\documentclass[10pt]{article}
\usepackage{multirow}
\usepackage[nottoc]{tocbibind}
\usepackage{geometry}
\usepackage[affil-it, auth-sc]{authblk}
\usepackage[utf8]{inputenc}
\usepackage{graphicx}
\usepackage{amsthm}
\usepackage{amssymb}
\usepackage{amsmath}
\usepackage{color, soul}
\usepackage{verbatim}
\usepackage{lineno}
\usepackage{todonotes}
\usepackage{subcaption}
\usepackage{ulem}
\usepackage[pdftex,colorlinks,bookmarks,linkcolor=darkred, citecolor=darkgreen]{hyperref}
\usepackage[ampersand]{easylist}
\usepackage{babel}
\usepackage{bm}
\usepackage{feynmf}
\usepackage[sort&compress,numbers]{natbib}
\usepackage{doi}
\usepackage{eso-pic}
\usepackage{caption}
\usepackage{floatrow}
\presetkeys{todonotes}{inline, color=blue!30, size=\small}{}
\definecolor{darkred}{rgb}{0.9, 0.0, 0.0}
\definecolor{darkgreen}{rgb}{0.0, 0.5, 0.0}
\allowdisplaybreaks

\def\fAt{f_{A3}}
\def\fAtbar{\bar{f}_{A3}}

\def\slash#1{#1\!\!\!/}
\newcommand{\nl}{\nonumber \\ }

\begin{document}

\AddToShipoutPictureFG*{\AtPageUpperLeft{\put(-60,-75){\makebox[\paperwidth][r]{FERMILAB-PUB-23-0822-LBNF-T,~LA-UR-22-22626}}}}

\title{\Large\bf Invariant amplitudes, unpolarized cross sections, and polarization asymmetries in (anti)neutrino-nucleon elastic scattering}

\author[1]{Kaushik Borah}
\affil[1]{Department of Physics and Astronomy, University of Kentucky, Lexington, KY 40506, USA \vspace{1.2mm}}

\author[2]{Minerba Betancourt}
\affil[2]{Fermilab, Batavia, IL 60510, USA
\vspace{1.2mm}}

\author[1,2]{Richard J.~Hill}

\author[2]{Thomas Junk}

\author[3]{Oleksandr Tomalak}
\affil[3]{Theoretical Division, Los Alamos National Laboratory, Los Alamos, NM 87545, USA \vspace{1.2mm}}

\date{\today}

\maketitle
\begin{abstract}
At leading order in weak and electromagnetic couplings, cross sections for (anti)neutrino-nucleon elastic scattering are determined by four nucleon form factors that depend on the momentum transfer $Q^2$. Including radiative corrections in the Standard Model and potential new physics contributions beyond the Standard Model, eight invariant amplitudes are possible, depending on both $Q^2$ and the (anti)neutrino energy $E_\nu$. We review the definition of these amplitudes and use them to compute both unpolarized and polarized observables including radiative corrections. We show that unpolarized accelerator neutrino cross-section measurements can probe new physics parameter space within the constraints inferred from precision beta decay measurements.
\end{abstract}

\newpage
\tableofcontents
\newpage

\section{Introduction}
\label{sec:introduction}

Next-generation neutrino oscillation experiments DUNE~\cite{Alion:2016uaj,Abi:2020evt} and Hyper-K~\cite{Hyper-Kamiokande:2016dsw} aim to discover charge-parity (CP) violation in the lepton sector and conclusively establish the ordering of neutrino masses. To achieve these goals, it is essential to control muon disappearance and electron appearance signals at the percent level which requires both theoretical and experimental progress in our understanding of neutrino-nucleus interactions~\cite{Formaggio:2013kya,Mosel:2016cwa,Alvarez-Ruso:2017oui}.

Vector form factors are known relatively well from electron scattering data~\cite{Bernauer:2010wm,Bernauer:2013tpr,Xiong:2019umf,Punjabi:2015bba,Ganichot:1972mb,Bosted:1989hy}, and the axial-vector form factor is the main source of uncertainty for neutrino interactions at the nucleon level. New experimental measurements~\cite{MINERvA:2023avz} and lattice-QCD evaluations~\cite{Gockeler:2003ay,Alexandrou:2018sjm,Yamazaki:2009zq,Chambers:2017tuf,Capitani:2015sba,Bruno:2014jqa,Bratt:2010jn,Kronfeld:2019nfb,Ishikawa:2018rew,Hasan:2019noy,RQCD:2019jai,Jang:2019vkm,Bali:2019yiy,Alexandrou:2020okk,Park:2021ypf,Djukanovic:2022wru,Jang:2023zts} of the axial-vector form factor have approached the precision of determinations from deuterium bubble-chamber data with (anti)neutrino beams~\cite{Mann:1973pr,Barish:1977qk,Miller:1982qi,Baker:1981su,Kitagaki:1983px,Meyer:2016oeg,Meyer:2022mix,Tomalak:2023pdi} and from pion electroproduction~\cite{Choi:1993vt,Bernard:1994pk,Blomqvist:1996tx,Liesenfeld:1999mv,Kamalov:2001qg,Gran:2006jn,Friscic:2015tga}. Future experimental measurements can potentially yield sub-percent uncertainties~\cite{Petti:2023abz}. Beyond unpolarized cross sections, a range of observables have been considered with polarization measurements using (anti)neutrino beams~\cite{Lee:1962jm,Florescu:1968zz,Dombey:1969wk,Pais:1971er,Cheng:1971mx,Tarrach:1974da,Oliver:1974de,Akhiezer:1974em,Kim:1978fx,Ridener:1984tb,Ridener:1986ey,Hagiwara:2003di,Hagiwara:2004gs,Graczyk:2004vg,Graczyk:2004uy,Bourrely:2004iy,Kuzmin:2004ke,Kuzmin:2004yb,Jachowicz:2004we,Jachowicz:2005np,Aoki:2005wb,Aoki:2005kc,Lava:2006fc,Meucci:2008zz,Bilenky:2013fra,Bilenky:2013iua,Fatima:2018tzs,Fatima:2018gjy,Fatima:2018wsy,Sobczyk:2019urm,Graczyk:2019xwg,Graczyk:2019opm,Tomalak:2020zlv,Fatima:2020pvv,Fatima:2021ctt,SajjadAthar:2022pjt}. In a previous publication~\cite{Tomalak:2020zlv} by one of the authors, and in a Snowmass 2021 white paper~\cite{Alvarez-Ruso:2022ctb}, we performed a detailed study, within the Standard Model, of how to access nucleon axial-vector structure from single-spin asymmetries in elastic charged-current (anti)neutrino-nucleon scattering. In particular, asymmetry measurements could determine the nucleon pseudoscalar form factor without the conventional ansatz of partially conserved axial current (PCAC) with pion-pole dominance~\cite{LlewellynSmith:1971uhs,Bernard:1998gv,Fuchs:2002zz,Kaiser:2003dr,Schindler:2006it,Lutz:2020dfi,Chen:2020wuq,Jang:2016kch,Gupta:2017dwj,Wright:1998gi,Winter:2011yp,MuCap:2015boo,MuCap:2012lei,Gorringe:2002xx,Hill:2017wgb,Choi:1993vt,Bernard:1994pk}. In order to interpret both unpolarized cross sections and polarization asymmetries with neutrino beams, and to understand the potential impact of new experimental data, we consider the general framework of invariant amplitudes for (anti)neutrino-nucleon elastic scattering. We compute both unpolarized and polarized observables including radiative corrections and compare predicted unpolarized cross sections and polarization asymmetries to constraints inferred from precision beta decay measurements.

The remainder of the paper is organized as follows. Section~\ref{sec:amplitudes_and_unpoalarized_cross_section} introduces the invariant amplitude decomposition for charged-current elastic (anti)neutrino-nucleon scattering and presents the expression for the unpolarized cross section. Section~\ref{sec:single_spin_asymmetries} provides expressions for all possible single-spin asymmetries and discusses the forward and backward limits. In Section~\ref{sec:observables}, we compare unpolarized cross sections and polarization asymmetries to constraints from precision beta decay measurements. We discuss the effects of radiative corrections on some observables of interest in Section~\ref{sec:radiative_corrections}. We present plots of cross sections, polarization asymmetries, and radiative corrections for illustrative (anti)neutrino energies in Appendix~\ref{sec:plots}.

\section{Amplitude decomposition and unpolarized cross section}
\label{sec:amplitudes_and_unpoalarized_cross_section}

\begin{figure}[tb]
\centering
\includegraphics[width=0.2\textwidth]{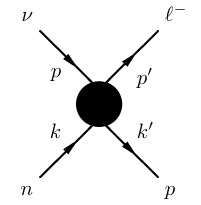}
\includegraphics[width=0.2\textwidth]{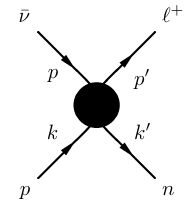}
\caption{Kinematics of charged-current elastic scattering for neutrino (left) and antineutrino (right). \label{fig:CCQE_kin}}
\end{figure}

In the Standard Model, the matrix element of the charged-current elastic process with a massless charged lepton can be expressed as\footnote{We use the shorthand notation $\bar{\ell}^- (\dots ) \nu_{\ell} = \bar{u}^{(\ell)}(p^\prime) (\dots ) u^{(\nu)}(p)$ and $\overline{\bar{\nu}}_\ell (\dots) \ell^+= \bar{v}^{(\nu)}(p) (\dots ) v^{(\ell)}(p^\prime)$ for the usual Dirac spinors with momentum assignment in Fig.~\ref{fig:CCQE_kin}.}
\begin{align}
T^{m_\ell = 0}_{\nu_\ell n \to \ell^- p} &= \sqrt{2}\mathrm{G}_\mathrm{F} V_{ud} \, \bar{\ell}^- \gamma^\mu \mathrm{P}_\mathrm{L} \nu_{\ell}\, \bar{p} \left(\gamma_\mu \left( {g}_M + {f}_A \gamma_5 \right) - \left( f_2 - 2 \fAt \gamma_5 \right) \frac{{K}_\mu}{M} \right) n \,, \nl
T^{m_\ell = 0}_{\bar{\nu}_\ell p \to \ell^+ n} &= \sqrt{2}\mathrm{G}_\mathrm{F} V^*_{ud} \, \overline{\bar{\nu}}_\ell \gamma^\mu \mathrm{P}_\mathrm{L} \ell^+ \, \bar{n} \left( \gamma_\mu \left( \bar{g}_M + \bar{f}_A \gamma_5 \right) - \left( \bar{f}_2 + 2  \fAtbar \gamma_5 \right) \frac{{K}_\mu}{M} \right) p , \label{eq:CCQE_amplitude}
\end{align}
with the averaged nucleon momentum $K_\mu = (k_\mu + k^\prime_\mu)/2$, and the averaged nucleon mass $M=(m_n+m_p)/2$. Following tree-level notations for the form factors, we define the electric and magnetic amplitudes $g_E$ and $g_M$ in terms of the amplitudes $f_1$ and $f_2$ as
\begin{align}
 g_E = f_1 - \tau f_2, \qquad g_M = f_1 + f_2.
\end{align}
Both the (anti)neutrino and the lepton are left-handed in the limit $m_\ell \to 0$ (we ignore the (anti)neutrino mass throughout, $m_{\nu_\ell}=0$). A similar decomposition describes neutral-current (anti)neutrino-nucleon elastic scattering in terms of four amplitudes, at arbitrary $m_\ell$. For nonvanishing lepton mass, the matrix element for charged currents contains four more invariant amplitudes\footnote{The form factors from Ref.~\cite{LlewellynSmith:1971uhs} contribute to our amplitudes as $f_1 = F_1^V$, $f_2 = \xi F_2^V$, $f_A = F_A$, $f_3 = F_3^V$, $f_{P} = F_P$, $\fAt = F_A^3$. At low energies, the invariant amplitudes reduce to local four-fermion couplings; conventionally normalized scalar and tensor interactions are~\cite{Gonzalez-Alonso:2018omy} $C_S \approx -\frac{m_\ell}{M} f_3 $ and $C_T \approx \frac{m_\ell}{2 M} f_T$.}
\begin{align}
T_{\nu_\ell n \to \ell^- p} &= T^{m_\ell = 0}_{\nu_\ell n \to \ell^- p} + \sqrt{2}\mathrm{G}_\mathrm{F} V_{ud} \frac{m_\ell}{M} \left[ \frac{{f}_{T}}{4} \, \bar{\ell}^-  \sigma^{\mu \nu} \mathrm{P}_\mathrm{L} \nu_{\ell}\, \bar{p} \sigma_{\mu \nu} n - \bar{\ell}^- \mathrm{P}_\mathrm{L} \nu_{\ell}\, \bar{p} \left( {f}_3 + f_P \gamma_5  - \frac{f_R}{4} \frac{\slash{P}}{M} \gamma_5 \right) n \right]  \,, \nl
T_{\bar{\nu}_\ell p \to \ell^+ n} &= T^{m_\ell = 0}_{\bar{\nu}_\ell p \to \ell^+ n} + \sqrt{2}\mathrm{G}_\mathrm{F} V^\star_{ud} \frac{m_\ell}{M} \left[ \frac{\bar{f}_{T}}{4} \overline{\bar{\nu}}_\ell \sigma^{\mu \nu} \mathrm{P}_\mathrm{R} \ell^+ \, \bar{n} \sigma_{\mu \nu}  p -\overline{\bar{\nu}}_\ell \mathrm{P}_\mathrm{R} \ell^+ \,
\bar{n} \left( \bar{f}_3 + \bar{f}_P \gamma_5 - \frac{\bar{f}_R}{4} \frac{\slash{P}}{M} \gamma_5 \right) p \right], \, \label{eq:CCQE_amplitudem}
\end{align}
which enter all observables with a factor of the lepton mass $m_\ell$. Here $P_\mu = (p_\mu + p^\prime_\mu)/2$ denotes the averaged lepton momentum. Invariant amplitudes are functions of two kinematic variables: the crossing-symmetric variable $\nu = E_\nu / M - \tau - r_\ell^2$, with $r_\ell = m_\ell/(2M)$, $\tau = Q^2/(4M^2)$, and the squared momentum transfer $Q^2 = - \left( p - p^\prime \right)^2 = - \left( k - k^\prime \right)^2$. For the antineutrino case, the invariant amplitudes are given by $\bar{f}_i \left( \nu+i0, Q^2 \right) = f_i \left( -\nu-i0, Q^2 \right)^*$. At tree level, $f_T$ and $f_{R}$ vanish, and each of the remaining amplitudes are real-valued functions of the squared momentum transfer $Q^2$ only. In the limit of isospin invariance, the amplitudes $~\fAt$ and $f_3$ also vanish at tree level. QED radiative corrections contribute to all eight invariant amplitudes~\cite{Tomalak:2021hec,Tomalak:2022xup}. The invariant amplitudes can receive corrections from new physics beyond the Standard Model. For example, the scalar interaction $f_3$ appears as a radiative correction in the minimally supersymmetric Standard Model (MSSM)~\cite{Ramsey-Musolf:2006evg,Hardy:2008gy} or at tree level if $R$-parity is violated~\cite{Yamanaka:2009hi,Kozela:2011mc}.

Using the representation in Eqs.~(\ref{eq:CCQE_amplitude})~and~(\ref{eq:CCQE_amplitudem}), the charged-current elastic cross section (without real radiation) in the laboratory frame is expressed in terms of invariant amplitudes as~\cite{LlewellynSmith:1971uhs}\footnote{We neglect the relative difference in nucleon masses, $(m_n-m_p)/(m_n+m_p)$, and electroweak power corrections suppressed by the $W$-boson mass $M_W$, i.e., the corrections of order $Q^2/M_W^2$; these effects contribute at the permille level.}
\begin{equation}
\frac{d\sigma}{dQ^2} (E_\nu, Q^2) = \frac{\mathrm{G}_\mathrm{F}^2 |V_{ud}|^2}{2\pi} \frac{M^2}{E_\nu^2} \left[ \left( \tau + r_\ell^2 \right)A(\nu,~Q^2) - \frac{\nu}{M^2} B(\nu,~Q^2) + \frac{\nu^2}{M^4} \frac{C(\nu,~Q^2)}{1+ \tau} \right] \,.\label{eq:xsection_CCQE}
\end{equation}
The quantities $A$, $B$, and $C$ are given by\footnote{The sign in front of $\fAt$ in $B$ differs from Ref.~\cite{LlewellynSmith:1971uhs} for antineutrino scattering, as noted in Refs.~\cite{Pais:1971er,Kuzmin:2007kr,Akbar:2015yda}.}
\begin{align}
A &= \tau | g_M |^2 - | g_E |^2 + (1+ \tau) | f_A |^2 - r_\ell^2 \left( | g_M |^2 + | f_A + 2 f_P |^2 - 4 \left( 1 + \tau \right) \left( |f_P|^2 + |f_3|^2\right)\right) - 4 \tau (1+ \tau) |\fAt|^2 \nonumber \\
& + \frac{r_\ell^2}{4} \left( \nu^2 + 1 + \tau - \left(1 + \tau + r_\ell^2 \right)^2 \right) |f_R|^2 - r_\ell^2 \left( 1 + 2 r_\ell^2 \right) |f_T|^2 - 2 r_\ell^2 \mathfrak{Re} \Big[ \Big( \left(  g_E + 2 g_M \right)  - 2 \left( 1 + \tau \right) \fAt \Big) f_T^\star \Big] \nonumber \\
&  - \eta r_\ell^2 \left( 1 + \tau + r_\ell^2 \right)  \mathfrak{Re}\Big [ f_A f_R^\star \Big] - 2 \eta r_\ell^4 \mathfrak{Re}\Big [ f_P f_R^\star \Big] \,, \\
B &= \mathfrak{Re} \Big[ 4 \eta \tau f^*_A g_M - 4 \eta r_\ell^2 \left( f_A - 2 \tau f_P \right)^*  \fAt + 4 r_\ell^2 g_E f_3^\star - 2 \eta r_\ell^2 \left( 3 f_A - 2 \tau \left( f_P - \eta f_3\right) \right) f_T^\star - r_\ell^4 \left( f_T + 2 \fAt  \right) f_R^\star  \Big]  \,,  \\
C &= \tau |g_M|^2 + |g_E|^2 + (1 + \tau) |f_A|^2 + 4 \tau (1 + \tau) |\fAt|^2 + 2 r_\ell^2 \left( 1 + \tau \right) |f_T|^2 + \eta r_\ell^2 \left( 1 + \tau \right)  \mathfrak{Re} \Big[ f_A f_R^\star \Big],
\end{align}
with $\eta = + 1$ for neutrino scattering and $\eta = - 1$ for antineutrino scattering.

In the following, we consider the limits of forward and backward charged lepton scattering, which correspond to the squared momentum transfers $Q_-^2$ and $Q_+^2$, respectively, where
\begin{equation}
Q^2_\pm = \frac{2 M E^2_\nu}{M+2 E_\nu} - 4M^2 \frac{M+E_\nu}{M+2 E_\nu} r_\ell^2 \pm \frac{4 M^2 E_\nu}{M+2 E_\nu} \sqrt{\left(  \frac{E_\nu}{2M}- r_\ell^2 \right)^2- r_\ell^2}.
\end{equation}
In the forward limit, up to terms that are suppressed by the lepton mass, the unpolarized cross section can be expressed as
\begin{equation}
    \frac{d\sigma}{dQ^2} (E_\nu, Q^2_-) =  \frac{\mathrm{G}_\mathrm{F}^2 |V_{ud}|^2}{2\pi} \left( |g_E|^2 + |f_A|^2 \right) + \mathrm{O} \left( m^2_\ell \right),
\end{equation}
and in the backward limit, the cross section can be expressed as
\begin{equation}
    \frac{d\sigma}{dQ^2} (E_\nu, Q^2_+) =  \frac{\mathrm{G}_\mathrm{F}^2 |V_{ud}|^2}{\pi} \frac{ | \left( M + E_\nu \right) f_A  - \eta E_\nu g_M|^2}{\left( M + 2 E_\nu \right)^2} + \mathrm{O} \left( m^2_\ell \right).
\end{equation}

\section{Single-spin asymmetries}
\label{sec:single_spin_asymmetries}

Single-spin asymmetries could provide new constraints on physics beyond the Standard Model, cf. Figs.~\ref{fig:pol_target} and \ref{fig:pol_lepton_tau} below. In this Section, we provide expressions for single-spin asymmetries in charged-current elastic (anti)neutrino-nucleon scattering, in terms of the invariant amplitudes defined above, and study limits of forward and backward scattering. The single-spin asymmetry is defined as the difference between the cross section $\sigma \left( S \right)$ with a definite spin four-vector $S$ of one initial- or final-state particle and the cross section with an opposite spin direction $\sigma \left( -S \right)$, divided by the sum of these cross sections:
\begin{equation}
\mathrm{T},\mathrm{R}, \mathrm{L} = \frac{d \sigma \left( S \right) - d \sigma \left( -S \right)}{d \sigma \left( S \right) + d \sigma \left( -S \right)},
\end{equation}
where $\mathrm{T}$, $\mathrm{R}$, and $\mathrm{L}$ denote target, recoil, and lepton asymmetries, respectively. We describe these observables in more detail below. It is convenient to express the asymmetries in a manner similar to that for the unpolarized cross section (\ref{eq:xsection_CCQE}), in terms of new structure-dependent functions that depend on the particle whose spin we are considering:
\begin{equation}
\mathrm{T}, \mathrm{R},\mathrm{L} = \frac{\left( \tau + r_\ell^2 \right)A^{\mathrm{T}, \mathrm{R},\mathrm{L}}(\nu, Q^2) -  \nu B^{\mathrm{T}, \mathrm{R},\mathrm{L}}(\nu, Q^2) + \frac{\nu^2}{1+\tau} C^{\mathrm{T}, \mathrm{R},\mathrm{L}}(\nu, Q^2)}{\left( \tau + r_\ell^2 \right)A(\nu, Q^2) -  \nu B(\nu, Q^2) + \frac{\nu^2}{1+\tau} C(\nu, Q^2)}. \label{eq:asymmetryABC}
\end{equation}

The target single-spin asymmetry $\mathrm{T}$ with a target spin four-vector $S^\mu$\footnote{We follow the same definitions, notations, and conventions as Ref.~\cite{Tomalak:2020zlv}. In particular, all spin four-vectors are normalized as $S^2 = - 1/M^2$.} is expressed in terms of the invariant amplitudes of Eqs.~(\ref{eq:CCQE_amplitude}) and~(\ref{eq:CCQE_amplitudem}) with the following structure-dependent functions:
\begin{align}
 A^\mathrm{T} &= \mathfrak{Re} \bigg [   \left(   f_A- \eta g_E \right) g^\star_M \left( k' \cdot S \right) - 2 \eta g_M g^\star_E \left( p' \cdot S \right) + 2 r_\ell^2 \left(  \frac{ \eta g_E - f_A + 2 \tau f_P }{\tau + r_\ell^2} \left(\left( k' \cdot S \right) + \left( p' \cdot S \right) \right) - f_P \left( k' \cdot S \right) \right) g^\star_M  \nonumber \\
 &+ 4 \frac{1+\tau}{\tau + r_\ell^2} \left( \tau \fAt - r_\ell^2 f_3\right) f_A^\star \left(\left( k' \cdot S \right) + \left( p' \cdot S \right) \right) + 2 \left( r_\ell^2 f_3 - \left( 1 + \tau \right) \fAt  \right) f^\star_A \left( k' \cdot S \right) - 2 \eta \fAt g^\star_E  \left( k' \cdot S \right) \nonumber \\
 &  + \eta  r_\ell^2 \left( \frac{2\left(  g_E - 2 \eta f_A + r_\ell^2 f_T  \right) - r_\ell^2 f_R}{\tau + r_\ell^2}\left(\left( k' \cdot S \right) + \left( p' \cdot S \right) \right) - \left(2 \fAt + \frac{f_R}{2} \right) \left( k' \cdot S \right) \right) f_T^\star - \eta r_\ell^2 \left( 1 + \tau \right)  f_3 f_R^\star   \left( p' \cdot S \right) \nonumber \\
 & - r_\ell^2 \left( \frac{\tau - r_\ell^2}{\tau+r_\ell^2} \left( k' \cdot S \right) + \frac{2\tau \left( p' \cdot S \right) }{\tau+r_\ell^2} \right)\left( \eta g_M + f_A - 2  f_P  \right) f_T^\star - \eta r_\ell^2 \left( \left( 1 + \tau \right) f_3 + \frac{r_\ell^2 g_M}{\tau + r_\ell^2} \right)  f_R^\star \left(\left( k' \cdot S \right) + \left( p' \cdot S \right) \right) \bigg ]  \nonumber \\
 &+ \frac{\rho_\perp}{\tau + r_\ell^2} \mathfrak{Im} \Big [  2 r_\ell^2 \left( g_M f^\star_3 - f_A f^\star_P \right) +  \eta f_A g^\star_E + 2 \tau \fAt g^\star_M + r_\ell^2 \left( \eta f_A +   g_M - 2 f_3  - 2 \fAt \right) f_T^\star  \Big] \nonumber \\
 & + \eta r_\ell^2 \mathfrak{Re} \bigg [  \left( 4 \eta f_{P} + r_\ell^2 f_R \right) f_3^\star - \frac{1}{2} g_M f^\star_R \bigg ] \left( k' \cdot S \right) + \rho_\perp \eta r_\ell^2\mathfrak{Im} \bigg [ \left( f_{P} - \frac{1}{2} \frac{f_A}{\tau + r_\ell^2}\right) f_R^\star \bigg ] , \\
 B^\mathrm{T} &= \mathfrak{Re} \bigg [\left( \eta |f_A|^2 - g_E f^\star_A +\eta \tau f_2	g^\star_M \right)  \left( k' \cdot S \right) - 2  f_A  g^\star_E \left( p' \cdot S \right) - r_\ell^2  \left( f_2 f^\star_A - 2  f_1 f^\star_P\right) \left( k' \cdot S \right) \nonumber \\
 & - 2 r_\ell^2 \left( \left(f_A - 2 \eta \fAt \right) f^\star_3 + \eta \fAt g^\star_M \right) \left( k' \cdot S \right) + 2 \tau \left( \eta g_M + f_A \right) \left( \fAt \right)^\star \left( k' \cdot S \right) + 4 \eta \tau \fAt g^\star_M \left( p' \cdot S \right) -  \frac{\eta r_\ell^4}{2} f_R f_T^\star \left( k' \cdot S \right)  \nonumber \\  
 & - \eta r_\ell^2 \left[ \left( g_M + f_2 + \eta f_A -2 f_3 \right) f_T^\star  \left( k' \cdot S \right)  + \frac{1}{2} g_E f_R^\star \left(\left( k' \cdot S \right) + 2 \left( p' \cdot S \right) \right) +   \left[ \left( \tau + r_\ell^2 \right) f_3 +\frac{r_\ell^2}{2} f_2 \right] f_R^\star \left( k' \cdot S \right) \right] \nonumber \\
 & +  \eta r_\ell^2\left( \frac{\tau }{2}  f_R + f_T \right) f_T^\star \left(  \left( k' \cdot S \right) + 2  \left( p' \cdot S \right) \right)   \bigg ]  -  \rho_\perp \mathfrak{Im} \bigg [  - 2 \eta \fAt f^\star_A + \frac{g_E g^\star_M}{1+\tau} - \frac{r_\ell^2}{2}\left( f_T + 2 \fAt \right) f_R^\star  \bigg ] \,, \\
 C^\mathrm{T} &=   \mathfrak{Re} \Big [  \left(  g_M - g_E \right) f^\star_A + 2 \eta \left(  g_E + \tau g_M \right) \left( \fAt \right)^\star  + \frac{ \eta r_\ell^2}{2} \left( 1 + \tau \right) \left(  f_2 + f_T  \right) f_R^\star \Big ]  \left( k' \cdot S \right), \label{eq:ABC_target3}
\end{align}
where $\rho_\perp= M^{-3} \varepsilon^{\mu \nu \lambda \rho} p_\mu p^\prime_\nu k_\lambda S_\rho$, $\eta=+1$ corresponds to neutrino scattering $\nu_\ell n \to \ell^- p$, and $\eta=-1$ corresponds to antineutrino scattering $\bar{\nu}_\ell p \to \ell^+ n$, with implied overline ($f_i\to \bar{f}_i$) for antineutrinos. It is worthwhile highlighting some special cases. To evaluate $\mathrm{T}_\mathrm{t}$, the asymmetry in which the target polarization is transverse to the beam direction with the spin vector in the scattering plane, we substitute $\left( k' \cdot S \right) = - \left( p' \cdot S \right) = (2M/E_\nu) \sqrt{\tau \nu^2 -  (1+\tau)  (\tau + r_\ell^2)^2 }$ and $\rho_\perp=0$. To evaluate $\mathrm{T}_\mathrm{l}$, the asymmetry in which the target polarization is along the beam direction, we substitute $\left( k' \cdot S \right) = - 2 \left[ \tau + (M/E_\nu) \left( \tau + r_\ell^2 \right)\right]$, $\left( p' \cdot S \right) = - \left( k' \cdot S \right) - E_\nu/M$, and $\rho_\perp=0$. To evaluate $\mathrm{T}_\perp$, the asymmetry in which the target polarization is transverse to the scattering plane, we substitute $\rho_\perp= 2\sqrt{ \tau \nu^2 - \left( 1 + \tau \right) ( \tau + r_\ell^2 )^2}$, $\left( k' \cdot S \right) =0$, and $\left( p' \cdot S \right) =0$. For non-orthogonal spin direction, $\rho_\perp$ is multiplied with $\cos \phi_\perp$, where $\phi_\perp$ is the angle between the spin vector and the vector $\vec{n} = \vec{p} \times \vec{p}^{\,\prime} / | \vec{p} \times \vec{p}^{\,\prime} |$ that is orthogonal to the scattering plane. The transverse target single-spin asymmetries $\mathrm{T_t}$ and $\mathrm{T}_\perp$ vanish at forward and backward angles. The longitudinal single-spin asymmetry is positive at forward scattering when the squared momentum transfer is $Q^2_\mathrm{-}$. Up to lepton-mass-suppressed terms, the asymmetry $\mathrm{T_l}$ reaches a maximum in absolute value at backward angles, when the squared momentum transfer is $Q^2_\mathrm{+}$. For these kinematic boundaries, the longitudinal target single-spin asymmetry is given by
\begin{equation}
\mathrm{T}_\mathrm{l} \left( Q^2_\mathrm{-} \right) = - \frac{2~ \mathfrak{Re} [ g_E f^\star_A ]}{ | g_E |^2 + |f_A|^2} + \mathrm{O} \left( m_\ell^2 \right), \qquad \mathrm{T}_\mathrm{l} \left( Q^2_\mathrm{+} \right)  = \eta + \mathrm{O} \left( m_\ell^2 \right).
\end{equation}

The structure-dependent parameters for the recoil single-spin asymmetry $\mathrm{R}$, with a recoil spin four-vector $S^\mu$, are expressed in terms of the invariant amplitudes of Eqs.~(\ref{eq:CCQE_amplitude}) and~(\ref{eq:CCQE_amplitudem}) as
\begin{align}
 A^\mathrm{R} &= \mathfrak{Re} [   \left(   f_A- \eta g_E \right) g^\star_M \left( k \cdot S \right) - 2 \eta g_M g^\star_E \left( p \cdot S \right) + 2 r_\ell^2 \left(  \frac{ \eta g_E + f_A - 2 \tau f_P }{\tau + r_\ell^2} \left( p\cdot S \right) - f_P \left( k \cdot S \right) \right) g^\star_M + 2  \eta \fAt g^\star_E \left( k \cdot S \right)\nonumber \\
 &- 4 \frac{1+\tau}{\tau + r_\ell^2} \left( \tau \fAt +  r_\ell^2 f_3\right) f_A^\star \left(\left( k \cdot S \right) + \left( p \cdot S \right) \right) +2 \left( \left( 1 + \tau \right)\frac{\tau - r_\ell^2}{\tau + r_\ell^2} \fAt +  r_\ell^2 \frac{\tau + 2 - r_\ell^2}{\tau + r_\ell^2} f_3\right) f^\star_A  \left( k \cdot S \right) \nonumber \\
  &  +  \eta r_\ell^2 \left( \frac{2 \left(   g_E + 2 \eta f_A + r_\ell^2 f_T  \right)}{\tau + r_\ell^2}\left( p \cdot S \right) +  \left( 2 \fAt - \frac{f_R}{2} \right) \left( k \cdot S \right)  - \left( \left( k \cdot S \right) +\frac{2\tau \left( p \cdot S \right) }{\tau+r_\ell^2} \right) \left( g_M -\eta f_A + 2 \eta f_P \right) \right) f_T^\star \nonumber \\
  & -  r_\ell^2 \left[ \left( 4 f_{P} + \eta r_\ell^2 f_R \right) f_3^\star +  \frac{\eta}{2} g_M f^\star_R \right] \left( k \cdot S \right)   + \frac{\eta r_\ell^4 \left( g_M + f_T \right) }{ \tau + r_\ell^2}   f_R^\star \left( p \cdot S \right) - \eta r_\ell^2  \left( 1 + \tau \right) f_3  f_R^\star \left(\left( k \cdot S \right) + 2 \left( p \cdot S \right) \right)  \bigg ]   \nonumber \\
  &+ \frac{\rho_\perp}{\tau + r_\ell^2} \mathfrak{Im} [ 2 r_\ell^2 \left( f_A f^\star_P + g_M f^\star_3 \right) + \eta f_A g^\star_E - 2  \tau \fAt g^\star_M  +   r_\ell^2 \left( \eta f_A + 2 \fAt -   \left( g_M + 2 f_3 \right) \right) f_T^\star ] \nonumber \\
  & - \rho_\perp \eta r_\ell^2\mathfrak{Im} \bigg [ \left( f_{P} - \frac{1}{2} \frac{f_A}{\tau + r_\ell^2}\right) f_R^\star \bigg ], \\
 B^\mathrm{R} &= \mathfrak{Re} [\left( \eta |f_A|^2 - g_E f^\star_A +\eta \tau f_2	g^\star_M \right)  \left( k \cdot S \right) - 2  f_A  g^\star_E \left( p \cdot S \right) + r_\ell^2  \left( f_2 f^\star_A - 2 f_1 f^\star_P\right) \left( k \cdot S \right) \nonumber \\
 & - 2 \eta  r_\ell^2 \left( \left(  \eta f_A + 2 \fAt \right) f^\star_3  +  \fAt g^\star_M \right) \left( k \cdot S \right) - 2 \tau \left( \eta g_M + f_A \right) \left( \fAt \right)^\star \left( k \cdot S \right) - 4  \eta \tau \fAt g^\star_M \left( p \cdot S \right) \nonumber \\
 & - \eta  r_\ell^2 \left[ \left( g_M + f_2 - \eta f_A + 2 f_3\right) f_T^\star  \left( k \cdot S \right)  + \frac{1}{2} g_E f_R^\star \left(\left( k \cdot S \right) + 2 \left( p \cdot S \right) \right) +   \left[ \left( \tau + r_\ell^2 \right) f_3 - \frac{r_\ell^2}{2} f_2 \right] f_R^\star \left( k \cdot S \right)  \right. \nonumber \\
 & \left. - \frac{ r_\ell^2}{2} \left( k \cdot S \right) f_R f_T^\star - \left( \frac{\tau }{2}  f_R - f_T \right) f_T^\star  \left( \left( k \cdot S \right) + 2 \left( p \cdot S \right) \right)   \right] \bigg ] -  \rho_\perp \mathfrak{Im} \bigg [ 2  \eta \fAt f^\star_A + \frac{g_E g^\star_M}{1+\tau} + \frac{r_\ell^2}{2}\left(  f_T + 2 \fAt \right) f_R^\star   \bigg ]\,, \\
 C^\mathrm{R} &=   \mathfrak{Re} \Big [ \left(  g_M - g_E \right) f^\star_A - 2  \eta \left(  g_E + \tau g_M \right) \left( \fAt \right)^\star + \frac{ \eta r_\ell^2}{2} \left( 1 + \tau \right) \left( f_2 +  f_T  \right) f_R^\star \Big ]  \left( k \cdot S \right). \label{eq:ABC_recoil3}
\end{align}
To evaluate $\mathrm{R}_\mathrm{t}$, the recoil nucleon spin asymmetry with the spin vector in the scattering plane and perpendicular to the recoiling nucleon's momentum, we substitute $\left( k \cdot S \right) = 0$, $\left( p \cdot S \right) = - \sqrt{ \tau \nu^2 -  (1+\tau)  (\tau + r_\ell^2)^2 }/\sqrt{\tau (1+\tau)}$, and $\rho_\perp=0$. To evaluate $\mathrm{R}_\mathrm{l}$, the recoil nucleon spin asymmetry with the spin vector in the scattering plane and parallel to the recoiling nucleon's momentum, we substitute $\left( k \cdot S \right) = 2  \sqrt{\tau (1+\tau)} $, $\left( p \cdot S \right) =   \left( \tau \nu -  (1+\tau)  (\tau + r_\ell^2) \right)/\sqrt{\tau (1+\tau)} $, and $\rho_\perp=0$. To evaluate $\mathrm{R}_\perp$, the asymmetry in which the recoil polarization is transverse to the scattering plane, we substitute $\rho_\perp = 2  \sqrt{ \tau \nu^2 - \left( 1 + \tau \right) ( \tau + r_\ell^2 )^2}$, $\left( k \cdot S \right) =0$, and $\left( p \cdot S \right) =0$. The transverse recoil single-spin asymmetries $\mathrm{R_t}$ and $\mathrm{R}_\perp$ vanish at forward and backward angles. The longitudinal single-spin asymmetry is positive at forward scattering when the squared momentum transfer is $Q^2_\mathrm{-}$. Up to lepton-mass-suppressed terms, the asymmetry $\mathrm{R_l}$ reaches a maximum in absolute value at backward angles, when the squared momentum transfer is $Q^2_\mathrm{+}$. For these kinematic boundaries, the longitudinal target single-spin asymmetry is given by
\begin{equation}
\mathrm{R}_\mathrm{l} \left( Q^2_\mathrm{-} \right) = - \frac{2 \mathfrak{Re} [ g_E f^\star_A ]}{  |g_E |^2 + |f_A|^2}  + \mathrm{O} \left( m_\ell^2 \right), \qquad \mathrm{R}_\mathrm{l} \left( Q^2_\mathrm{+} \right) = - \eta + \mathrm{O} \left( m_\ell^2 \right).
\end{equation}

The structure-dependent factors for the lepton single-spin asymmetry $\mathrm{L}$, with a recoil spin four-vector $S^\mu$, are expressed in terms of the invariant amplitudes of Eqs.~(\ref{eq:CCQE_amplitude}) and~(\ref{eq:CCQE_amplitudem}) as
\begin{align}
   \left( \tau + r_\ell^2 \right) A^\mathrm{L}  &=  \mathfrak{Re} [ - \eta A \left( p \cdot r_\ell S \right) + 2 \left( \tau + r_\ell^2 \right) \left(  f_A g^\star_M -\left( f_A - 2 \tau f_P \right) \left( \fAt \right)^\star + \eta  g_E f^\star_3 \right) \left( p + 2 k \cdot r_\ell S \right) \nonumber \\
   &-  2 \eta r_\ell^2   \left( |g_M|^2 + |f_A + 2 f_P|^2 - 4 \left( 1 + \tau \right) \left(  |f_P|^2 + |f_3|^2 \right) + 2 g_M  f_T^\star \right) \left( p \cdot r_\ell S \right)  \nonumber \\
   &+ \left( \tau + r_\ell^2 \right)  \left(  \left( f_A - 2r_\ell^2 \left( f_P - \eta f_3 \right) \right)  f_T^\star + \left( \frac{ \tau + r_\ell^2}{2} \eta g_M -r_\ell^2 \fAt  - \eta \frac{r_\ell^2}{2} f_T \right) f_R^\star \right) \left( p + 2 k \cdot r_\ell S \right)  \nonumber \\
   &- r_\ell^2 \left( 2 \left( 1 + \tau + r_\ell^2 \right)  f_A + 4  r_\ell^2 f_P + \eta\frac{\left( \tau + r_\ell^2 \right)^2 + \tau +2 r_\ell^2}{2} f_R \right) f_R^\star\left( p \cdot r_\ell S \right) -  2 \eta r_\ell^2 |f_T|^2 \left( p \cdot r_\ell S \right) ] \nonumber \\
   &+  \eta r_\ell \rho_\perp \mathfrak{Im} \bigg [ 2 g_E f^\star_3  -2 \fAt f^\star_A - 4  \tau f_P \left( \fAt \right)^\star +  \left( f_A + 2 r_\ell^2 \left( f_P - \eta f_3 \right) \right) f_T^\star \bigg] \nonumber \\
   &+  \eta r_\ell \rho_\perp \mathfrak{Im} \bigg [  \frac{\tau + r_\ell^2}{2} g_M f_R^\star -  \frac{r_\ell^2}{2} \left( f_T + 2 \fAt \right) f_R^\star \bigg ]\,, \label{eq:ABC_lepton1}\\
   B^\mathrm{L}  &=  \mathfrak{Re} [-2 f_A g^\star_M \left( p \cdot r_\ell S \right) + \frac{\eta C}{1+\tau}  \left( p + 2 k \cdot r_\ell S \right) + \left( 2 \eta g_E f^\star_3 - 2 \left( f_A - 2 \tau f_P \right) \left( \fAt\right)^\star - r_\ell^2 \fAt f_R^\star  \right) \left( p \cdot r_\ell S \right) \nonumber \\
   & + \left( \left( f_A + 2 r_\ell^2 \left( f_P - \eta f_3 \right) \right) \left( p \cdot r_\ell S \right) +  \eta \left( \tau -r_\ell^2\right) \left(f_2 + 2 \fAt \right) \left( p + 2 k \cdot r_\ell S \right) \right) f_T^\star - \eta \frac{r_\ell^2}{2}  f_{T}  f_R^\star \left( p \cdot r_\ell S \right) \nonumber \\
   &+ \eta \left( \frac{\tau-r^2_\ell}{2} f_A f_R^\star- 4 r_\ell^2 |f_T|^2 \right)  \left( p + 2 k \cdot r_\ell S \right)  + \eta \frac{\tau-r^2_\ell}{2}  g_M f_R^\star \left( p \cdot r_\ell S \right) ] \nonumber \\
   &+  r_\ell \rho_\perp \mathfrak{Im} [ \left( f_2 + 2 \fAt \right) f_T^\star + \frac{\eta}{2} f_A  f_R^\star ]\,,\label{eq:ABC_lepton2} \\
   C^\mathrm{L}  &= \eta \left( 1 + \tau \right) \mathfrak{Re} [  \left(  f_2 + 2 \fAt \right) f_T^\star + \frac{1}{2}\left( \eta f_A + r_\ell^2 f_{R} \right) f_R^\star] \left( p \cdot r_\ell S \right)\,. \label{eq:ABC_lepton3}
\end{align}
To evaluate $\mathrm{L}_\mathrm{t}$, the lepton spin asymmetry with the spin vector in the scattering plane and perpendicular to the lepton momentum, we substitute $\left( k \cdot r_\ell S \right) = 0 $, $\left( p \cdot r_\ell S \right) = 2 r \sqrt{ \tau \nu^2 -  (1+\tau)  (\tau + r_\ell^2)^2 }/\sqrt{\left( \nu + r_\ell^2 - \tau \right)^2-4r_\ell^2}$, and $\rho_\perp=0$. To evaluate $\mathrm{L}_\mathrm{l}$, the lepton spin asymmetry with the spin vector in the scattering plane and parallel to the lepton momentum, we substitute $2 \left( k \cdot r_\ell S \right) = \sqrt{\left( \nu + r_\ell^2 - \tau \right)^2-4r_\ell^2}$, $ \left( p \cdot r_\ell S \right) =  - \left(\left(r_\ell^2 - \tau \right) \nu + \left( \tau + r_\ell^2 \right)^2\right)/\sqrt{\left( \nu + r_\ell^2 - \tau \right)^2-4r_\ell^2}$, and $\rho_\perp=0$. To evaluate $\mathrm{L}_\perp$, the asymmetry in which the lepton polarization is transverse to the scattering plane, we substitute $\rho_\perp = 2 \sqrt{ \tau \nu^2 - \left( 1 + \tau \right) ( \tau + r_\ell^2 )^2}$, $\left( k \cdot r_\ell S \right) =0$, and $\left( p \cdot r_\ell S \right) =0$. The transverse lepton single-spin asymmetries $\mathrm{L_t}$ and $\mathrm{L}_\perp$ vanish at forward and backward angles. Up to lepton-mass-suppressed terms, the longitudinal single-spin asymmetry reaches its extremum reflecting the chiral nature of the weak interaction, i.e., $\mathrm{L_l} = -\eta + \mathrm{O} \left( m_\ell^2 \right)$.

For all asymmetries (target, recoil, and lepton), only imaginary parts of the structure amplitudes contribute to the asymmetries when the spin vector is transverse to the scattering plane. These asymmetries vanish at tree level in the Standard Model.

\section{Phenomenology}
\label{sec:observables}

It is interesting to consider the potential for accelerator neutrino cross section measurements to constrain new physics beyond the Standard Model. In this Section, we consider possible short-distance interactions that could give rise to non-standard contributions to the invariant amplitudes appearing in Eqs.~(\ref{eq:CCQE_amplitude}) and~(\ref{eq:CCQE_amplitudem}). As a benchmark for comparison, we plot observables including the region allowed by constraints on such contact interactions inferred from precision beta decays. In making this comparison, we assume the Standard Model scaling of the amplitudes with lepton mass given in Eqs.~(\ref{eq:CCQE_amplitude}) and (\ref{eq:CCQE_amplitudem}), i.e., assume $m_\ell$-independent BSM contributions to $f_i$. This scaling naturally appears when the BSM contributions can be identified with modifications to the tree-level electroweak form factors associated with quark currents. For example, $f_3$ and $f_{A3}$ are identified with second-class form factors~\cite{Weinberg:1958ut,LlewellynSmith:1971uhs,Day:2012gb}. This scaling with lepton mass can also be motivated by the assumption of ``minimal flavor violation"~\cite{Cirigliano:2005ck}. It is also interesting to consider the constraints of accelerator (anti)neutrinos, with muon flavor (anti)neutrinos, independent of such assumptions, which we describe in our accompanying paper~\cite{Tomalak:2024yvq}. In this case, the parameter space is essentially unconstrained by beta decay since the charged muon mass is kinematically inaccessible to nuclear beta decays.

For the benchmark Standard Model prediction in numerical illustrations, we employ the vector form factors from Ref.~\cite{Borah:2020gte} and the axial-vector form factor from Ref.~\cite{Meyer:2016oeg}. For the pseudoscalar form factor, we use a standard ansatz (partially conserved axial-vector current and the assumption of pion pole dominance): $F_P(Q^2) = 2M^2 F_A(Q^2)/\left(m_\pi^2 + Q^2\right)$, with the charged pion mass $m_\pi$.

\begin{table}[t]
\begin{center}
\begin{tabular}{ |c|c|c| } \hline
parameter & value & reference \\ \hline
$\mathfrak{Re}f_3(0)$ & $0.0 \pm 1.8$ & \cite{Hardy:2020qwl} \\
$\mathfrak{Im}f_3(0)$ & $ 13 \pm 54$ & \cite{Gonzalez-Alonso:2018omy} \\
$\mathfrak{Re}f_{T}(0)$ & $-9.3\pm 10.3$ & \cite{Gonzalez-Alonso:2018omy} \\
$\mathfrak{Im}f_{T}(0)$ & $-1.9\pm 15.4$ & \cite{Gonzalez-Alonso:2018omy} \\ 
$\mathfrak{Re}\fAt(0)$ & $ 0 \pm 0.075$ & \cite{Day:2012gb} \\ \hline
\end{tabular}
\caption{Beta decay constraints on the real and imaginary parts of amplitudes at $Q^2=0$. \label{tab:beta}}
\end{center}
\end{table}

\subsection{Beta decay constraints}

Neglecting the isospin-breaking corrections, the amplitudes $f_T$, $f_3$, and $f_{A3}$ vanish at tree level in the Standard Model. Focusing on these interactions, we consider the substitution
\begin{align} \label{eq:fp5_constraint}
   f_i (\nu, Q^2)  &=  \frac{\mathfrak{Re} f_i \left( 0 \right) + i \mathfrak{Im} f_i \left( 0 \right)}{\left( 1 + \frac{Q^2}{\Lambda^2} \right)^2} \,.
\end{align}
For simplicity, we have neglected (anti)neutrino energy dependence and have assumed the dipole form for $Q^2$ dependence~\cite{Holstein:1984ga,AHRENS1988284,Day:2012gb} with the parameter $\Lambda \approx 1~\mathrm{GeV}$ describing a typical hadronic form-factor scale. The amplitudes $g_M$, $f_2$, and $f_A$ appear at tree level in the Standard Model and are unsuppressed by the lepton mass; they are thus well-constrained by (electronic) beta decay measurements and we do not consider new physics constraints on them here.  The amplitude $f_P$ is suppressed by the lepton mass but receives a large and uncertain (induced pseudoscalar) tree-level contribution, and we do not consider new physics constraints on it. The amplitudes $f_{T}$ and $f_3$ are suppressed by the lepton mass and relatively poorly constrained by beta decay; these amplitudes will be our focus. We also consider the amplitude $f_{A3}$, which is unsuppressed by the lepton mass but recoil-suppressed in low-energy beta decays. Finally, we mention $f_R$, which is suppressed by both the lepton mass and recoil and is essentially unconstrained by beta decay.

Numerous constraints have been placed on the normalization of elastic (anti)neutrino-nucleon interactions at low energies~\cite{Adler:1975he,Monsay:1977gw,Masuda:1979zz,Oka:1979bw,Dominguez:1979xp,Bardin:1981cq,Holstein:1984ga,Grenacs:1985da,AHRENS1988284,Shiomi:1996np,Govaerts:2000ps,Wilkinson:2000gx,Minamisono:2001cd,Hardy:2004id,Severijns:2006dr,Sumikama:2008zz,Minamisono:2011zz,Hardy:2014qxa,Holstein:2014fha,Gonzalez-Alonso:2018omy,Fatima:2018tzs,Falkowski:2020pma,Falkowski:2021vdg}. The best low-energy constraint for the real part of the scalar amplitude $f_3$ comes from precise measurements of beta decay rates: $\mathfrak{Re}f_3(0)/f_1(0)= 0.0 \pm 1.8$~\cite{Hardy:2020qwl} (see also Refs.~\cite{Day:2012gb,Hardy:2004id,Hardy:2008gy,Hardy:2014qxa}). The imaginary part is constrained by the triple-correlation coefficient in neutron decay:\footnote{The triple correlation is with respect to the neutron polarization, electron momentum, and electron spin, cf. Eq.~(1) of Ref.~\cite{Kozela:2011mc}.}~$\mathfrak{Im}f_3(0) =  13 \pm 54$~\cite{Gonzalez-Alonso:2018omy} (see also Ref.~\cite{Kozela:2011mc}). For the tensor amplitude $f_T$, we take constraints from the fit to beta decay data in Ref.~\cite{Gonzalez-Alonso:2018omy} (see also Refs.~\cite{Lee:1956qn,Jackson:1957zz,Severijns:2006dr,Cirigliano:2009wk,Bhattacharya:2011qm,Cirigliano:2012ab,Naviliat-Cuncic:2013ylu,Gonzalez-Alonso:2016etj,Falkowski:2017pss,Gonzalez-Alonso:2018omy,Alioli:2018ljm,Falkowski:2019xoe,Falkowski:2020pma,Falkowski:2021bkq,Cirigliano:2023nol,Dawid:2024wmp}): $\mathfrak{Re}f_{T}(0) = -9.3\pm10.3$, $\mathfrak{Im}f_{T}(0) = -1.9\pm 15.4$. For the amplitude $\fAt$, which vanishes at tree level, we take the compilation of the experimental data from Ref.~\cite{Day:2012gb}: $|\mathfrak{Re}\fAt(0)| < 0.075$. In Table~\ref{tab:beta}, we collect the beta decay constraints on $\mathfrak{Re} f_i$ and $\mathfrak{Im} f_i$ that we use for the plots to follow.

\subsection{Unpolarized observables}

\begin{figure}[htb]
\centering
\includegraphics[width=0.4\textwidth]{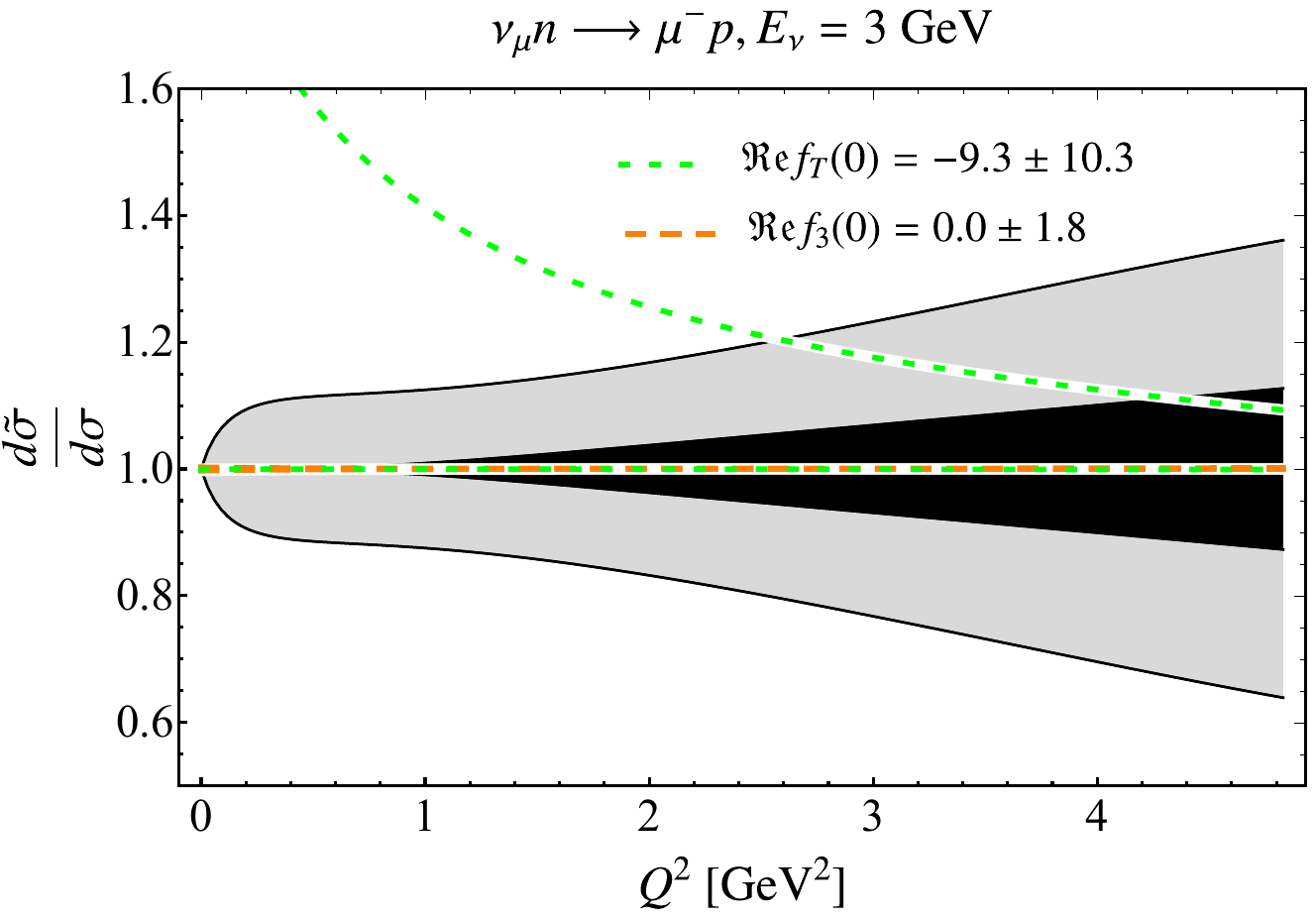}
\includegraphics[width=0.4\textwidth]{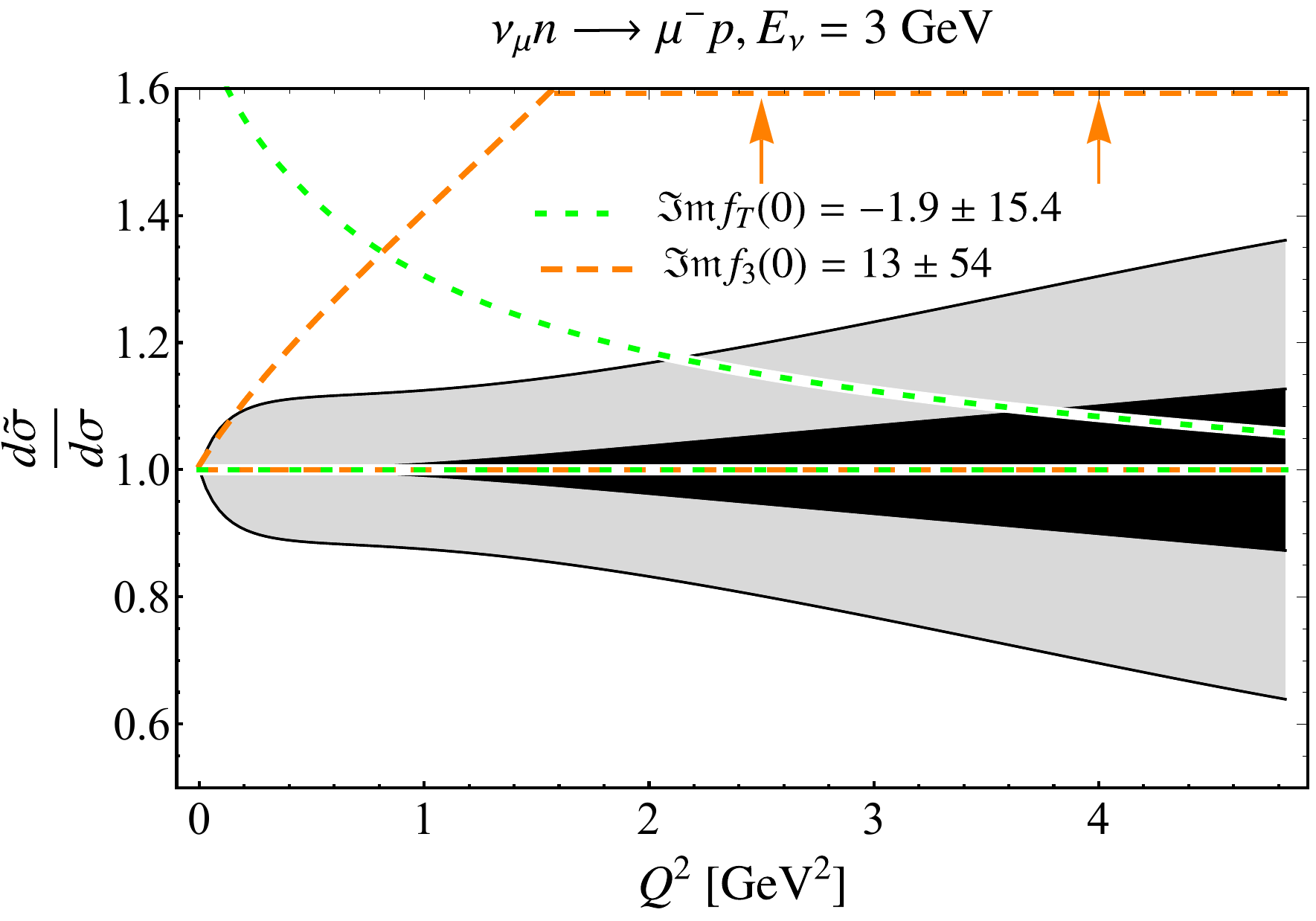}
\caption{Ratio of unpolarized muon neutrino cross section to the tree-level result as a function of the squared momentum transfer $Q^2$ at fixed muon neutrino energy $E_\nu = 3$~GeV. The dark black and light gray bands correspond to vector and axial-vector uncertainty, respectively. Orange dashed and green dotted lines represent allowed regions for $f_3$ and $f_T$, respectively, as described in the text. The left and right plots show results for one extra real-valued amplitude and one extra imaginary-valued amplitude, respectively. \label{fig:unpol}}
\end{figure}

Figure~\ref{fig:unpol} presents allowed regions for the muon-flavor charged-current elastic neutrino-neutron unpolarized cross section, for an illustrative $3\,{\rm GeV}$ neutrino energy. The cross section is evaluated with an additional amplitude of Eq.~(\ref{eq:fp5_constraint}) and is displayed as a ratio to the benchmark Standard Model cross section. We compare the uncertainties from the vector and axial-vector form factors to the envelope of possible values of the unpolarized cross section obtained by including one additional amplitude within the errors from Table~\ref{tab:beta}. The effects of $\mathfrak{Re}f_3$ and $\mathfrak{Re}\fAt$ are within the current uncertainties on the cross sections. However, the plots indicate that constraints on $\mathfrak{Re}f_{T}$, $\mathfrak{Im}f_{T}$, and $\mathfrak{Im}f_3$ can be significantly improved using accelerator neutrino cross-section data~\cite{Tomalak:2024yvq}. We present additional plots in Figs.~\ref{fig:nu_xsec_ratio_SCFFRe}-\ref{fig:antinu_xsec_ratio_SCFFIm} of Appendix~\ref{app:unpol_plots} for several beam energies between $300~\mathrm{MeV}$ and $3\,{\rm GeV}$, for both neutrinos and antineutrinos.

Although precise experimental data on tau (anti)neutrino cross sections is currently lacking for $\mathrm{GeV}$ energies~\cite{DONuT:2007bsg,OPERA:2015wbl,IceCube:2019dqi,DsTau:2019wjb}, in anticipation of future data we repeat our study of unpolarized cross sections for tau (anti)neutrinos with energies above the tau production threshold. Results are displayed in Figs.~\ref{fig:nu_xsec_ratio_SCFFtauRe}-\ref{fig:antinu_xsec_ratio_SCFFFtauIm} of Appendix~\ref{app:unpol_plots}. As expected from the charged lepton mass dependence in Eq.~(\ref{eq:CCQE_amplitudem}), cross sections with tau flavor are even more sensitive to the amplitudes $f_3$ and $f_T$ than cross sections with muon flavor, including sensitivity to values of $\mathfrak{Re}f_3$ that are allowed by beta decay constraints.

\subsection{Polarization observables}

\begin{figure}[htb]
\centering
\includegraphics[width=0.4\textwidth]{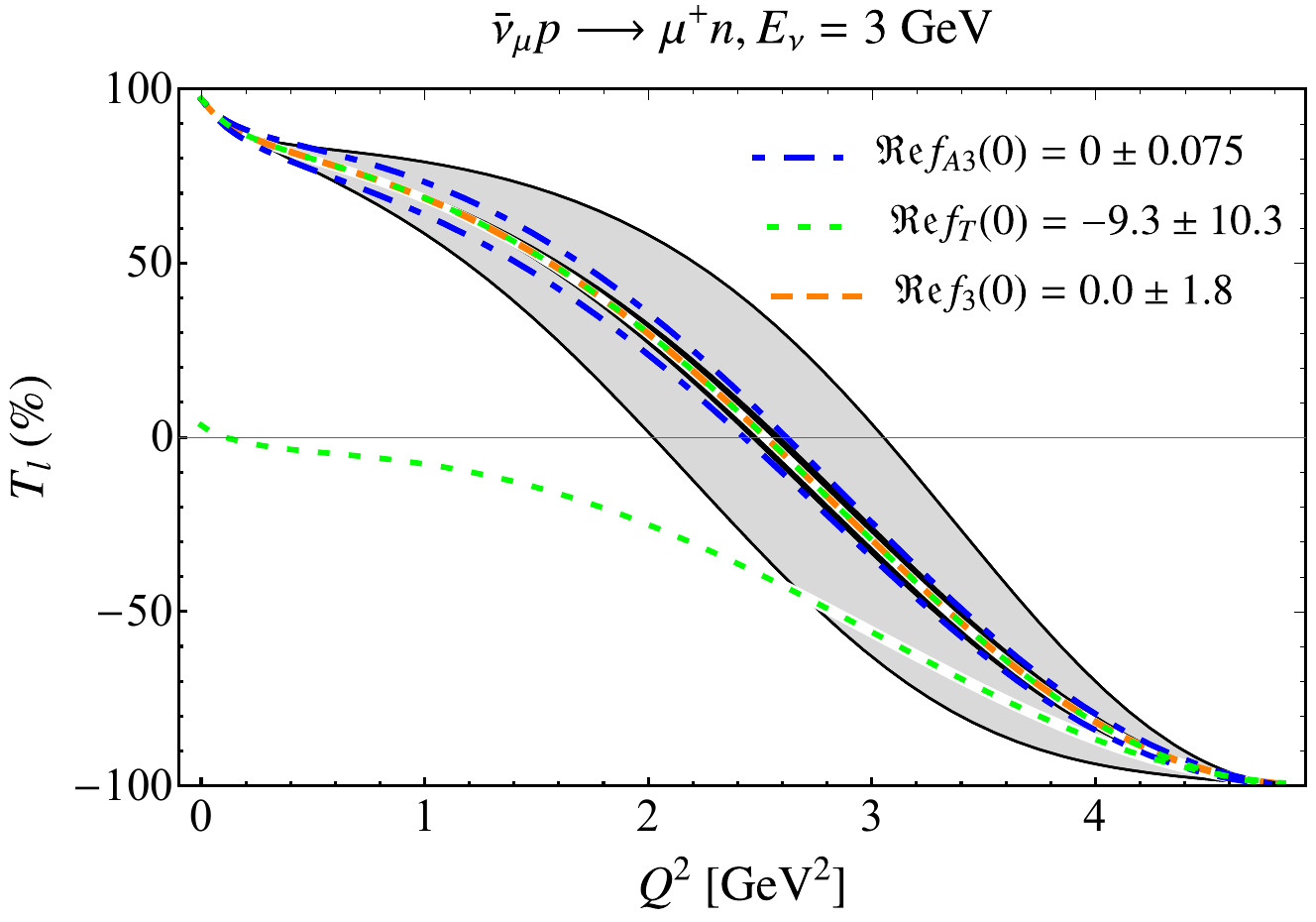}
\includegraphics[width=0.4\textwidth]{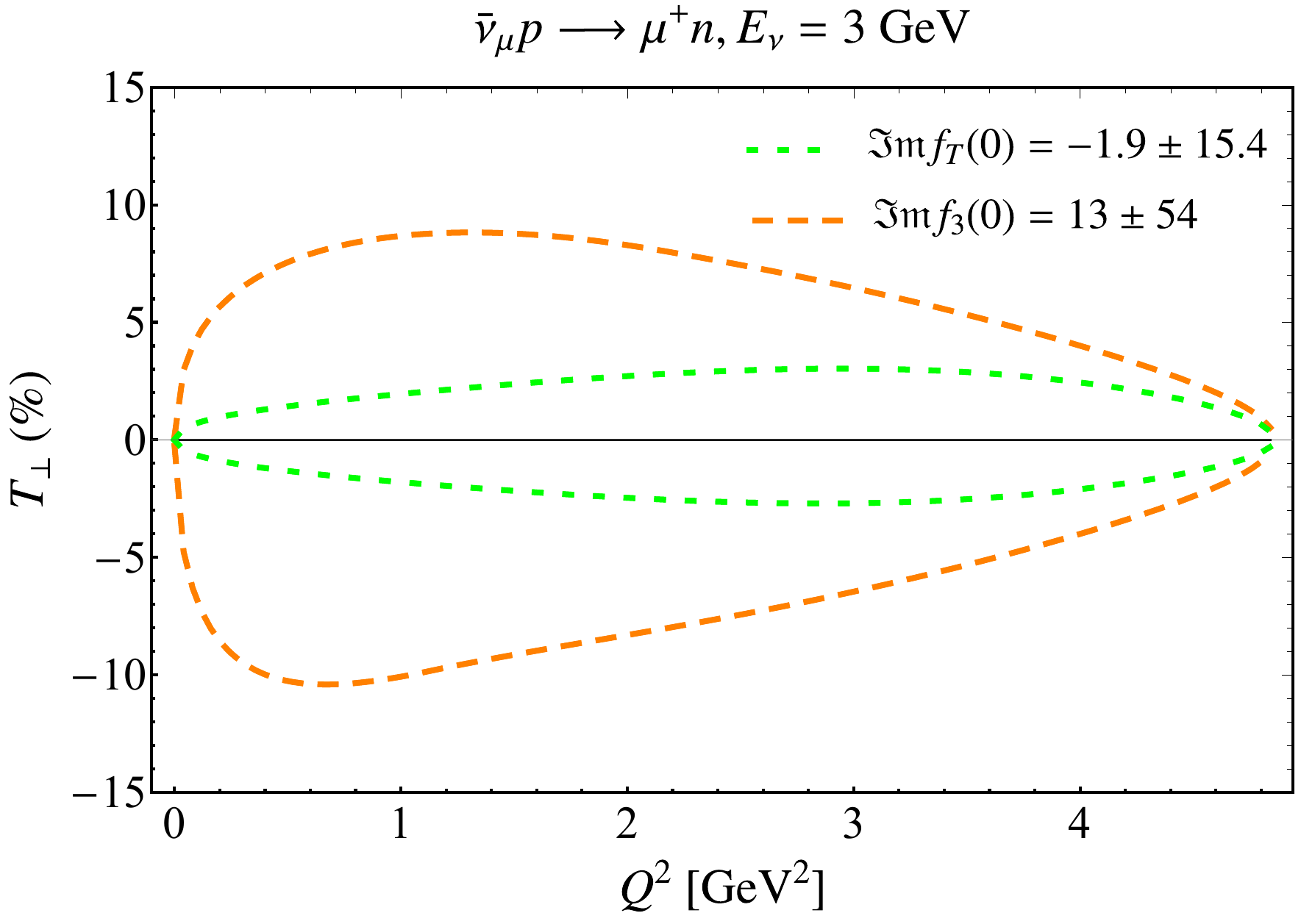}
\caption{Polarization observables $T_l$ with one extra real-valued amplitude (left) and $T_\perp$ with one extra imaginary-valued amplitude (right) compared to the tree-level result at fixed muon antineutrino energies $E_\nu = 3$~GeV. The dark black and light gray bands correspond to vector and axial-vector uncertainties, respectively. Orange dashed, green dotted, and blue dashed-dotted lines represent allowed regions for $f_3$, $f_T$, and $\fAt$, respectively, as described in the text. \label{fig:pol_target}}
\end{figure}

\noindent
We note that the contribution of the scalar amplitude $f_3$ and the tensor amplitude $f_T$ to the (anti)neutrino-nucleon elastic unpolarized charged-current cross sections is always suppressed by the lepton mass, which causes relatively small effects in the scattering of electron and muon (anti)neutrinos. Consequently, it is interesting to study the contributions of these amplitudes to spin-dependent observables, where mass suppression is not necessarily present. In Fig.~\ref{fig:pol_target} as an illustrative example, we consider the asymmetry predicted for an illustrative $3\,{\rm GeV}$ muon antineutrino beam on a polarized proton target, including the effect of possible new physics encoded in the amplitudes in Eq.~(\ref{eq:fp5_constraint}). Figures~\ref{fig:nu_Tt_SCFF}-\ref{fig:antinu_LT_SCFF} of Appendix~\ref{app:pol_plots} present the complete set of polarization-transfer observables $T_t,T_l,T_\perp,R_t,R_l,R_\perp,L_t,L_l,L_\perp$ for the scattering of muon (anti)neutrinos with energies between $300\,{\rm MeV}$ and $3\,{\rm GeV}$. In each case, we present the envelope of possible asymmetries after varying the amplitudes $\mathfrak{Re}f_3$, $\mathfrak{Re}\fAt$, $\mathfrak{Re}f_{T}$, $\mathfrak{Im}f_3$, and $\mathfrak{Im}f_{T}$ within the constraints of Table~\ref{tab:beta}. The transverse nucleon spin asymmetries $T_t$ and $R_t$ both for neutrinos and antineutrinos, as well as the longitudinal asymmetries $T_l$ and $R_l$, are sensitive to new parameter space for the tensor coupling $\mathfrak{Re}f_{T}$, and also exhibit sensitivity to $\mathfrak{Re}\fAt$ at neutrino energies above $\sim 1~\mathrm{GeV}$. The asymmetry $R_l$ both for neutrinos and antineutrinos and the target asymmetry $T_l$ for antineutrinos provide sensitivity above current uncertainties on vector and axial-vector form factors to the amplitude $\mathfrak{Re}f_{T}$ only. The target nucleon asymmetry $T_l$ in neutrino scattering is sensitive also to the amplitude $\mathfrak{Re}f_3$ near the muon production threshold, cf. Fig.~\ref{fig:nu_Tl_SCFF}. The lepton-spin asymmetries $L_t$ and $L_l$ are sensitive to the amplitudes $\mathfrak{Re}f_{T}$ and $\mathfrak{Re}f_3$ both in the scattering of neutrinos and antineutrinos. The transverse polarization asymmetries $T_\perp,~R_\perp,~L_\perp$ receive small Standard Model contributions but potentially large contributions from $\mathfrak{Im}f_{T}$ and $\mathfrak{Im}f_3$ and are thus sensitive to these quantities.

\begin{figure}[b]
\centering
\includegraphics[width=0.32\textwidth]{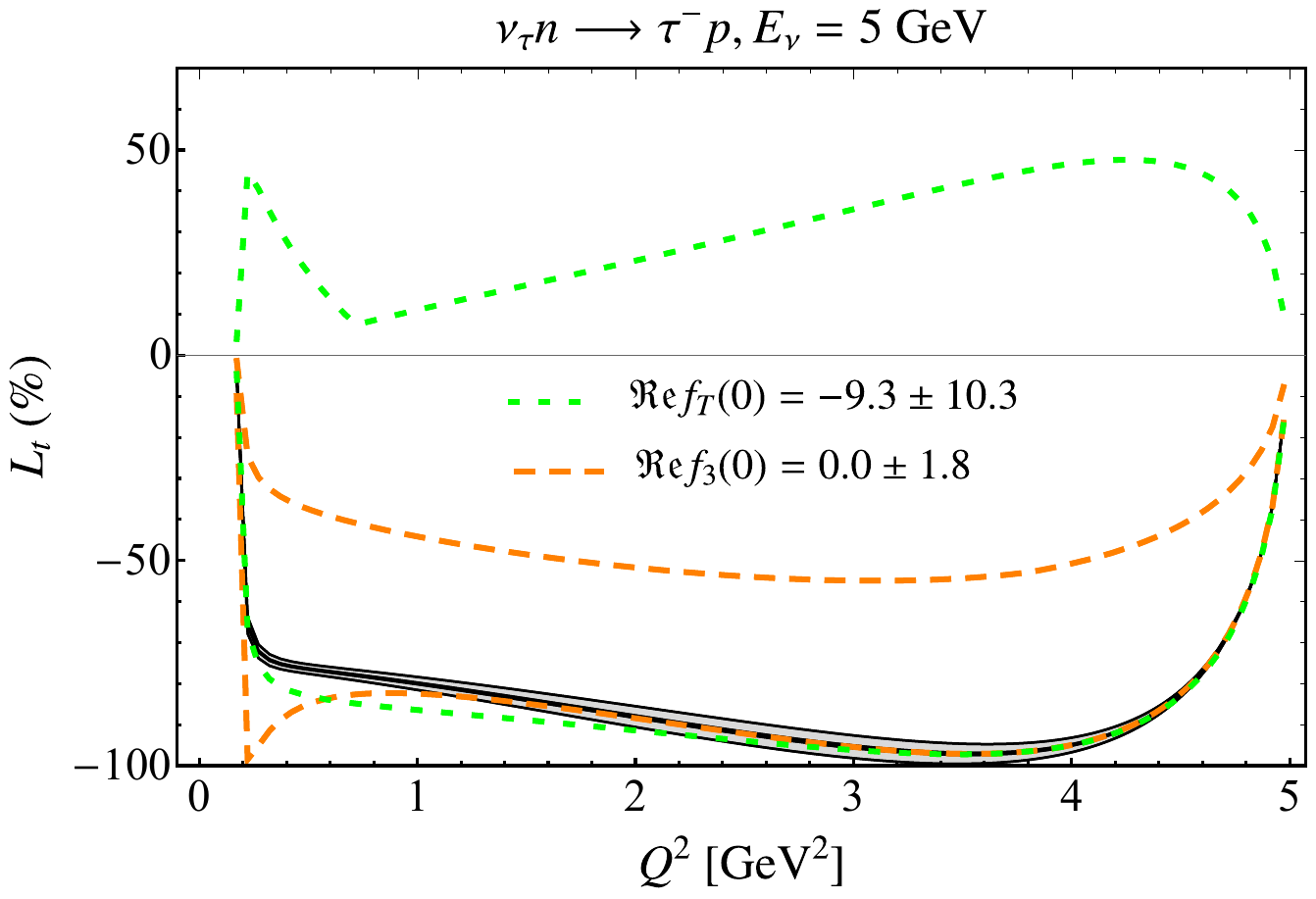}
\includegraphics[width=0.32\textwidth]{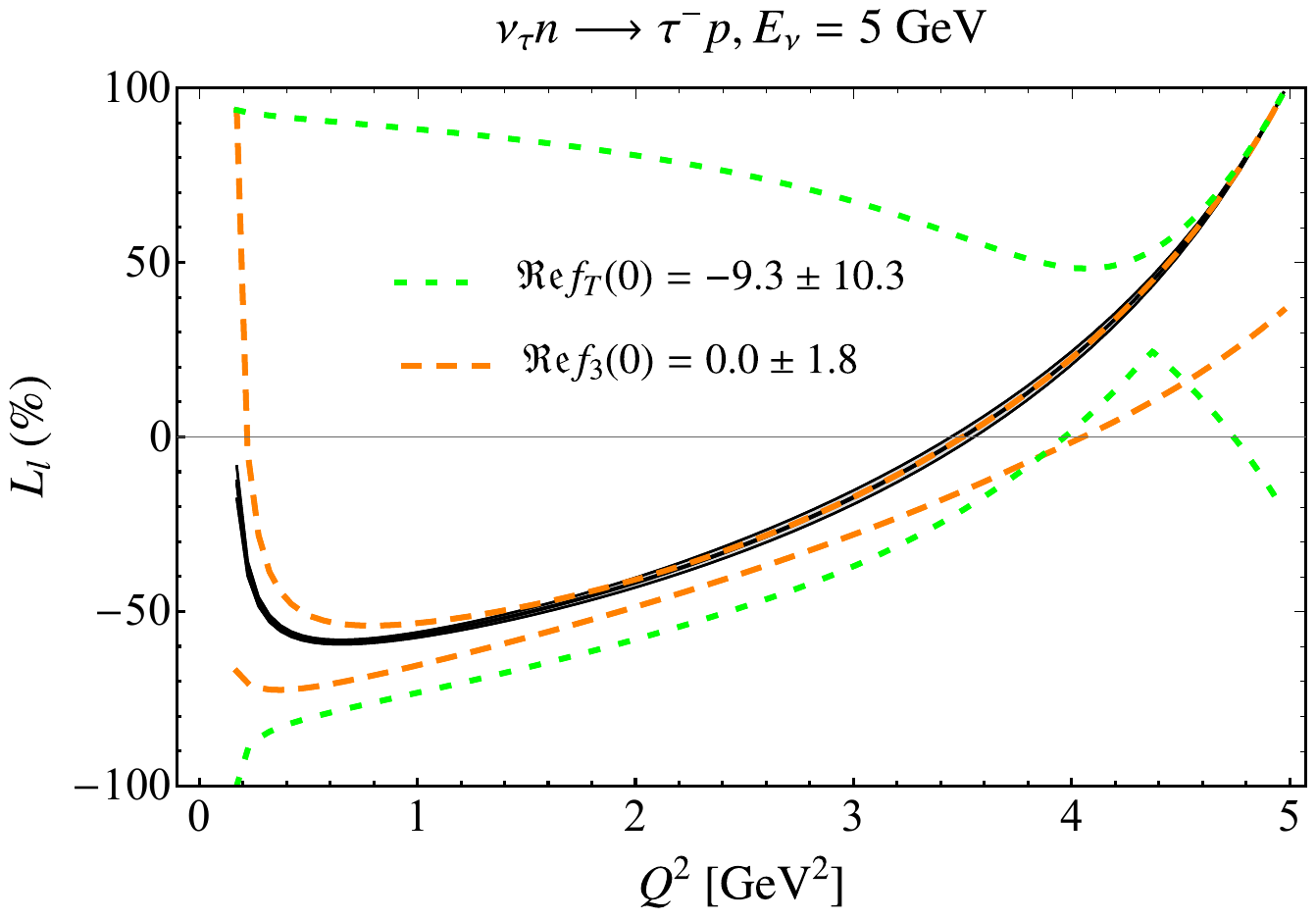}
\includegraphics[width=0.32\textwidth]{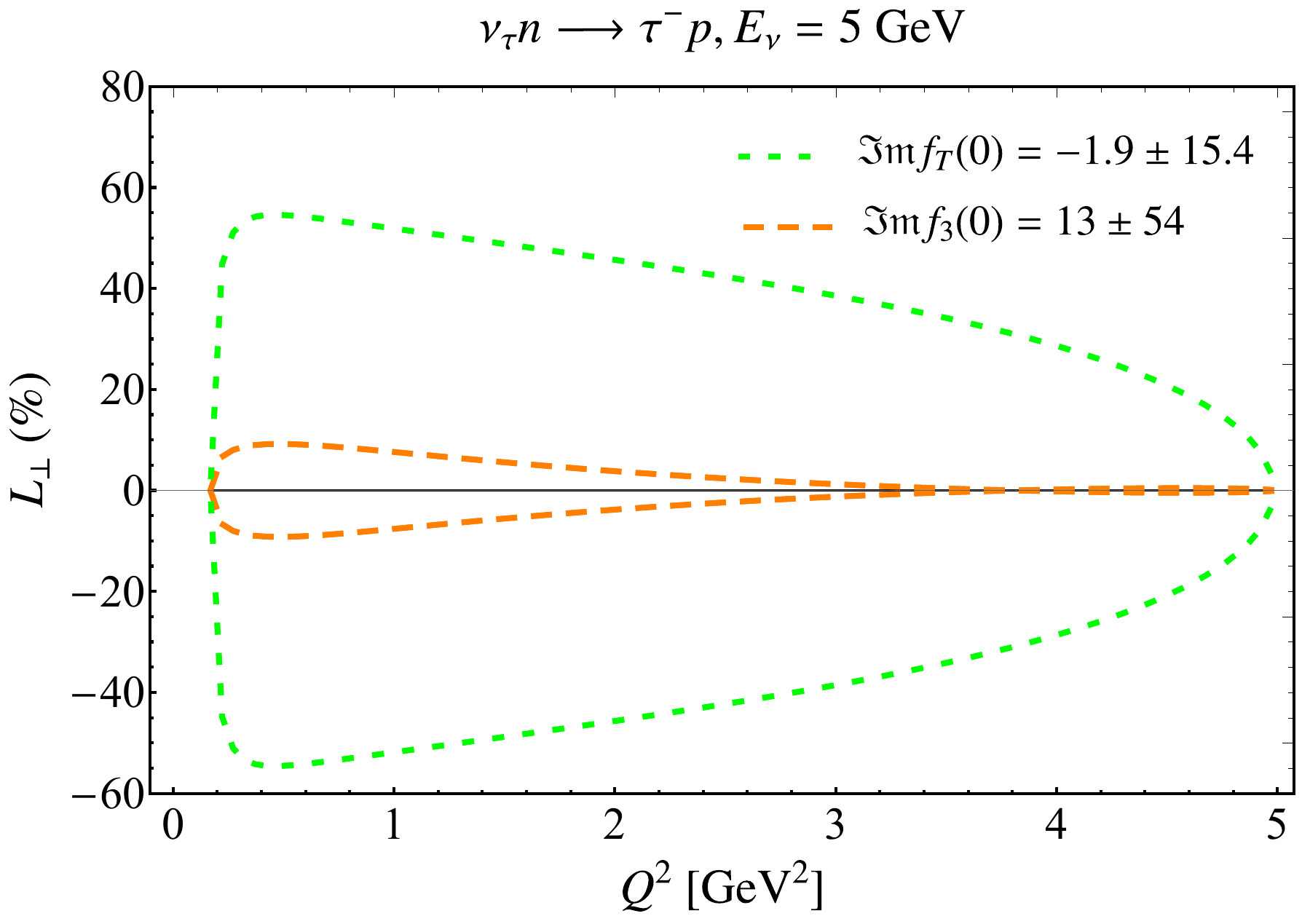}
\caption{Polarization observables $L_t$, $L_l$ with one extra real-valued amplitude, and $L_\perp$ with one extra imaginary-valued amplitude, compared to the tree-level result at fixed tau neutrino energy $E_\nu = 5$~GeV.} \label{fig:pol_lepton_tau}
\end{figure}

We repeat our study for tau flavor polarization asymmetries. As an illustration, Fig.~\ref{fig:pol_lepton_tau} considers the lepton asymmetries that can be determined by measuring the polarization of the final-state tau lepton. For neutrinos with $E_\nu=5\,{\rm GeV}$, the results demonstrate that asymmetries differ significantly from Standard Model predictions in large regions of parameter space that are allowed by beta decay constraints. Figures~\ref{fig:nu_Tt_SCFF_tau}-\ref{fig:nu_Lt_SCFF_tau} of Appendix~\ref{app:taupol_plots} present results for all single-spin asymmetries over a range of (anti)neutrino beam energies between $5\,{\rm GeV}$ and $15\,{\rm GeV}$, above the tau production threshold. For tau neutrinos, all polarization observables with a reference spin in the scattering plane are sensitive to the amplitudes $\mathfrak{Re}f_{T}$ and $\mathfrak{Re}f_3$. Contrary to the unpolarized cross section, the transverse spin asymmetries $T_t$ and $R_t$ in both neutrino and antineutrino scattering are also sensitive to $\mathfrak{Re}\fAt$. The transverse polarization asymmetries $T_\perp,R_\perp,L_\perp$, with spin direction orthogonal to the scattering plane, are sensitive to $\mathfrak{Im}f_{T}$ and $\mathfrak{Im}f_3$, in both neutrino and antineutrino scattering.

\section{Radiative corrections to unpolarized cross sections and single-spin asymmetries} \label{sec:radiative_corrections}

\begin{figure}[bt]
\centering
\includegraphics[width=0.4\textwidth]{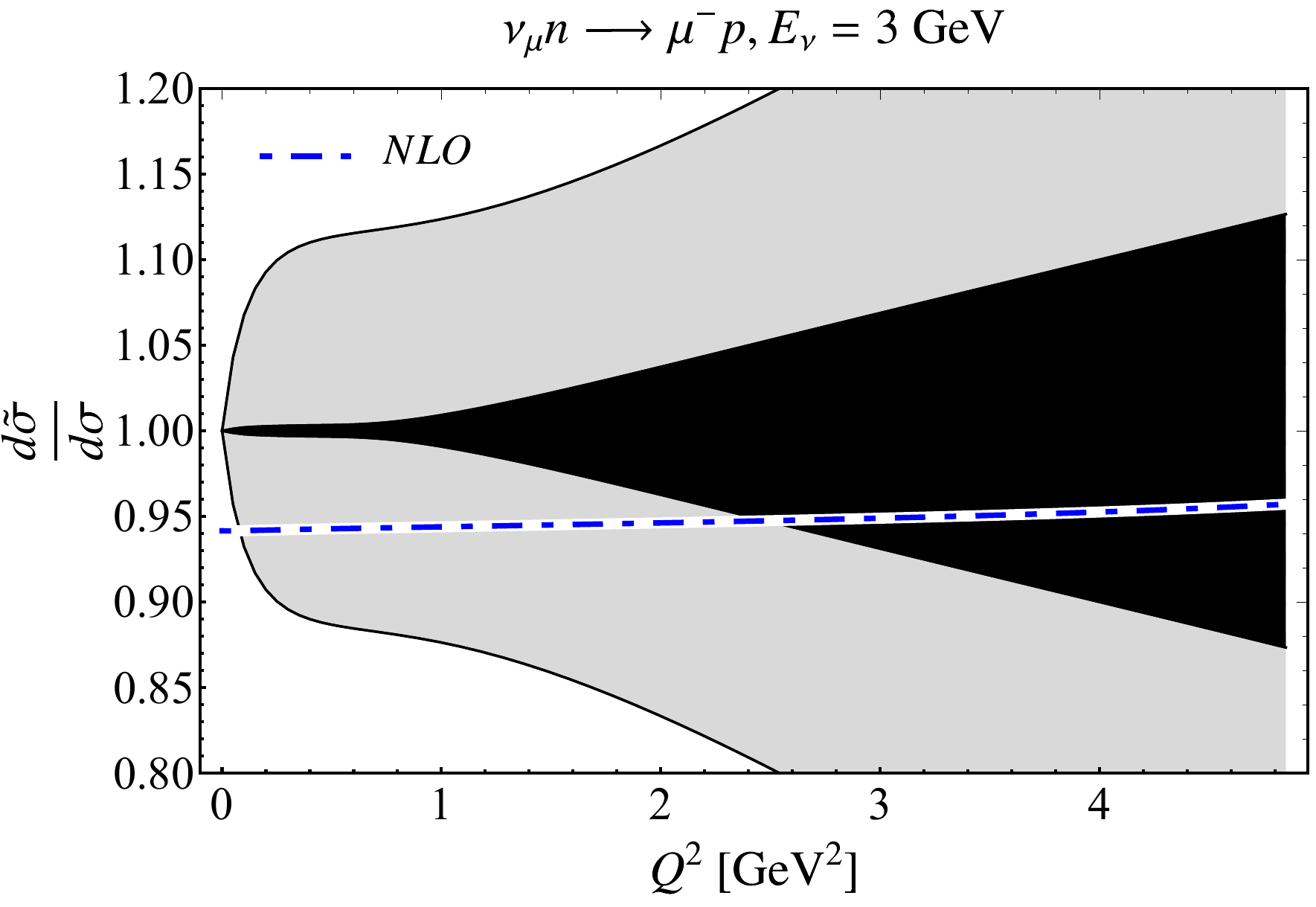}
\caption{Radiative corrections to unpolarized cross section in Fig.~\ref{fig:unpol}. The blue dashed-dotted line represents the radiatively-corrected result.
\label{fig:radcor}}
\end{figure}

In this Section, we study the effects of radiative corrections on the unpolarized cross sections and polarization observables considered above. We take virtual contributions to the amplitudes in Eqs.~(\ref{eq:CCQE_amplitude}) and~(\ref{eq:CCQE_amplitudem}) from the recent calculation of Refs.~\cite{Tomalak:2021hec,Tomalak:2022xup}. For the muon and tau charged-current events considered in this paper, we include the radiation of one real photon with energy below $\Delta E = 10~\mathrm{MeV}$ and label this cross section as next-to-leading order (NLO).

In Fig.~\ref{fig:radcor}, we compare the effect of radiative corrections to the theoretical cross-section uncertainties that are obtained by propagating errors from the vector and axial-vector nucleon form factors, cf. Fig.~\ref{fig:unpol}. Over most of the kinematic range, the effect is larger than the current vector form factor uncertainty~\cite{Borah:2020gte} but smaller than the current axial-vector form factor uncertainty~\cite{Meyer:2016oeg}. The effects of radiative corrections are at or below the current level of experimental uncertainties but will become increasingly important with more precise data. These corrections are also critical for the determination of flavor ratios~\cite{Tomalak:2021hec}. In Appendix~\ref{app:radcorplots}, we present results over a range of energies, for unpolarized cross sections of both neutrinos and antineutrinos, cf. Figs.~\ref{fig:nu_xsec_ratio_radcorrRe}-\ref{fig:antinu_xsec_ratio_radcorrFtauRe}. Results are displayed as a ratio of the cross sections evaluated with radiative corrections to the benchmark tree-level cross section, for the illustrative energies considered in Appendix~\ref{app:unpol_plots}. The results reproduce those from Refs.~\cite{Tomalak:2021hec,Tomalak:2022xup} for the corresponding kinematics and muon (anti)neutrino flavor. At very low squared momentum transfers, radiative corrections for muon (anti)neutrino flavor typically exceed the tree-level cross-section uncertainty, determined mainly by the axial-vector charge according to Figs.~\ref{fig:radcor},~\ref{fig:nu_xsec_ratio_radcorrRe}, and~\ref{fig:antinu_xsec_ratio_radcorrRe}. The corrections are positive at hundreds of MeV incoming (anti)neutrino energies and become negative for energies above $500~\mathrm{MeV}$.  For tau (anti)neutrino scattering, cf.~Figs.~\ref{fig:nu_xsec_ratio_radcorrtauRe} and~\ref{fig:antinu_xsec_ratio_radcorrFtauRe}, the radiative corrections are well below uncertainties from the axial-vector form factor and reach vector form-factor uncertainties only at lowest squared momentum transfers.

For the first time, we investigate the effects of radiative corrections on the polarization asymmetries in neutrino scattering $T_t, T_l, R_t, R_l, L_t, L_l$ for muon and tau flavors.  We present these observables in Figs.~\ref{fig:nu_TTT_radcorr}-\ref{fig:antinu_LTT_radcorr_tau} of Appendix~\ref{app:radcorplots}. Radiative corrections to these asymmetries are below the uncertainties from the axial-vector form factor but have the same order of magnitude as vector form-factor errors. Besides the motivated model for the real part of amplitudes in Refs.~\cite{Tomalak:2021hec,Tomalak:2022xup}, we consider also imaginary parts of all invariant amplitudes in the same model as an estimate of the order-of-magnitude for possible contributions from imaginary parts. The asymmetries with a spin vector transverse to the scattering plane vanish at tree level and are generated by the imaginary parts of invariant amplitudes. Both for muon and tau flavor, the transverse spin asymmetries $T_\perp, R_\perp$ are typically at permille level. The asymmetry $L_\perp$ is below permille and can reach $\sim 0.1~\%$ at some kinematics.

\section{Conclusions and Outlook}

In this paper, we present the general decomposition of (anti)neutrino-nucleon charged-current elastic scattering amplitudes. We provide expressions for the unpolarized cross section and single-spin asymmetries in terms of these amplitudes and investigate both the forward and the backward limits for these observables. We provide numerical results for all observables for relevant fixed (anti)neutrino energies as a function of the squared momentum transfer considering the uncertainty of tree-level nucleon form factors and also accounting for radiative corrections. QED contributions to muon (anti)neutrino-nucleon charged-current elastic scattering cross sections are of the order of the theoretical uncertainty, while the radiative corrections to the scattering of tau (anti)neutrinos are much smaller than the theoretical error from current knowledge of the nucleon form factors. The radiative corrections to the single-spin asymmetries largely cancel between the numerator and denominator resulting in negligible effects within the uncertainties of tree-level predictions.

Assuming the dipole form for the $Q^2$ dependence and $\nu$ independence for possible new physics contributions to the invariant amplitudes, and Standard Model scaling with the lepton mass, we examine the parameter space allowed by beta decay constraints on unpolarized cross sections and polarization transfer observables. Surprisingly, we find that the available constraints on both the real and the imaginary parts of the tensor interaction coefficient, as well as constraints on the imaginary part of the scalar coefficient, can be significantly improved with current data on the unpolarized antineutrino-hydrogen and neutrino-deuterium scattering cross sections. We present an analysis of the recent antineutrino-hydrogen measurement by the MINERvA Collaboration in an accompanying publication~\cite{Tomalak:2024yvq}. Relatively imprecise first measurements of single-spin asymmetries with (anti)neutrino beams would be sensitive to new parameter space for both the real and the imaginary parts of the scalar and tensor couplings.

\section*{Acknowledgments}
We thank Susan Gardner, Kevin McFarland, and Ryan Plestid for useful discussions. This work is supported by the US Department of Energy through the Los Alamos National Laboratory and by LANL’s Laboratory Directed Research and Development (LDRD/PRD) program under projects 20210968PRD4, 20210190ER, and 20240127ER. Los Alamos National Laboratory is operated by Triad National Security, LLC, for the National Nuclear Security Administration of U.S. Department of Energy (Contract No. 89233218CNA000001). This work was supported by the U.S. Department of Energy, Office of Science, Office of High Energy Physics, under Award DE-SC0019095. R.J.H. acknowledges support from a Fermilab Intensity Frontier Fellowship. K.B. acknowledges support from the Visiting Scholars Award Program of the Universities Research Association and the Fermilab Neutrino Physics Center Fellowship Program. Fermilab is operated by Fermi Research Alliance, LLC under Contract No. DE-AC02-07CH11359 with the United States Department of Energy. FeynCalc~\cite{Mertig:1990an,Shtabovenko:2016sxi}, LoopTools~\cite{Hahn:1998yk}, and Mathematica~\cite{Mathematica} were extremely useful in this work.

\newpage

\begin{appendix}

\section{Cross section, asymmetry, and radiative corrections plots \label{sec:plots}}

In this Appendix, we collect illustrative plots showing Standard Model uncertainties and constraints on new physics extrapolated from beta decay measurements. Section~\ref{app:unpol_plots} considers unpolarized cross sections for muon flavor and tau flavor (anti)neutrinos. Sections~\ref{app:pol_plots} and~\ref{app:taupol_plots} consider polarization asymmetries for muon and tau flavor (anti)neutrinos, respectively. Section~\ref{app:radcorplots} shows the impact of radiative corrections on each of the above observables.

\subsection{Unpolarized cross sections \label{app:unpol_plots}}

In this Section, we consider the unpolarized cross section with one extra real- or imaginary-valued amplitude, $f_i (\nu, Q^2)  =  [\mathfrak{Re} f_i \left( 0 \right) + i \mathfrak{Im} f_i \left( 0 \right)]/\left( 1 + \frac{Q^2}{\Lambda^2} \right)^2$, for illustrative muon (tau) neutrino and antineutrino energies $E_\nu = 300$~MeV, $600$~MeV, $1$~GeV, and $3$~GeV ($E_\nu = 5$~GeV, $7$~GeV, $10$~GeV, and $15$~GeV), and $\Lambda = 1$~GeV.
We vary the amplitude normalizations within the ranges from Table~\ref{tab:beta}, and compare to the uncertainty from vector and axial-vector form factors from Sec.~\ref{sec:observables}. 
The effect from the extra amplitude $\mathfrak{Re}\fAt(0)$ from Table~\ref{tab:beta} cannot be distinguished from zero on plots with the unpolarized cross section.

\begin{figure}[H]
\centering
\includegraphics[width=0.4\textwidth]{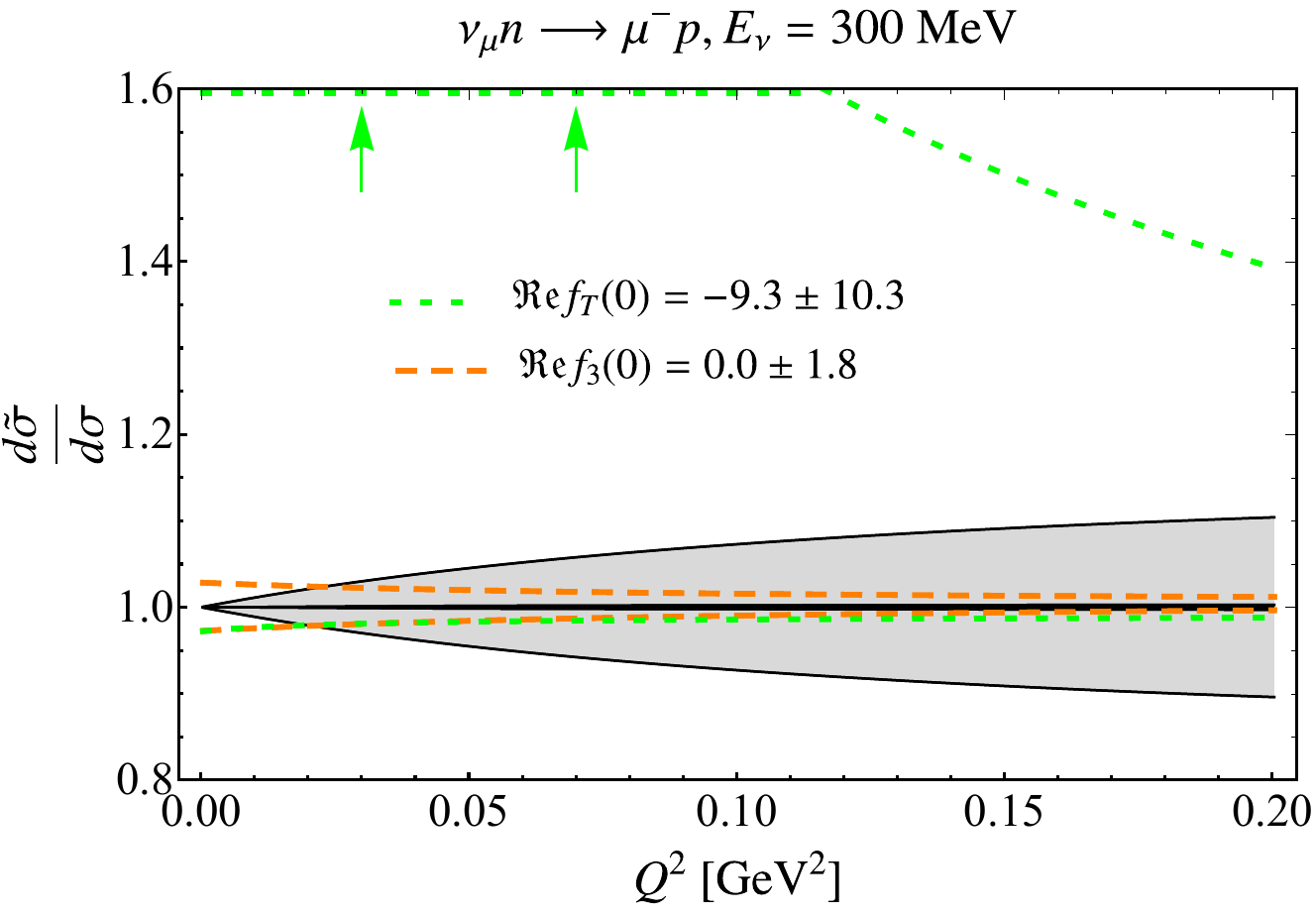}
\includegraphics[width=0.4\textwidth]{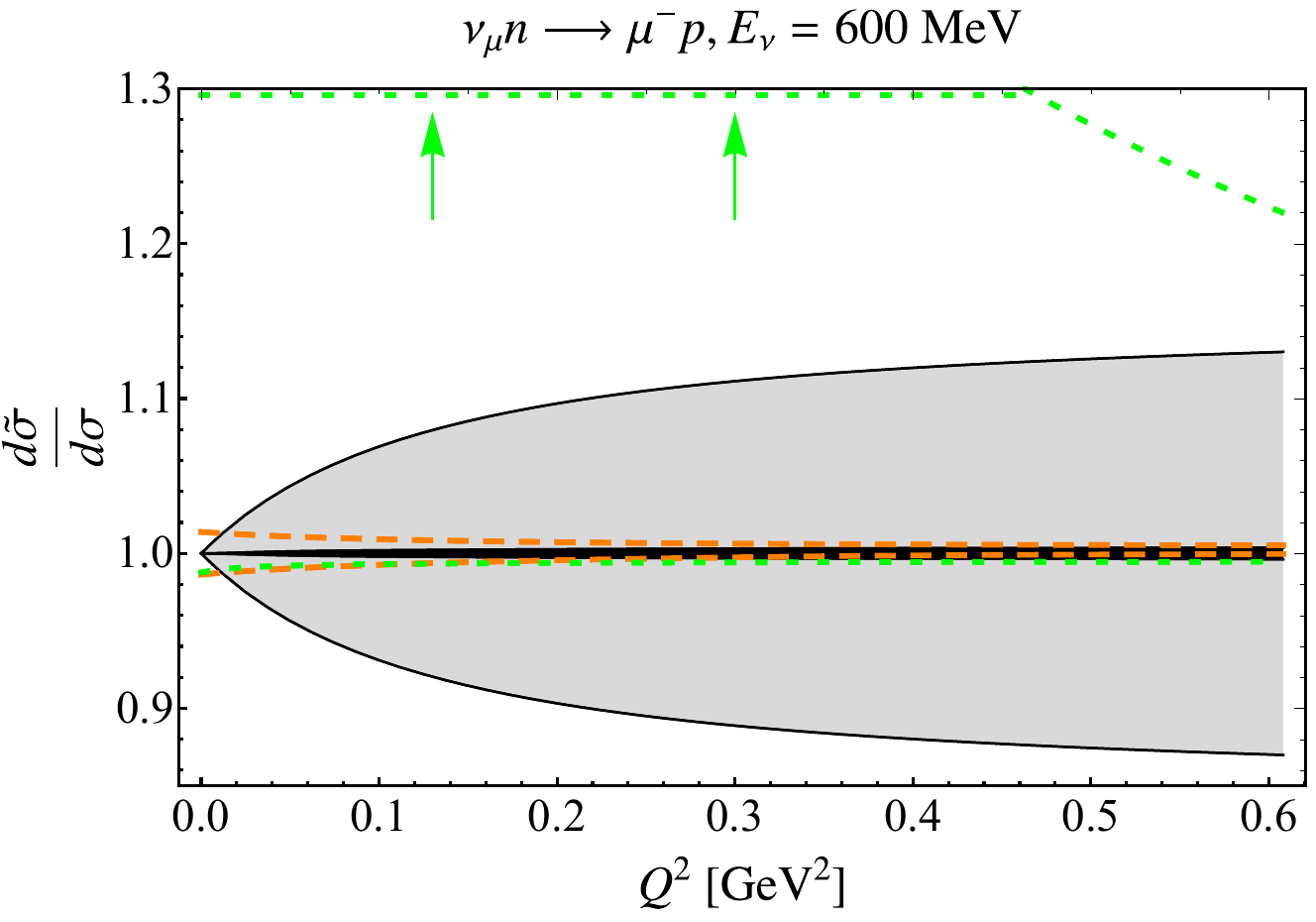}
\includegraphics[width=0.4\textwidth]{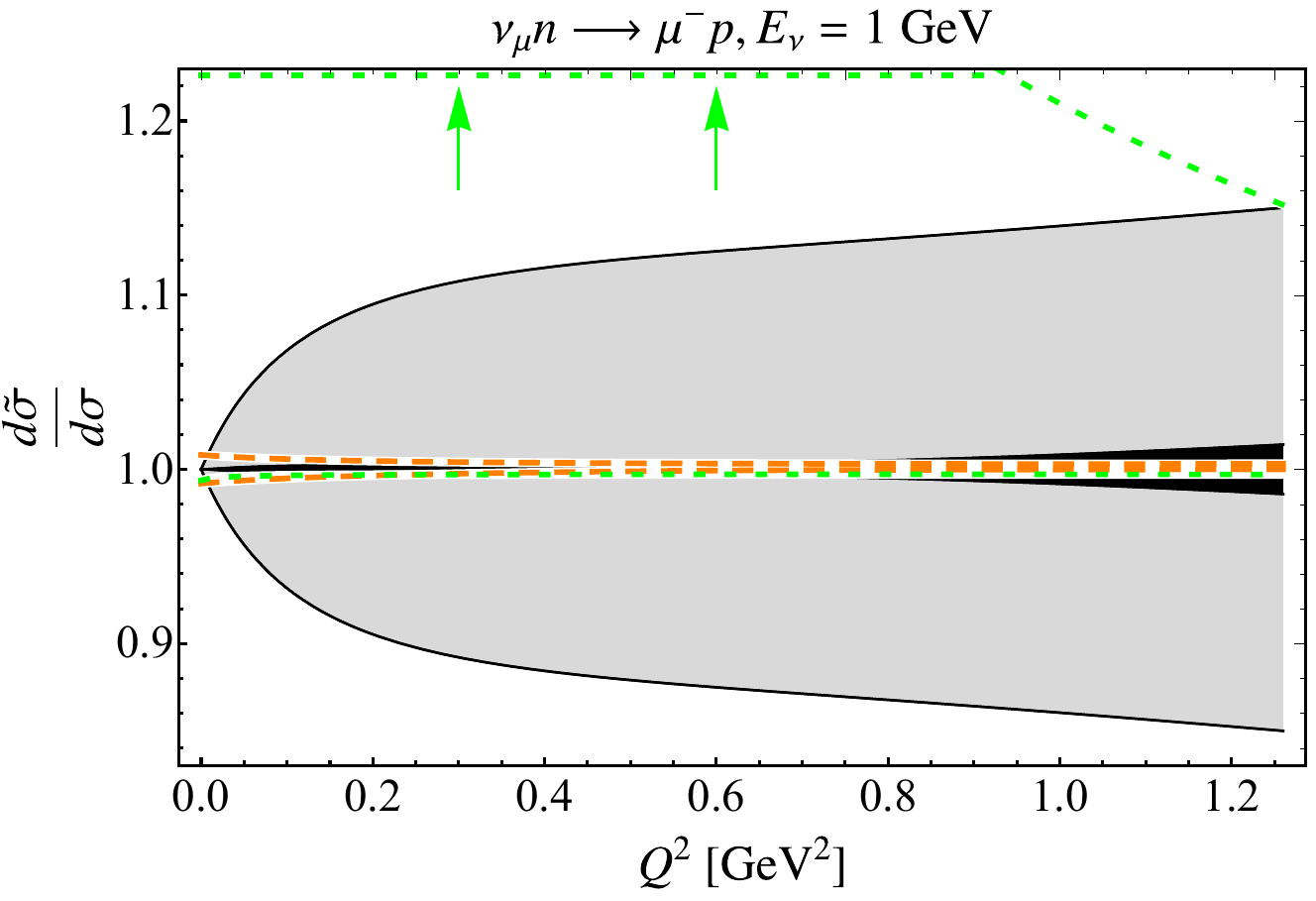}
\includegraphics[width=0.4\textwidth]{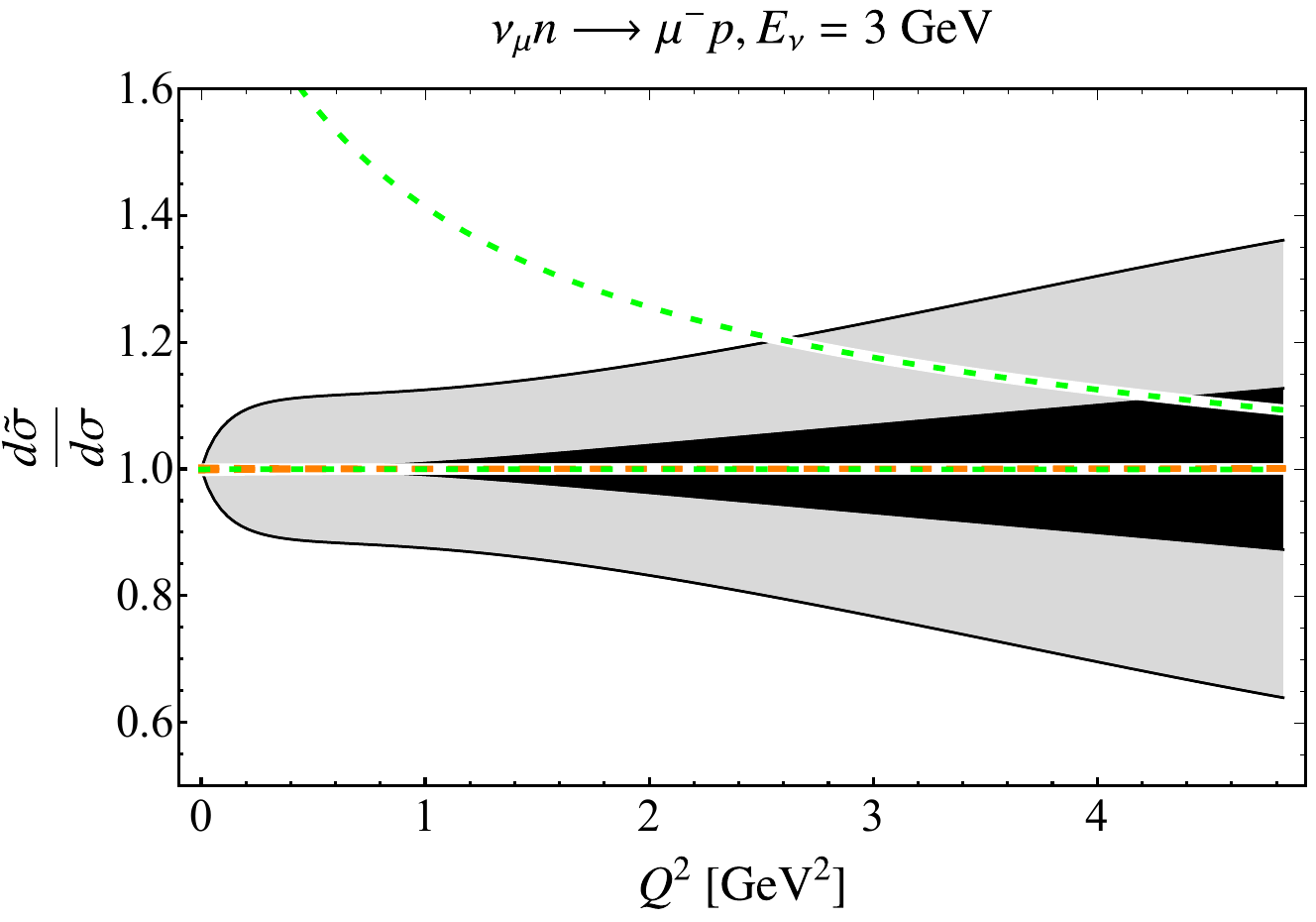}
\caption{Ratio of unpolarized muon neutrino cross section with one extra real-valued amplitude to the tree-level result as a function of the squared momentum transfer $Q^2$ at fixed muon neutrino energies $E_\nu = 300$~MeV, $600$~MeV, $1$~GeV, and $3$~GeV. The dark black and light gray bands correspond to vector and axial-vector uncertainty, respectively. Orange dashed and green dotted lines represent allowed regions for $f_3$ and $f_T$, respectively, as described in the text. \label{fig:nu_xsec_ratio_SCFFRe}}
\end{figure}

\begin{figure}[H]
\centering
\includegraphics[width=0.4\textwidth]{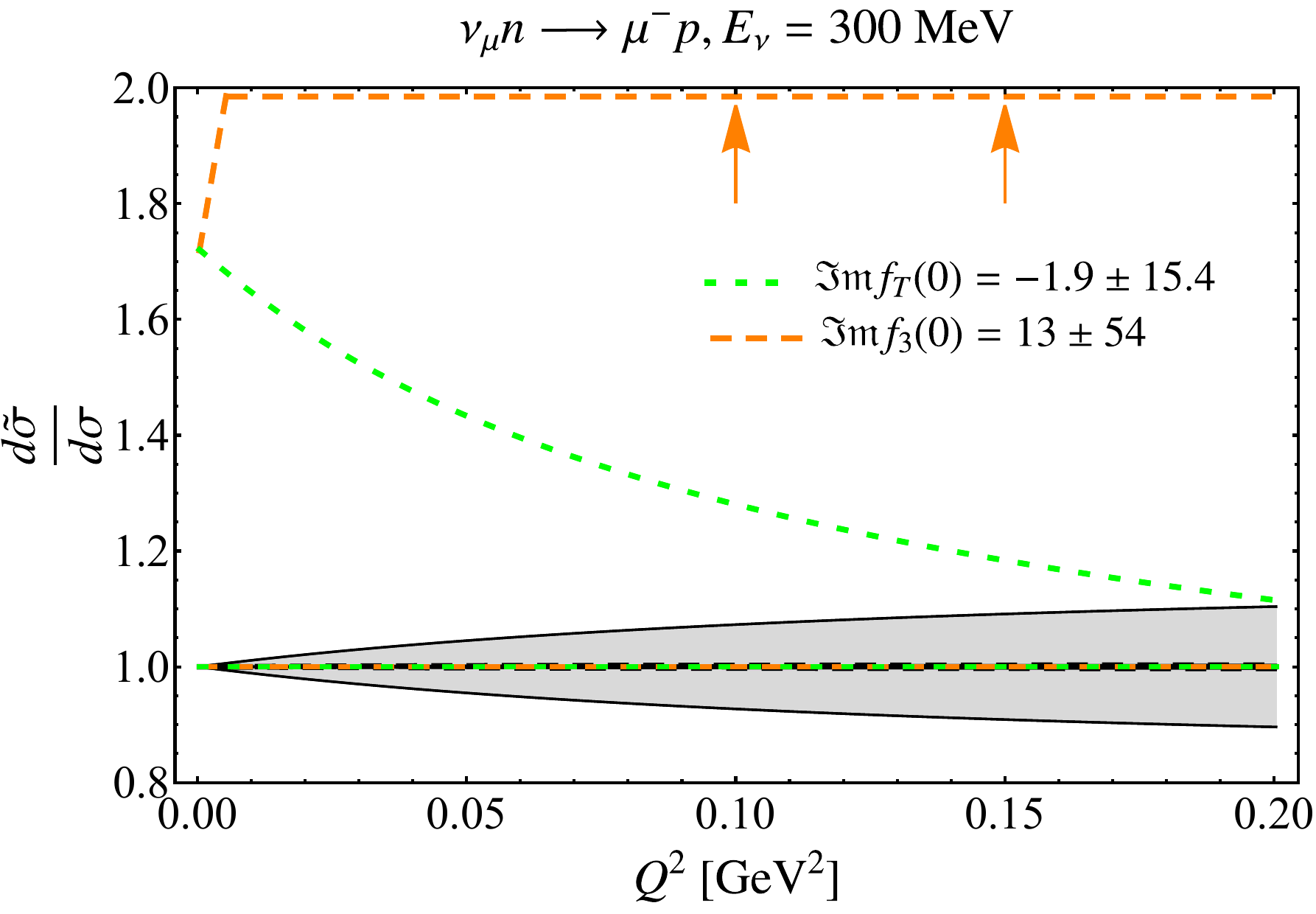}
\includegraphics[width=0.4\textwidth]{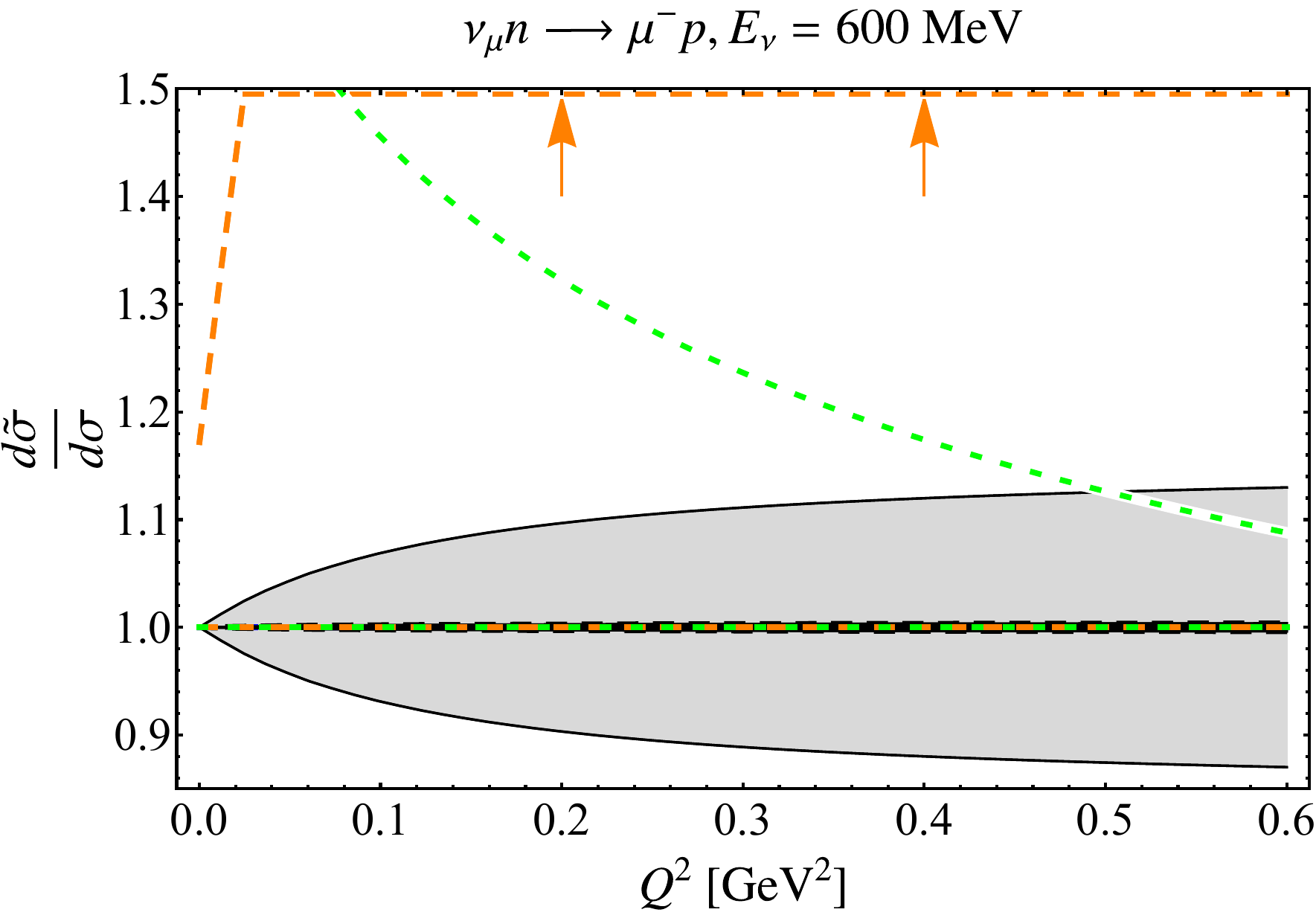}
\includegraphics[width=0.4\textwidth]{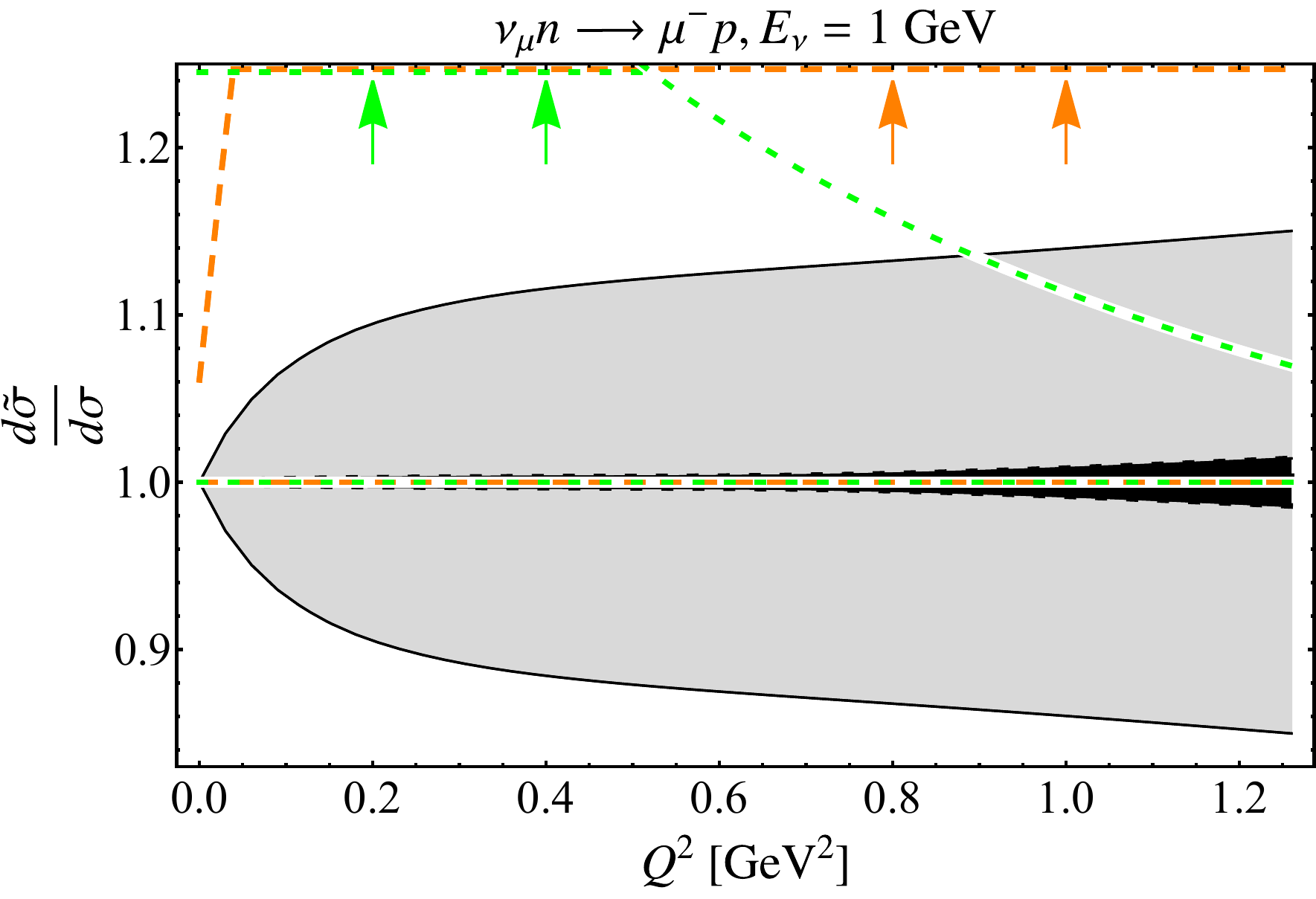}
\includegraphics[width=0.4\textwidth]{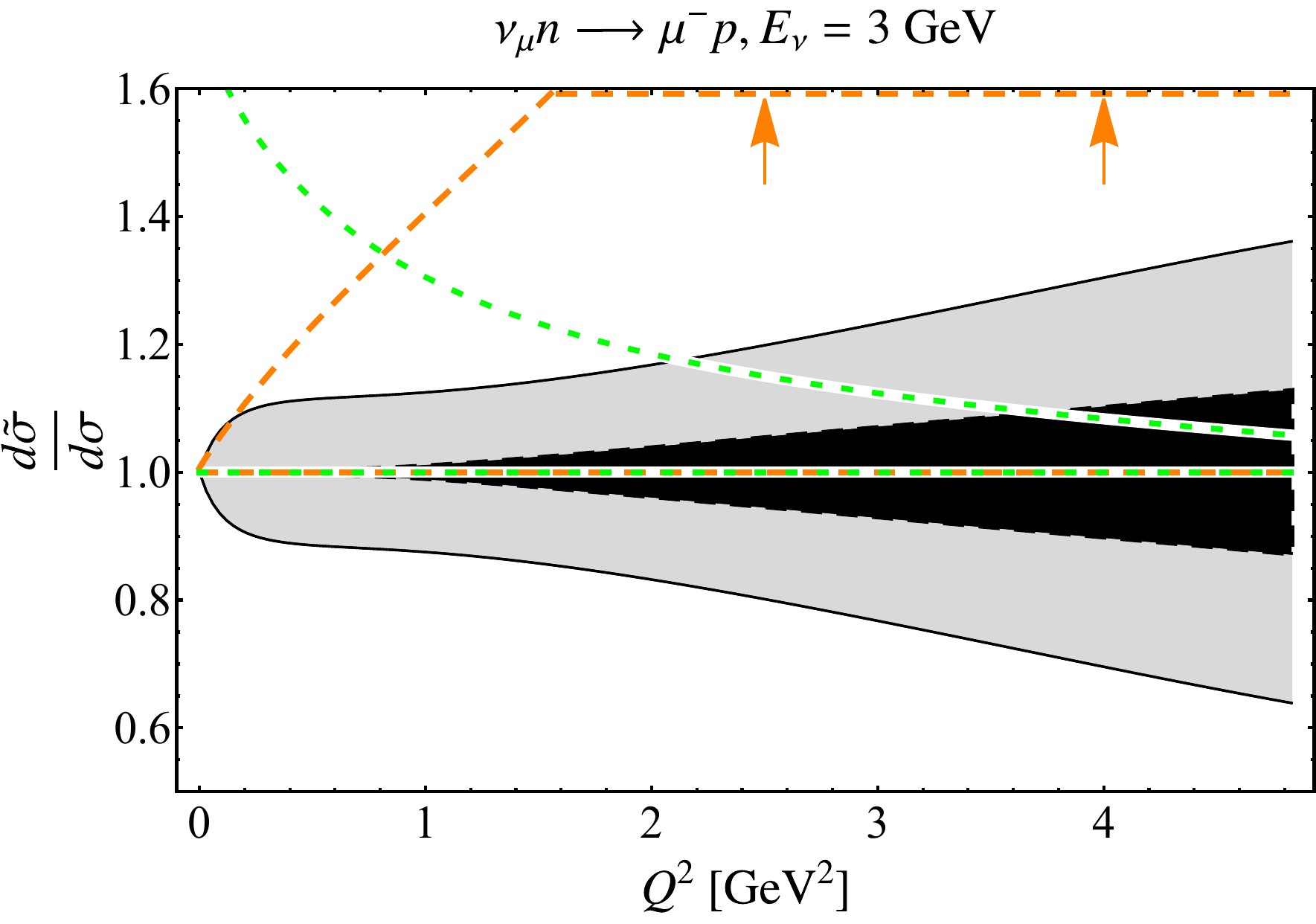}
\caption{Same as Fig.~\ref{fig:nu_xsec_ratio_SCFFRe}, but for imaginary amplitudes. \label{fig:nu_xsec_ratio_SCFFIm}}
\end{figure}

\begin{figure}[H]
\centering
\includegraphics[width=0.4\textwidth]{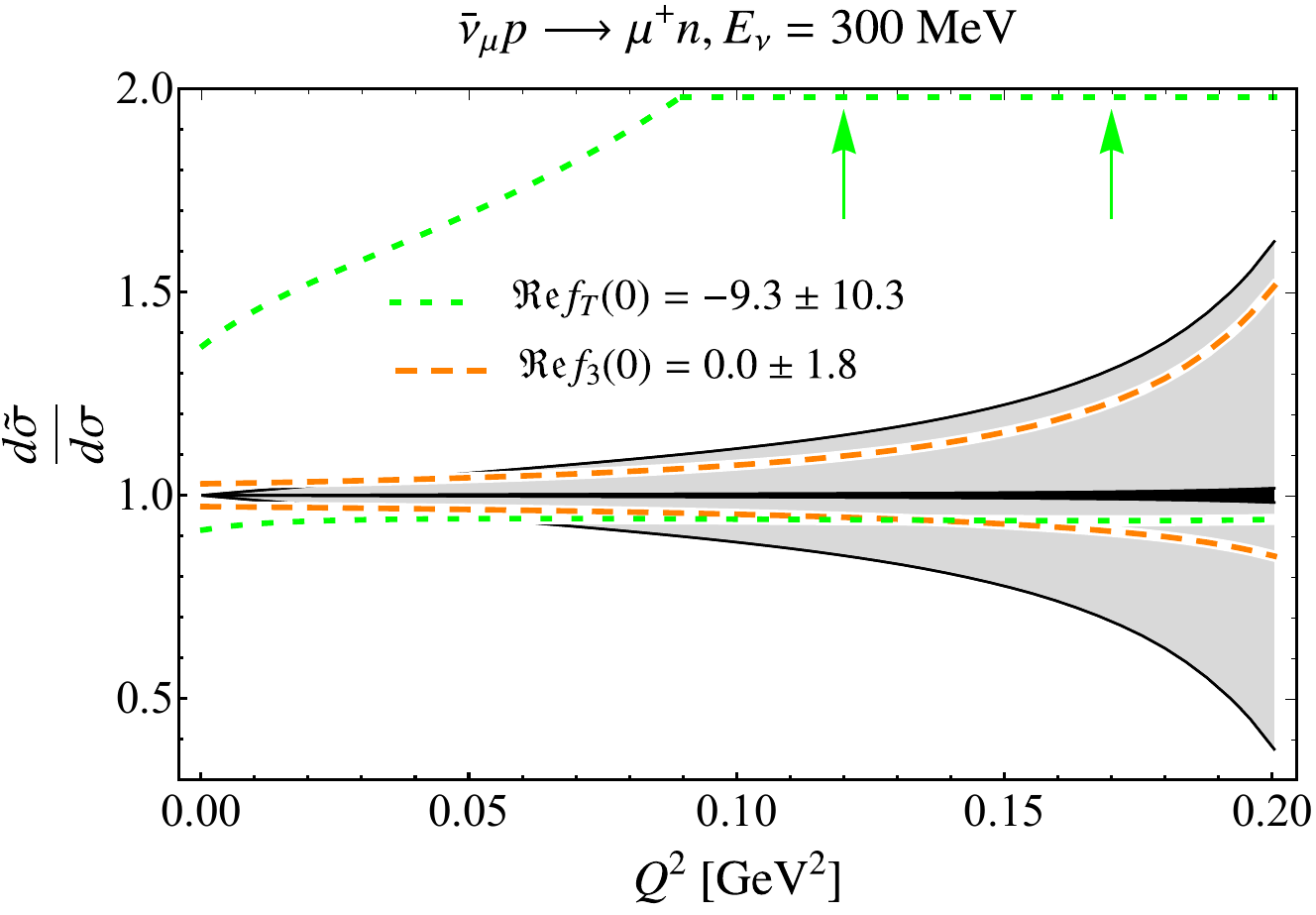}
\includegraphics[width=0.4\textwidth]{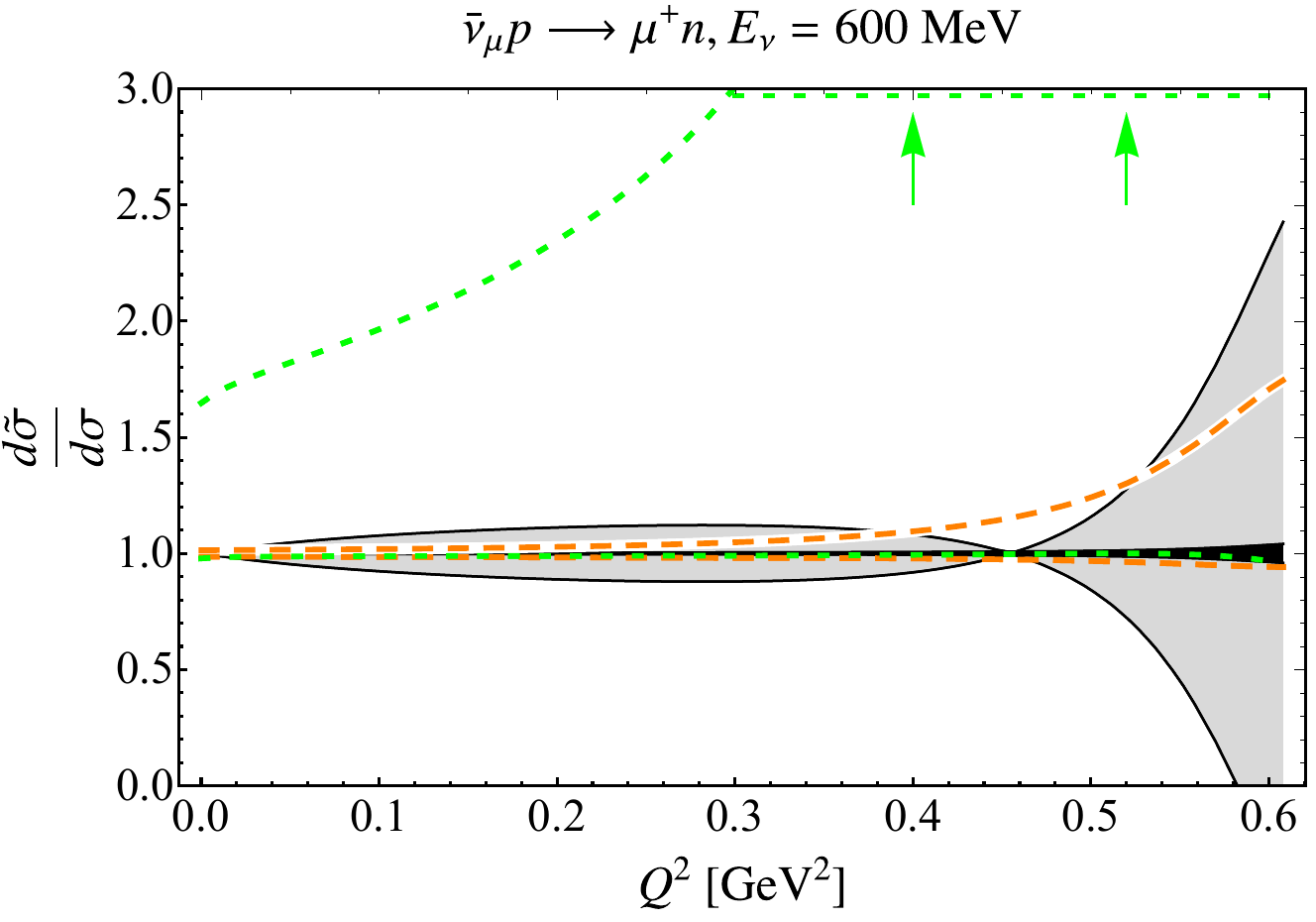}
\includegraphics[width=0.4\textwidth]{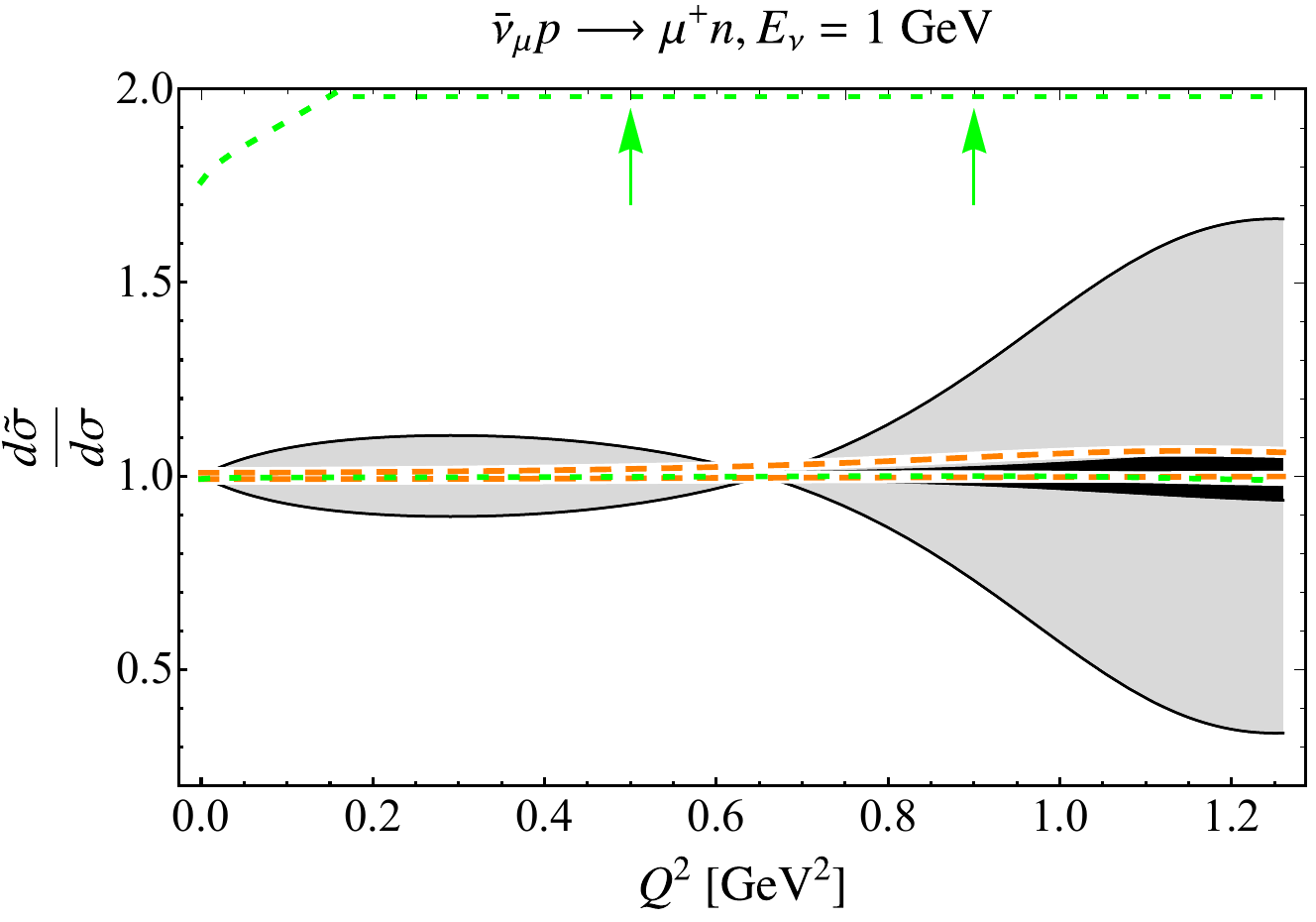}
\includegraphics[width=0.4\textwidth]{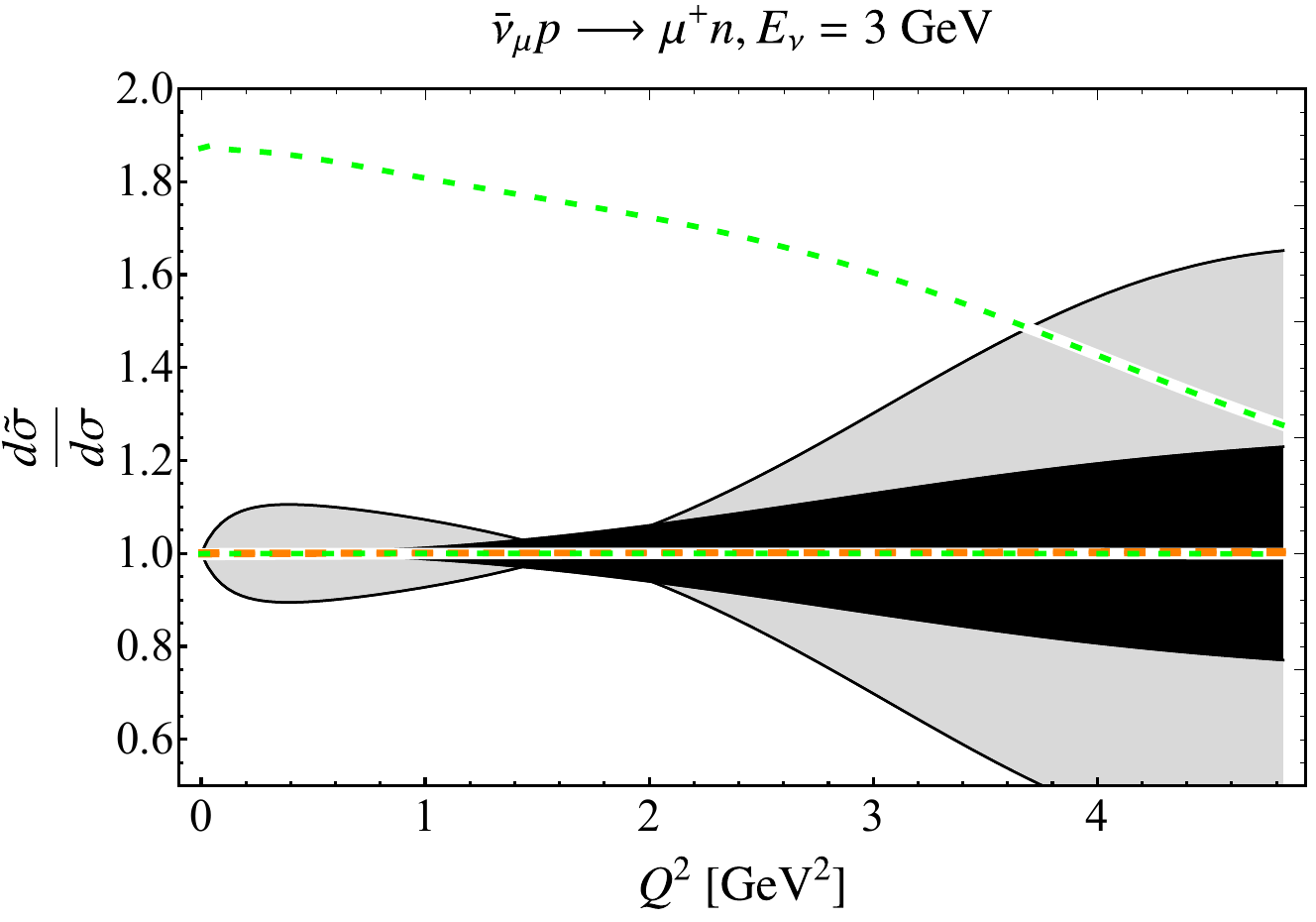}
\caption{Same as Fig.~\ref{fig:nu_xsec_ratio_SCFFRe} but for antineutrinos. \label{fig:antinu_xsec_ratio_SCFFRe}}
\end{figure}

\begin{figure}[H]
\centering
\includegraphics[width=0.4\textwidth]{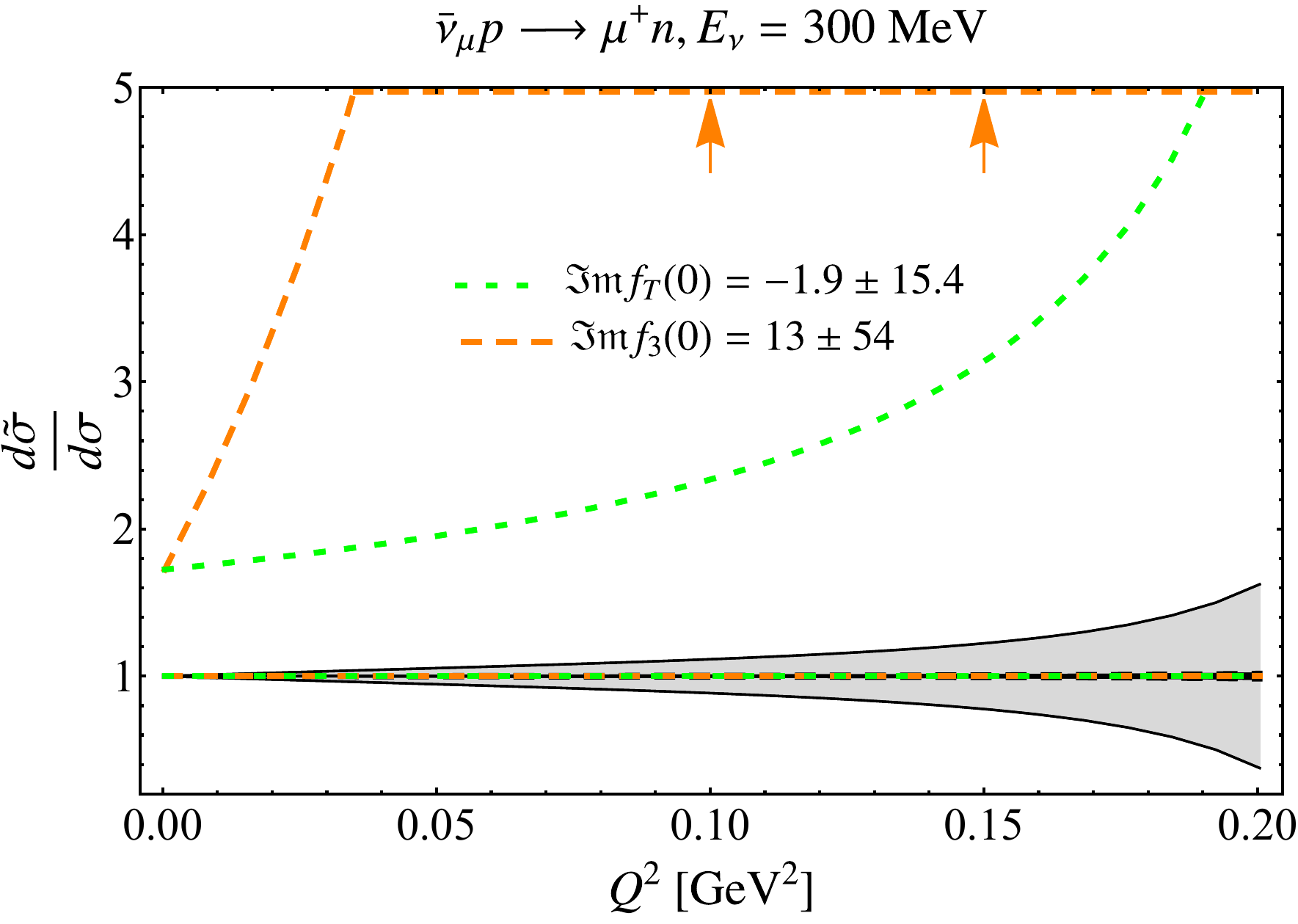}
\includegraphics[width=0.4\textwidth]{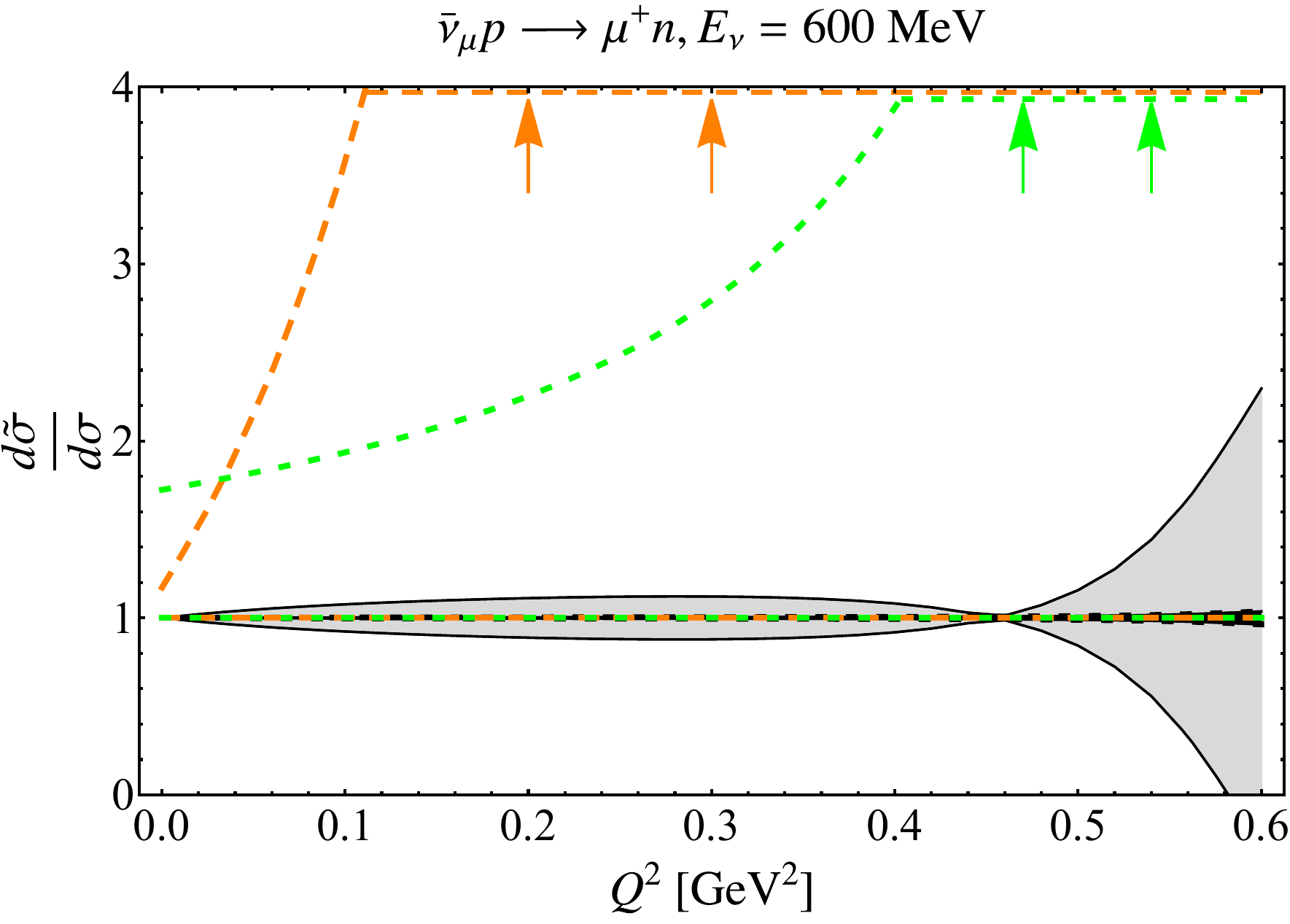}
\includegraphics[width=0.4\textwidth]{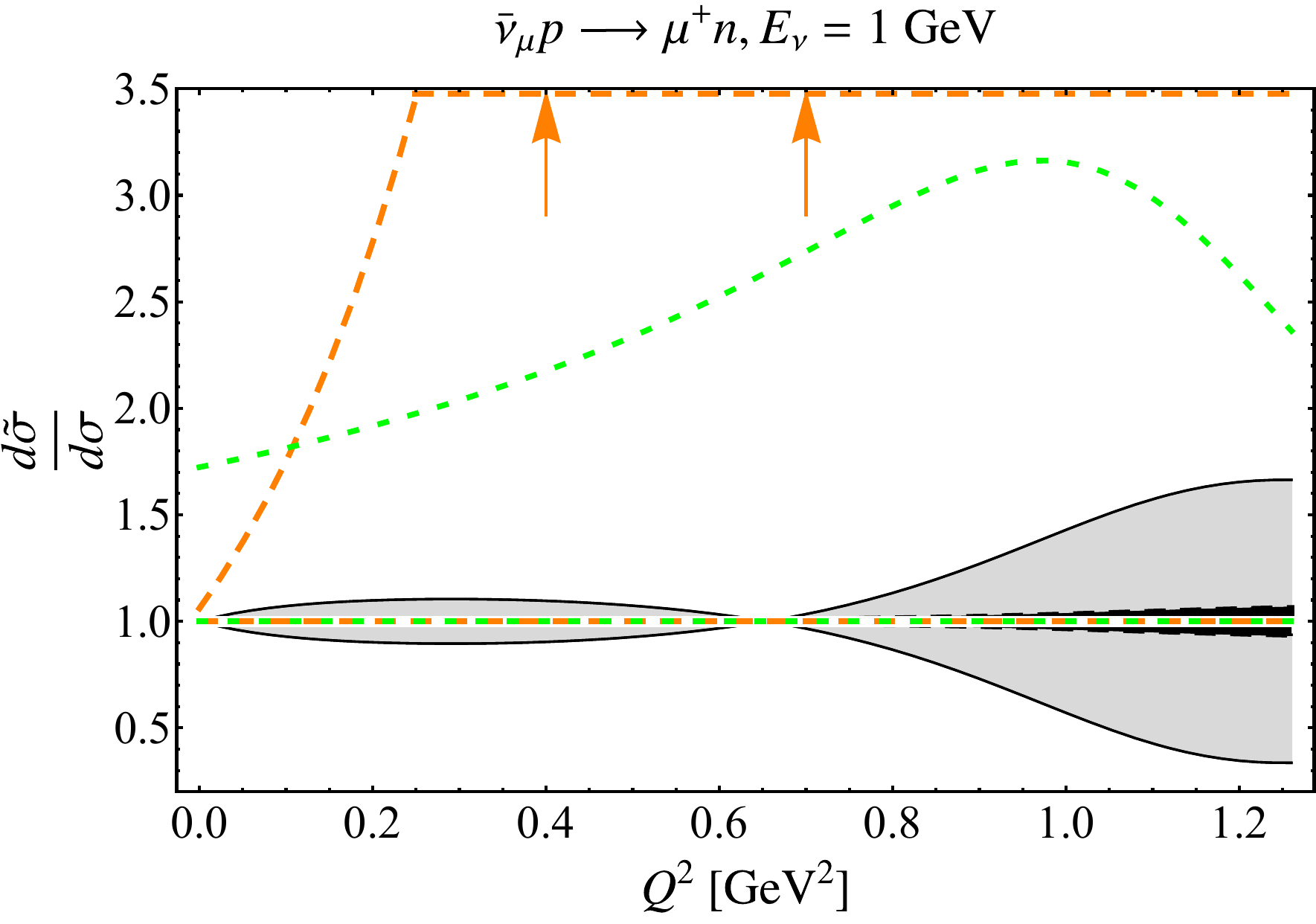}
\includegraphics[width=0.4\textwidth]{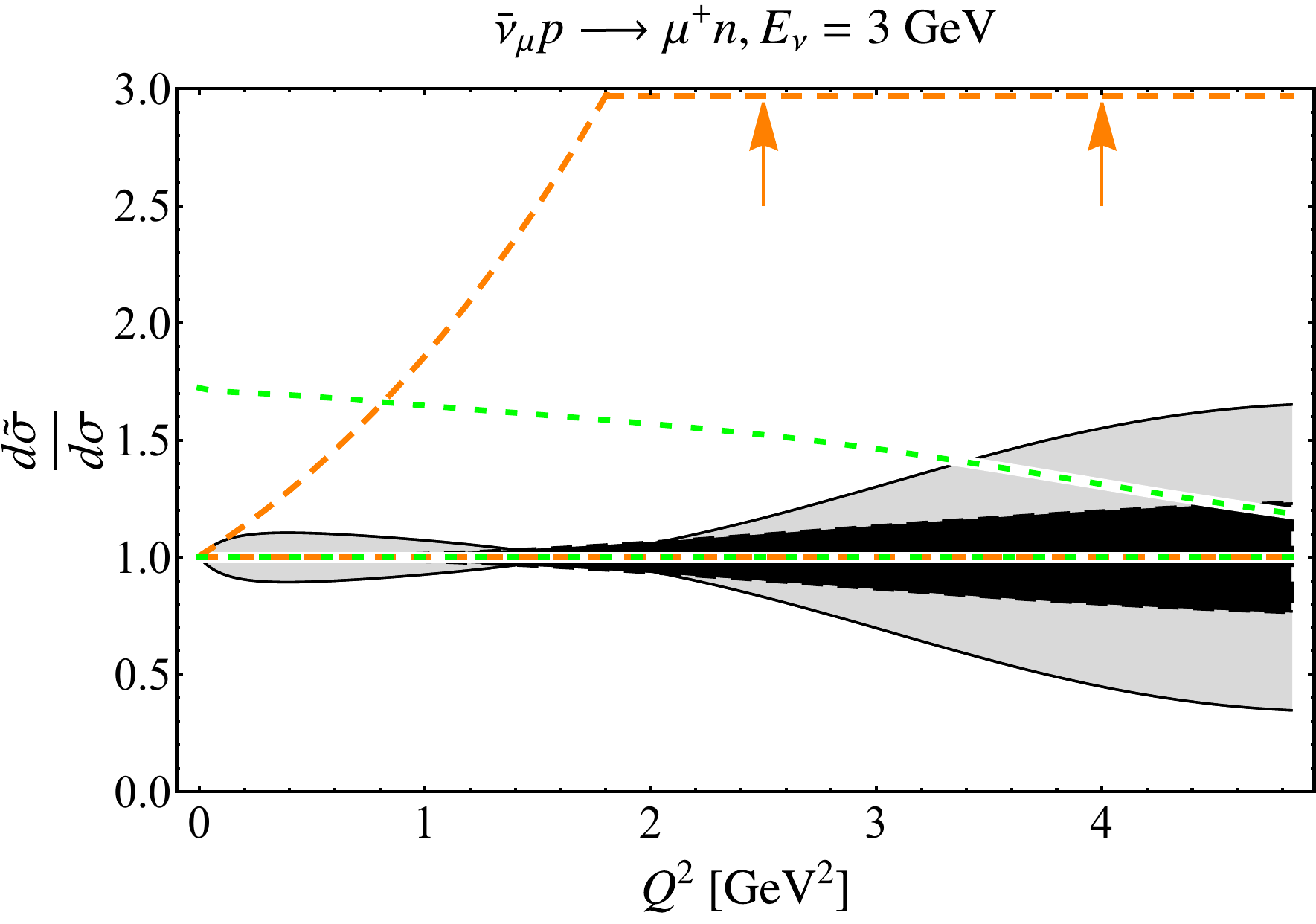}
\caption{Same as Fig.~\ref{fig:nu_xsec_ratio_SCFFIm} but for antineutrinos. \label{fig:antinu_xsec_ratio_SCFFIm}}
\end{figure}

\begin{figure}[H]
\centering
\includegraphics[width=0.4\textwidth]{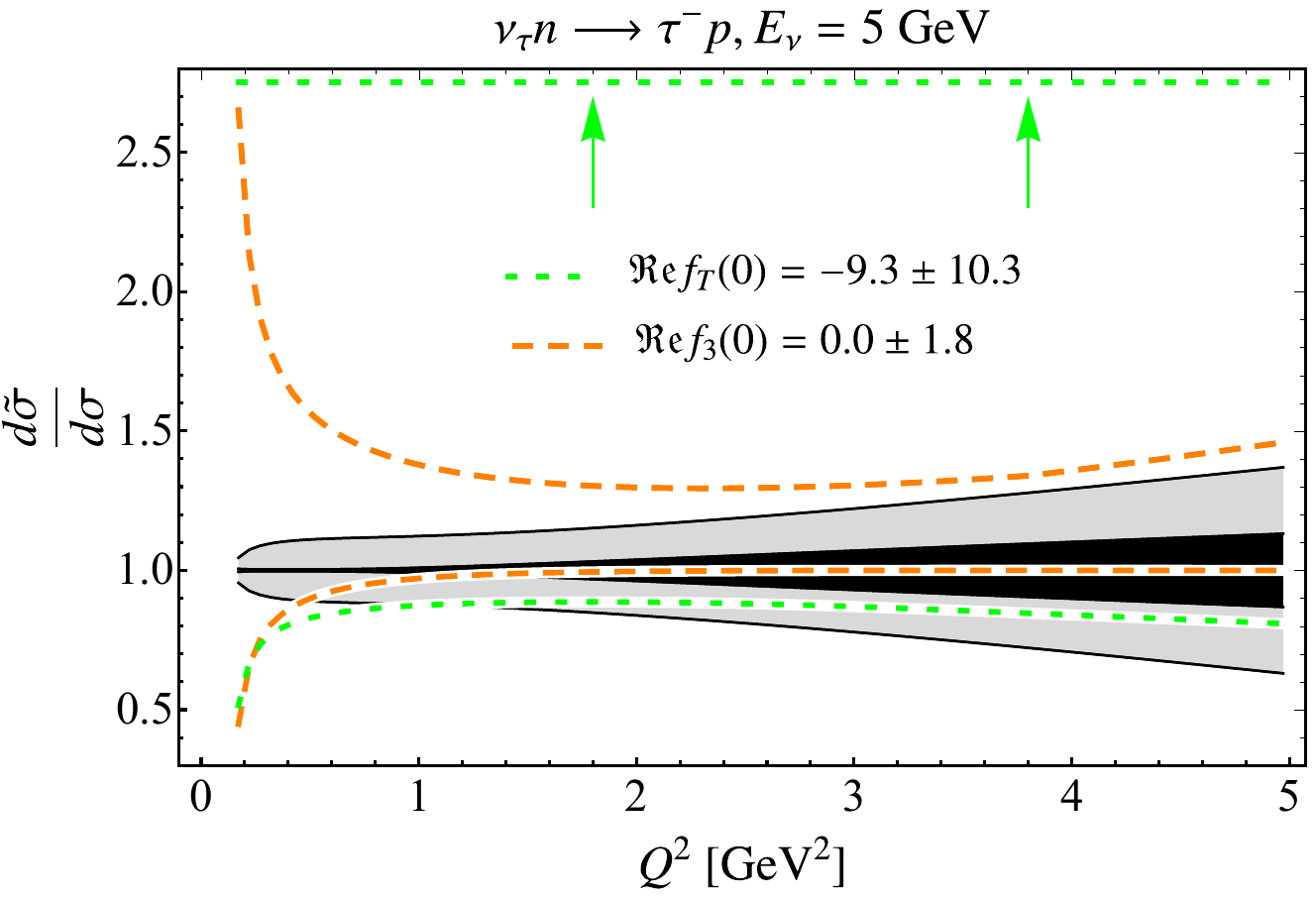}
\includegraphics[width=0.4\textwidth]{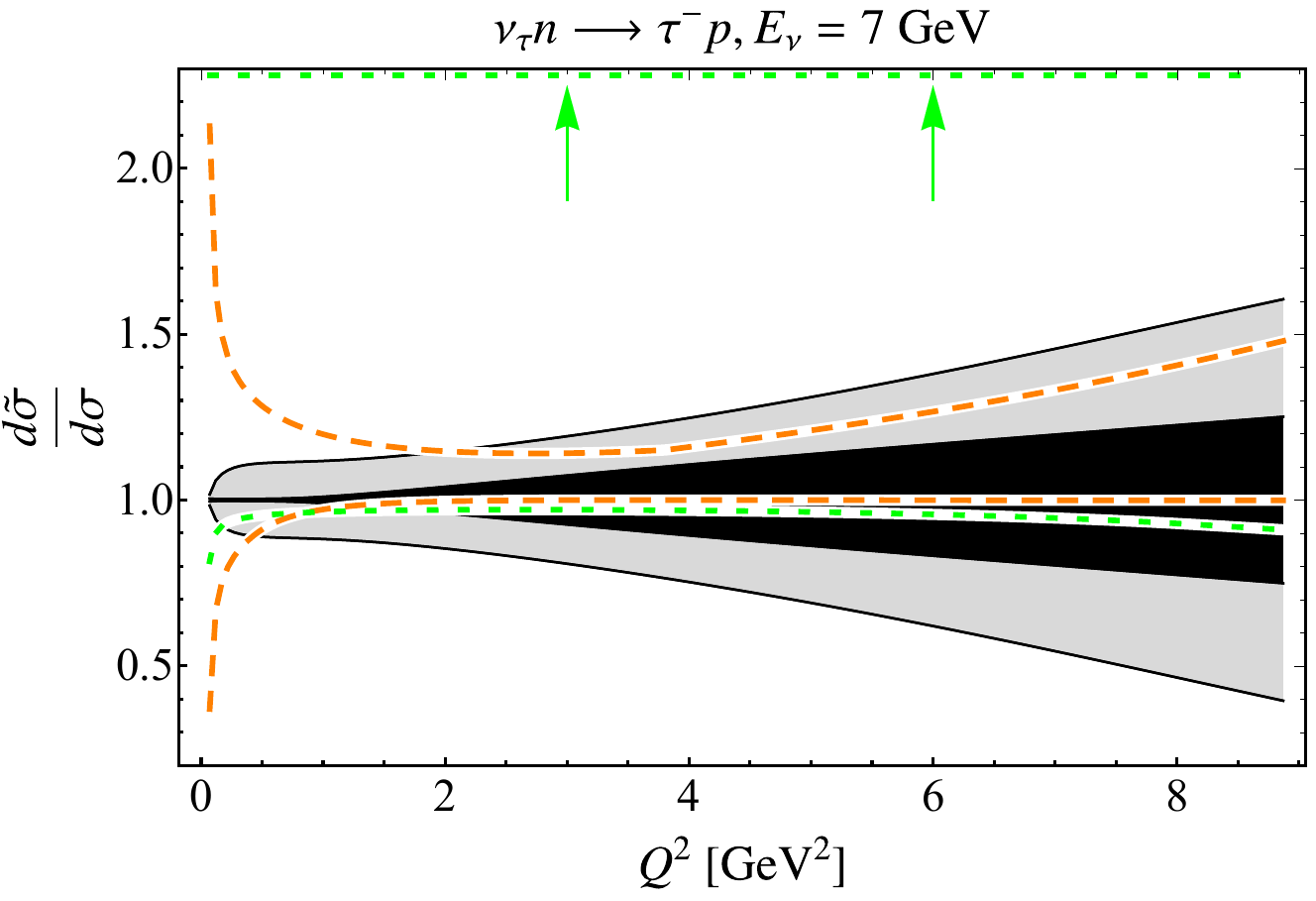}
\includegraphics[width=0.4\textwidth]{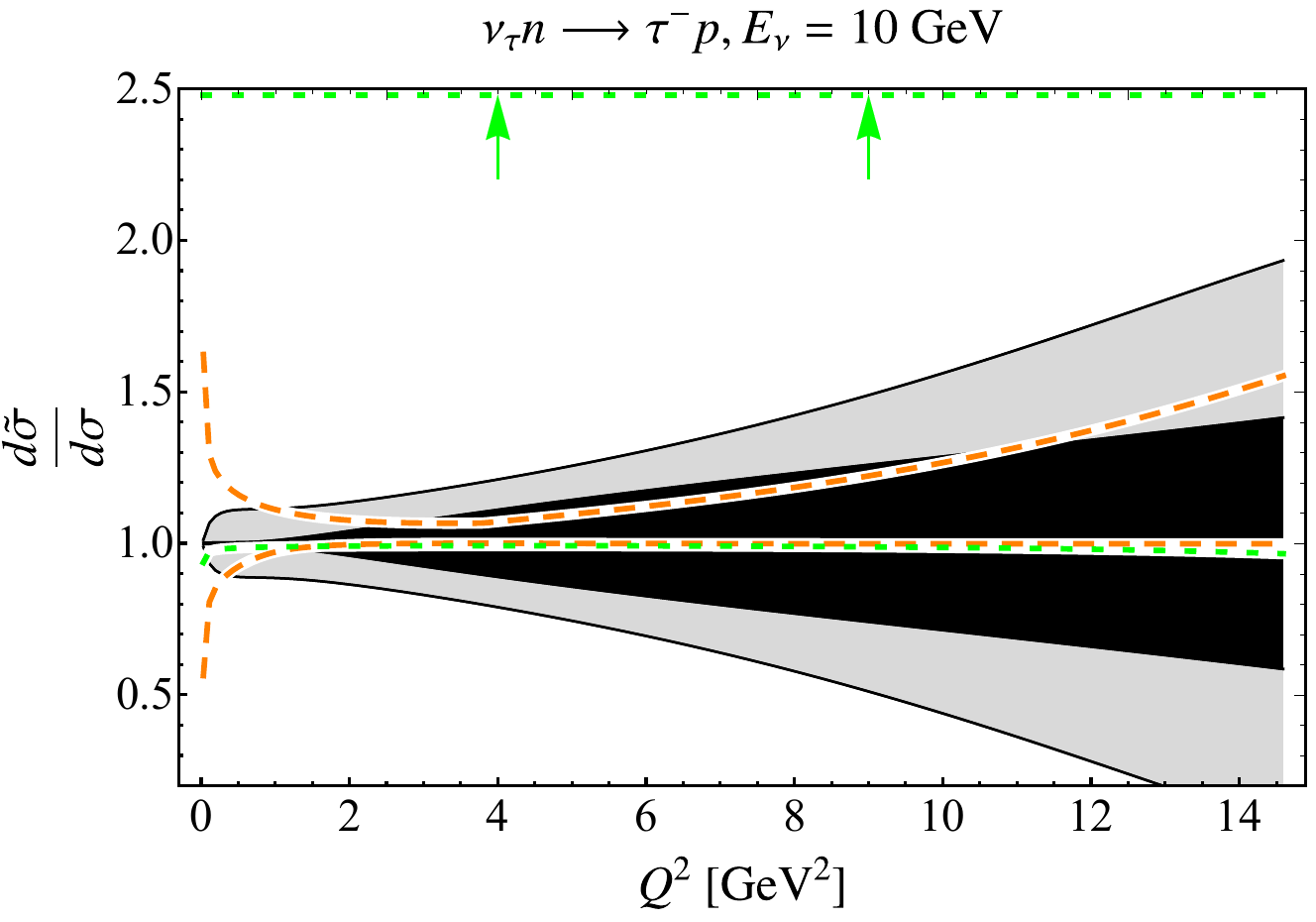}
\includegraphics[width=0.4\textwidth]{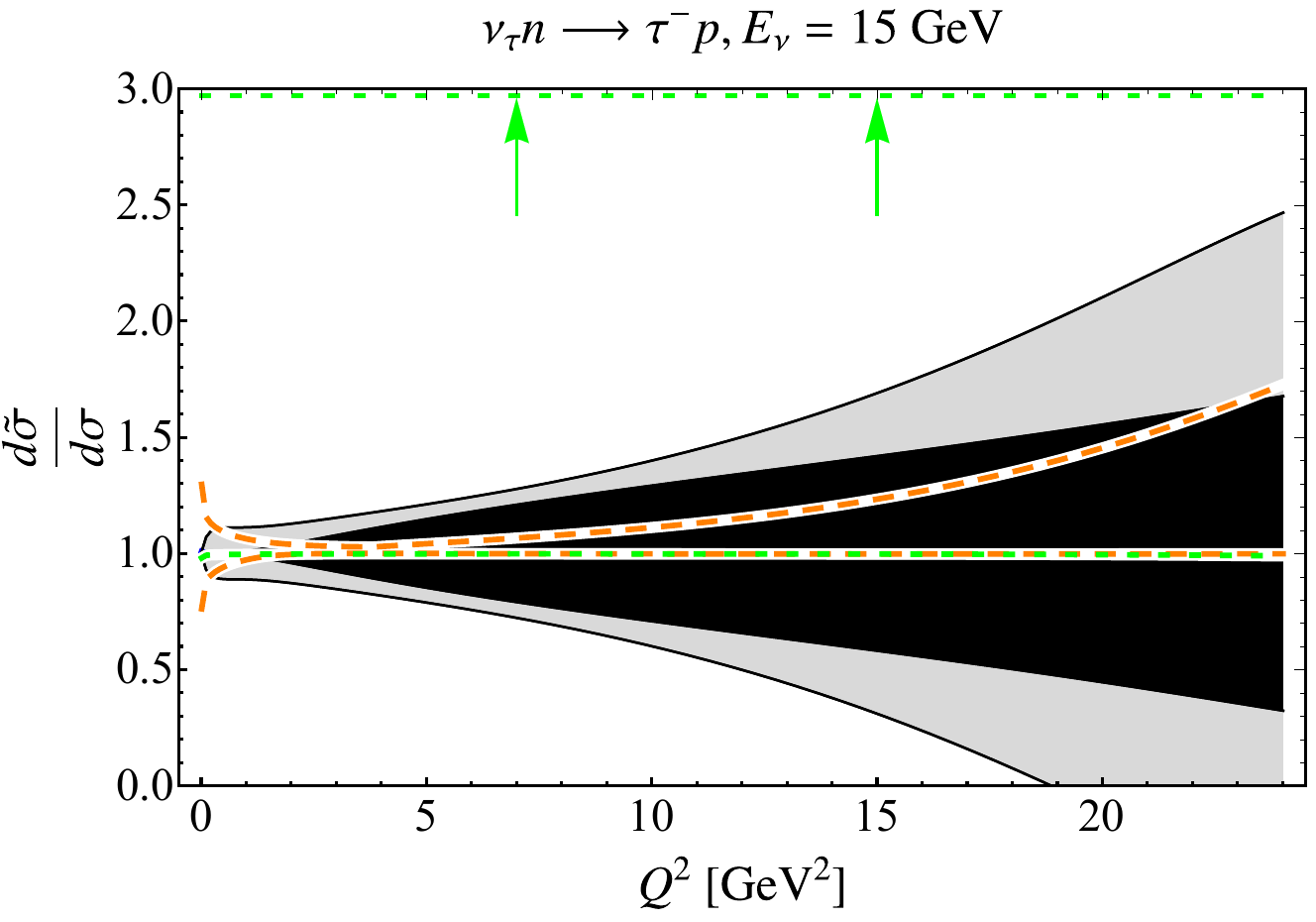}
\caption{Same as Fig.~\ref{fig:nu_xsec_ratio_SCFFRe} but for real amplitudes in the scattering of tau neutrinos at fixed beam energies $E_\nu = 5$~GeV, $7$~GeV, $10$~GeV, and $15$~GeV. \label{fig:nu_xsec_ratio_SCFFtauRe}}
\end{figure}

\begin{figure}[H]
\centering
\includegraphics[width=0.4\textwidth]{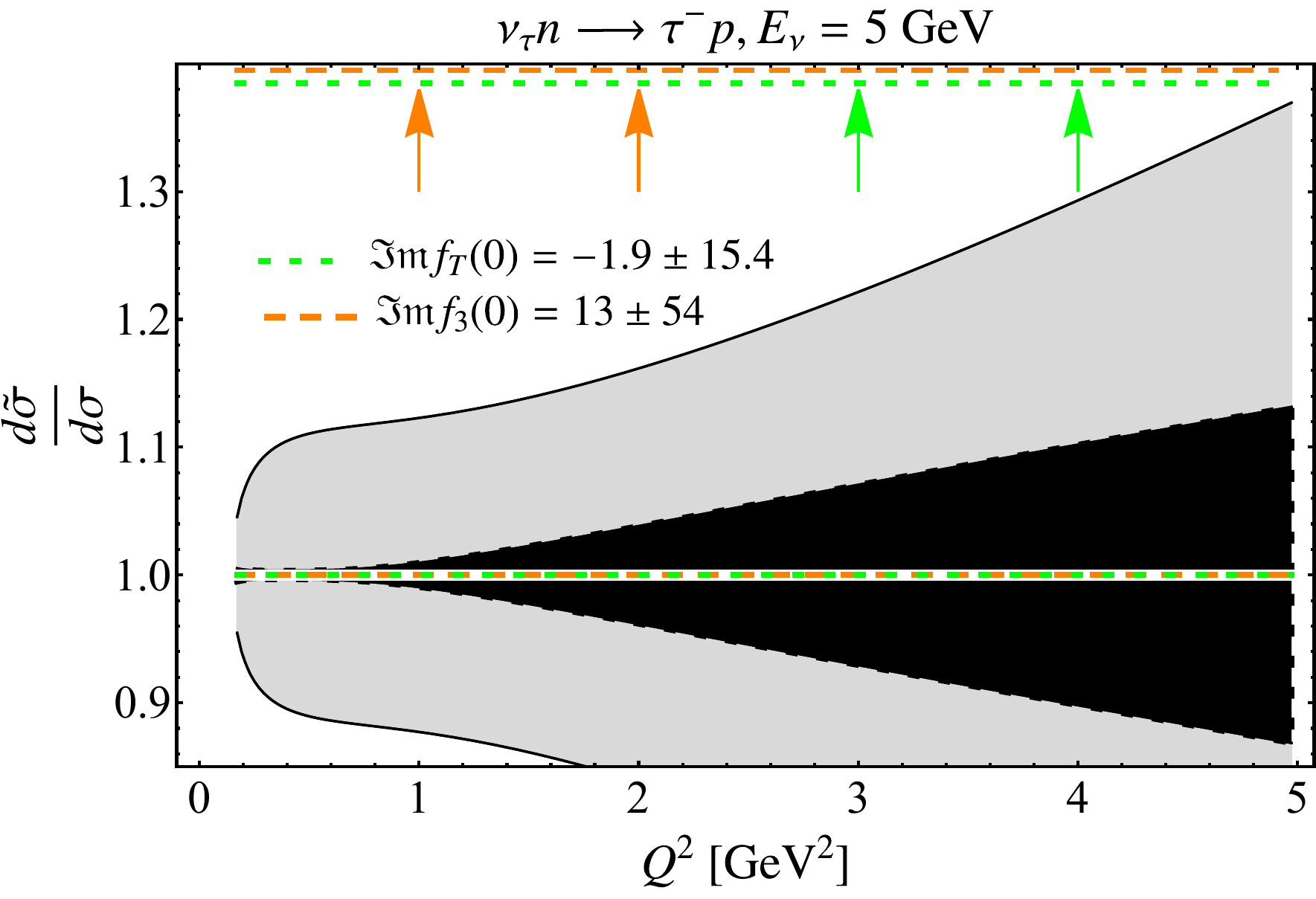}
\includegraphics[width=0.4\textwidth]{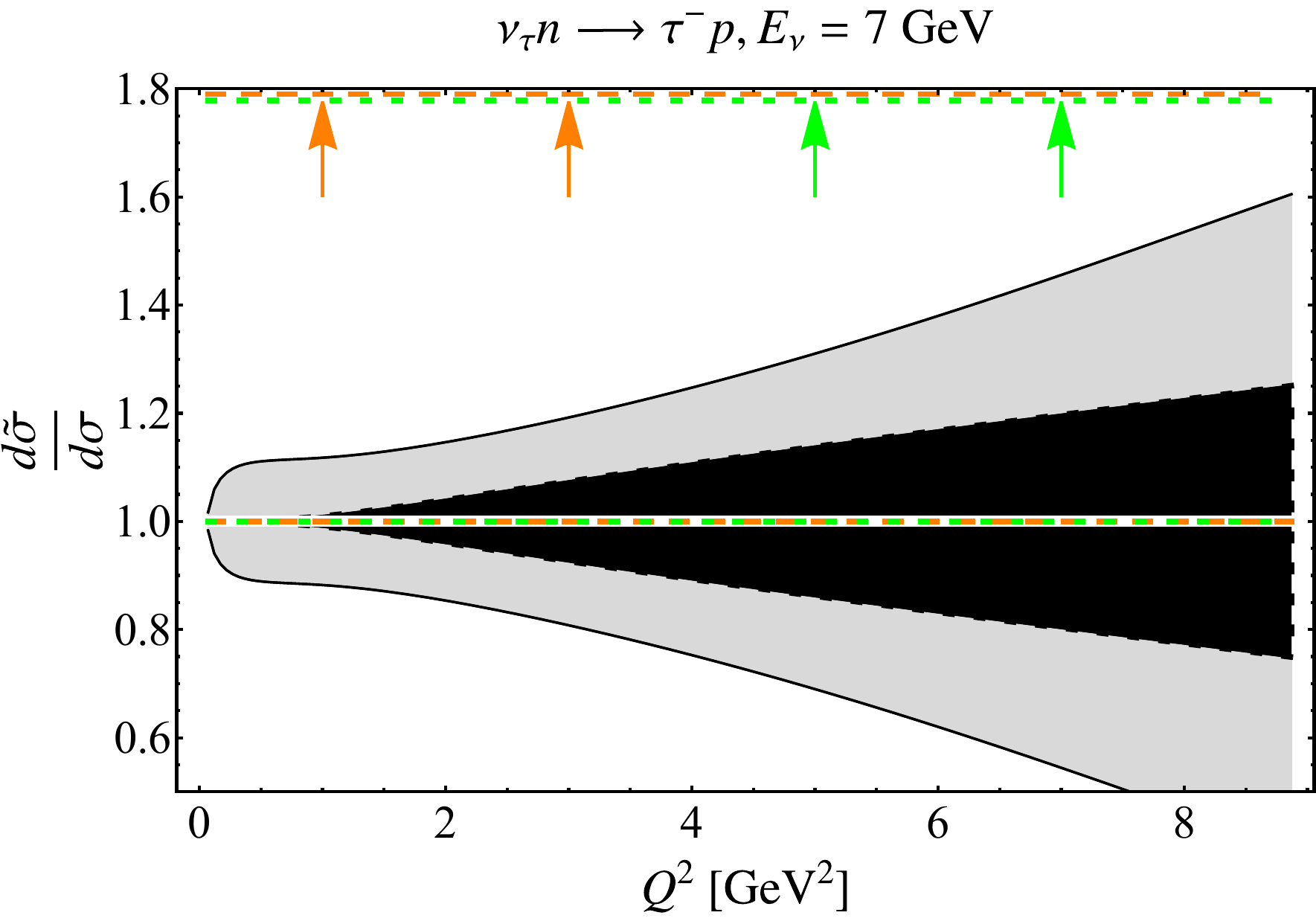}
\includegraphics[width=0.4\textwidth]{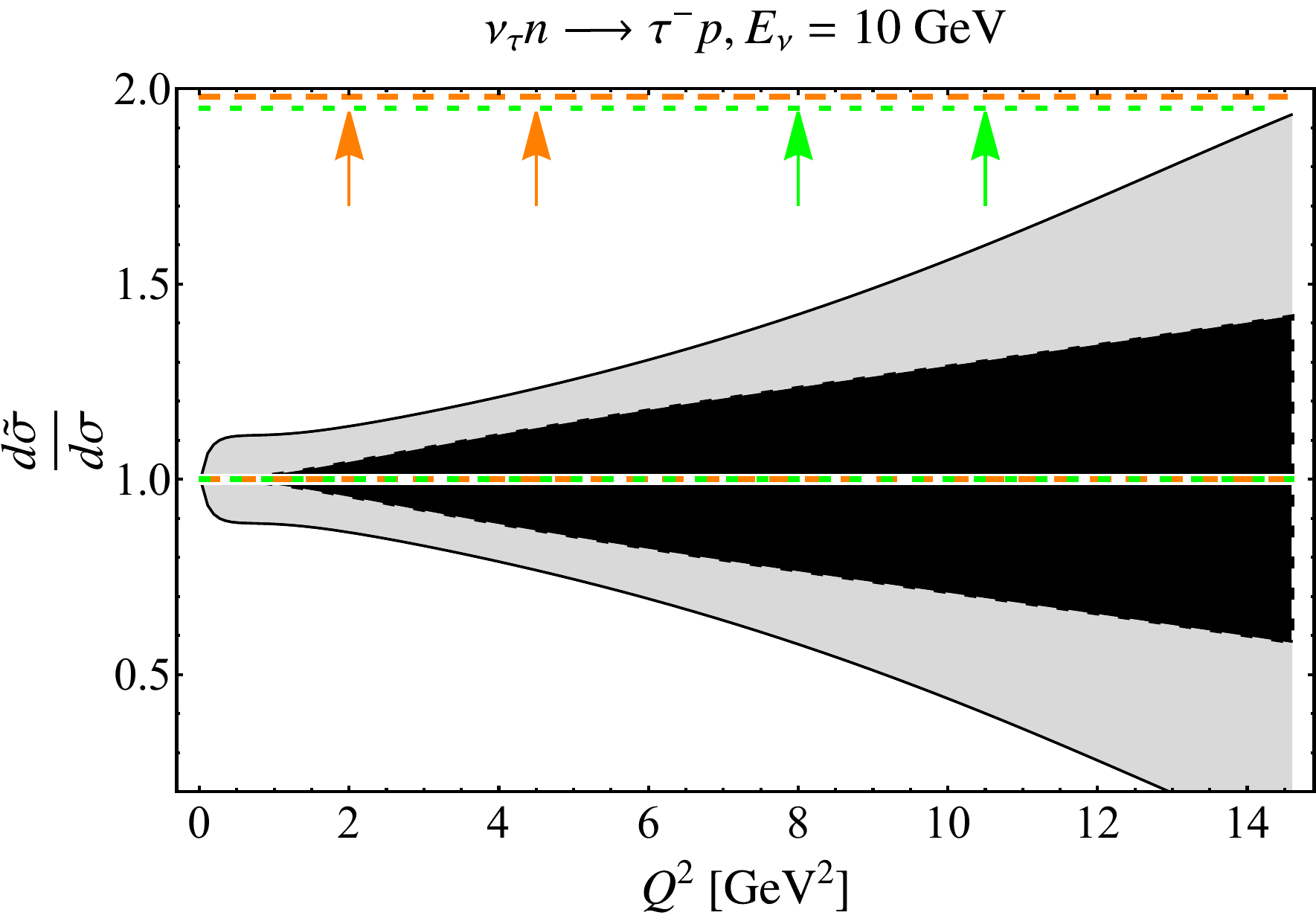}
\includegraphics[width=0.4\textwidth]{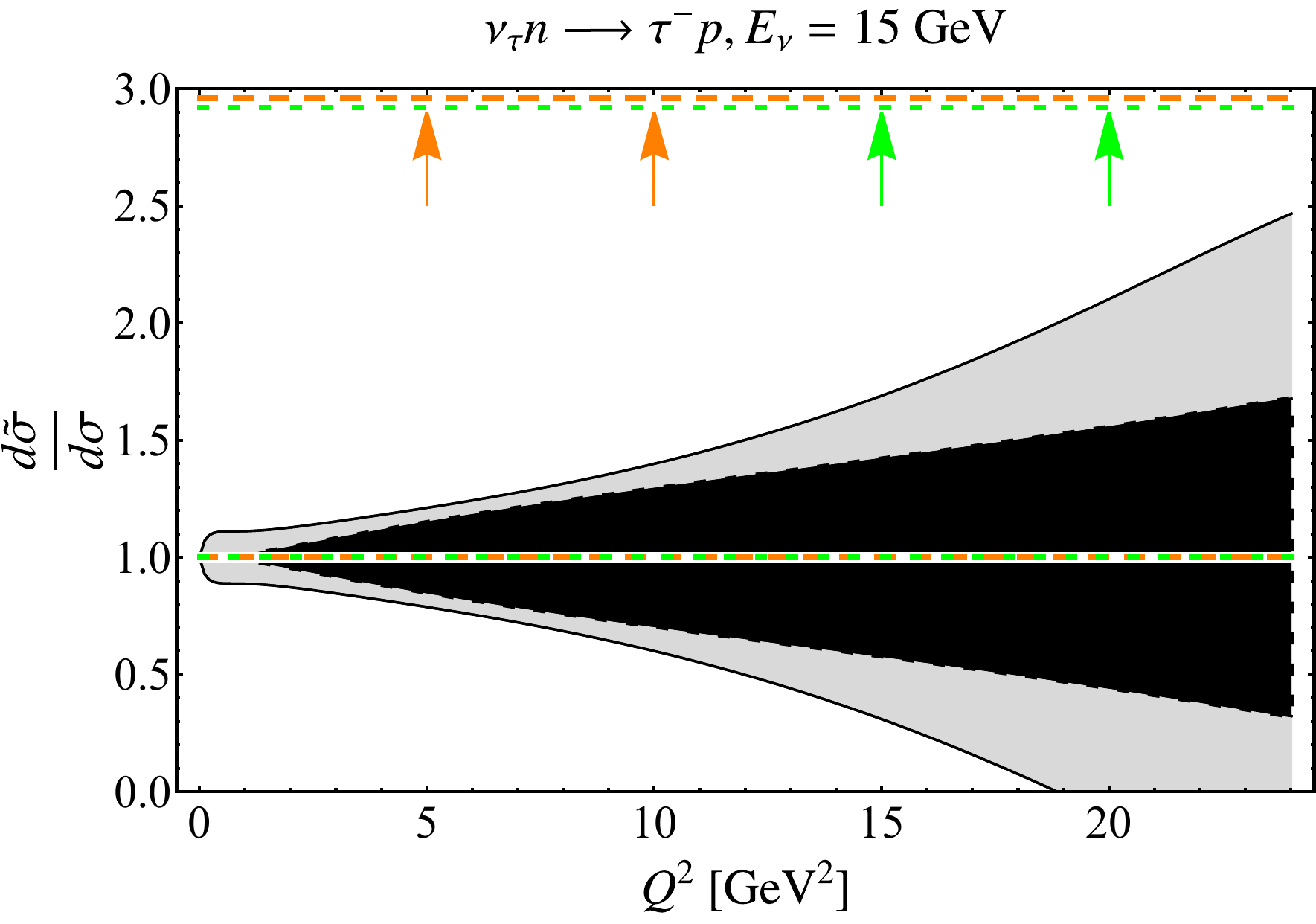}
\caption{Same as Fig.~\ref{fig:nu_xsec_ratio_SCFFtauRe} but for imaginary amplitudes. \label{fig:nu_xsec_ratio_SCFFtauIm}}
\end{figure}

\begin{figure}[H]
\centering
\includegraphics[width=0.4\textwidth]{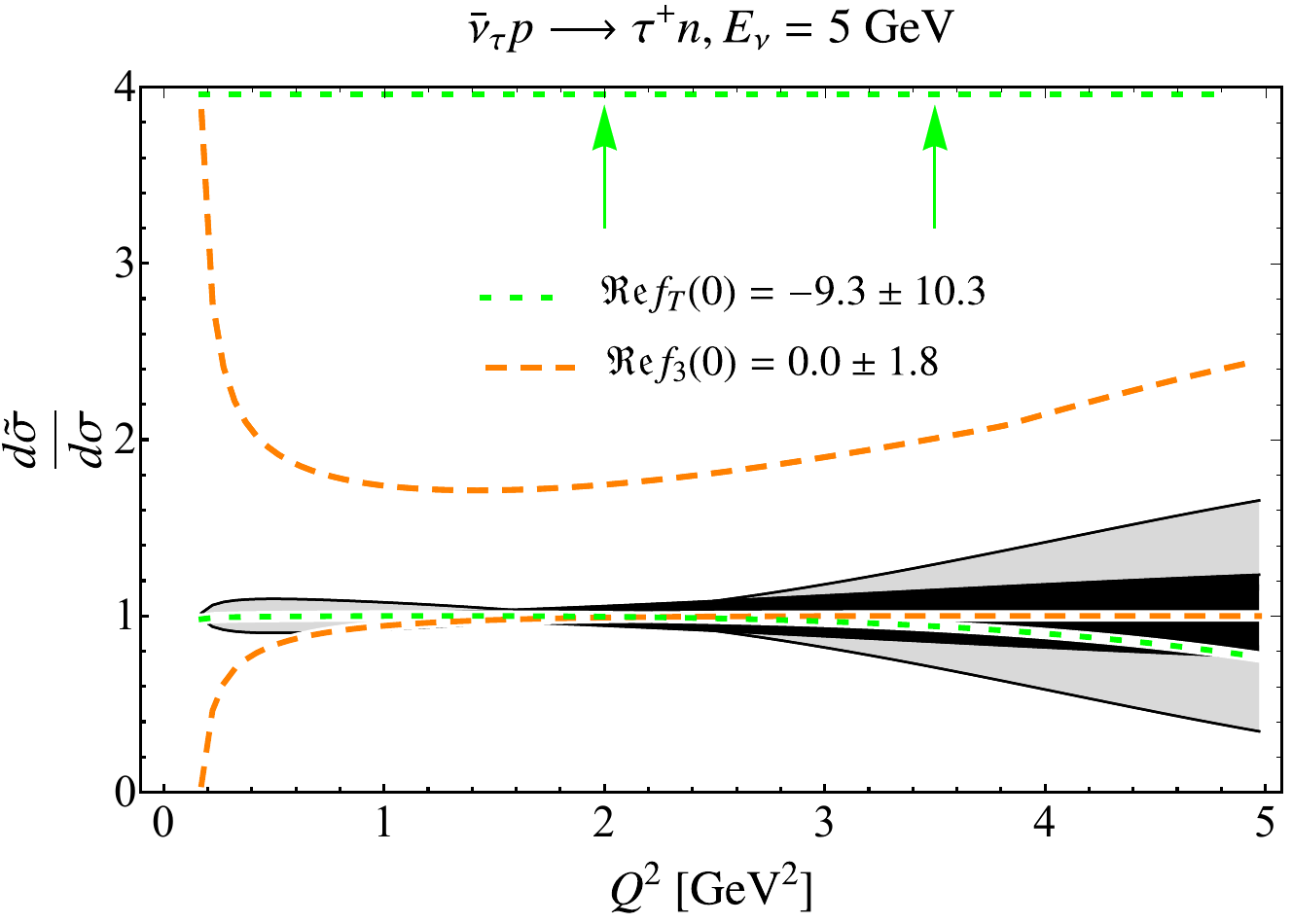}
\includegraphics[width=0.4\textwidth]{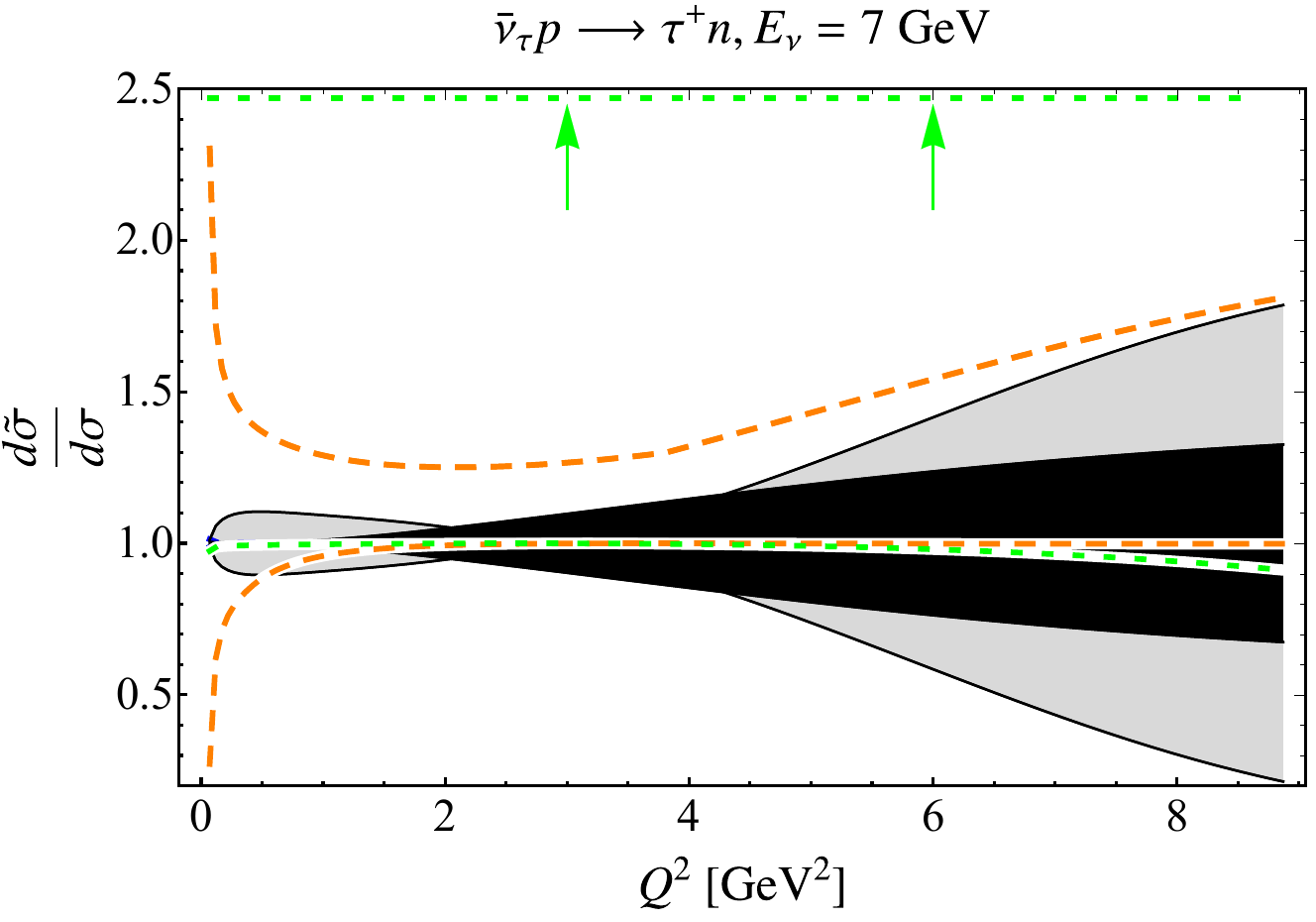}
\includegraphics[width=0.4\textwidth]{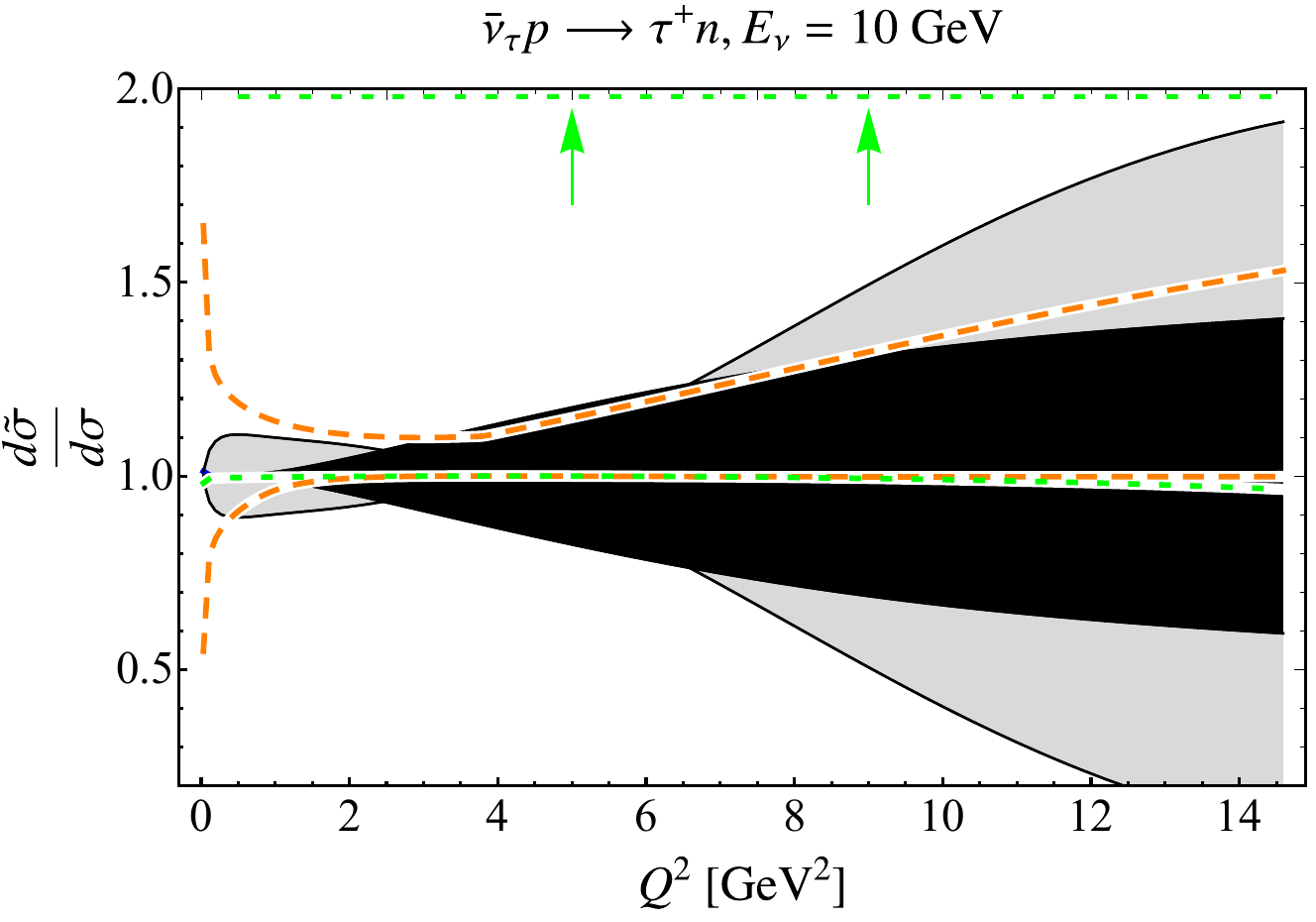}
\includegraphics[width=0.4\textwidth]{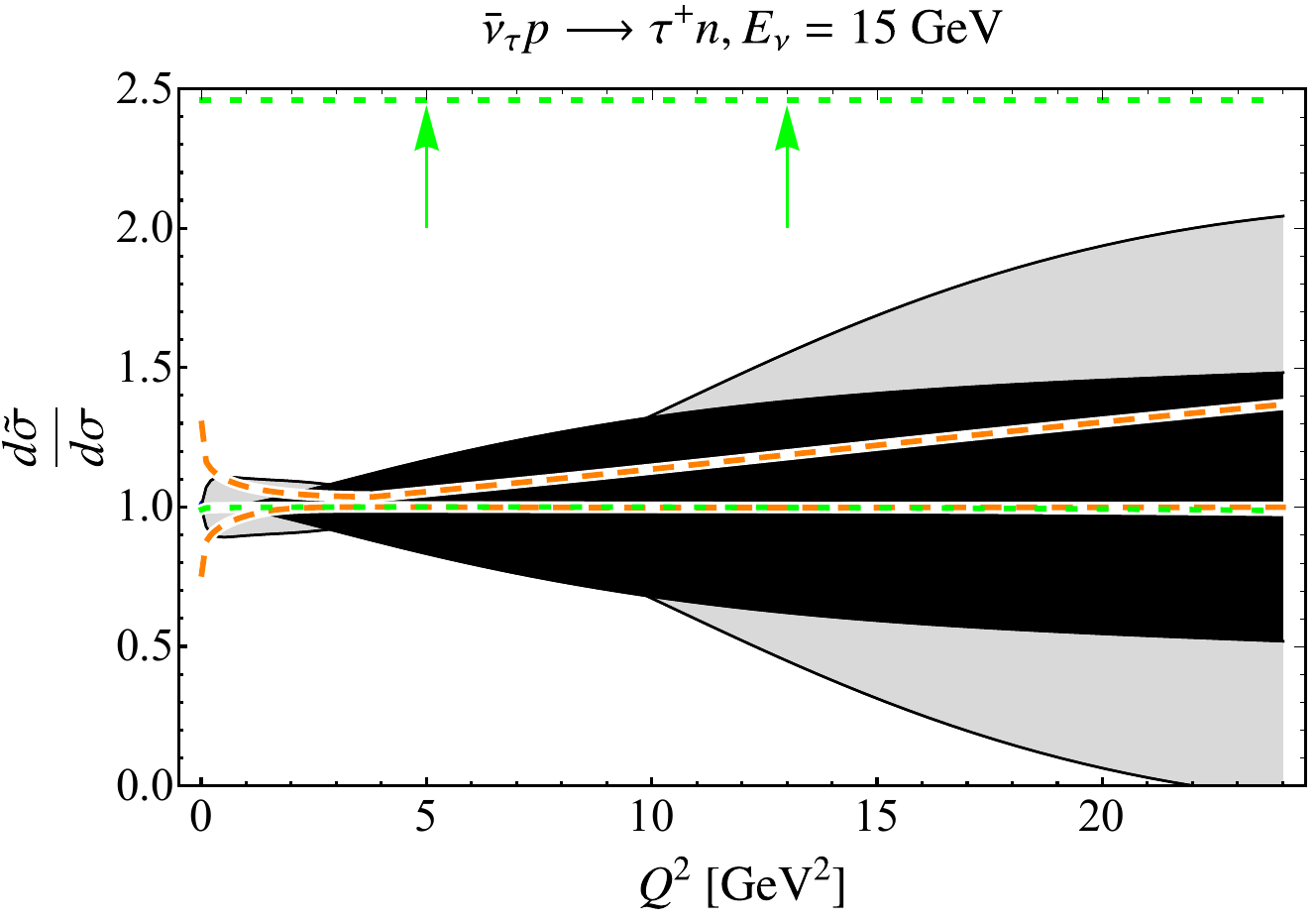}
\caption{Same as Fig.~\ref{fig:nu_xsec_ratio_SCFFtauRe} but for antineutrinos. \label{fig:antinu_xsec_ratio_SCFFFtauRe}}
\end{figure}

\begin{figure}[H]
\centering
\includegraphics[width=0.4\textwidth]{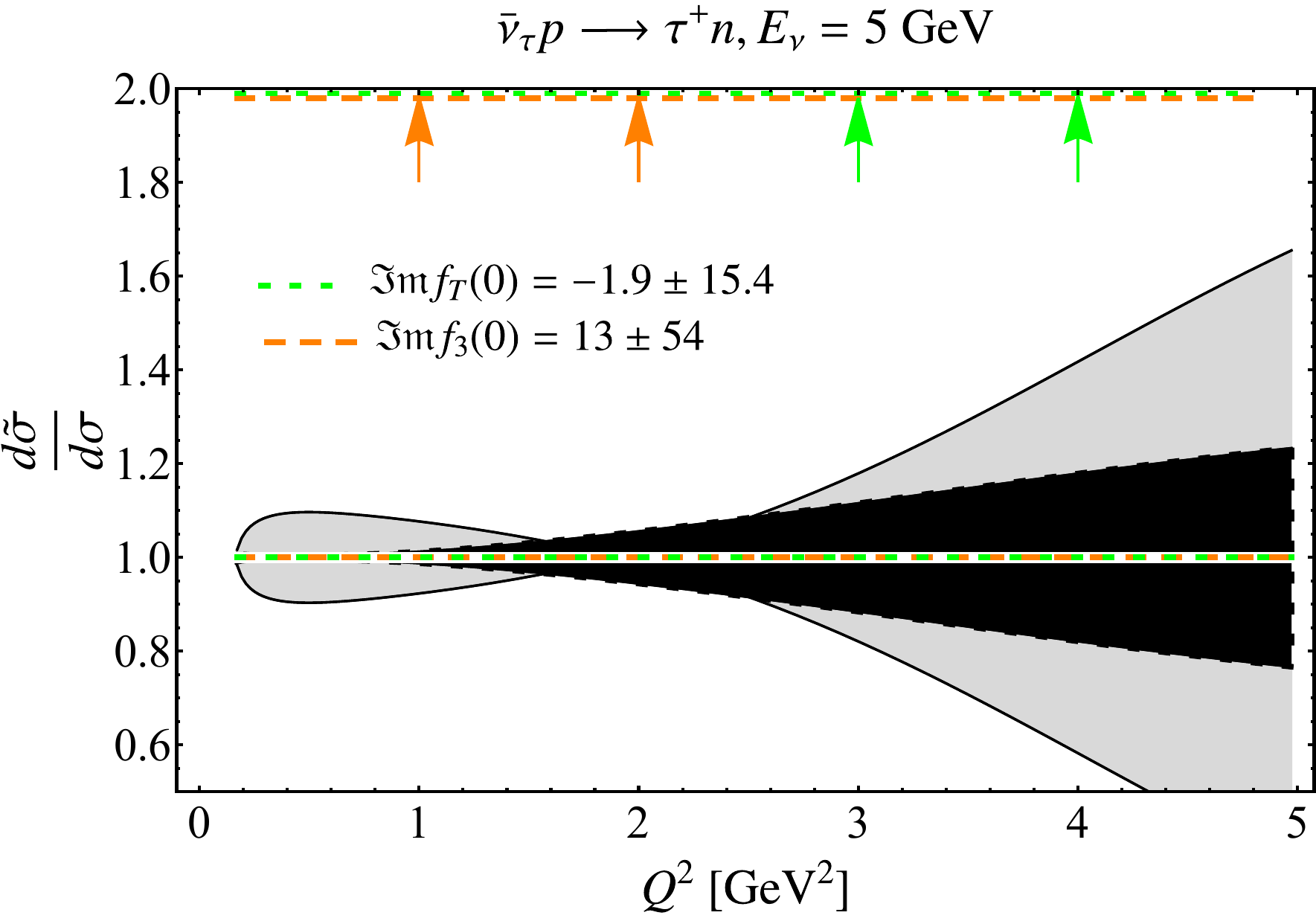}
\includegraphics[width=0.4\textwidth]{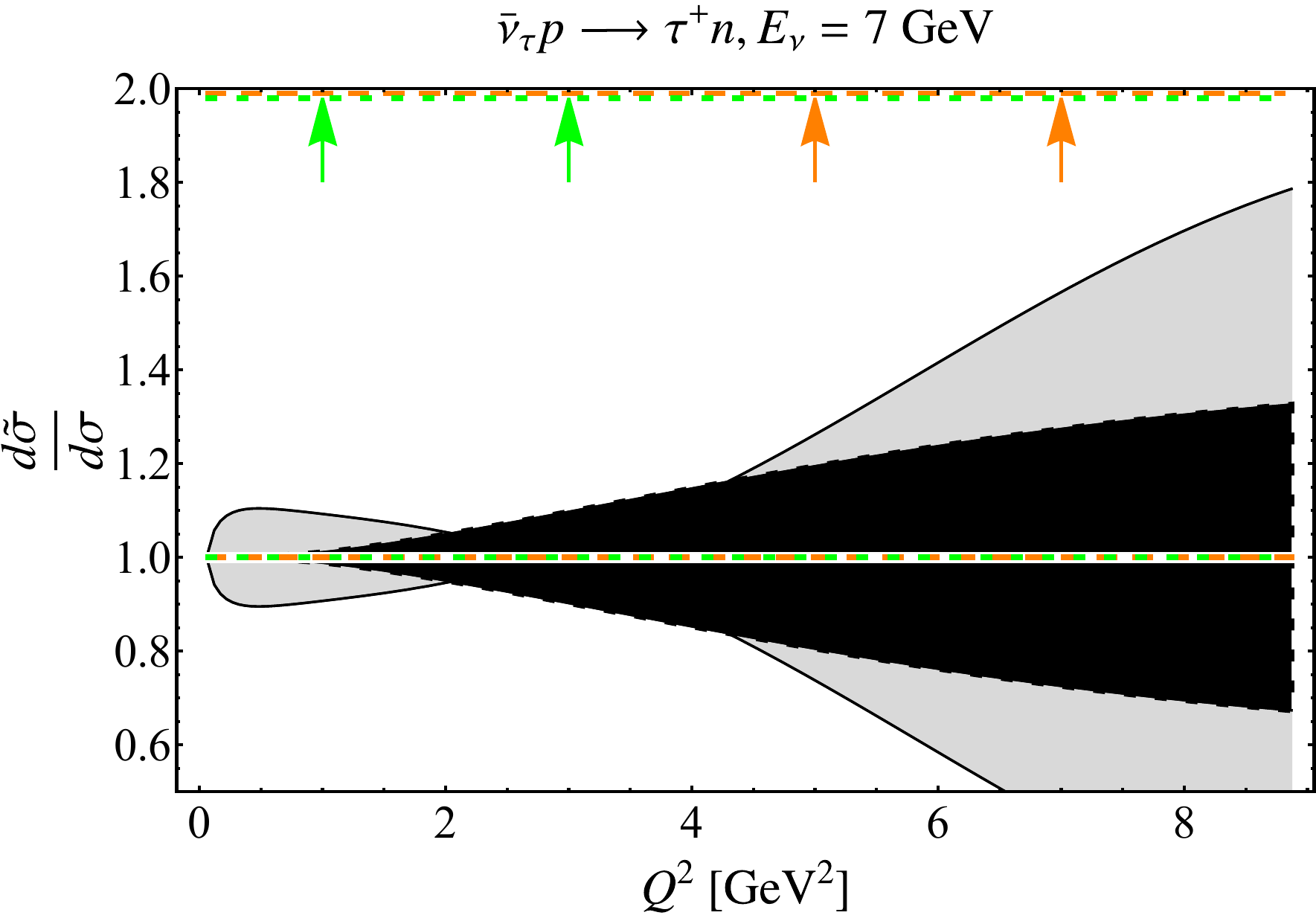}
\includegraphics[width=0.4\textwidth]{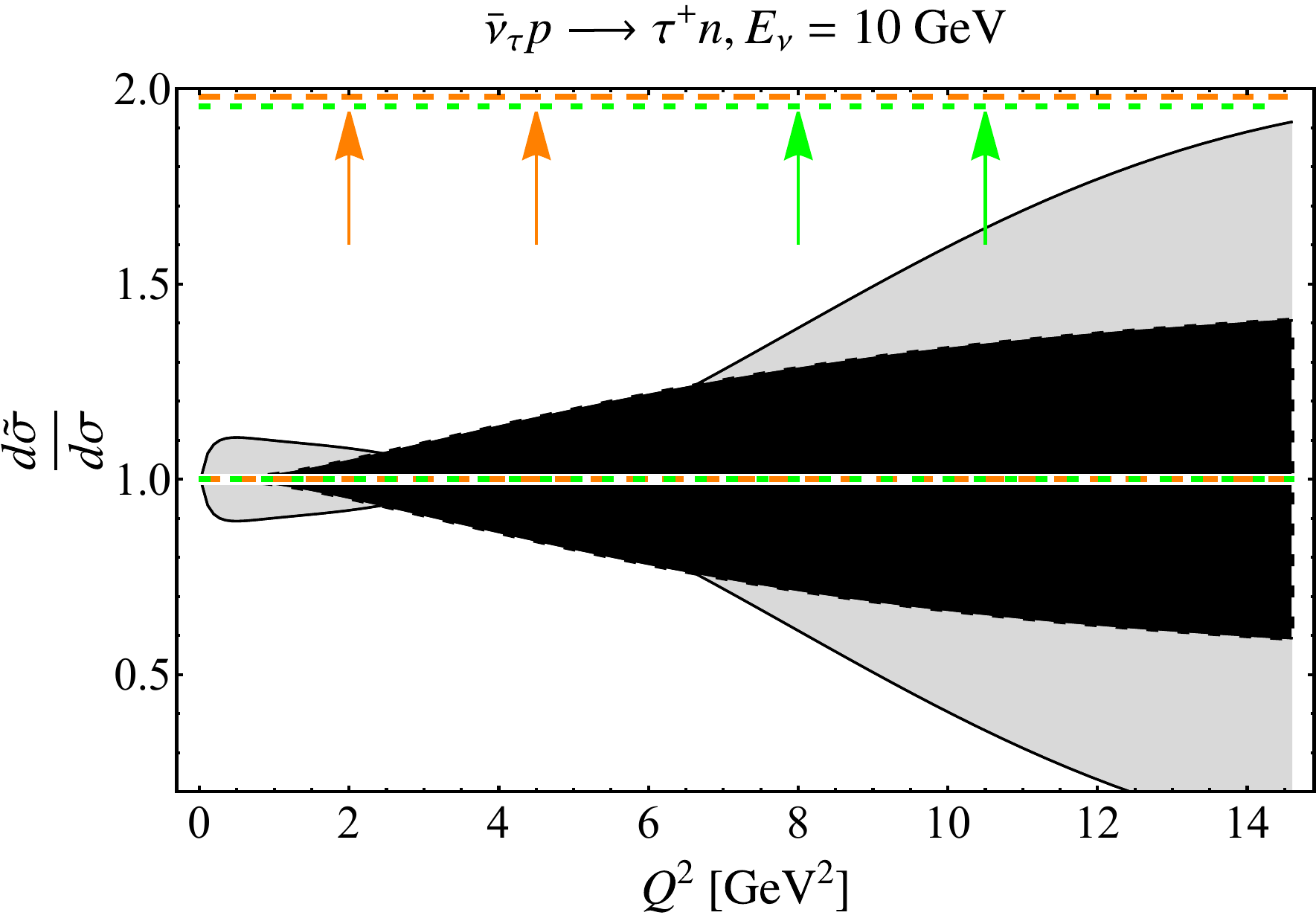}
\includegraphics[width=0.4\textwidth]{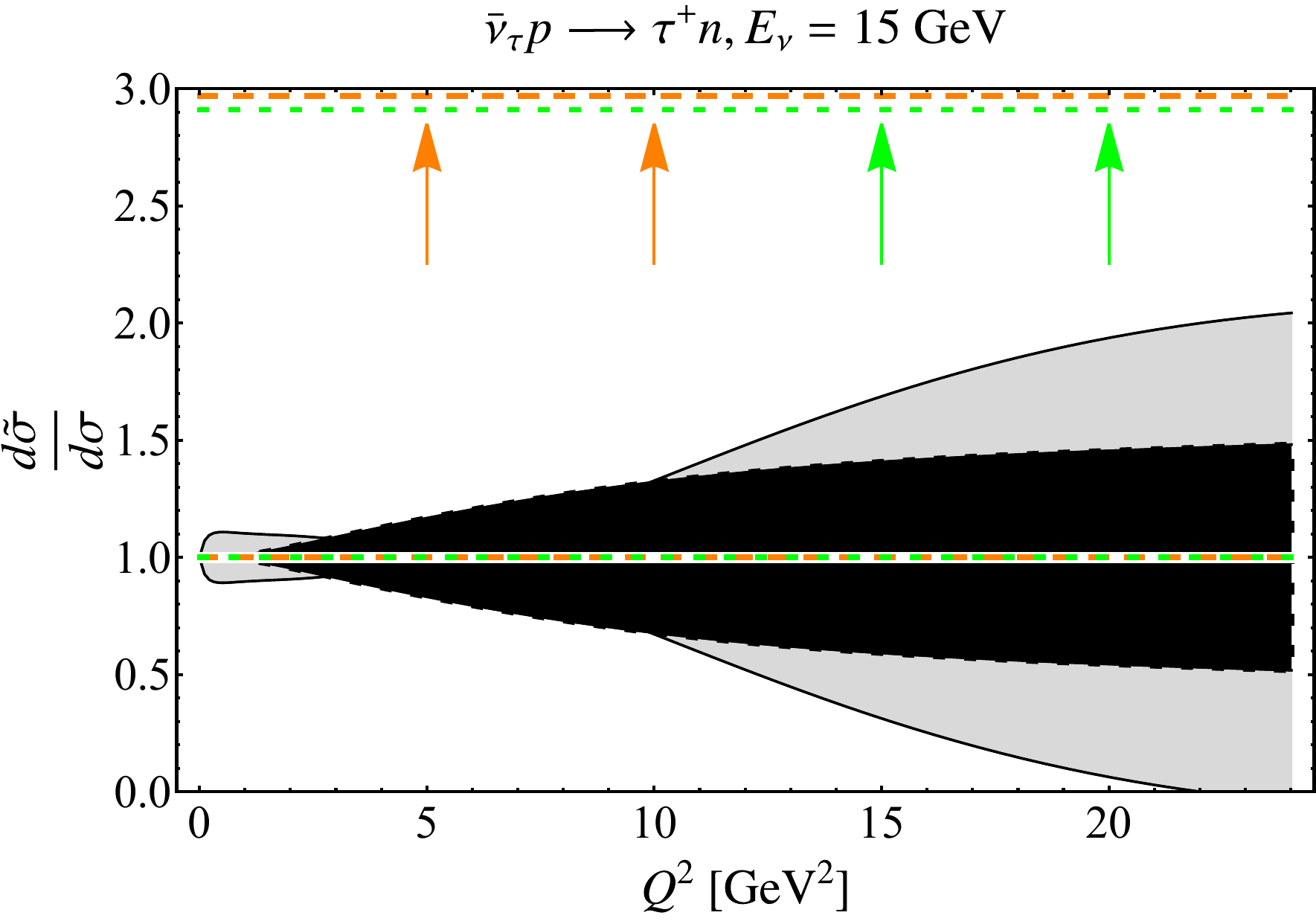}
\caption{Same as Fig.~\ref{fig:nu_xsec_ratio_SCFFtauIm} but for antineutrinos. \label{fig:antinu_xsec_ratio_SCFFFtauIm}}
\end{figure}

\newpage

\subsection{Polarization asymmetries, muon (anti)neutrino \label{app:pol_plots}}

In this Section, we present all independent single-spin asymmetries for muon neutrinos and antineutrinos, with one extra real- or imaginary-valued amplitude $f_i (\nu, Q^2)  =  [\mathfrak{Re} f_i \left( 0 \right) + i \mathfrak{Im} f_i \left( 0 \right)]/\left( 1 + \frac{Q^2}{\Lambda^2} \right)^2$, for illustrative neutrino energies $E_\nu = 300$~MeV, $600$~MeV, $1$~GeV, and $3$~GeV, and $\Lambda = 1$~GeV. We vary the amplitude normalizations within the ranges from Table~\ref{tab:beta}, and compare to the uncertainty from vector and axial-vector form factors from Sec.~\ref{sec:observables}. 

\begin{figure}[H]
\centering
\includegraphics[width=0.4\textwidth]{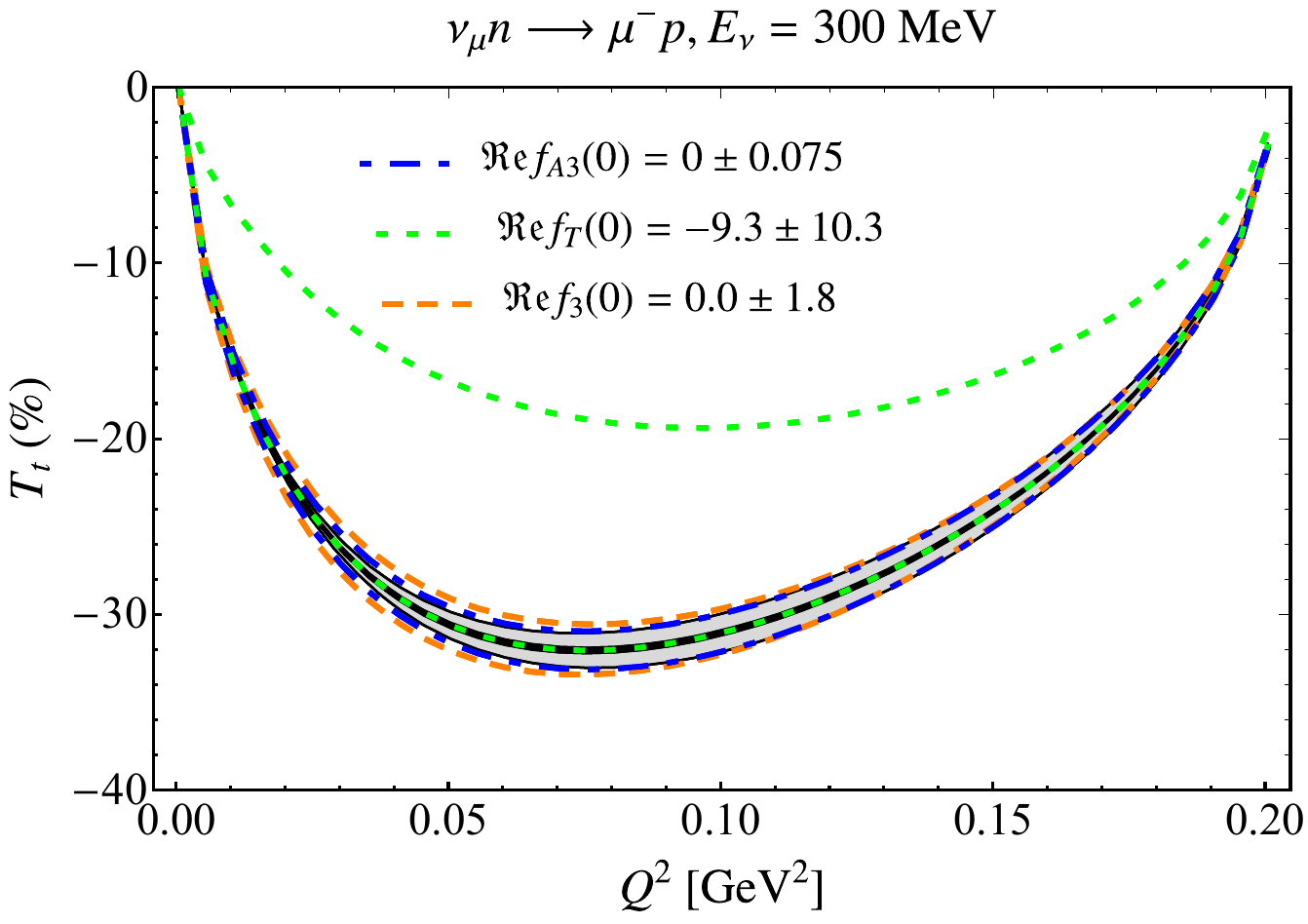}
\includegraphics[width=0.4\textwidth]{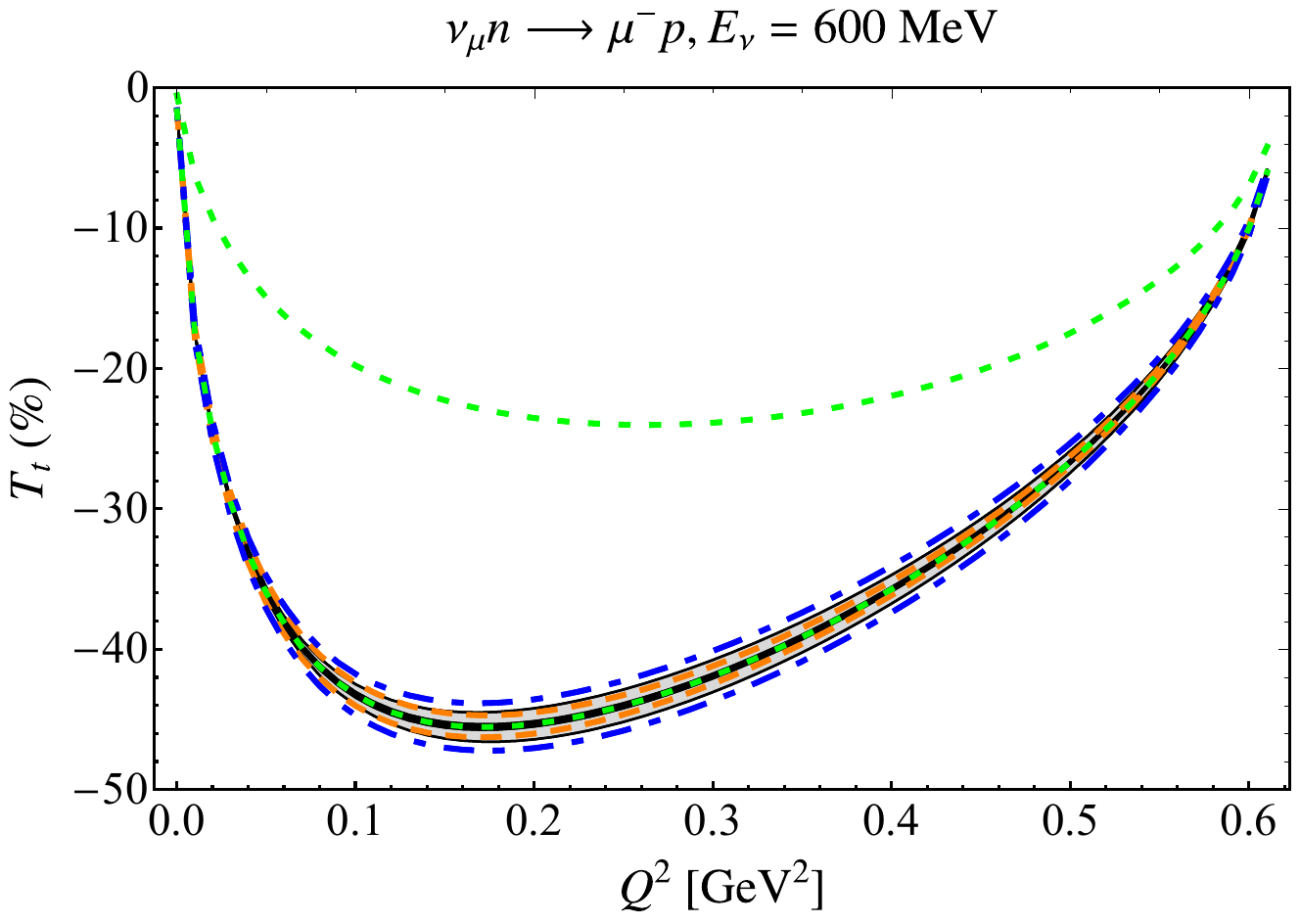}
\includegraphics[width=0.4\textwidth]{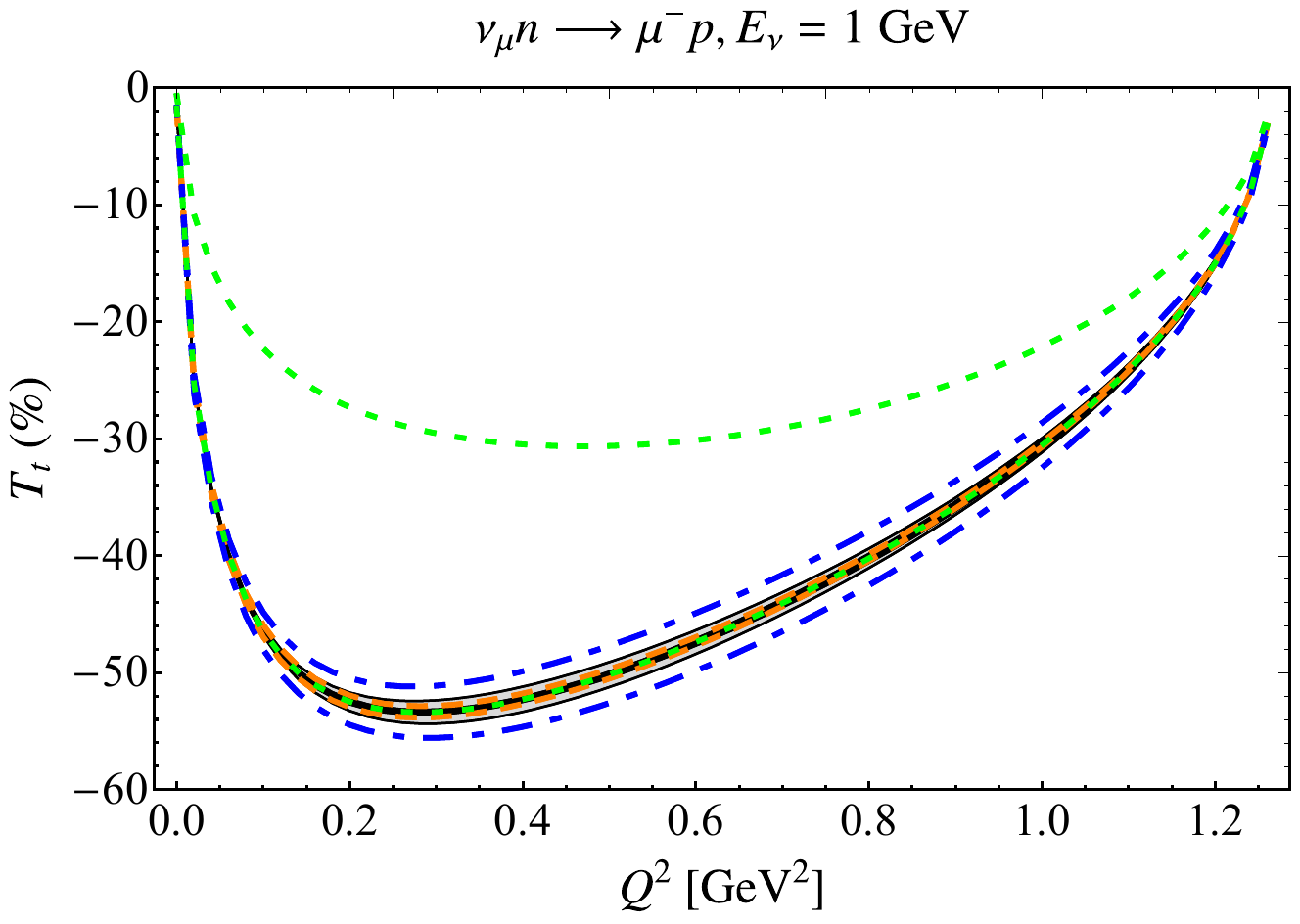}
\includegraphics[width=0.4\textwidth]{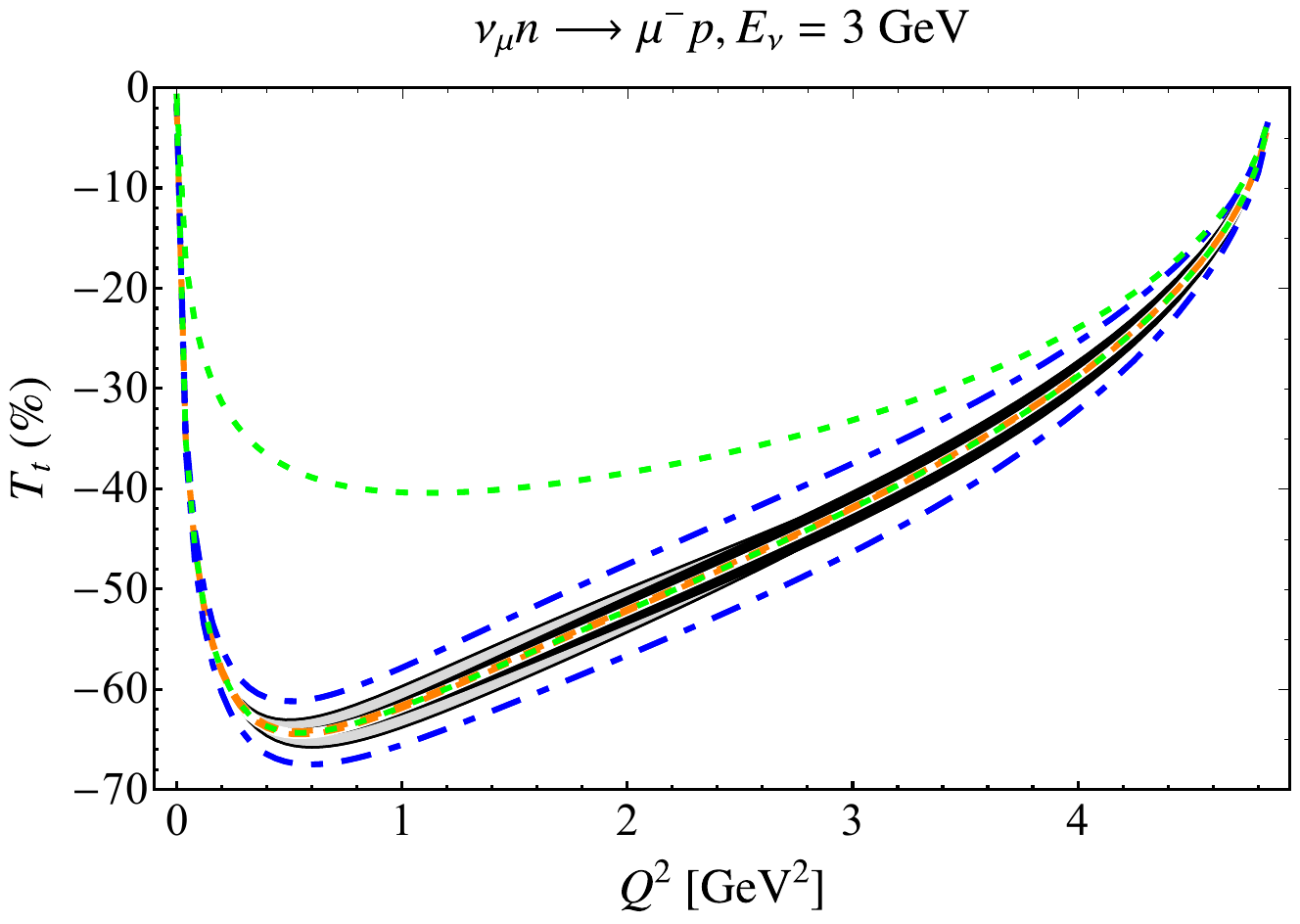}
\caption{Transverse polarization observable, $T_t$, with one extra real-valued amplitude, compared to the tree-level result at fixed muon neutrino energies $E_\nu = 300$~MeV, $600$~MeV, $1$~GeV, and $3$~GeV. The dark black and light gray bands correspond to vector and axial-vector uncertainty, respectively. Orange dashed, green dotted, and blue dashed-dotted lines represent allowed regions for $f_3$, $f_T$, and $\fAt$, respectively, as described in the text. \label{fig:nu_Tt_SCFF}}
\end{figure}

\begin{figure}[H]
\centering
\includegraphics[width=0.4\textwidth]{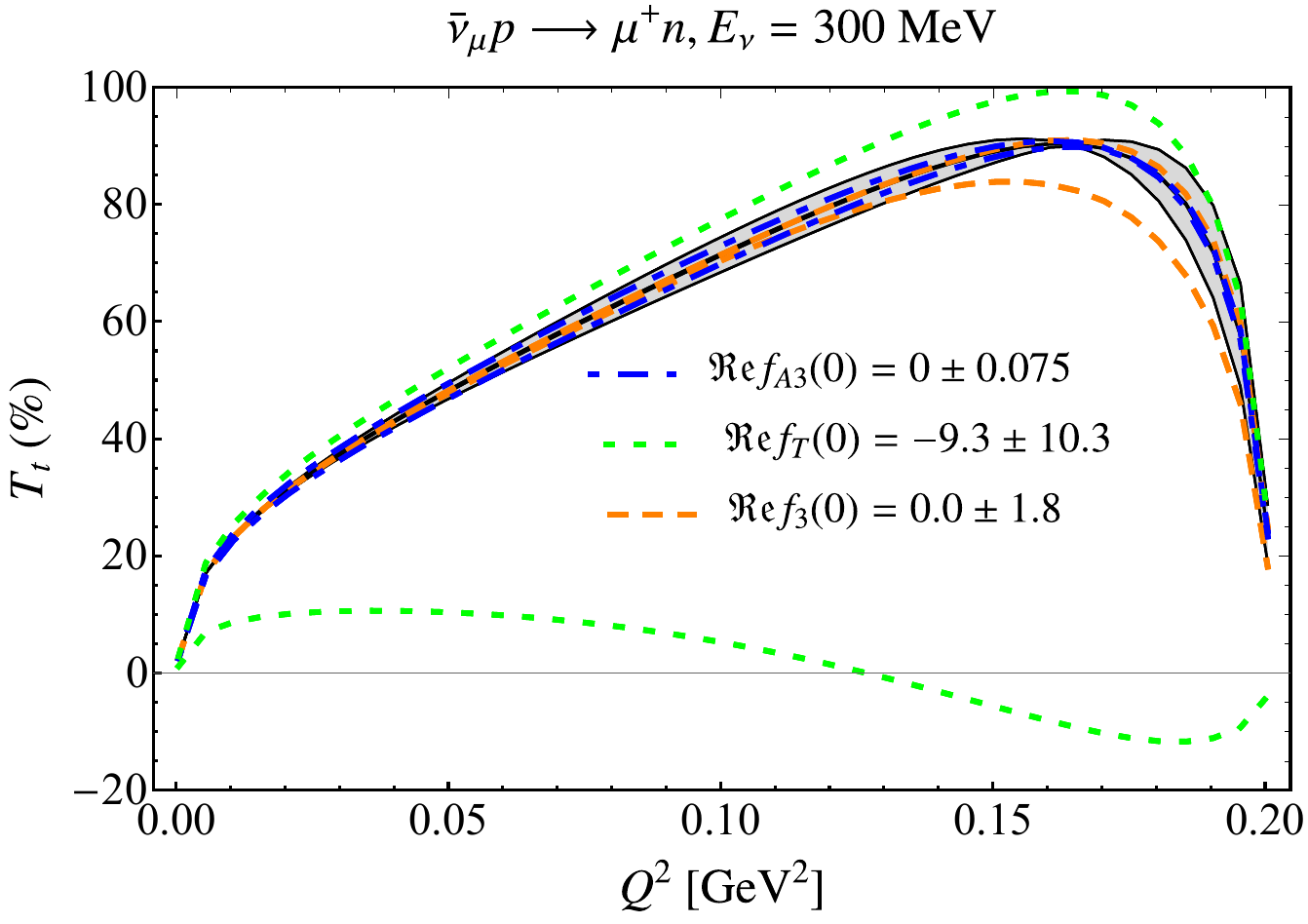}
\includegraphics[width=0.4\textwidth]{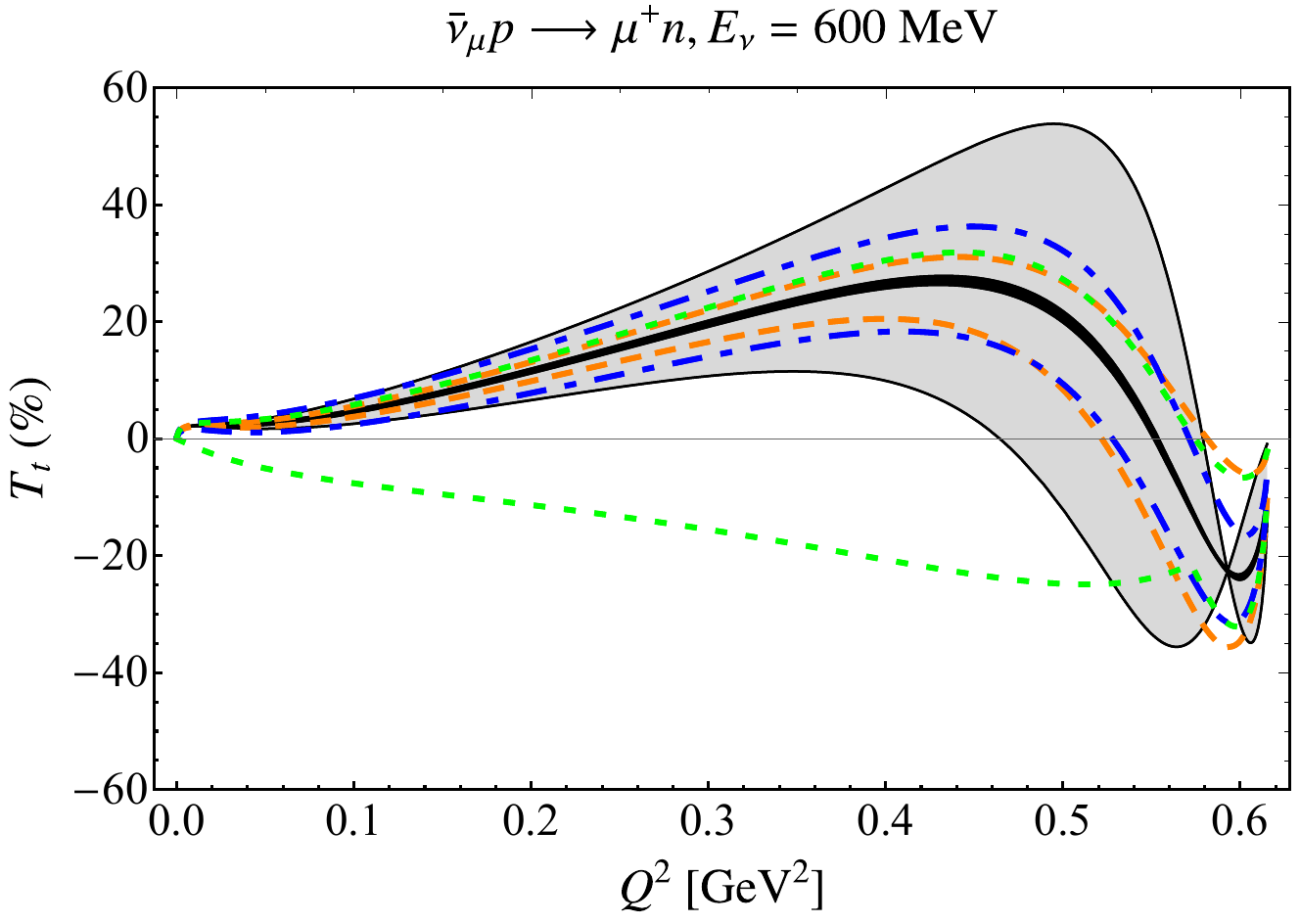}
\includegraphics[width=0.4\textwidth]{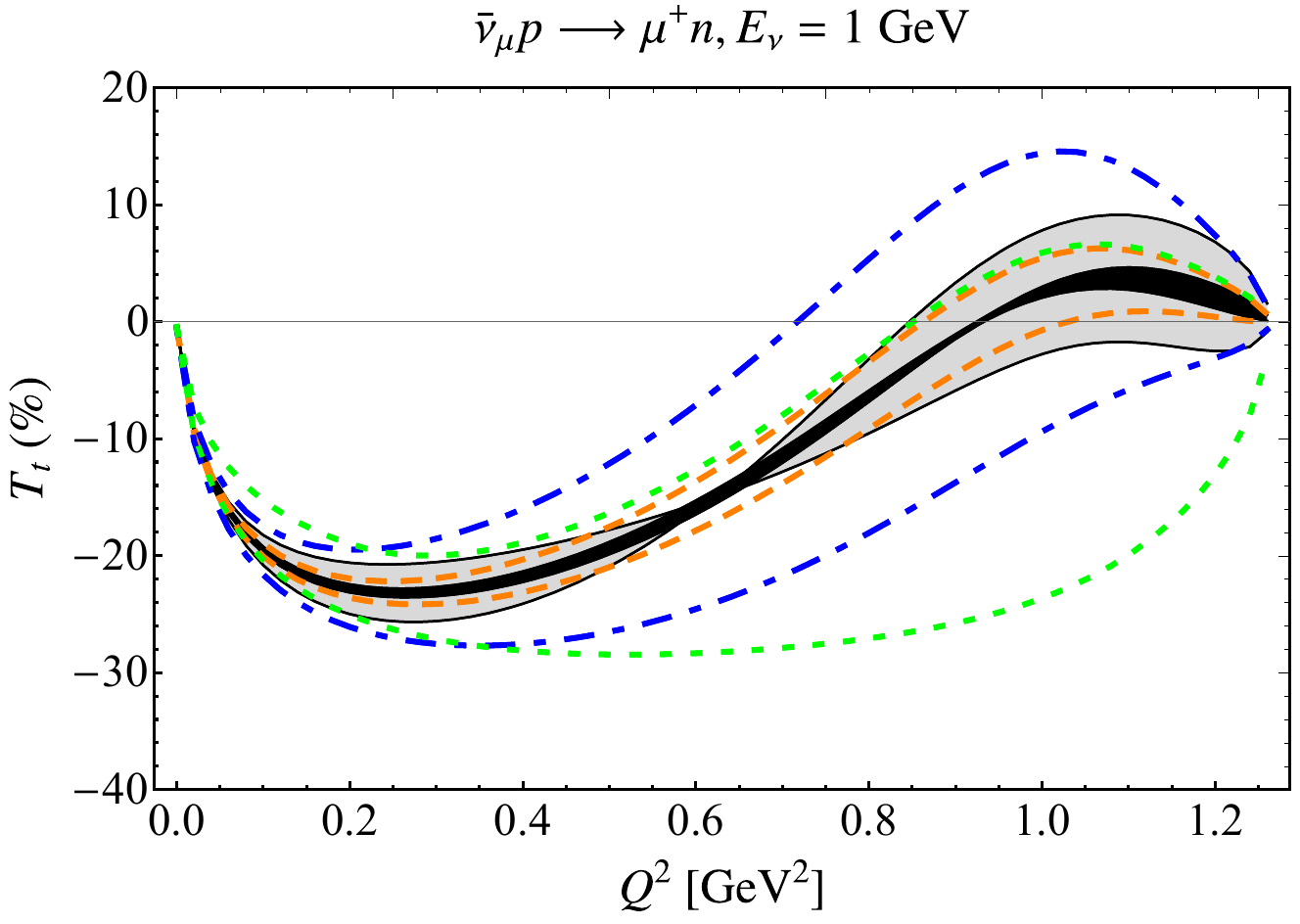}
\includegraphics[width=0.4\textwidth]{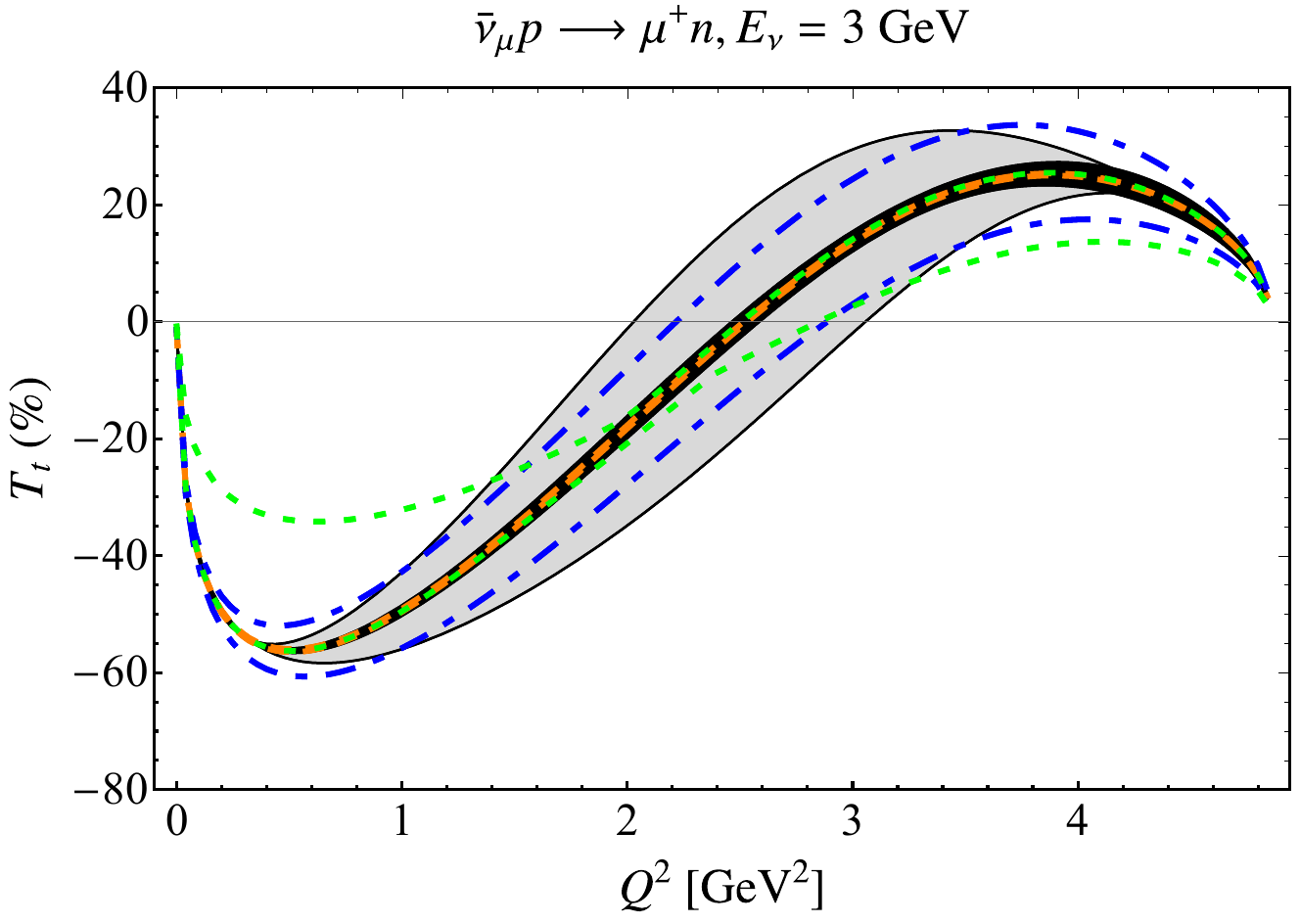}
\caption{Same as Fig.~\ref{fig:nu_Tt_SCFF} but for antineutrinos. \label{fig:antinu_Tt_SCFF}}
\end{figure}

\begin{figure}[H]
\centering
\includegraphics[width=0.4\textwidth]{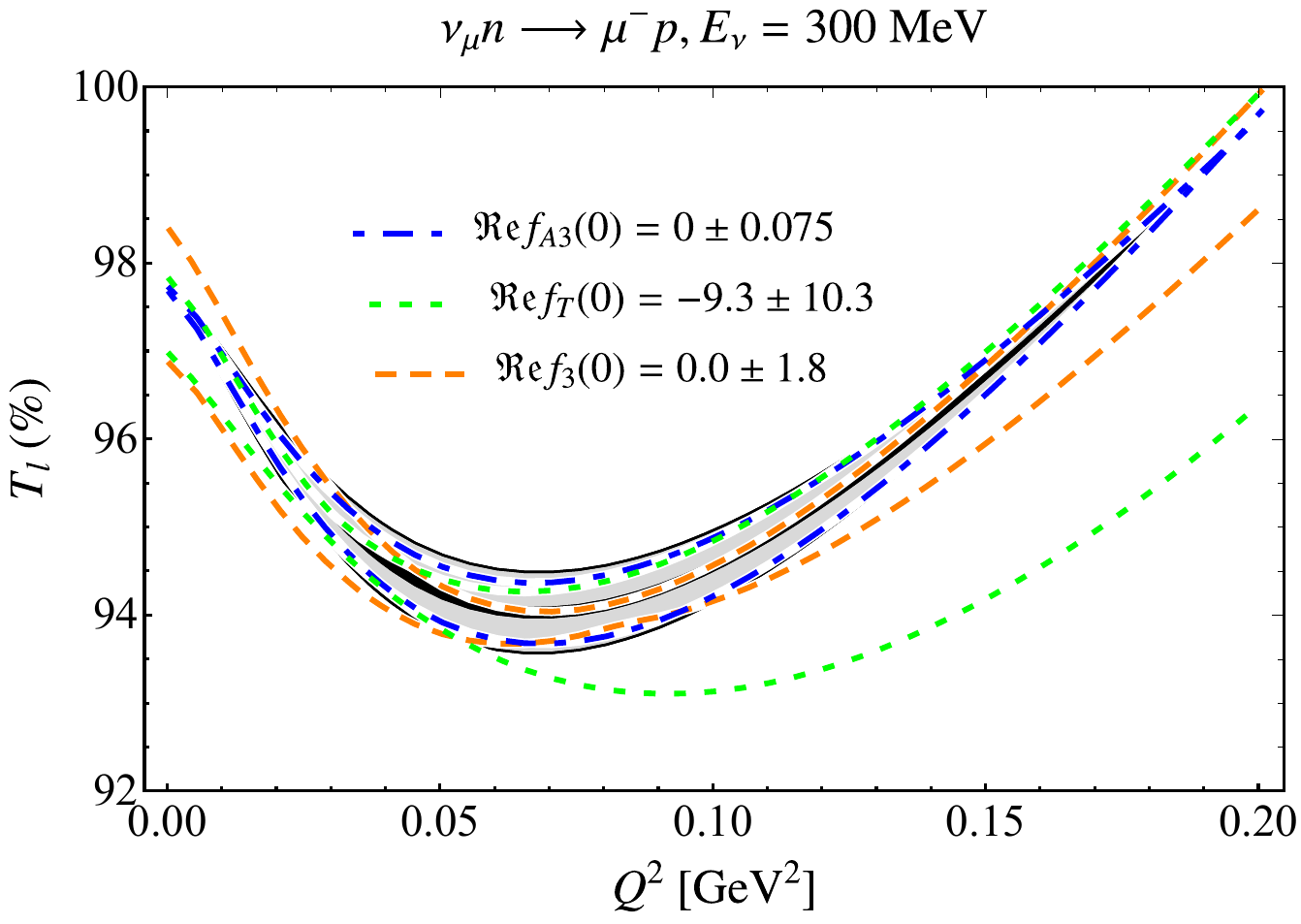}
\includegraphics[width=0.4\textwidth]{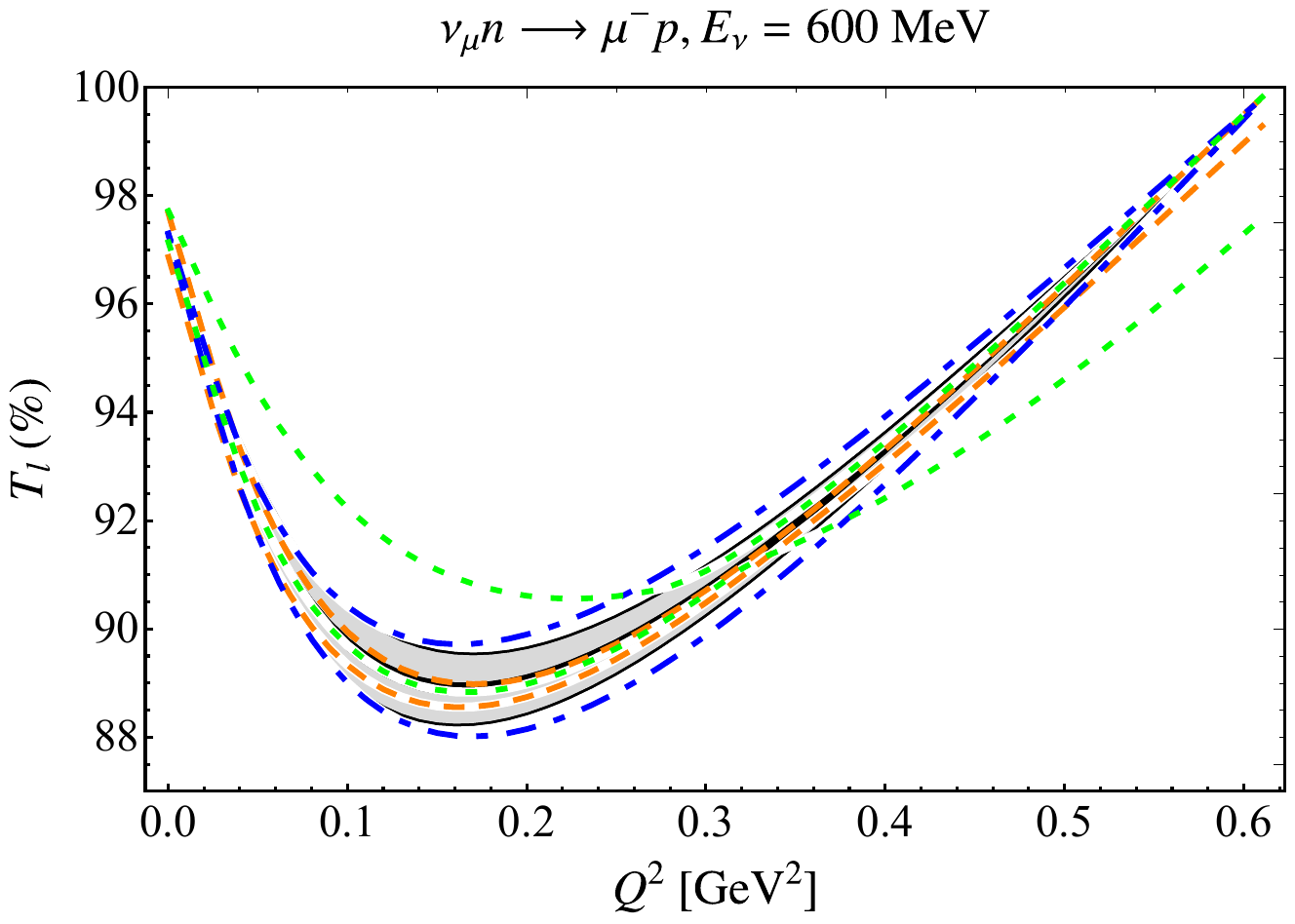}
\includegraphics[width=0.4\textwidth]{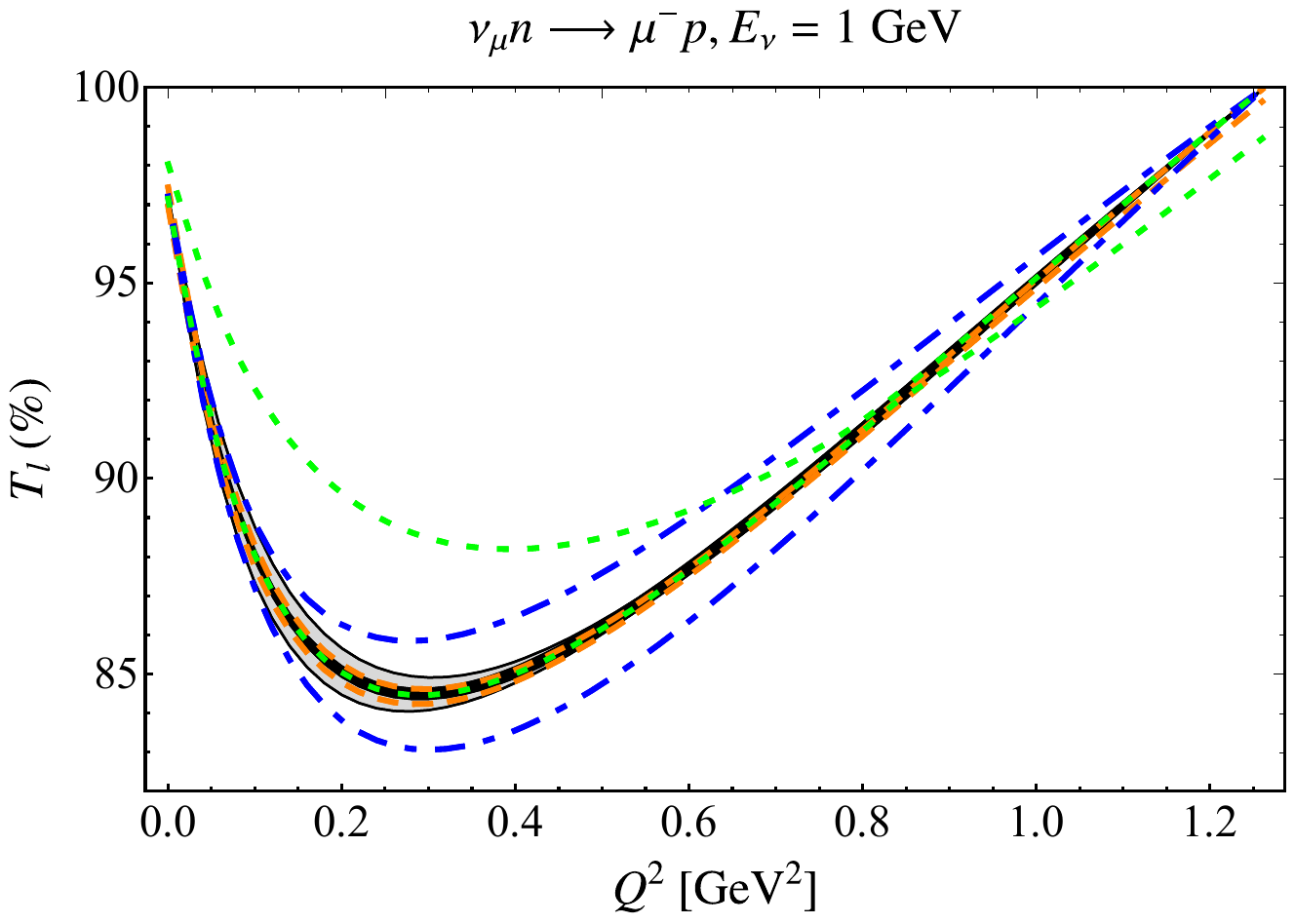}
\includegraphics[width=0.4\textwidth]{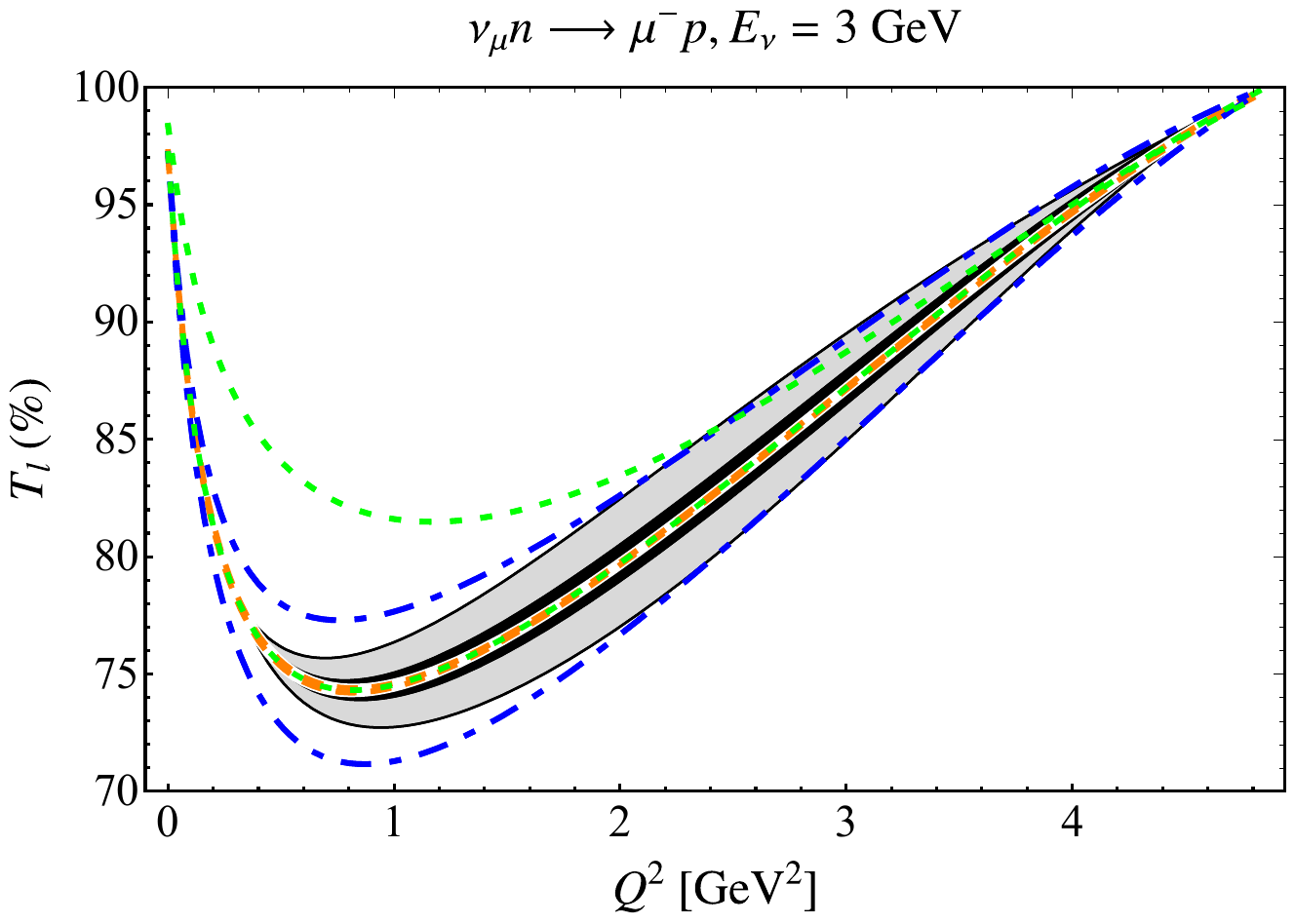}
\caption{Same as Fig.~\ref{fig:nu_Tt_SCFF} but for the longitudinal polarization observable $T_l$. \label{fig:nu_Tl_SCFF}}
\end{figure}

\begin{figure}[H]
\centering
\includegraphics[width=0.4\textwidth]{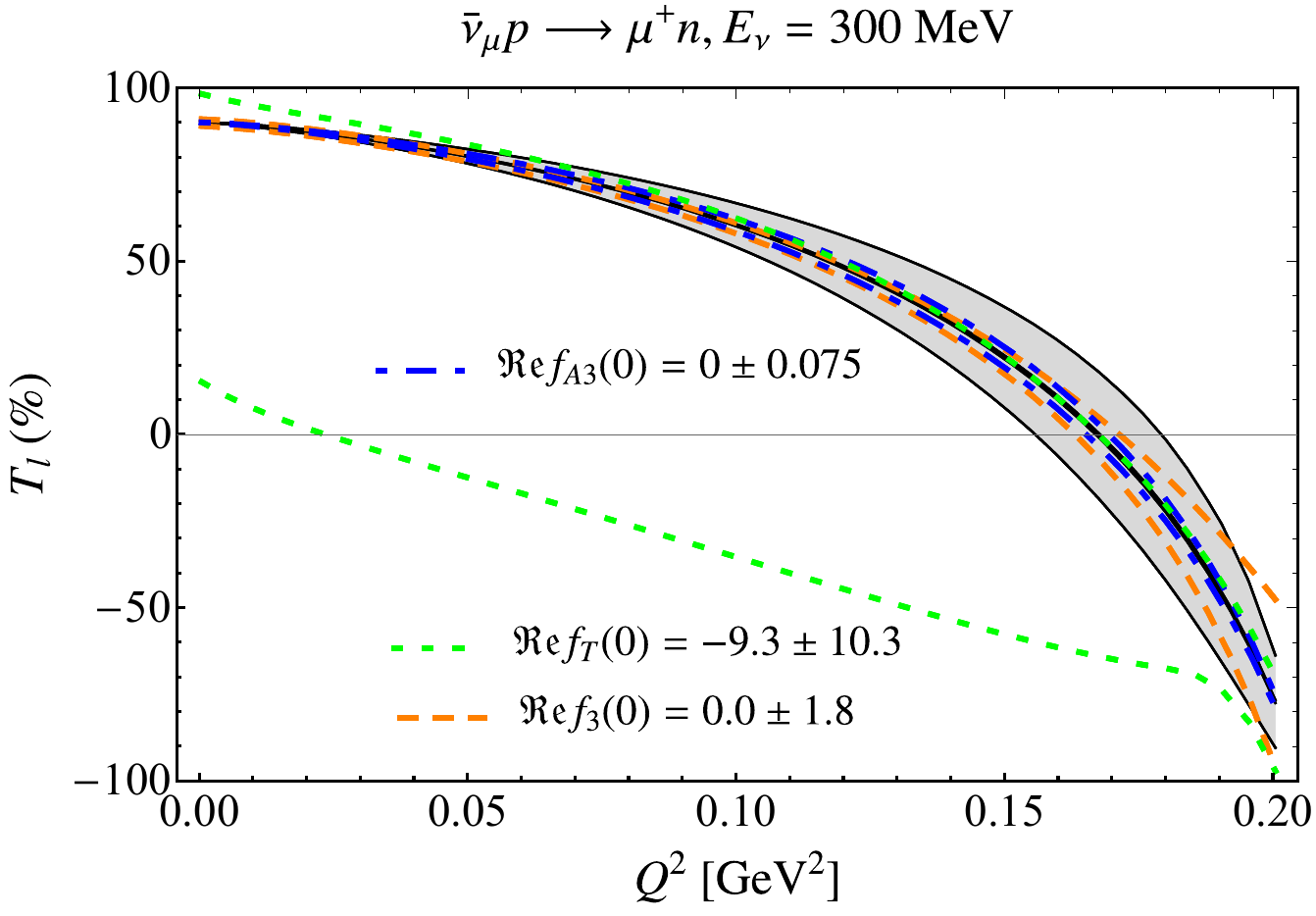}
\includegraphics[width=0.4\textwidth]{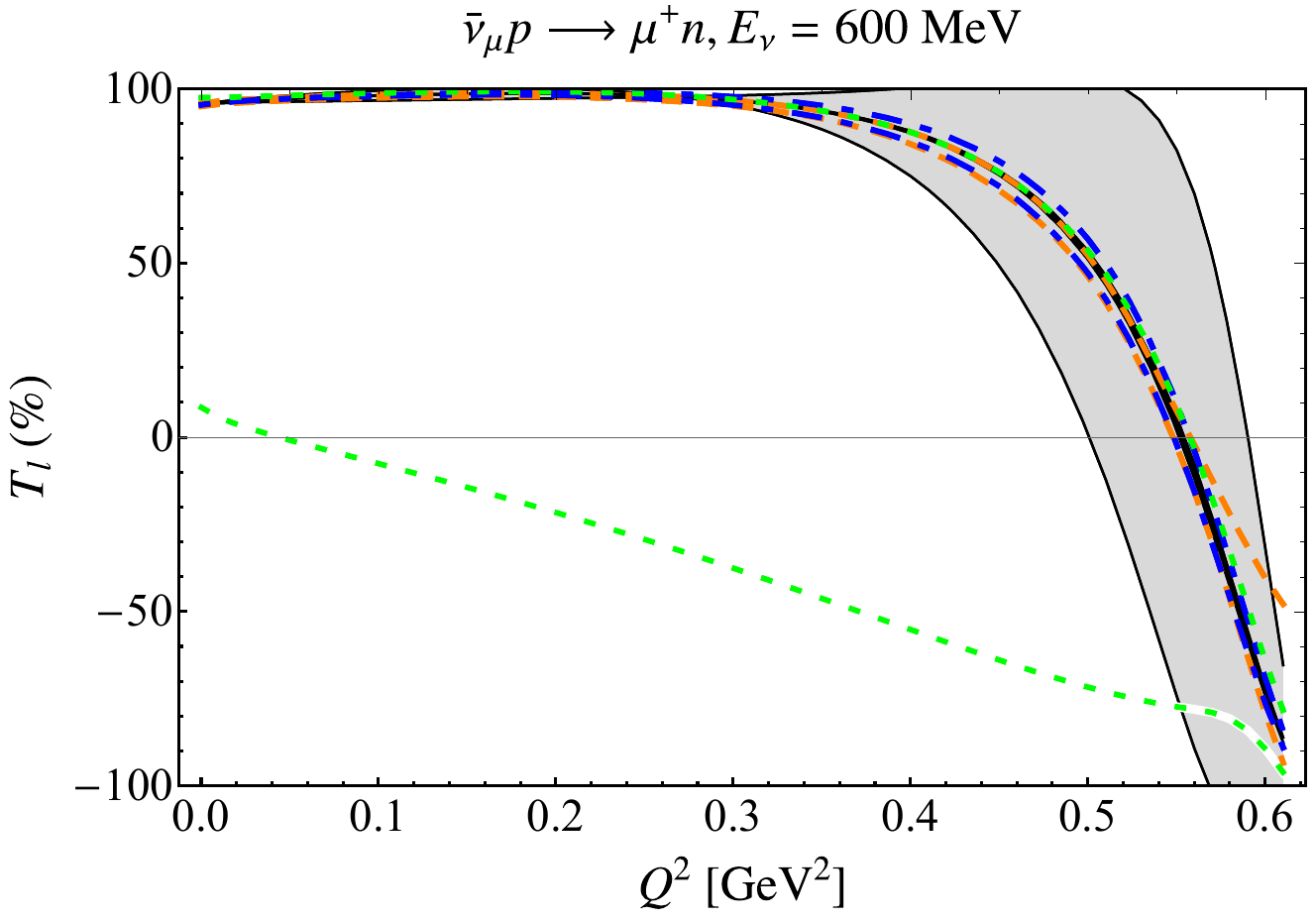}
\includegraphics[width=0.4\textwidth]{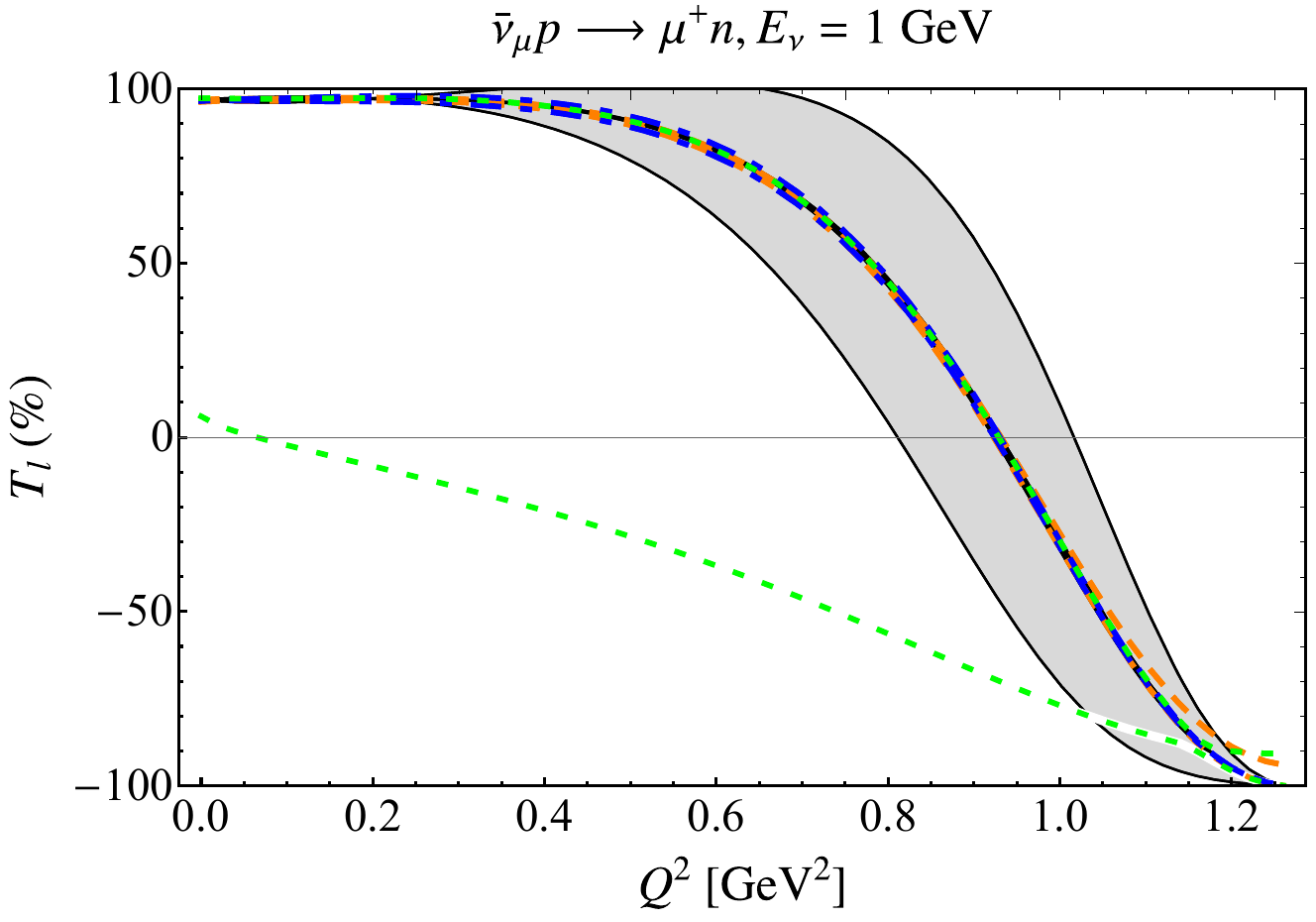}
\includegraphics[width=0.4\textwidth]{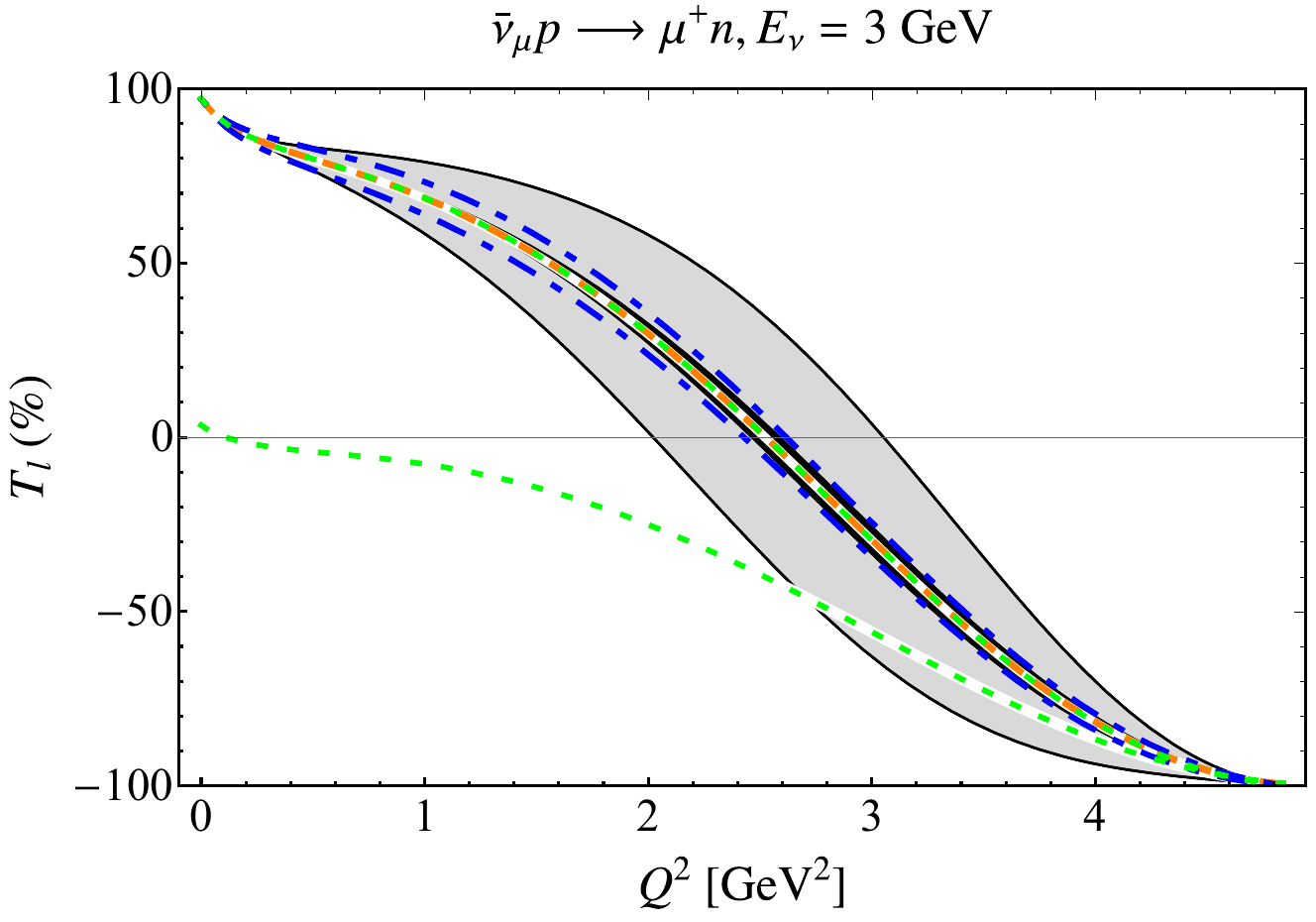}
\caption{Same as Fig.~\ref{fig:antinu_Tt_SCFF} but for the longitudinal polarization observable $T_l$. \label{fig:antinu_Tl_SCFF}}
\end{figure}

\begin{figure}[H]
\centering
\includegraphics[width=0.4\textwidth]{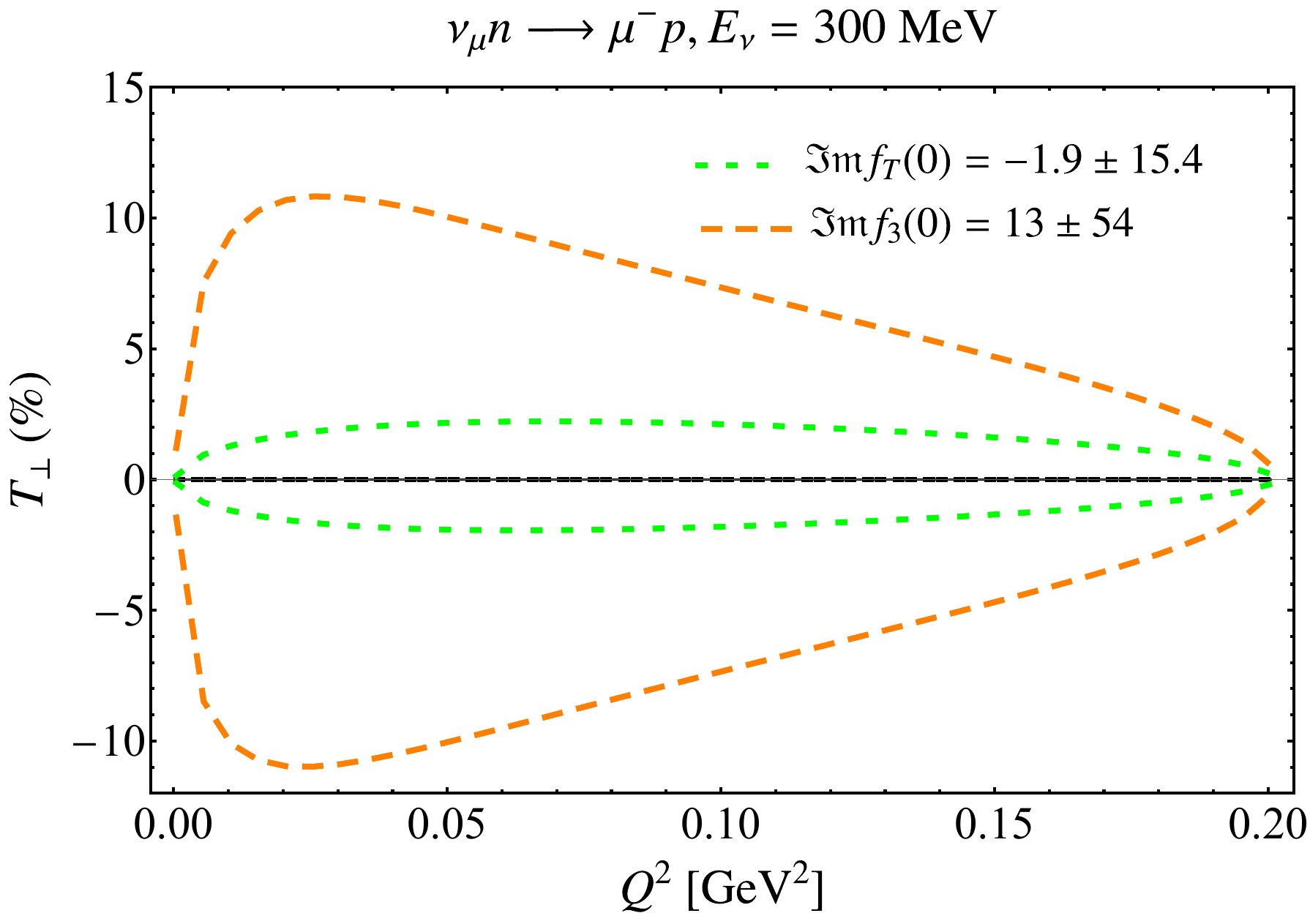}
\includegraphics[width=0.4\textwidth]{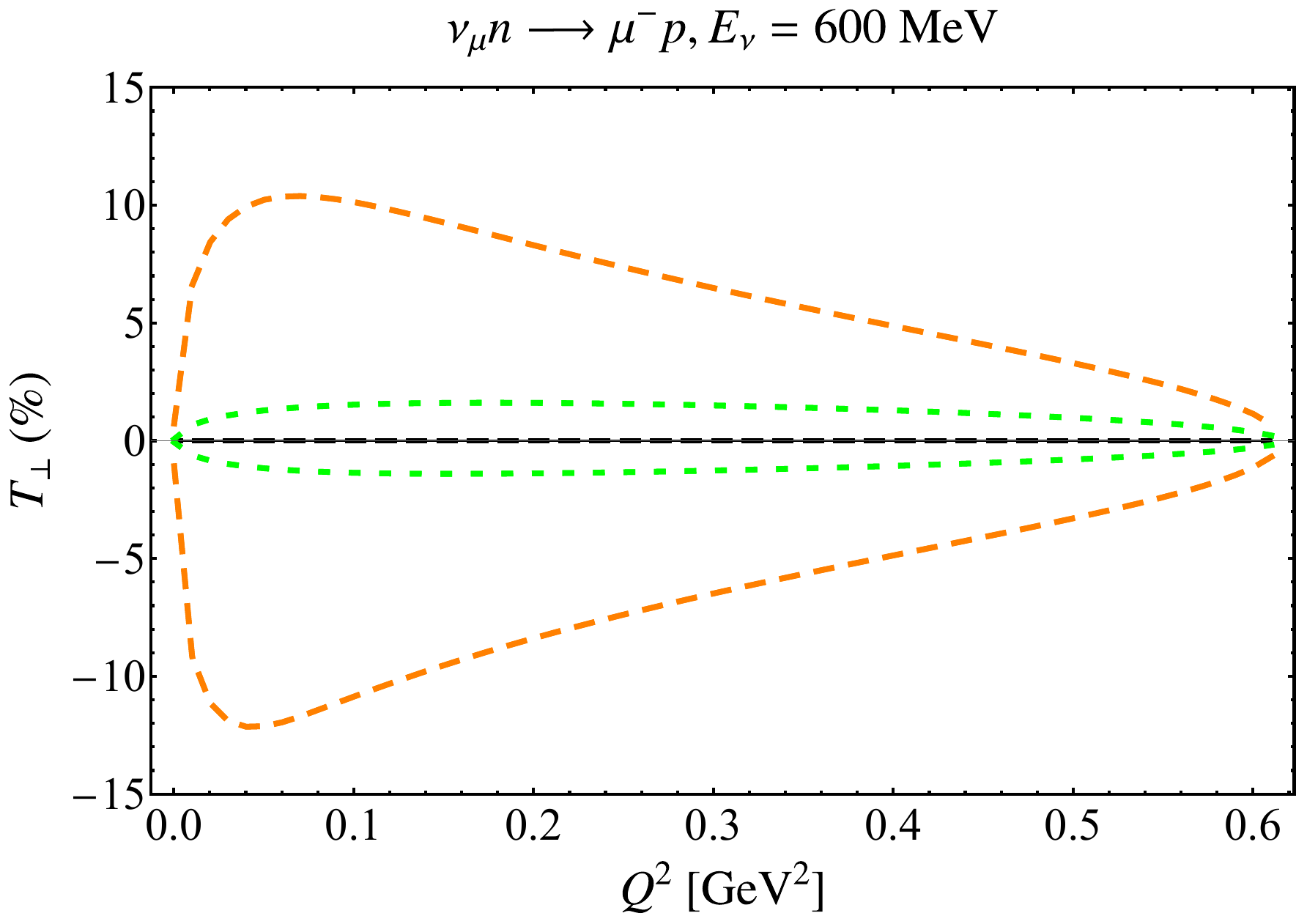}
\includegraphics[width=0.4\textwidth]{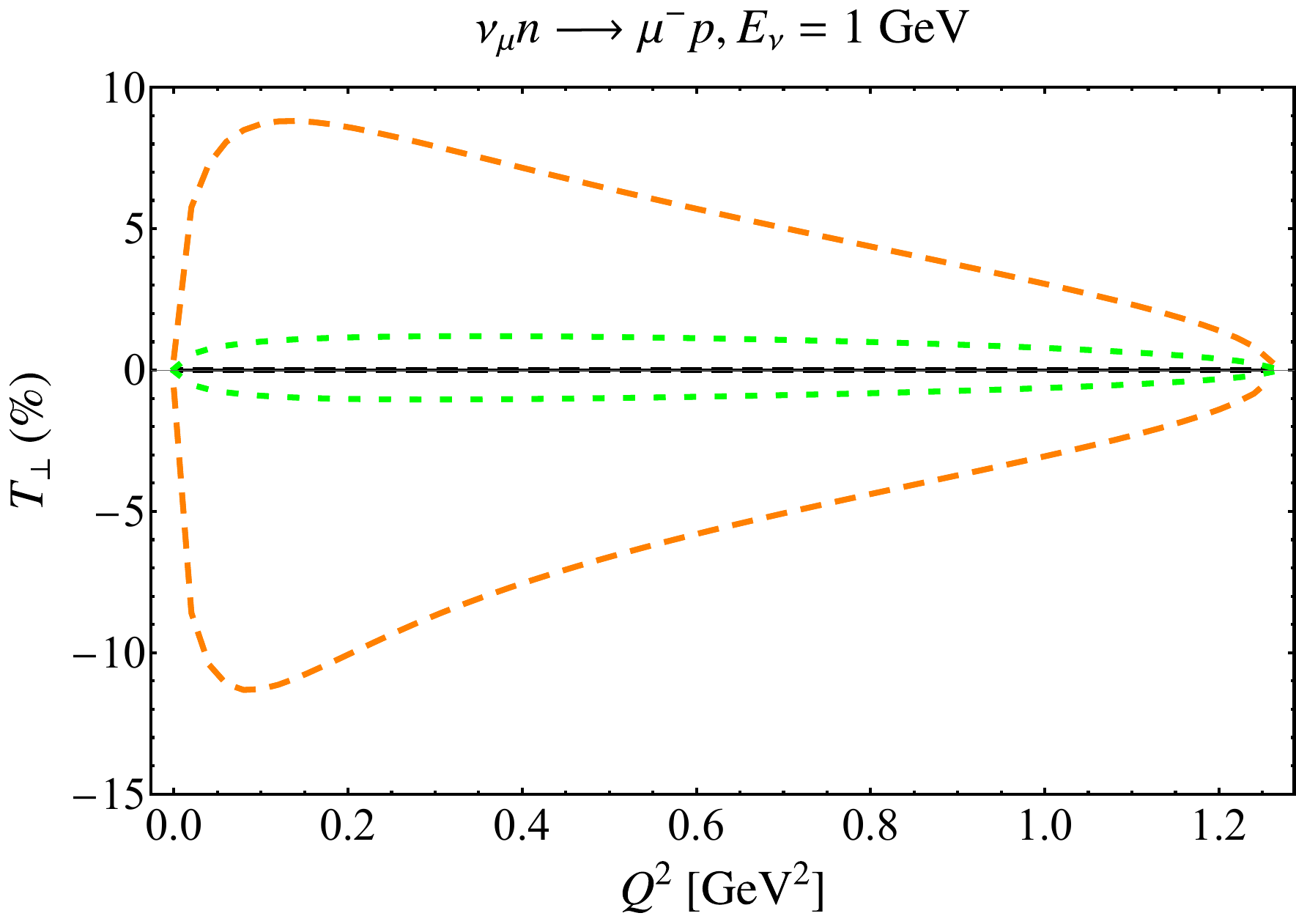}
\includegraphics[width=0.4\textwidth]{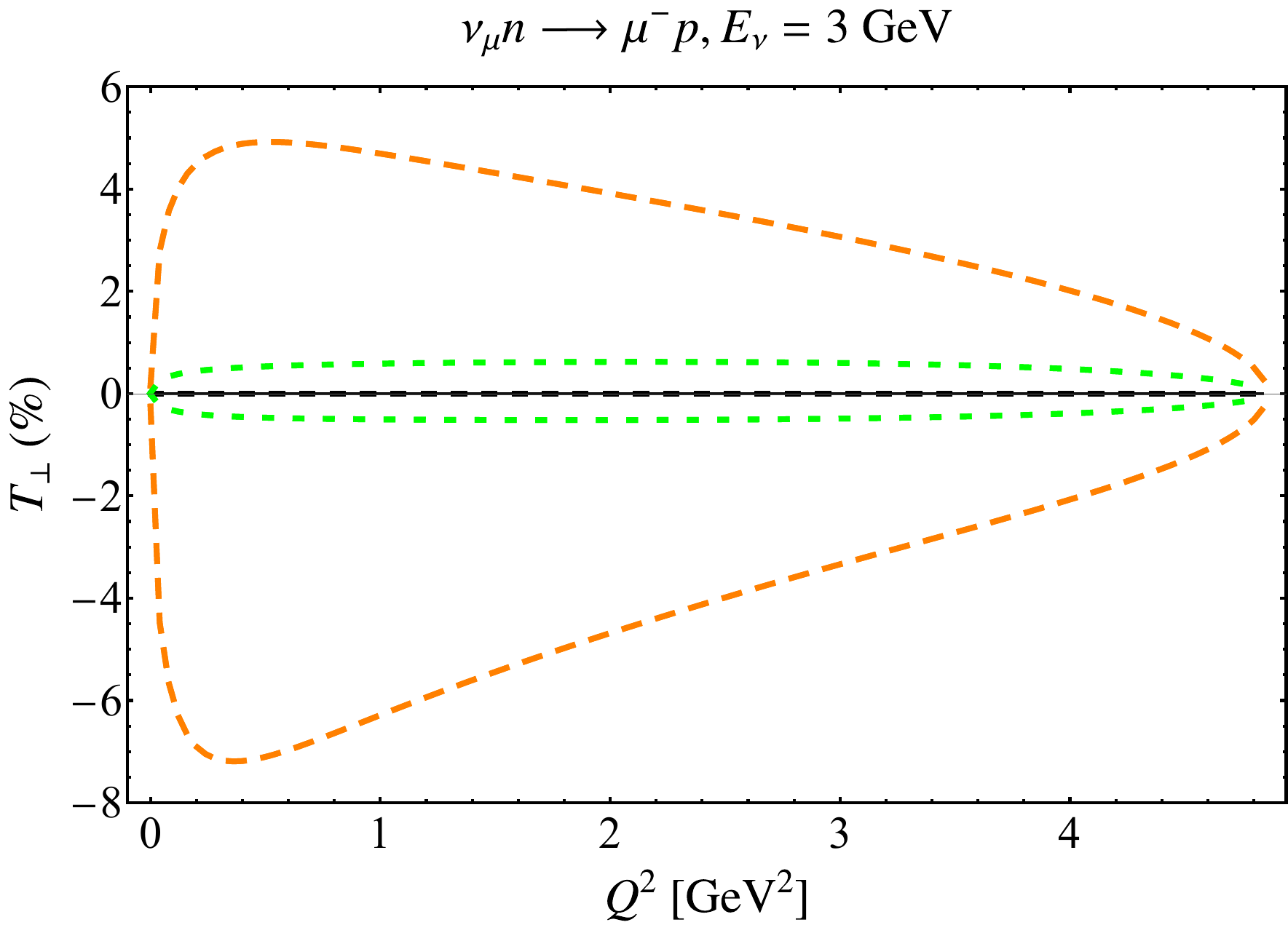}
\caption{Same as Fig.~\ref{fig:nu_Tt_SCFF} but for the transverse polarization observable $T_\perp$ and imaginary amplitudes. \label{fig:nu_TT_SCFF}}
\end{figure}

\begin{figure}[H]
\centering
\includegraphics[width=0.4\textwidth]{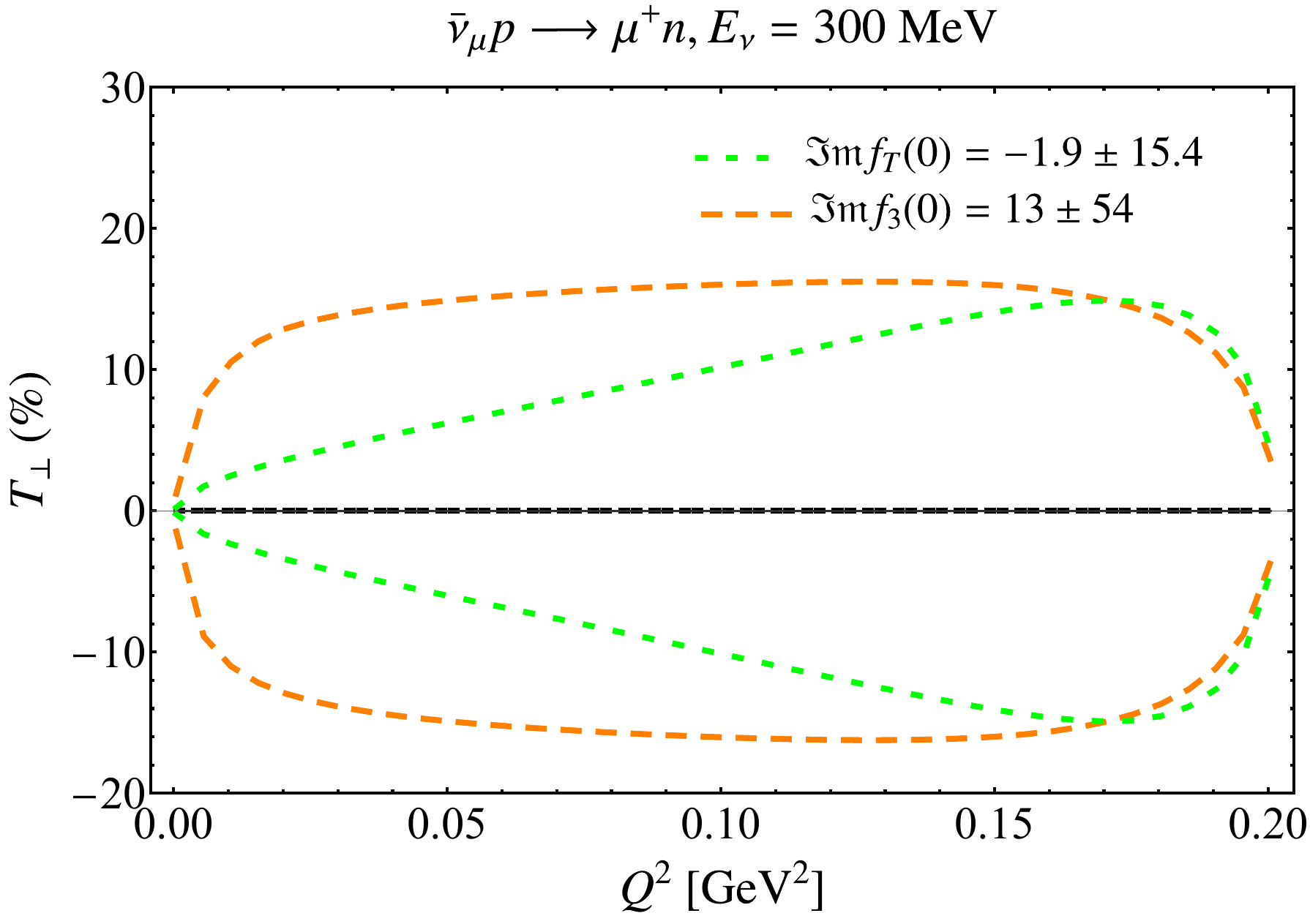}
\includegraphics[width=0.4\textwidth]{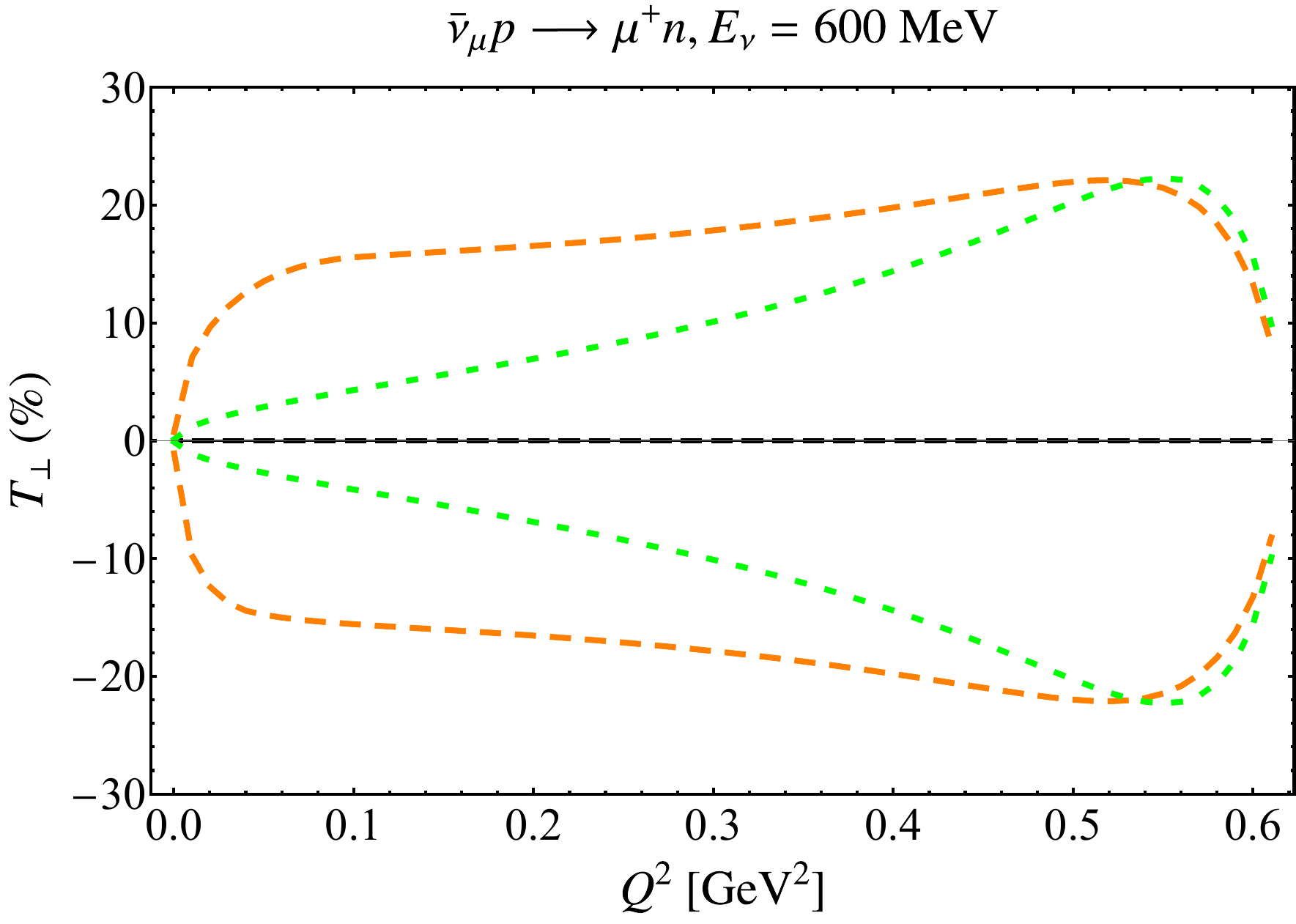}
\includegraphics[width=0.4\textwidth]{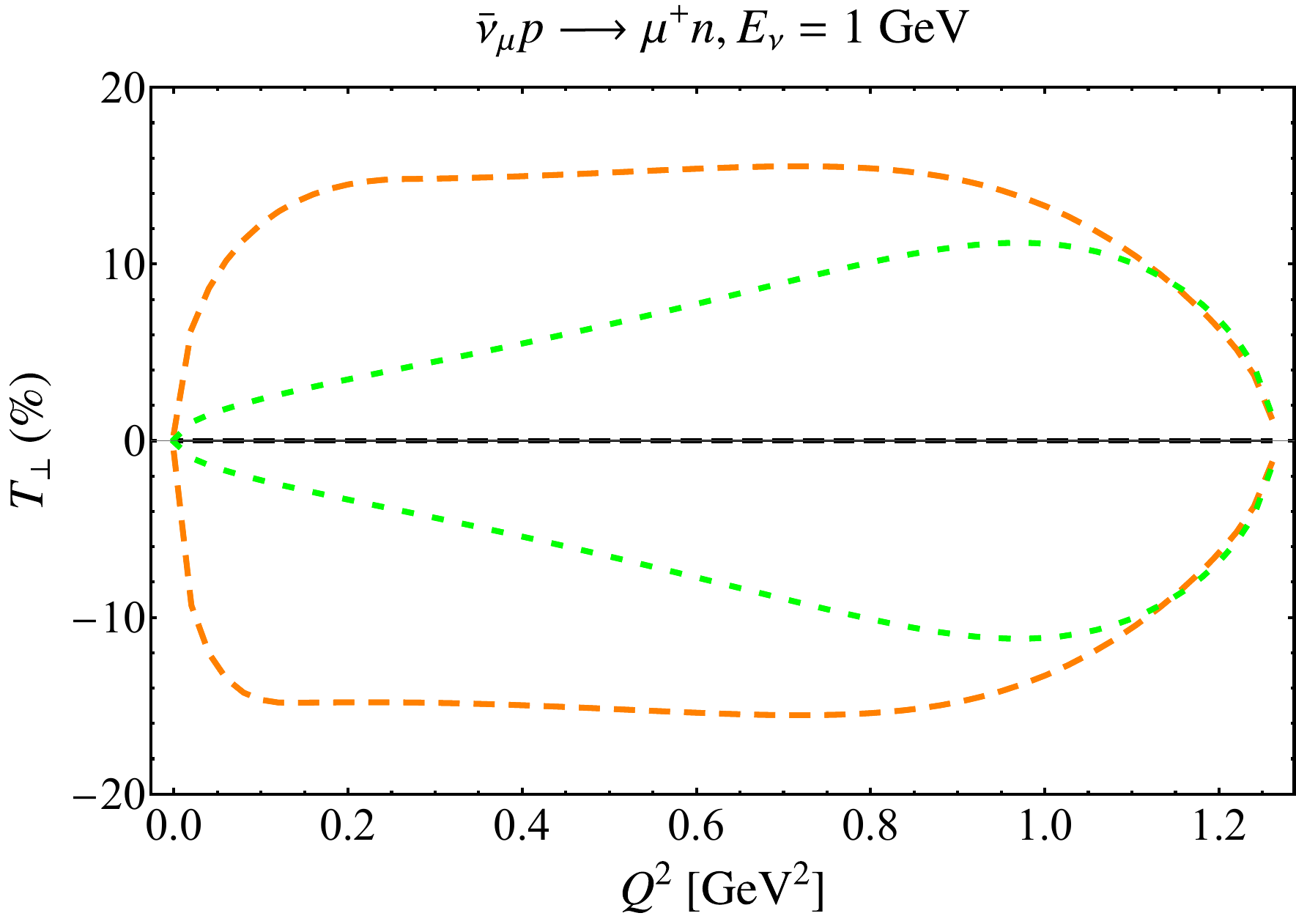}
\includegraphics[width=0.4\textwidth]{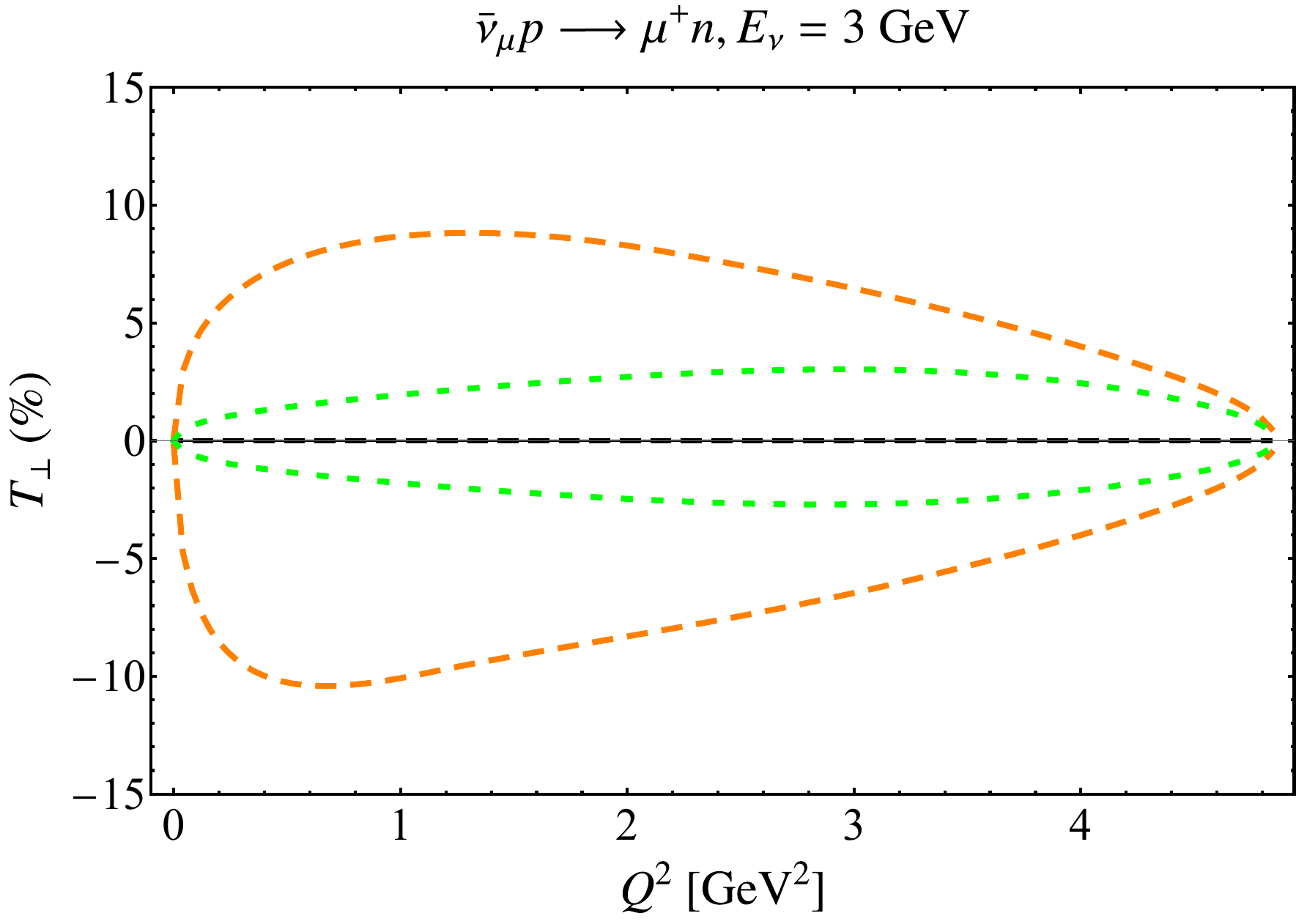}
\caption{Same as Fig.~\ref{fig:antinu_Tt_SCFF} but for the transverse polarization observable $T_\perp$ and imaginary amplitudes. \label{fig:antinu_TT_SCFF}}
\end{figure}

\begin{figure}[H]
\centering
\includegraphics[width=0.4\textwidth]{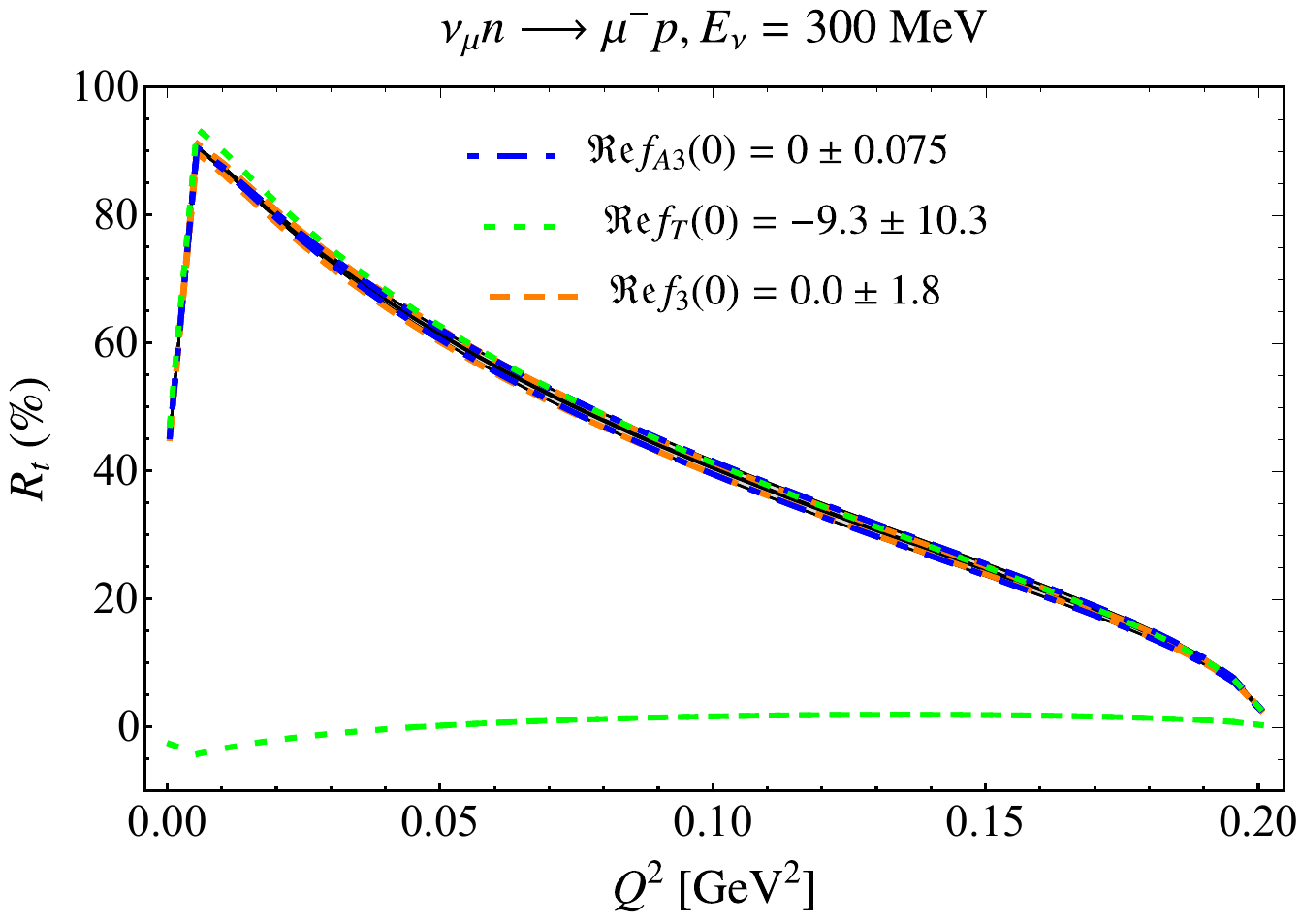}
\includegraphics[width=0.4\textwidth]{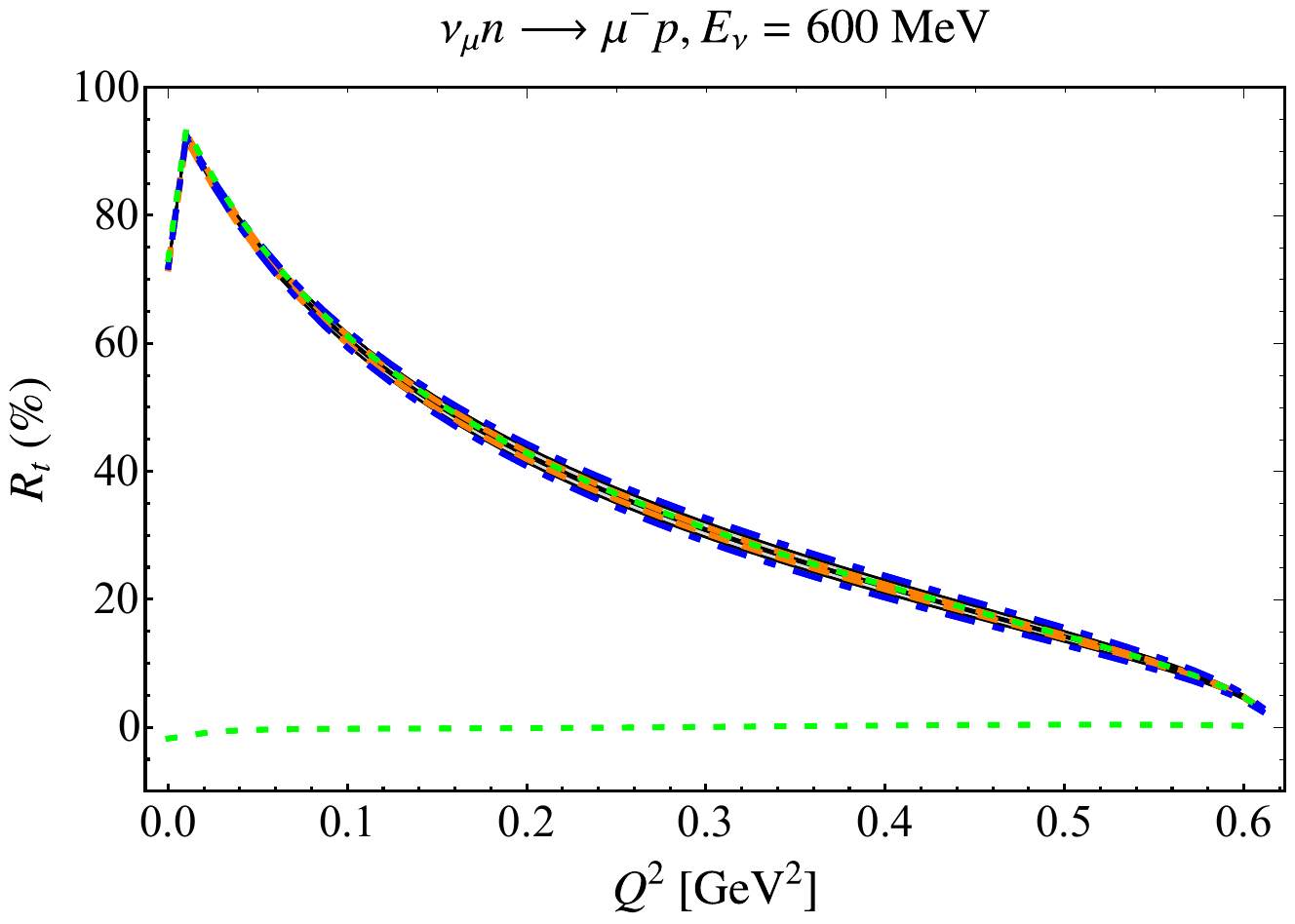}
\includegraphics[width=0.4\textwidth]{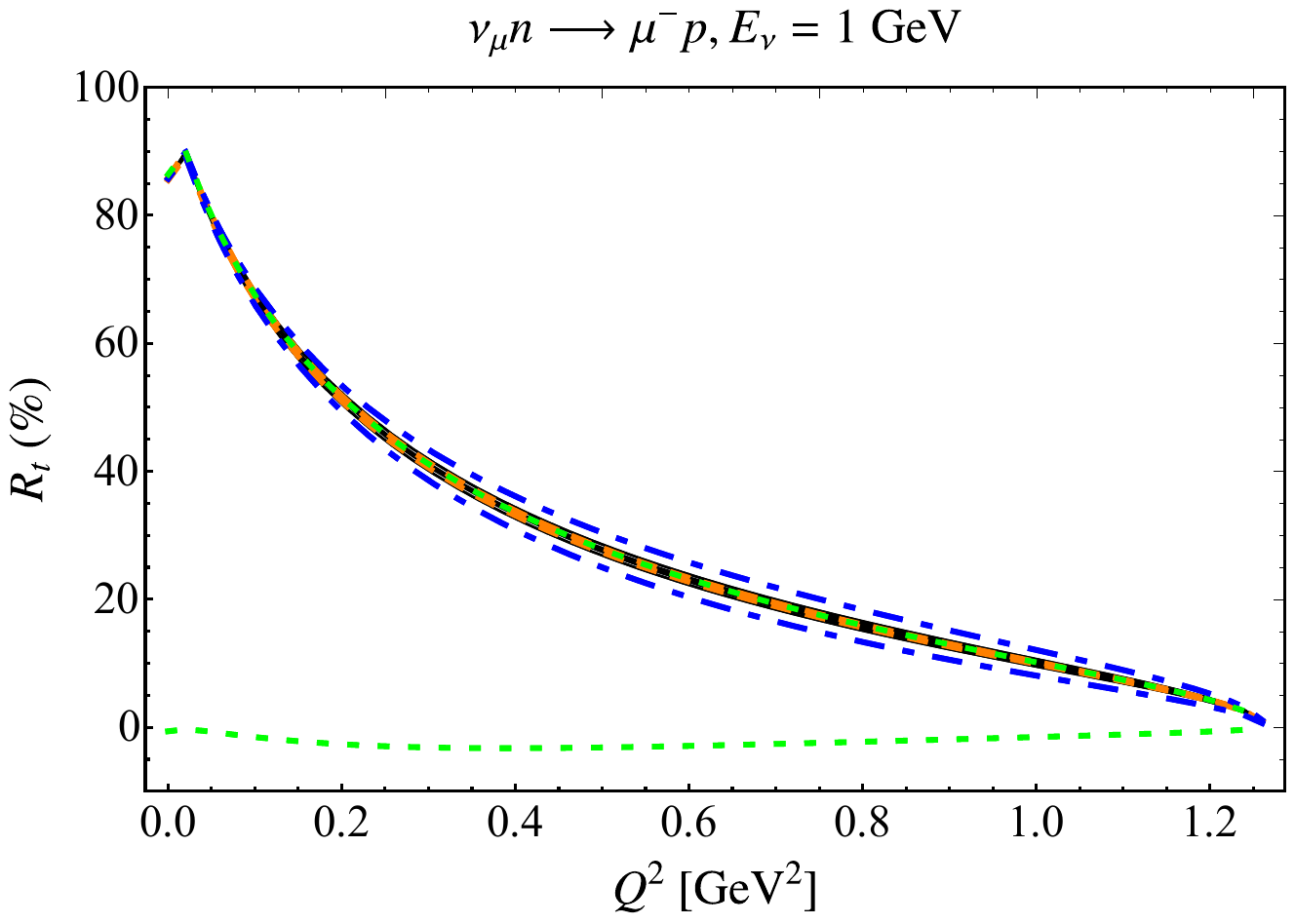}
\includegraphics[width=0.4\textwidth]{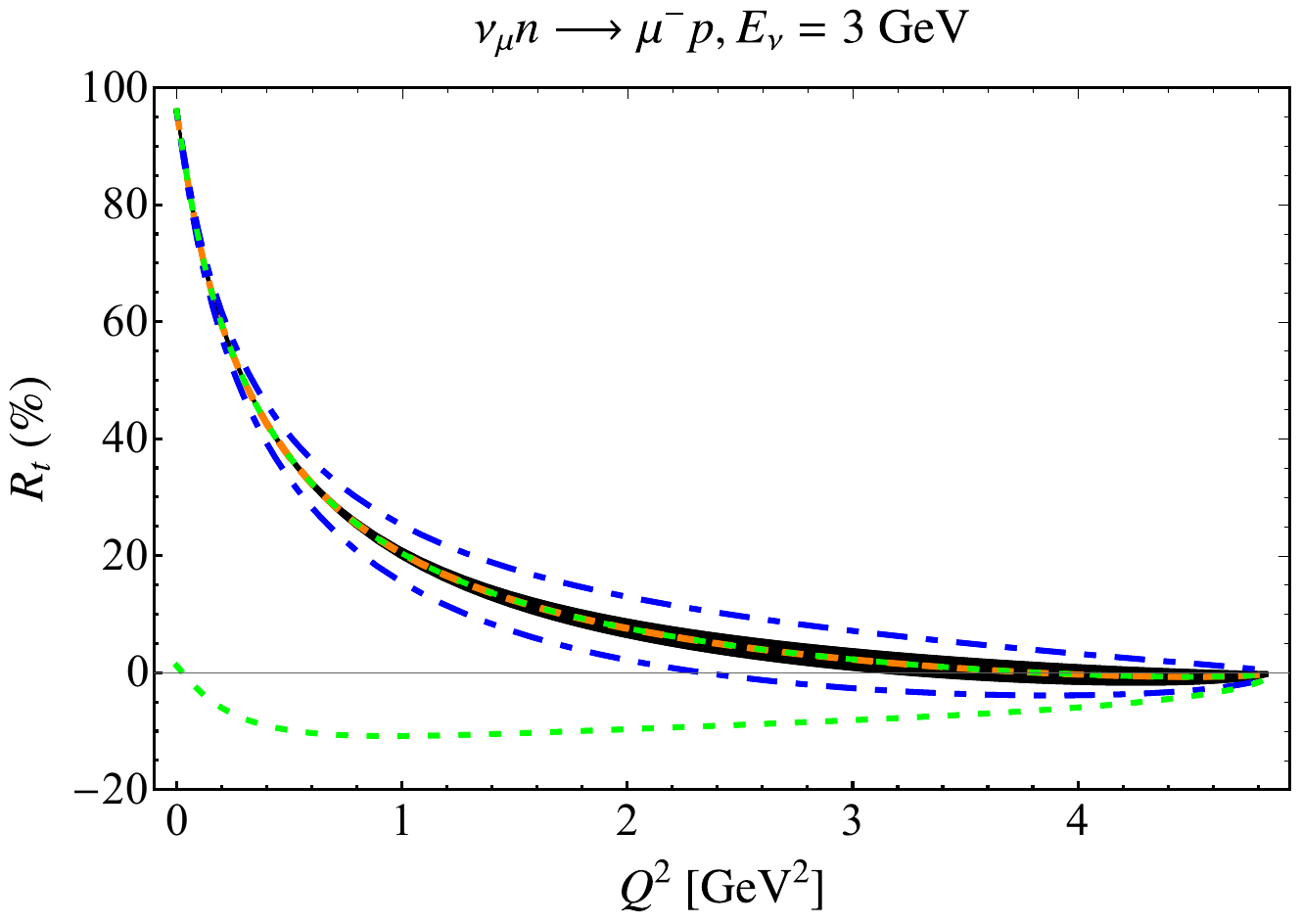}
\caption{Same as Fig.~\ref{fig:nu_Tt_SCFF} but for the transverse polarization observable $R_t$. \label{fig:nu_Rt_SCFF}}
\end{figure}

\begin{figure}[H]
\centering
\includegraphics[width=0.4\textwidth]{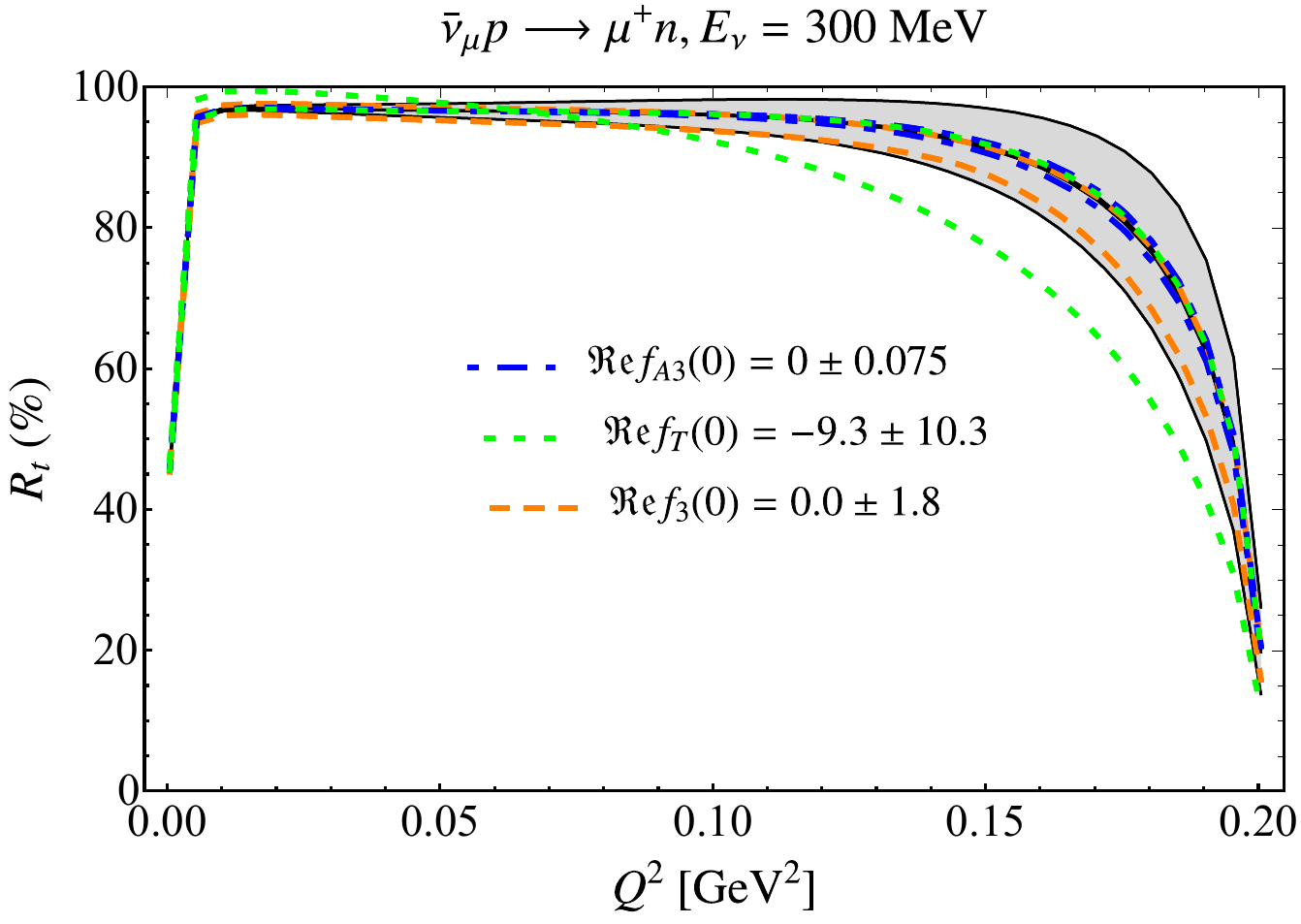}
\includegraphics[width=0.4\textwidth]{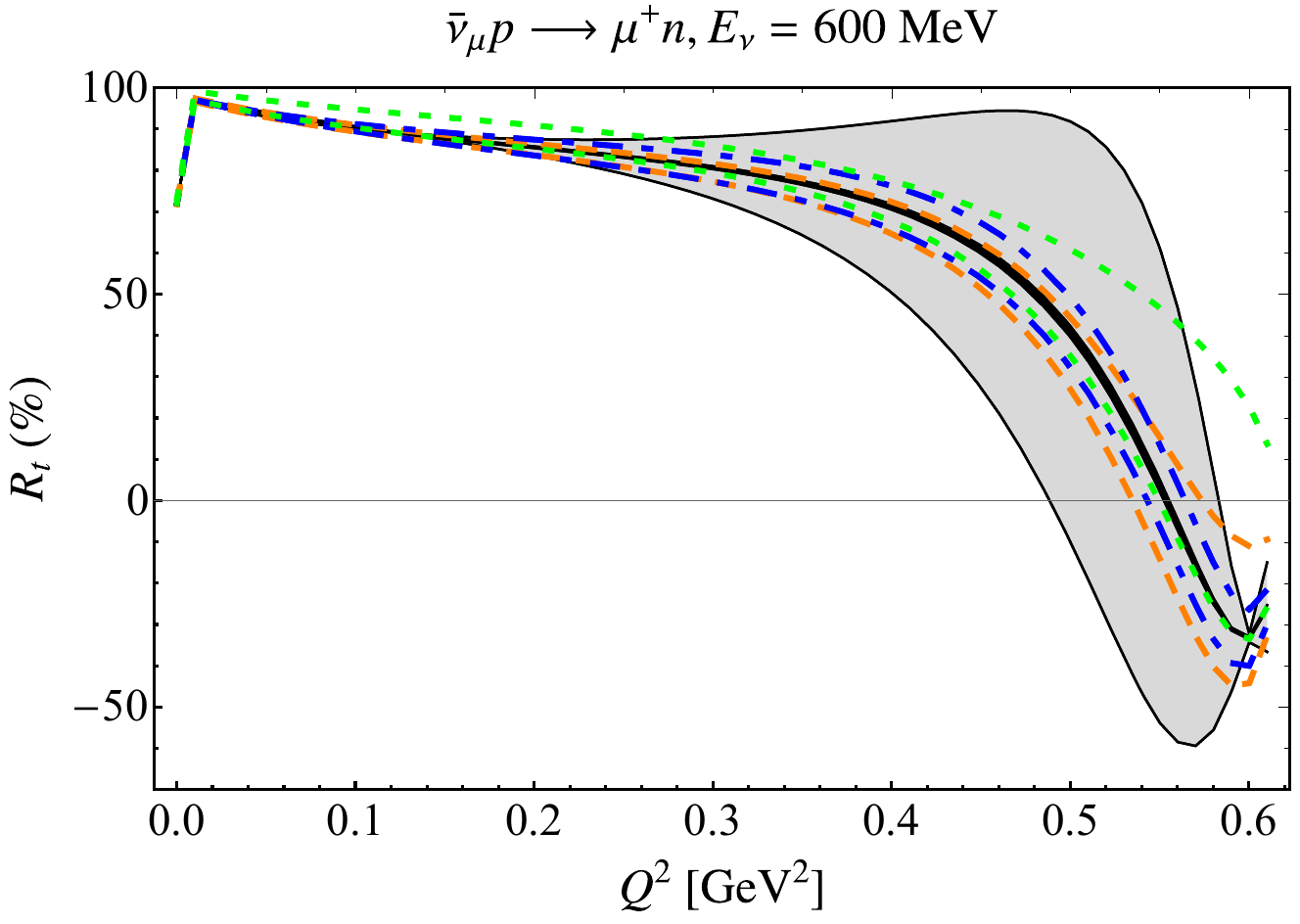}
\includegraphics[width=0.4\textwidth]{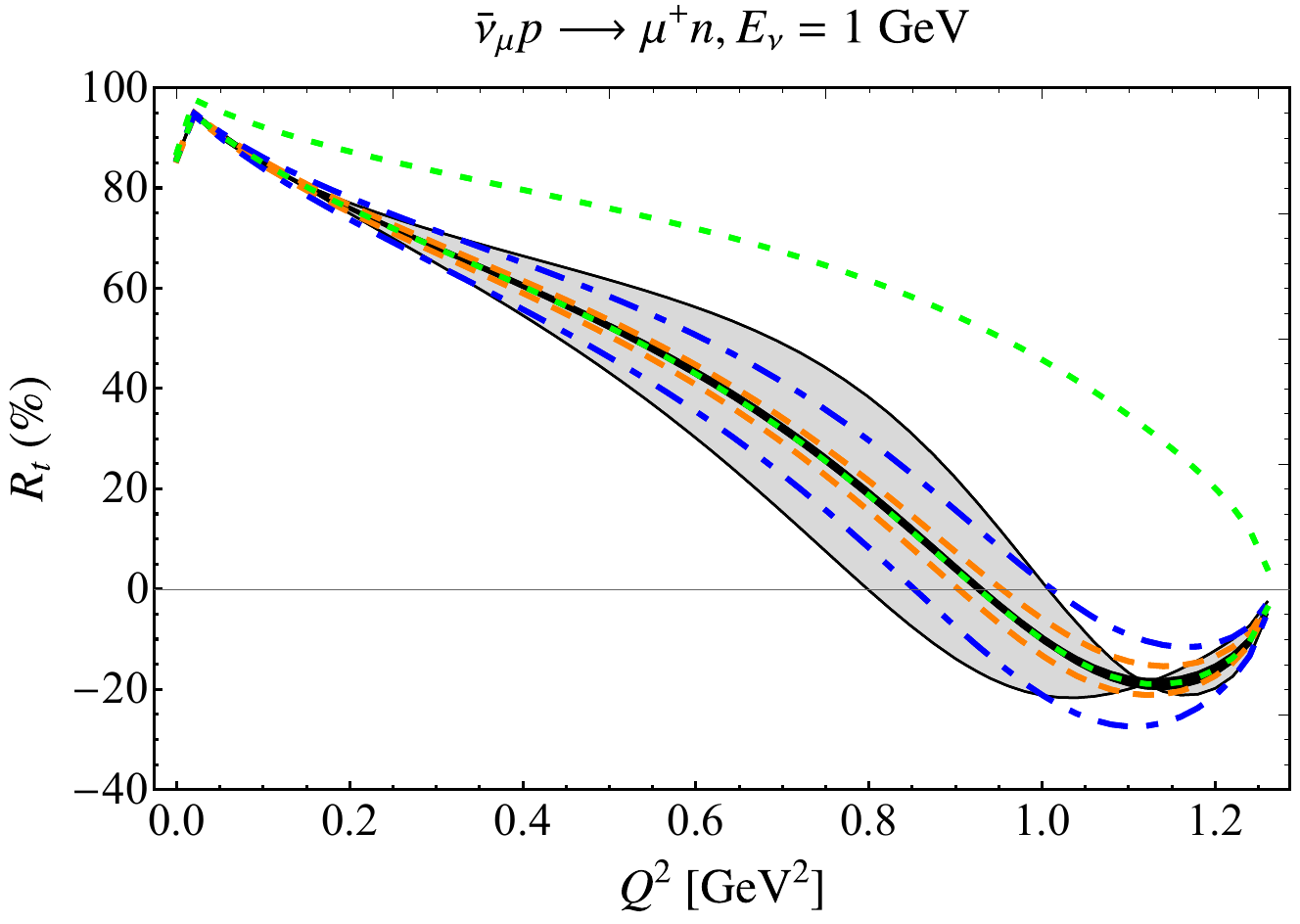}
\includegraphics[width=0.4\textwidth]{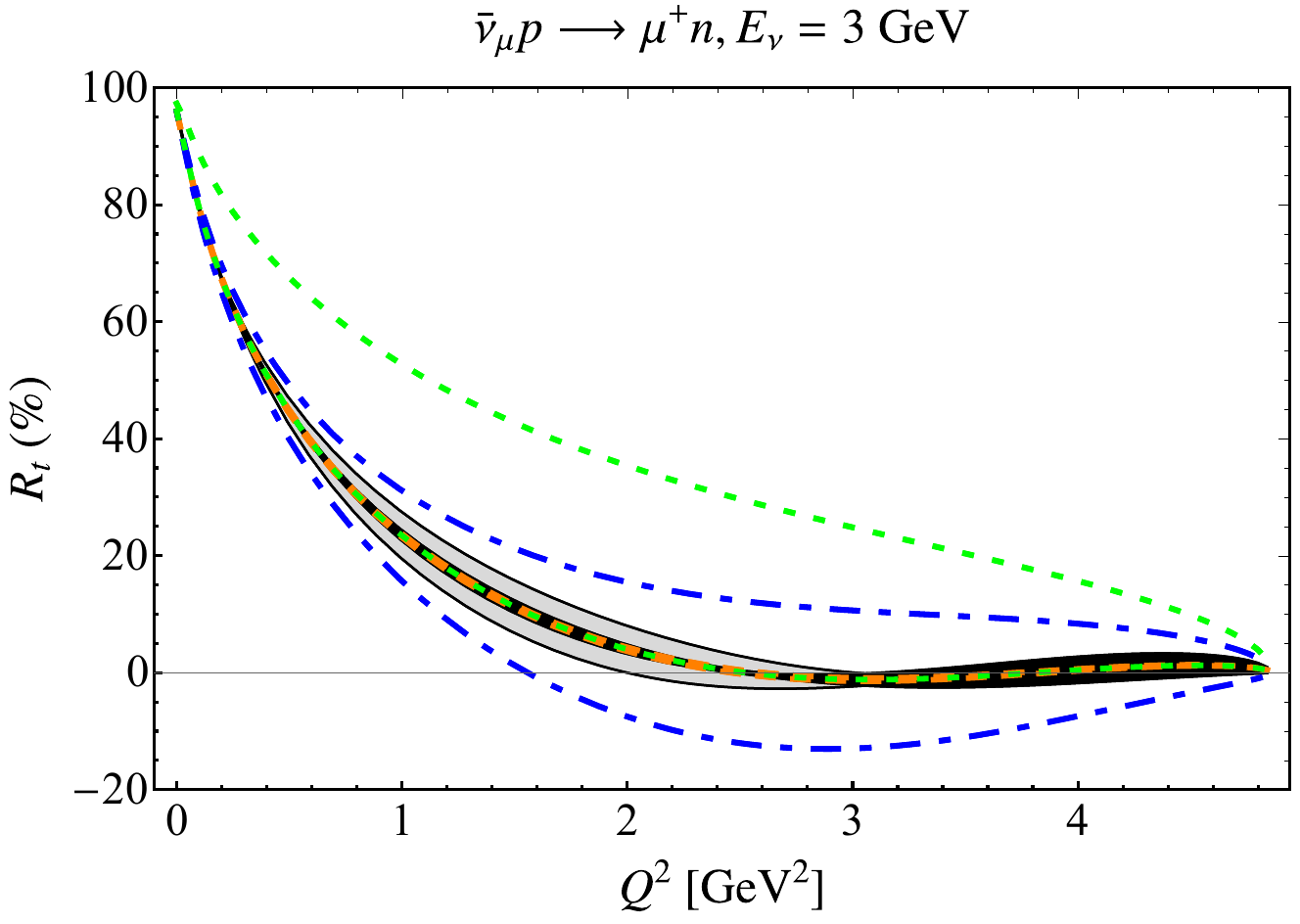}
\caption{Same as Fig.~\ref{fig:antinu_Tt_SCFF} but for the transverse polarization observable $R_t$. \label{fig:antinu_Rt_SCFF}}
\end{figure}

\begin{figure}[H]
\centering
\includegraphics[width=0.4\textwidth]{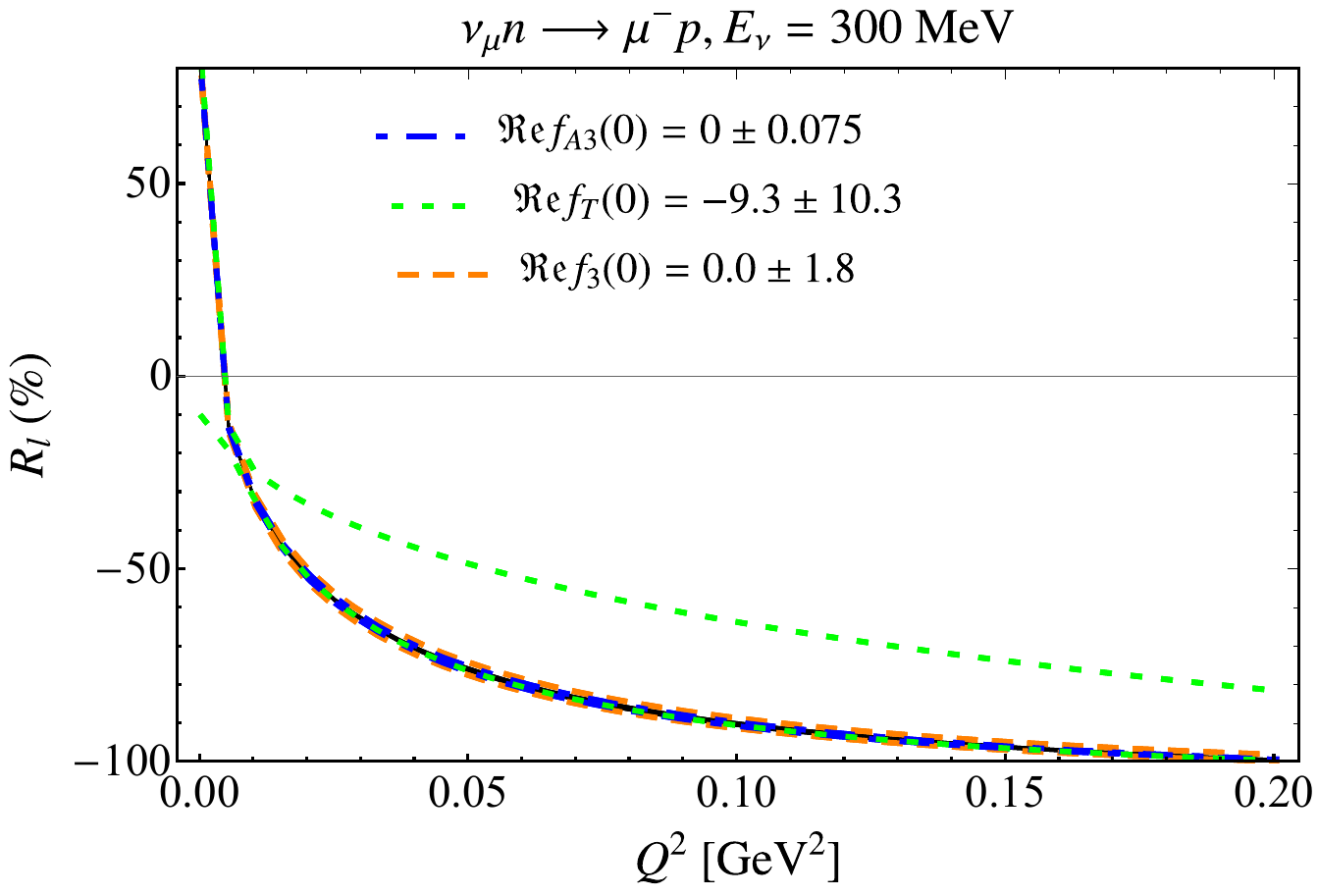}
\includegraphics[width=0.4\textwidth]{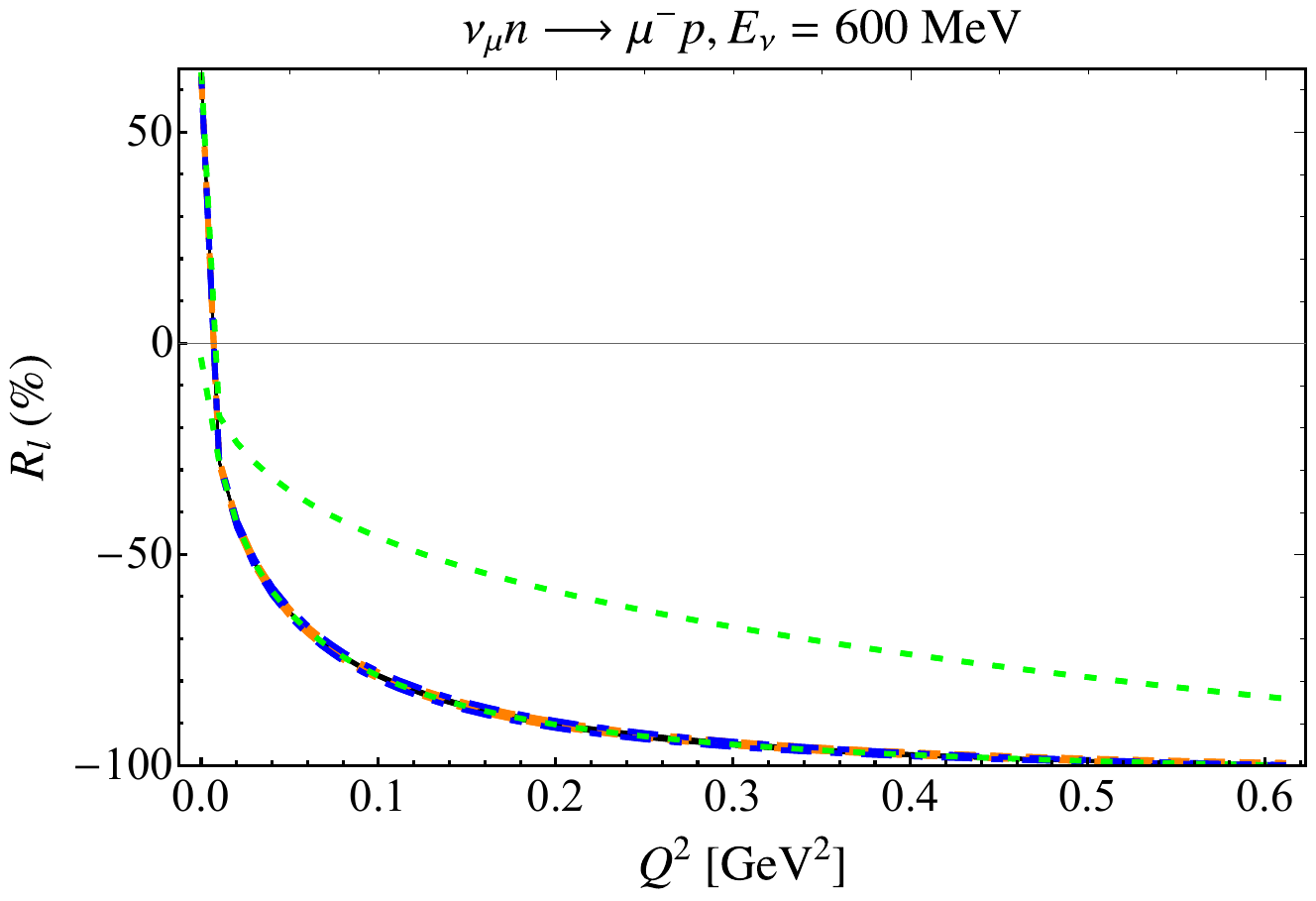}
\includegraphics[width=0.4\textwidth]{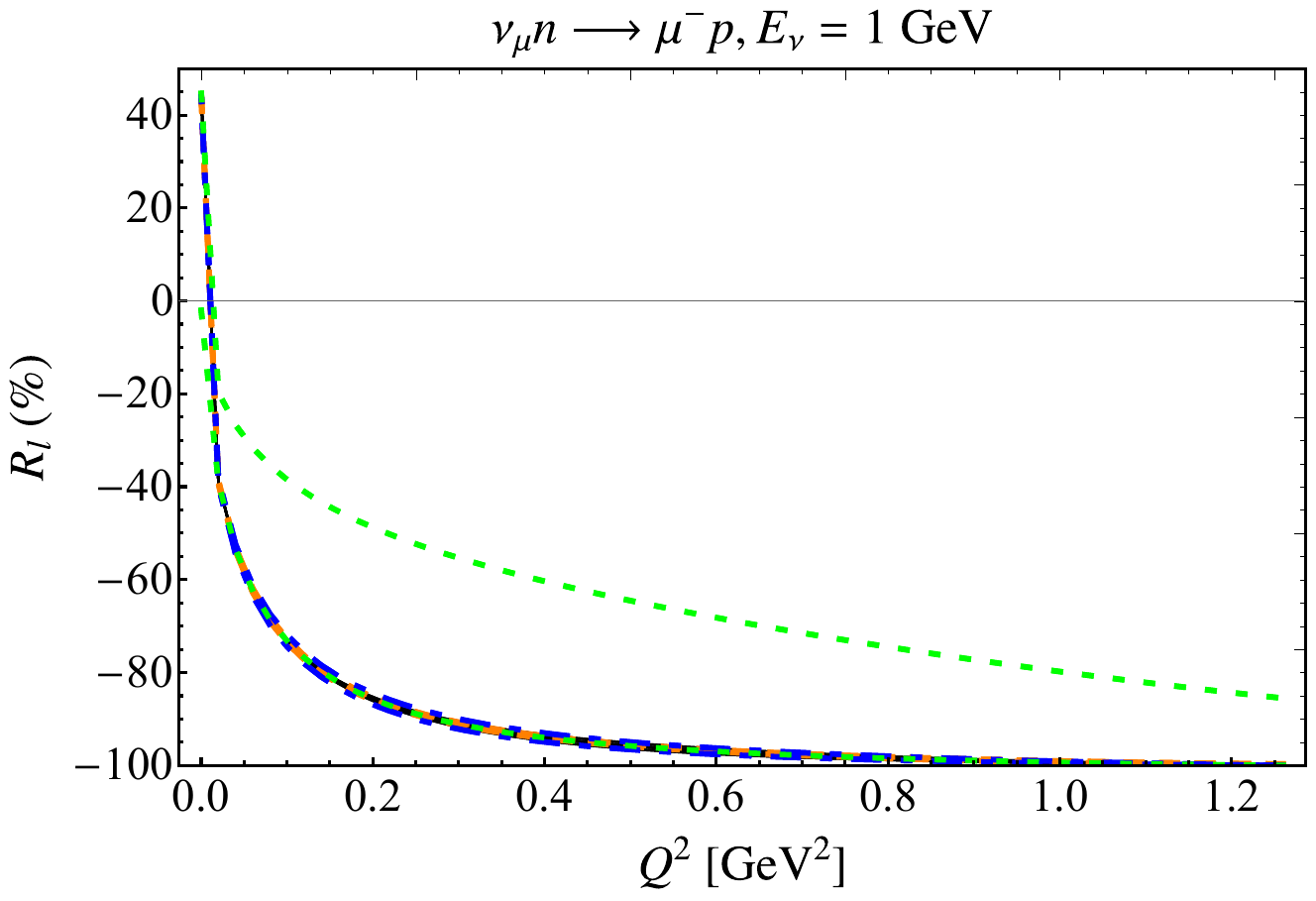}
\includegraphics[width=0.4\textwidth]{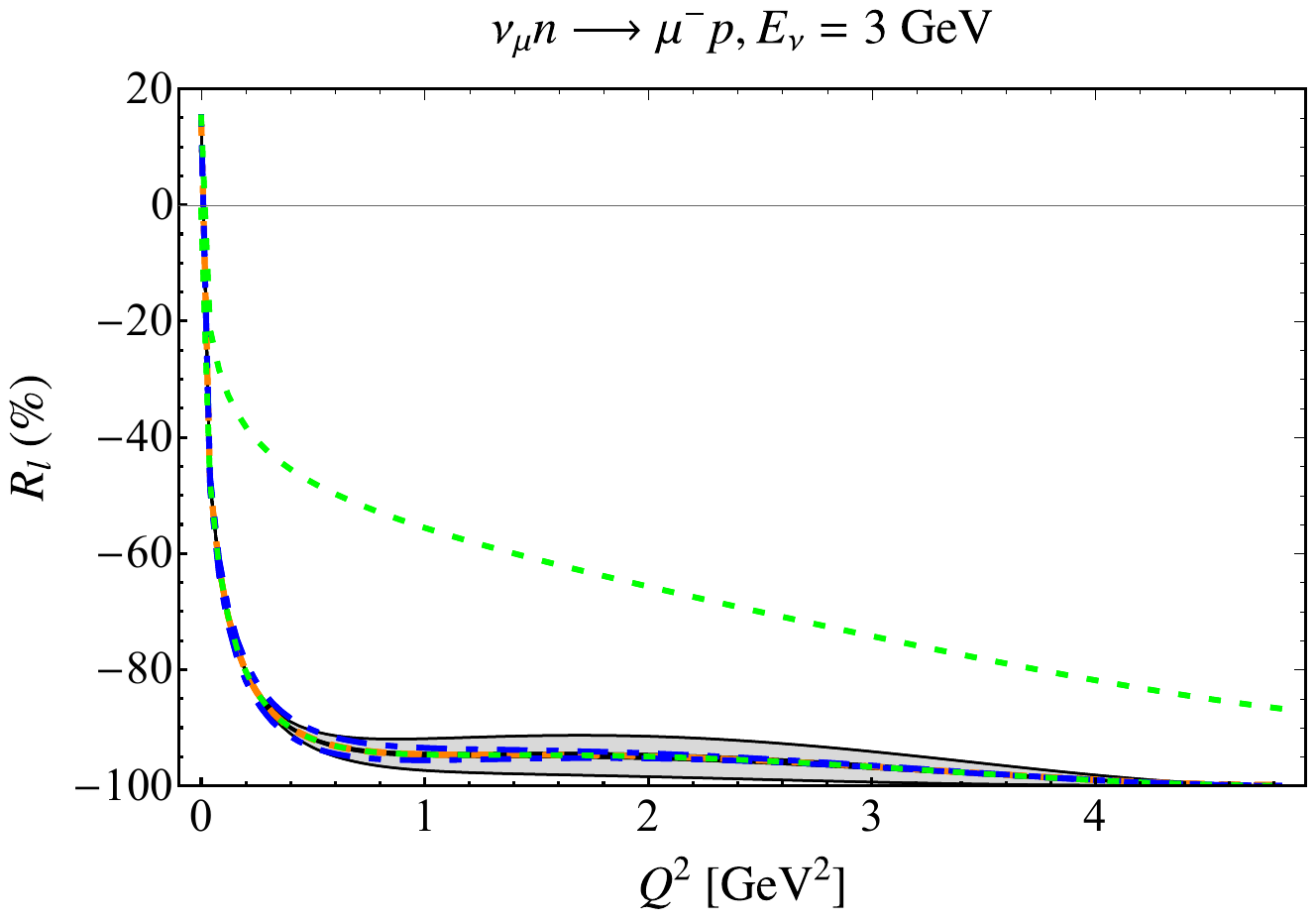}
\caption{Same as Fig.~\ref{fig:nu_Tt_SCFF} but for the longitudinal polarization observable $R_l$. \label{fig:nu_Rl_SCFF}}
\end{figure}

\begin{figure}[H]
\centering
\includegraphics[width=0.4\textwidth]{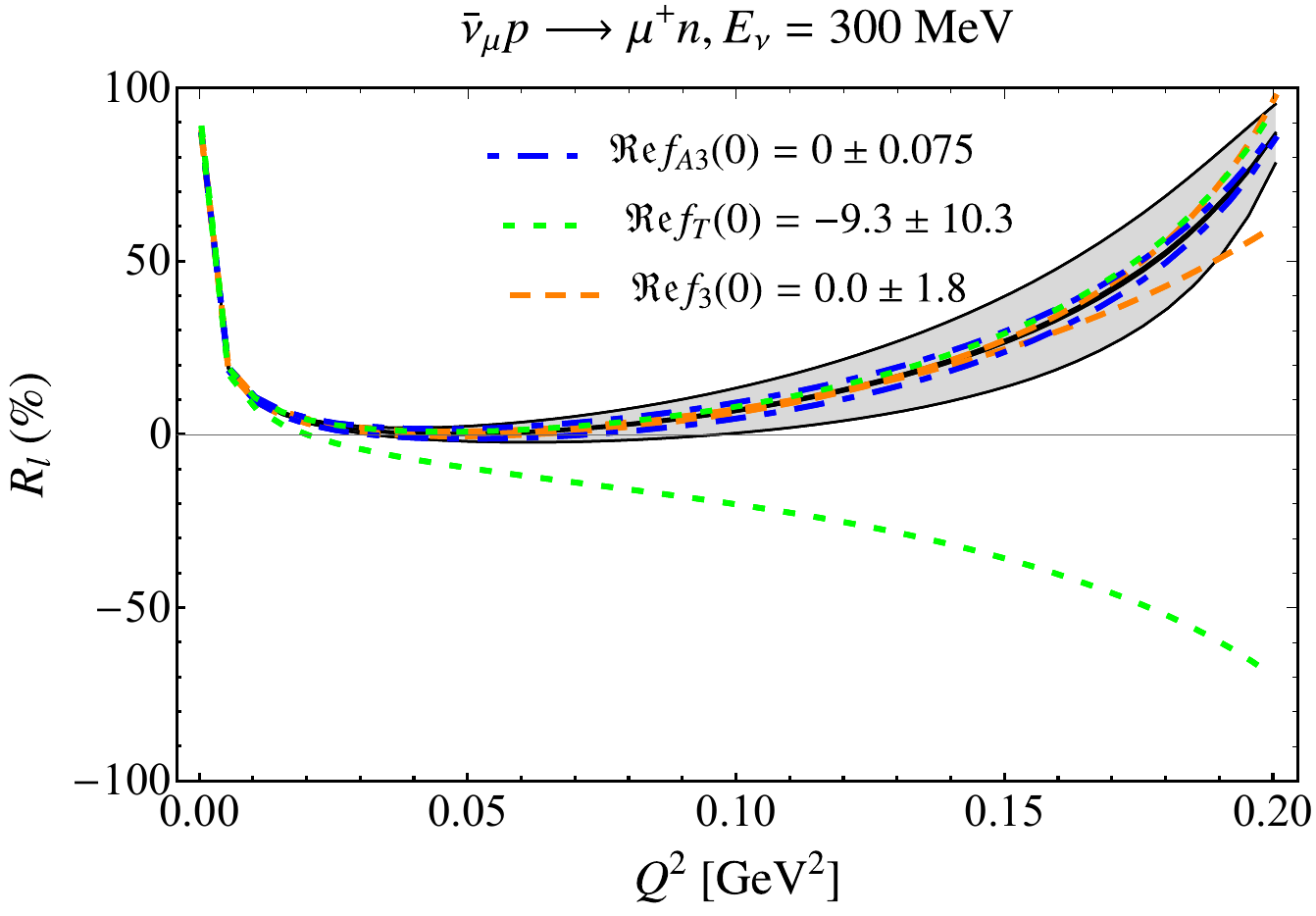}
\includegraphics[width=0.4\textwidth]{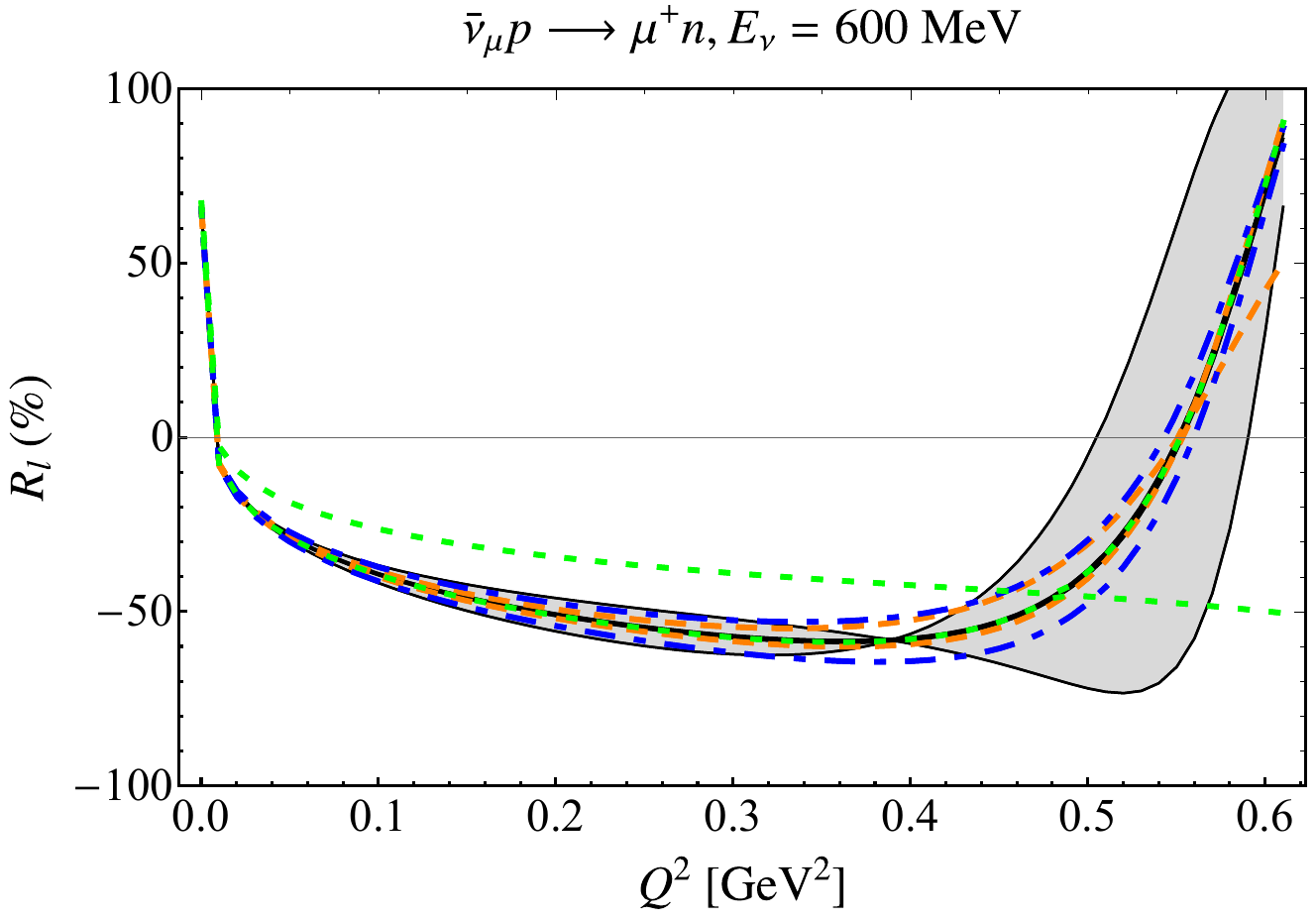}
\includegraphics[width=0.4\textwidth]{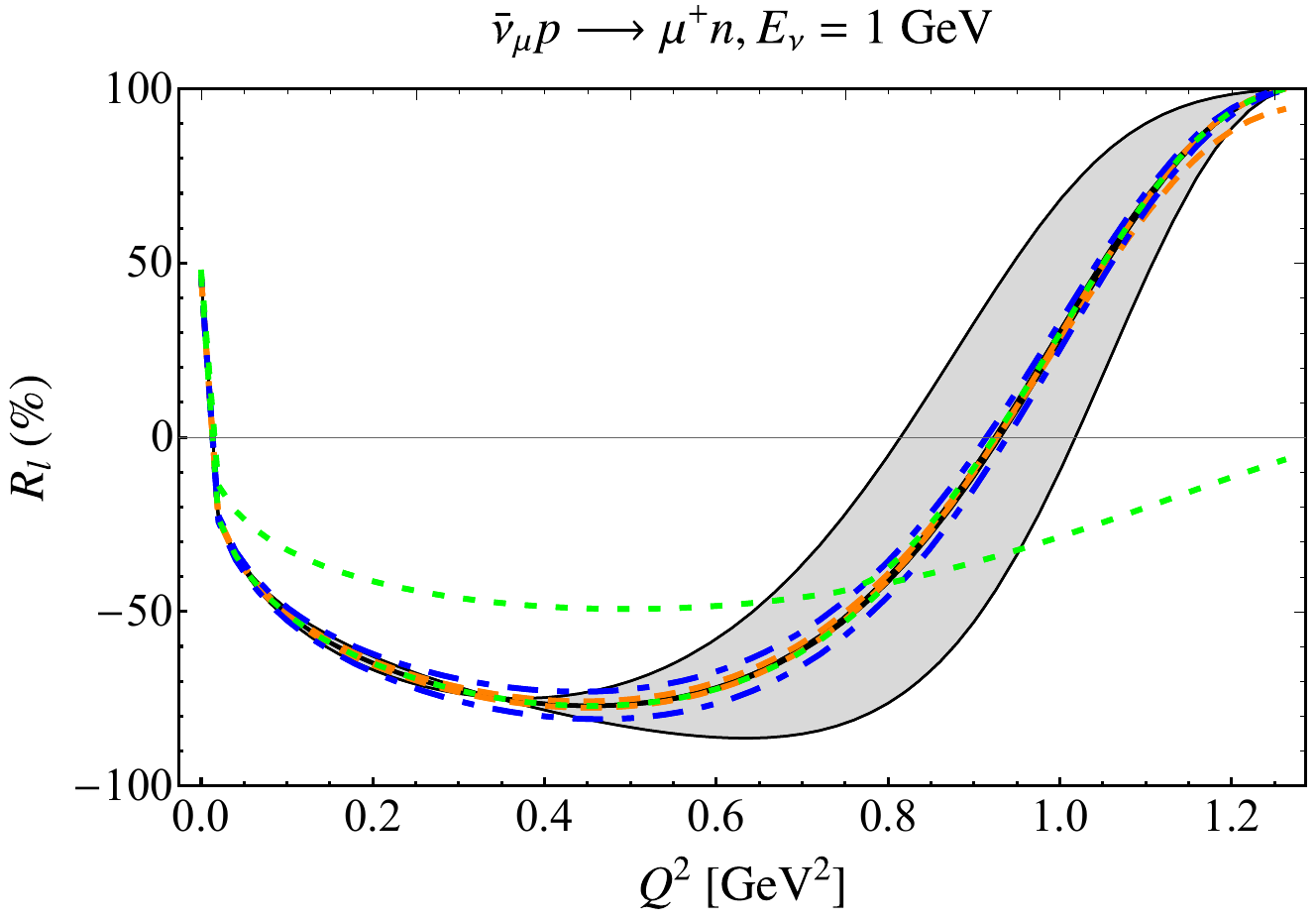}
\includegraphics[width=0.4\textwidth]{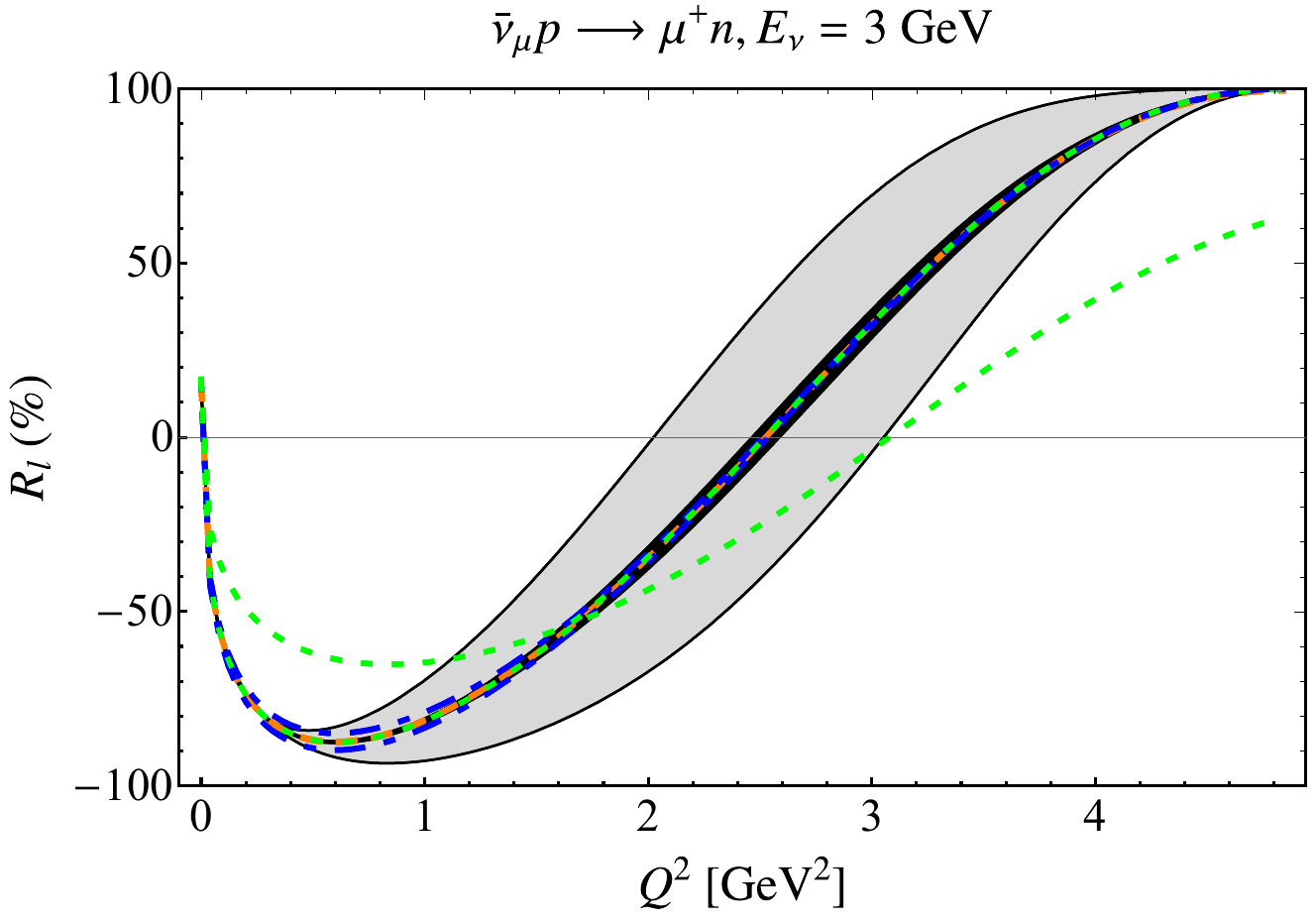}
\caption{Same as Fig.~\ref{fig:antinu_Tt_SCFF} but for the longitudinal polarization observable $R_l$. \label{fig:antinu_Rl_SCFF}}
\end{figure}

\begin{figure}[H]
\centering
\includegraphics[width=0.4\textwidth]{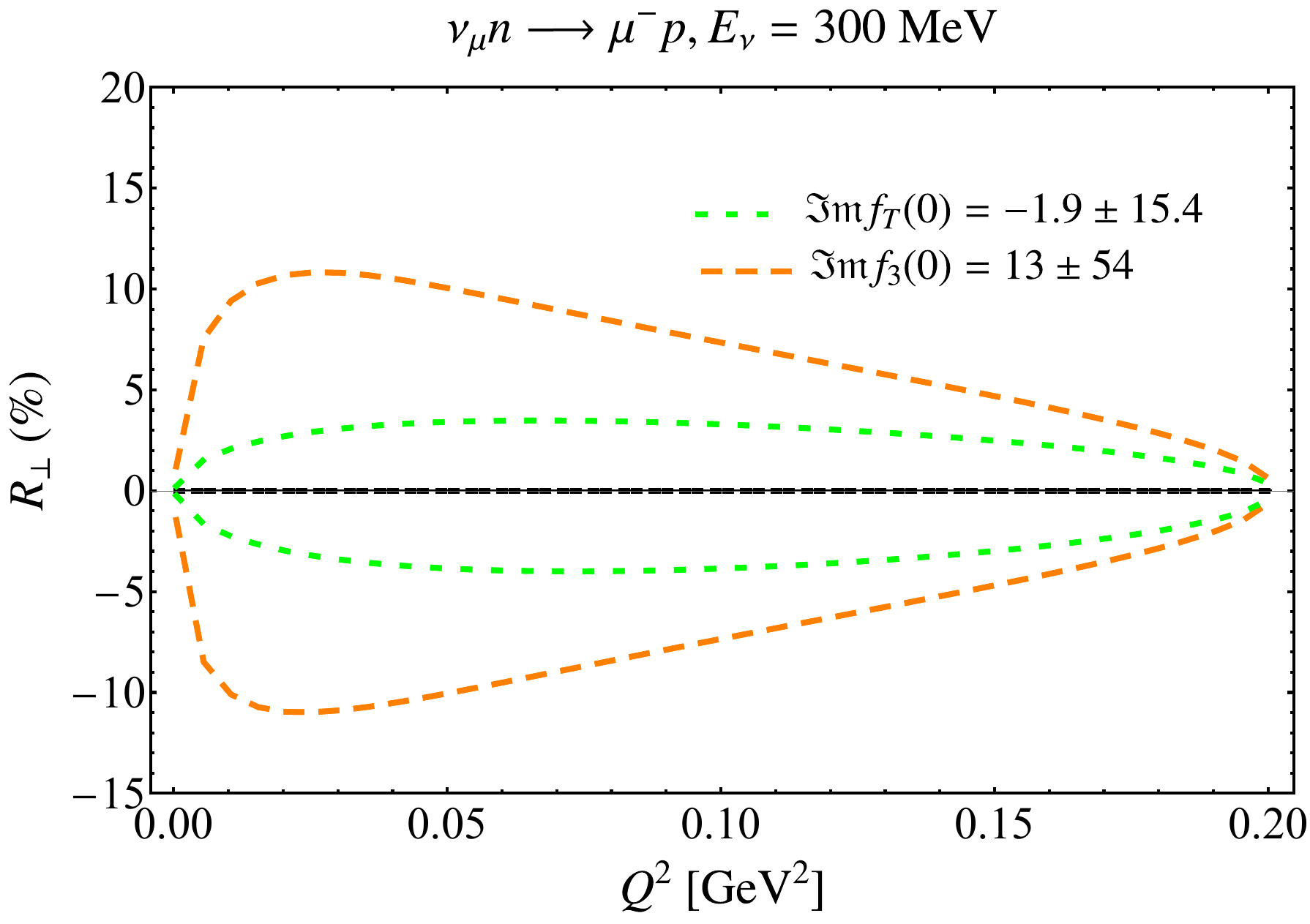}
\includegraphics[width=0.4\textwidth]{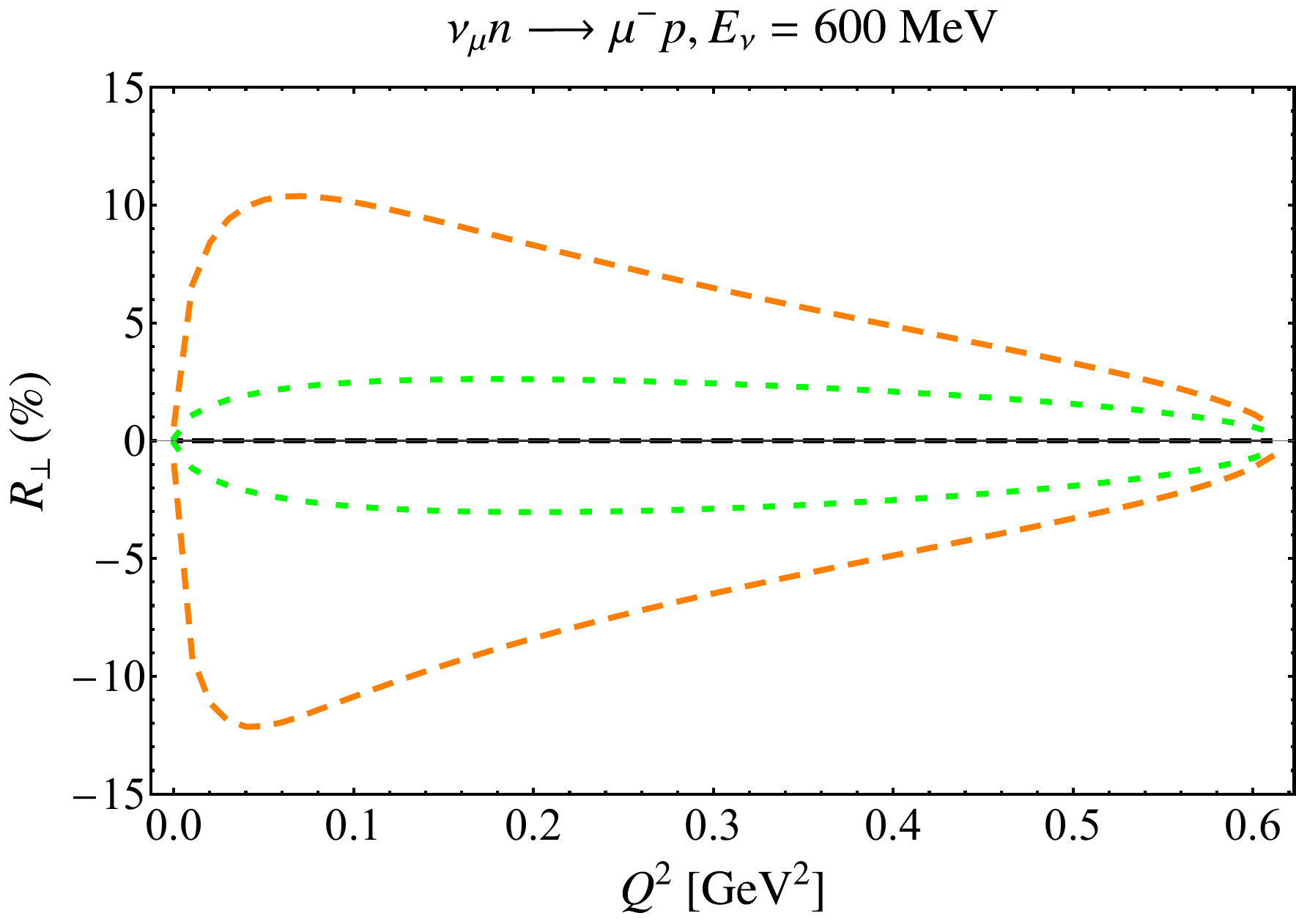}
\includegraphics[width=0.4\textwidth]{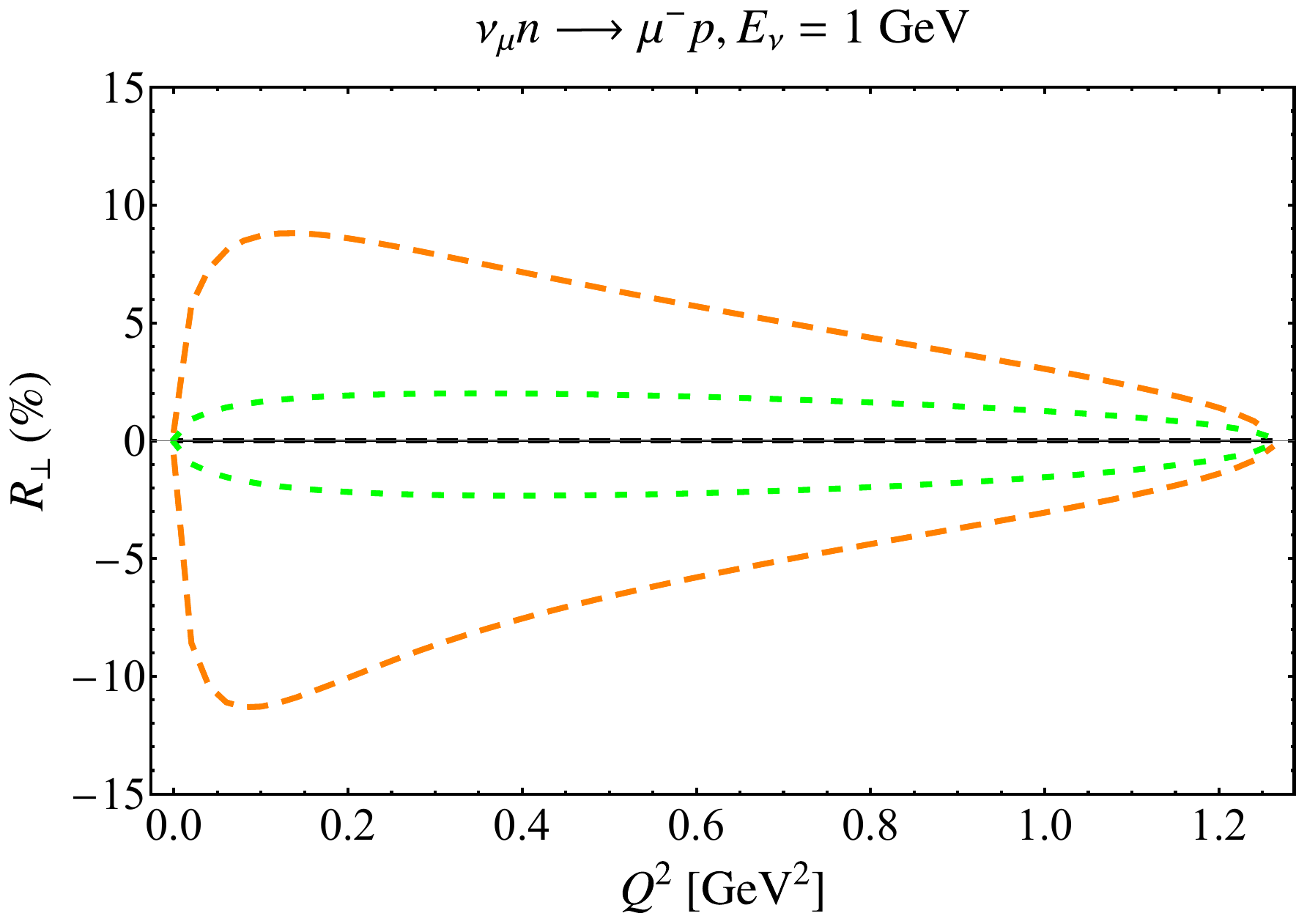}
\includegraphics[width=0.4\textwidth]{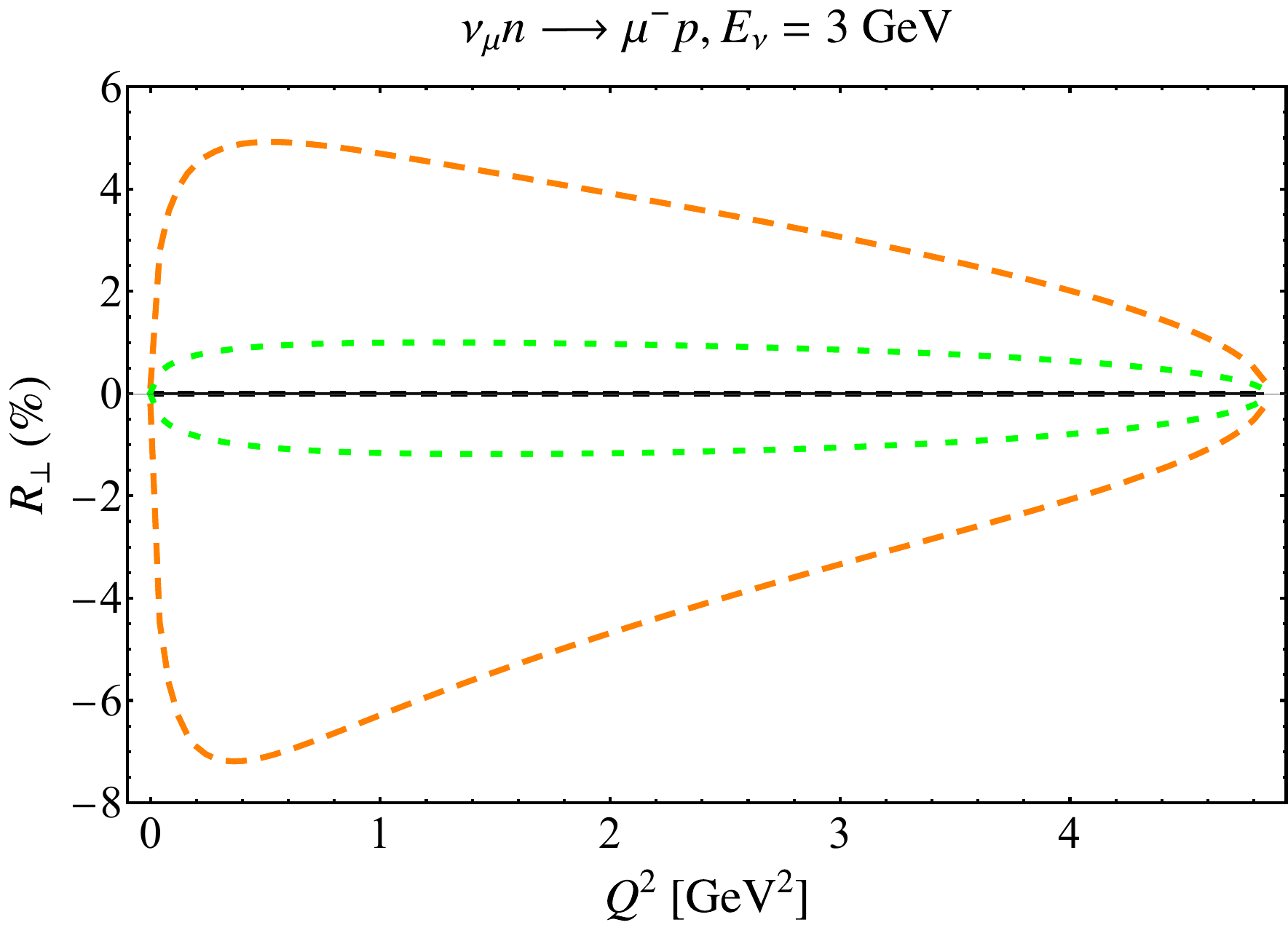}
\caption{Same as Fig.~\ref{fig:nu_Tt_SCFF} but for the transverse polarization observable $R_\perp$ and imaginary amplitudes. \label{fig:nu_RT_SCFF}}
\end{figure}

\begin{figure}[H]
\centering
\includegraphics[width=0.4\textwidth]{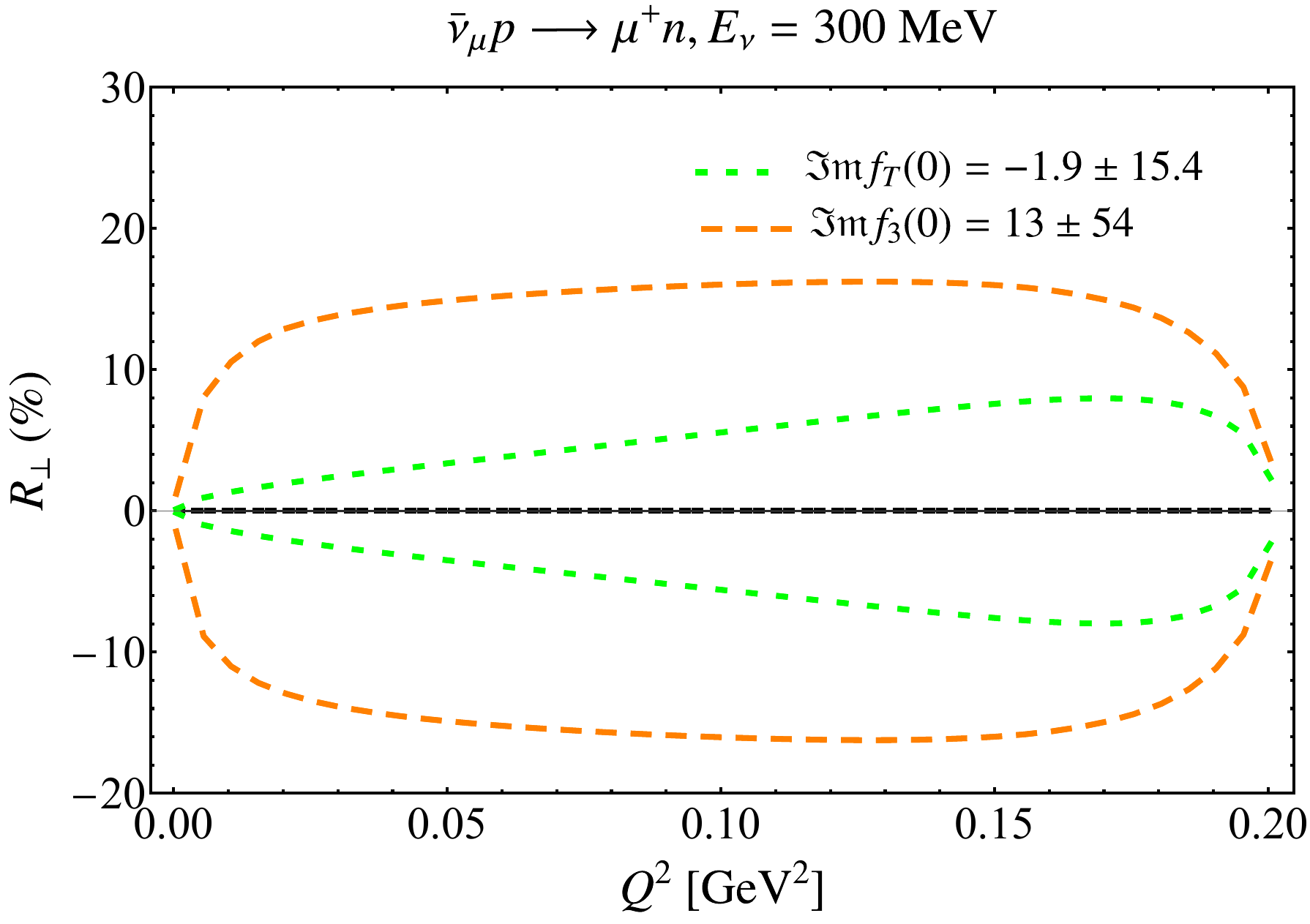}
\includegraphics[width=0.4\textwidth]{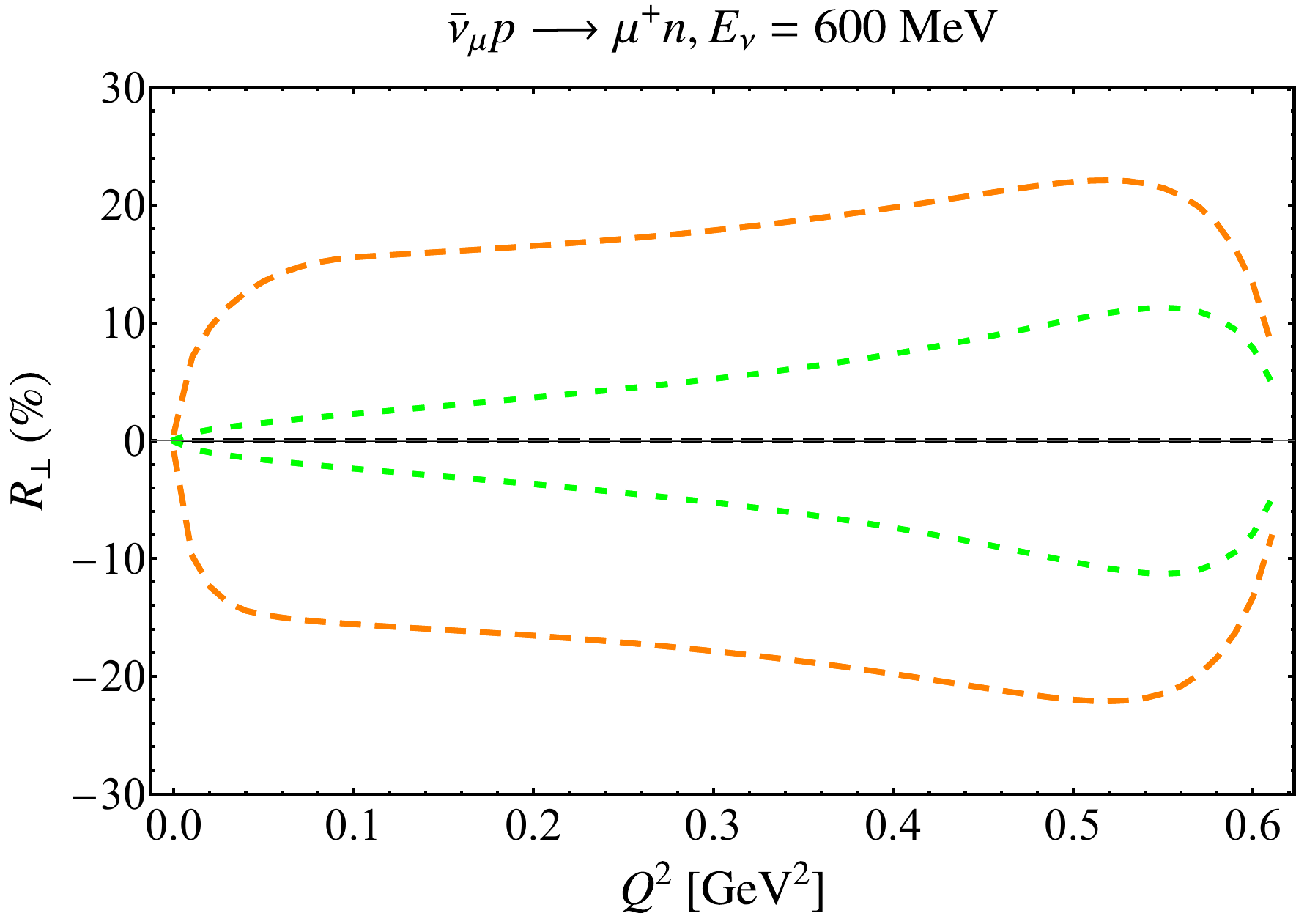}
\includegraphics[width=0.4\textwidth]{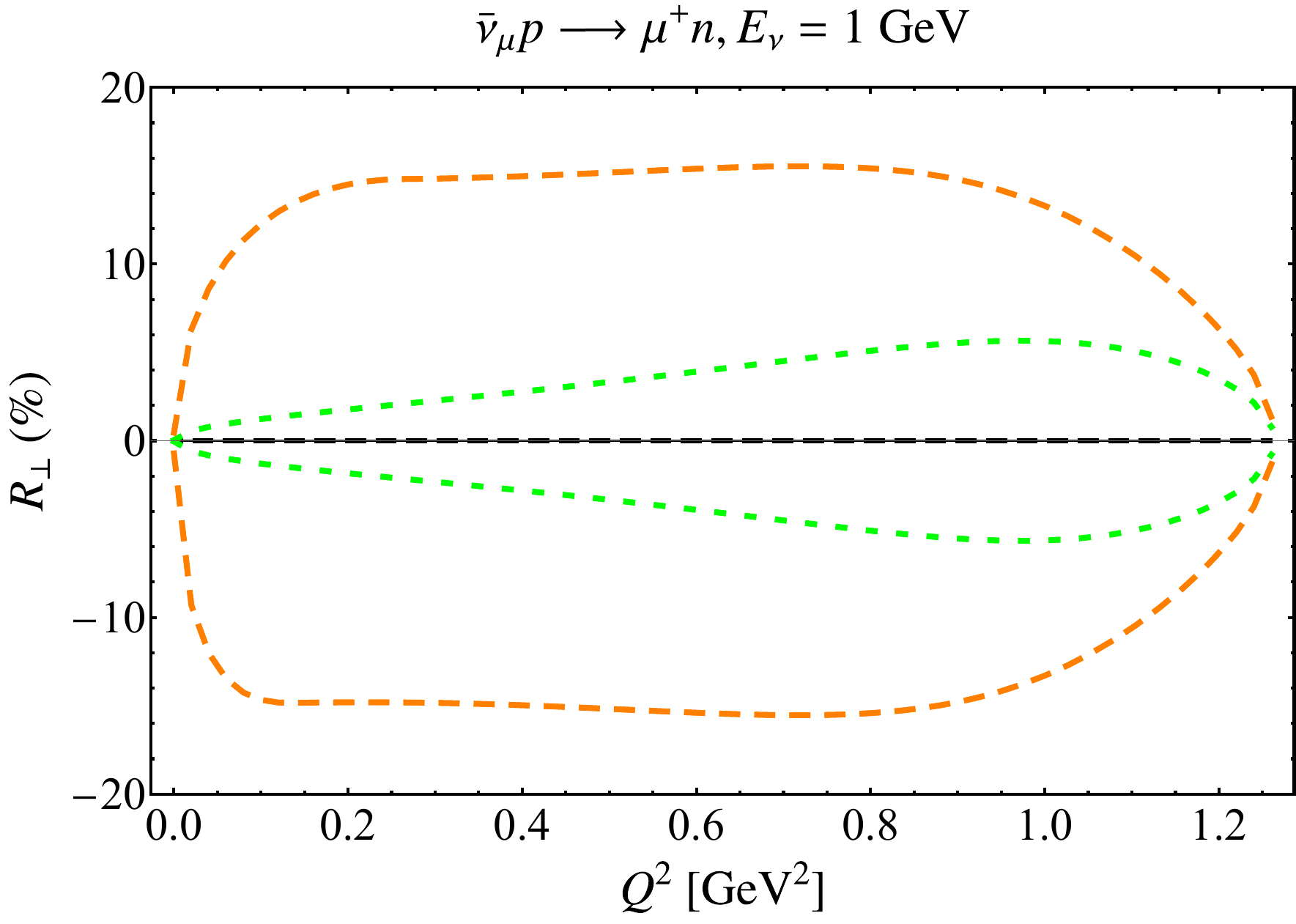}
\includegraphics[width=0.4\textwidth]{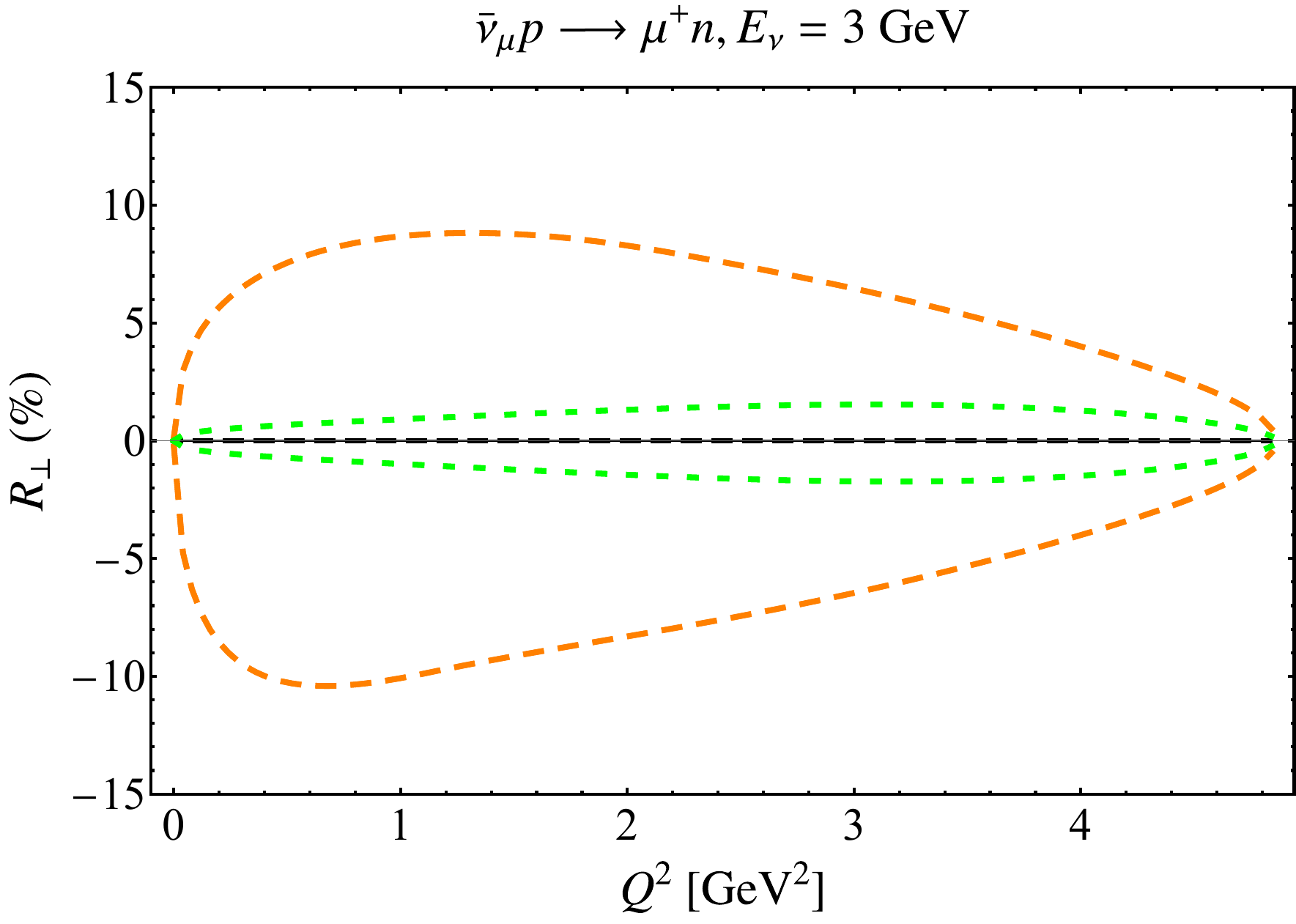}
\caption{Same as Fig.~\ref{fig:antinu_Tt_SCFF} but for the transverse polarization observable $R_\perp$ and imaginary amplitudes. \label{fig:antinu_RT_SCFF}}
\end{figure}

\begin{figure}[H]
\centering
\includegraphics[width=0.4\textwidth]{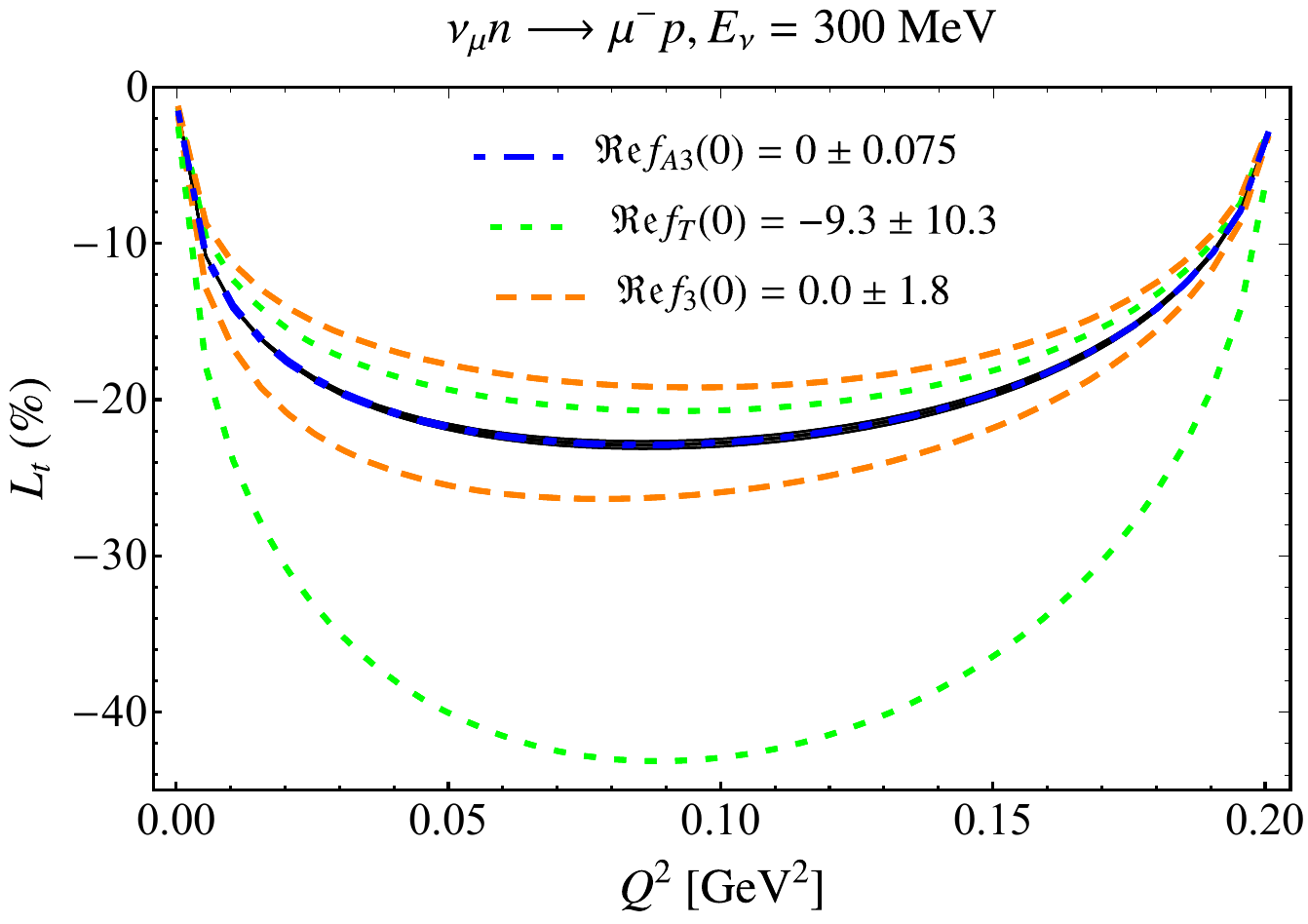}
\includegraphics[width=0.4\textwidth]{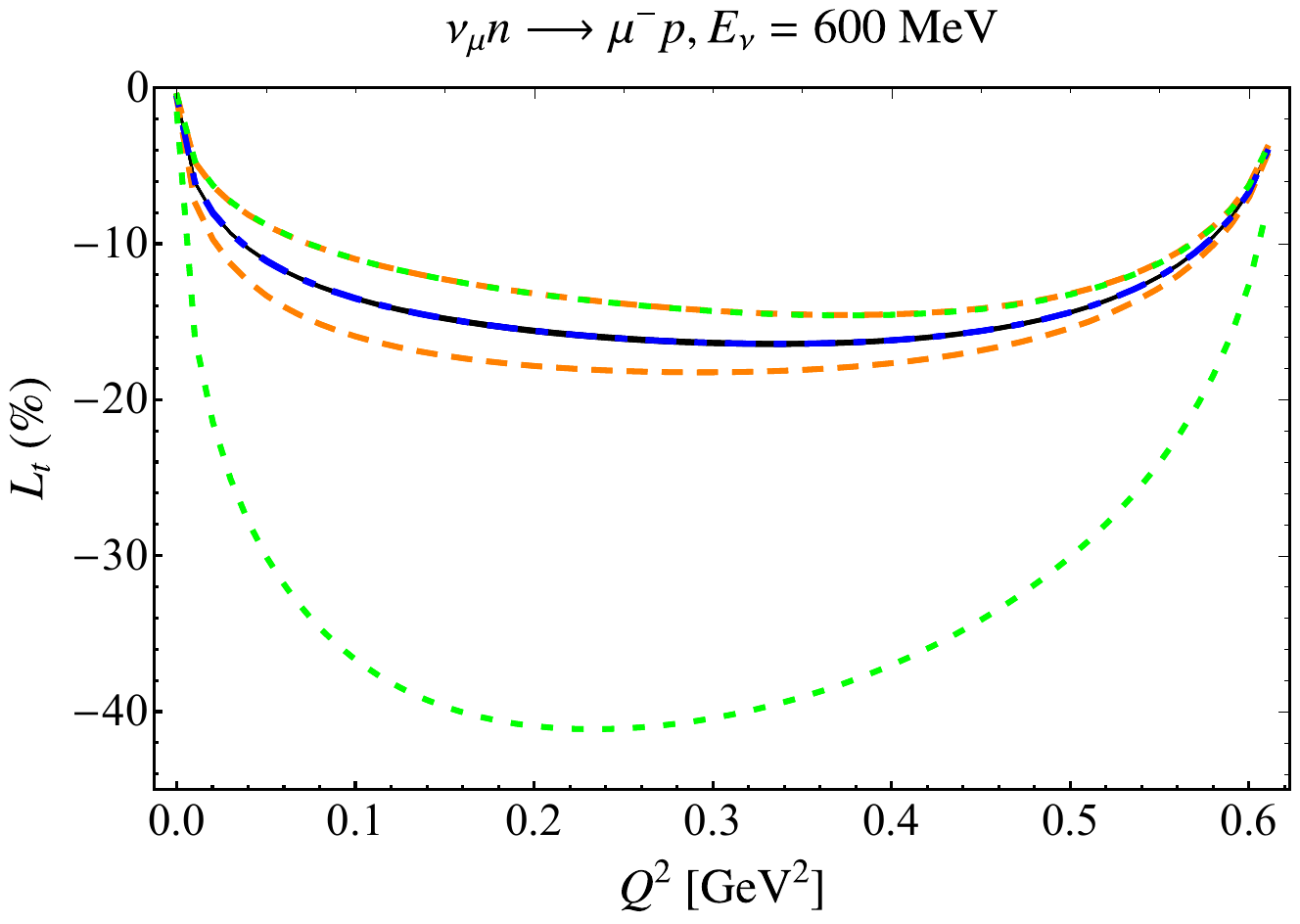}
\includegraphics[width=0.4\textwidth]{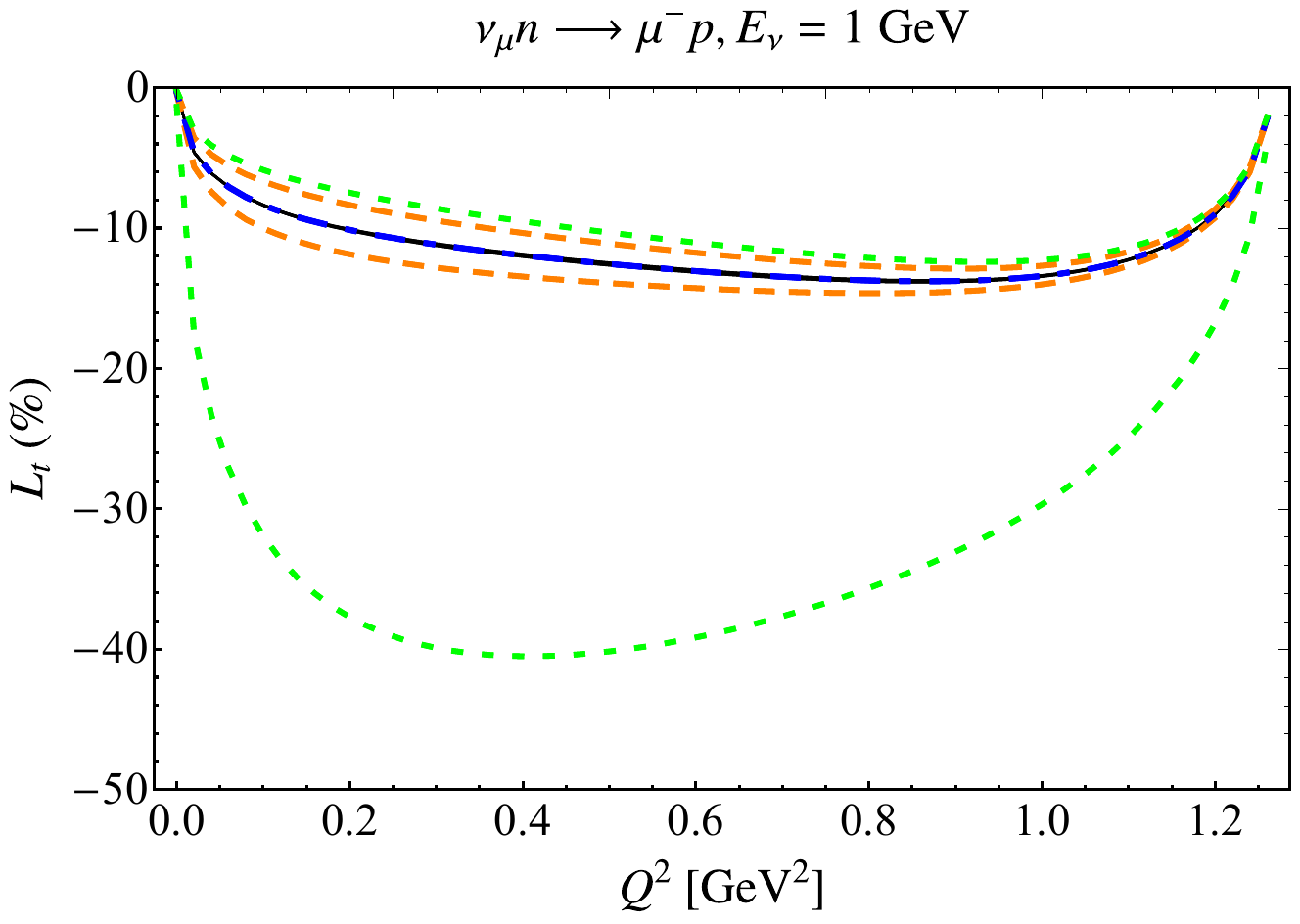}
\includegraphics[width=0.4\textwidth]{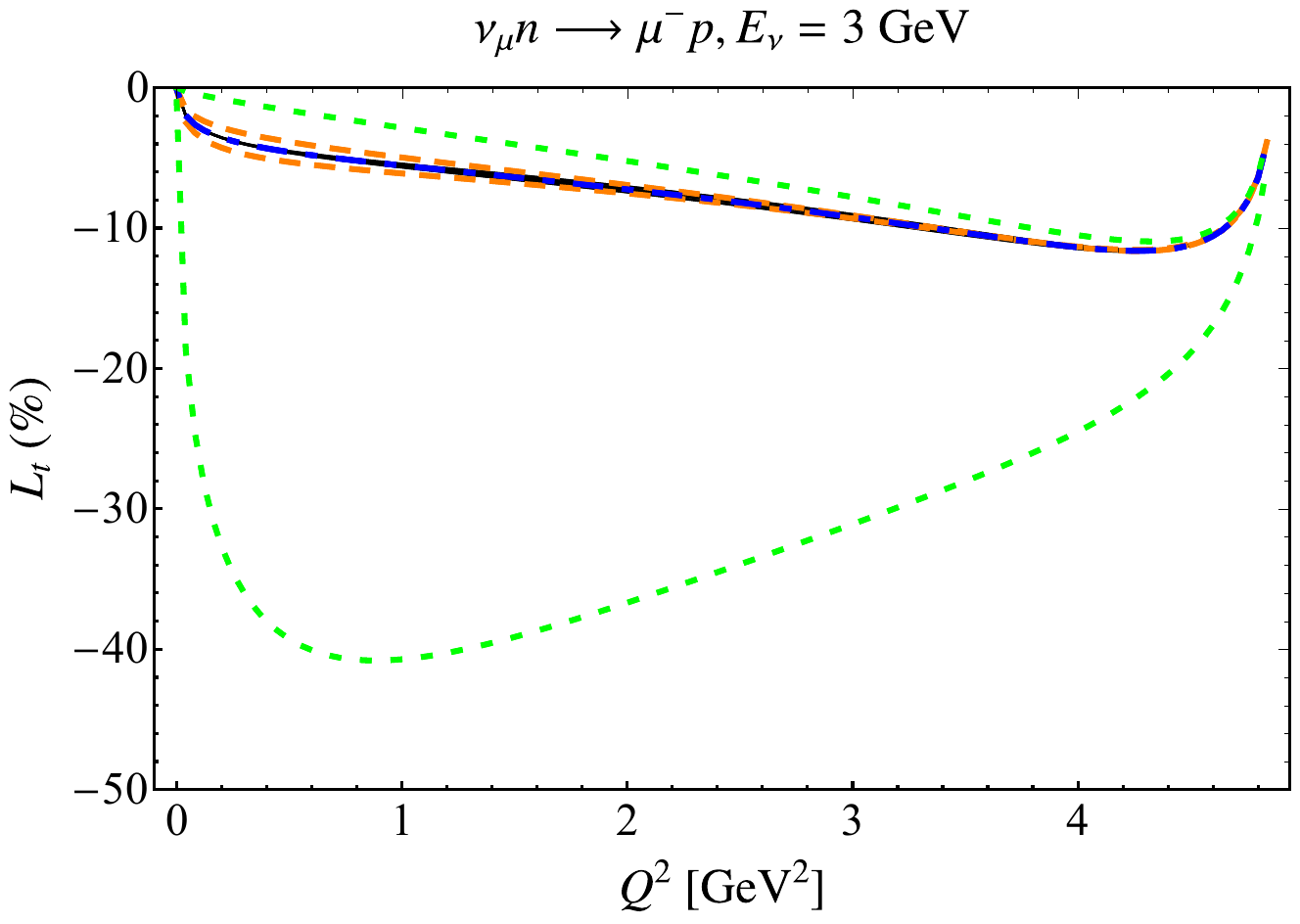}
\caption{Same as Fig.~\ref{fig:nu_Tt_SCFF} but for the transverse polarization observable $L_t$. \label{fig:nu_Lt_SCFF}}
\end{figure}

\begin{figure}[H]
\centering
\includegraphics[width=0.4\textwidth]{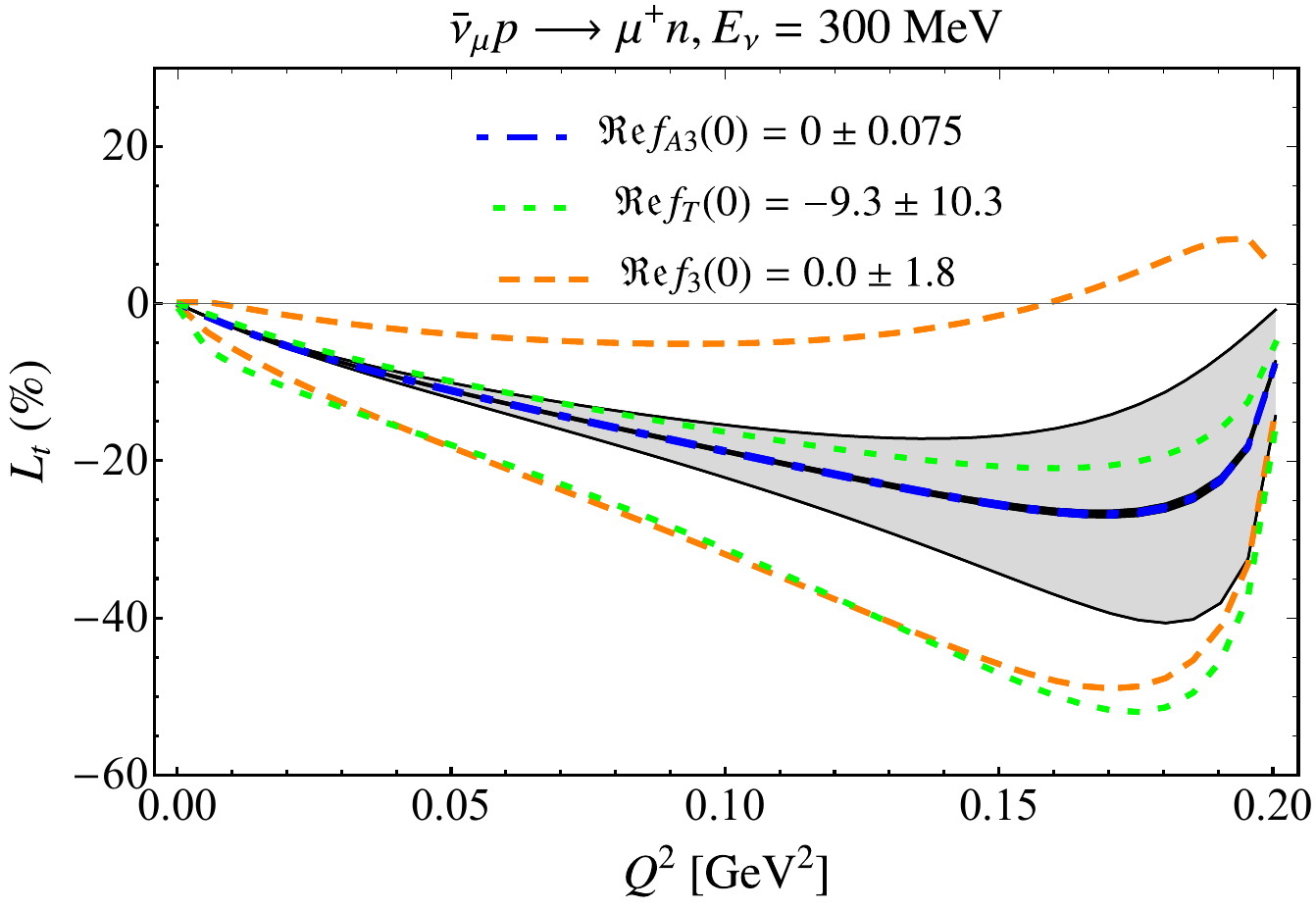}
\includegraphics[width=0.4\textwidth]{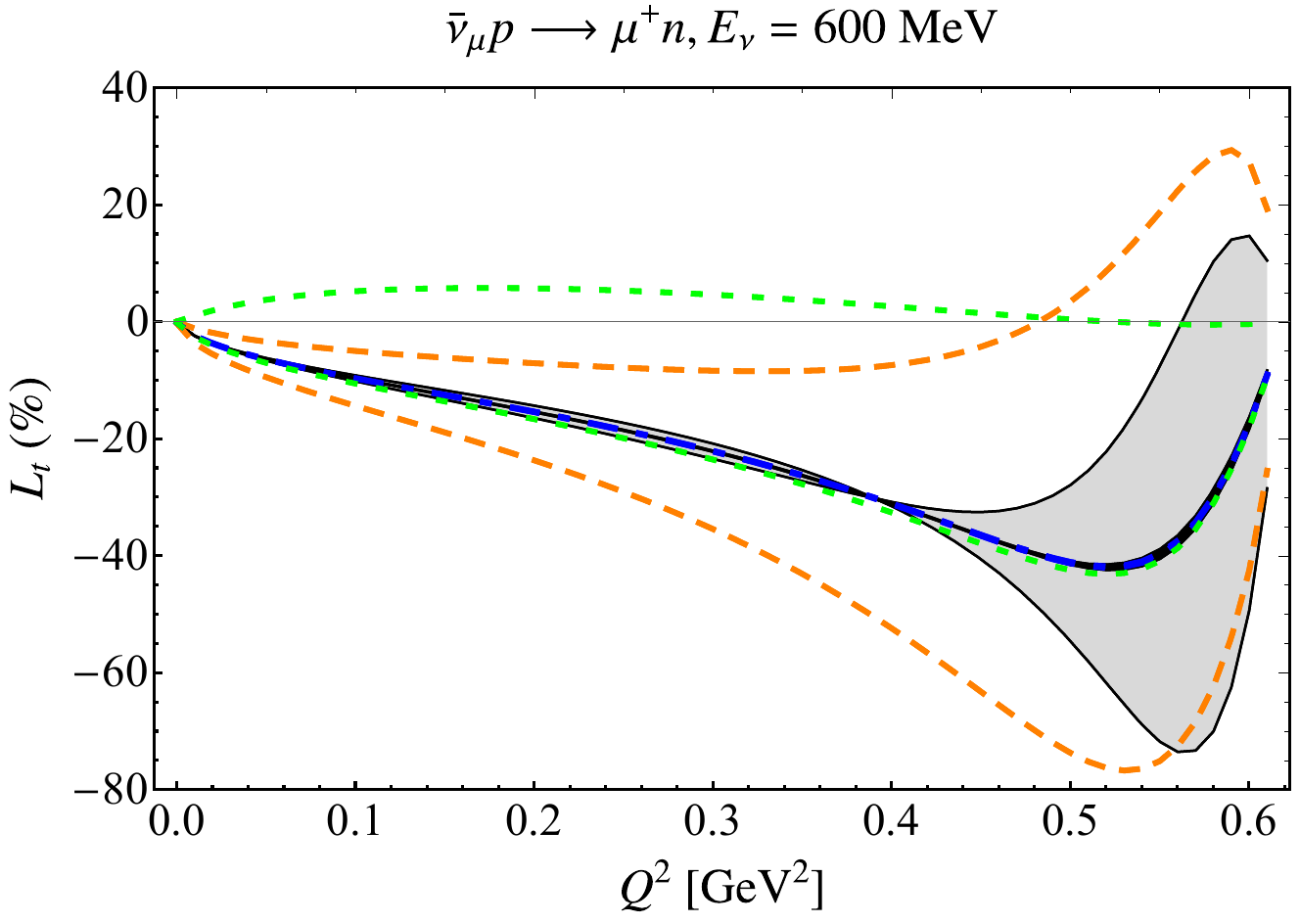}
\includegraphics[width=0.4\textwidth]{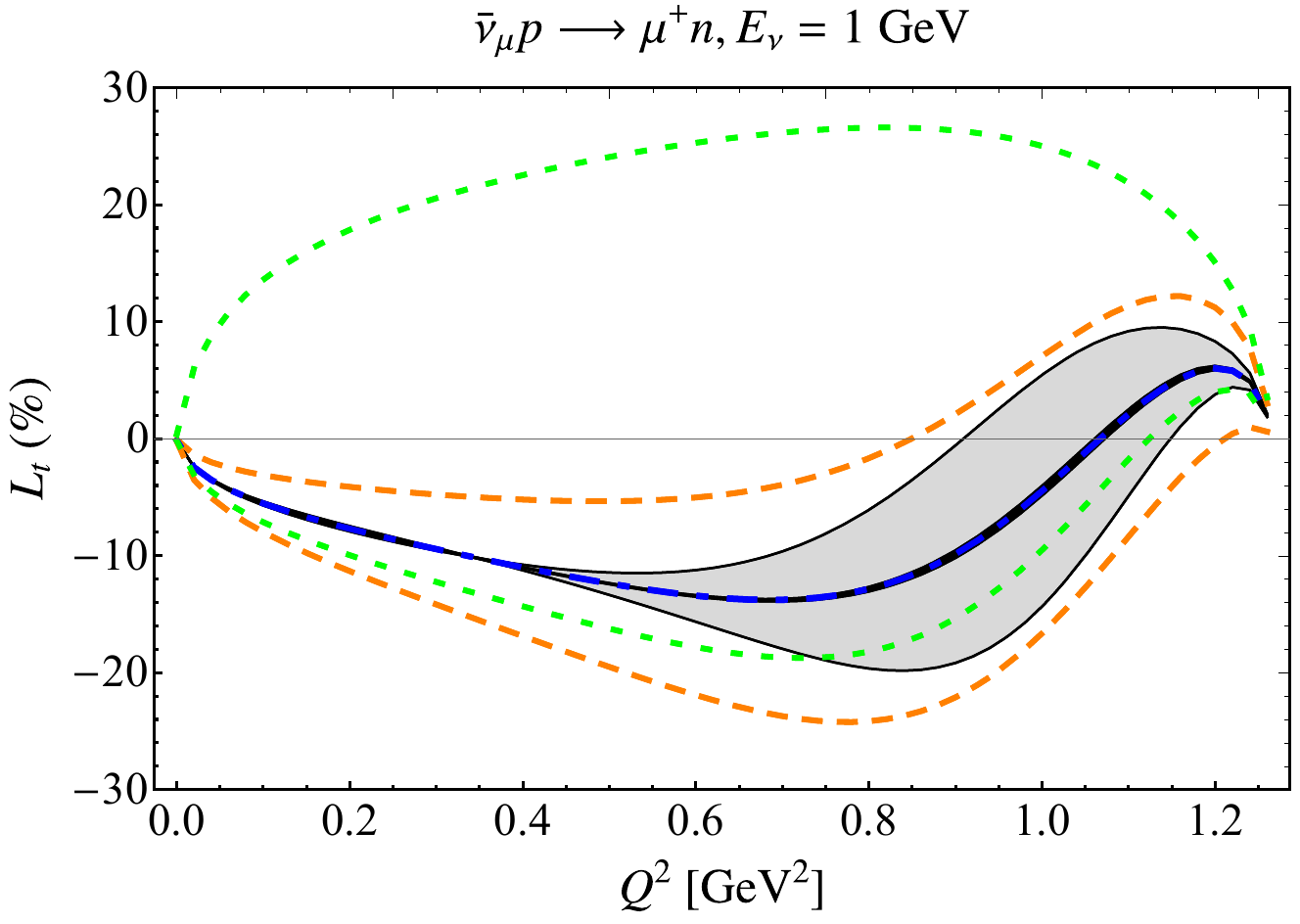}
\includegraphics[width=0.4\textwidth]{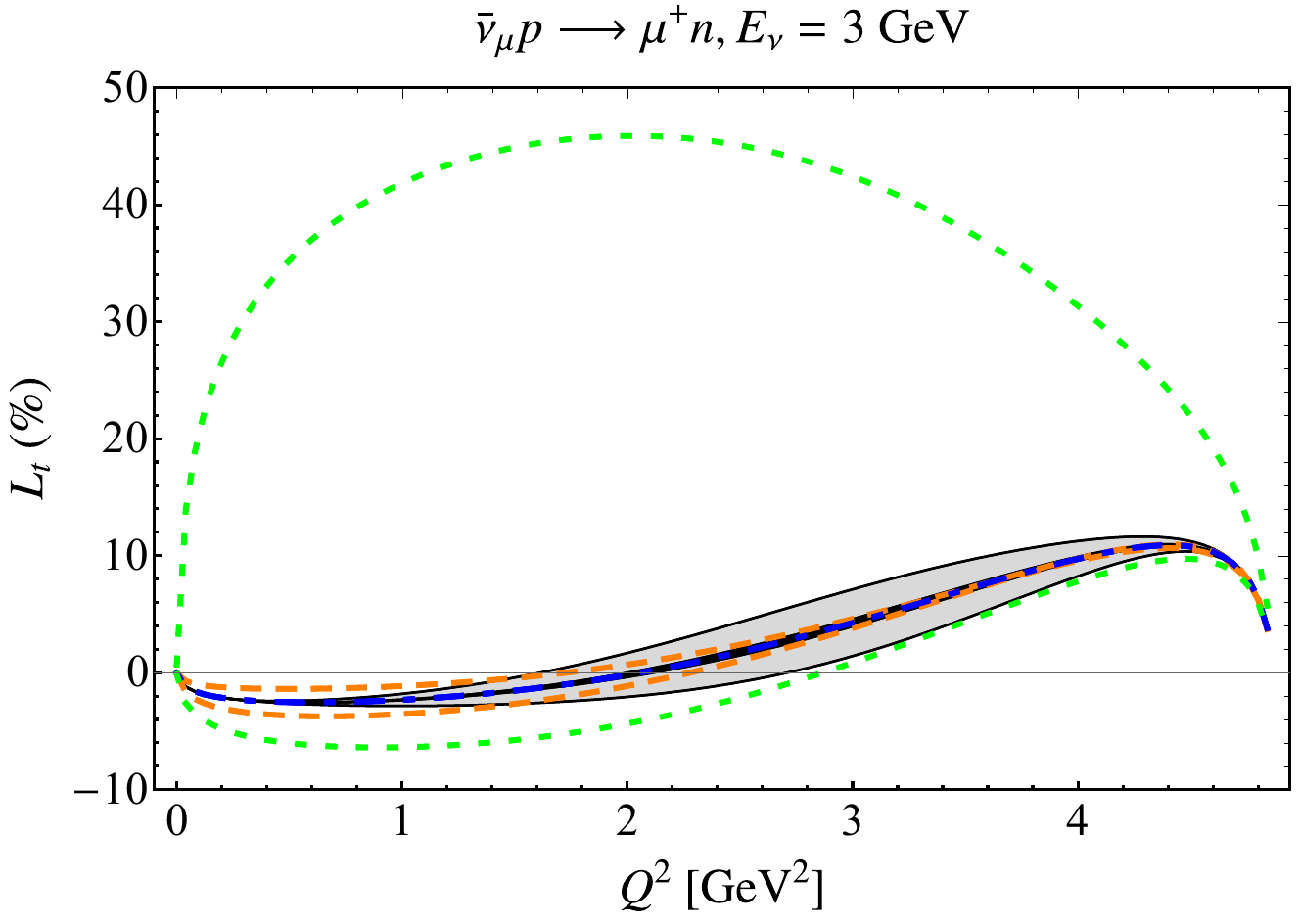}
\caption{Same as Fig.~\ref{fig:antinu_Tt_SCFF} but for the transverse polarization observable $L_t$. \label{fig:antinu_Lt_SCFF}}
\end{figure}

\begin{figure}[H]
\centering
\includegraphics[width=0.4\textwidth]{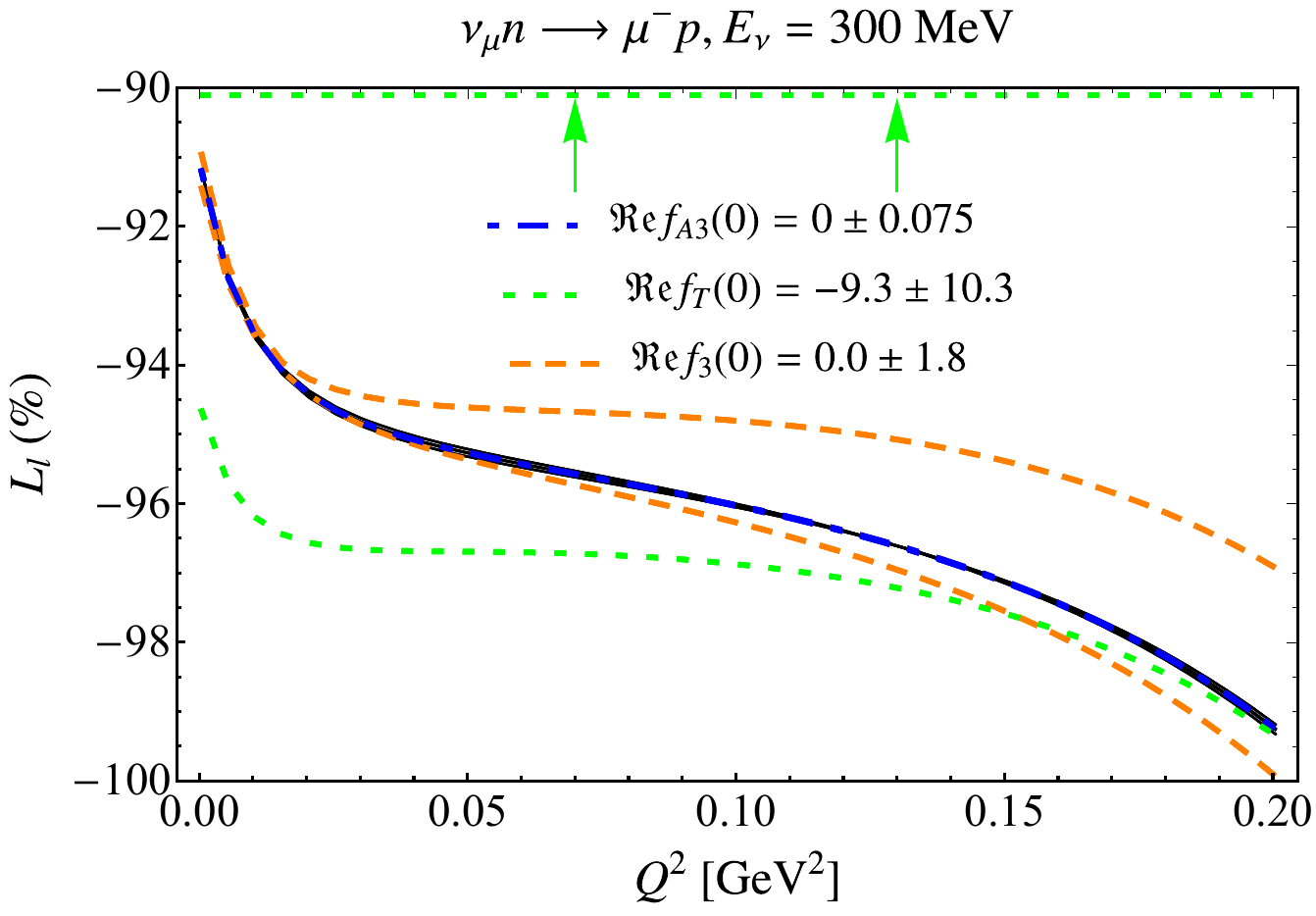}
\includegraphics[width=0.4\textwidth]{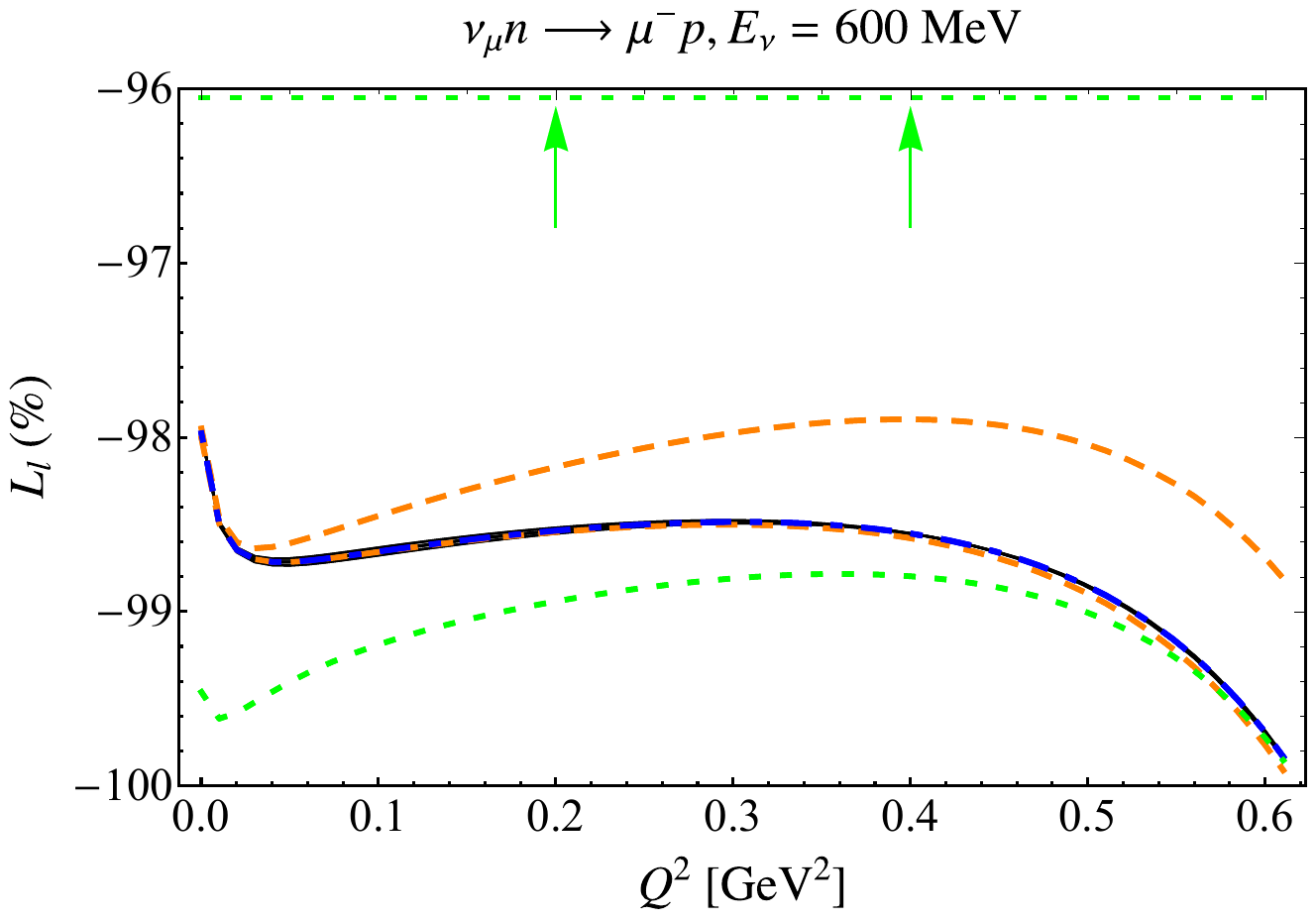}
\includegraphics[width=0.4\textwidth]{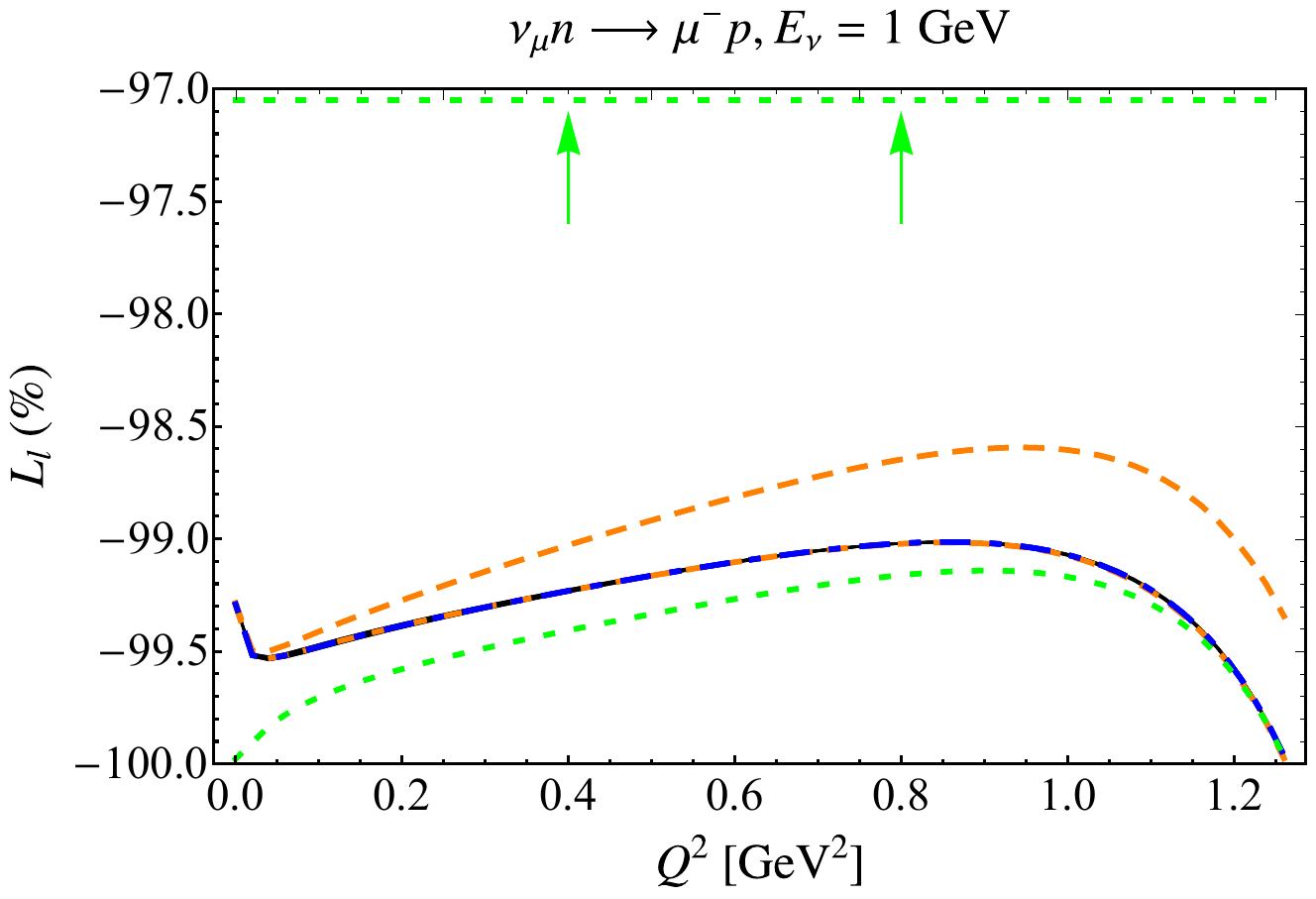}
\includegraphics[width=0.4\textwidth]{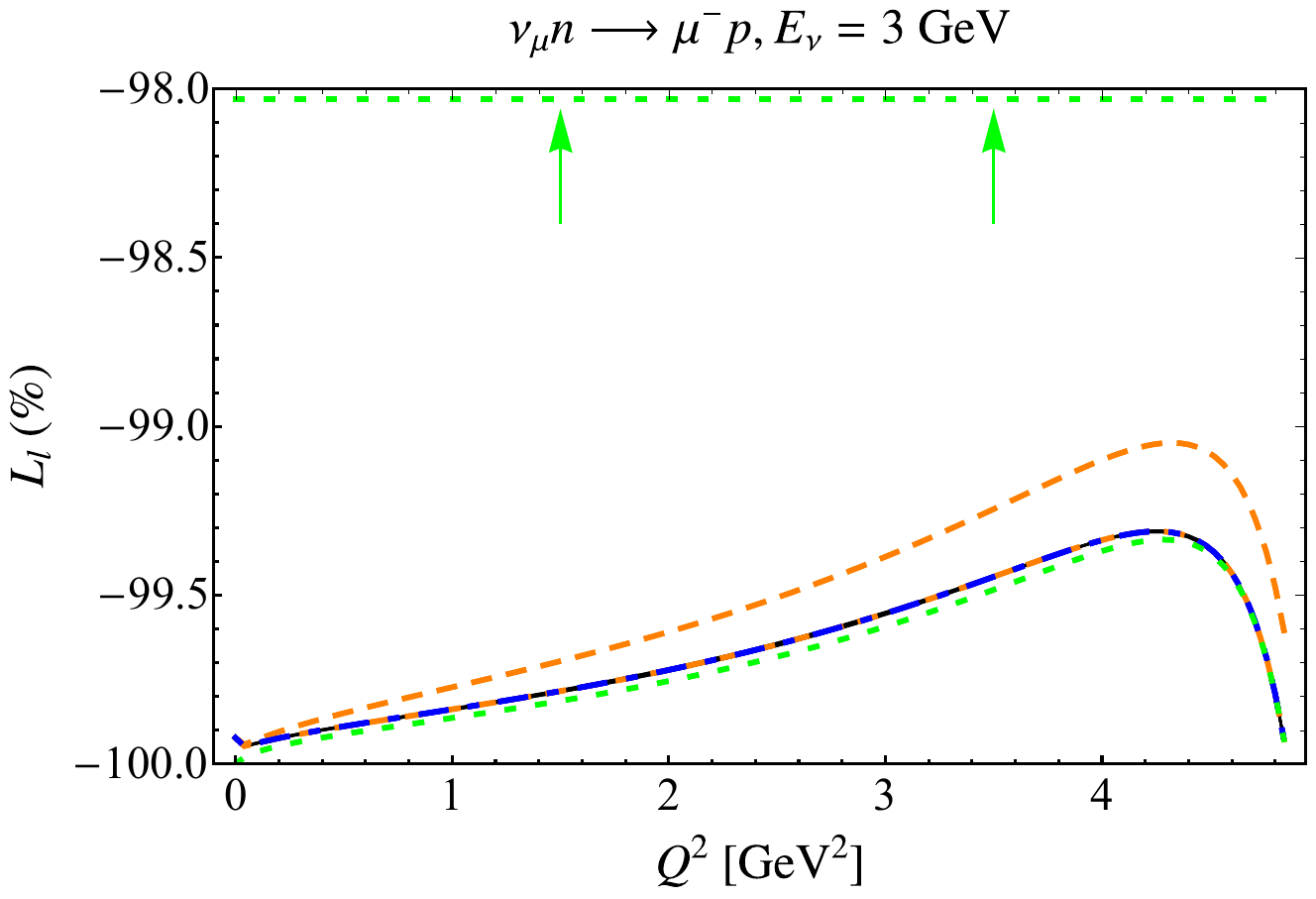}
\caption{Same as Fig.~\ref{fig:nu_Tt_SCFF} but for the longitudinal polarization observable $L_l$. \label{fig:nu_Ll_SCFF}}
\end{figure}

\begin{figure}[H]
\centering
\includegraphics[width=0.4\textwidth]{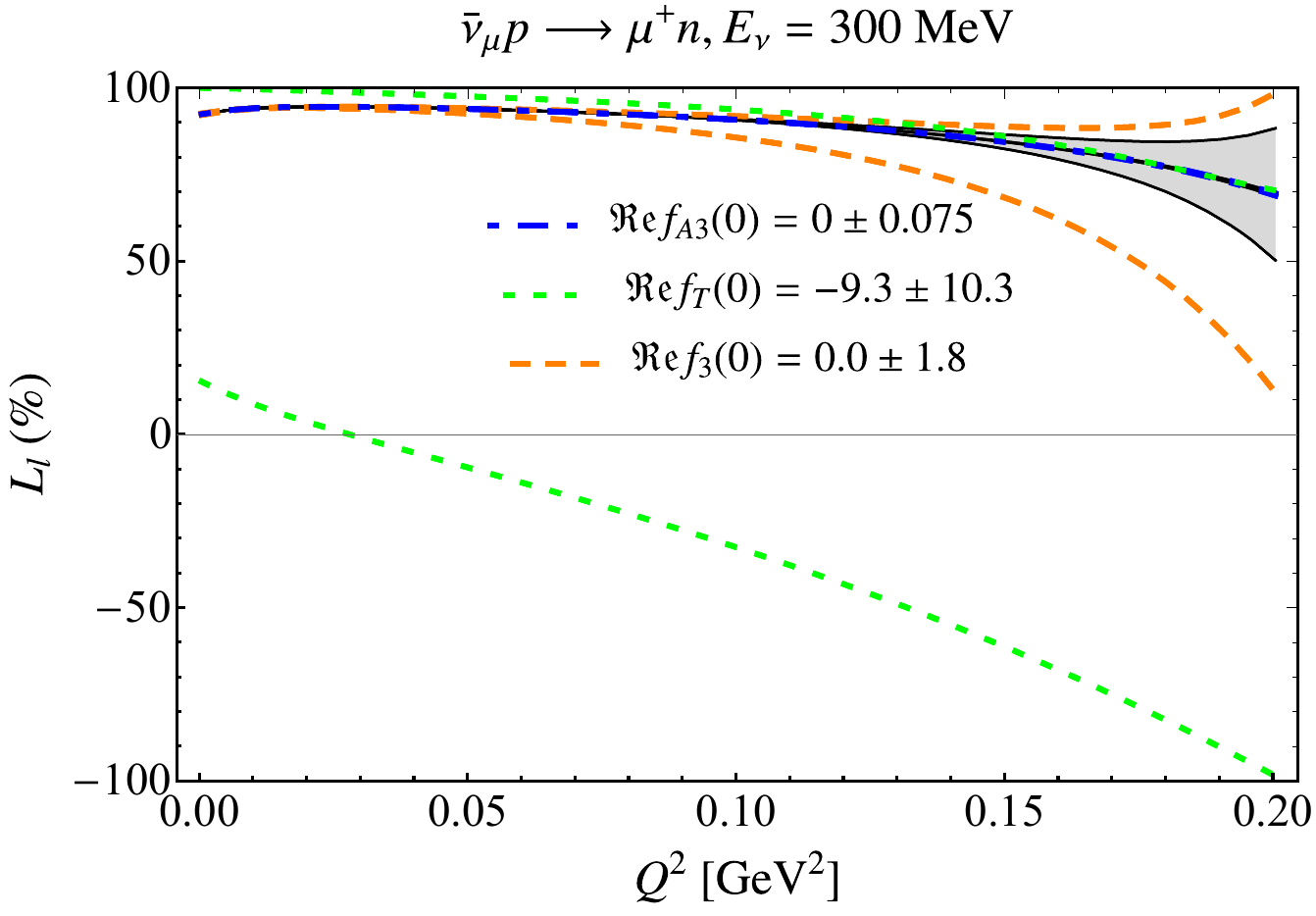}
\includegraphics[width=0.4\textwidth]{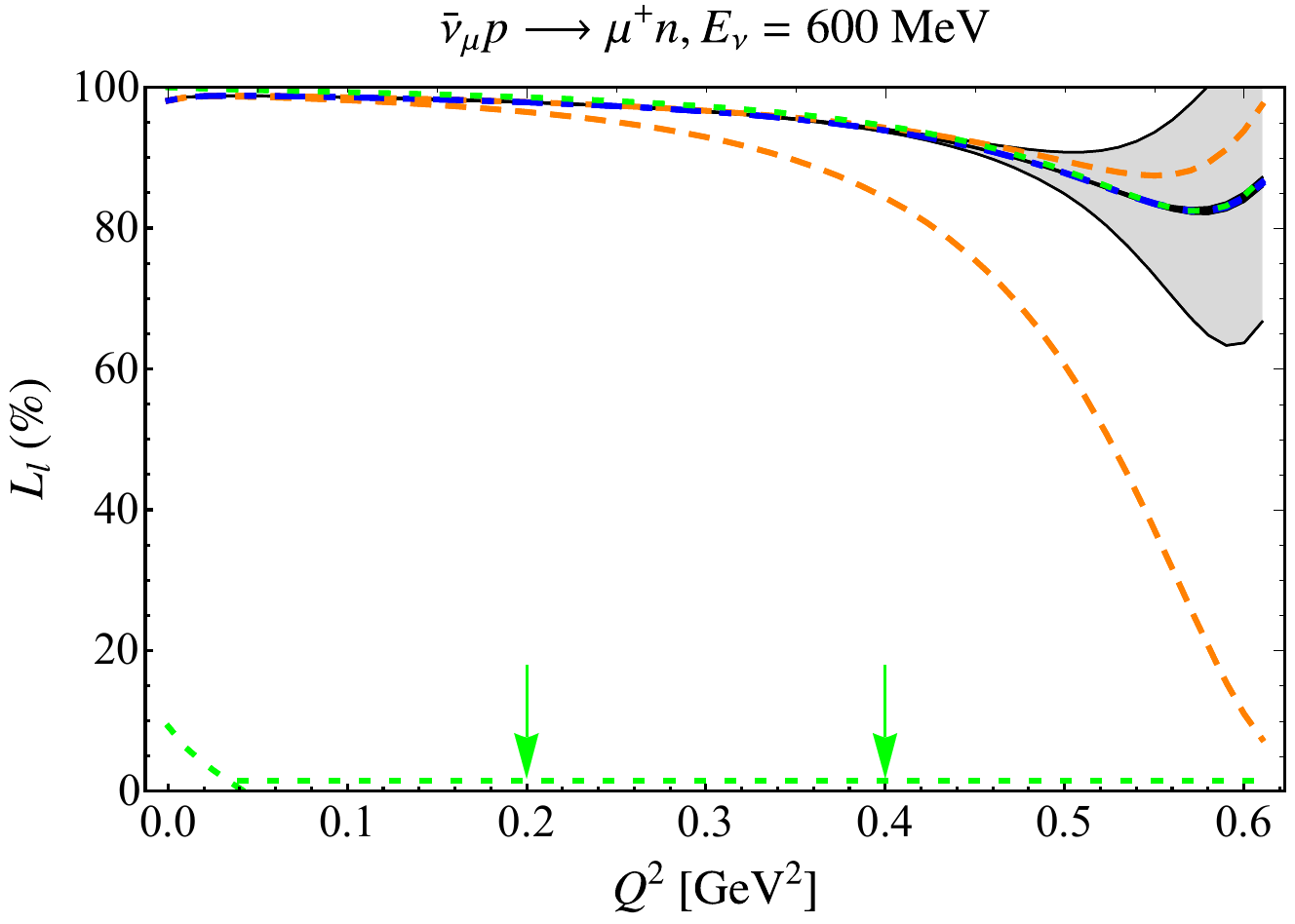}
\includegraphics[width=0.4\textwidth]{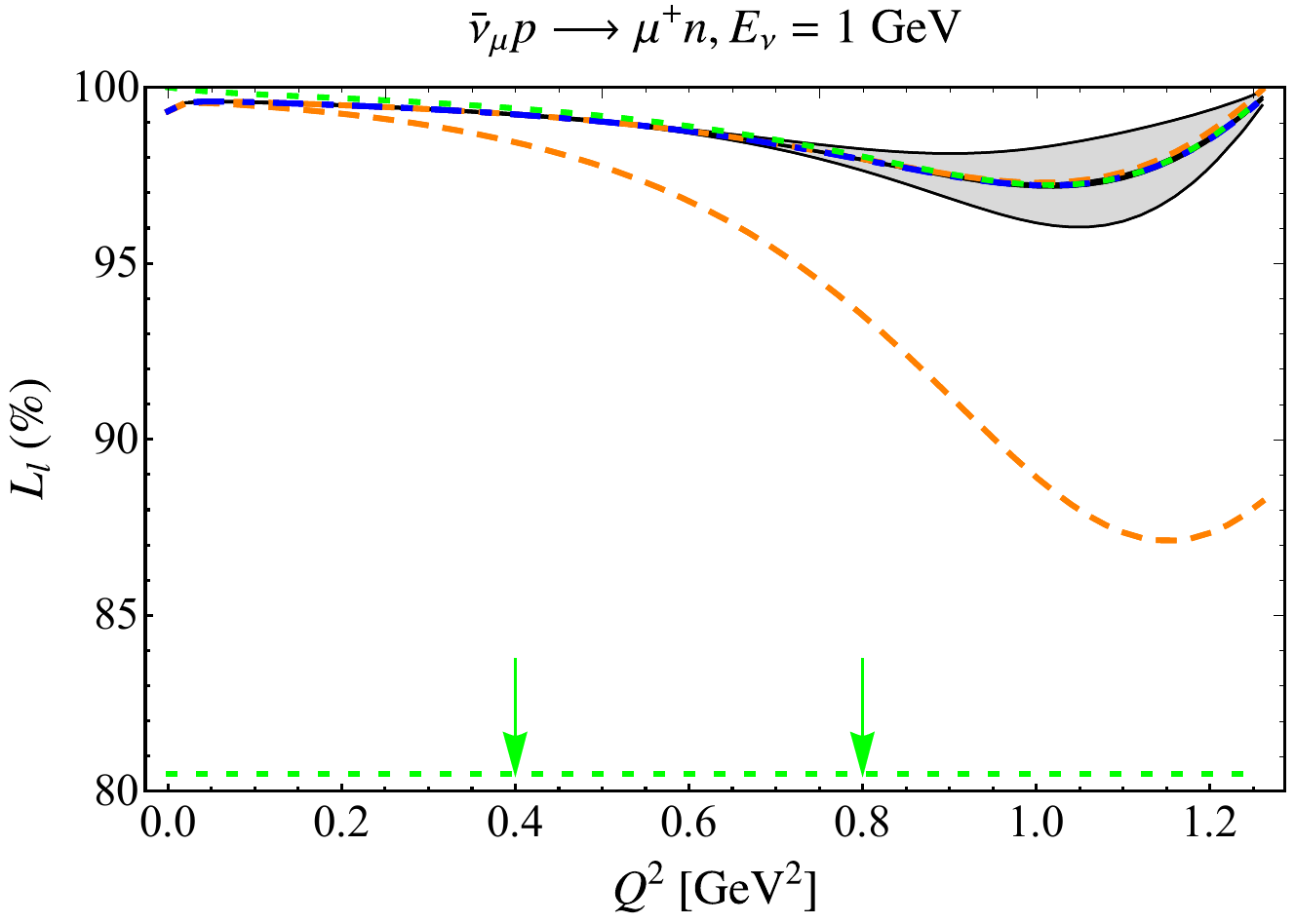}
\includegraphics[width=0.4\textwidth]{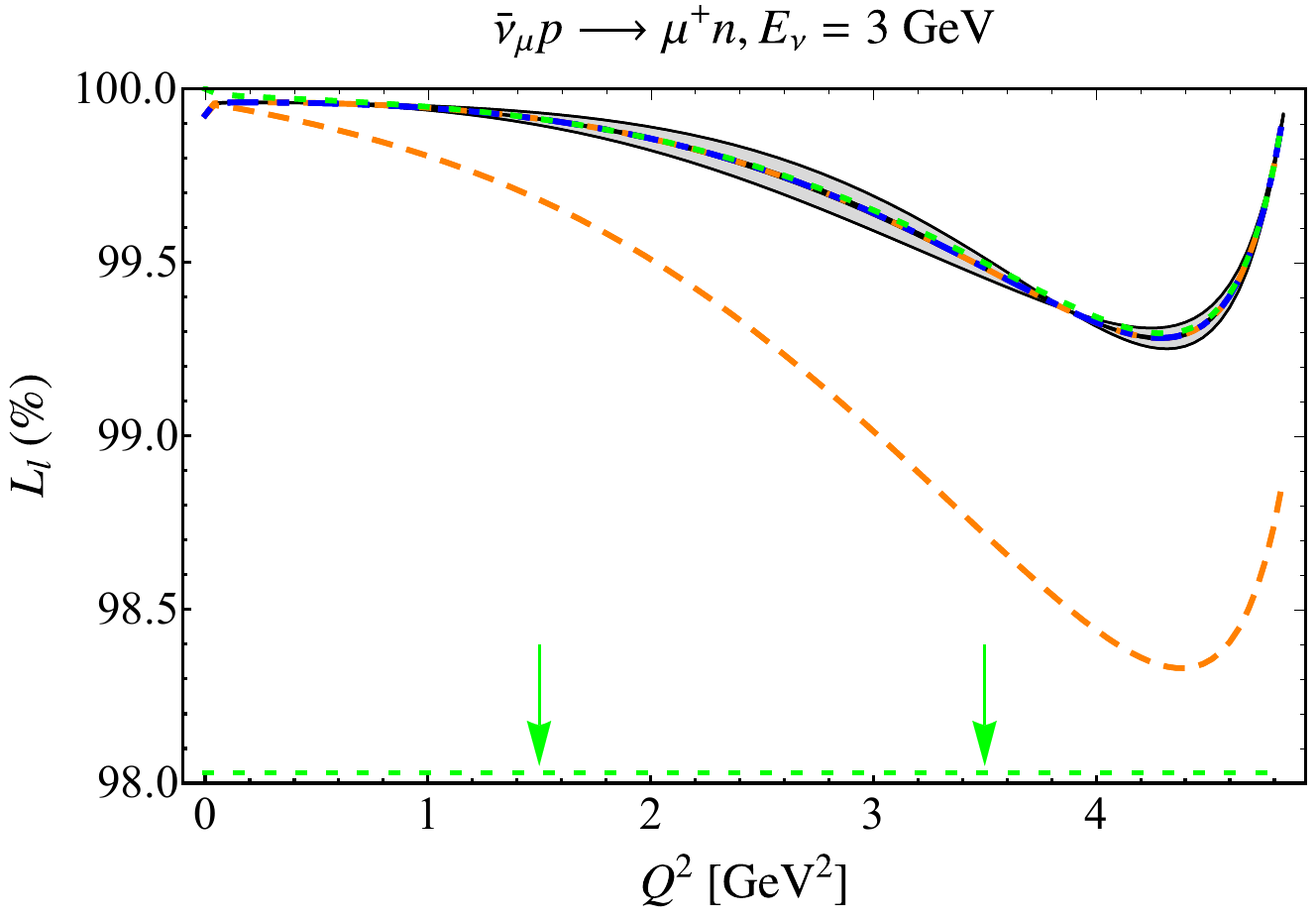}
\caption{Same as Fig.~\ref{fig:antinu_Tt_SCFF} but for the longitudinal polarization observable $L_l$. \label{fig:antinu_Ll_SCFF}}
\end{figure}

\begin{figure}[H]
\centering
\includegraphics[width=0.4\textwidth]{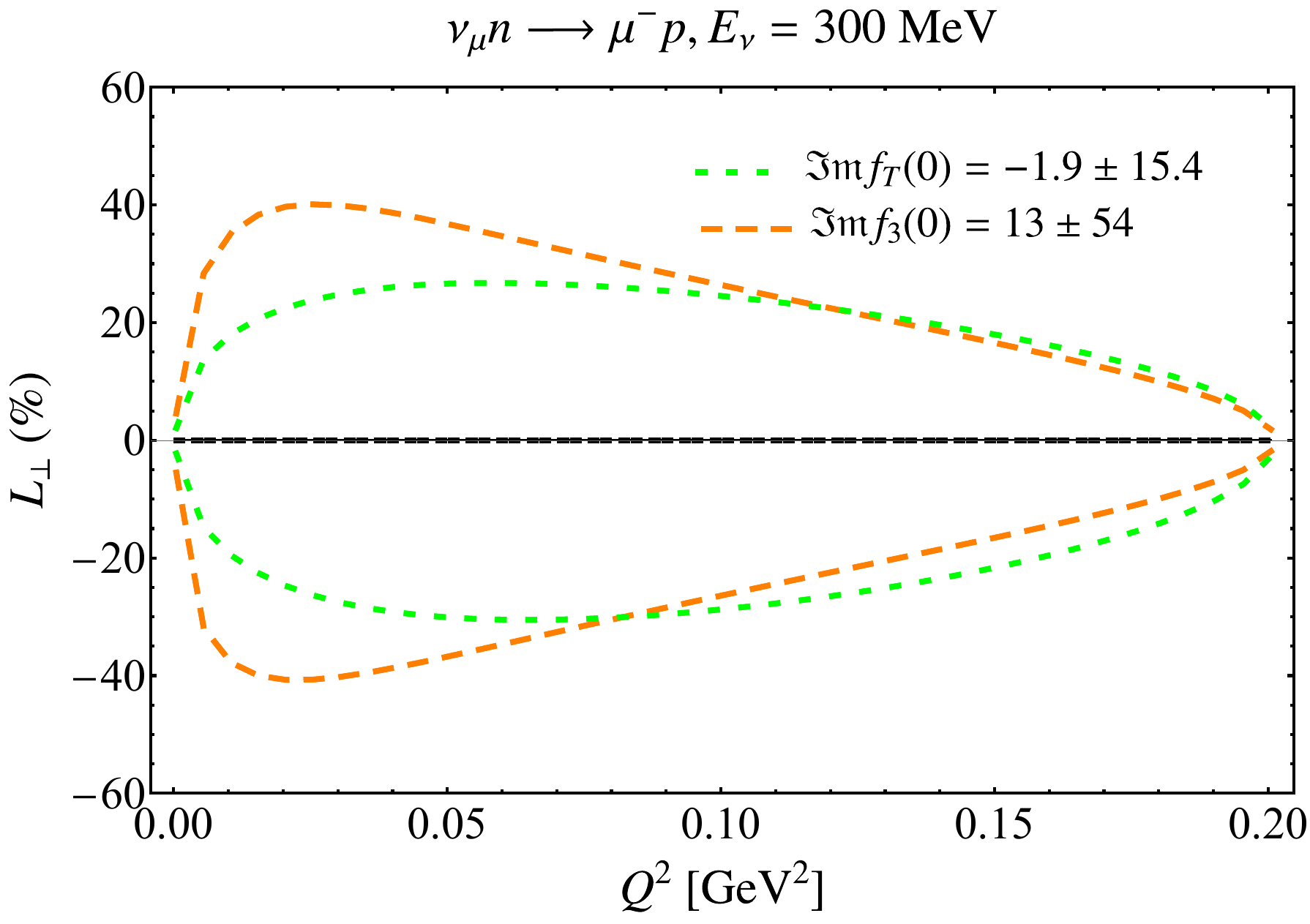}
\includegraphics[width=0.4\textwidth]{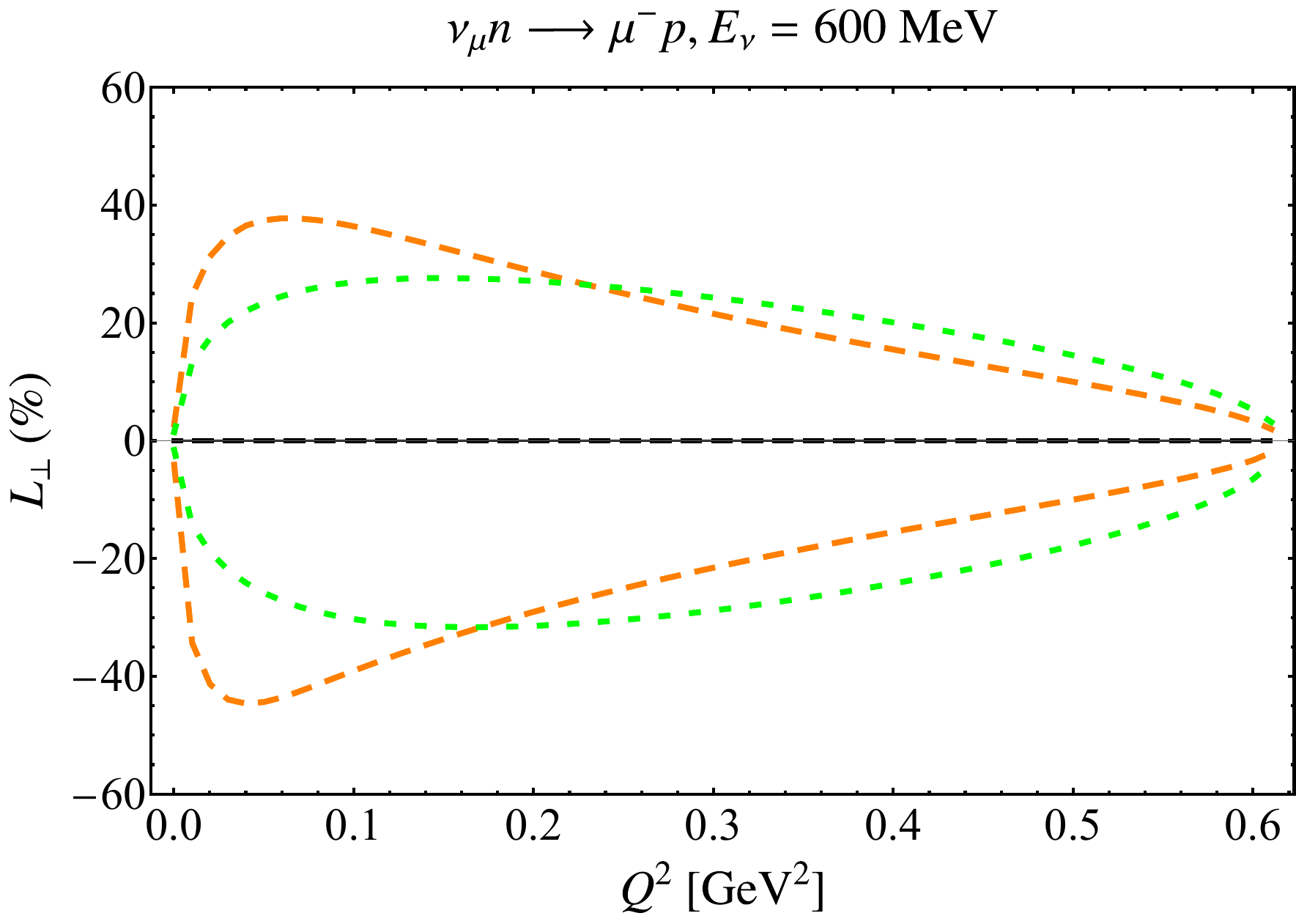}
\includegraphics[width=0.4\textwidth]{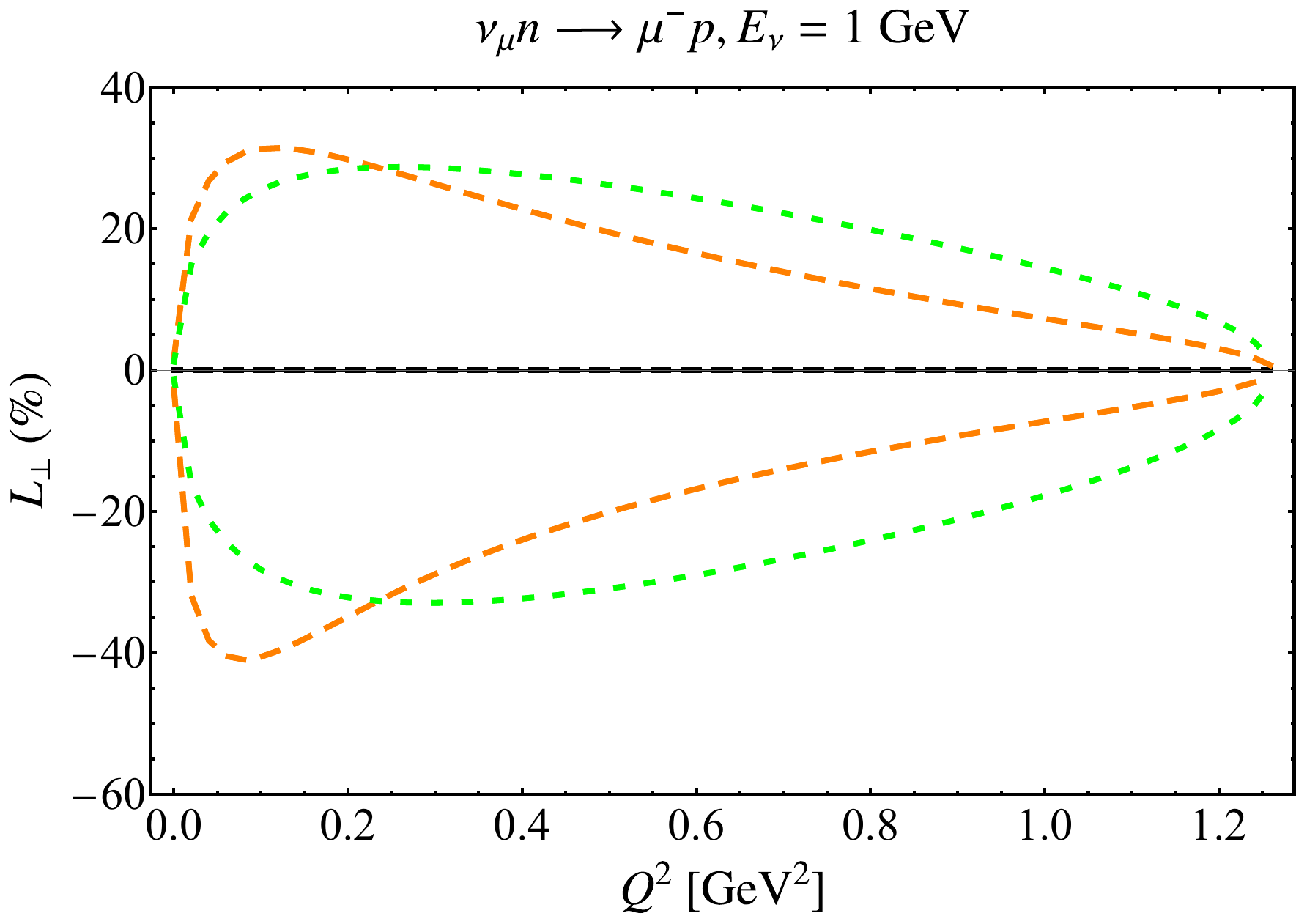}
\includegraphics[width=0.4\textwidth]{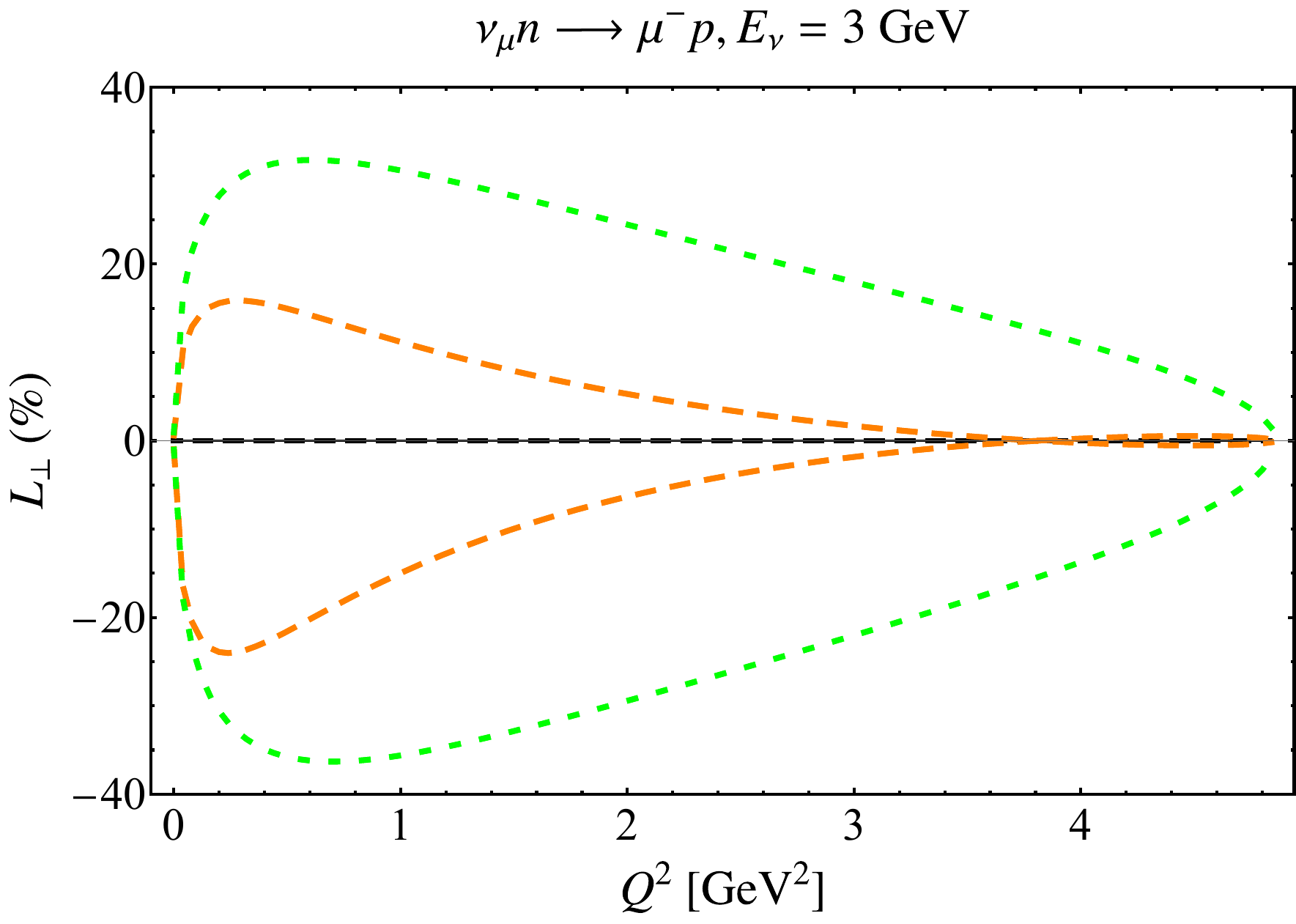}
\caption{Same as Fig.~\ref{fig:nu_Tt_SCFF} but for the transverse polarization observable $L_\perp$ and imaginary amplitudes. \label{fig:nu_LT_SCFF}}
\end{figure}

\begin{figure}[H]
\centering
\includegraphics[width=0.4\textwidth]{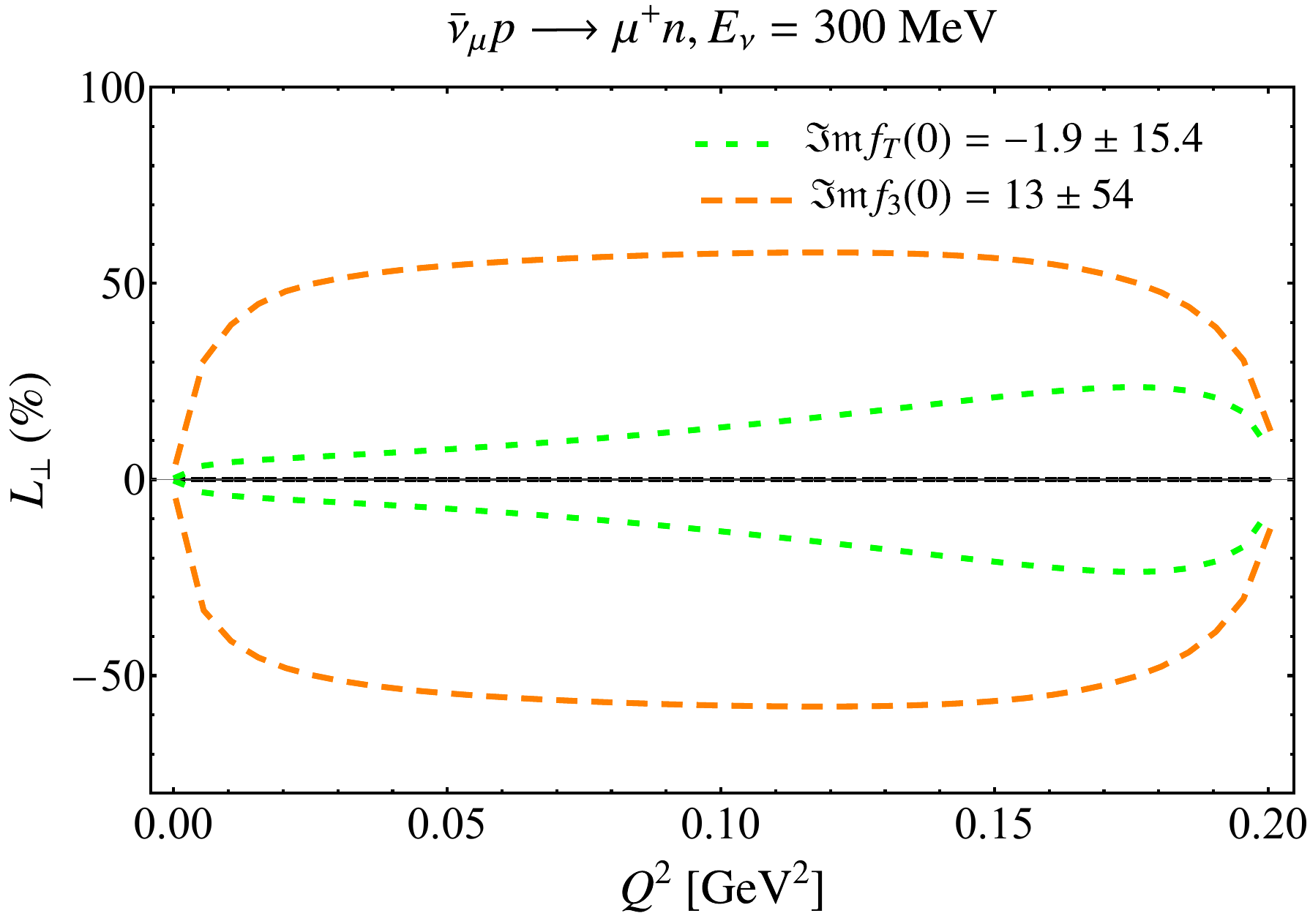}
\includegraphics[width=0.4\textwidth]{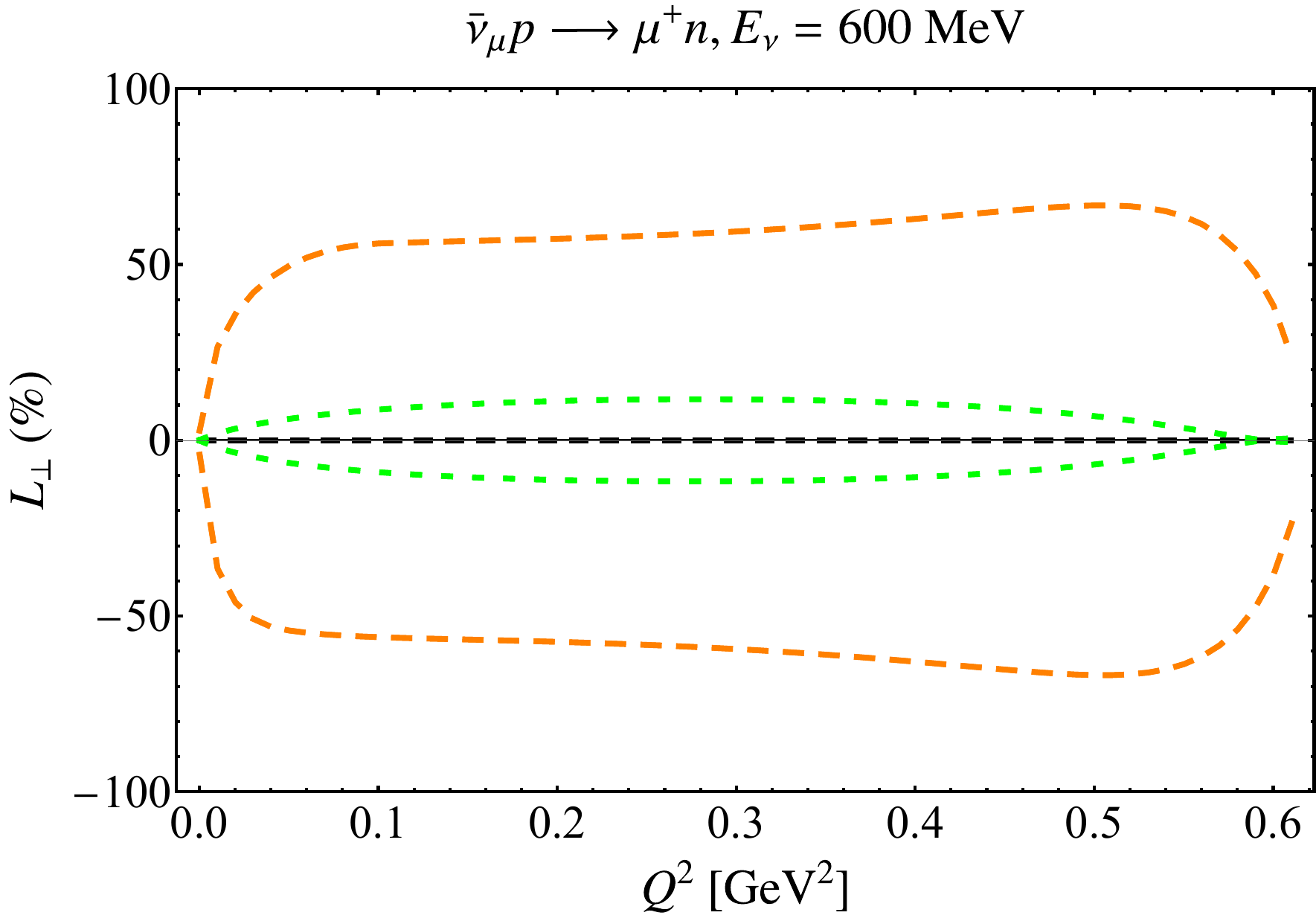}
\includegraphics[width=0.4\textwidth]{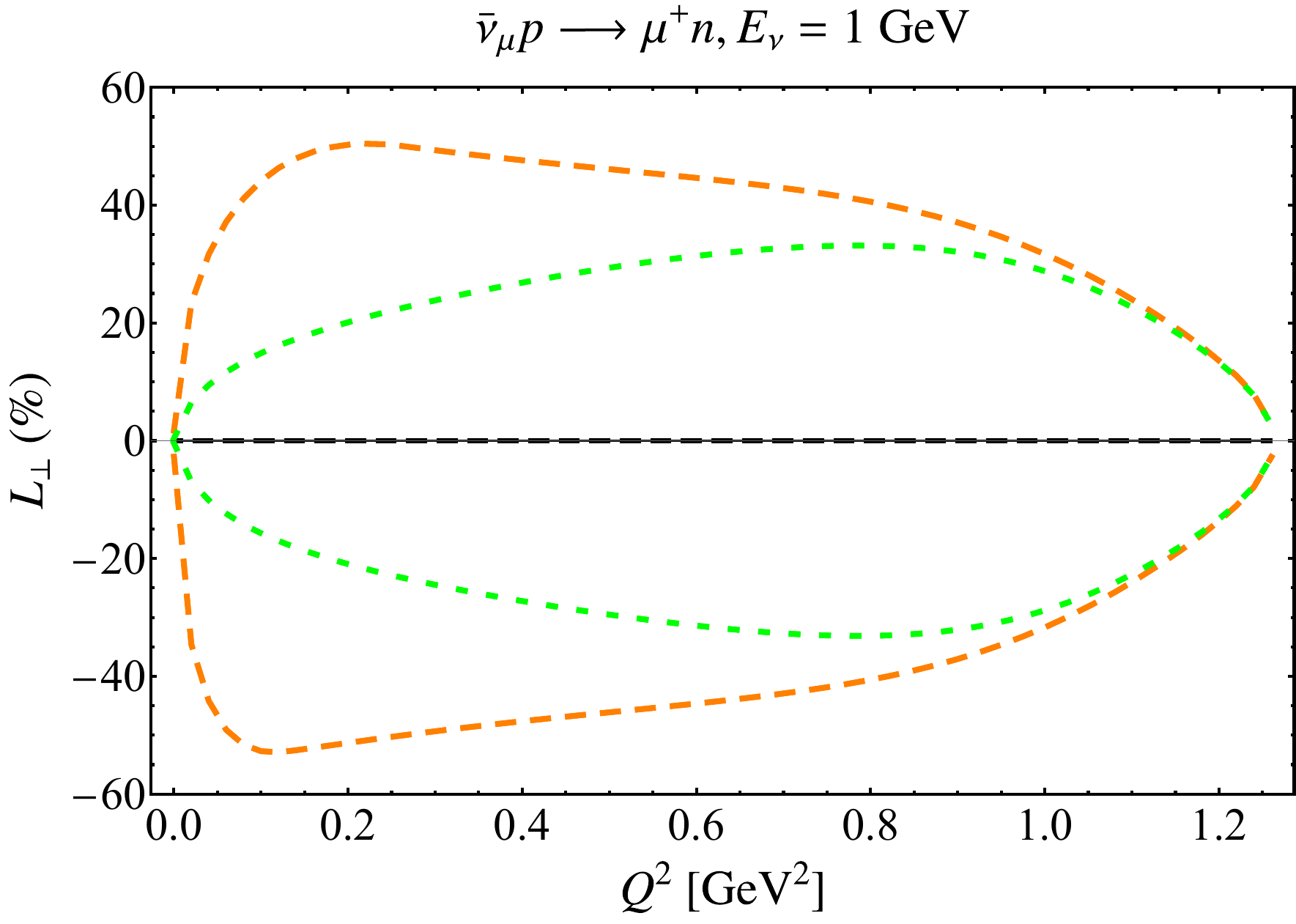}
\includegraphics[width=0.4\textwidth]{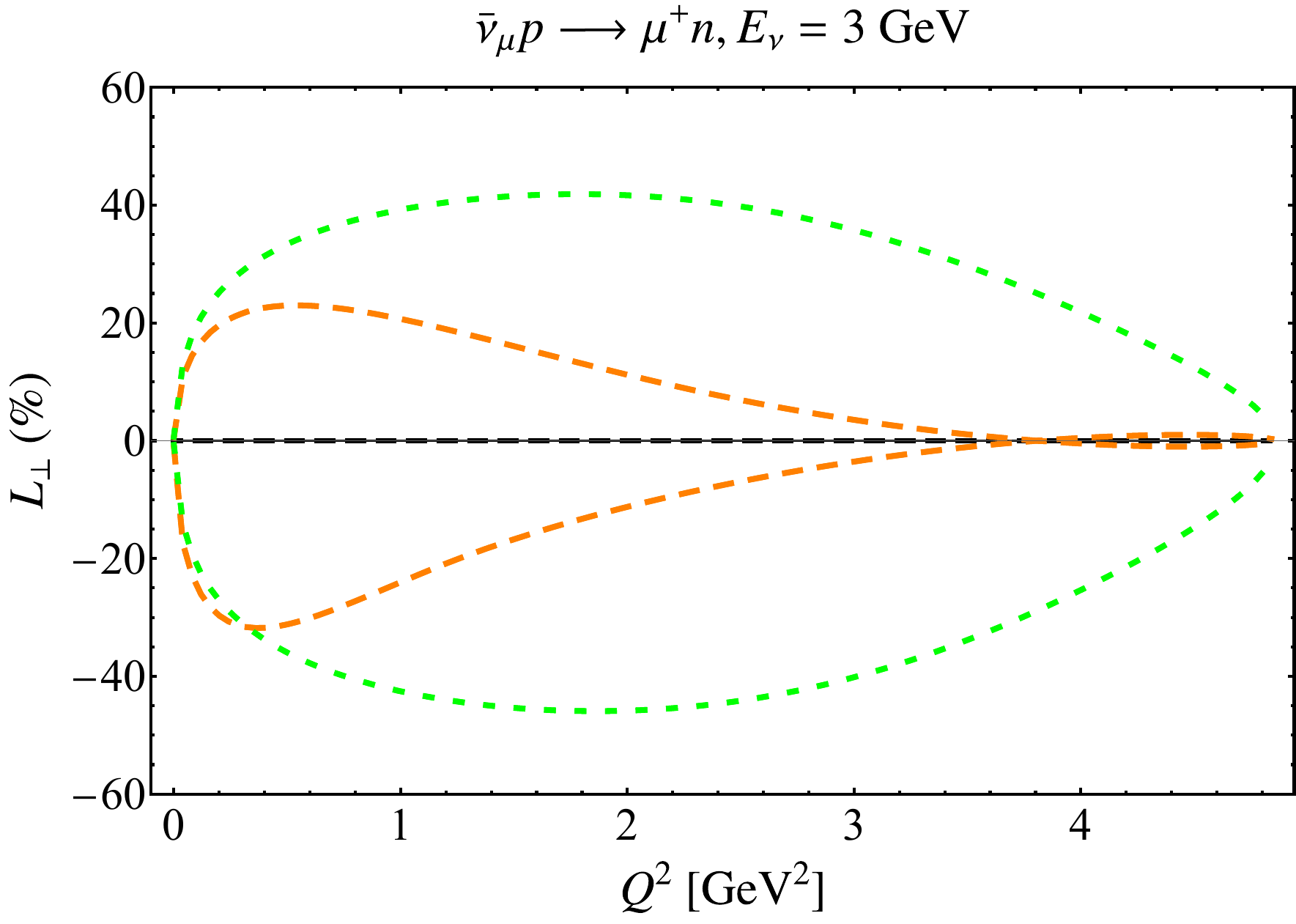}
\caption{Same as Fig.~\ref{fig:antinu_Tt_SCFF} but for the transverse polarization observable $L_\perp$ and imaginary amplitudes. \label{fig:antinu_LT_SCFF}}
\end{figure}

\newpage

\subsection{Polarization asymmetries, tau (anti)neutrino \label{app:taupol_plots}}

In this Section, we present all independent single-spin asymmetries for tau neutrinos and antineutrinos, with one extra real- or imaginary-valued amplitude $f_i (\nu, Q^2)  =  [\mathfrak{Re} f_i \left( 0 \right) + i \mathfrak{Im} f_i \left( 0 \right)]/\left( 1 + \frac{Q^2}{\Lambda^2} \right)^2$, for illustrative neutrino energies $E_\nu = 5$~GeV, $7$~GeV, $10$~GeV, and $15$~GeV, and $\Lambda = 1$~GeV. 
We vary the amplitude normalizations within the ranges from Table~\ref{tab:beta}, and compare to the uncertainty from vector and axial-vector form factors from Sec.~\ref{sec:observables}. 

\begin{figure}[H]
\centering
\includegraphics[width=0.4\textwidth]{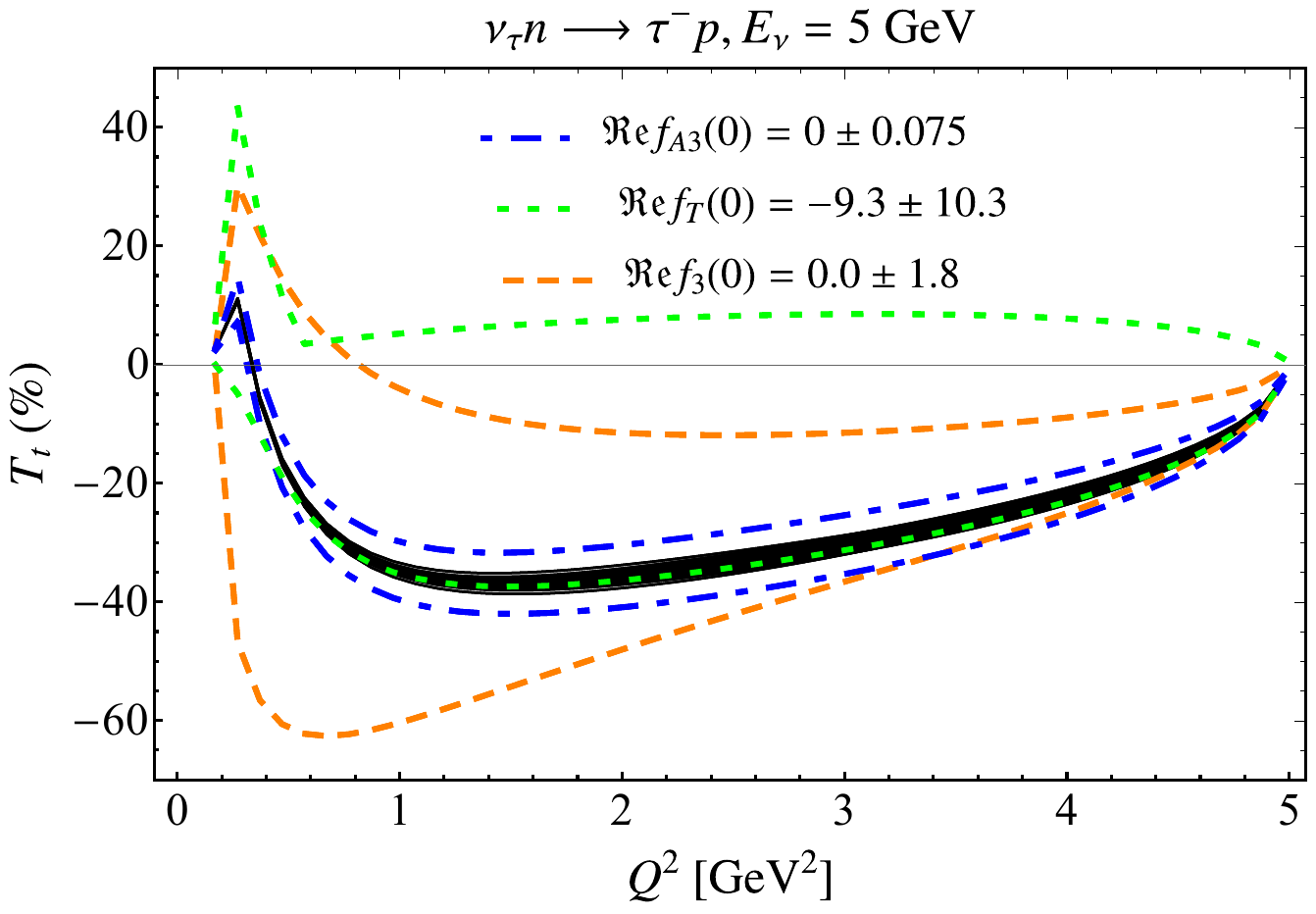}
\includegraphics[width=0.4\textwidth]{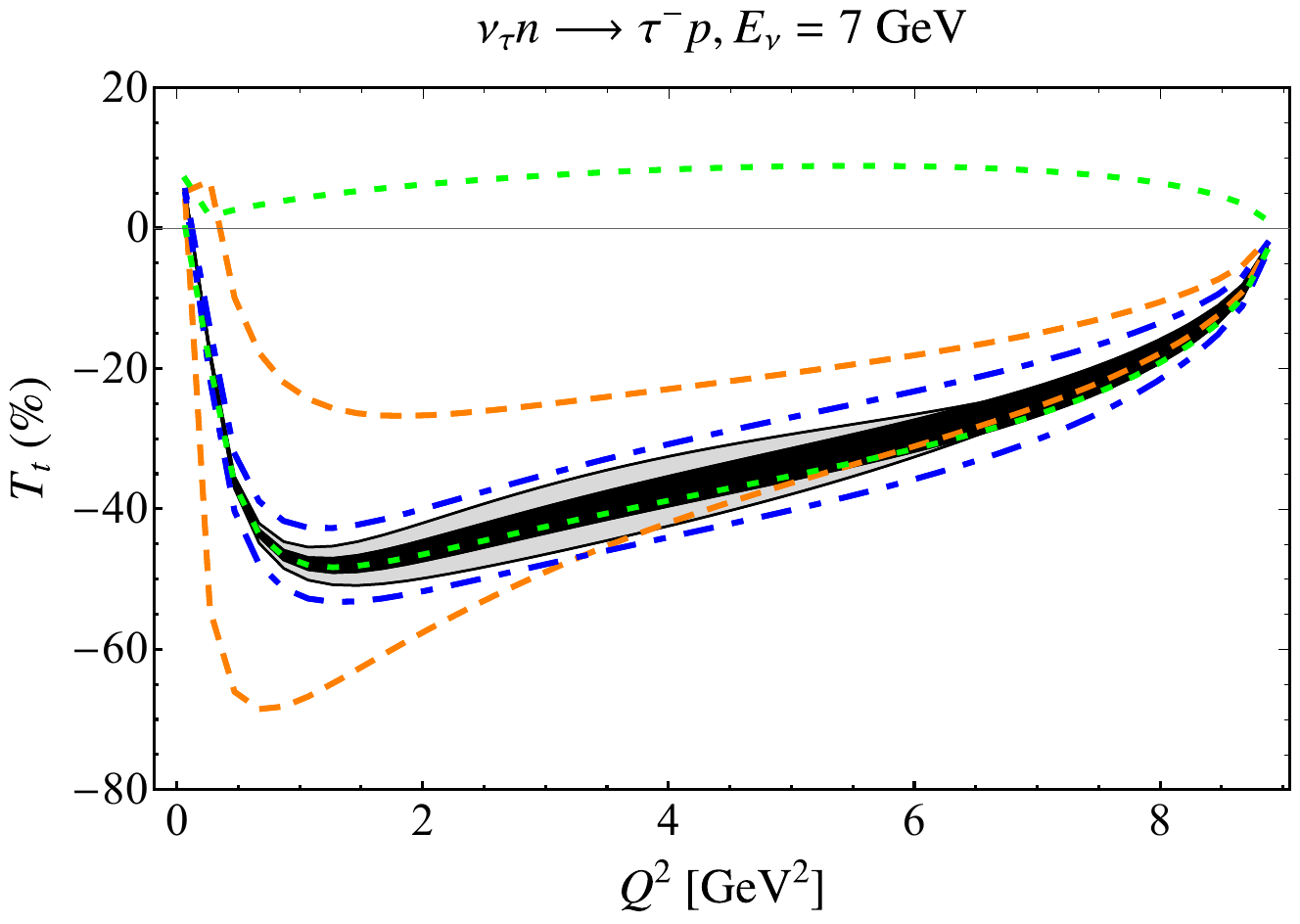}
\includegraphics[width=0.4\textwidth]{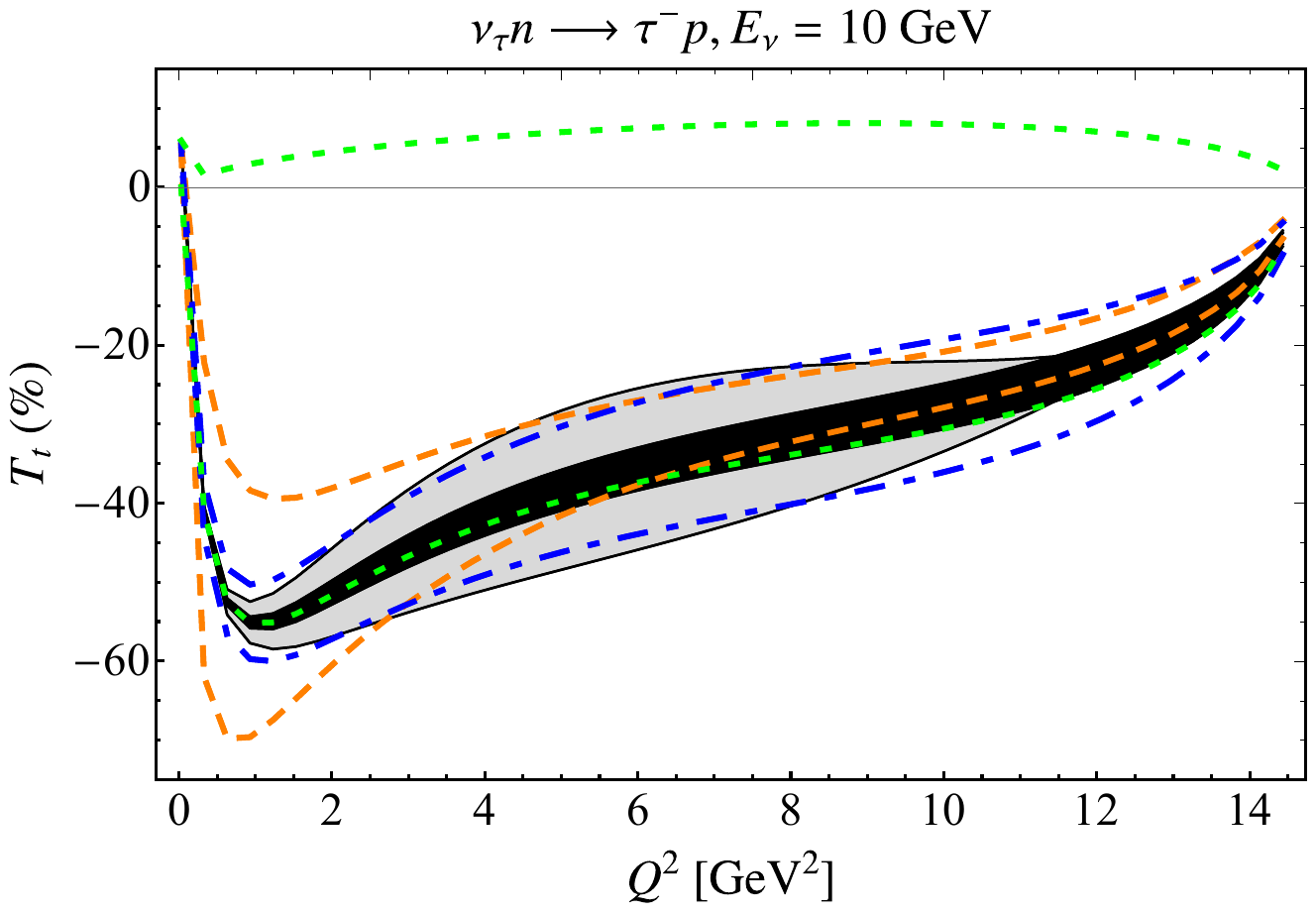}
\includegraphics[width=0.4\textwidth]{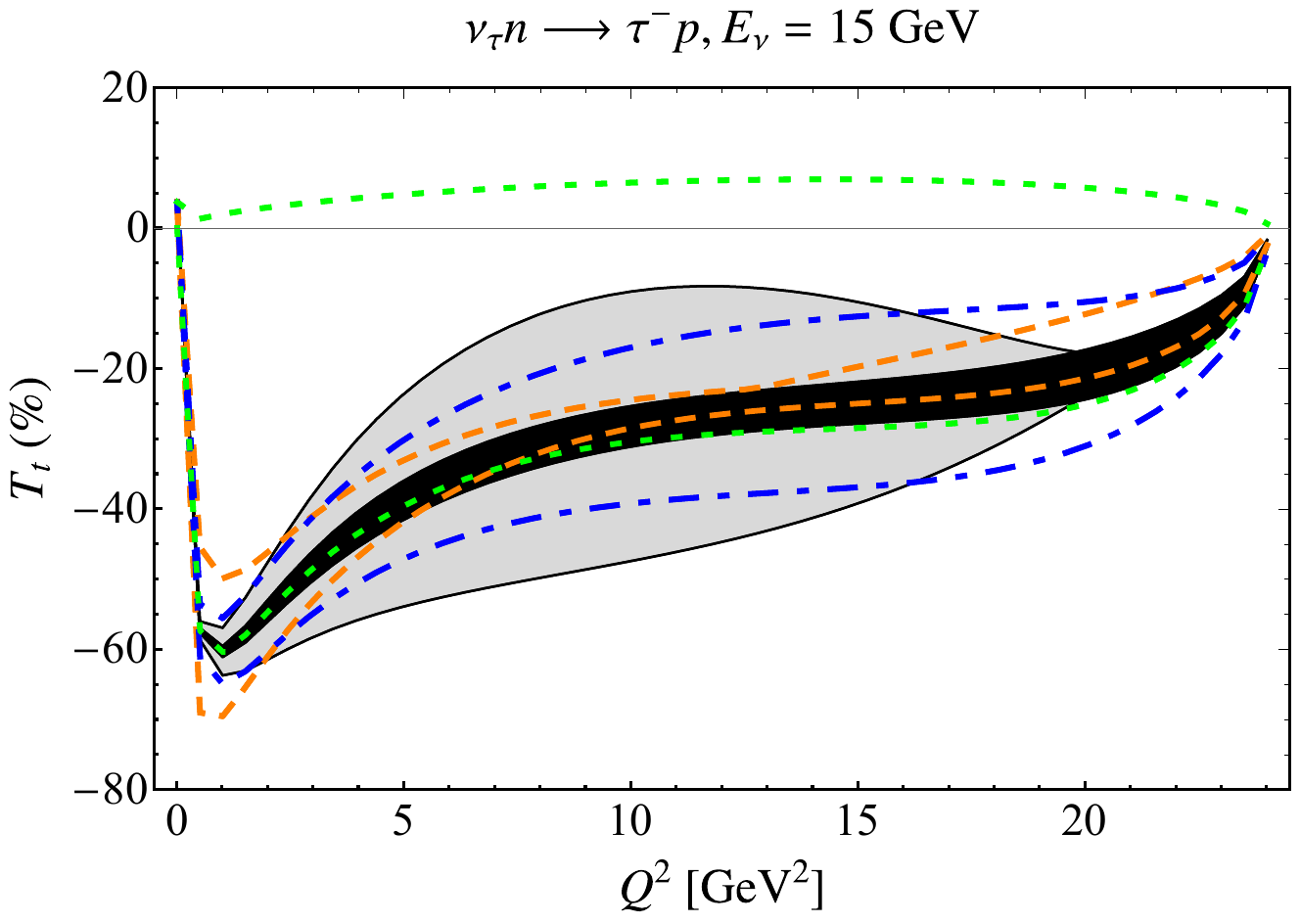}
\caption{Transverse polarization observable, $T_t$, with one extra real-valued amplitude, compared to the tree-level result at fixed tau neutrino energies $E_\nu = 5$~GeV, $7$~GeV, $10$~GeV, and $15$~GeV. The dark black and light gray bands correspond to vector and axial-vector uncertainty, respectively. Orange dashed, green dotted, and blue dashed-dotted lines represent allowed regions for $f_3$, $f_T$, and $\fAt$, respectively, as described in the text.\label{fig:nu_Tt_SCFF_tau}}
\end{figure}

\begin{figure}[H]
\centering
\includegraphics[width=0.4\textwidth]{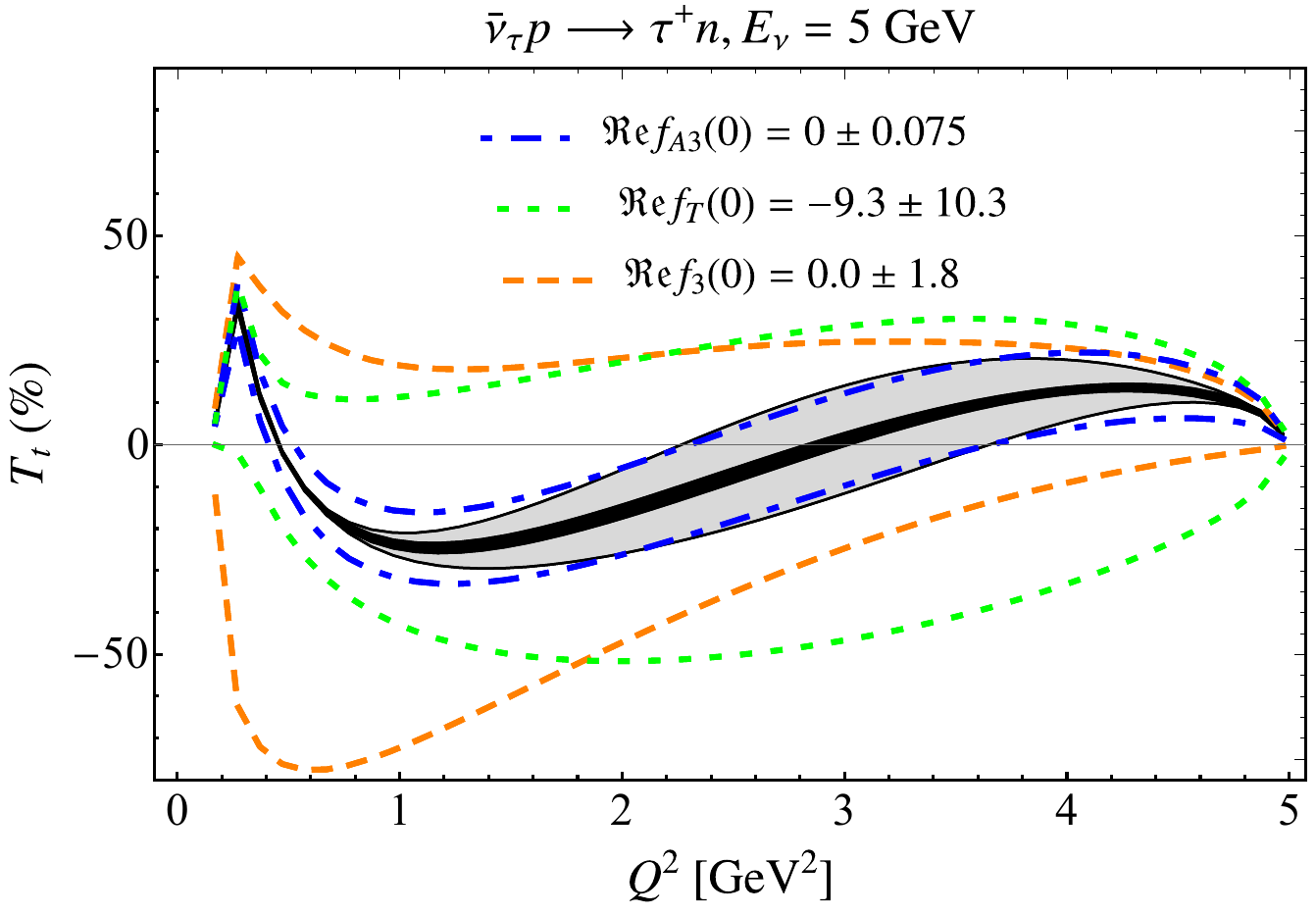}
\includegraphics[width=0.4\textwidth]{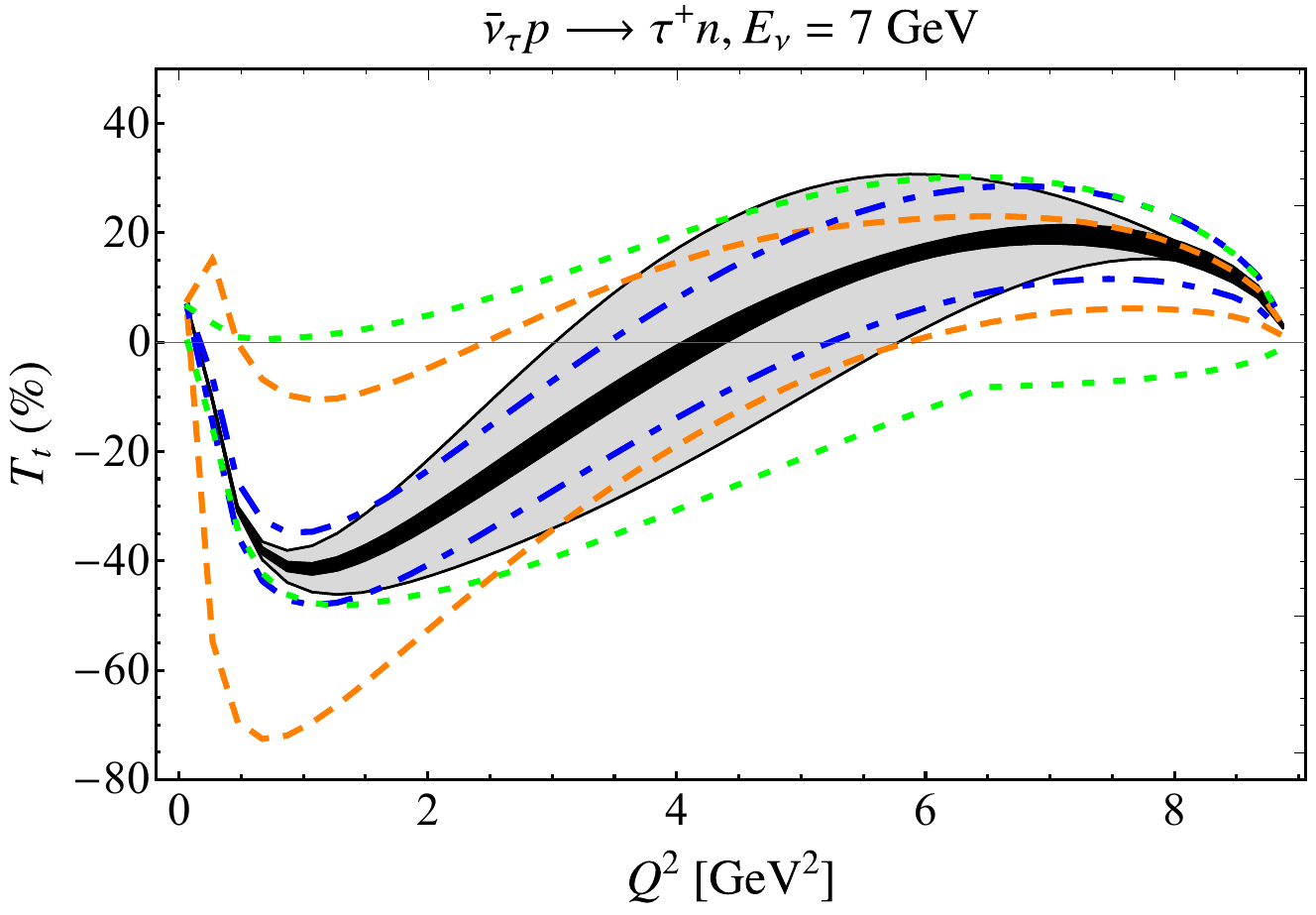}
\includegraphics[width=0.4\textwidth]{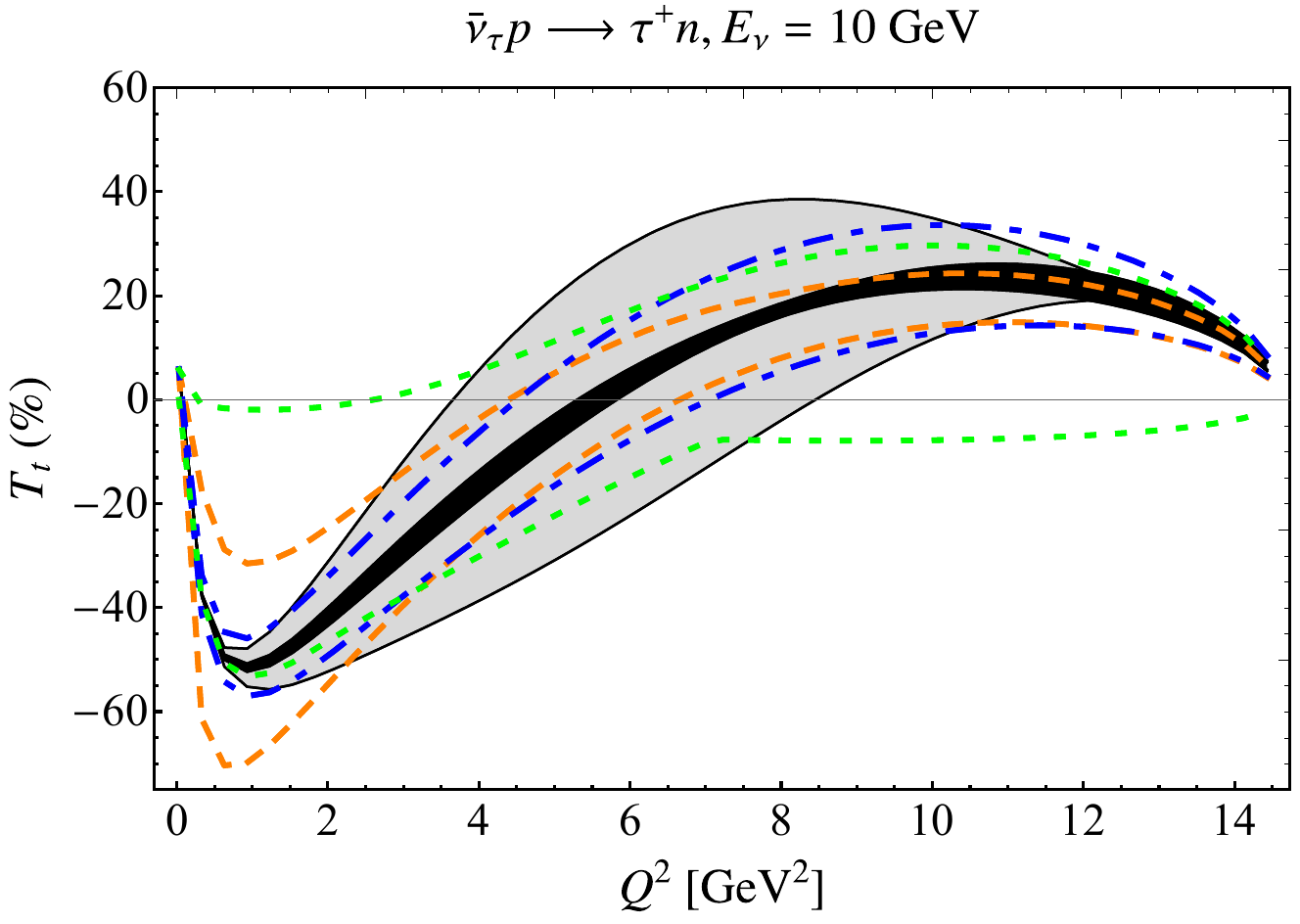}
\includegraphics[width=0.4\textwidth]{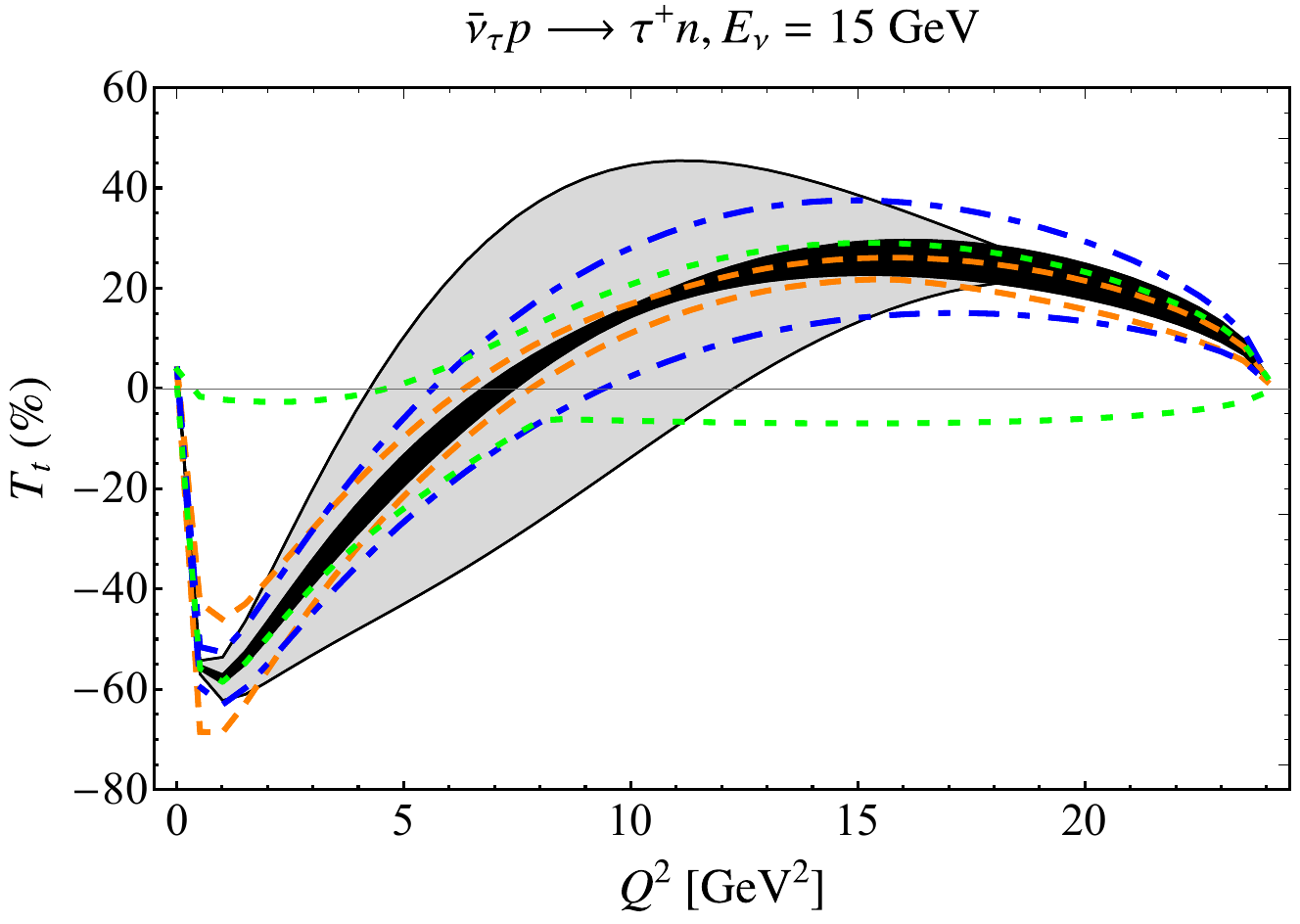}
\caption{Same as Fig.~\ref{fig:nu_Tt_SCFF_tau} but for antineutrinos. \label{fig:antinu_Tt_SCFF_tau}}
\end{figure}

\begin{figure}[H]
\centering
\includegraphics[width=0.4\textwidth]{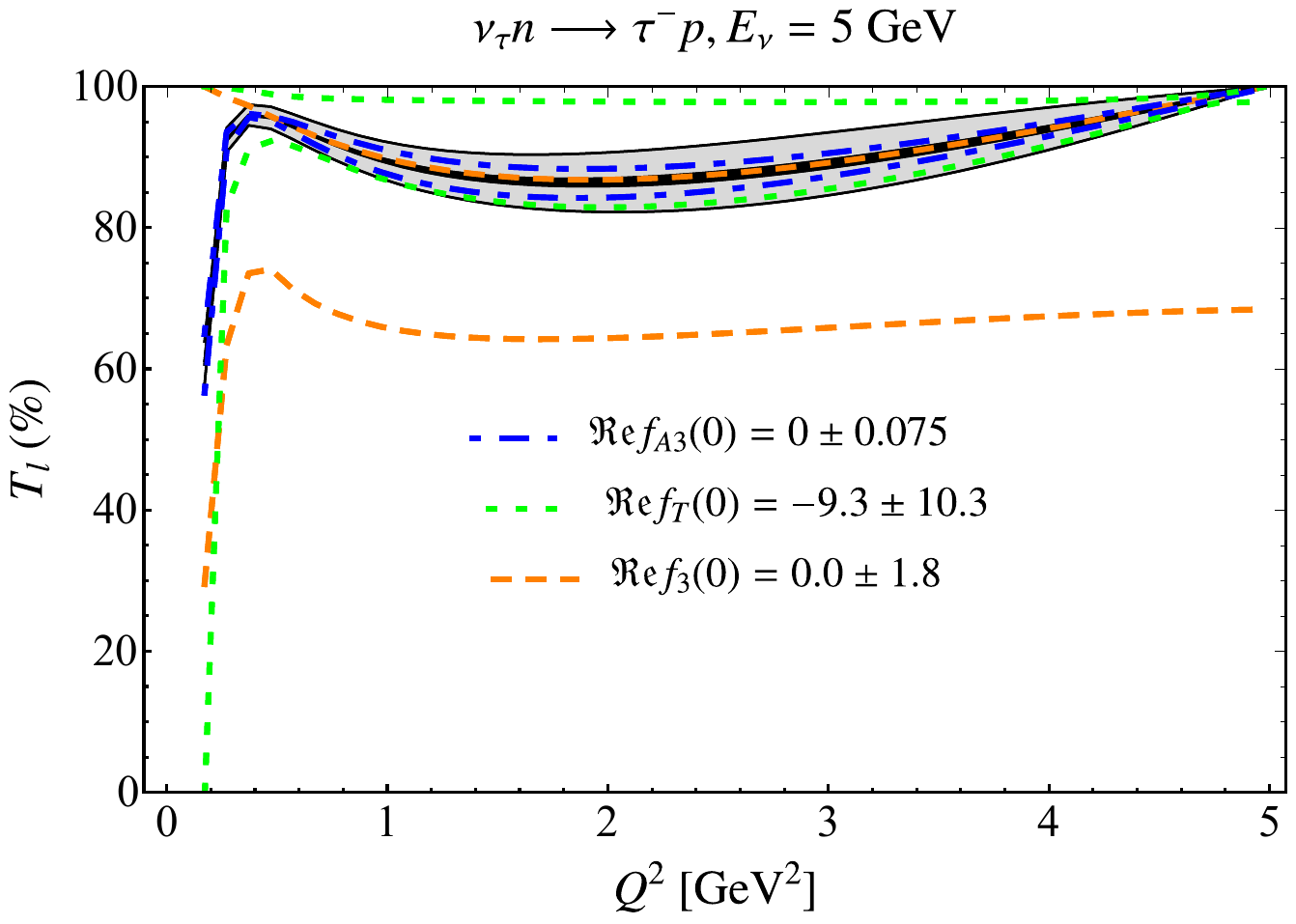}
\includegraphics[width=0.4\textwidth]{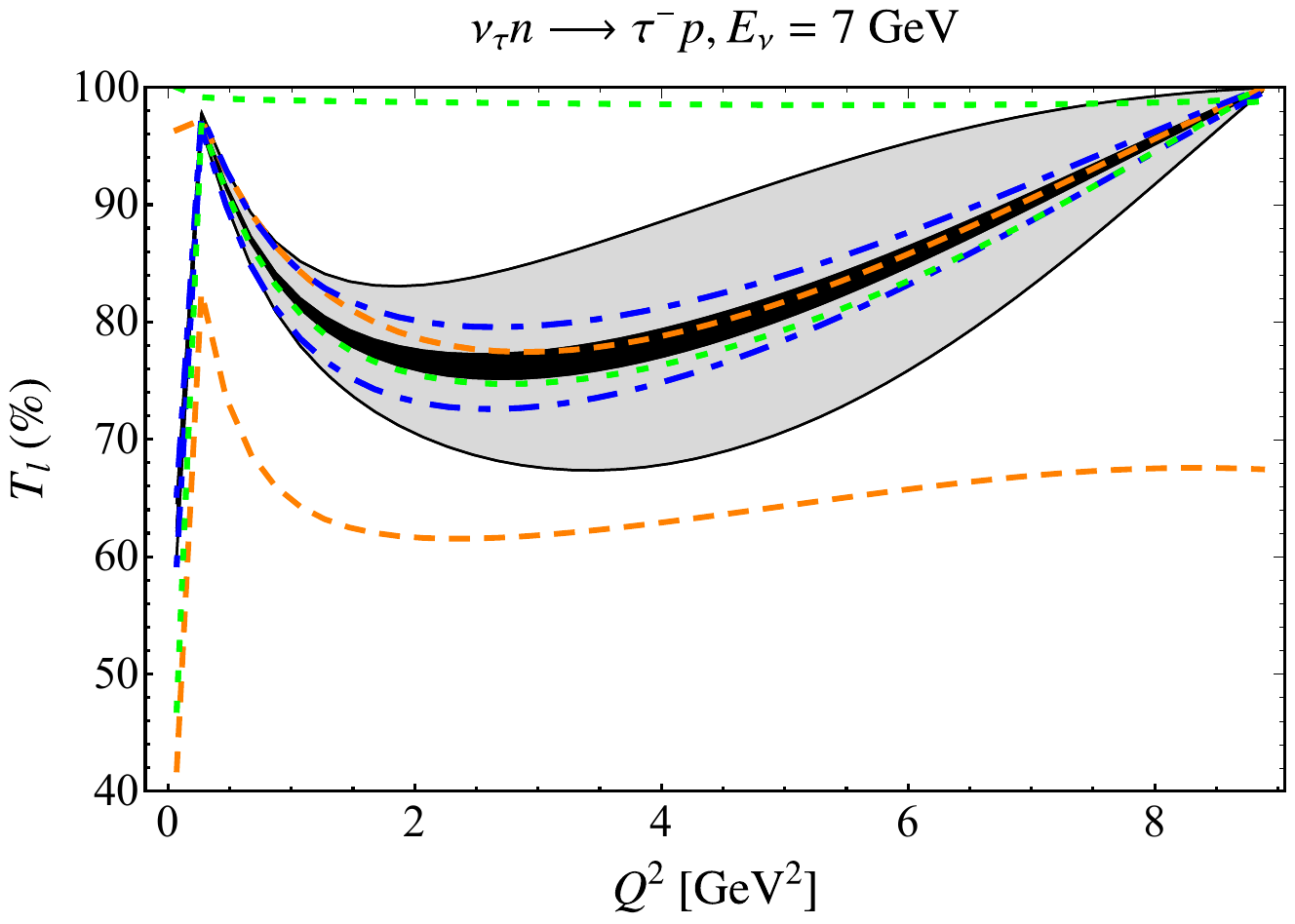}
\includegraphics[width=0.4\textwidth]{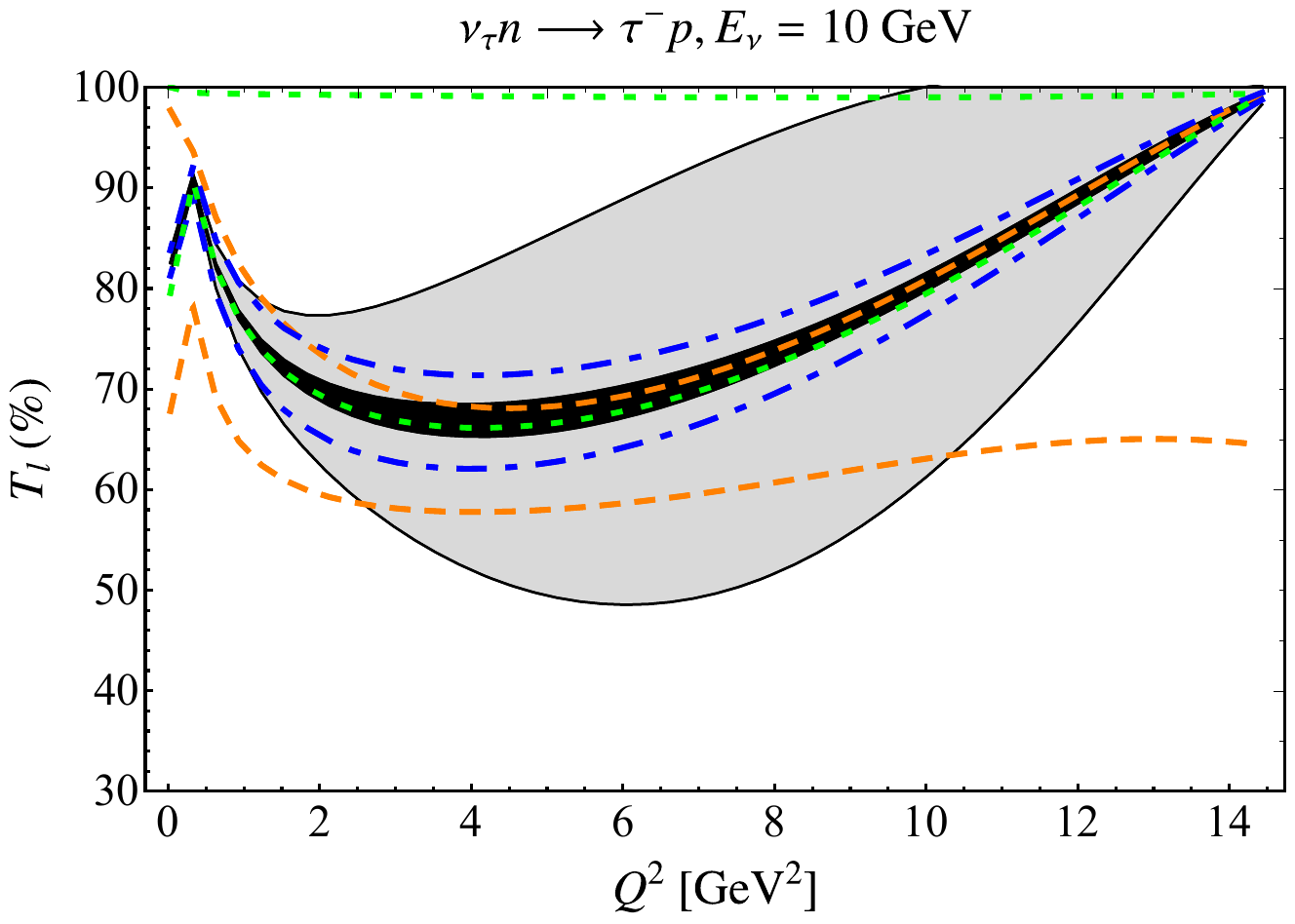}
\includegraphics[width=0.4\textwidth]{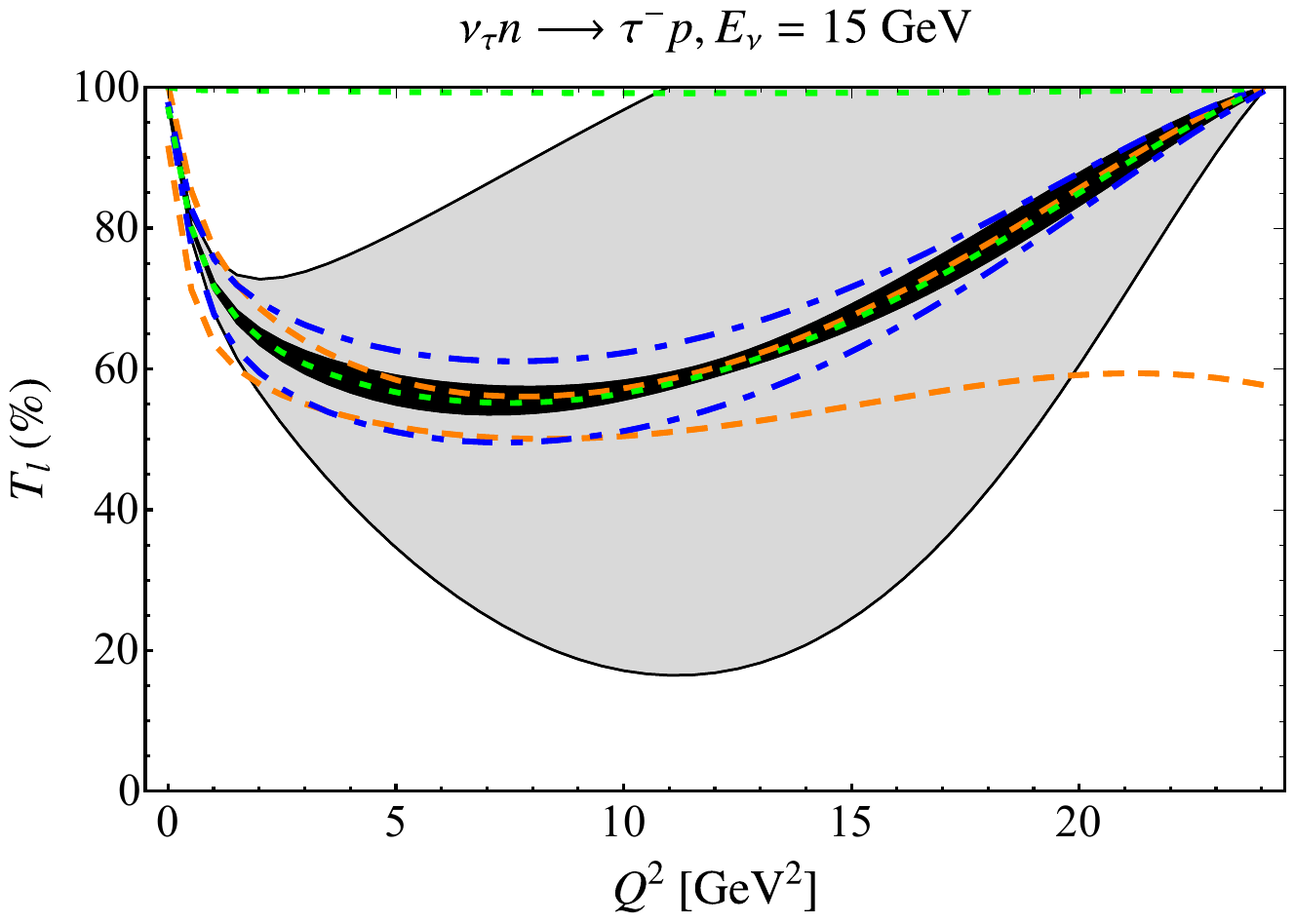}
\caption{Same as Fig.~\ref{fig:nu_Tt_SCFF_tau} but for the longitudinal polarization observable $T_l$. \label{fig:nu_Tl_SCFF_tau}}
\end{figure}

\begin{figure}[H]
\centering
\includegraphics[width=0.4\textwidth]{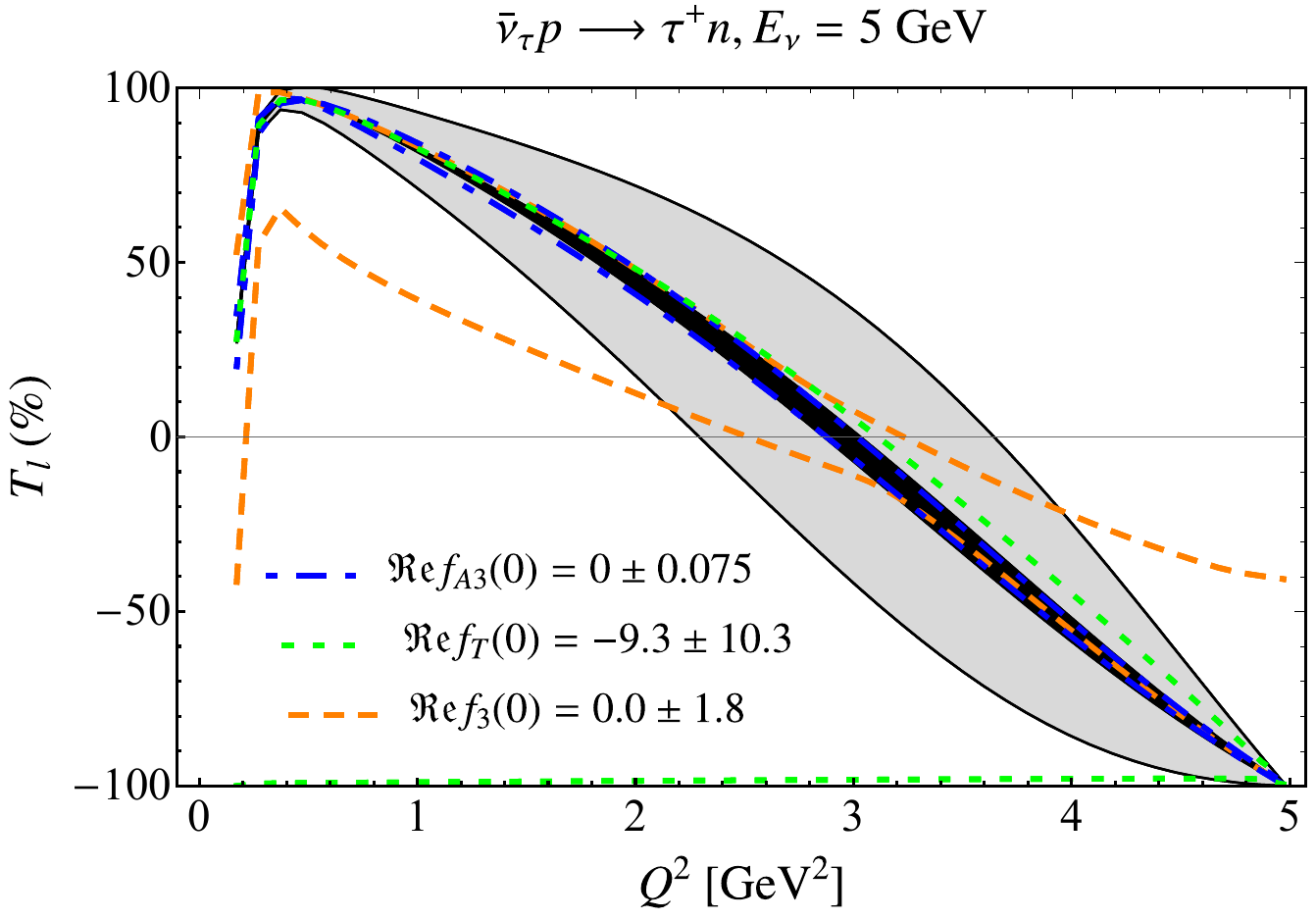}
\includegraphics[width=0.4\textwidth]{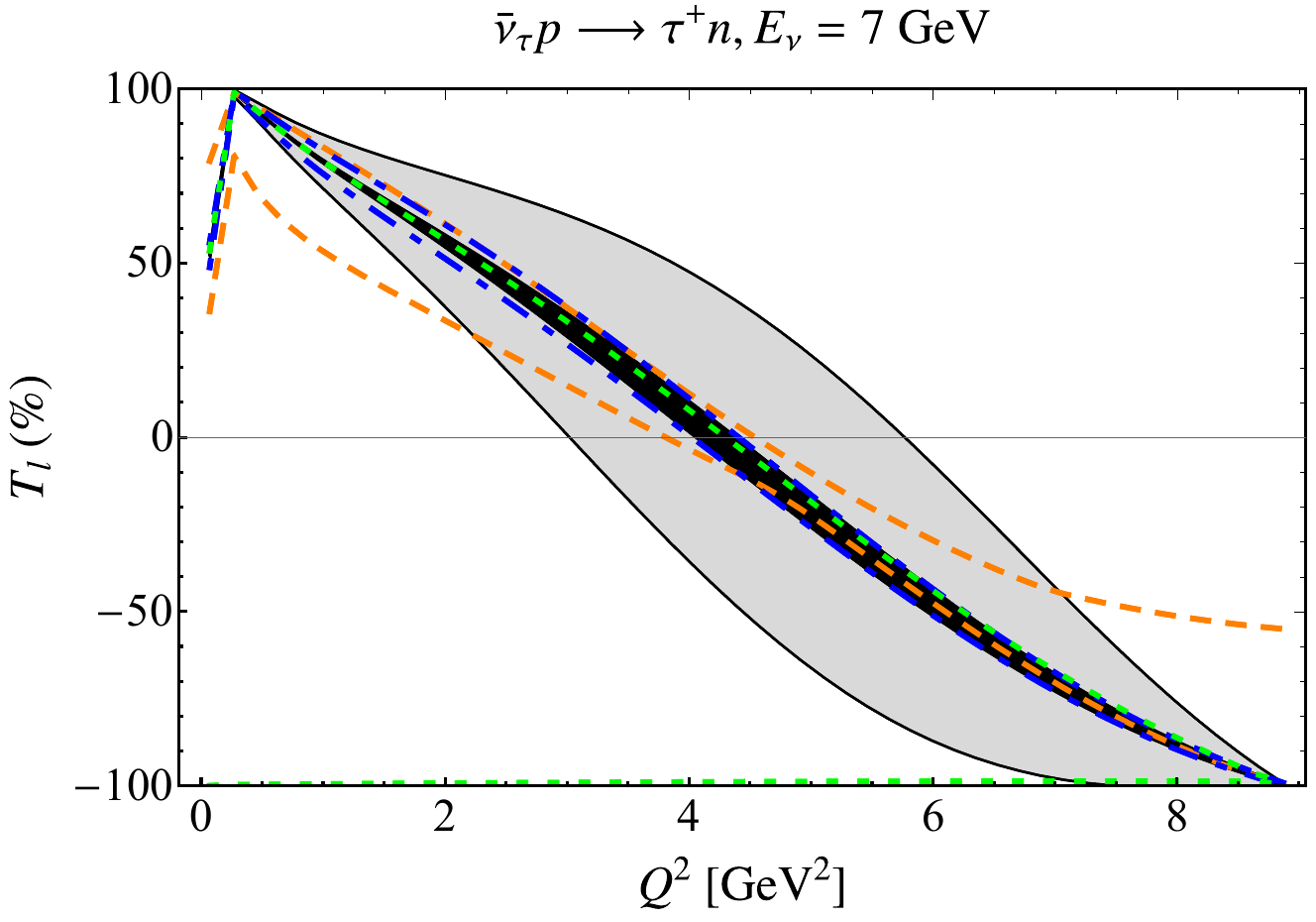}
\includegraphics[width=0.4\textwidth]{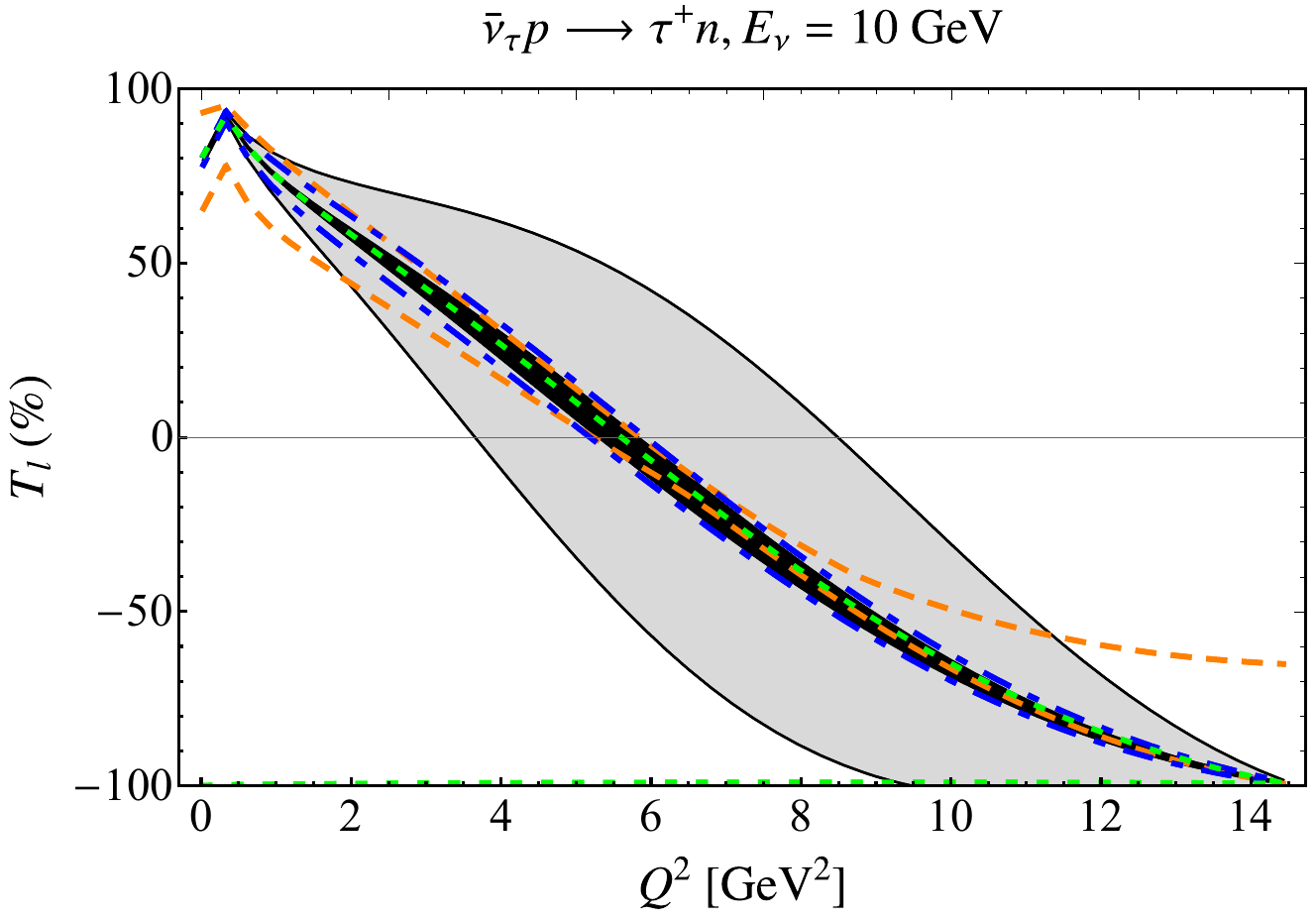}
\includegraphics[width=0.4\textwidth]{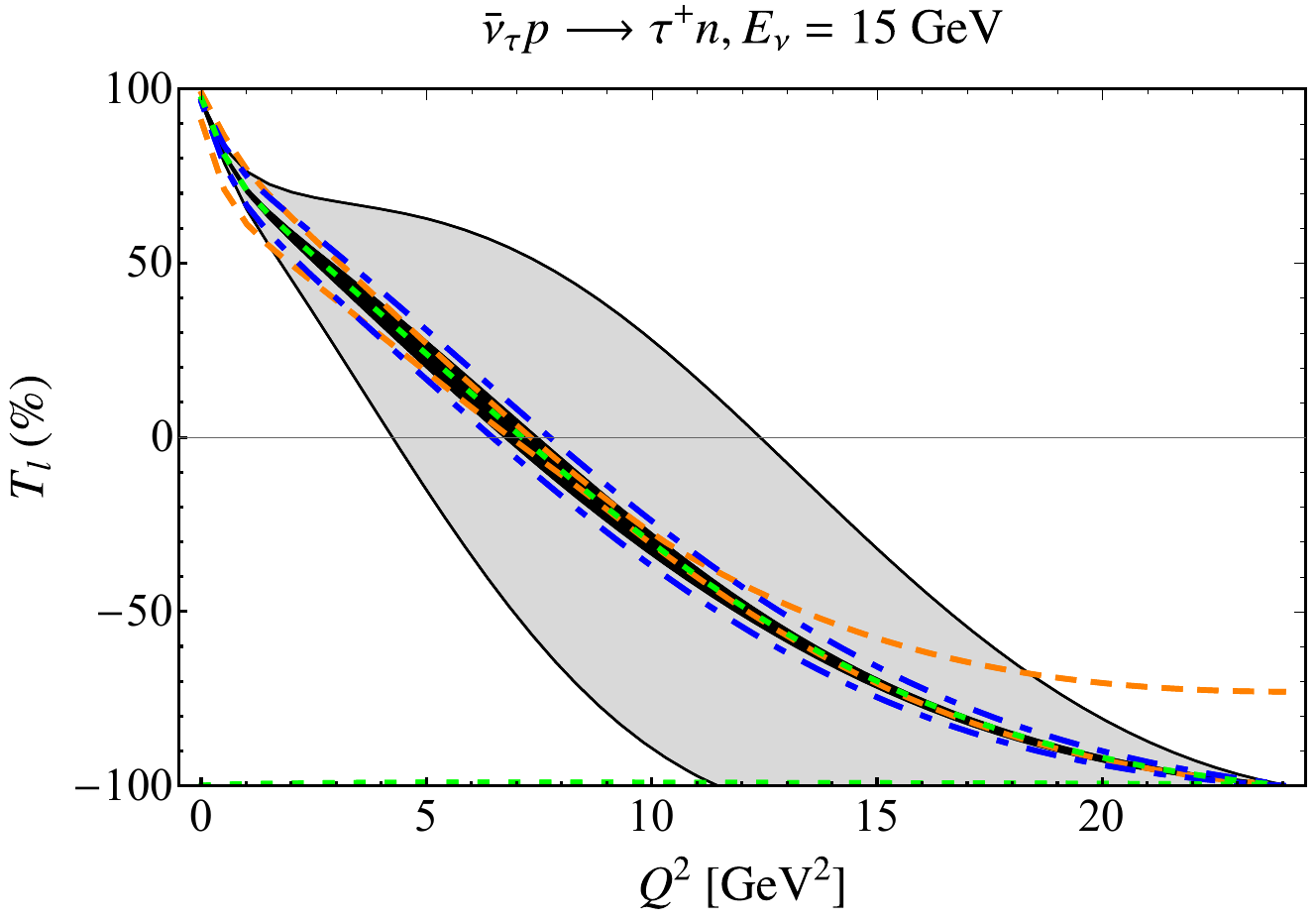}
\caption{Same as Fig.~\ref{fig:antinu_Tt_SCFF_tau} but for the longitudinal polarization observable $T_l$. \label{fig:antinu_Tl_SCFF_tau}}
\end{figure}

\begin{figure}[H]
\centering
\includegraphics[width=0.4\textwidth]{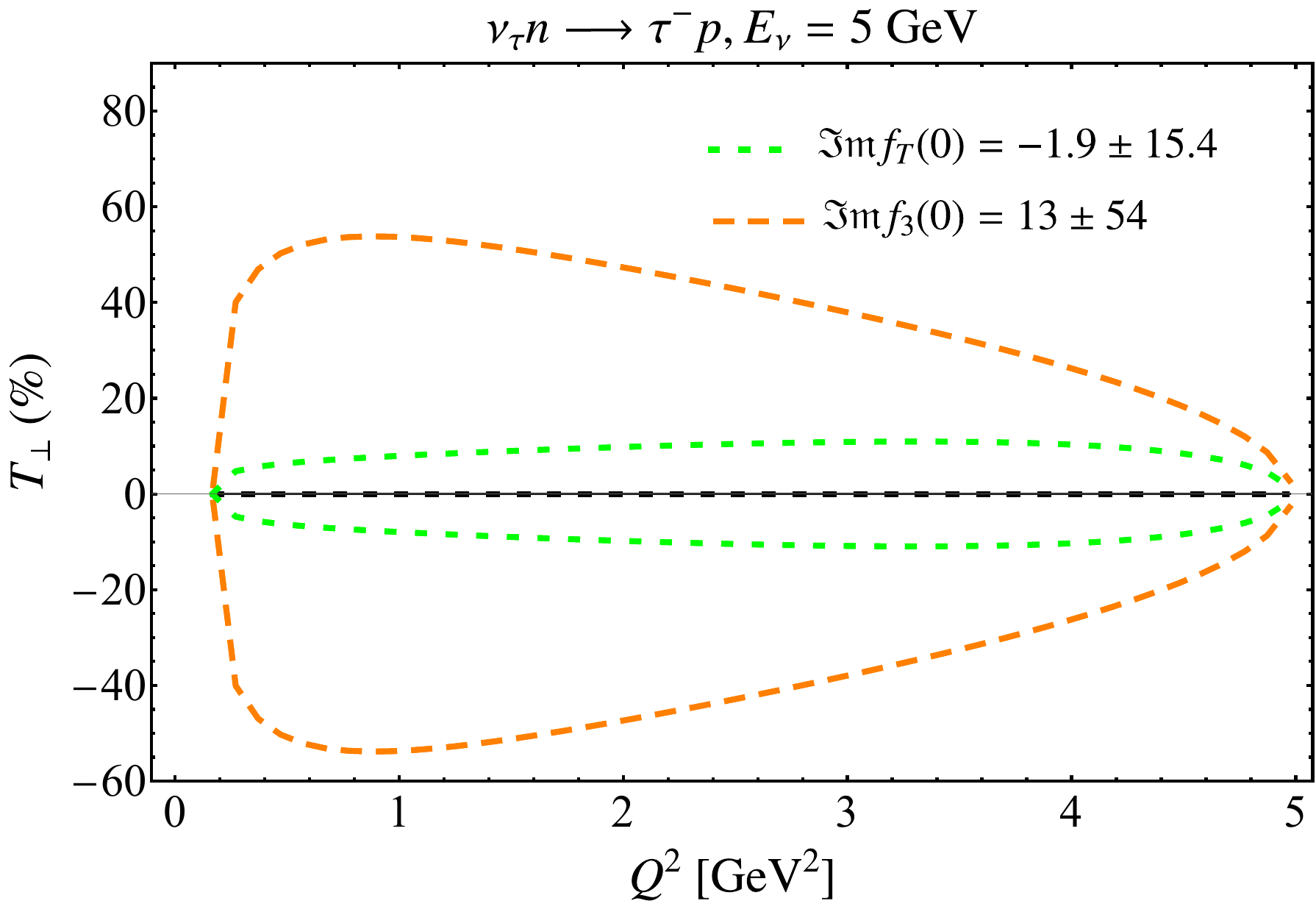}
\includegraphics[width=0.4\textwidth]{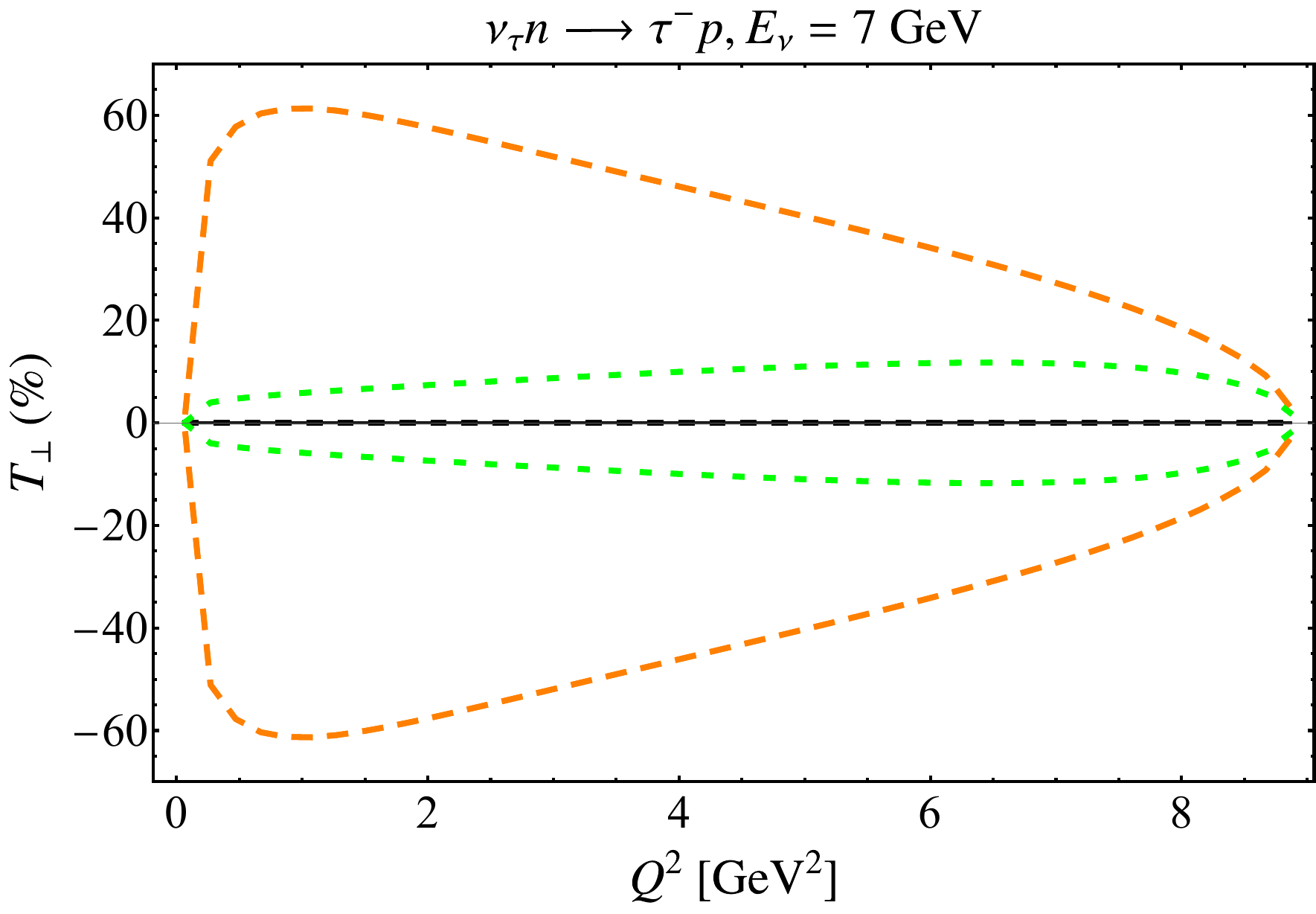}
\includegraphics[width=0.4\textwidth]{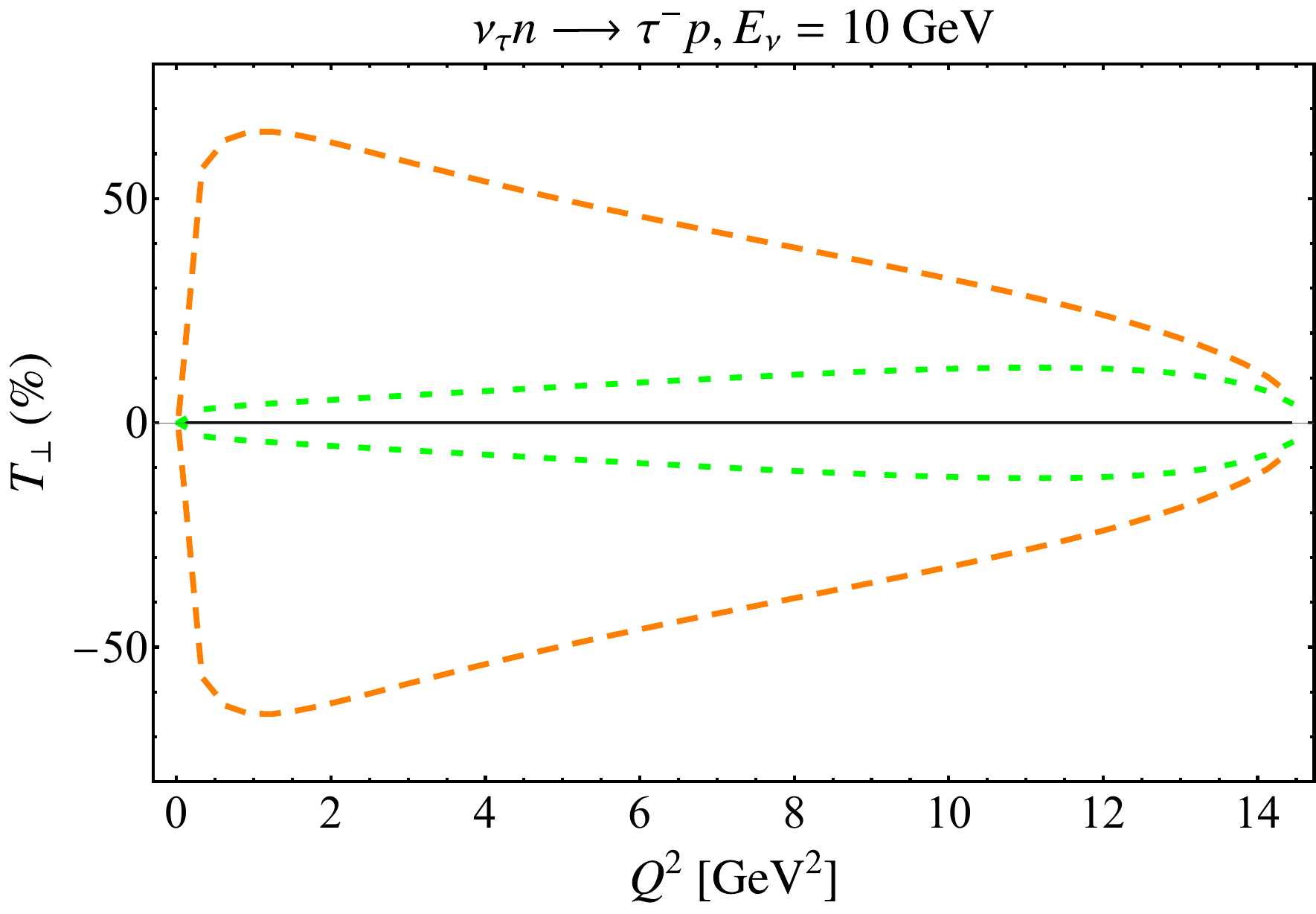}
\includegraphics[width=0.4\textwidth]{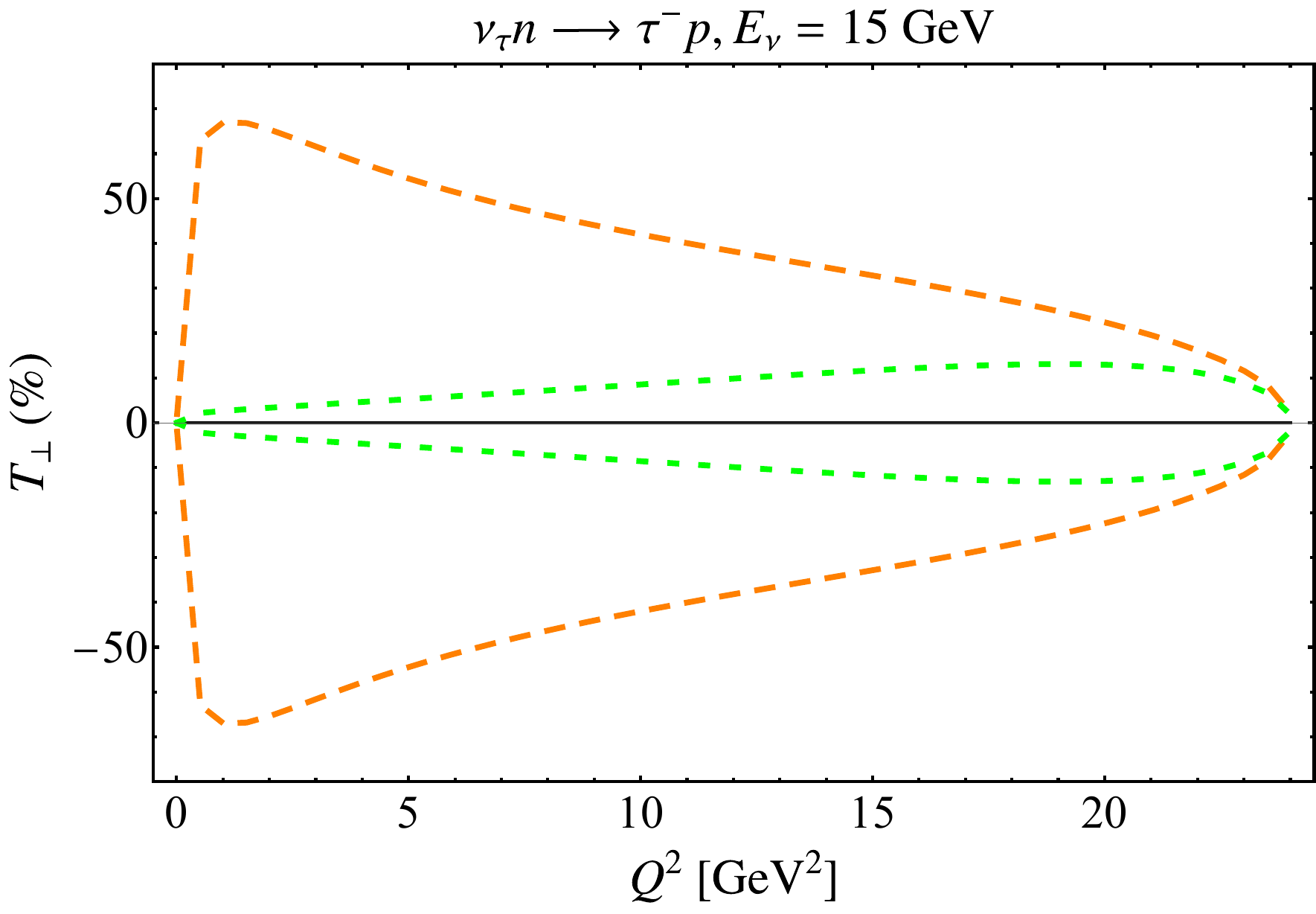}
\caption{Same as Fig.~\ref{fig:nu_Tt_SCFF_tau} but for the transverse polarization observable $T_\perp$ and imaginary amplitudes. \label{fig:nu_TT_SCFF_tau}}
\end{figure}

\begin{figure}[H]
\centering
\includegraphics[width=0.4\textwidth]{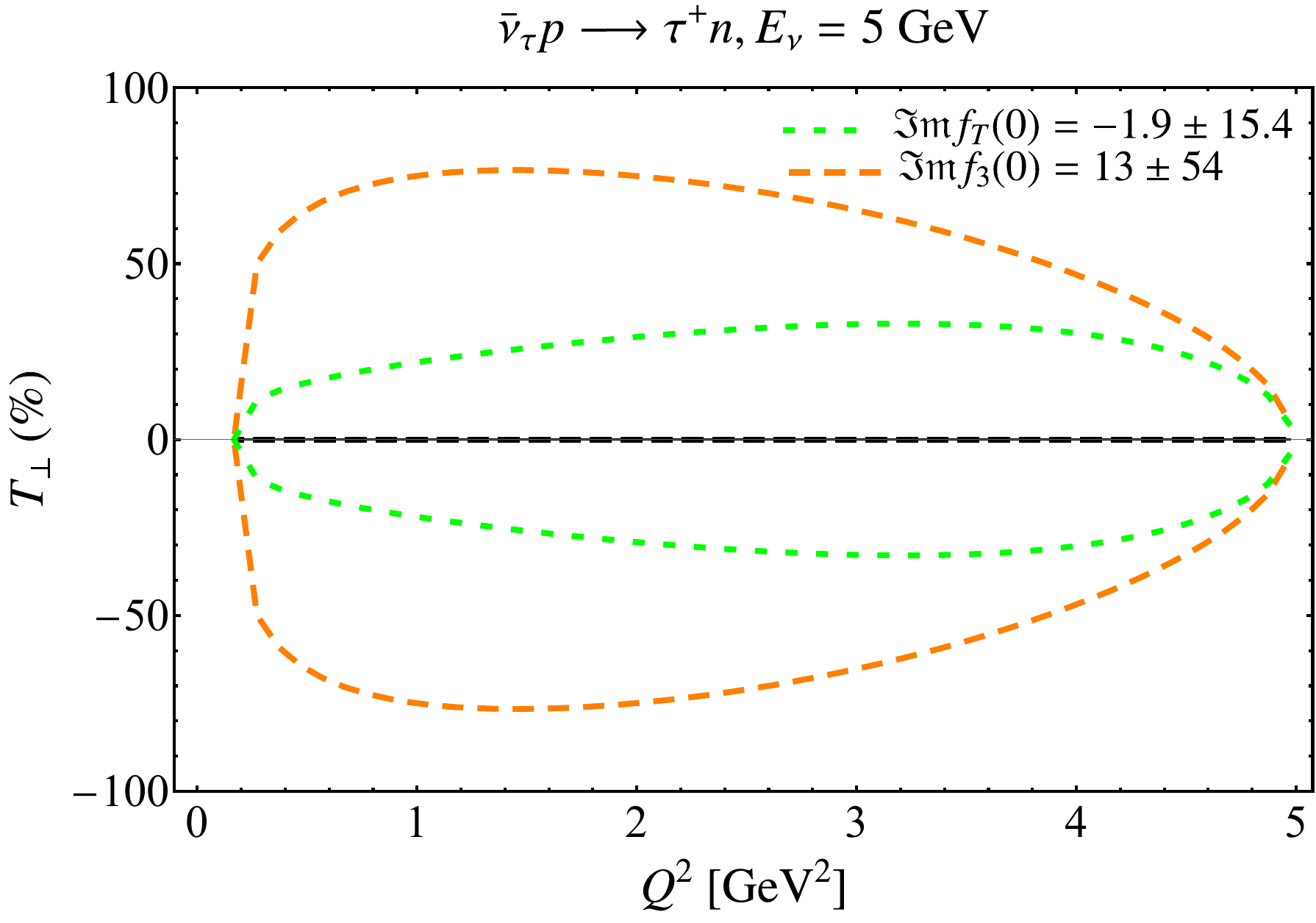}
\includegraphics[width=0.4\textwidth]{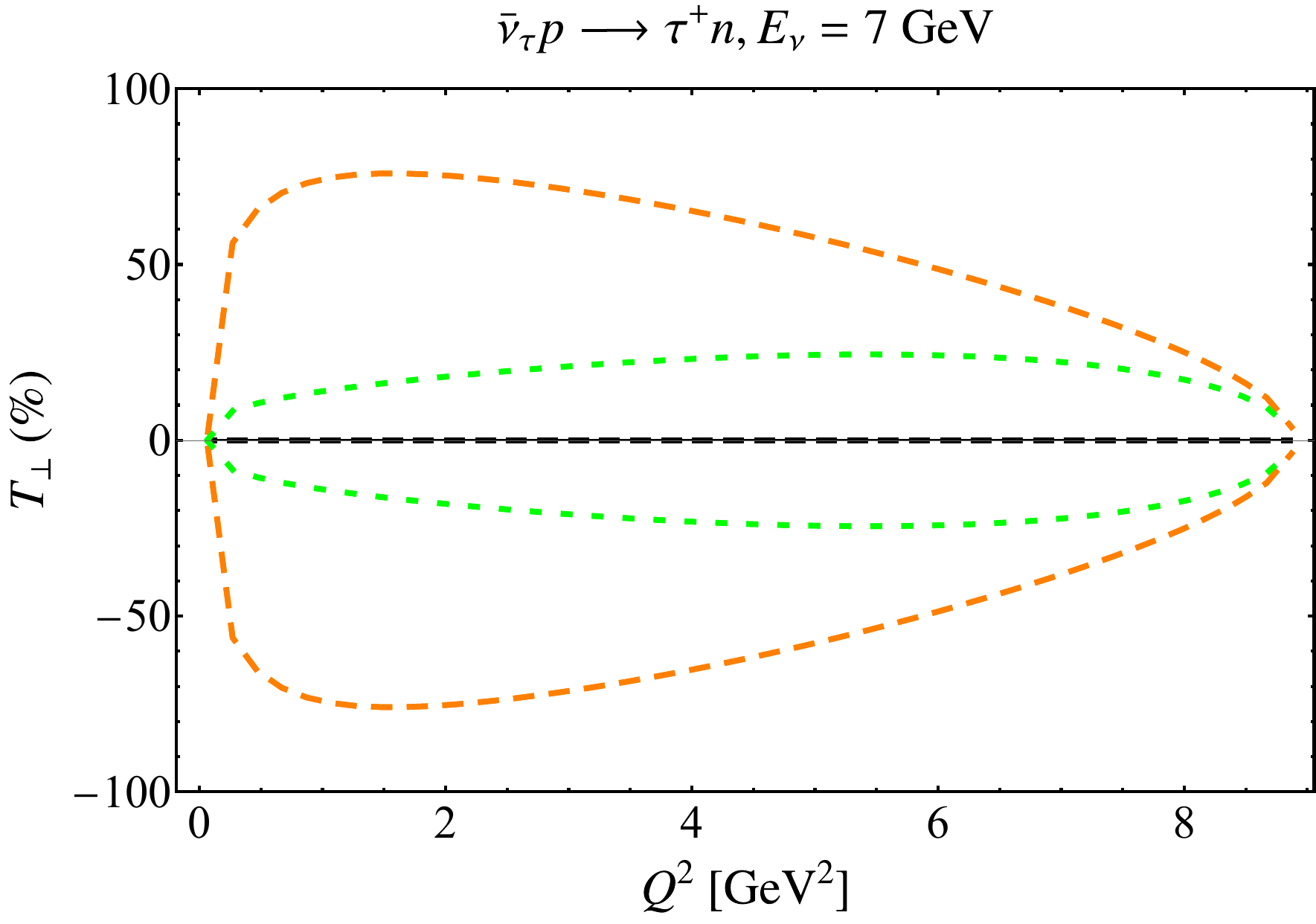}
\includegraphics[width=0.4\textwidth]{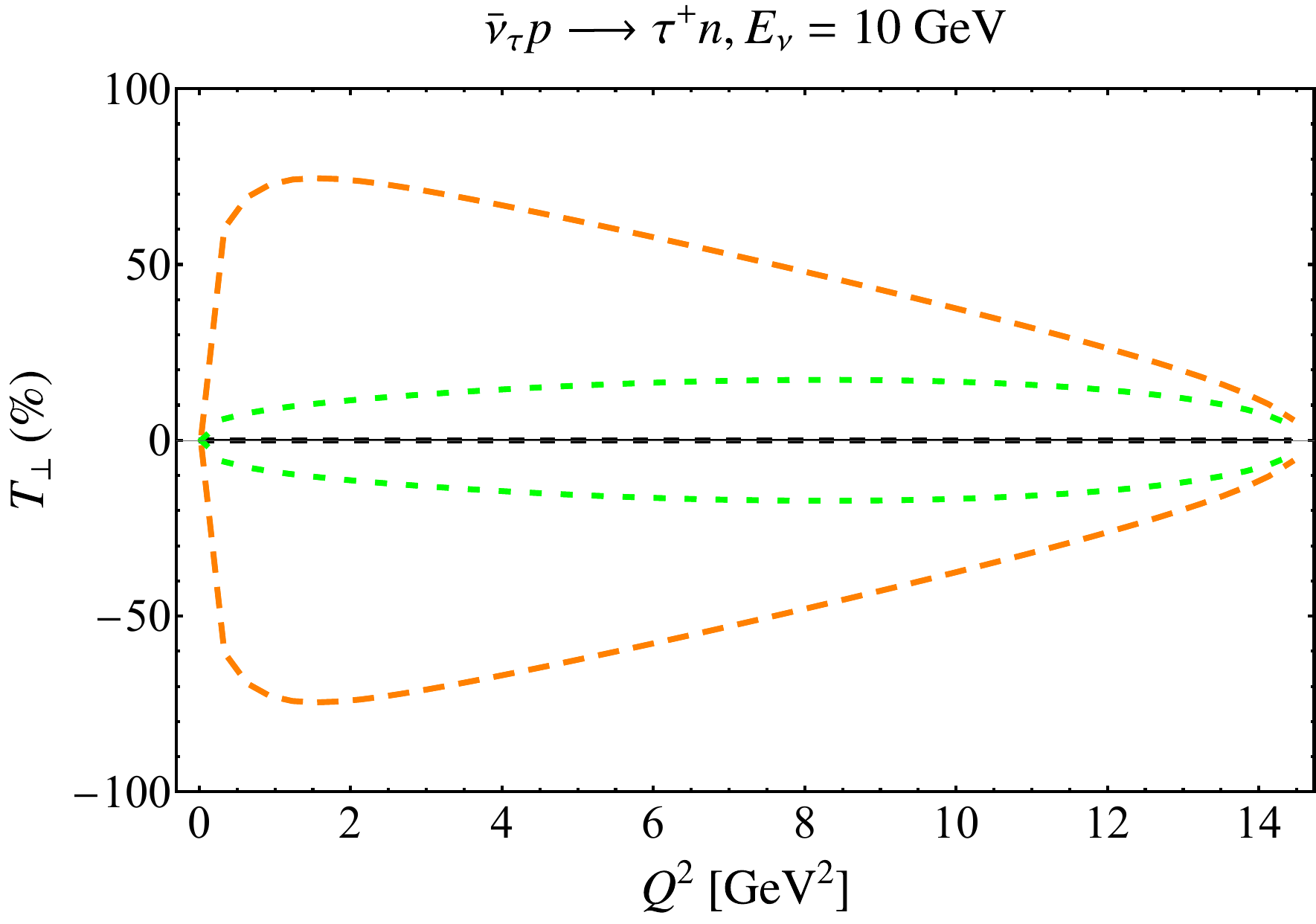}
\includegraphics[width=0.4\textwidth]{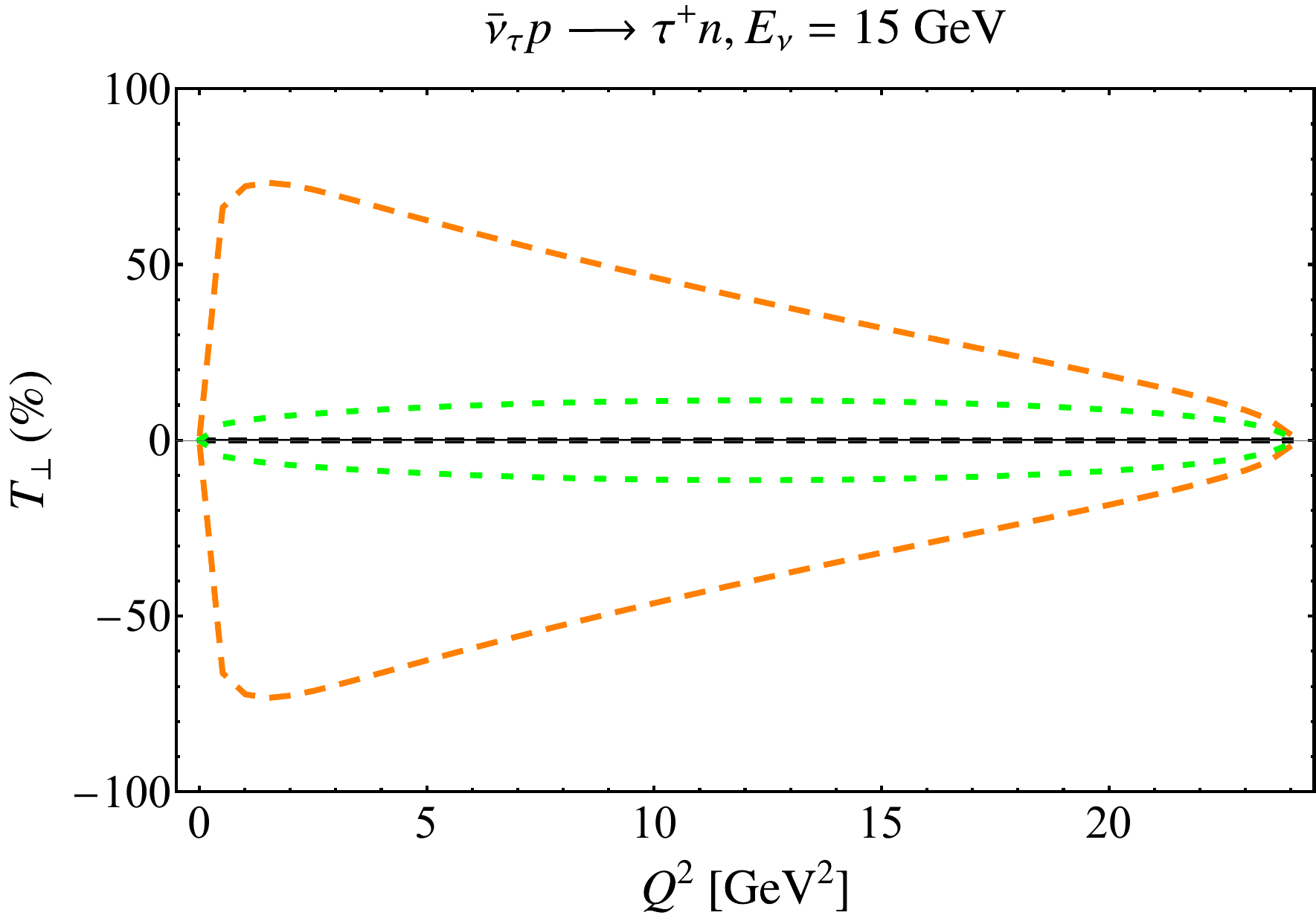}
\caption{Same as Fig.~\ref{fig:antinu_Tt_SCFF_tau} but for the transverse polarization observable $T_\perp$ and imaginary amplitudes. \label{fig:antinu_TT_SCFF_tau}}
\end{figure}

\begin{figure}[H]
\centering
\includegraphics[width=0.4\textwidth]{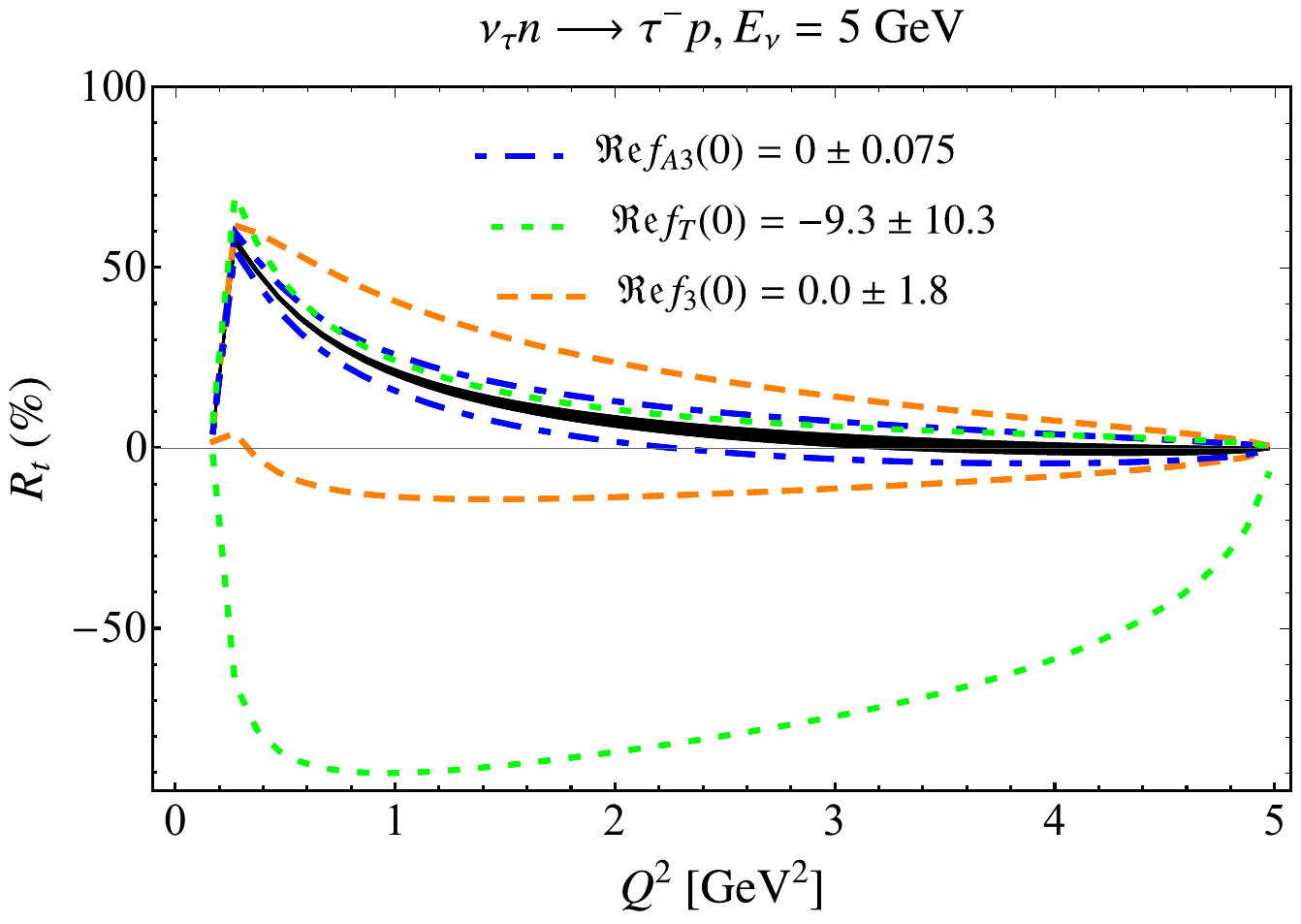}
\includegraphics[width=0.4\textwidth]{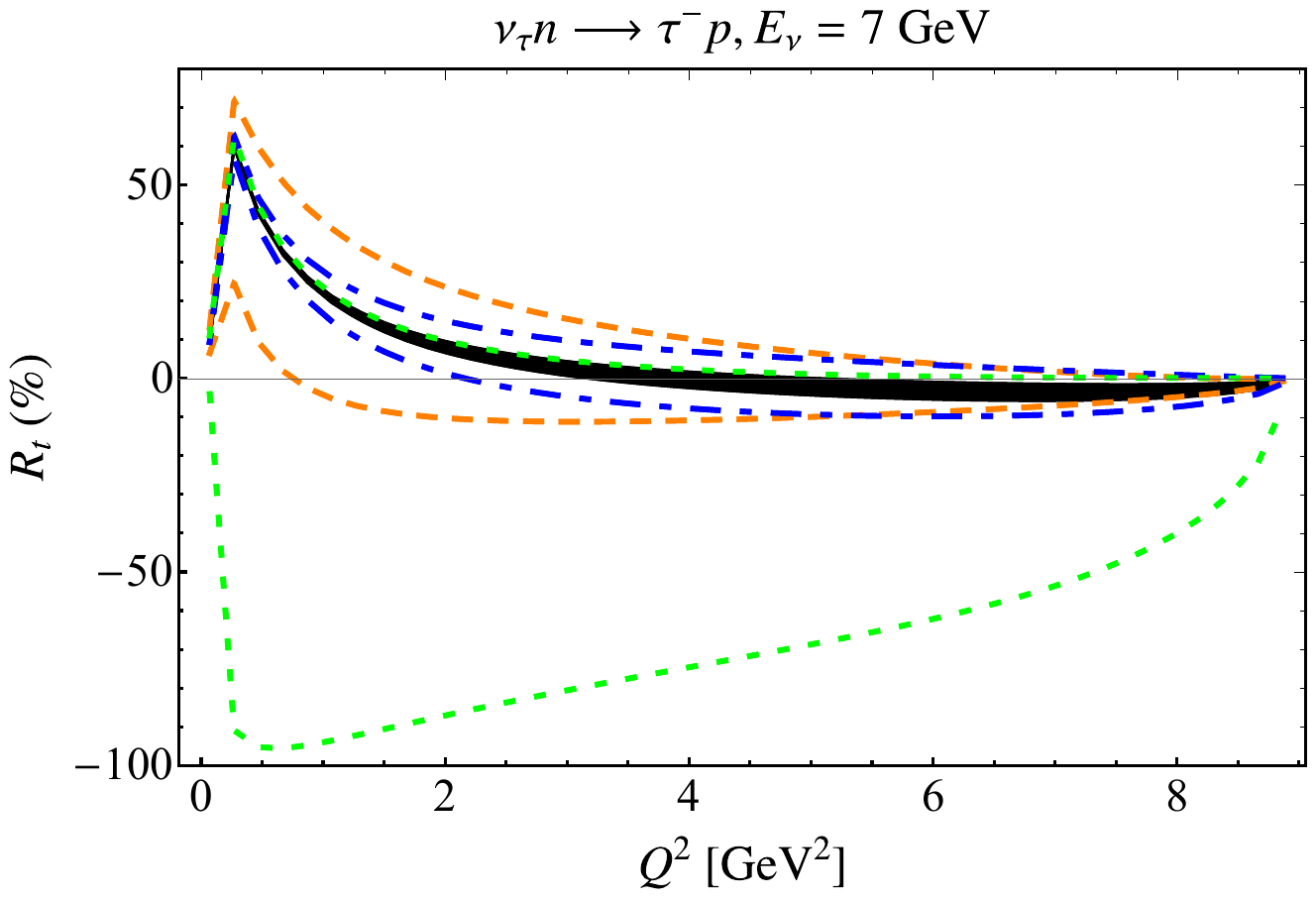}
\includegraphics[width=0.4\textwidth]{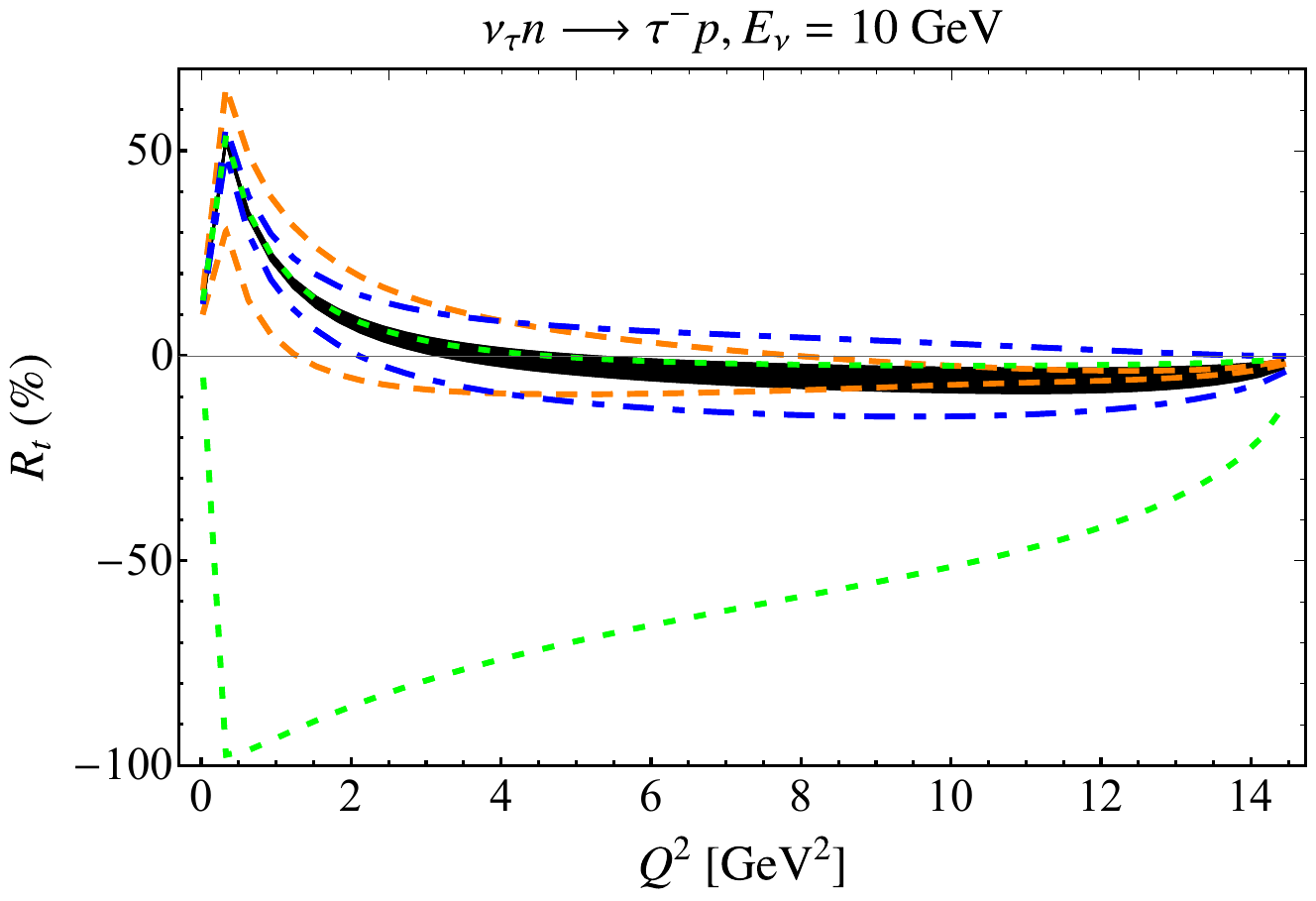}
\includegraphics[width=0.4\textwidth]{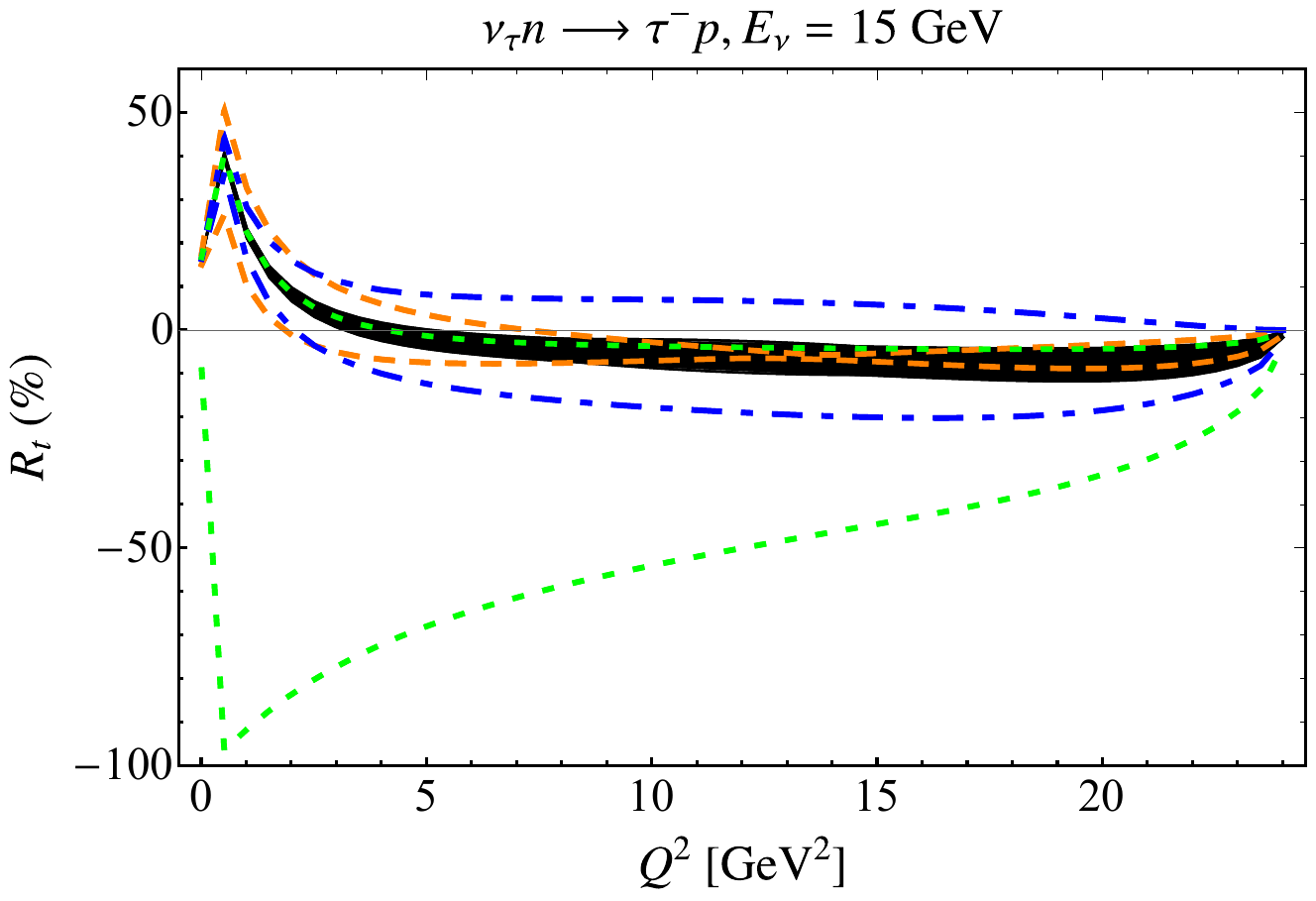}
\caption{Same as Fig.~\ref{fig:nu_Tt_SCFF_tau} but for the transverse polarization observable $R_t$. \label{fig:nu_Rt_SCFF_tau}}
\end{figure}

\begin{figure}[H]
\centering
\includegraphics[width=0.4\textwidth]{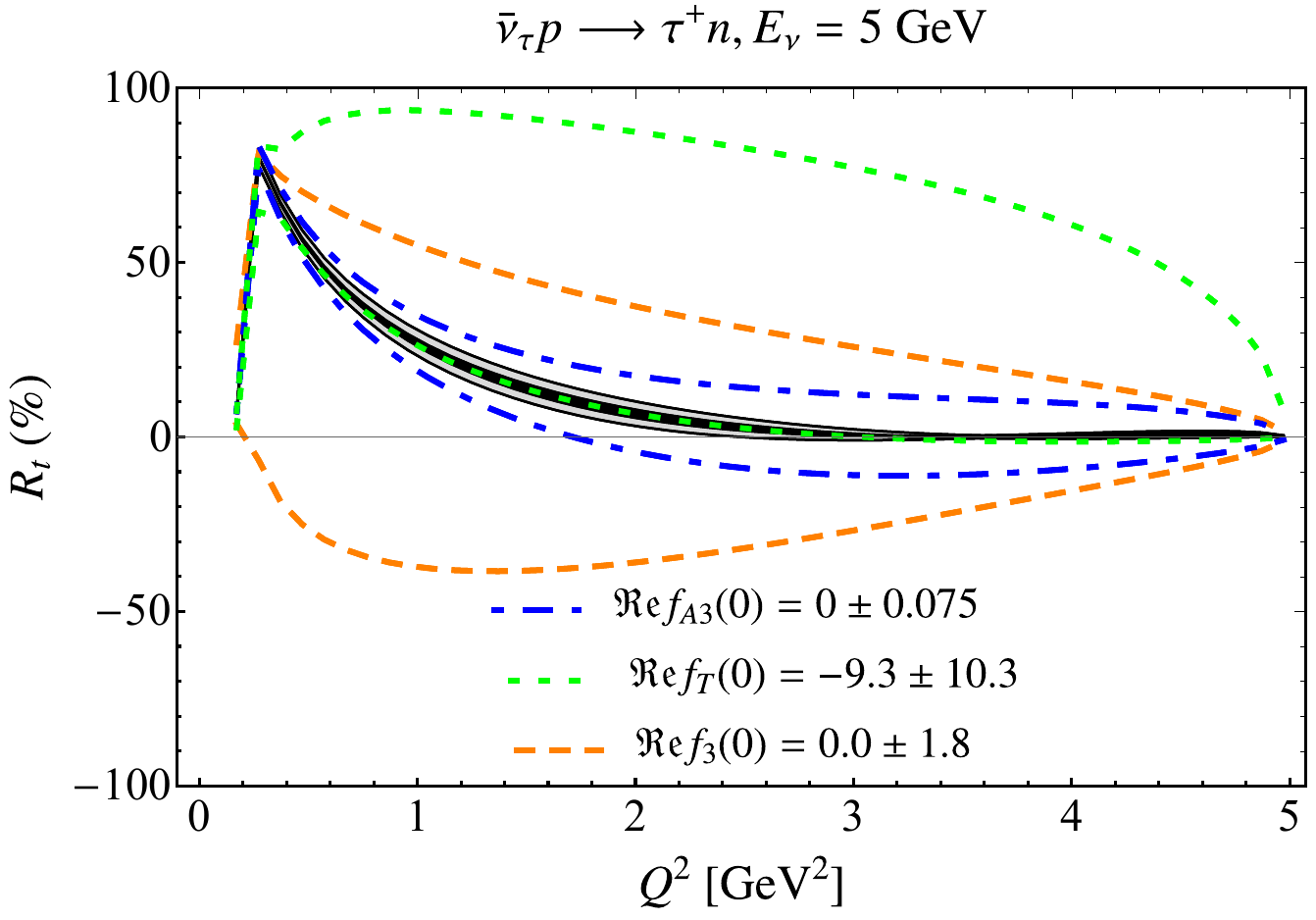}
\includegraphics[width=0.4\textwidth]{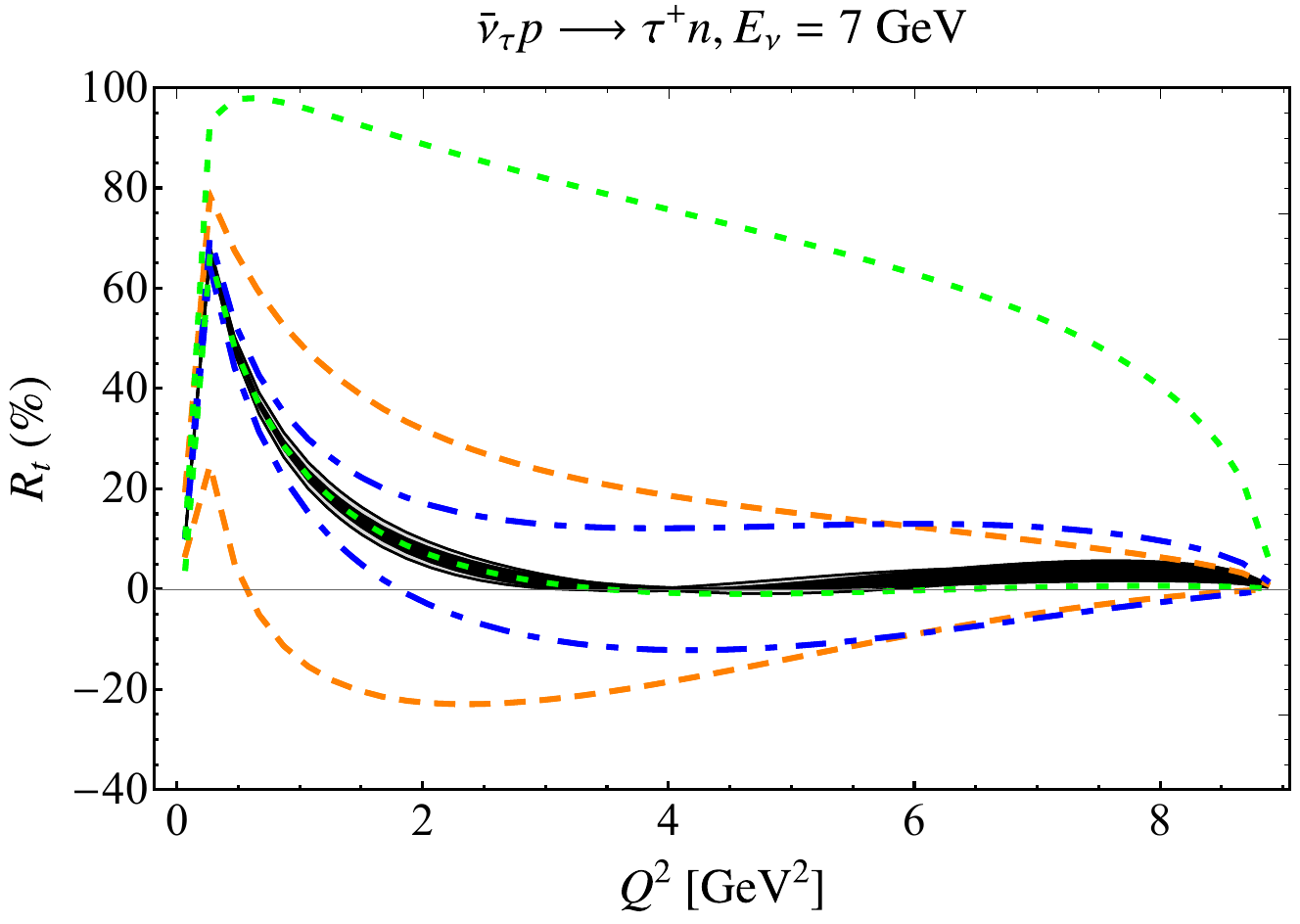}
\includegraphics[width=0.4\textwidth]{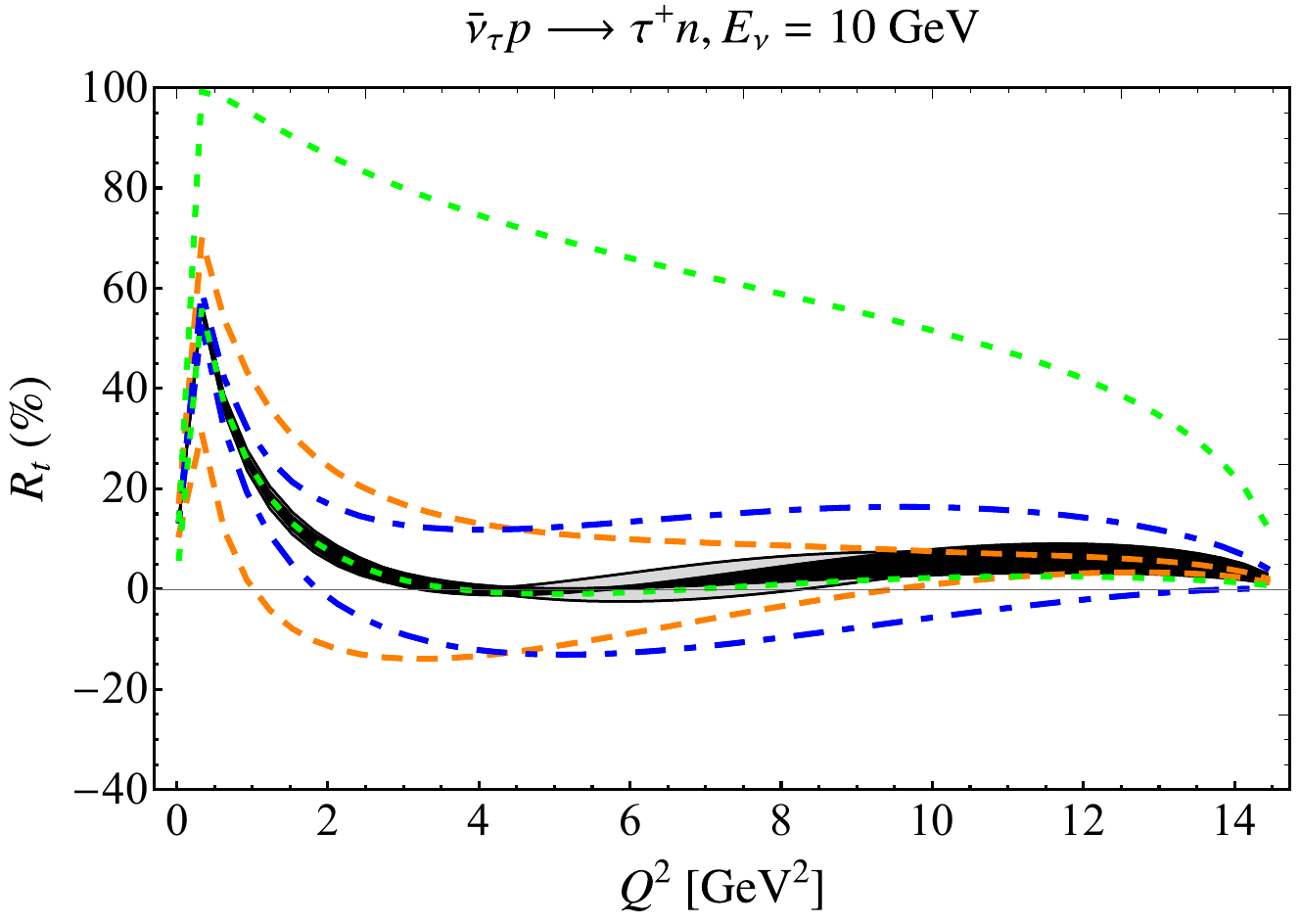}
\includegraphics[width=0.4\textwidth]{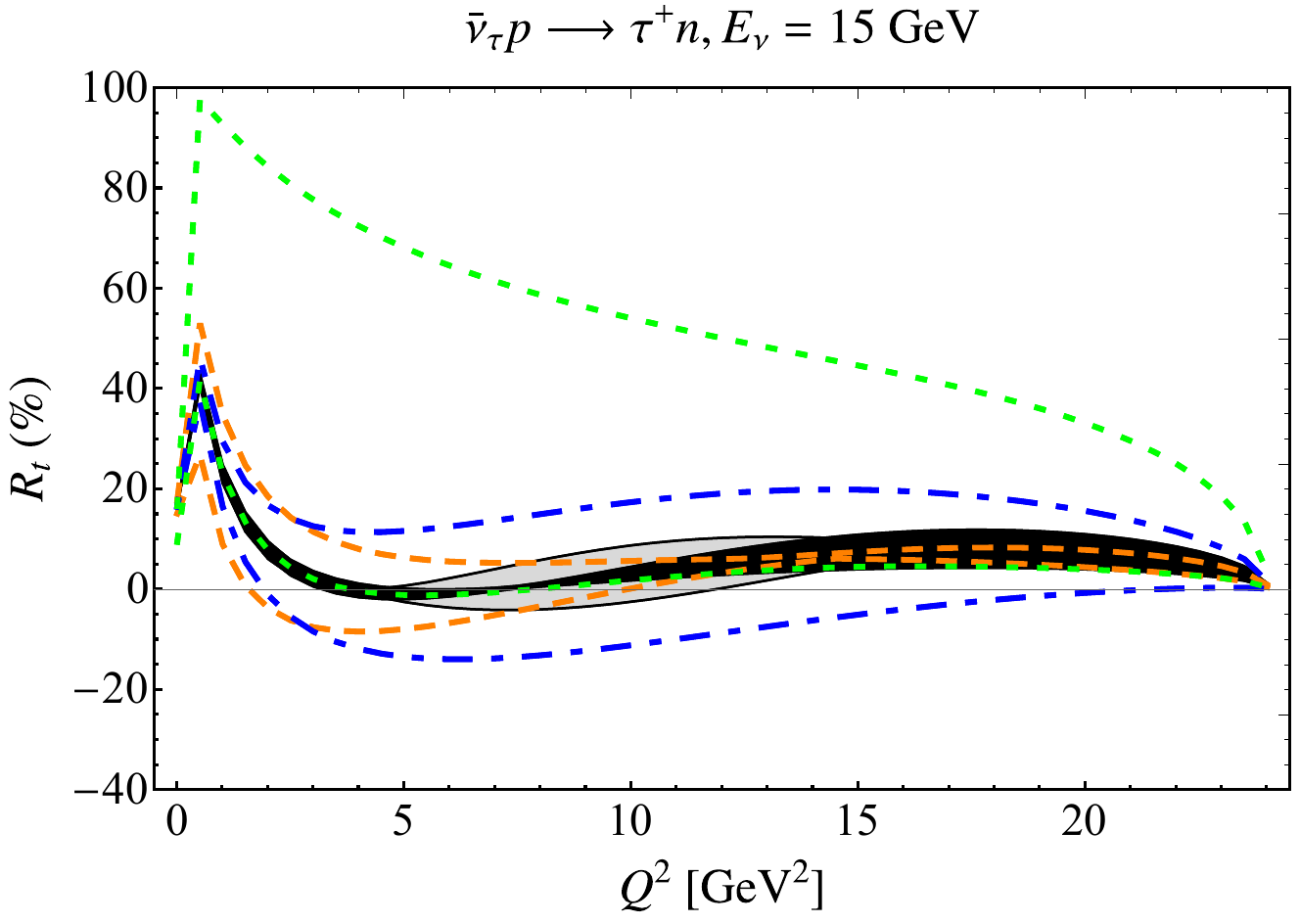}
\caption{Same as Fig.~\ref{fig:antinu_Tt_SCFF_tau} but for the transverse polarization observable $R_t$. \label{fig:antinu_Rt_SCFF_tau}}
\end{figure}

\begin{figure}[H]
\centering
\includegraphics[width=0.4\textwidth]{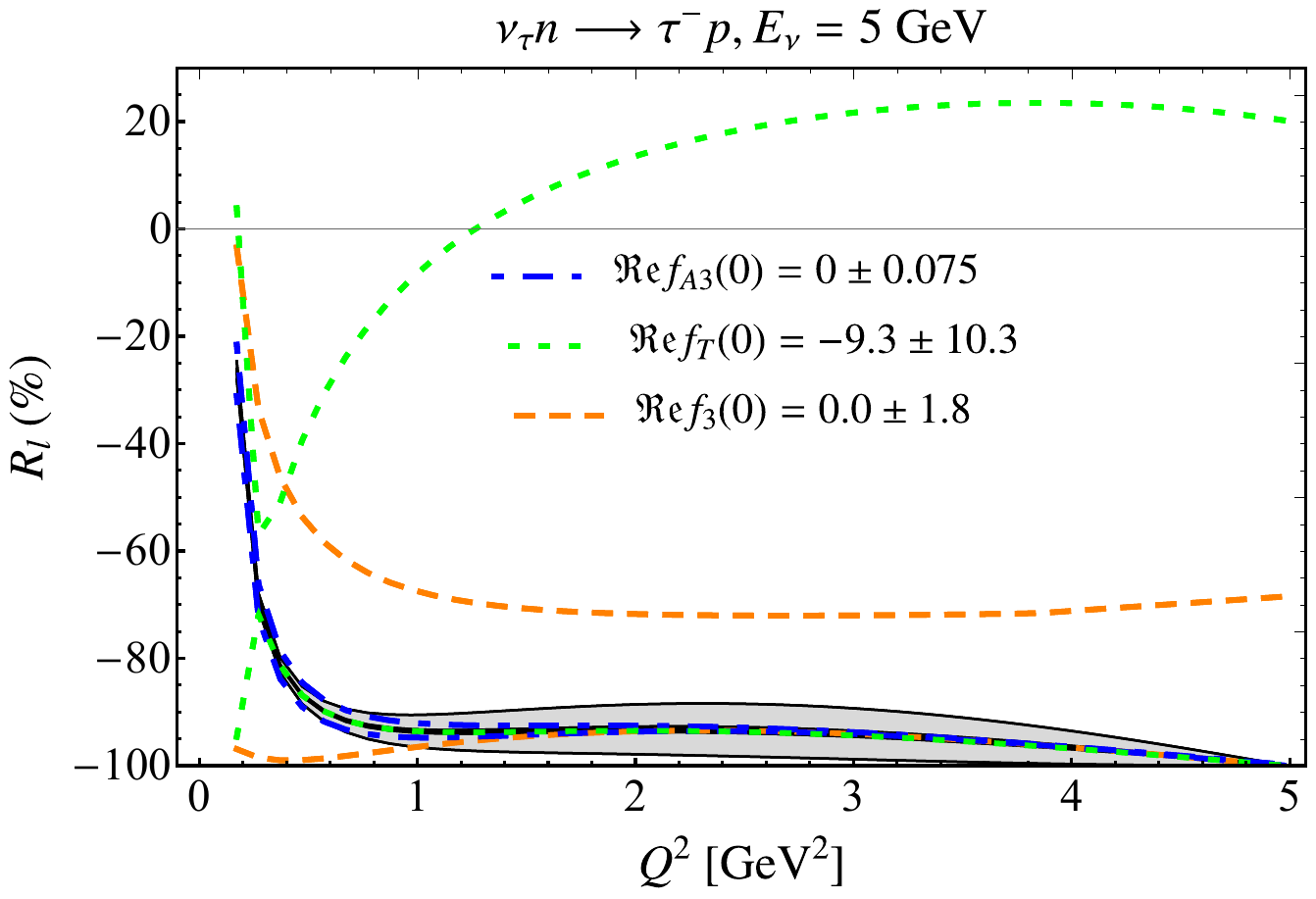}
\includegraphics[width=0.4\textwidth]{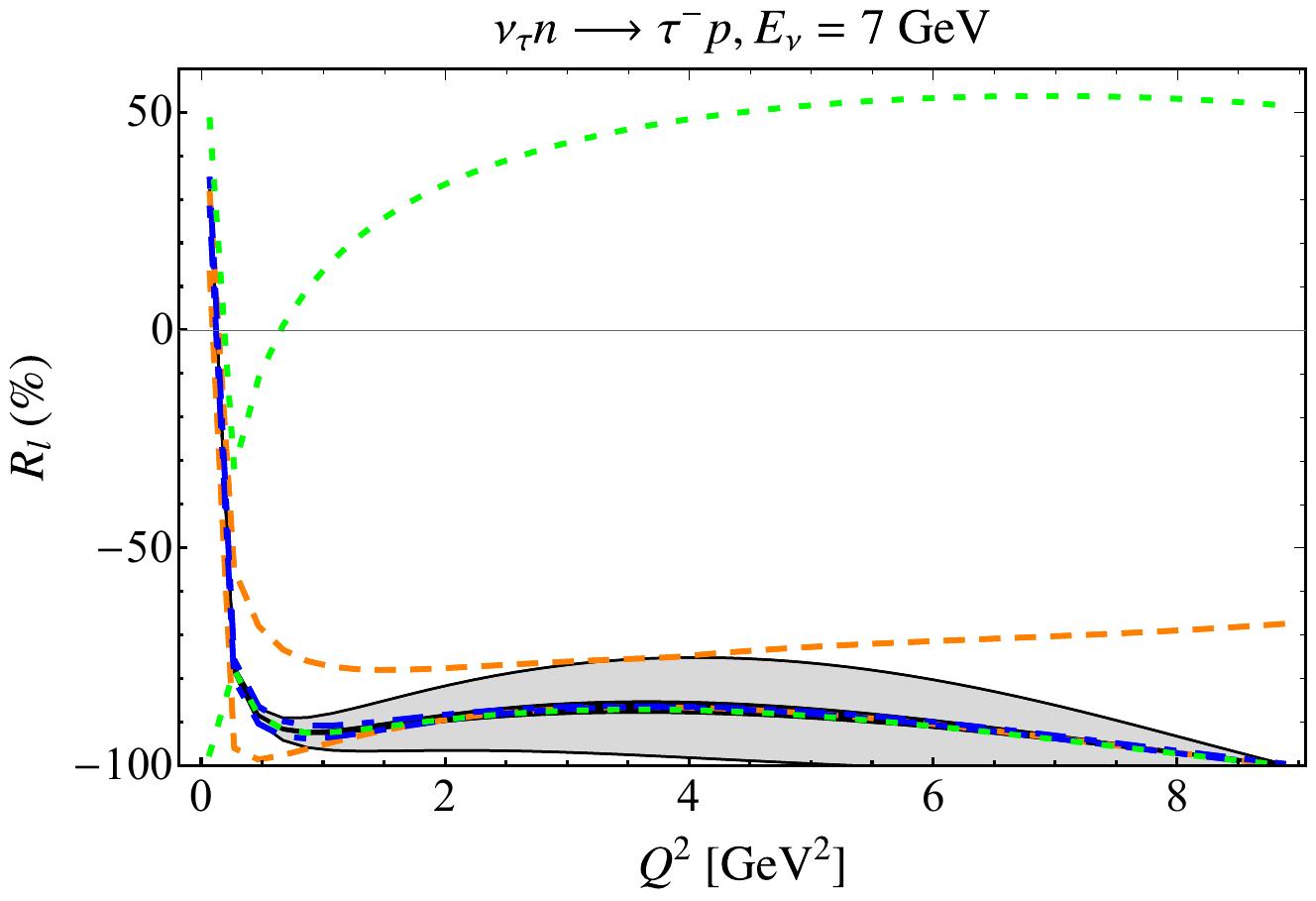}
\includegraphics[width=0.4\textwidth]{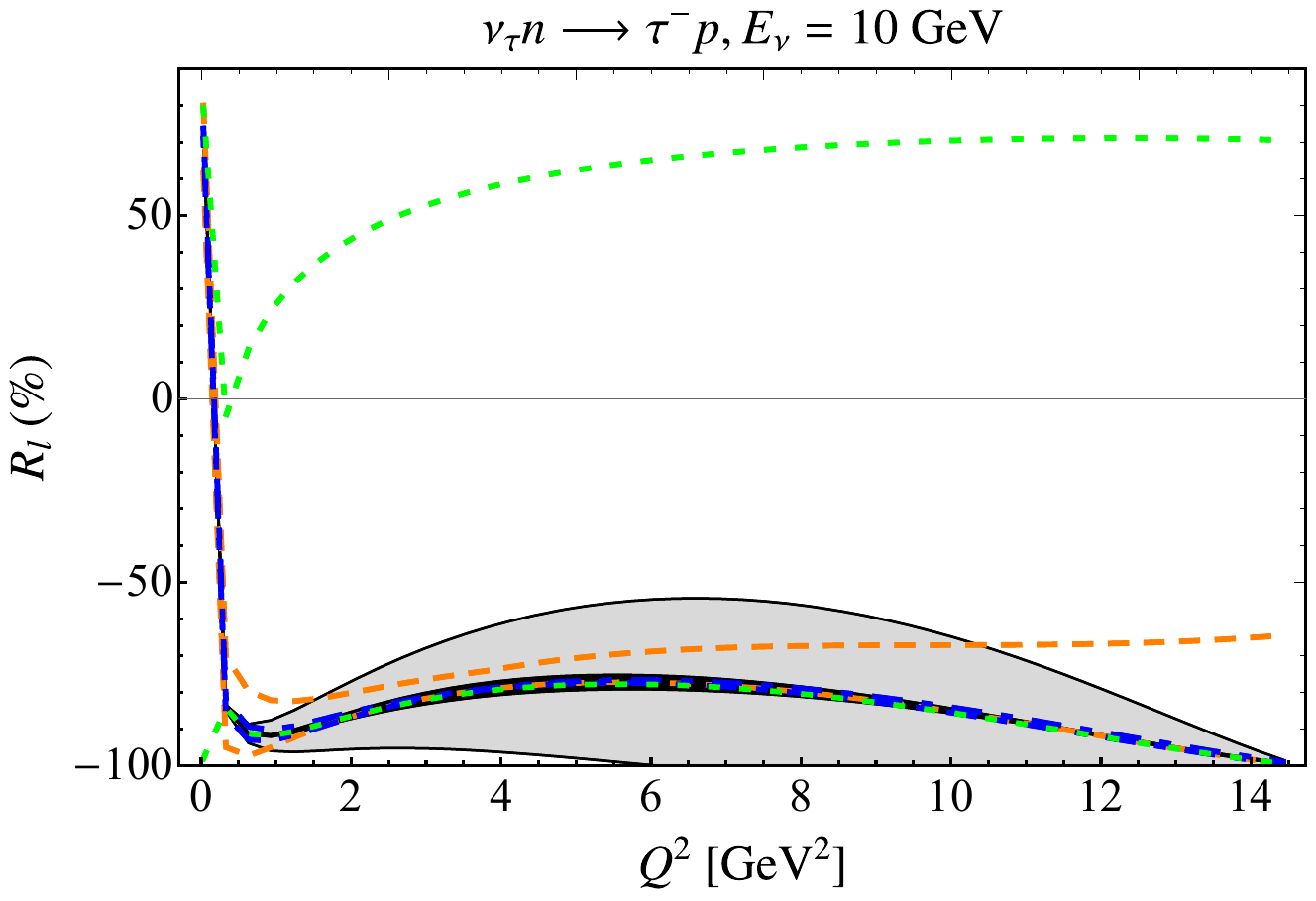}
\includegraphics[width=0.4\textwidth]{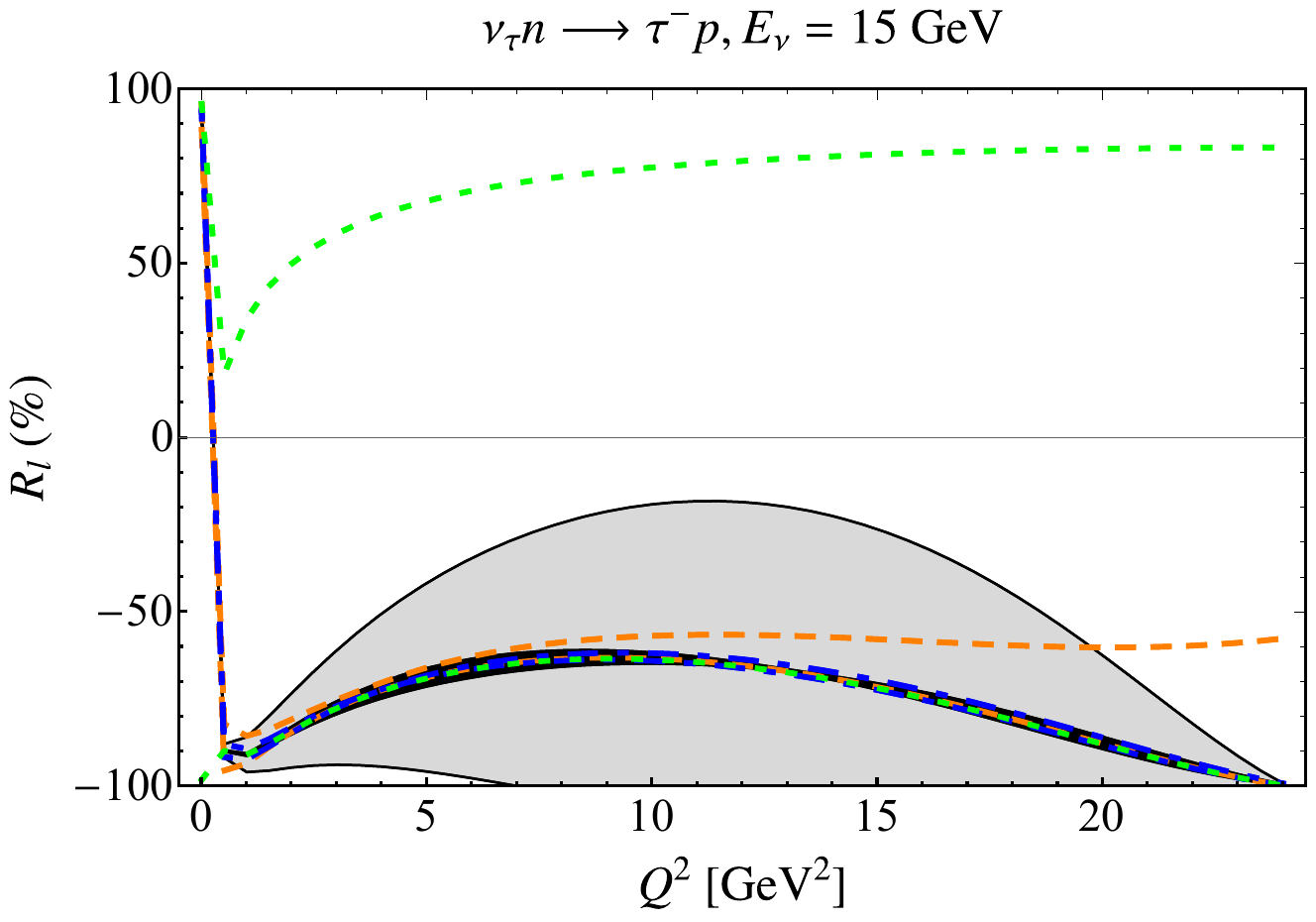}
\caption{Same as Fig.~\ref{fig:nu_Tt_SCFF_tau} but for the longitudinal polarization observable $R_l$. \label{fig:nu_Rl_SCFF_tau}}
\end{figure}

\begin{figure}[H]
\centering
\includegraphics[width=0.4\textwidth]{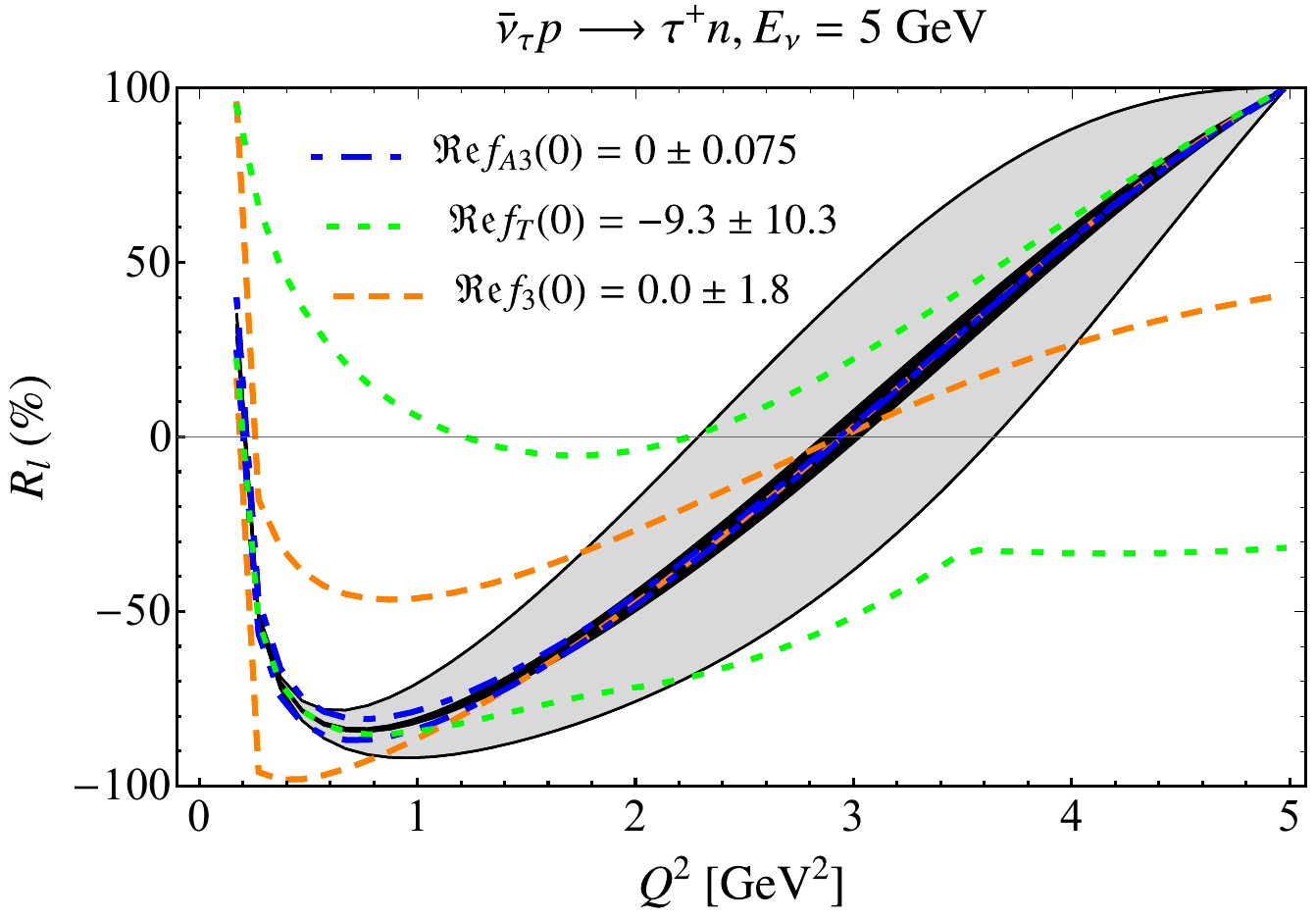}
\includegraphics[width=0.4\textwidth]{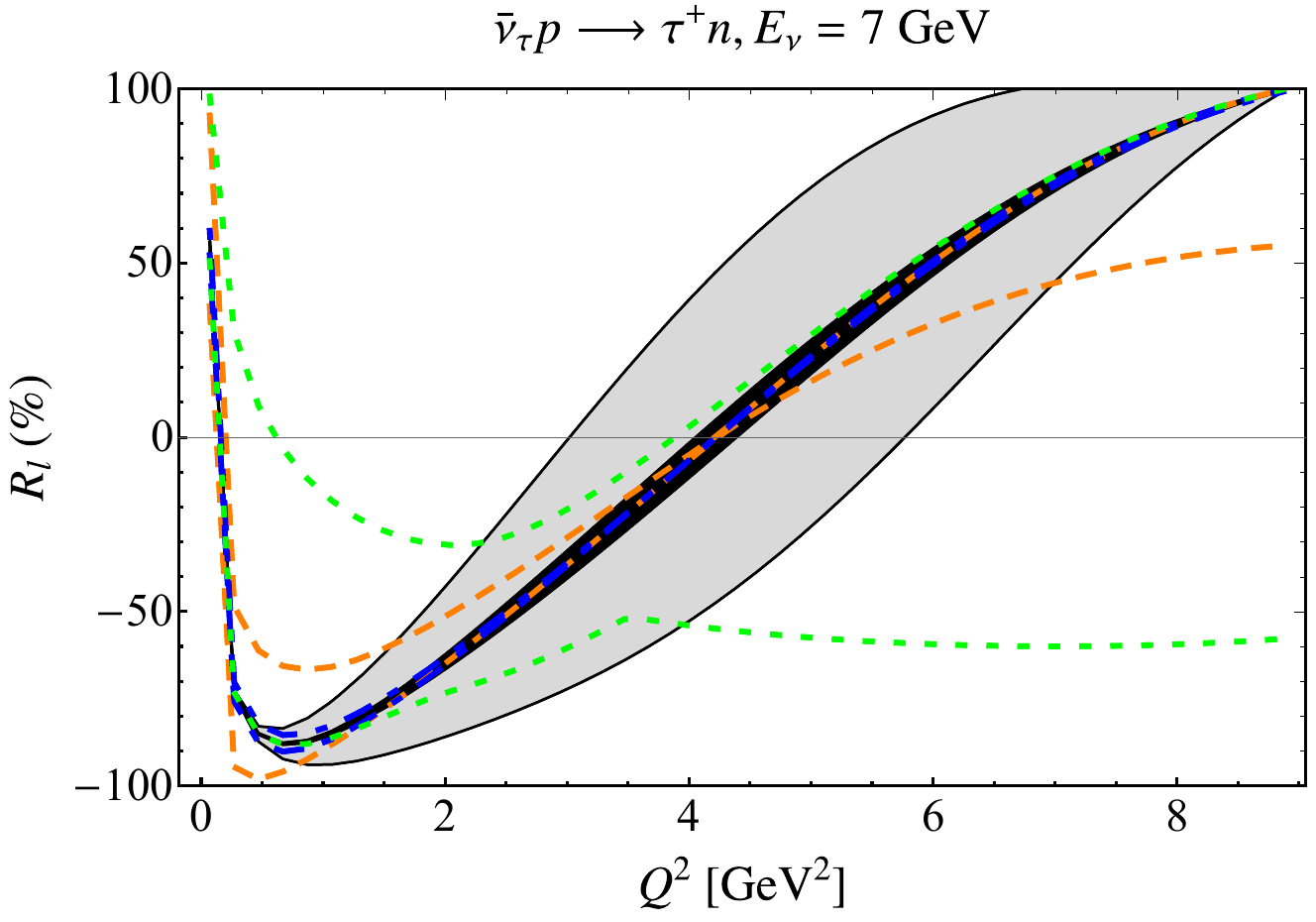}
\includegraphics[width=0.4\textwidth]{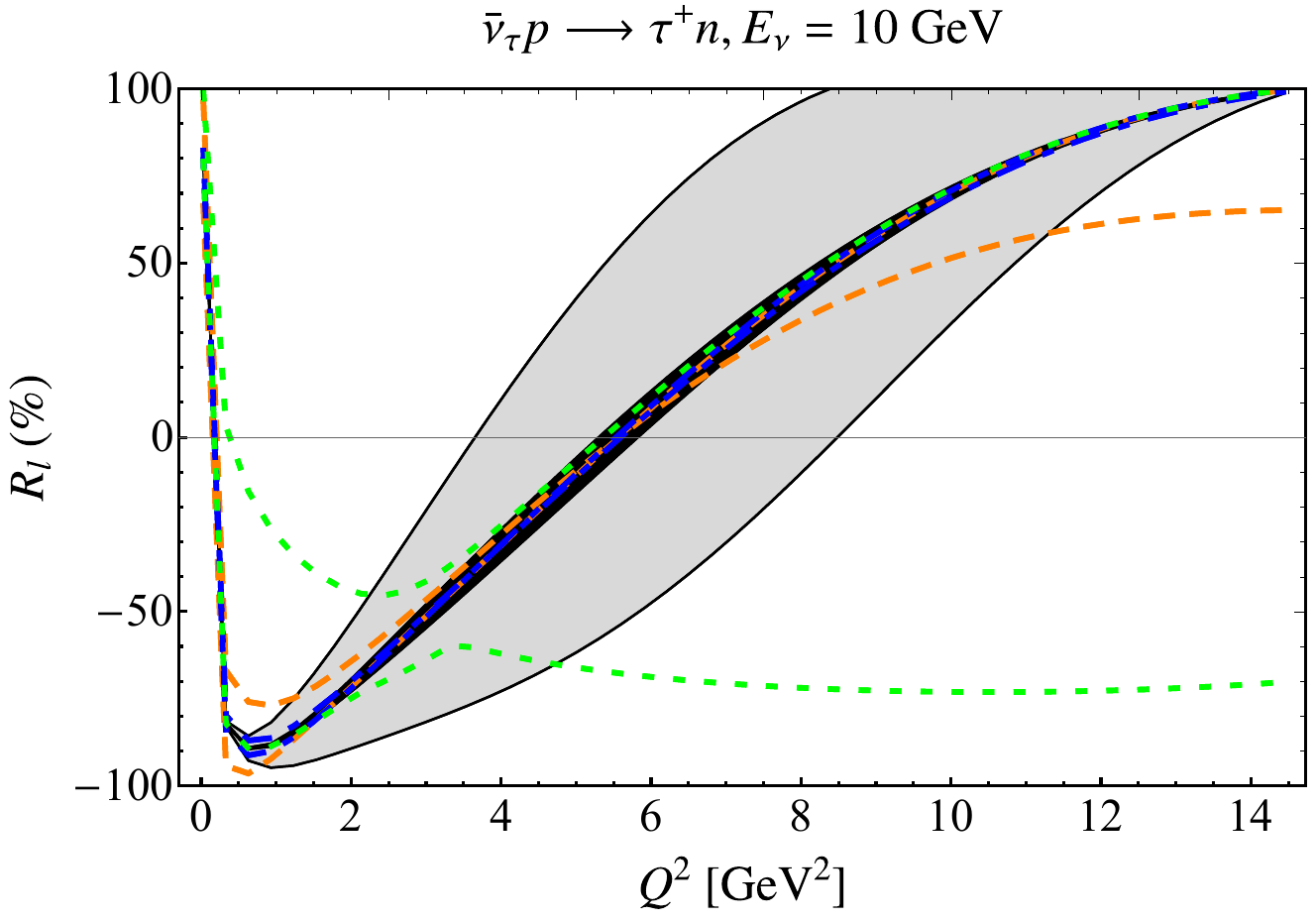}
\includegraphics[width=0.4\textwidth]{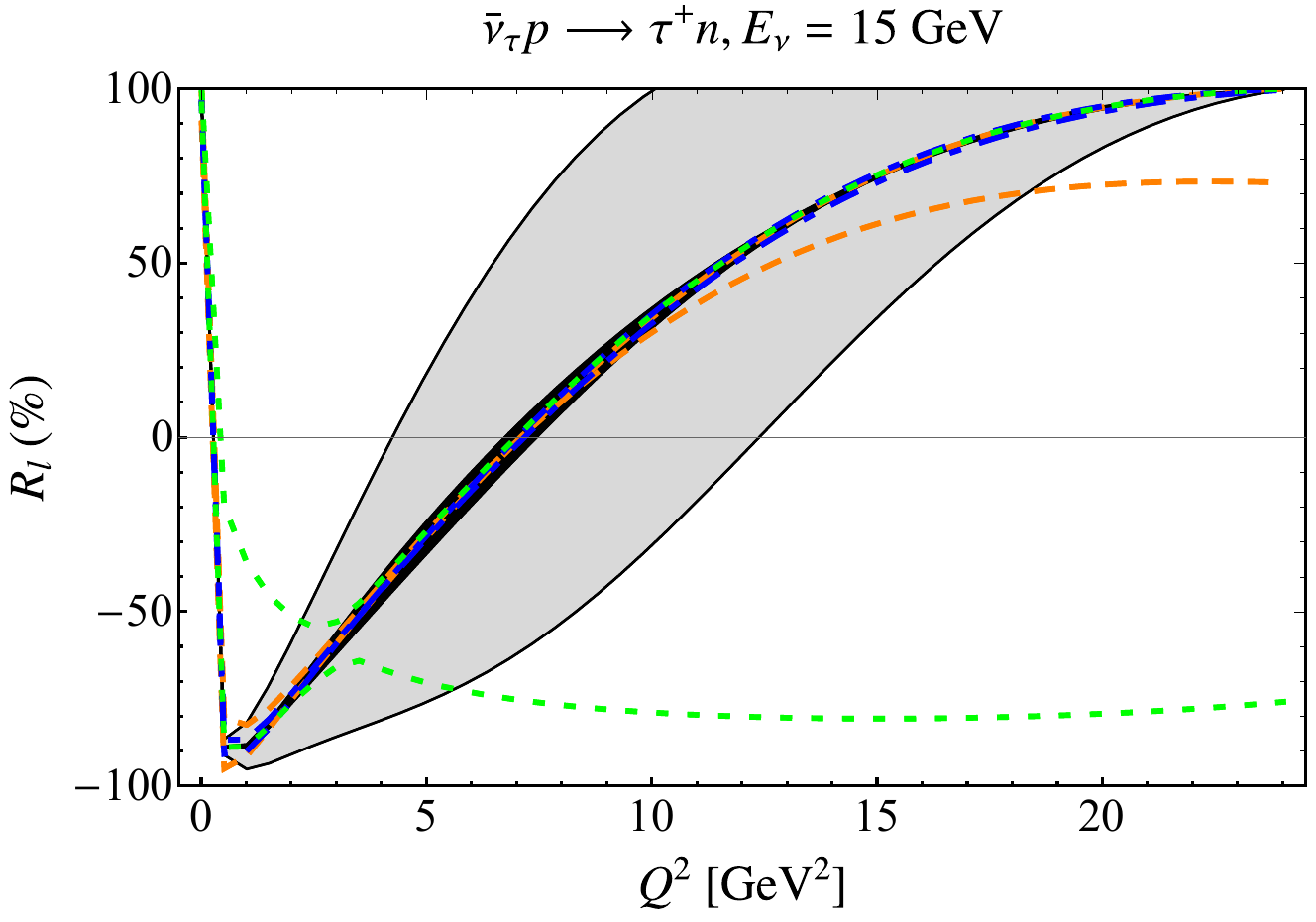}
\caption{Same as Fig.~\ref{fig:antinu_Tt_SCFF_tau} but for the longitudinal polarization observable $R_l$. \label{fig:antinu_Rl_SCFF_tau}}
\end{figure}

\begin{figure}[H]
\centering
\includegraphics[width=0.4\textwidth]{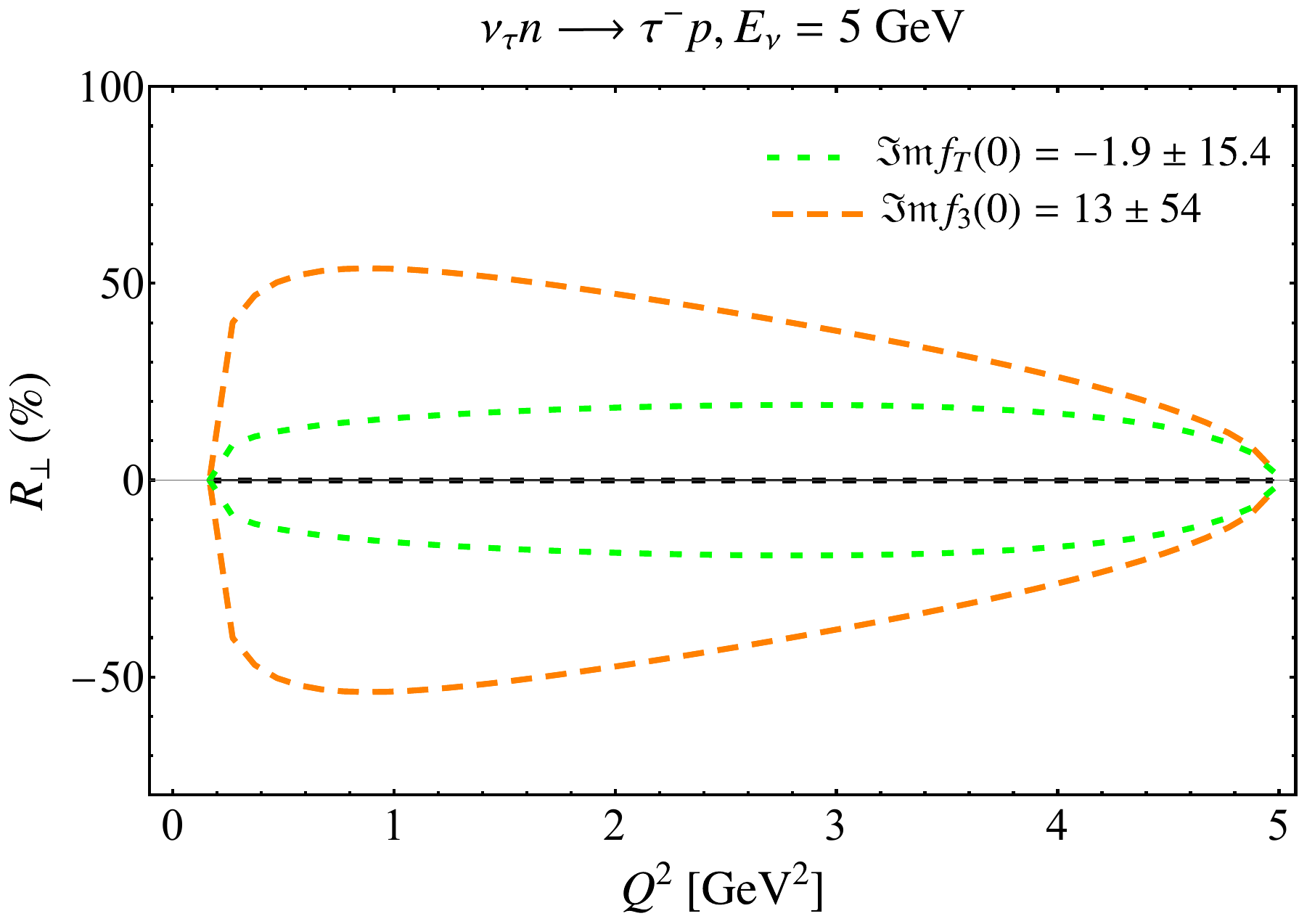}
\includegraphics[width=0.4\textwidth]{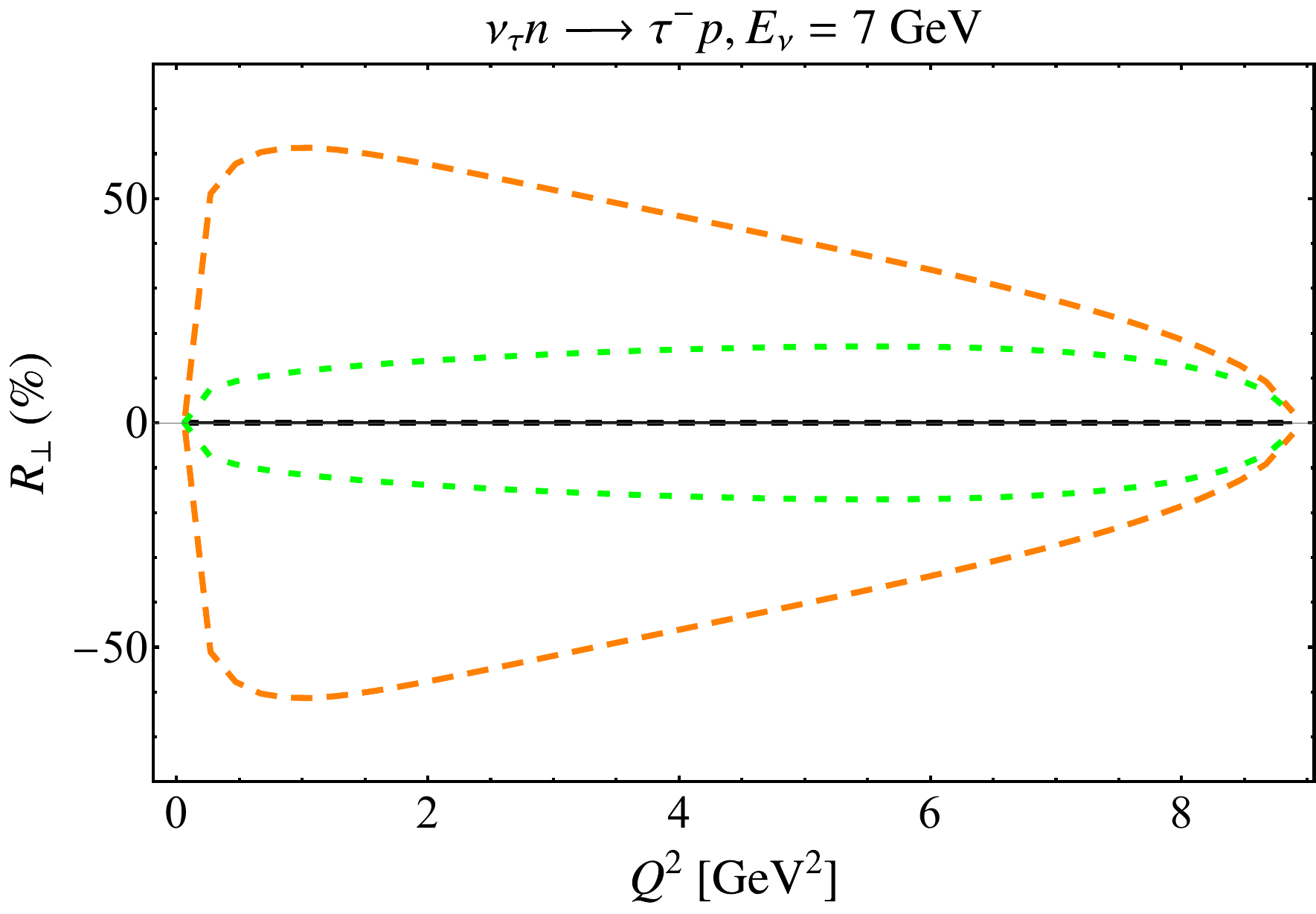}
\includegraphics[width=0.4\textwidth]{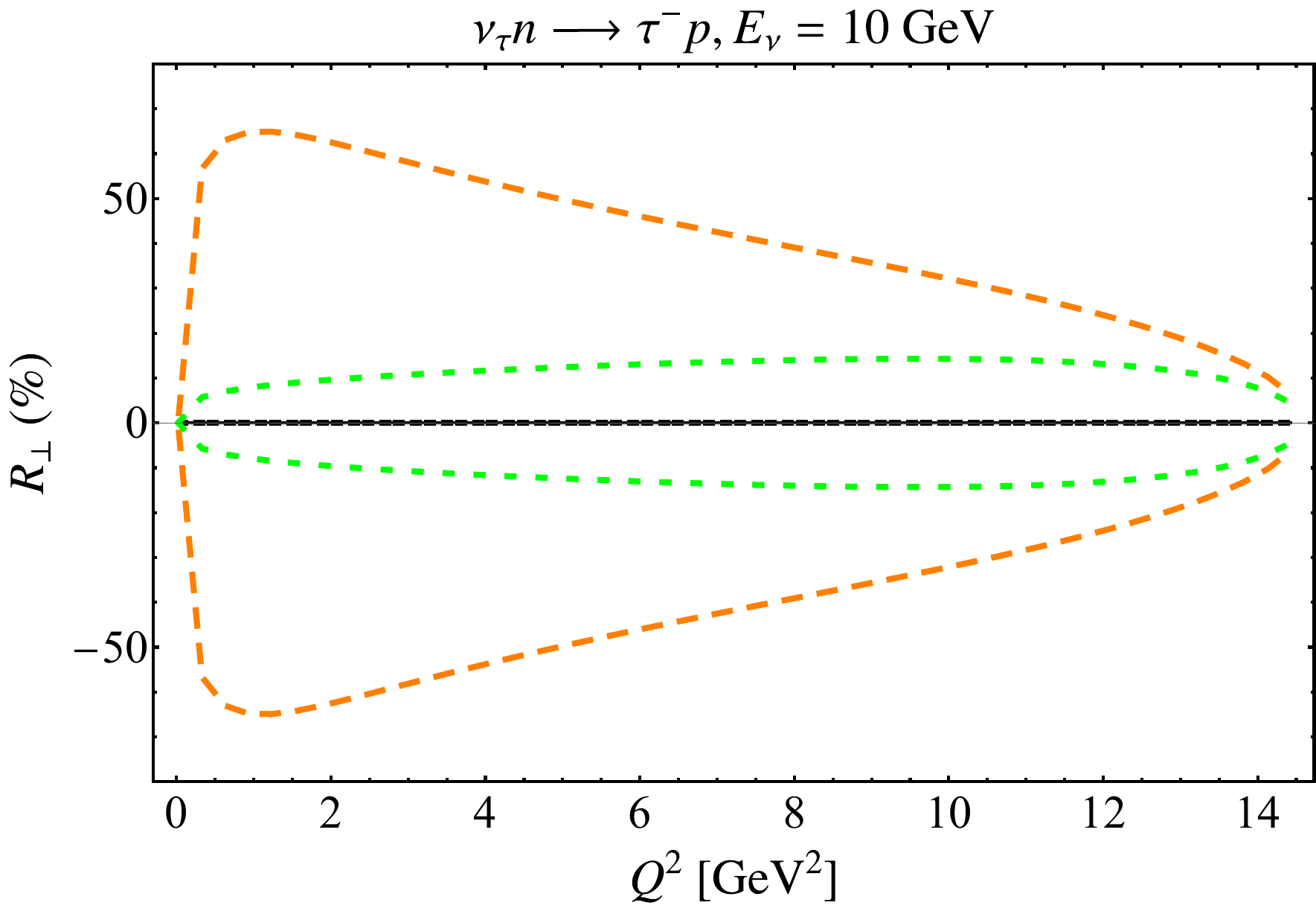}
\includegraphics[width=0.4\textwidth]{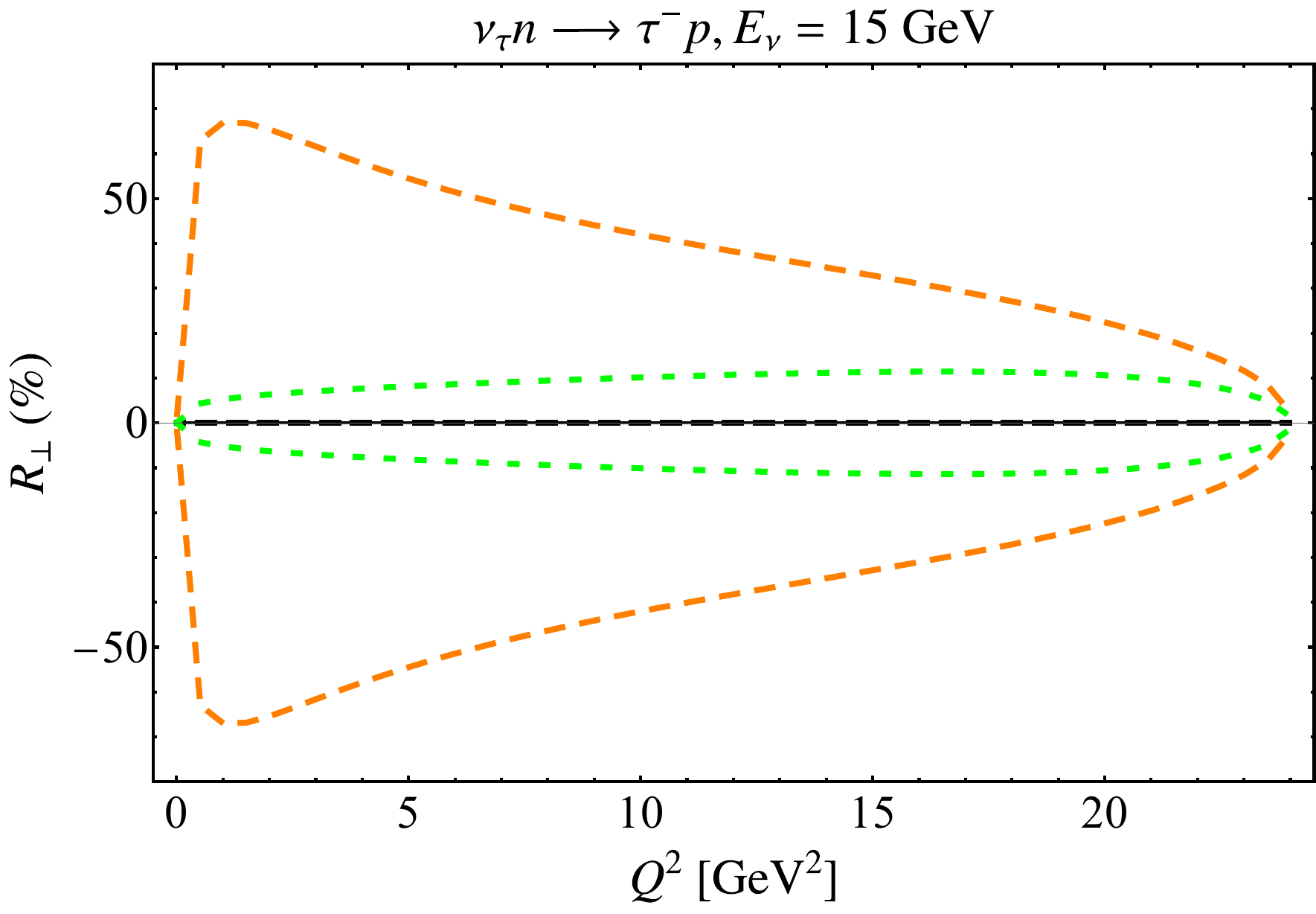}
\caption{Same as Fig.~\ref{fig:nu_Tt_SCFF_tau} but for the transverse polarization observable $R_\perp$ and imaginary amplitudes. \label{fig:nu_RTT_SCFF_tau}}
\end{figure}

\begin{figure}[H]
\centering
\includegraphics[width=0.4\textwidth]{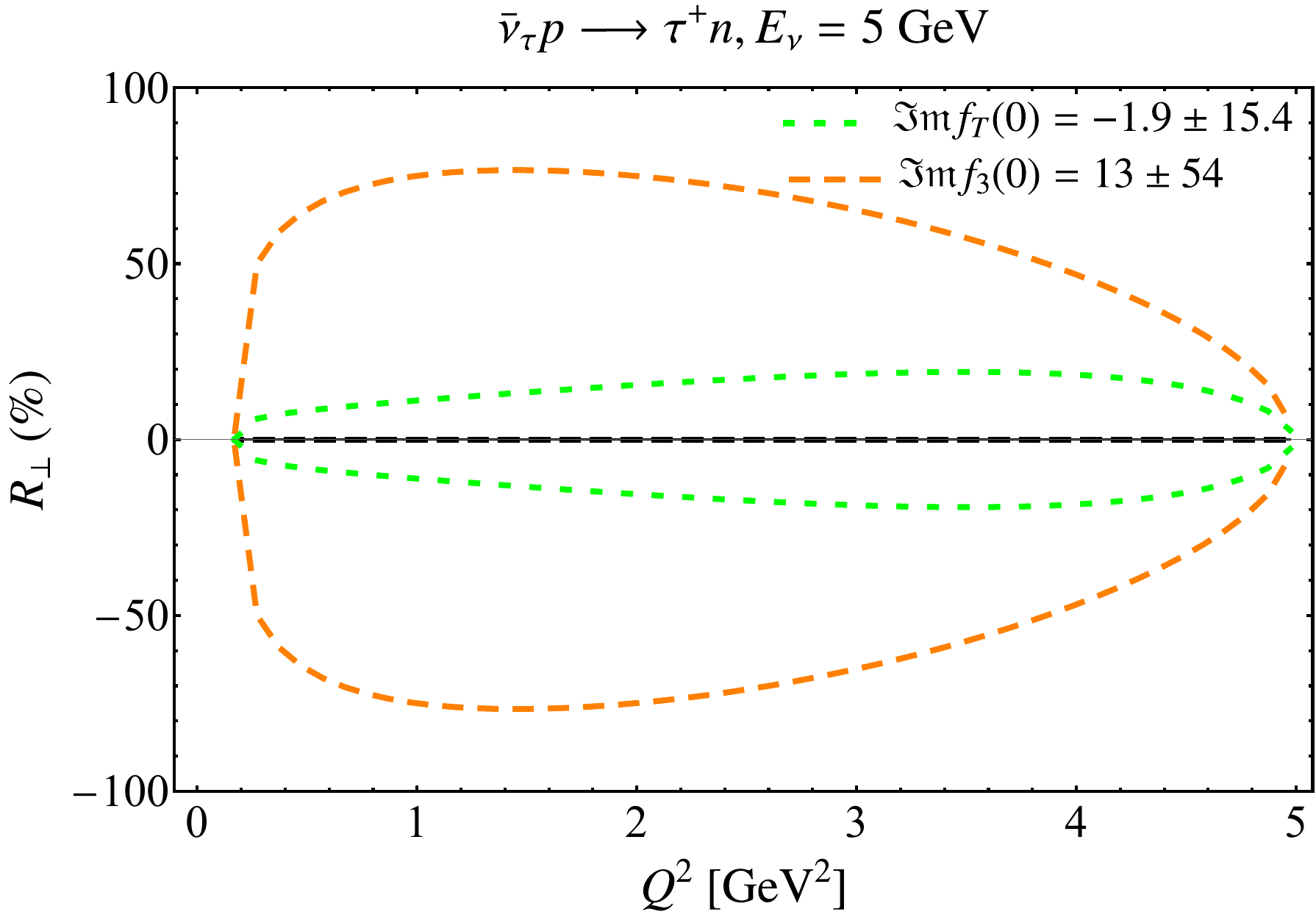}
\includegraphics[width=0.4\textwidth]{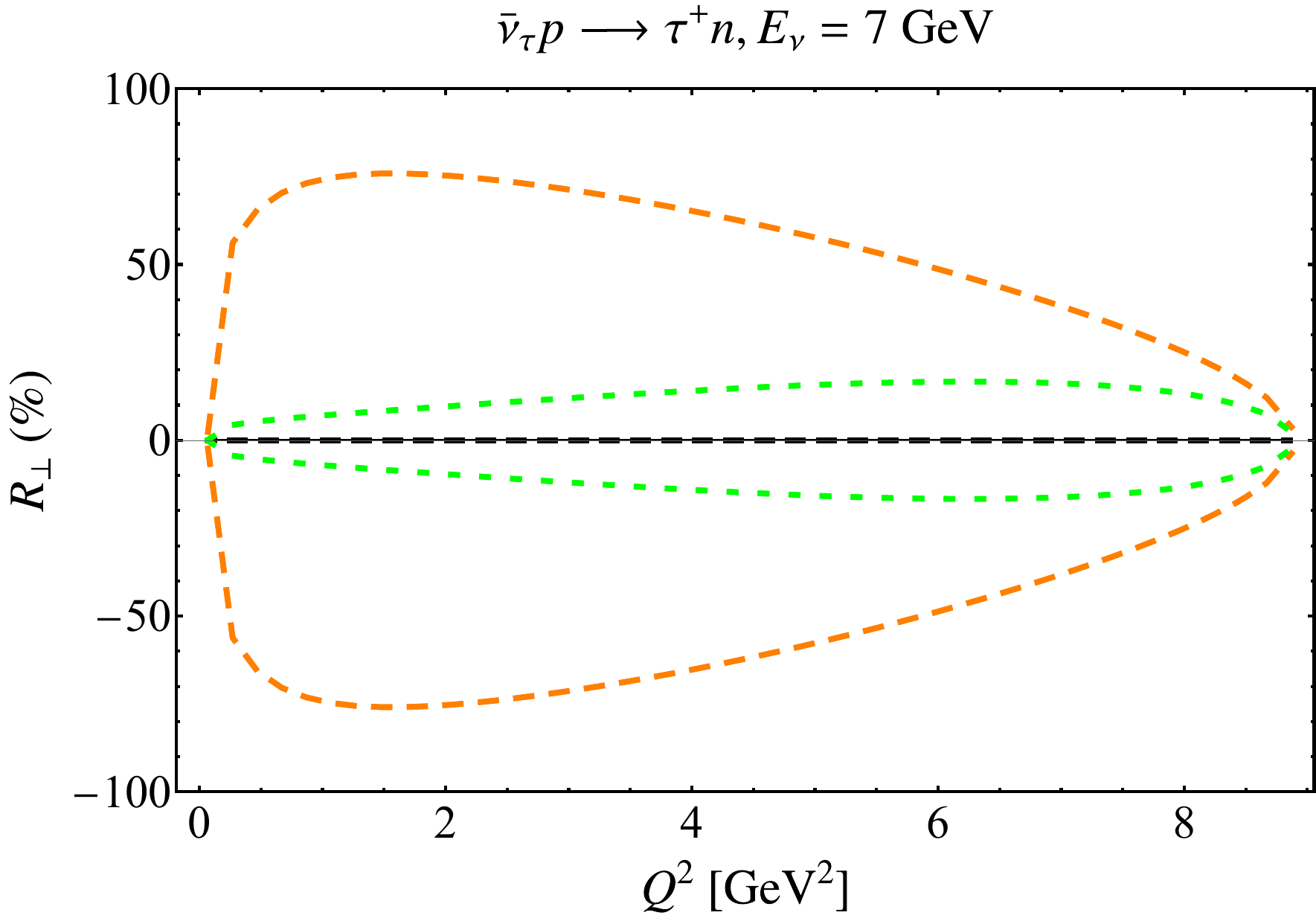}
\includegraphics[width=0.4\textwidth]{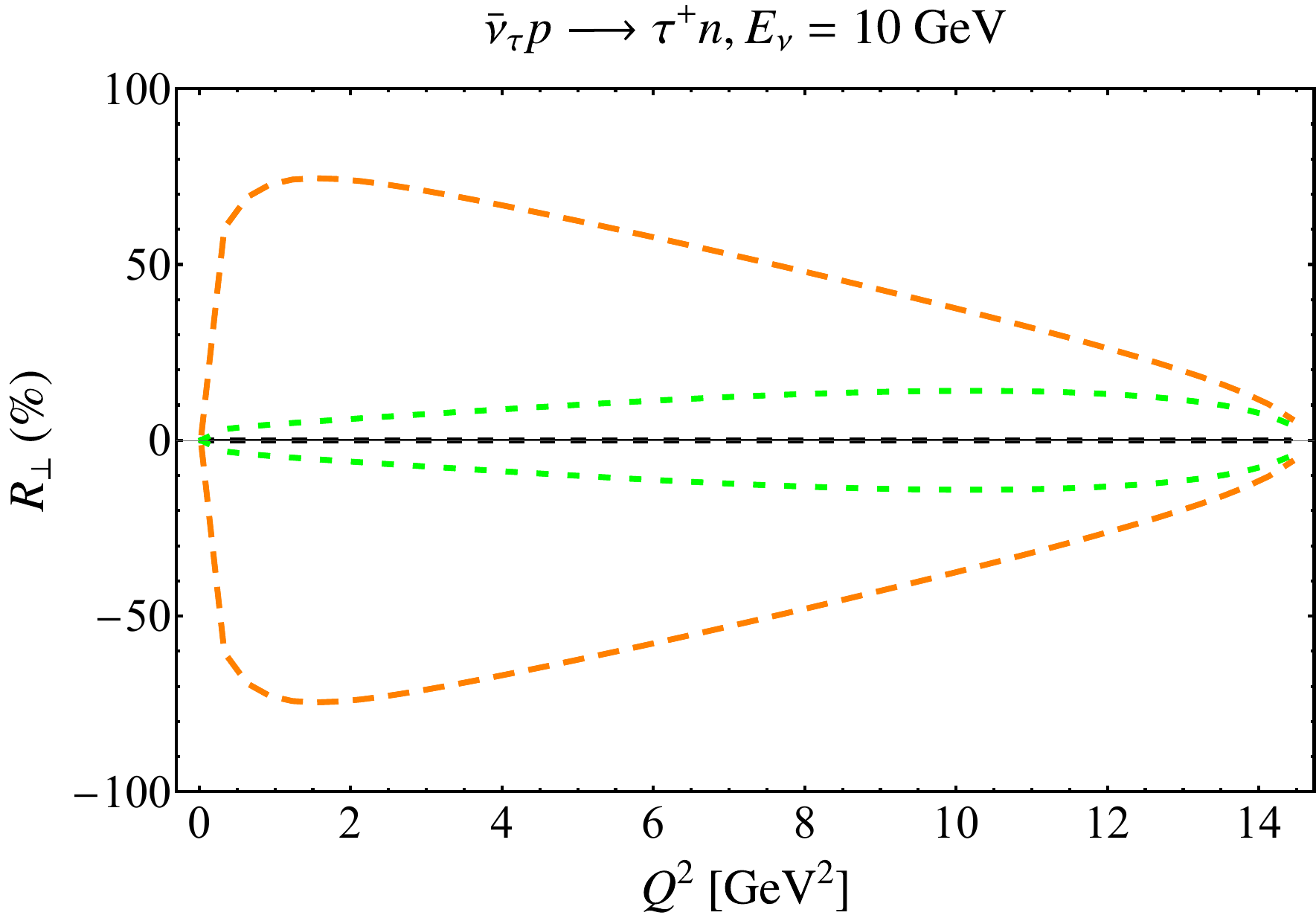}
\includegraphics[width=0.4\textwidth]{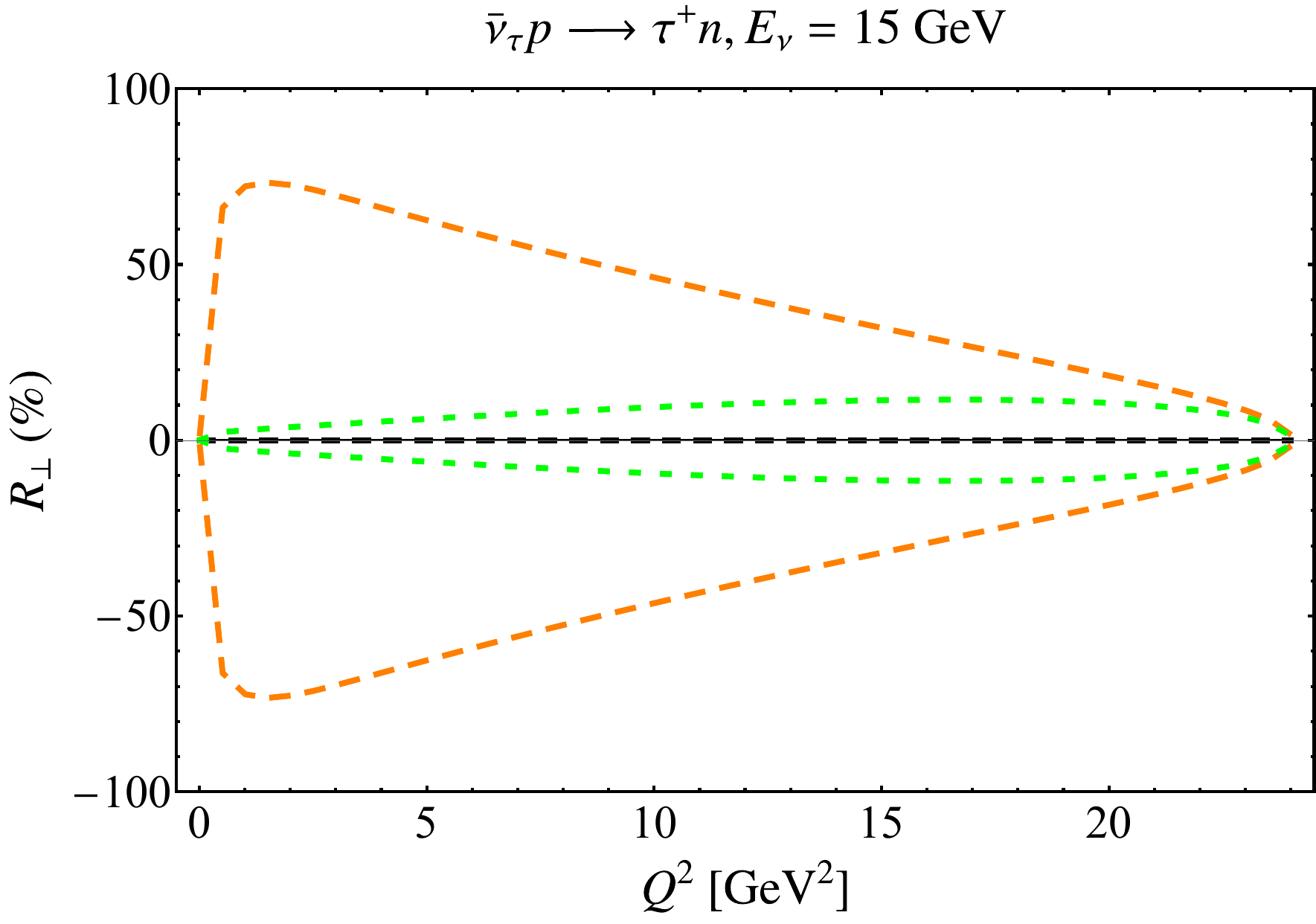}
\caption{Same as Fig.~\ref{fig:antinu_Tt_SCFF_tau} but for the transverse polarization observable $R_\perp$ and imaginary amplitudes. \label{fig:antinu_RTT_SCFF_tau}}
\end{figure}

\begin{figure}[H]
\centering
\includegraphics[width=0.4\textwidth]{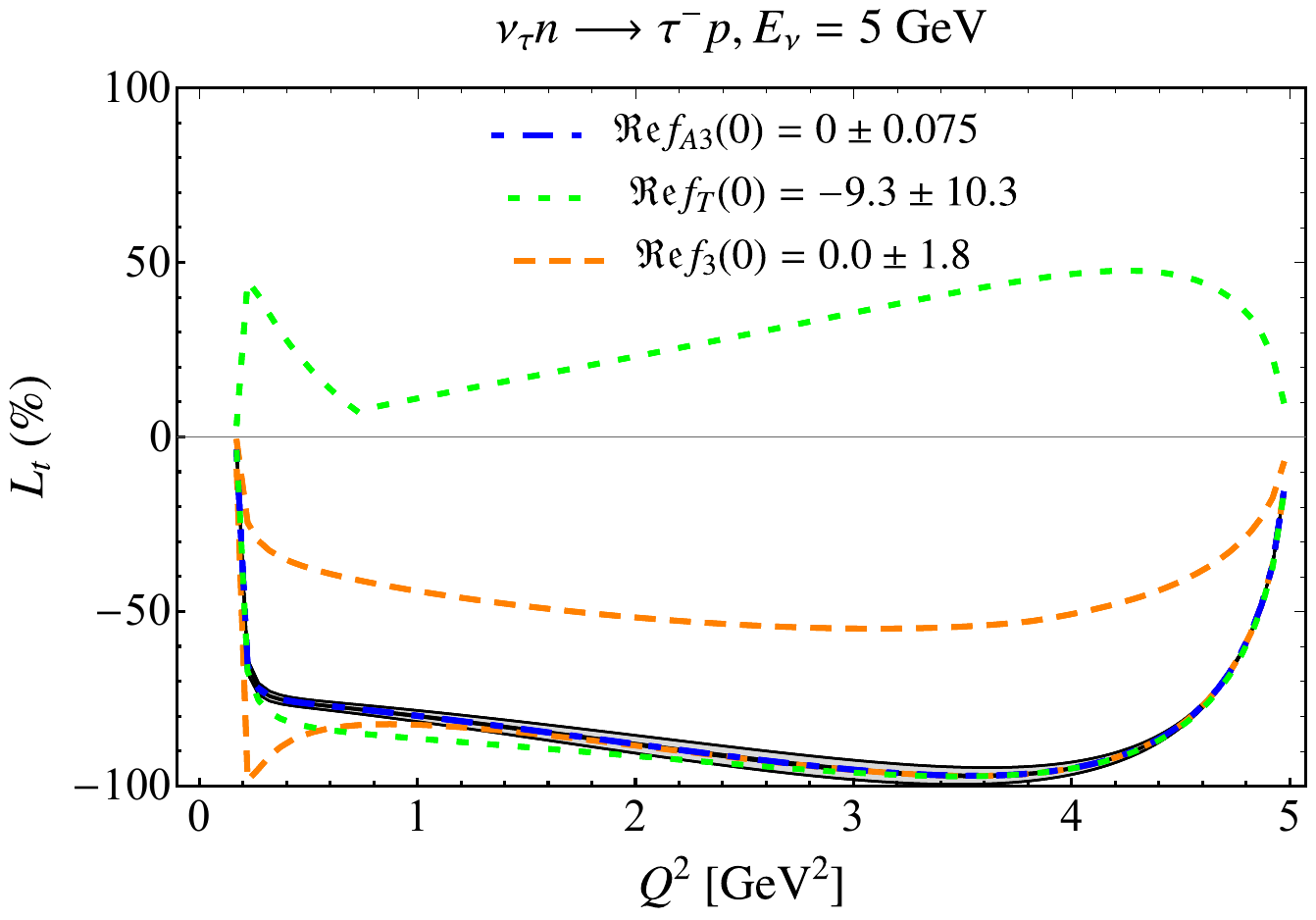}
\includegraphics[width=0.4\textwidth]{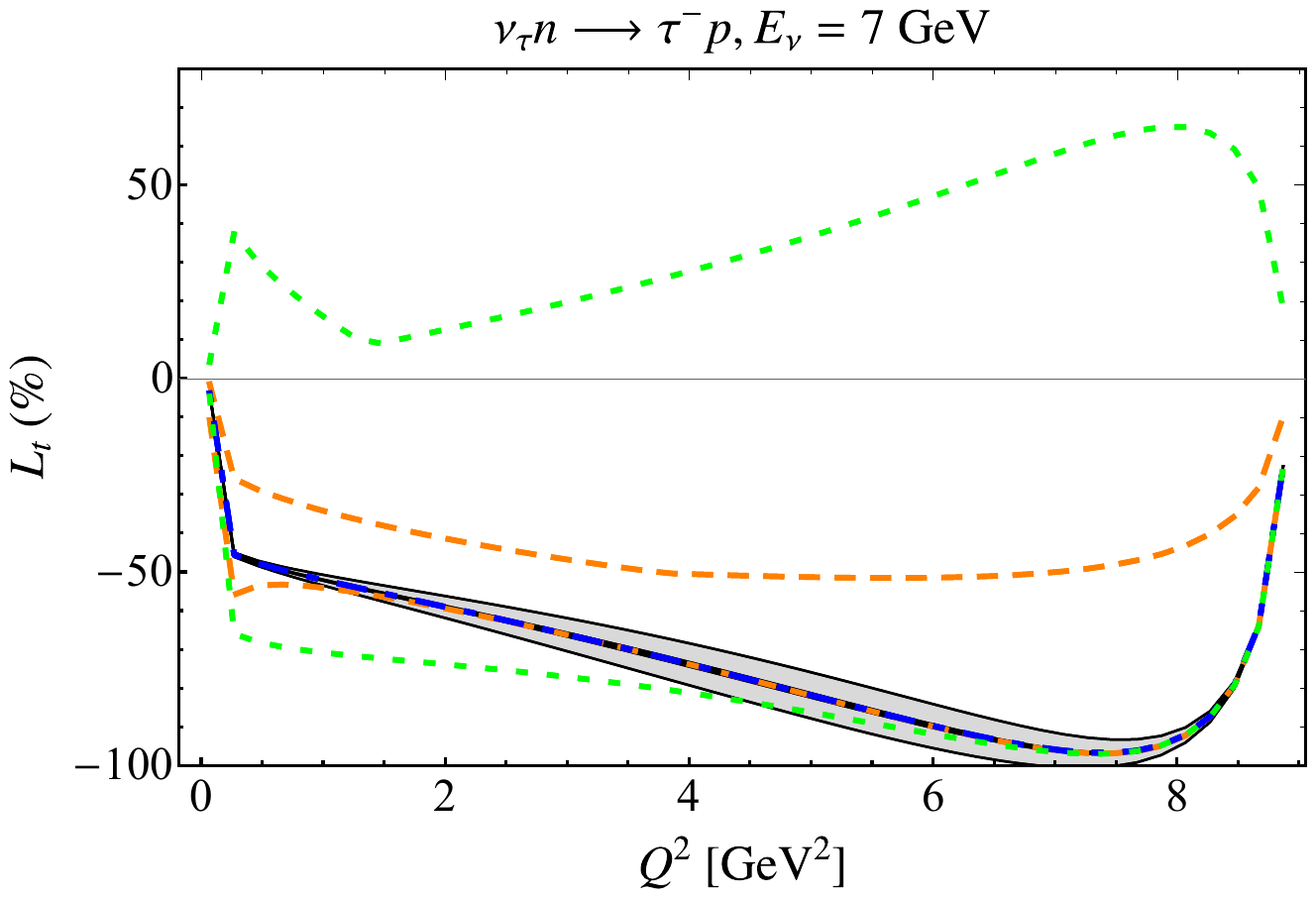}
\includegraphics[width=0.4\textwidth]{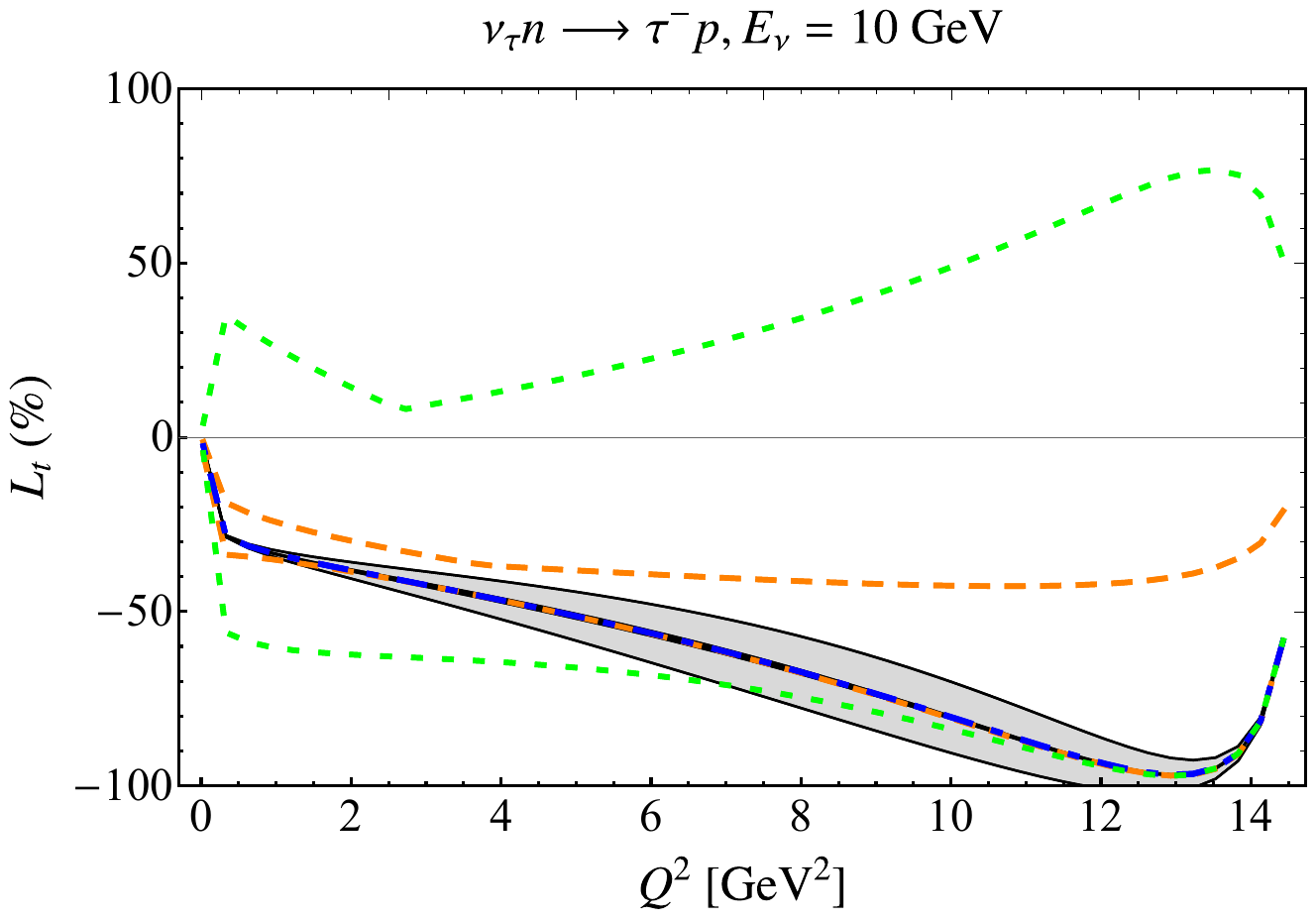}
\includegraphics[width=0.4\textwidth]{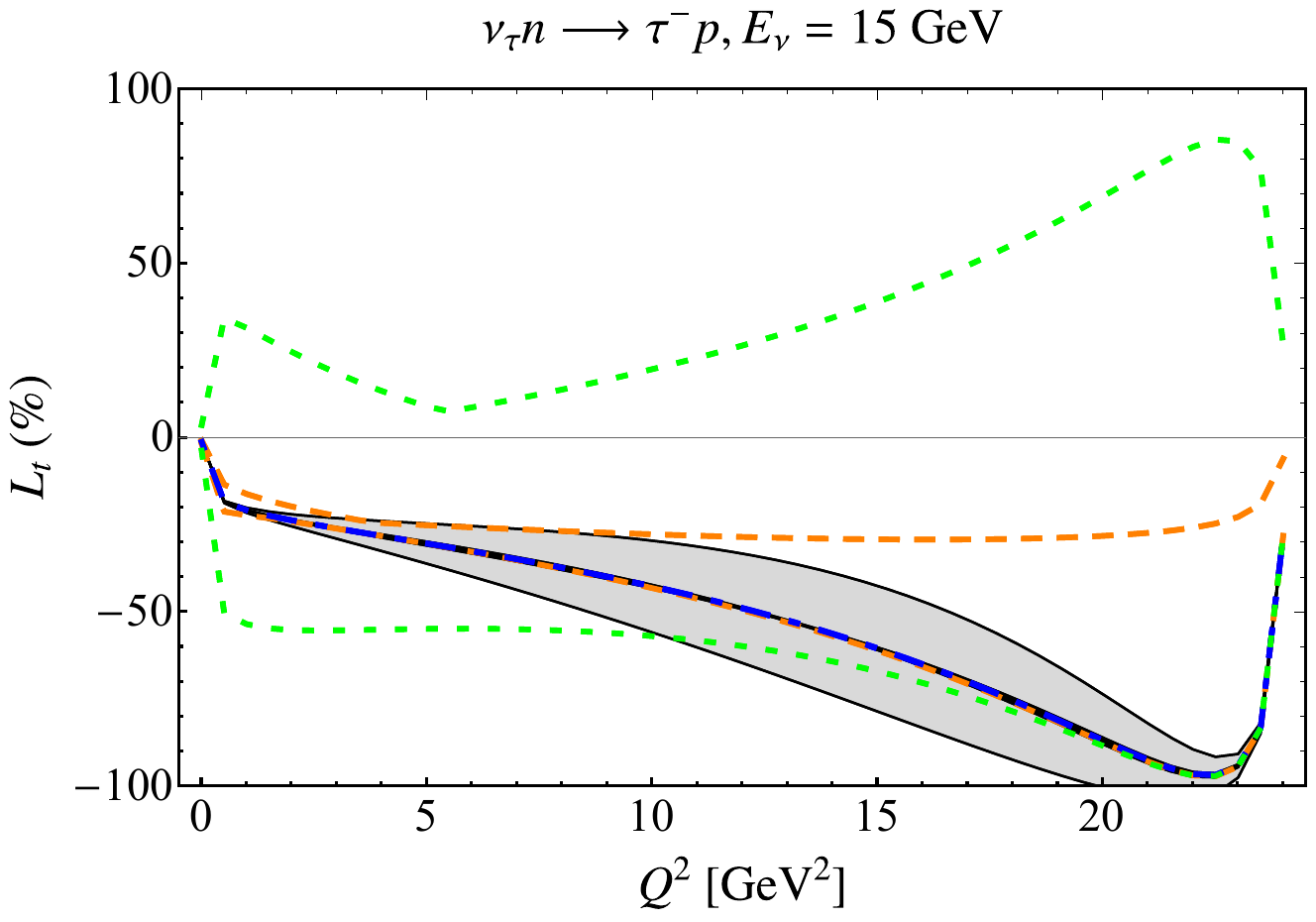}
\caption{Same as Fig.~\ref{fig:nu_Tt_SCFF_tau} but for the transverse polarization observable $L_t$. \label{fig:nu_Lt_SCFF_tau}}
\end{figure}

\begin{figure}[H]
\centering
\includegraphics[width=0.4\textwidth]{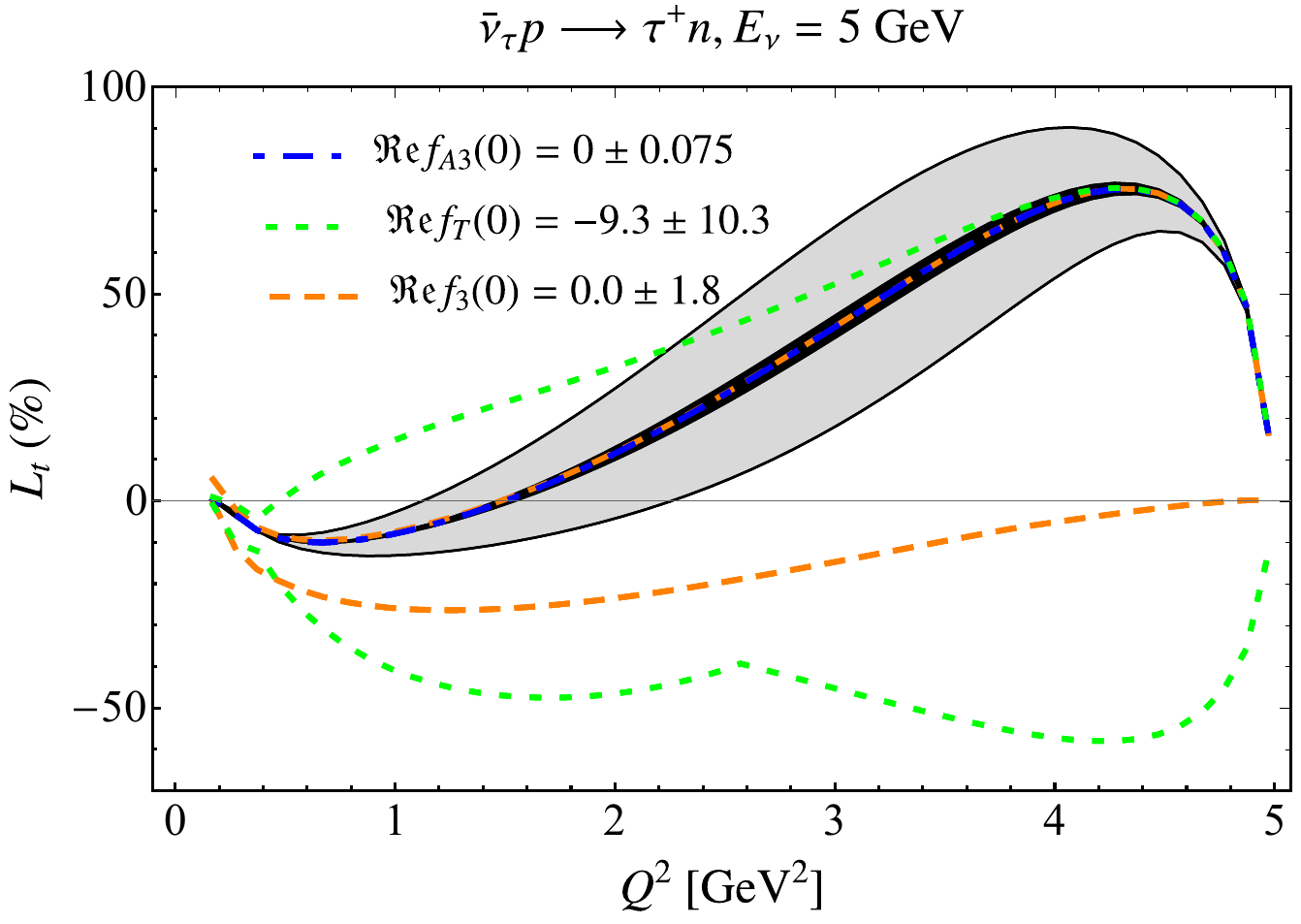}
\includegraphics[width=0.4\textwidth]{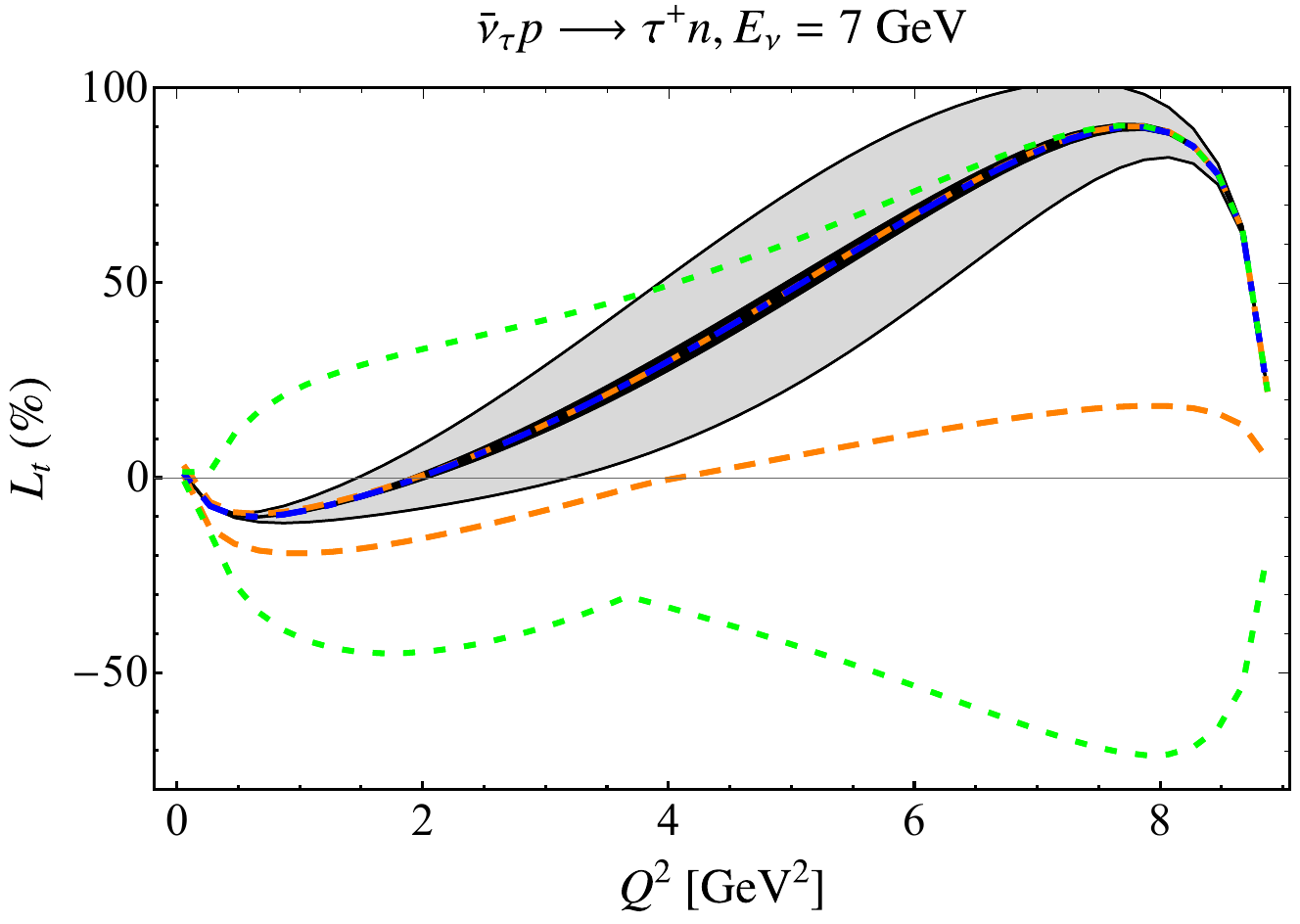}
\includegraphics[width=0.4\textwidth]{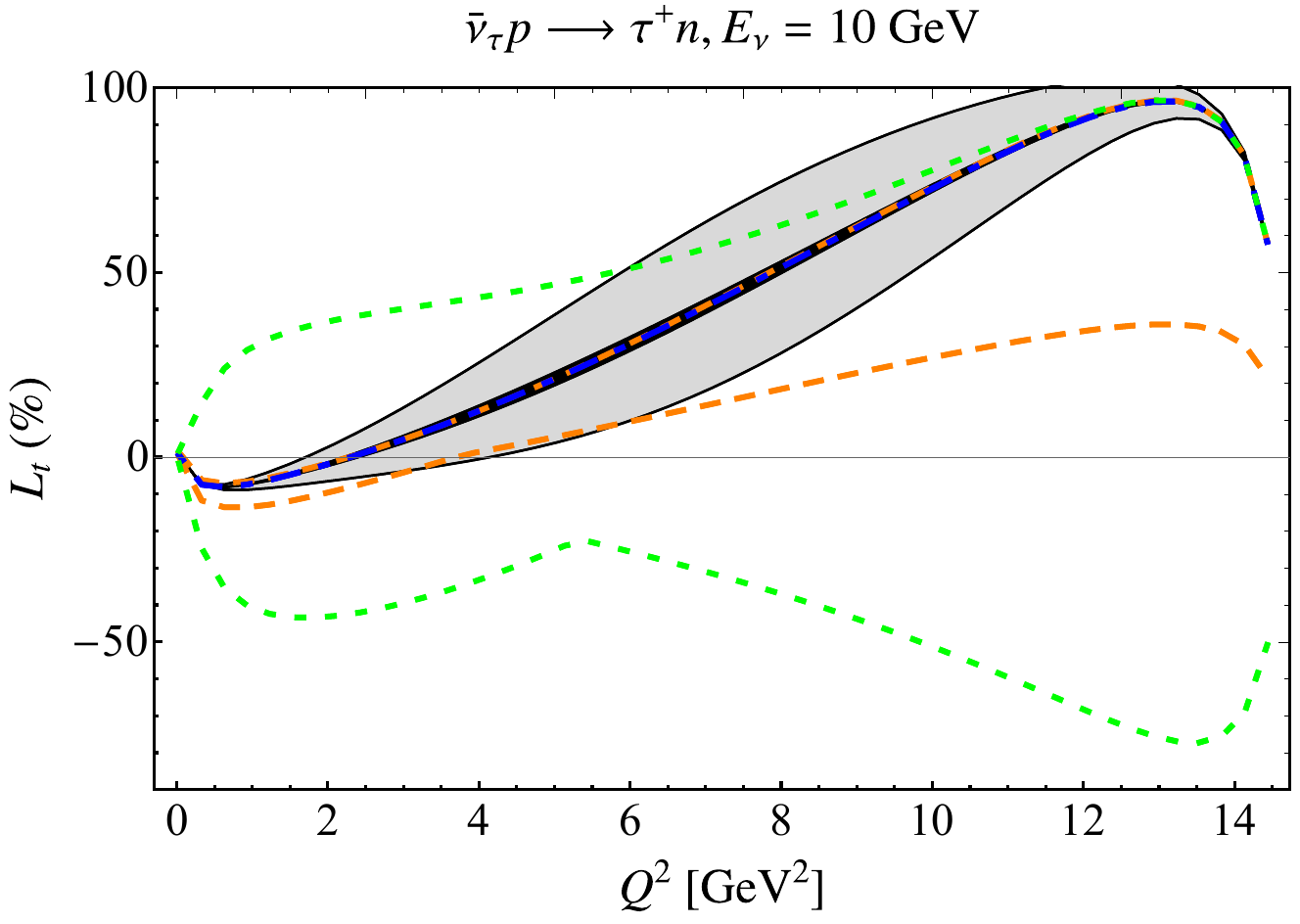}
\includegraphics[width=0.4\textwidth]{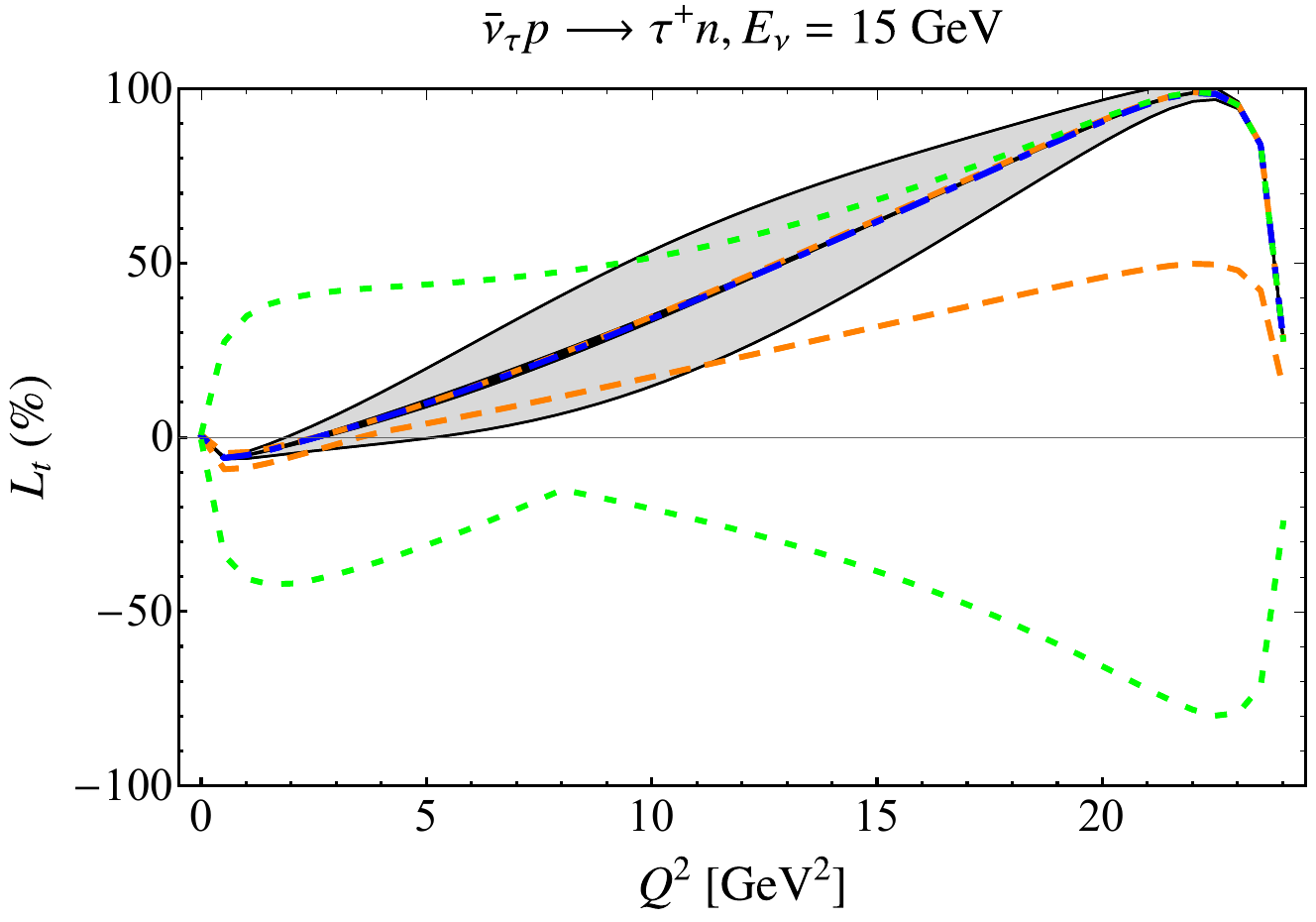}
\caption{Same as Fig.~\ref{fig:nu_Tt_SCFF_tau} but for the transverse polarization observable $L_t$. \label{fig:antinu_Lt_SCFF_tau}}
\end{figure}

\begin{figure}[H]
\centering
\includegraphics[width=0.4\textwidth]{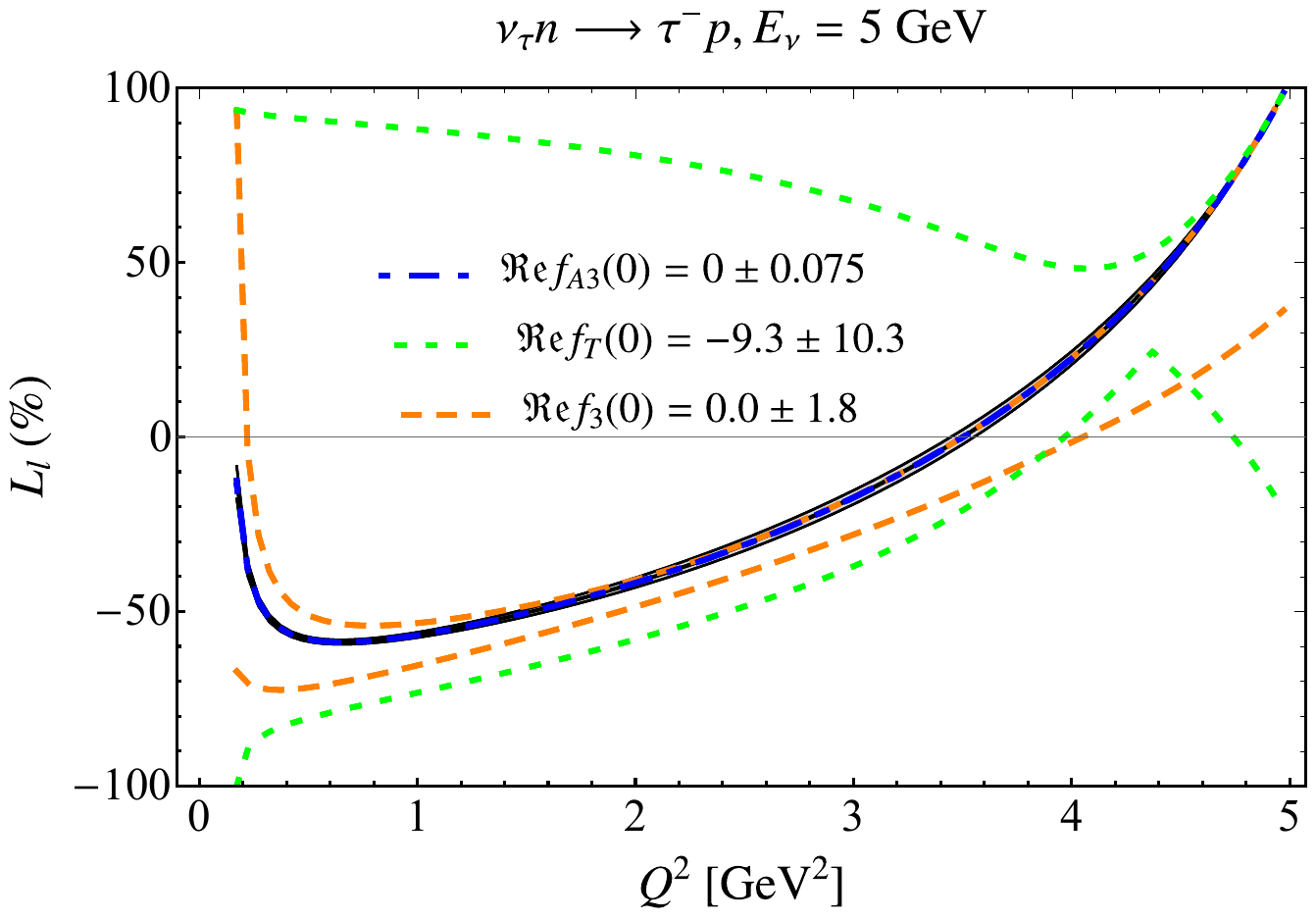}
\includegraphics[width=0.4\textwidth]{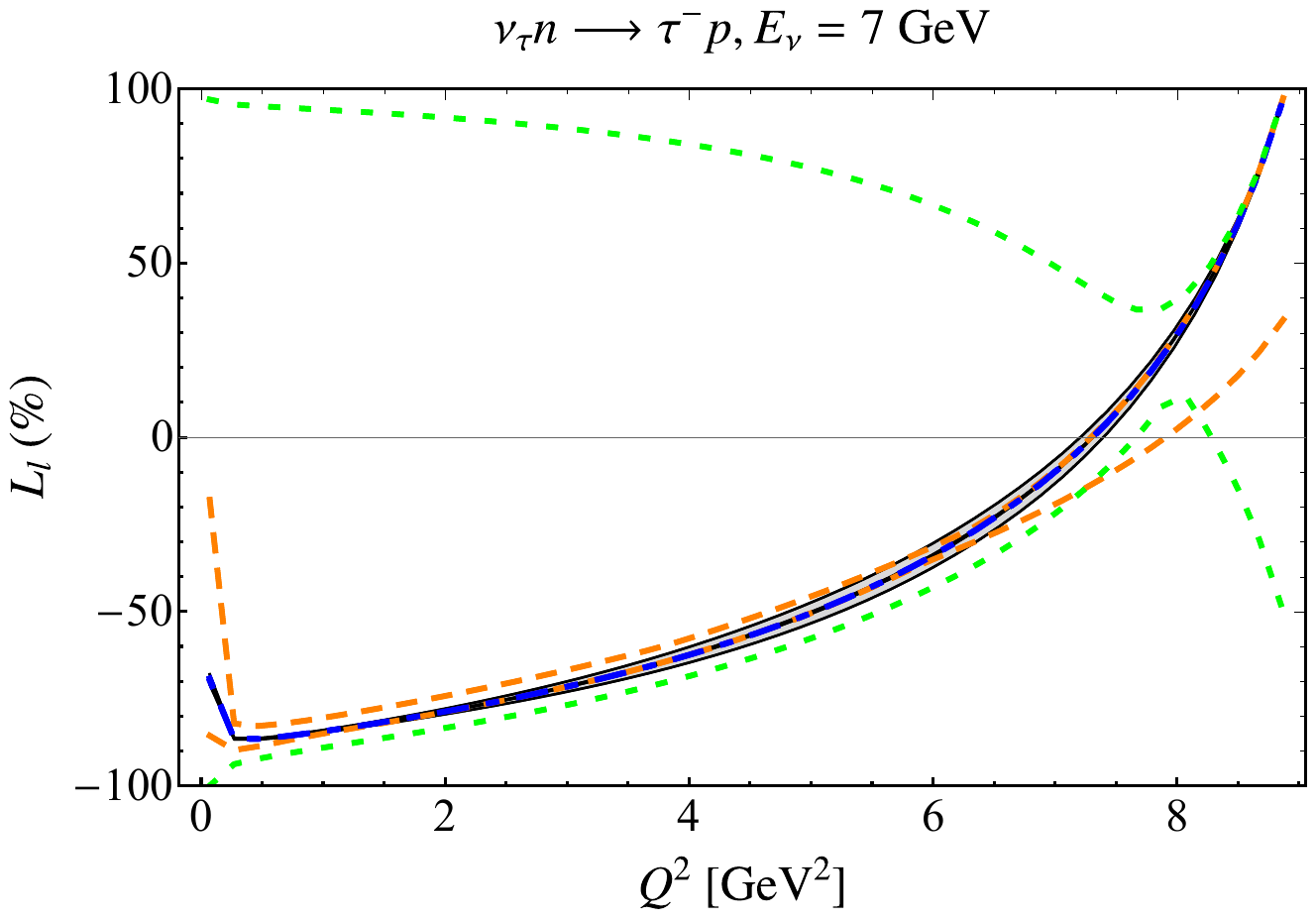}
\includegraphics[width=0.4\textwidth]{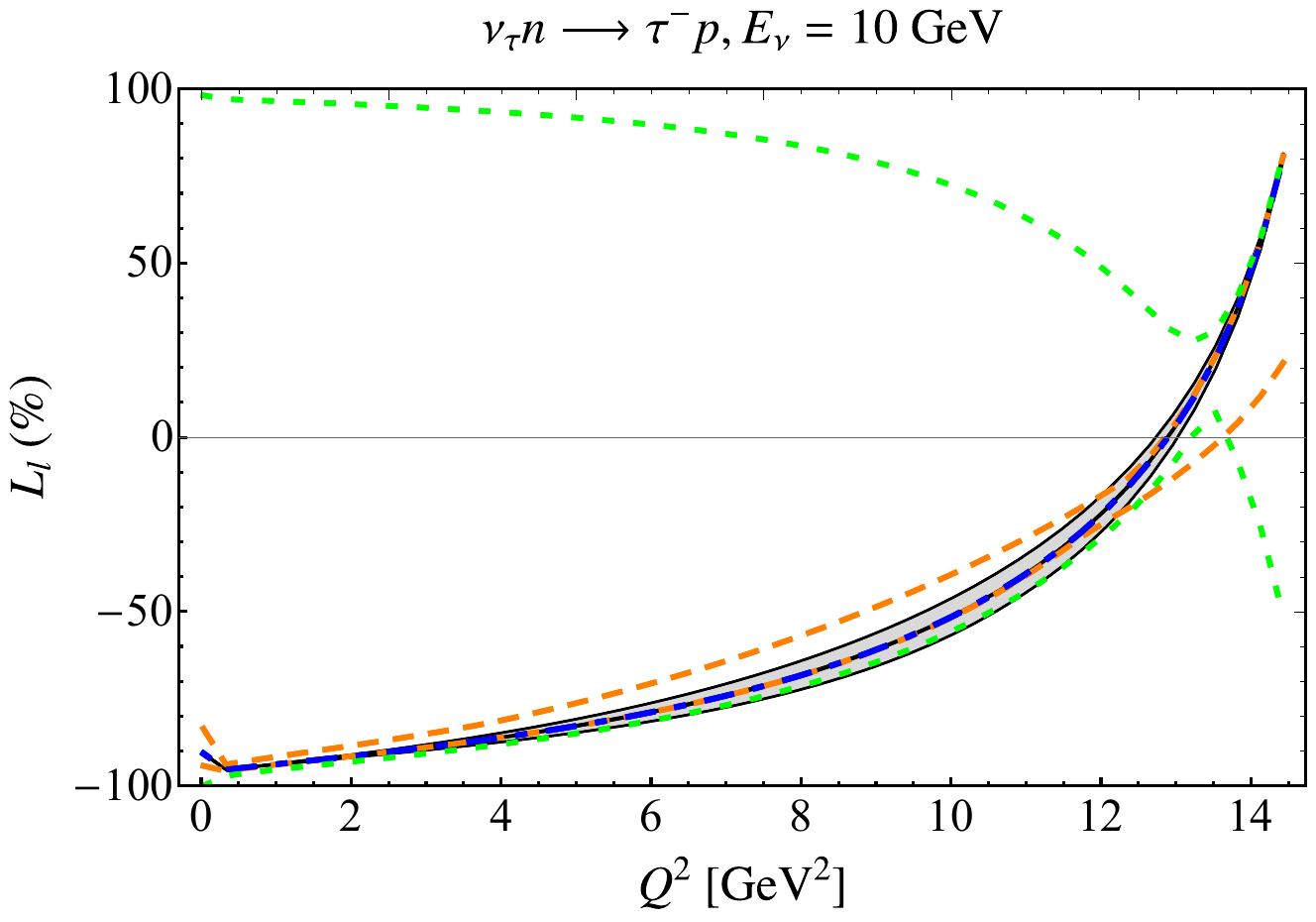}
\includegraphics[width=0.4\textwidth]{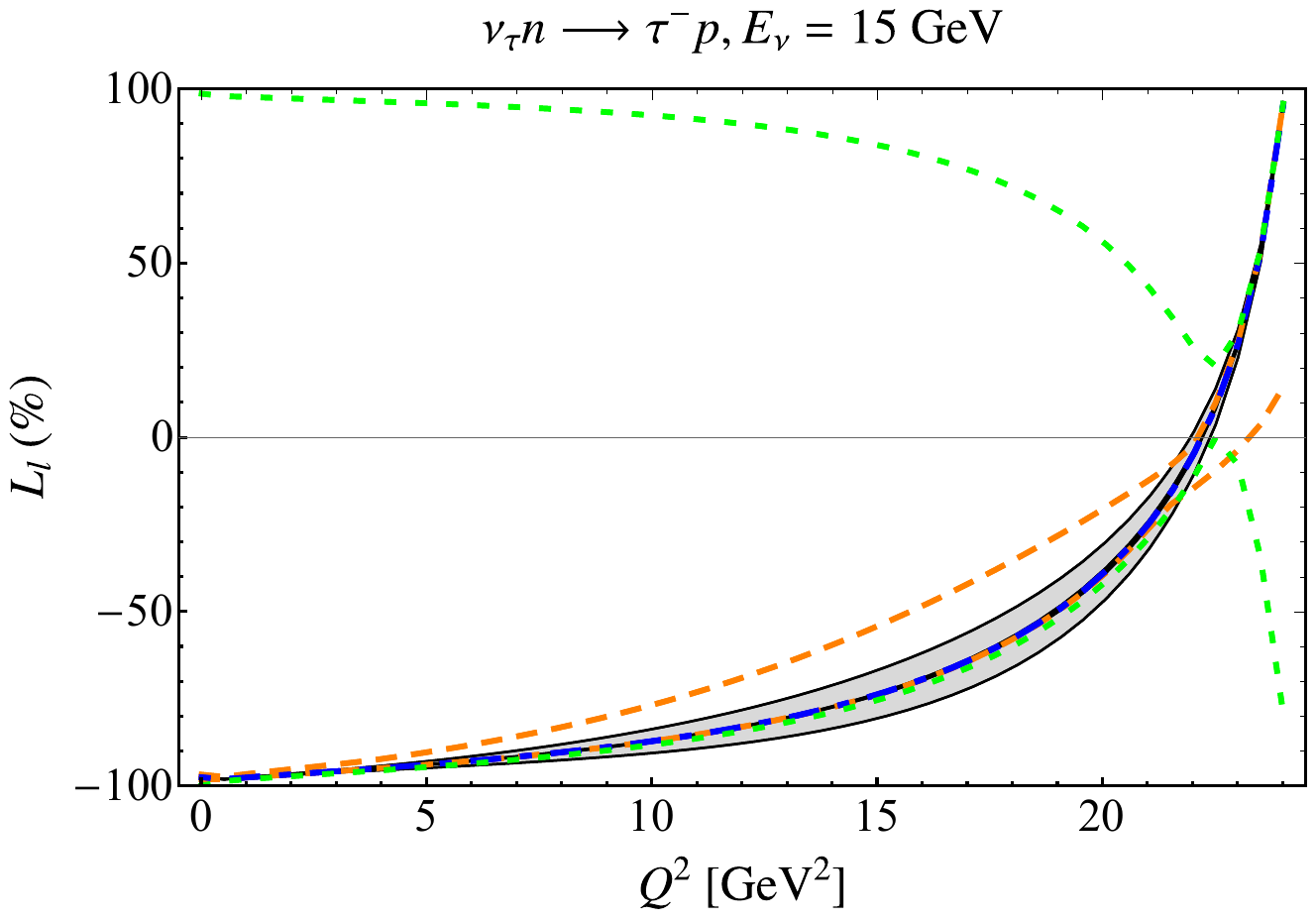}
\caption{Same as Fig.~\ref{fig:nu_Tt_SCFF_tau} but for the longitudinal polarization observable $L_l$. \label{fig:nu_Ll_SCFF_tau}}
\end{figure}

\begin{figure}[H]
\centering
\includegraphics[width=0.4\textwidth]{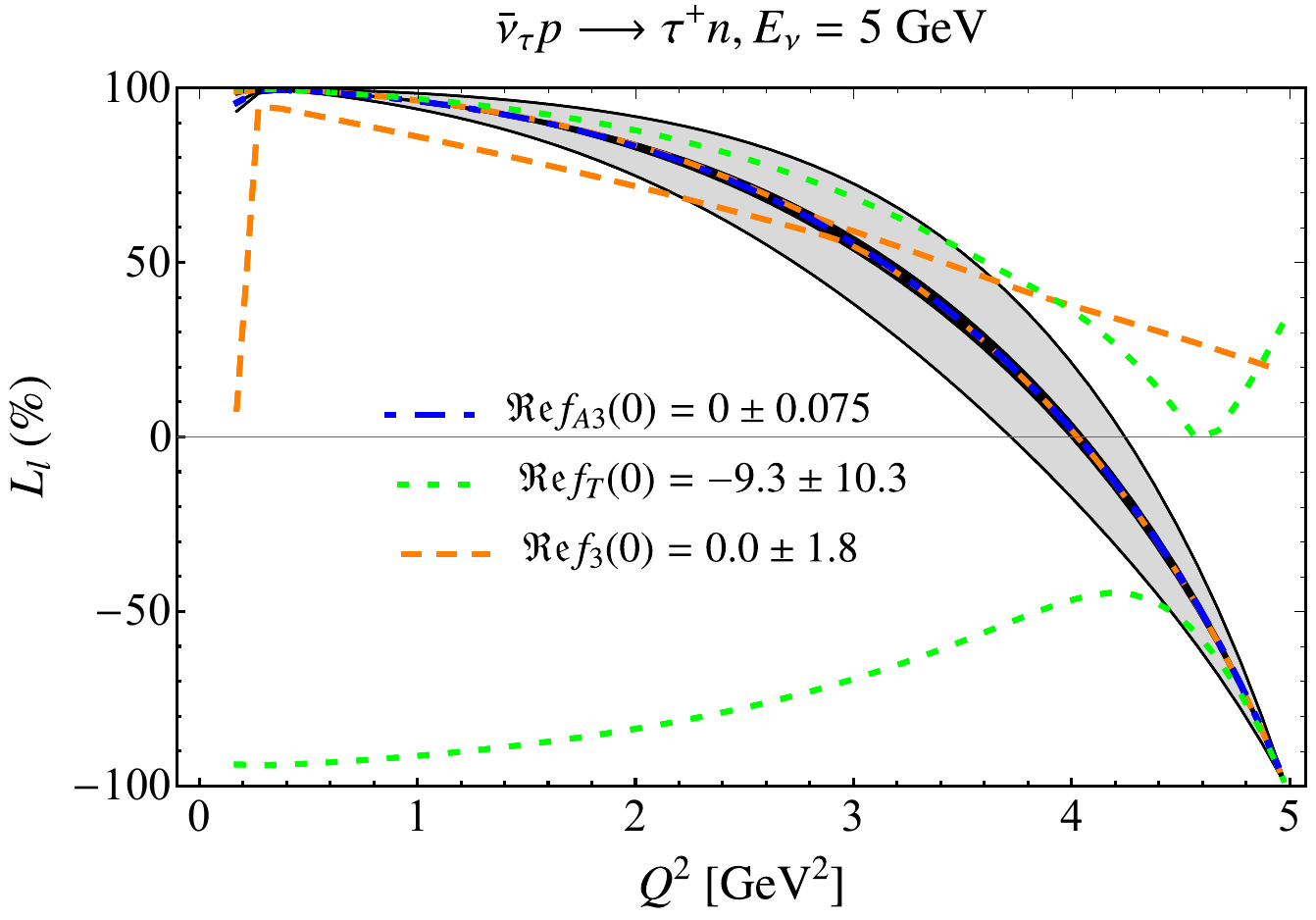}
\includegraphics[width=0.4\textwidth]{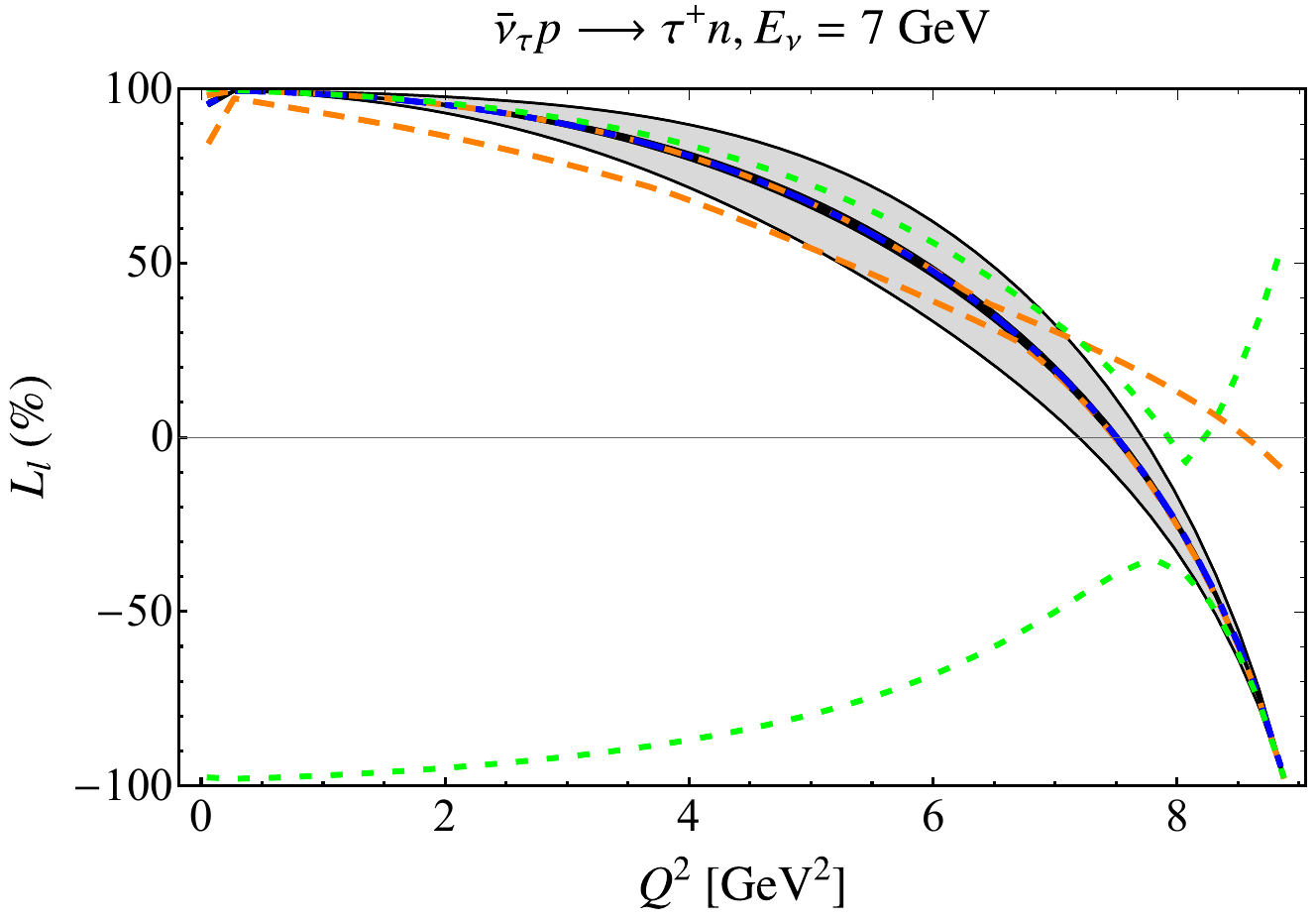}
\includegraphics[width=0.4\textwidth]{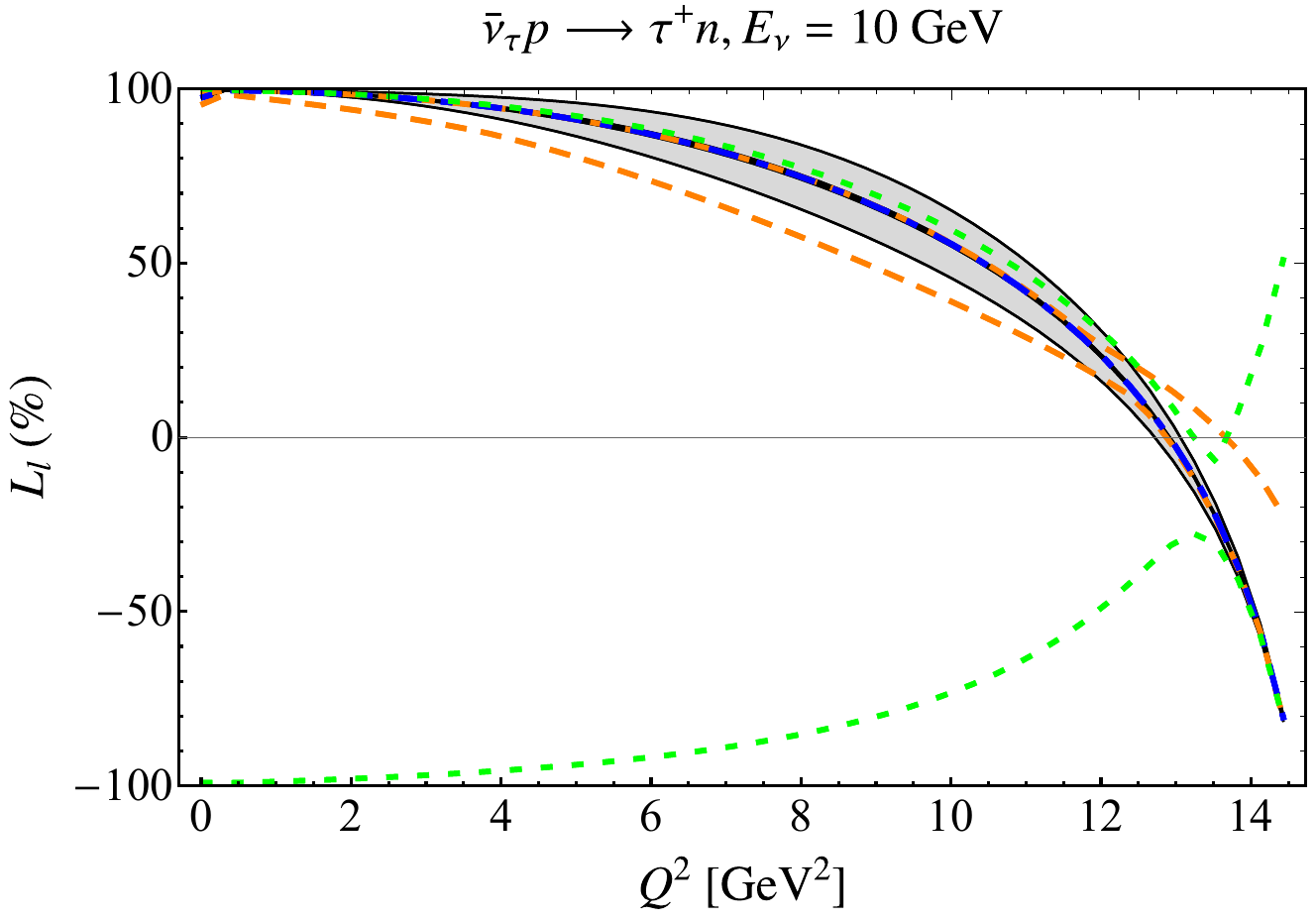}
\includegraphics[width=0.4\textwidth]{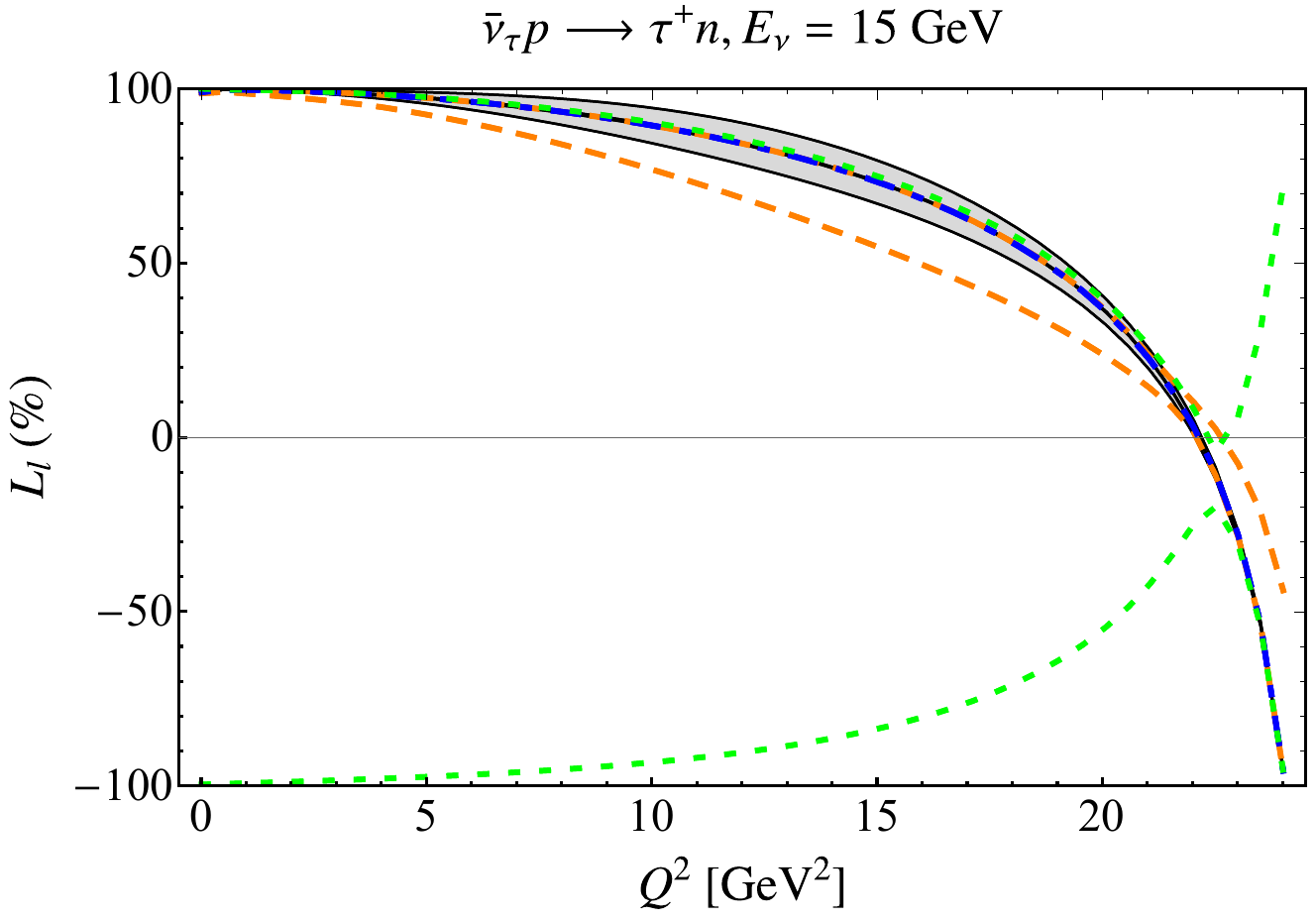}
\caption{Same as Fig.~\ref{fig:antinu_Tt_SCFF_tau} but for the longitudinal polarization observable $L_l$. \label{fig:antinu_Ll_SCFF_tau}}
\end{figure}

\begin{figure}[H]
\centering
\includegraphics[width=0.4\textwidth]{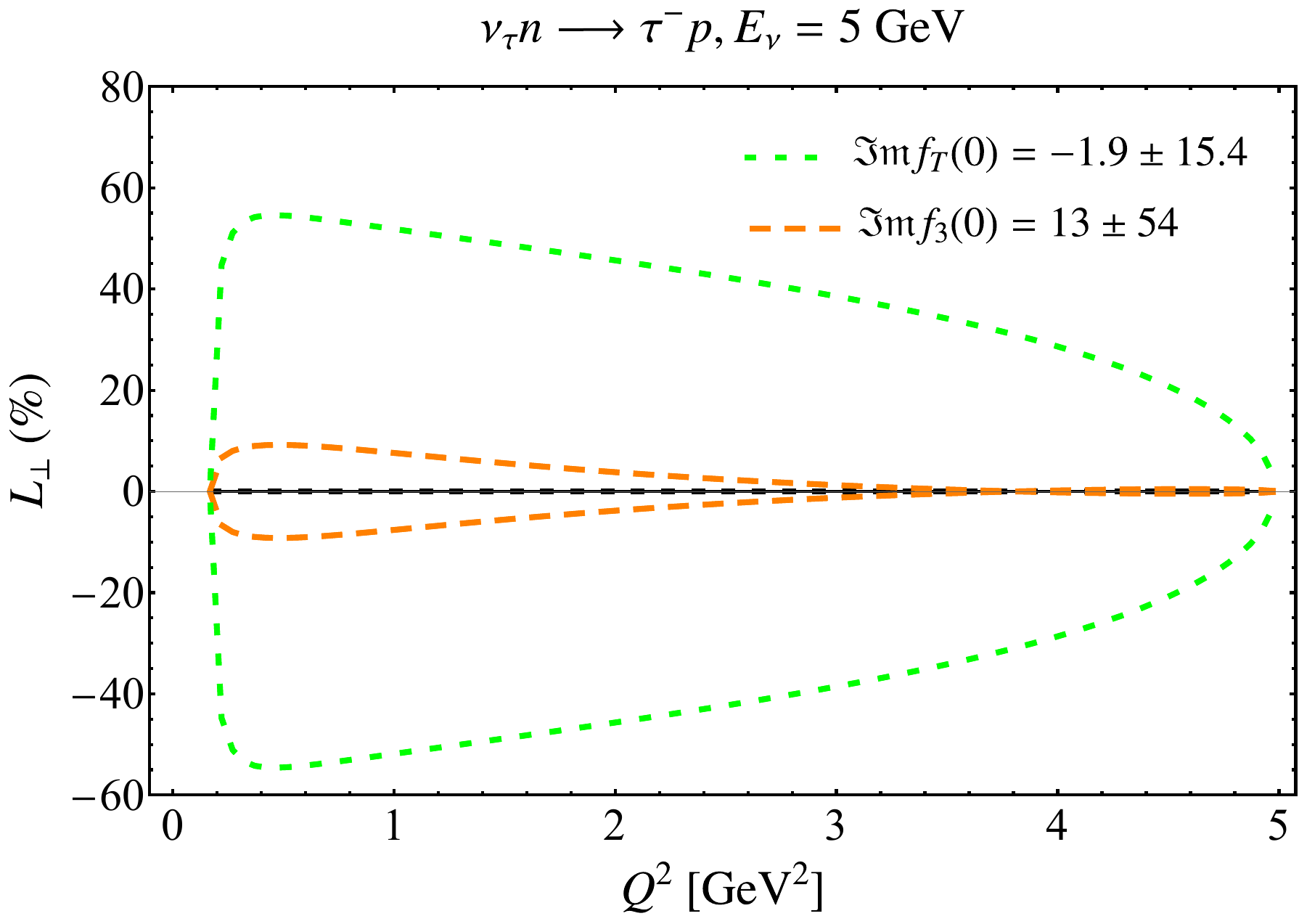}
\includegraphics[width=0.4\textwidth]{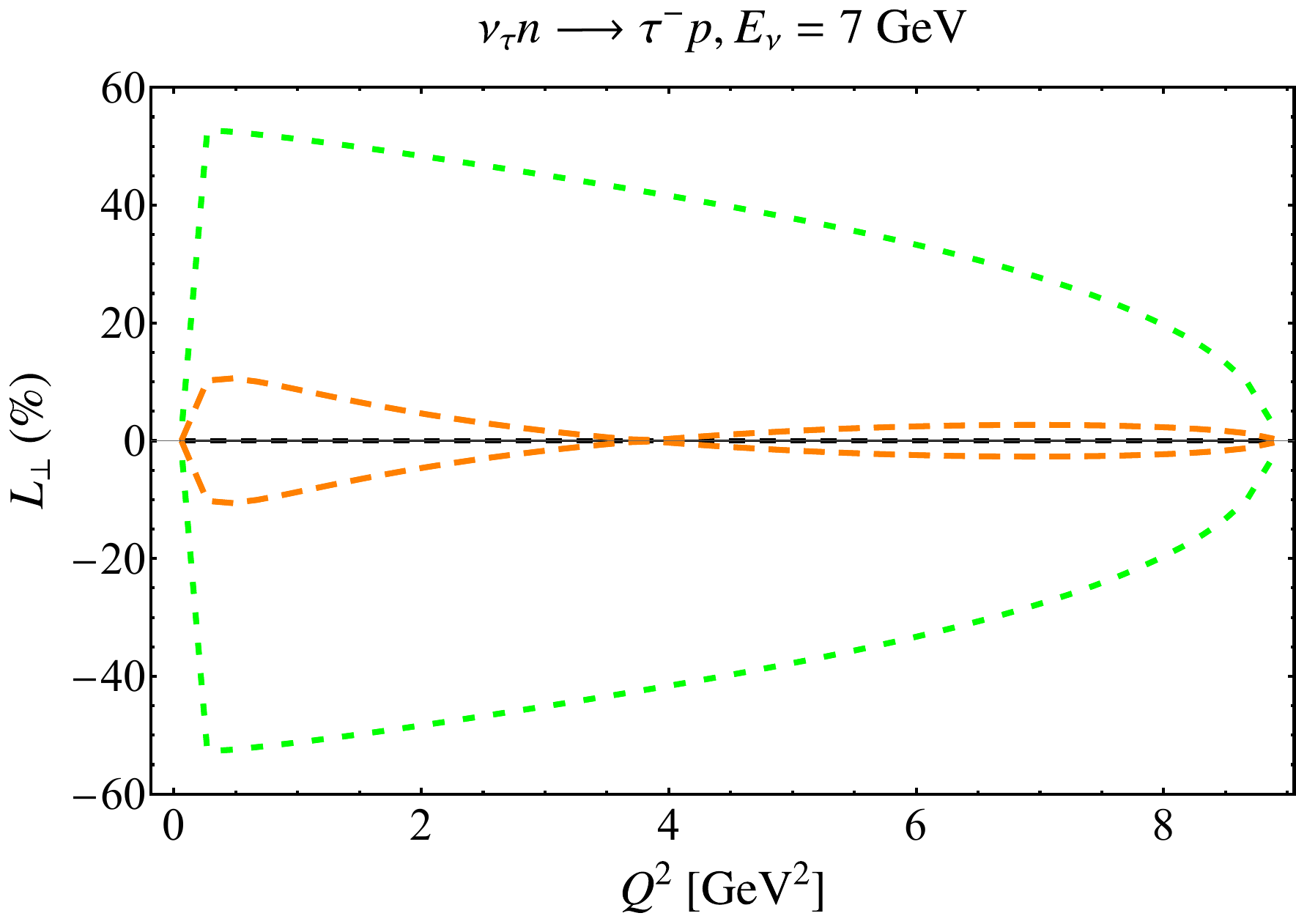}
\includegraphics[width=0.4\textwidth]{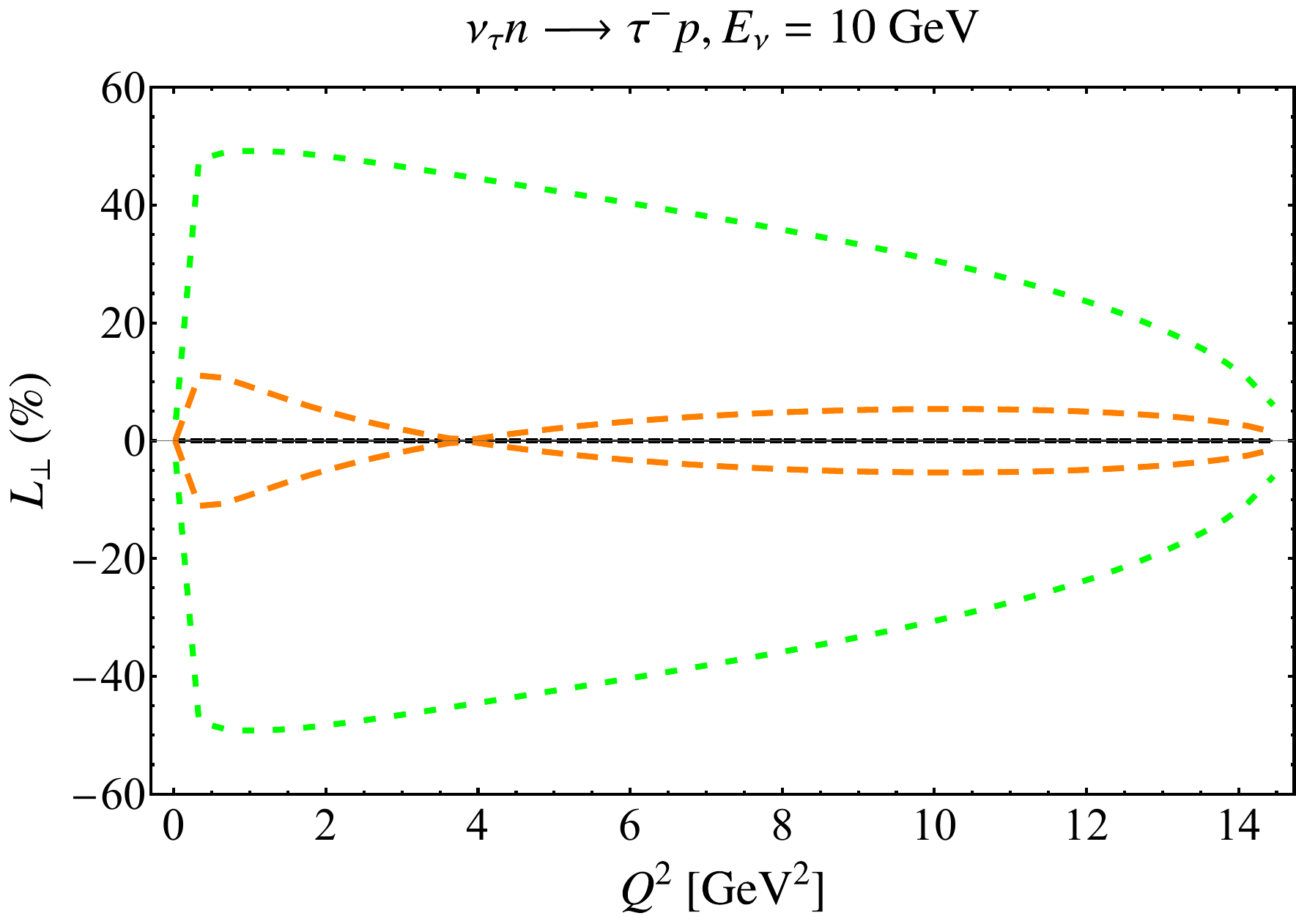}
\includegraphics[width=0.4\textwidth]{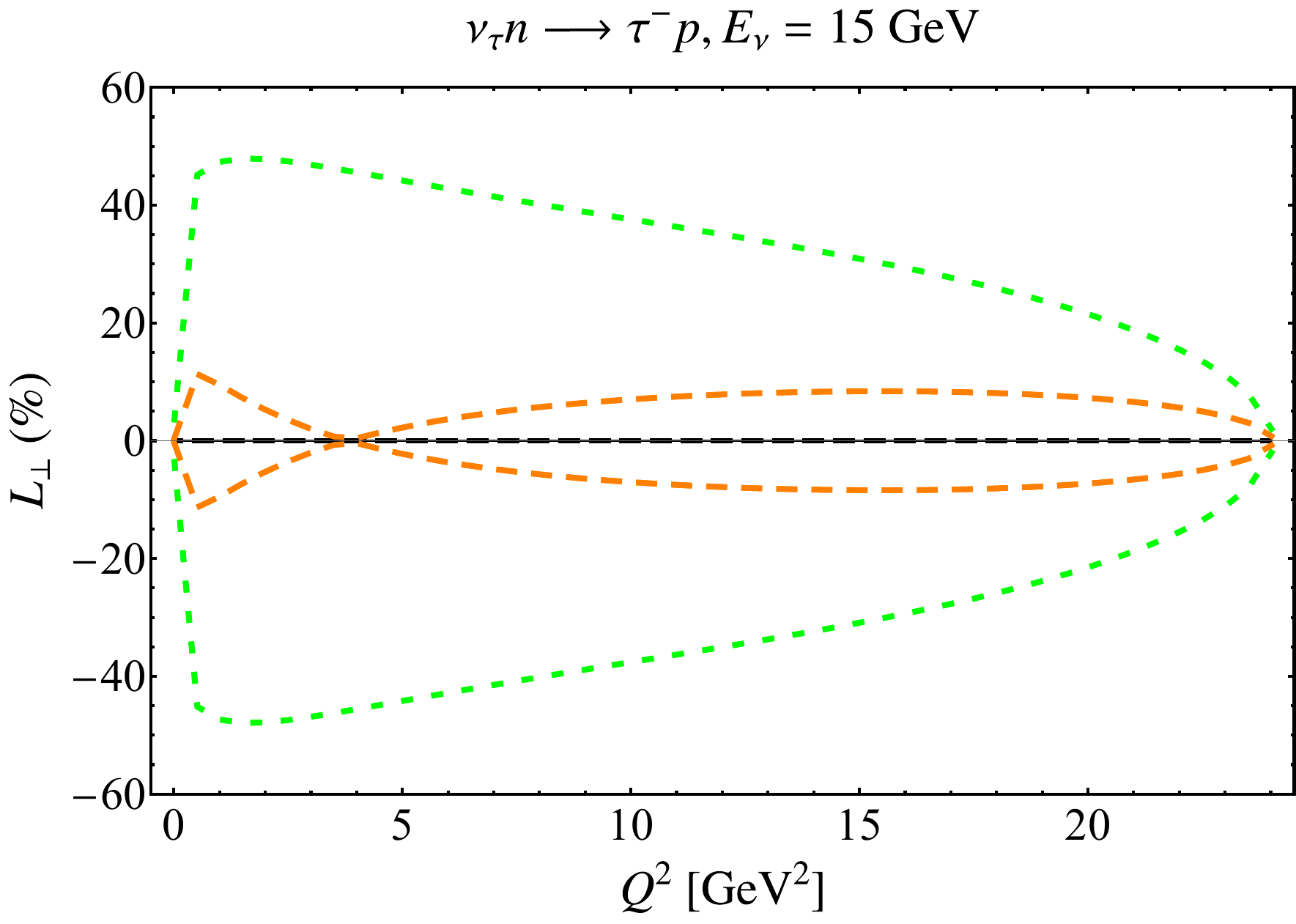}
\caption{Same as Fig.~\ref{fig:nu_Tt_SCFF_tau} but for the transverse polarization observable $L_\perp$ and imaginary amplitudes. \label{fig:nu_LTT_SCFF_tau}}
\end{figure}

\begin{figure}[H]
\centering
\includegraphics[width=0.4\textwidth]{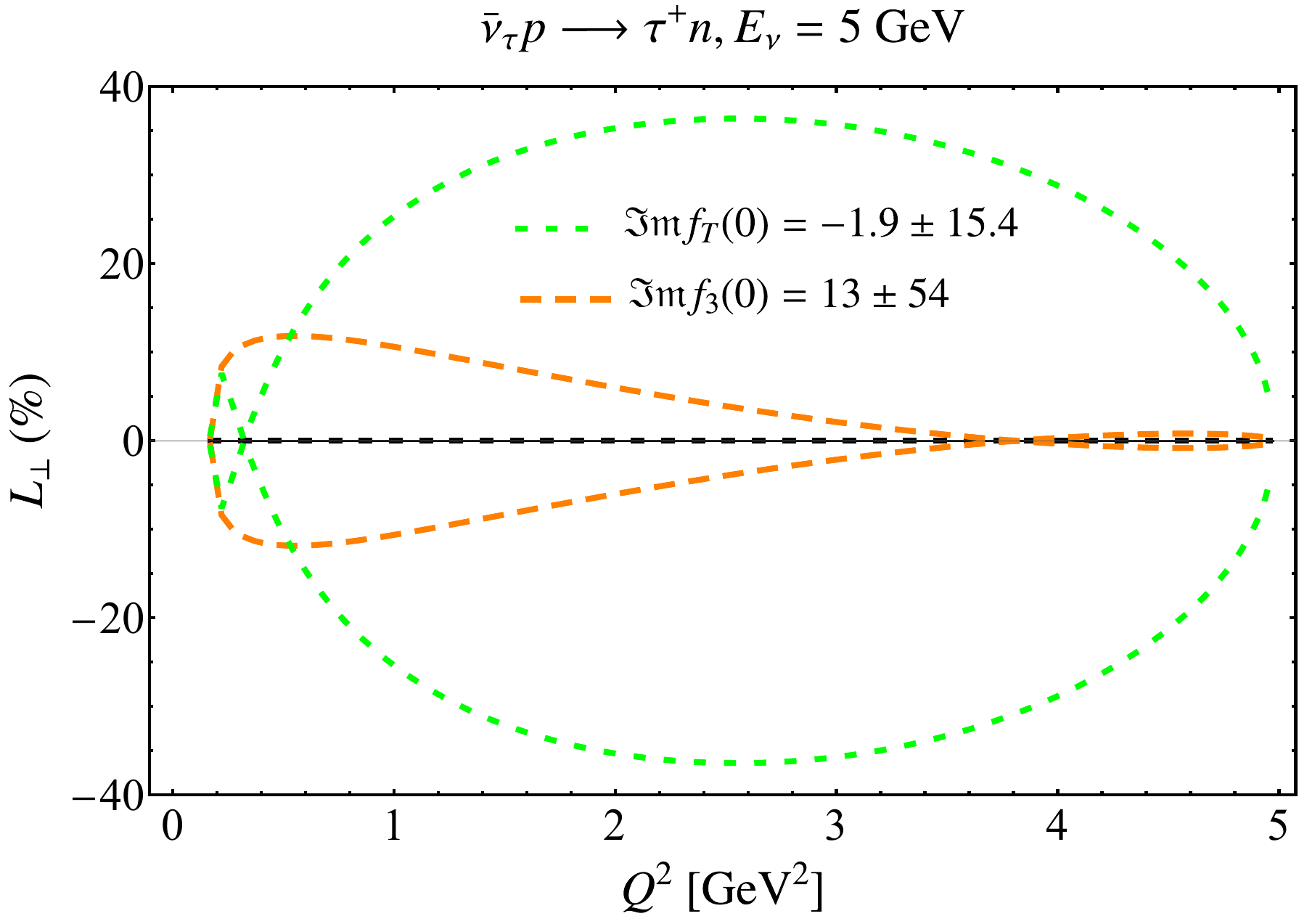}
\includegraphics[width=0.4\textwidth]{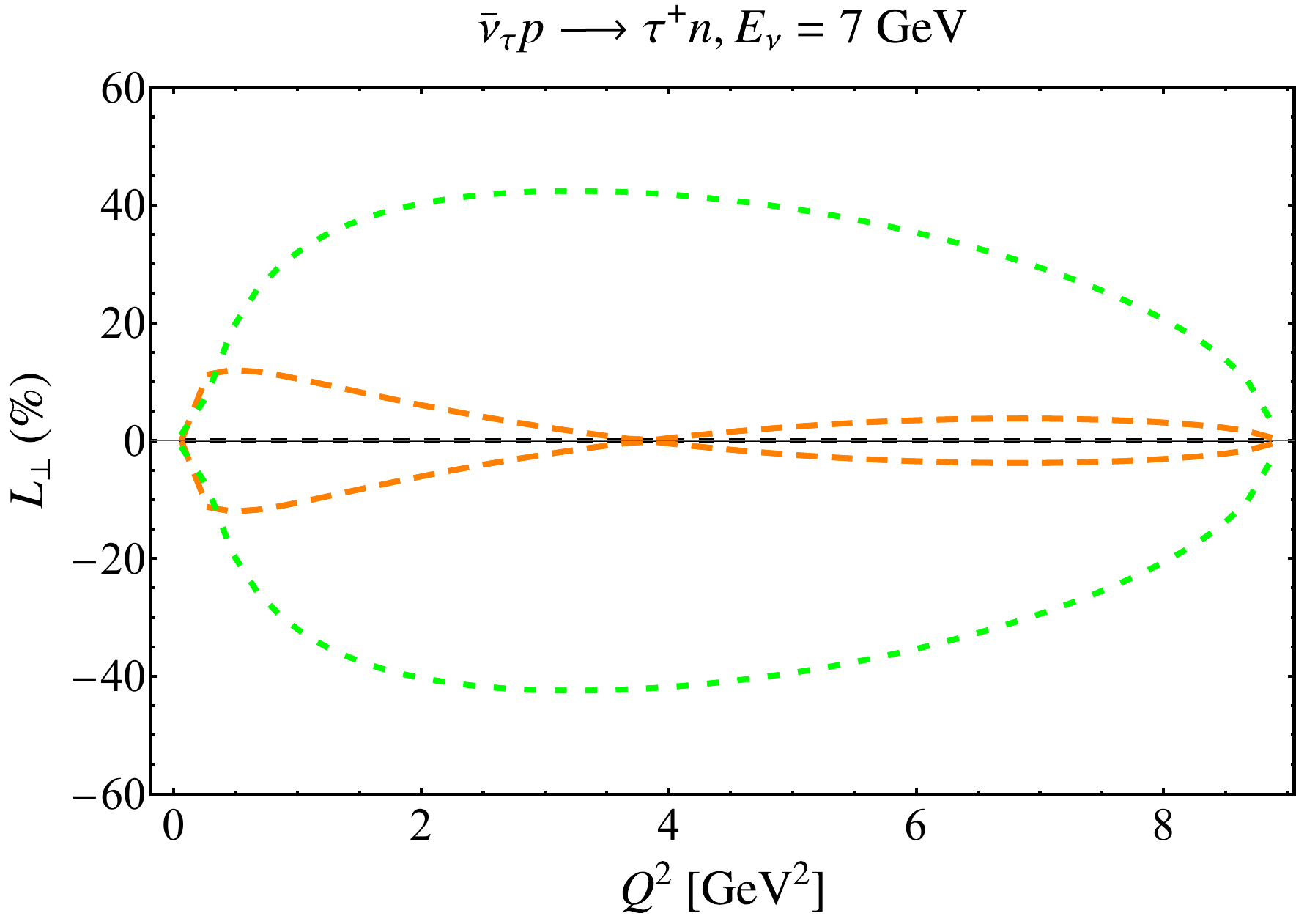}
\includegraphics[width=0.4\textwidth]{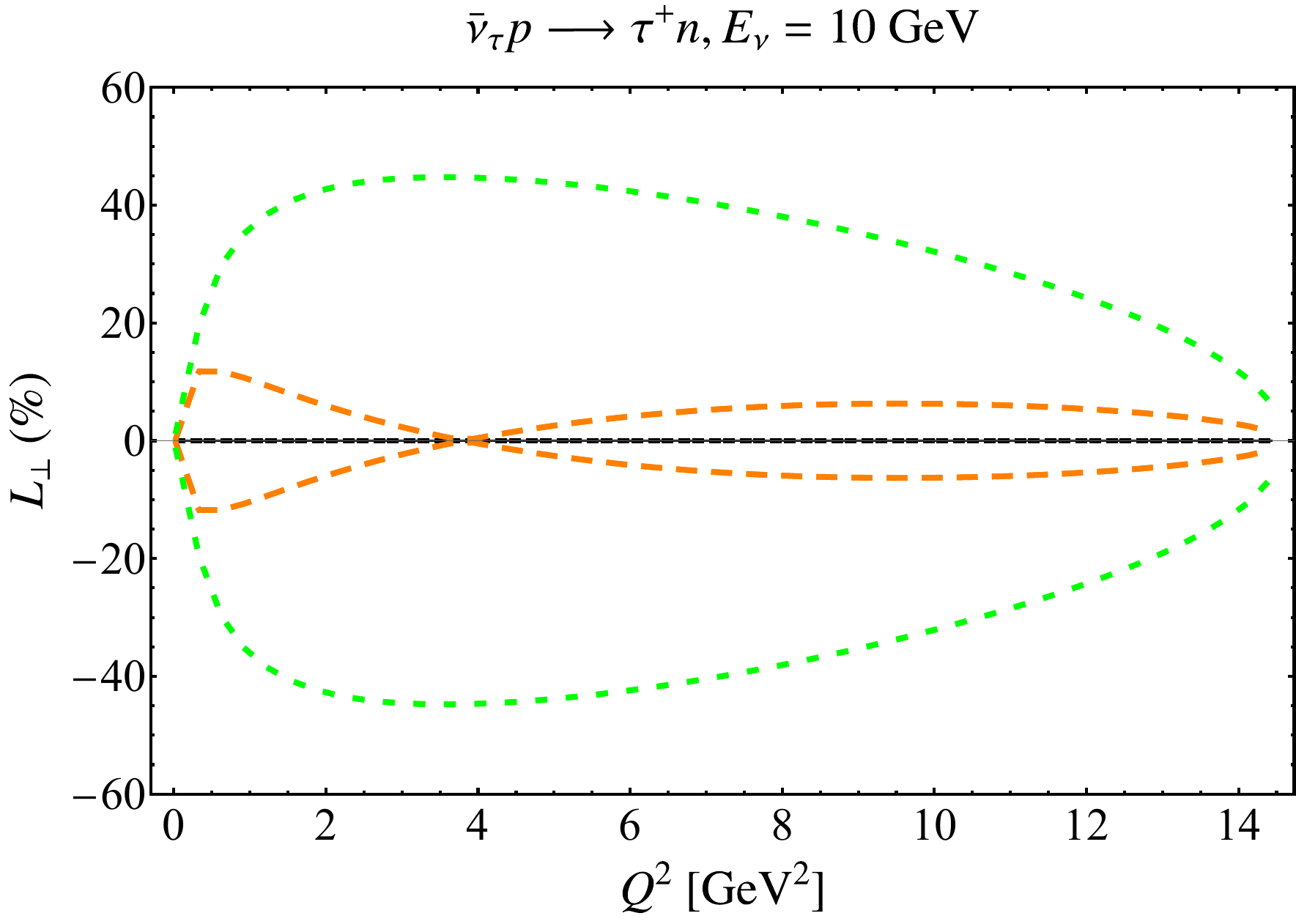}
\includegraphics[width=0.4\textwidth]{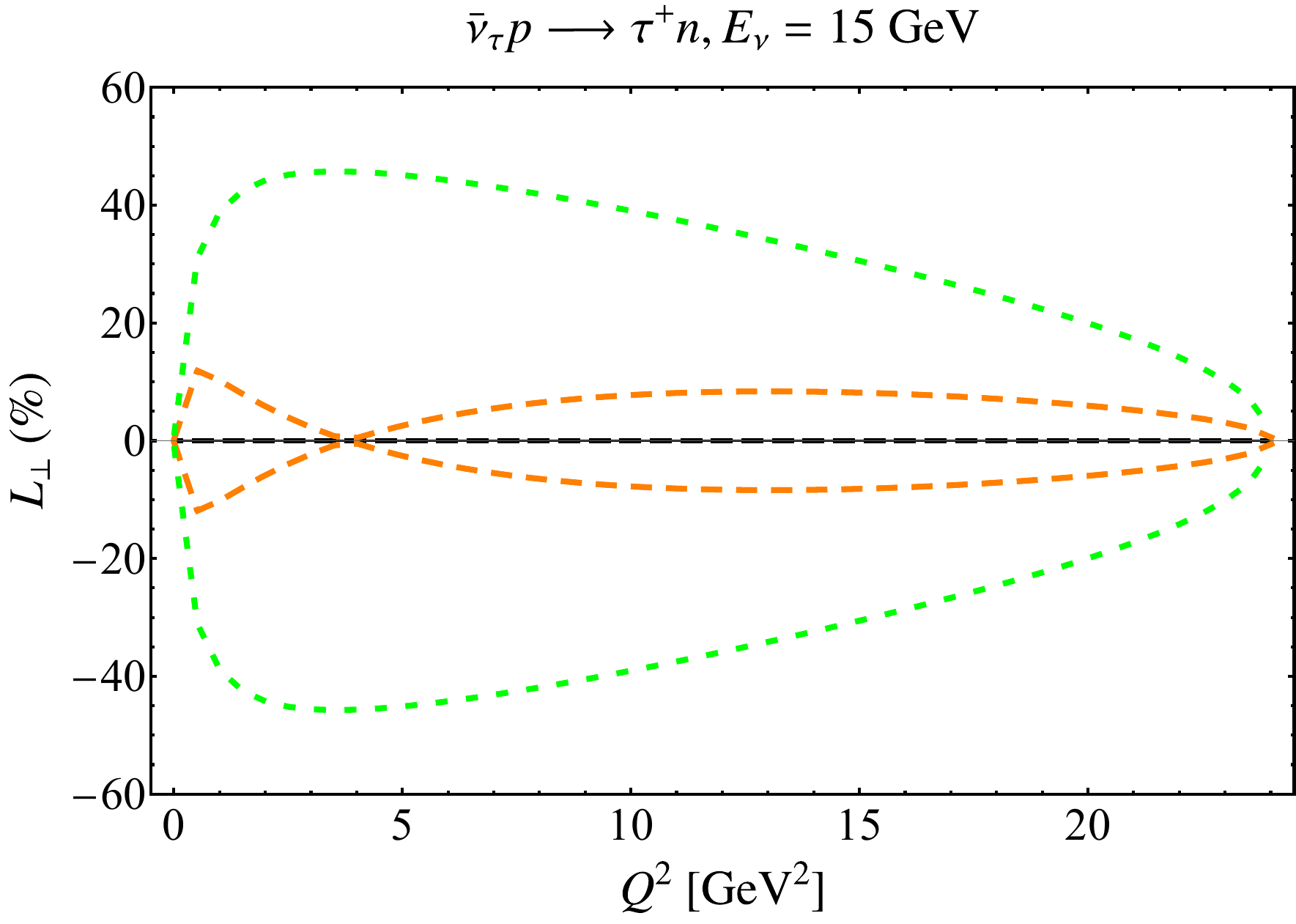}
\caption{Same as Fig.~\ref{fig:antinu_Tt_SCFF_tau} but for the transverse polarization observable $L_\perp$ and imaginary amplitudes. \label{fig:antinu_LTT_SCFF_tau}}
\end{figure}

\newpage

\subsection{Radiative corrections \label{app:radcorplots}}

In this Section, we consider radiative corrections to the observables illustrated in Secs.~\ref{app:unpol_plots}, \ref{app:pol_plots}, and \ref{app:taupol_plots}.

\subsubsection{Radiative corrections to unpolarized cross sections}

In this Section, we present unpolarized cross sections including radiative corrections as described in Sec.~\ref{sec:radiative_corrections}. We consider muon (tau) neutrino and antineutrino energies $E_\nu = 300$~MeV, $600$~MeV, $1$~GeV, and $3$~GeV ($E_\nu = 5$~GeV, $7$~GeV, $10$~GeV, and $15$~GeV), and compare to the uncertainty from vector and axial-vector form factors from Sec.~\ref{sec:observables}.

\begin{figure}[H]
\centering
\includegraphics[width=0.4\textwidth]{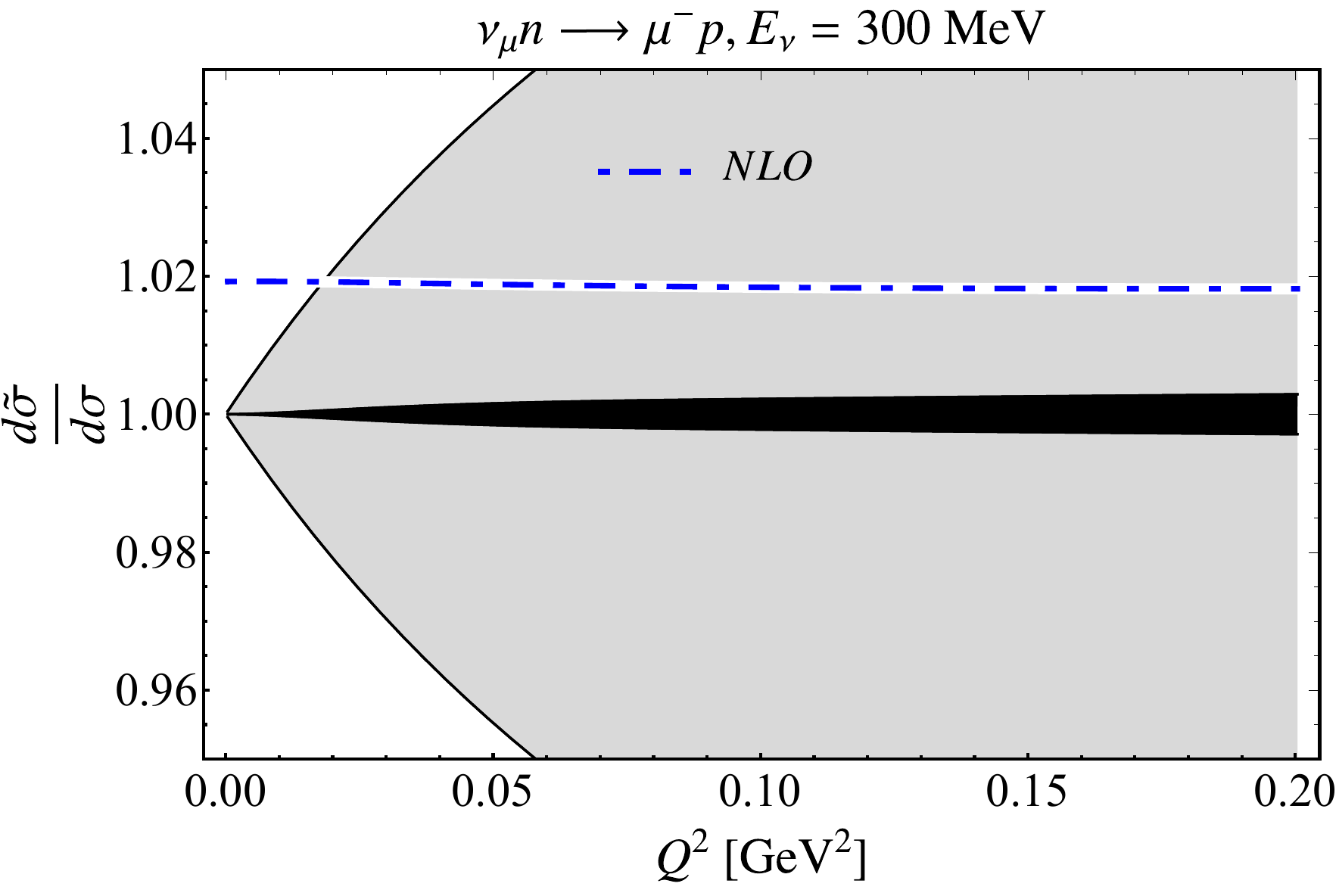}
\includegraphics[width=0.4\textwidth]{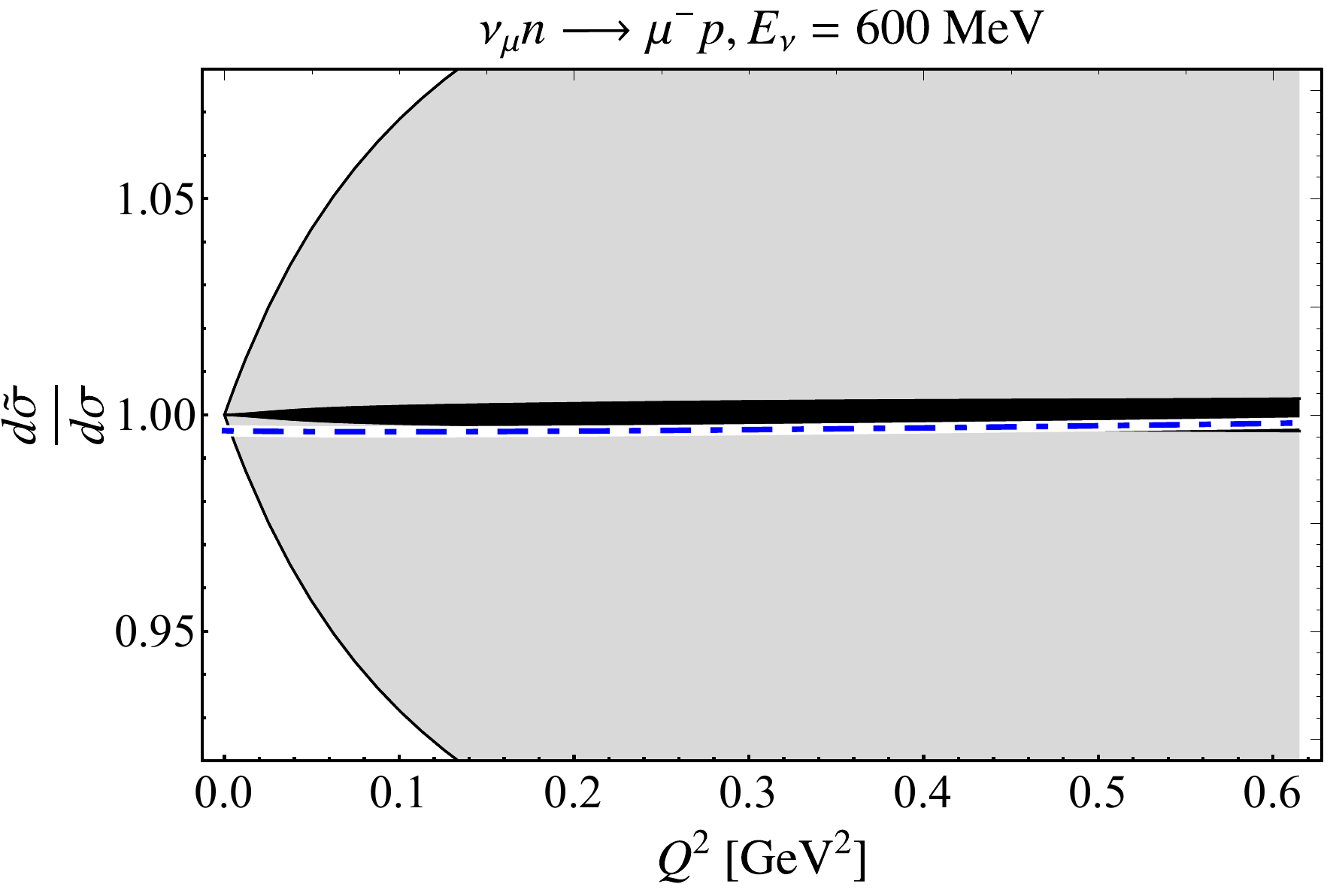}
\includegraphics[width=0.4\textwidth]{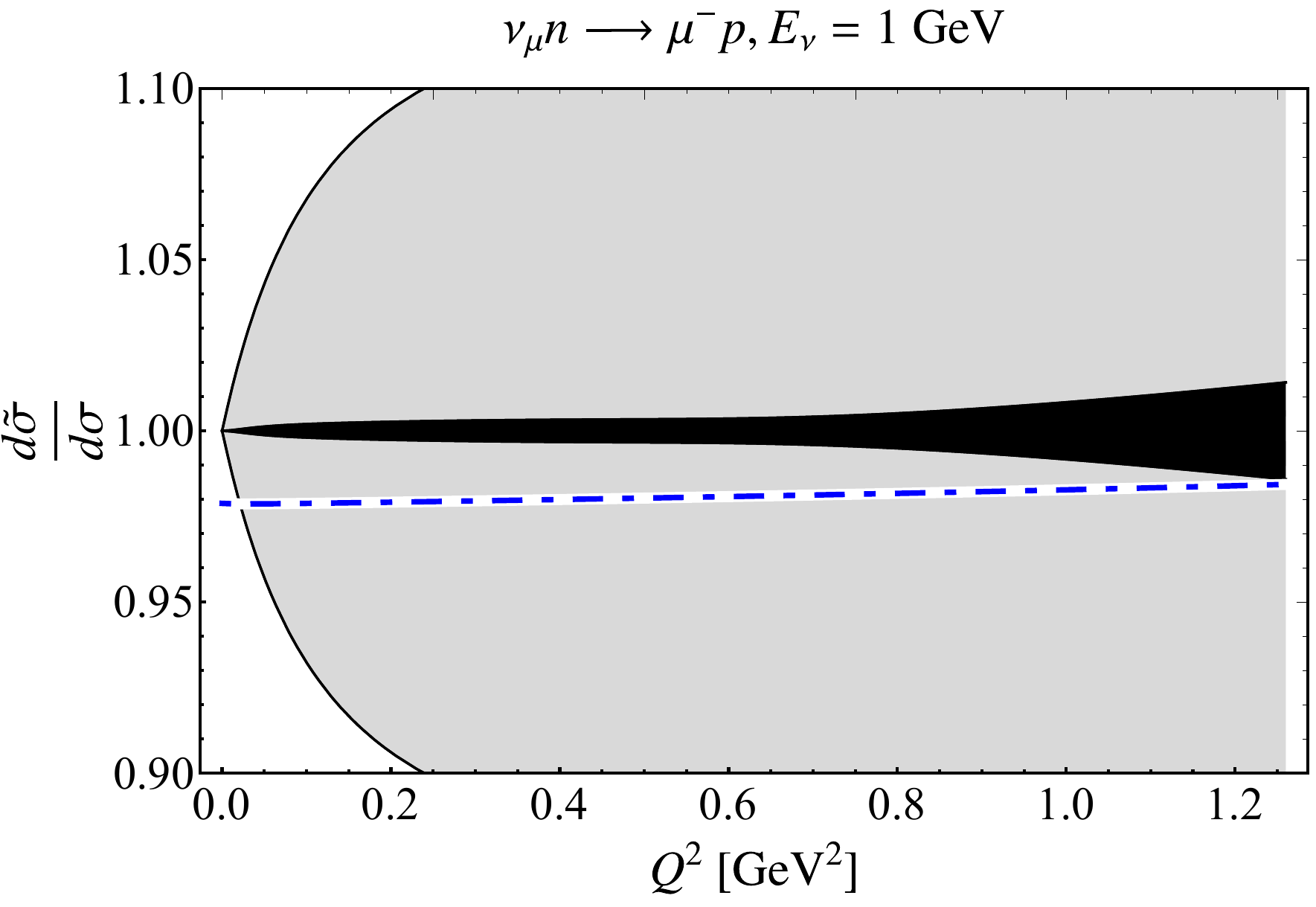}
\includegraphics[width=0.4\textwidth]{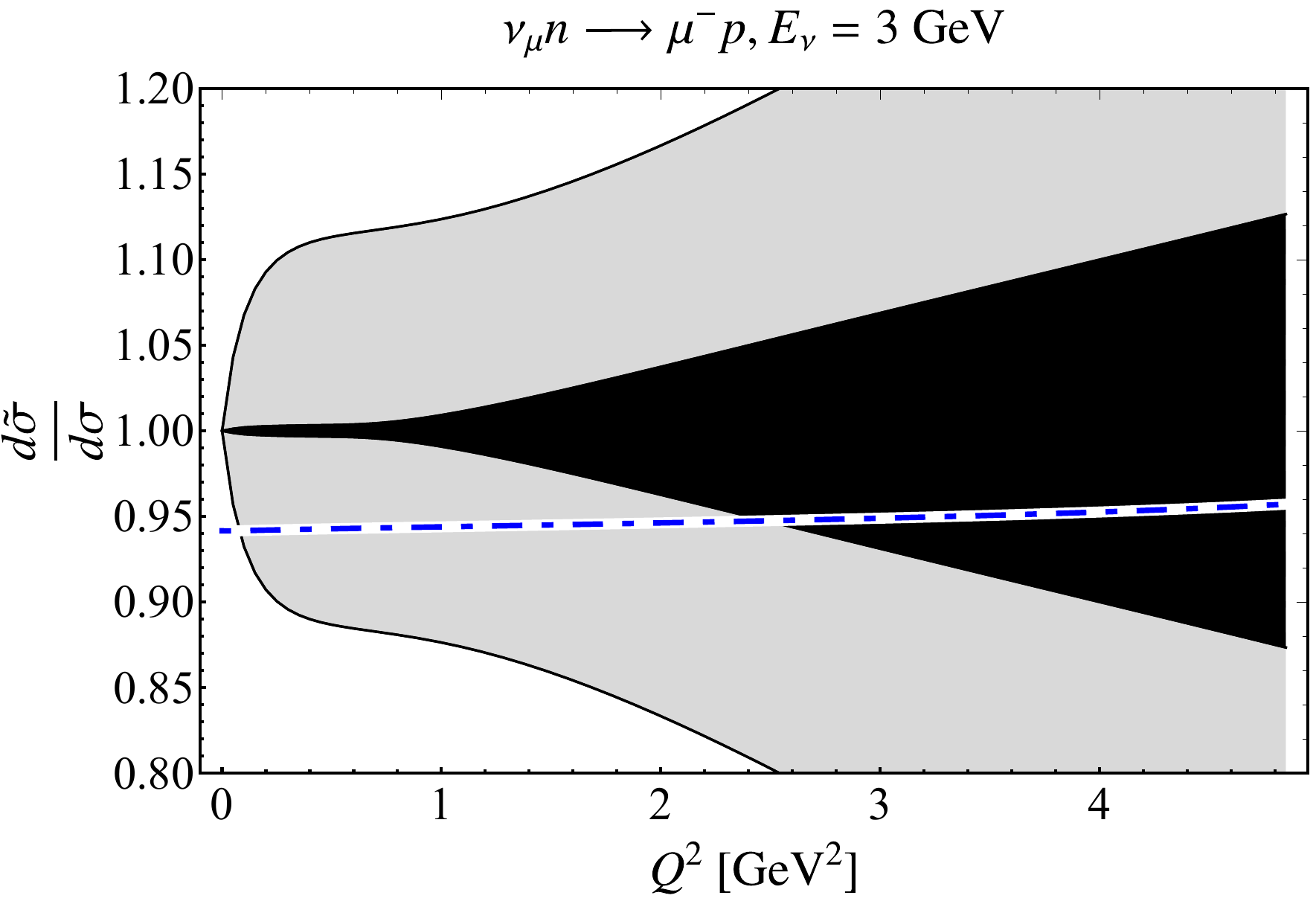}
\caption{Ratio of the unpolarized cross section including radiative corrections to the tree-level result as a function of the squared momentum transfer $Q^2$ at fixed muon neutrino energies $E_\nu = 300$~MeV, $600$~MeV, $1$~GeV, and $3$~GeV is illustrated. The radiatively-corrected cross section, which includes virtual contributions and one real photon of energy below $10~\mathrm{MeV}$, is shown by the blue dashed-dotted line. The dark black and light gray bands correspond to vector and axial-vector form factor uncertainty. \label{fig:nu_xsec_ratio_radcorrRe}}
\end{figure}

\begin{figure}[H]
\centering
\includegraphics[width=0.4\textwidth]{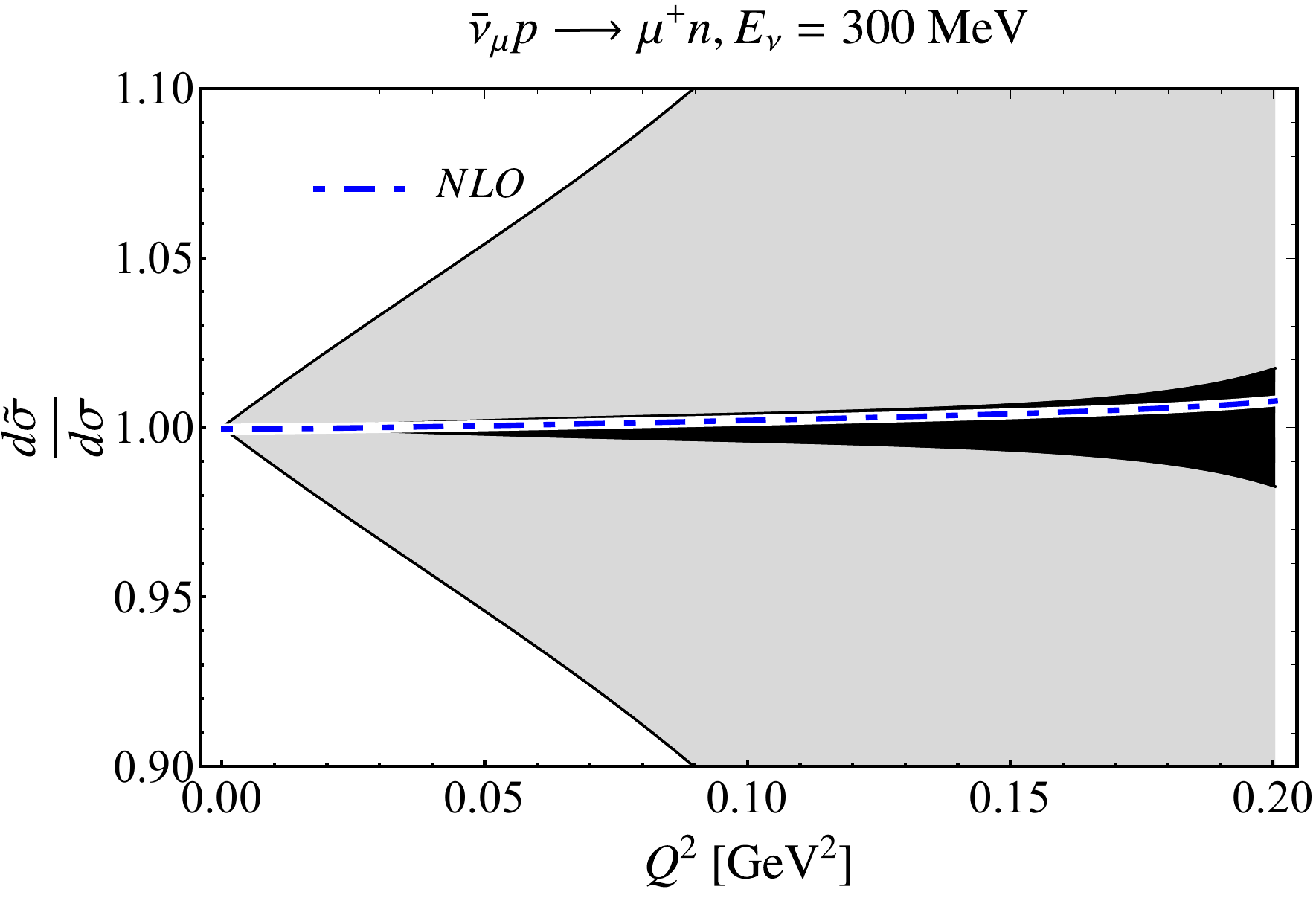}
\includegraphics[width=0.4\textwidth]{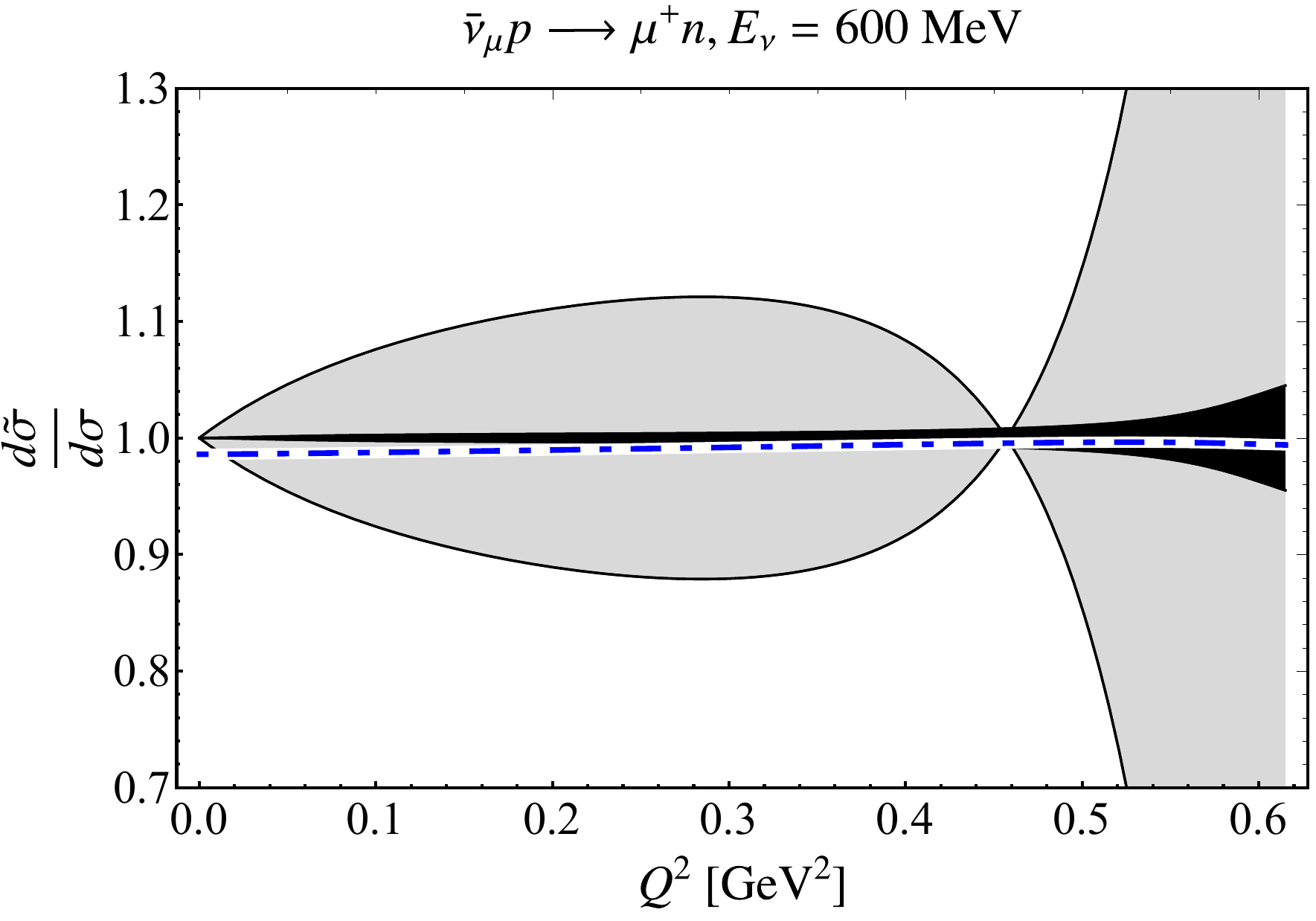}
\includegraphics[width=0.4\textwidth]{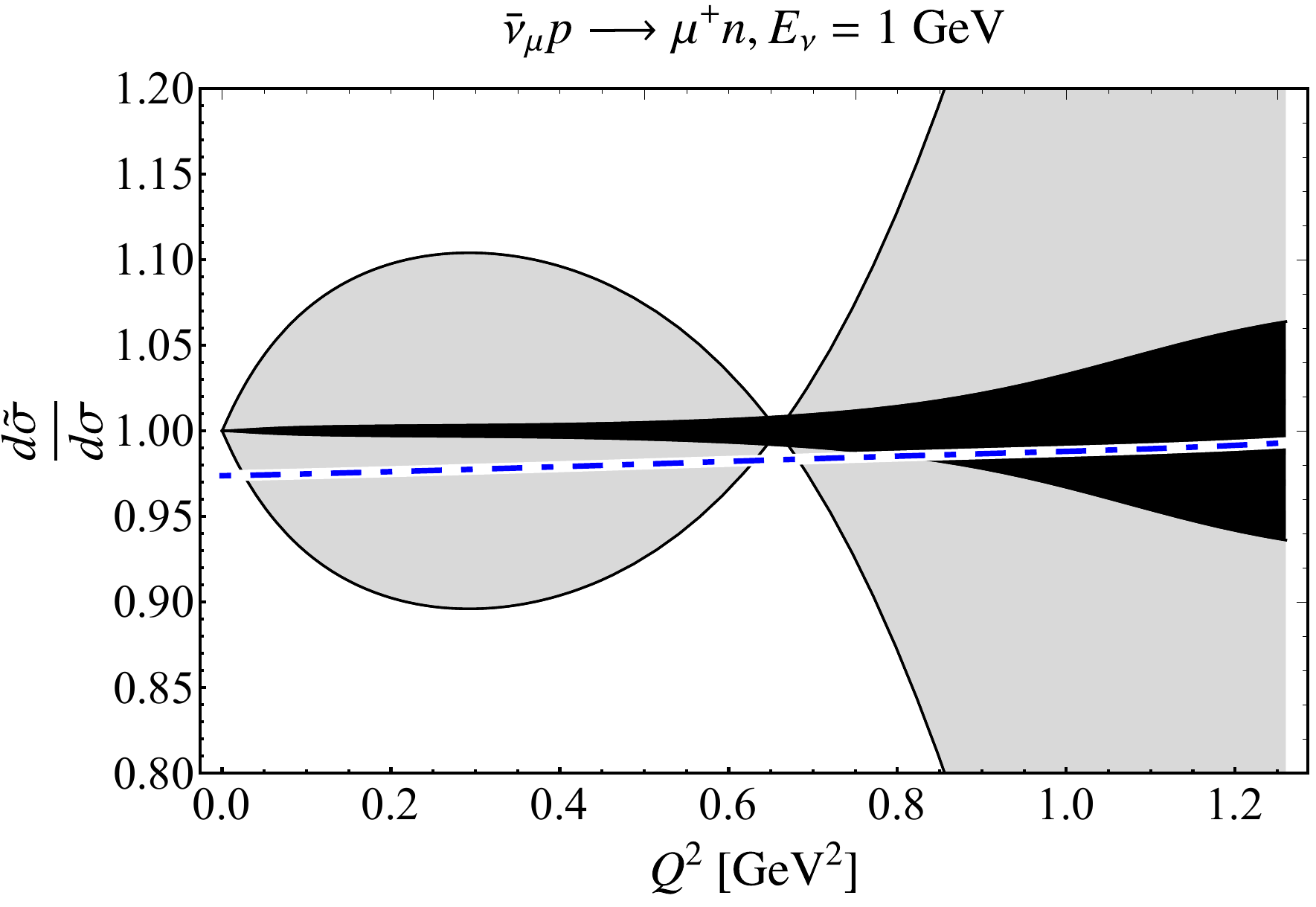}
\includegraphics[width=0.4\textwidth]{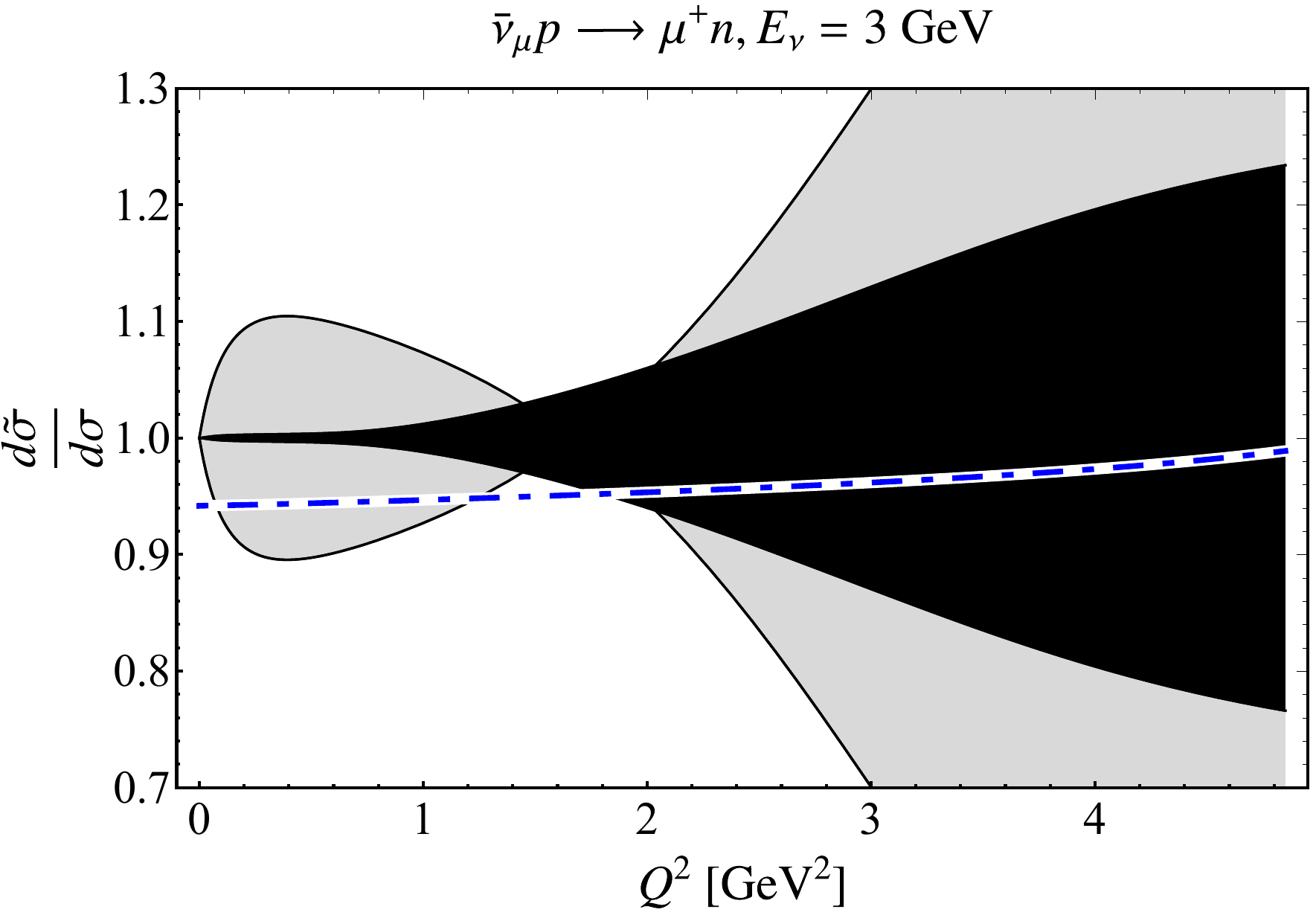}
\caption{Same as Fig.~\ref{fig:nu_xsec_ratio_radcorrRe} but for antineutrinos. \label{fig:antinu_xsec_ratio_radcorrRe}}
\end{figure}

\begin{figure}[H]
\centering
\includegraphics[width=0.4\textwidth]{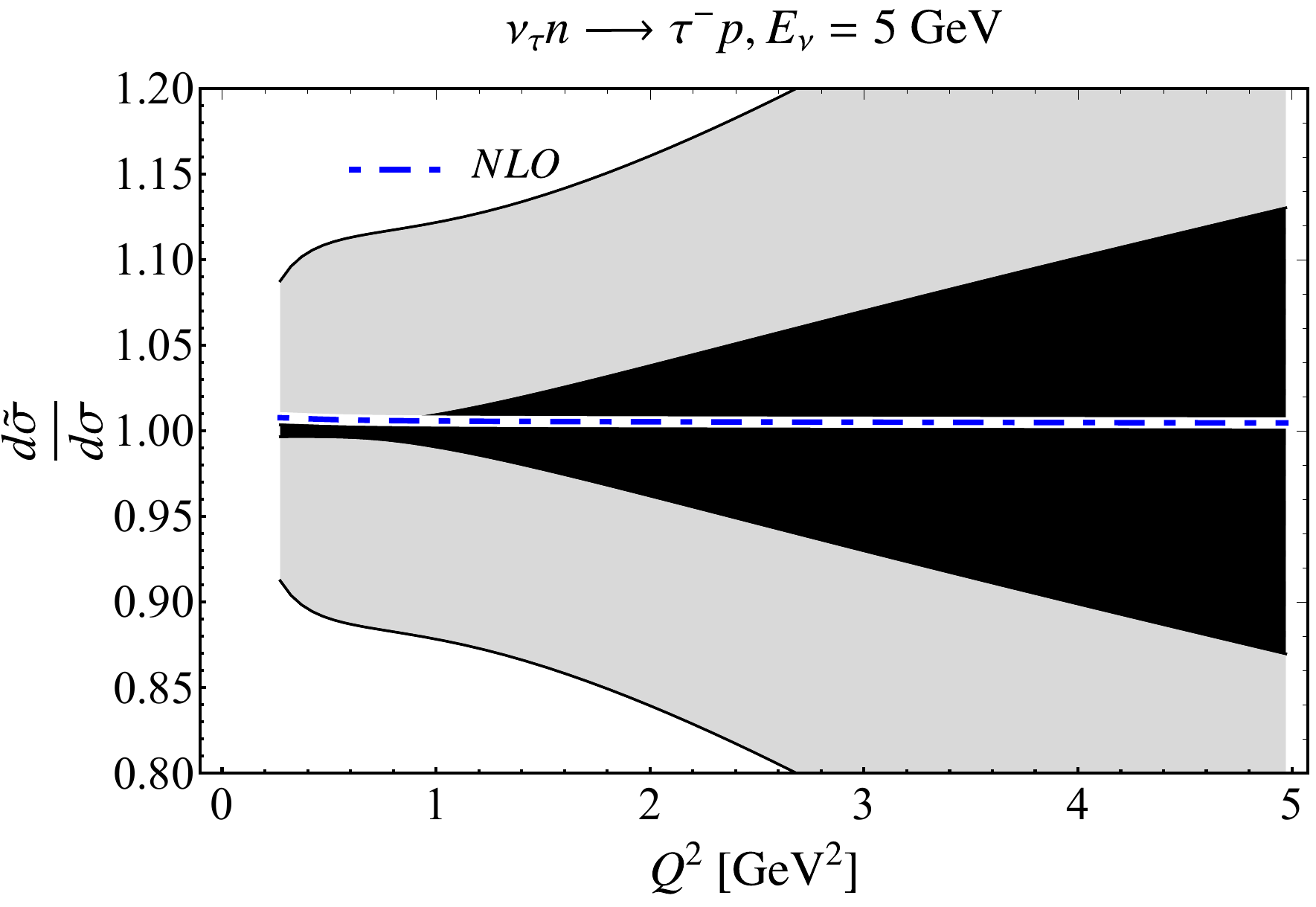}
\includegraphics[width=0.4\textwidth]{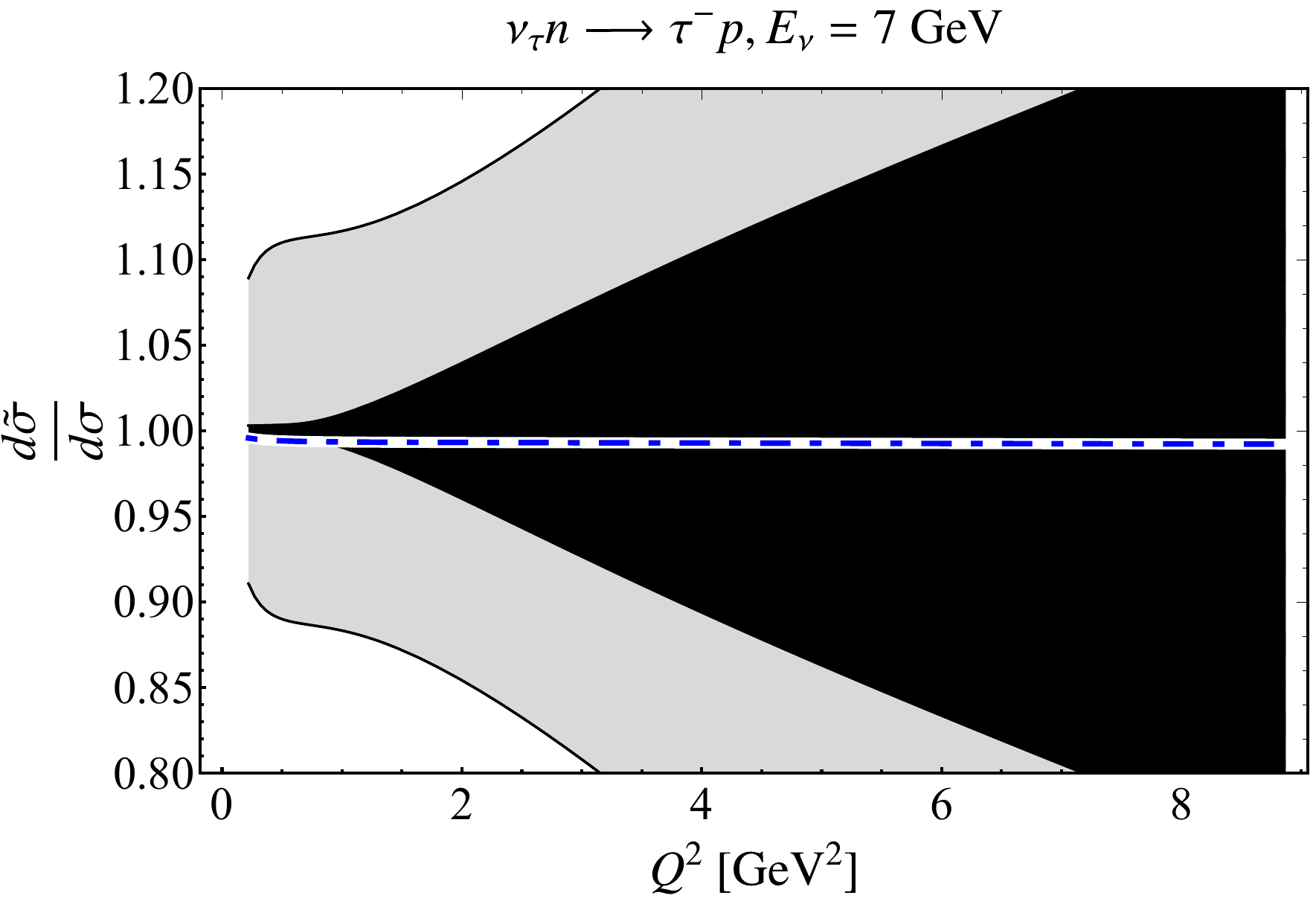}
\includegraphics[width=0.4\textwidth]{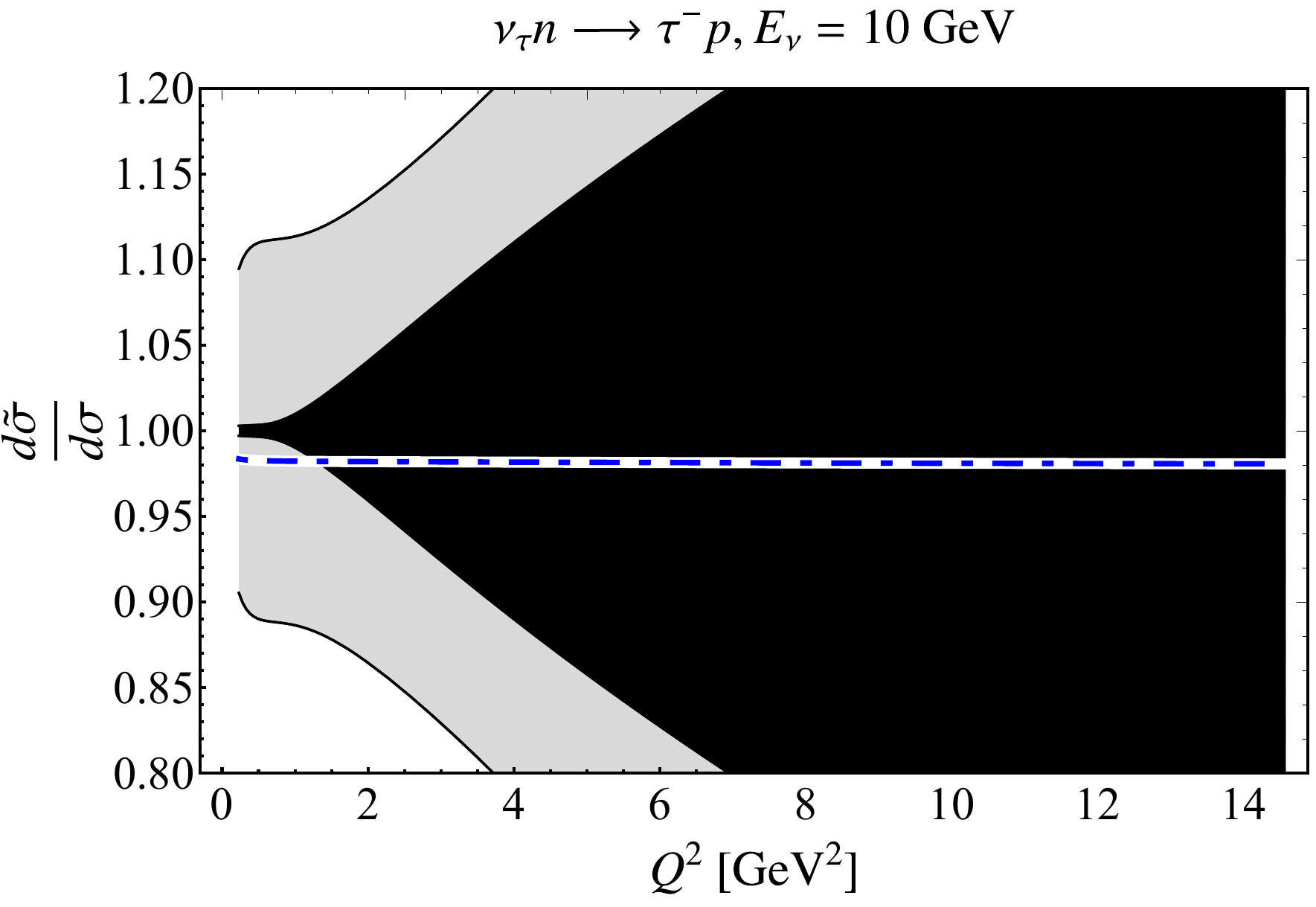}
\includegraphics[width=0.4\textwidth]{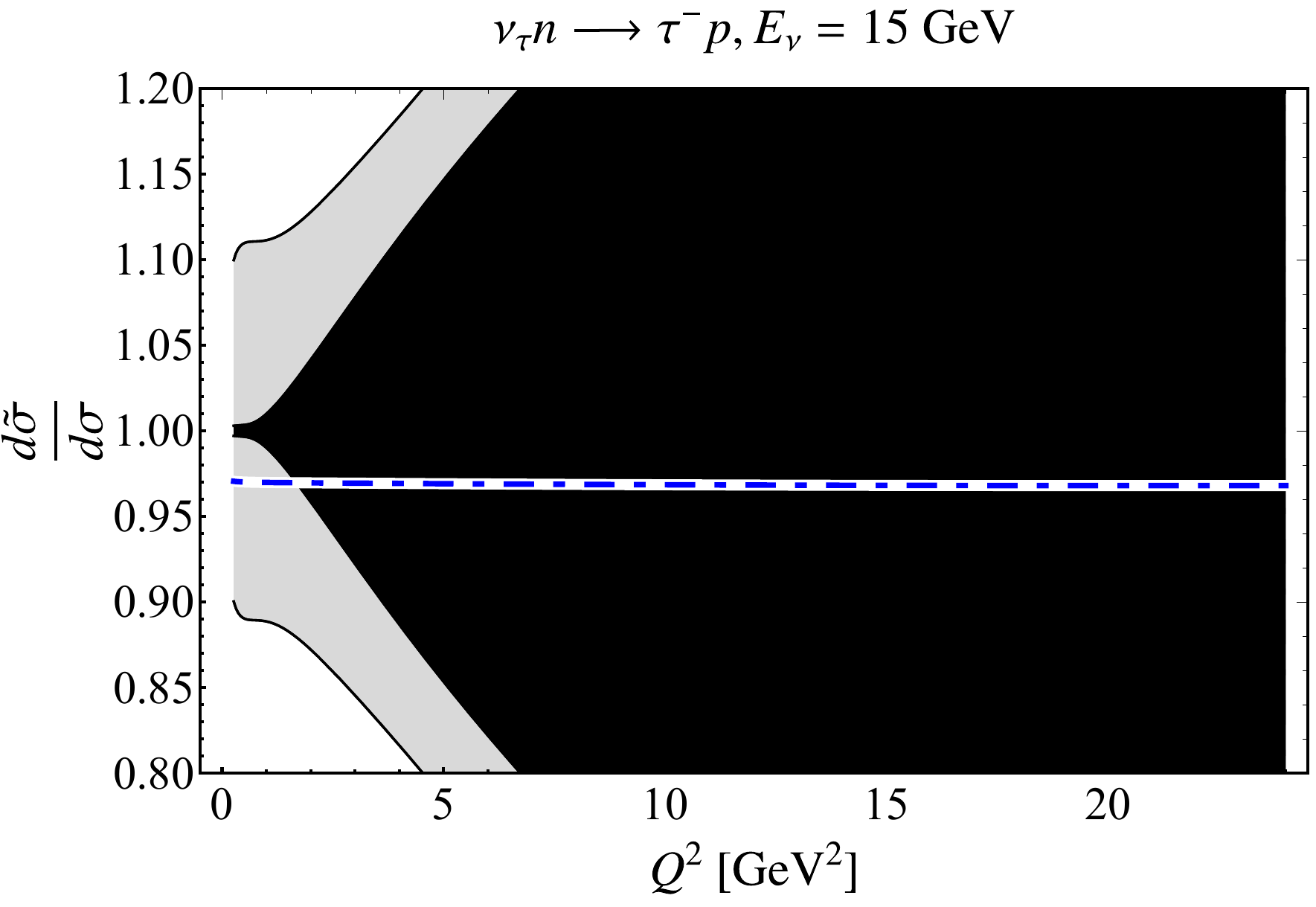}
\caption{Same as Fig.~\ref{fig:nu_xsec_ratio_radcorrRe} but for tau neutrinos at fixed beam energies $E_\nu = 5$~GeV, $7$~GeV, $10$~GeV, and $15$~GeV.\label{fig:nu_xsec_ratio_radcorrtauRe}}
\end{figure}

\begin{figure}[H]
\centering
\includegraphics[width=0.4\textwidth]{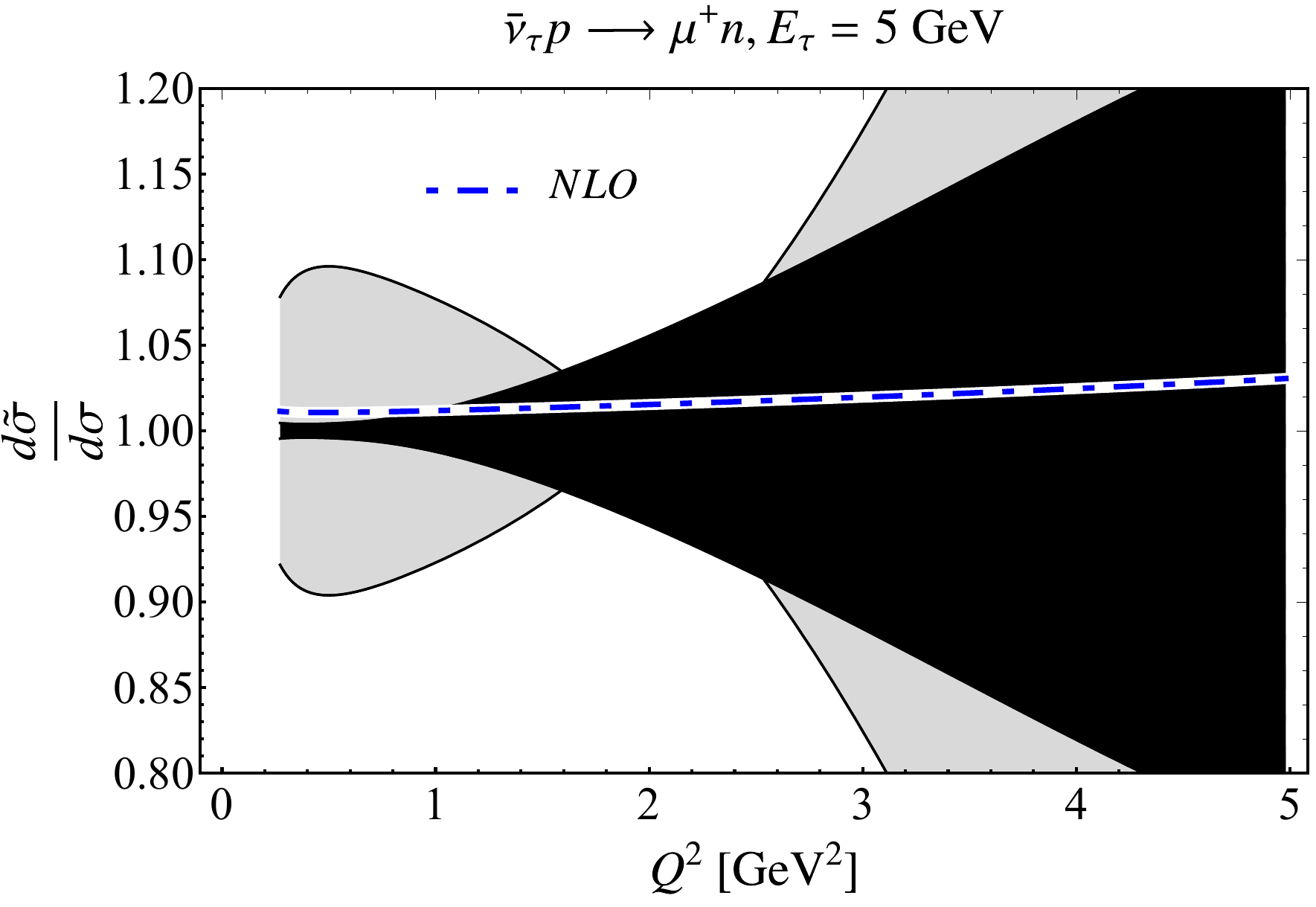}
\includegraphics[width=0.4\textwidth]{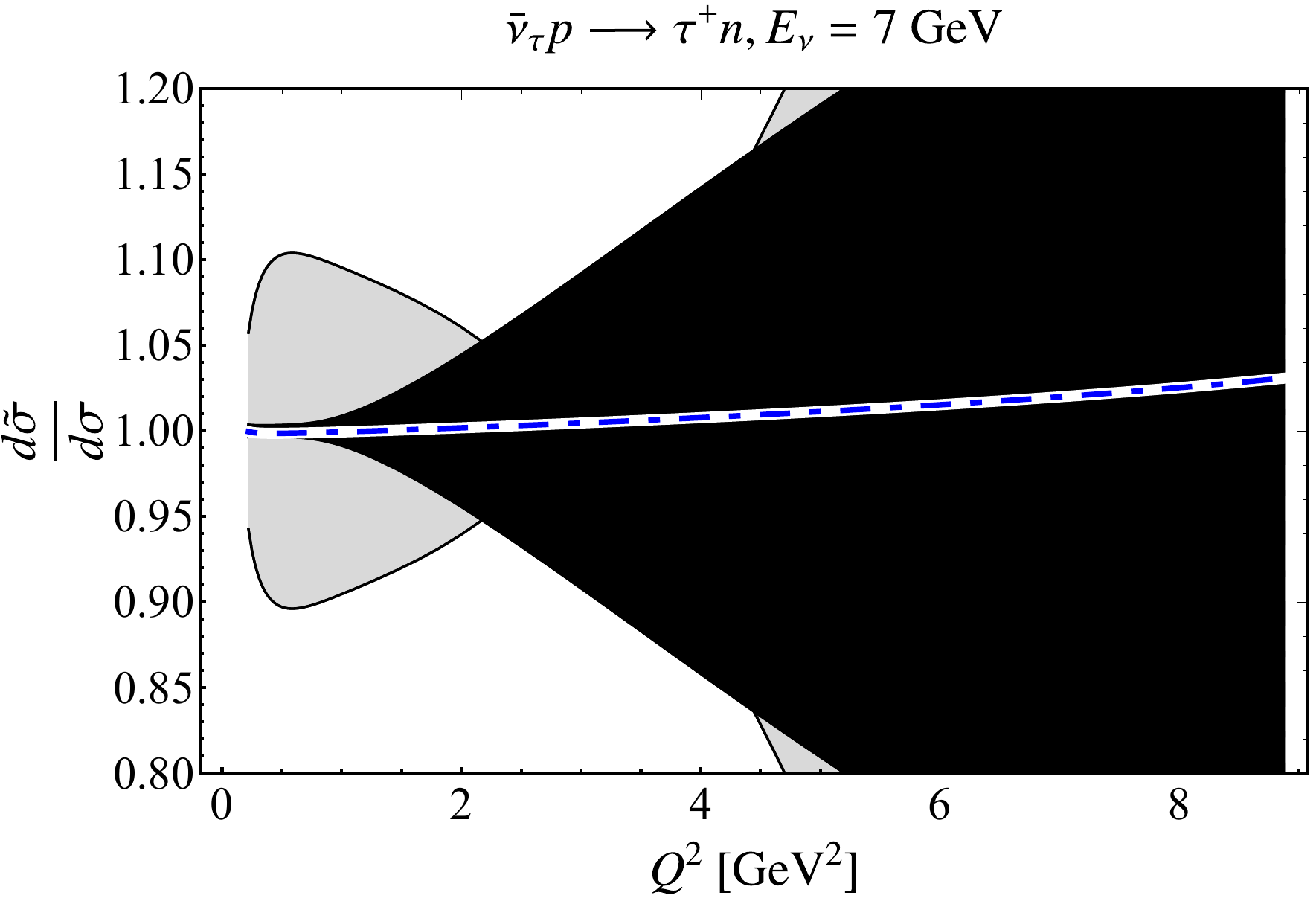}
\includegraphics[width=0.4\textwidth]{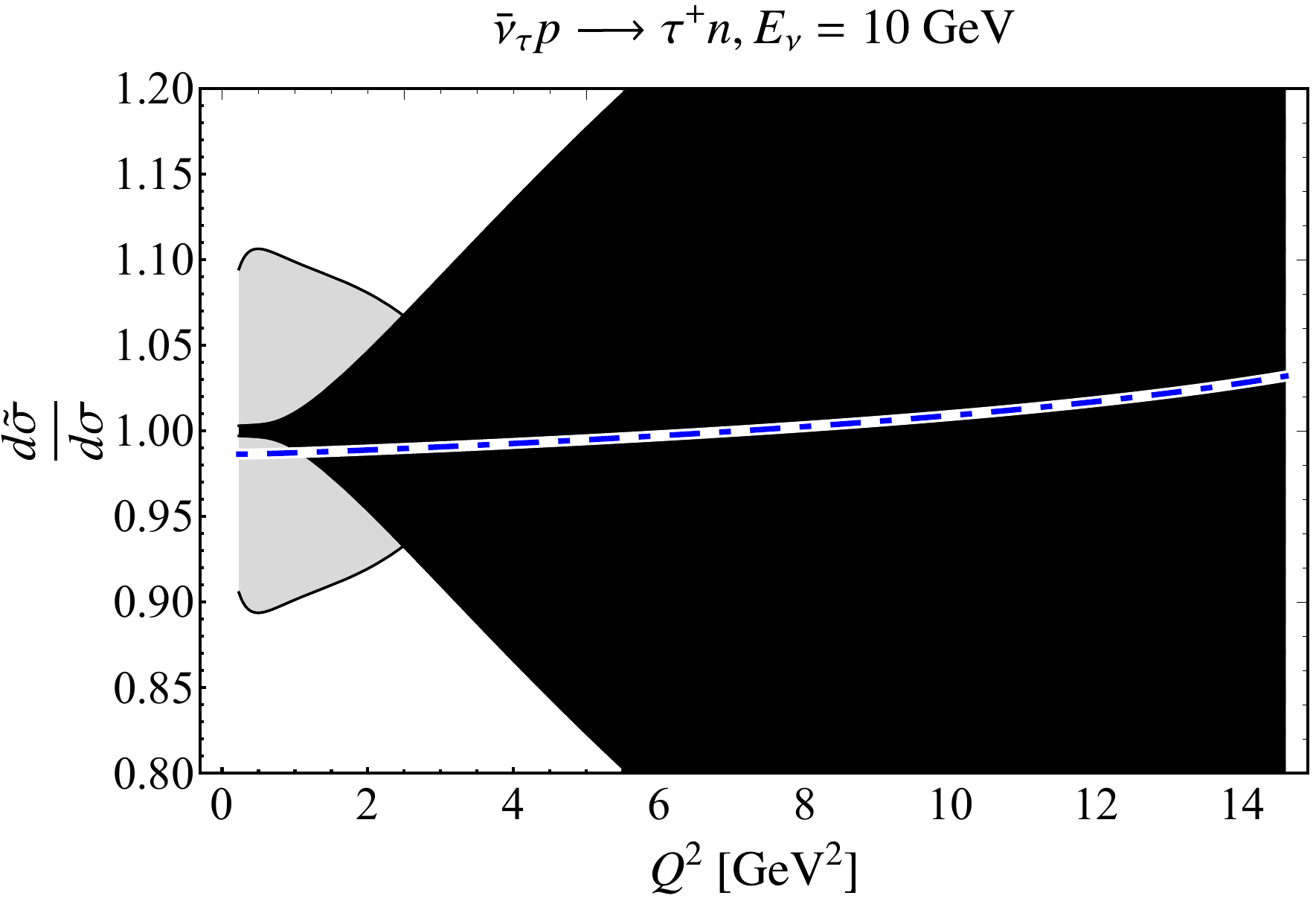}
\includegraphics[width=0.4\textwidth]{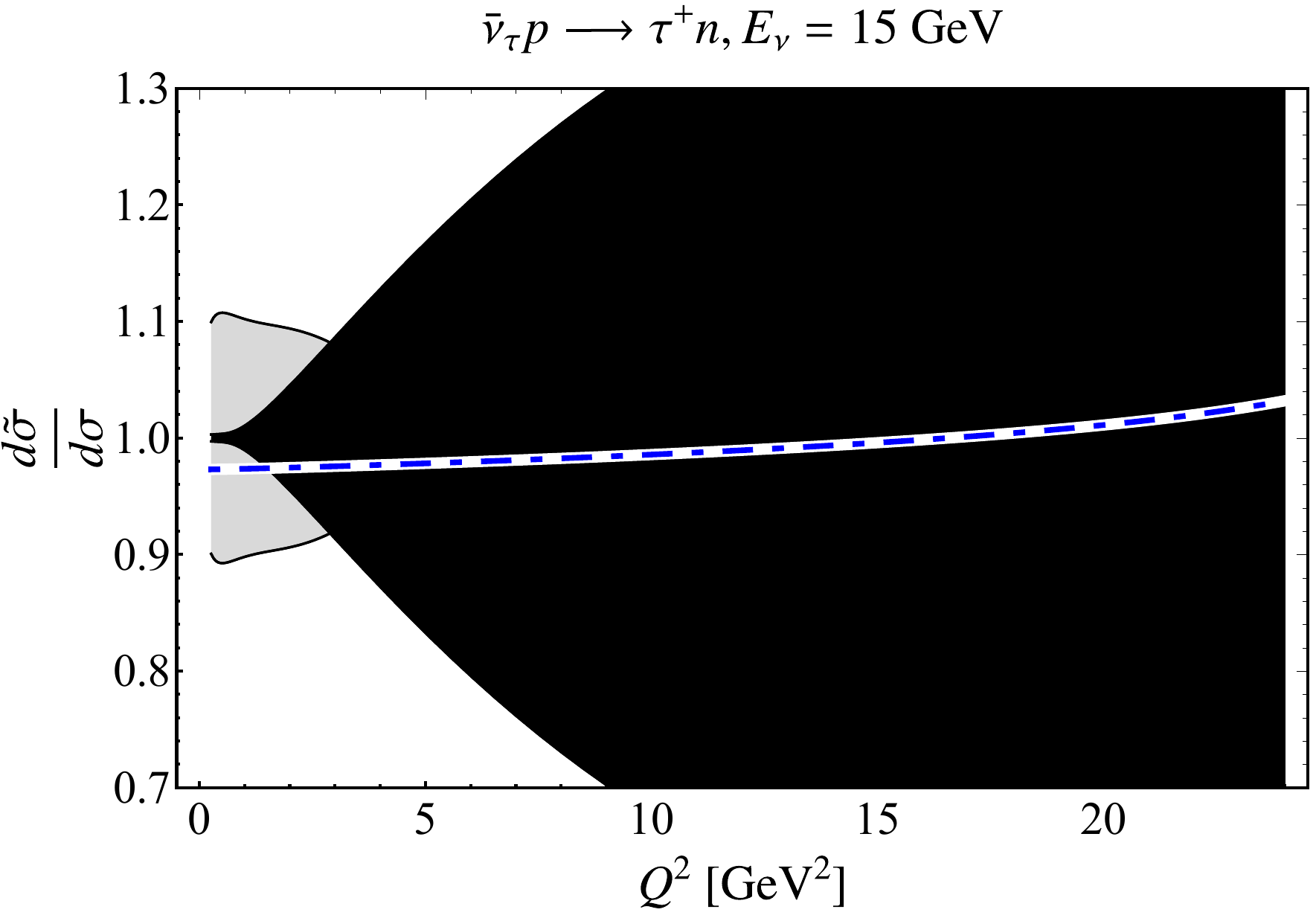}
\caption{Same as Fig.~\ref{fig:nu_xsec_ratio_radcorrtauRe} but for antineutrinos.\label{fig:antinu_xsec_ratio_radcorrFtauRe}}
\end{figure}

\subsubsection{Radiative corrections to polarization asymmetries, muon (anti)neutrino}

In this Section, we present single-spin asymmetries for muon neutrinos and antineutrinos, including radiative corrections as described in Sec.~\ref{sec:radiative_corrections}. We consider neutrino energies $E_\nu = 300$~MeV, $600$~MeV, $1$~GeV, and $3$~GeV, and compare to the uncertainty from vector and axial-vector form factors from Sec.~\ref{sec:observables}. 

\begin{figure}[H]
\centering
\includegraphics[width=0.4\textwidth]{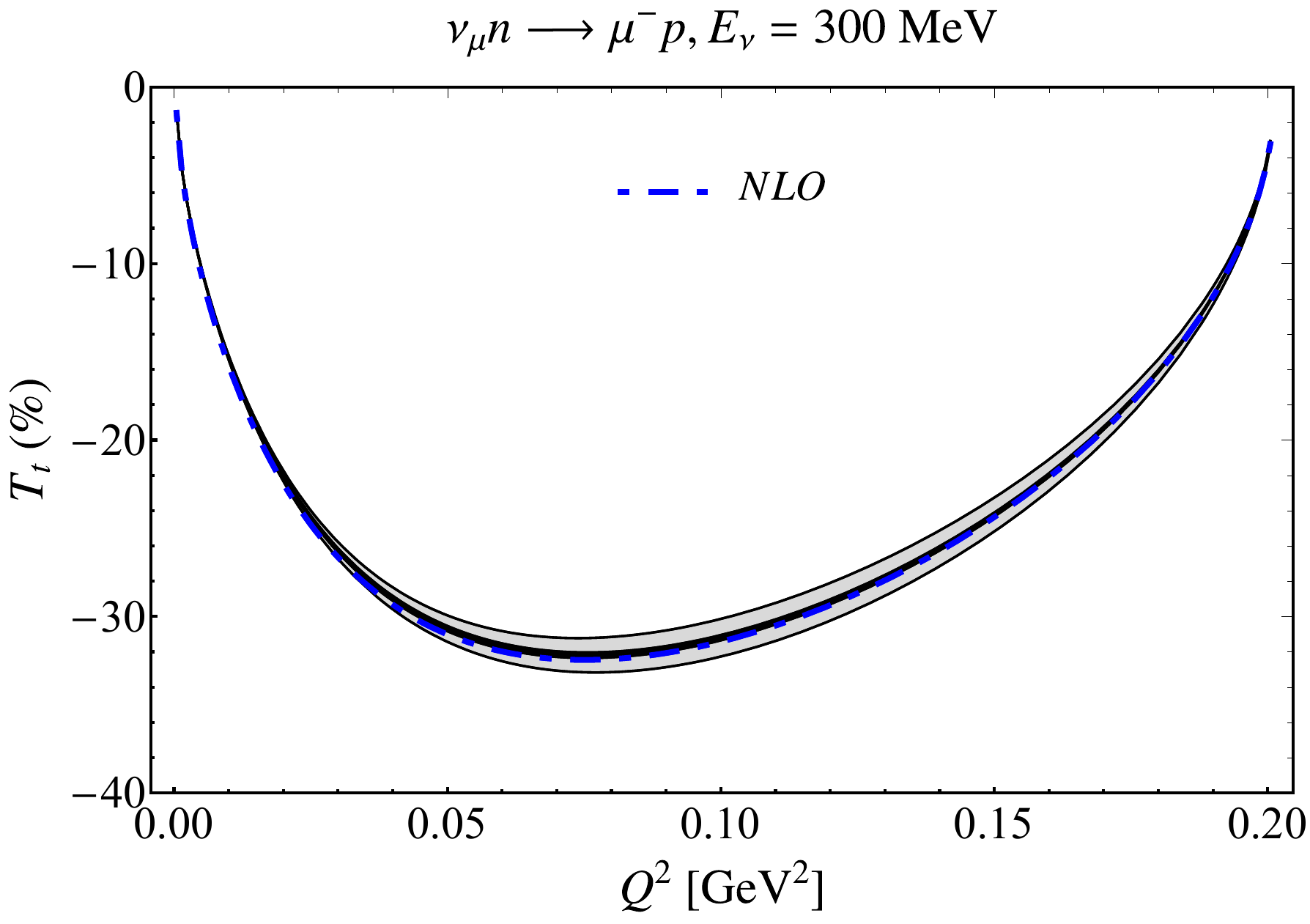}
\includegraphics[width=0.4\textwidth]{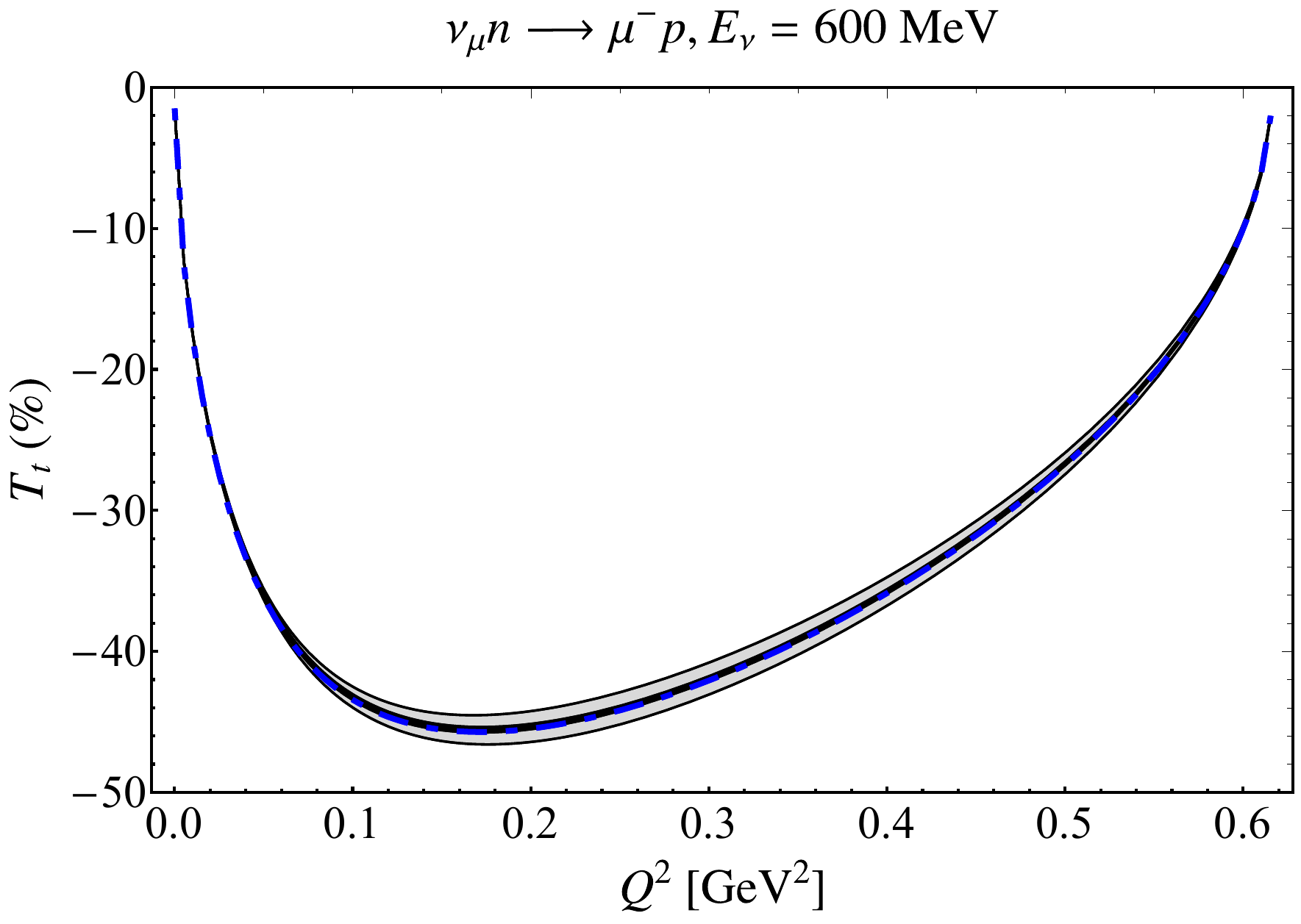}
\includegraphics[width=0.4\textwidth]{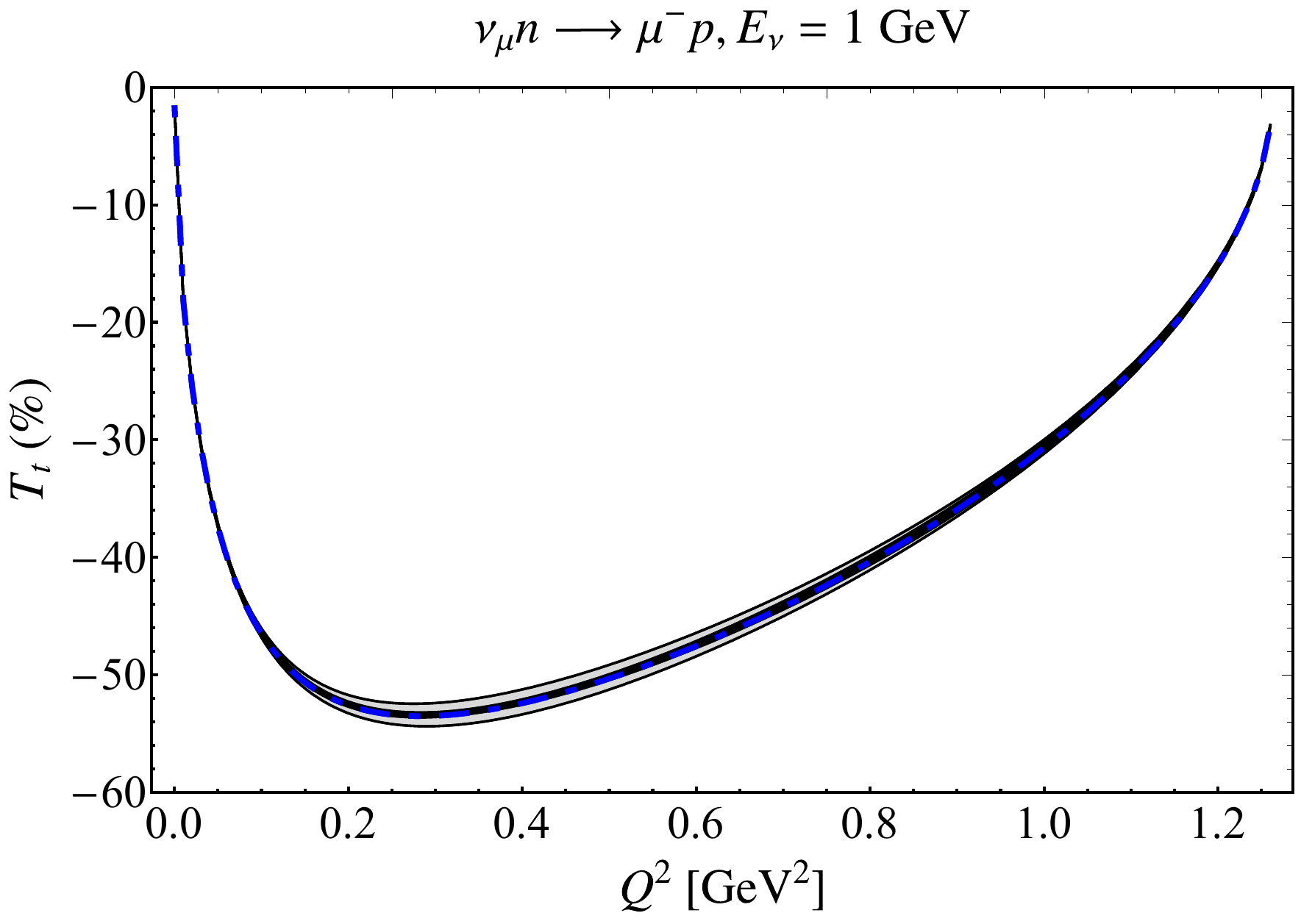}
\includegraphics[width=0.4\textwidth]{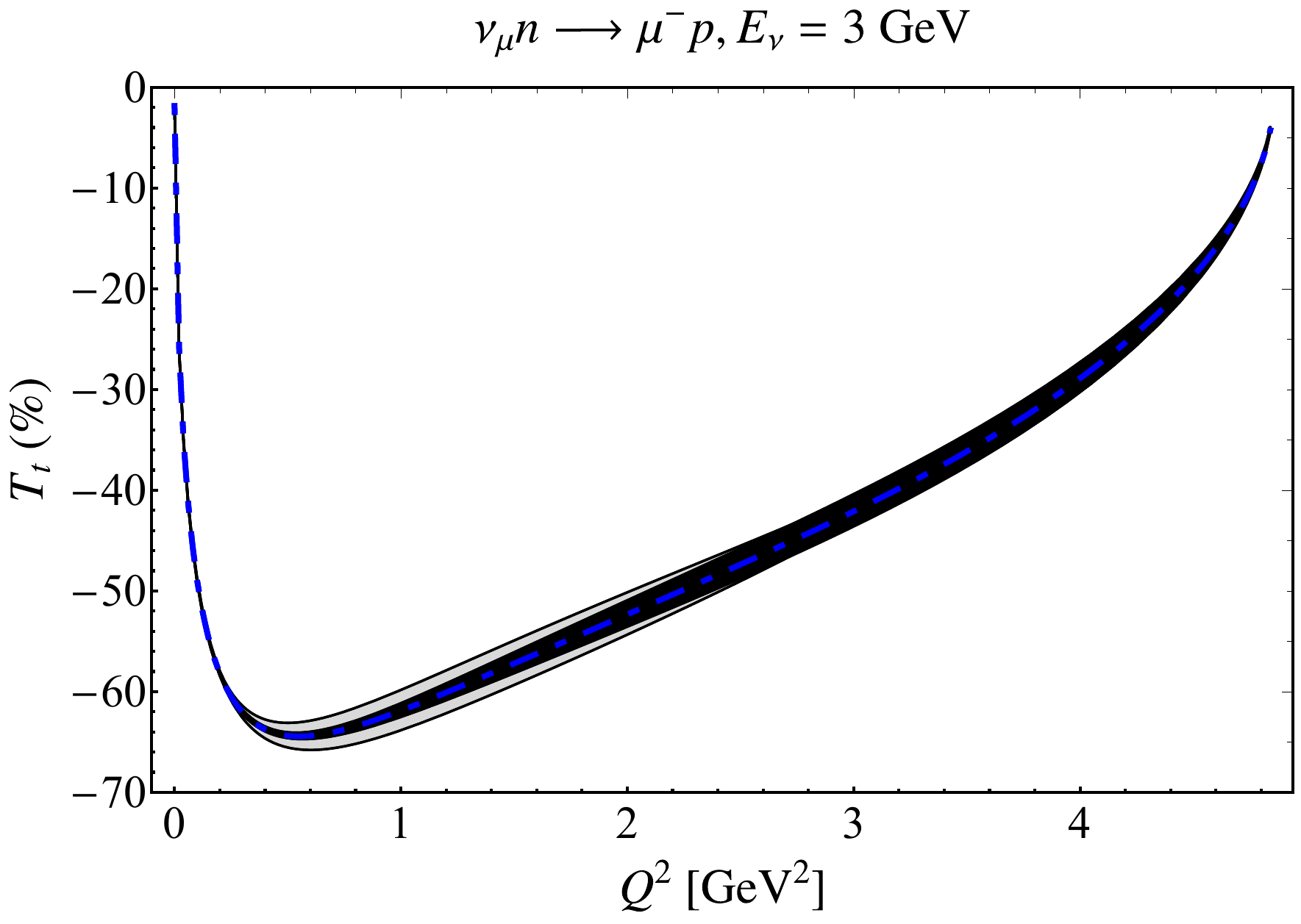}
\caption{Radiative correction to the transverse polarization observable $T_t$, at fixed muon neutrino energies $E_\nu = 300$~MeV, $600$~MeV, $1$~GeV, and $3$~GeV is illustrated. Radiatively-corrected observable, which includes virtual contributions and one real photon of energy below $10~\mathrm{MeV}$, is shown by the blue dashed-dotted line. The dark black and light gray bands correspond to vector and axial-vector form factor uncertainty. \label{fig:nu_Tt_radcorr}}
\end{figure}

\begin{figure}[H]
\centering
\includegraphics[width=0.4\textwidth]{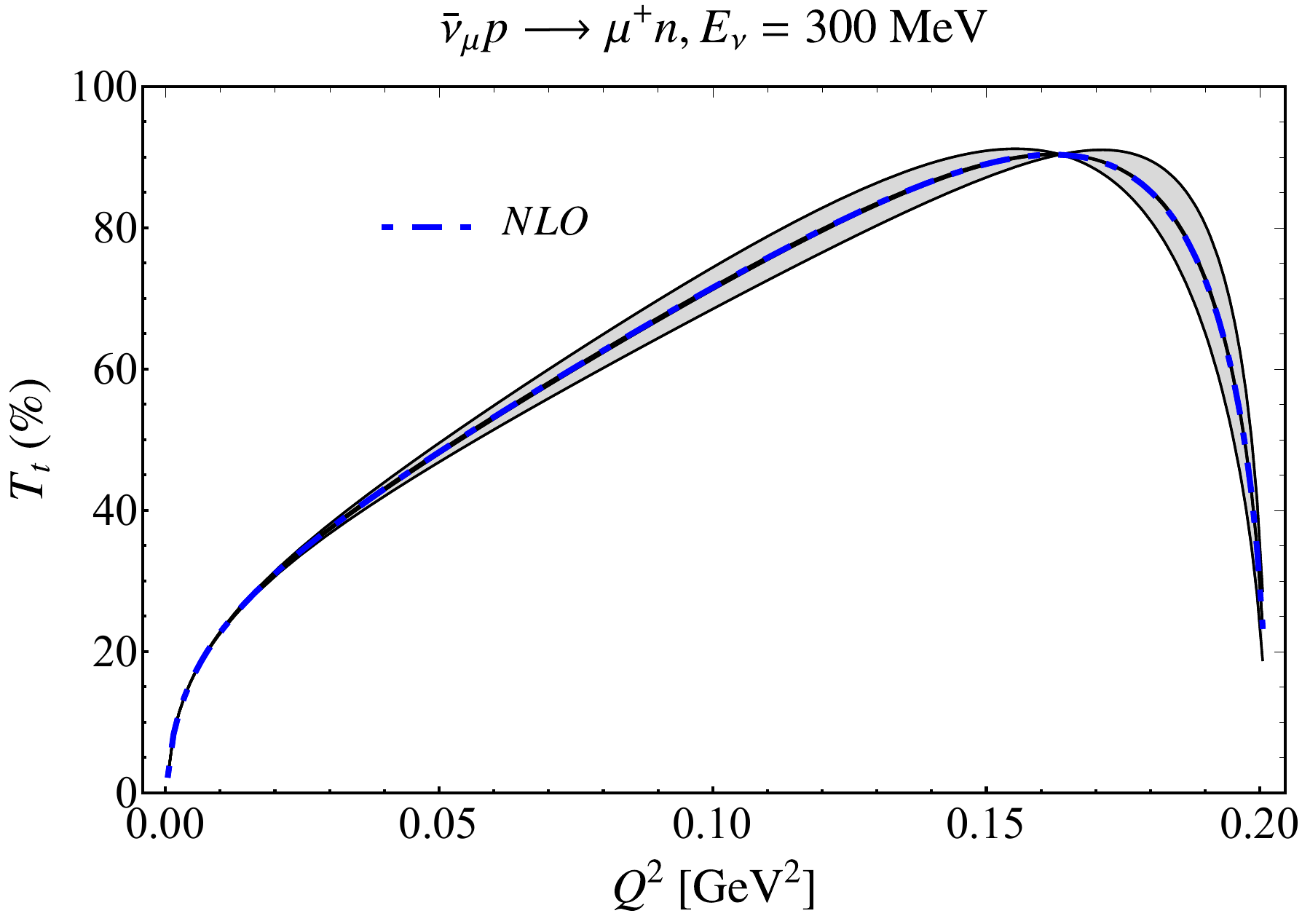}
\includegraphics[width=0.4\textwidth]{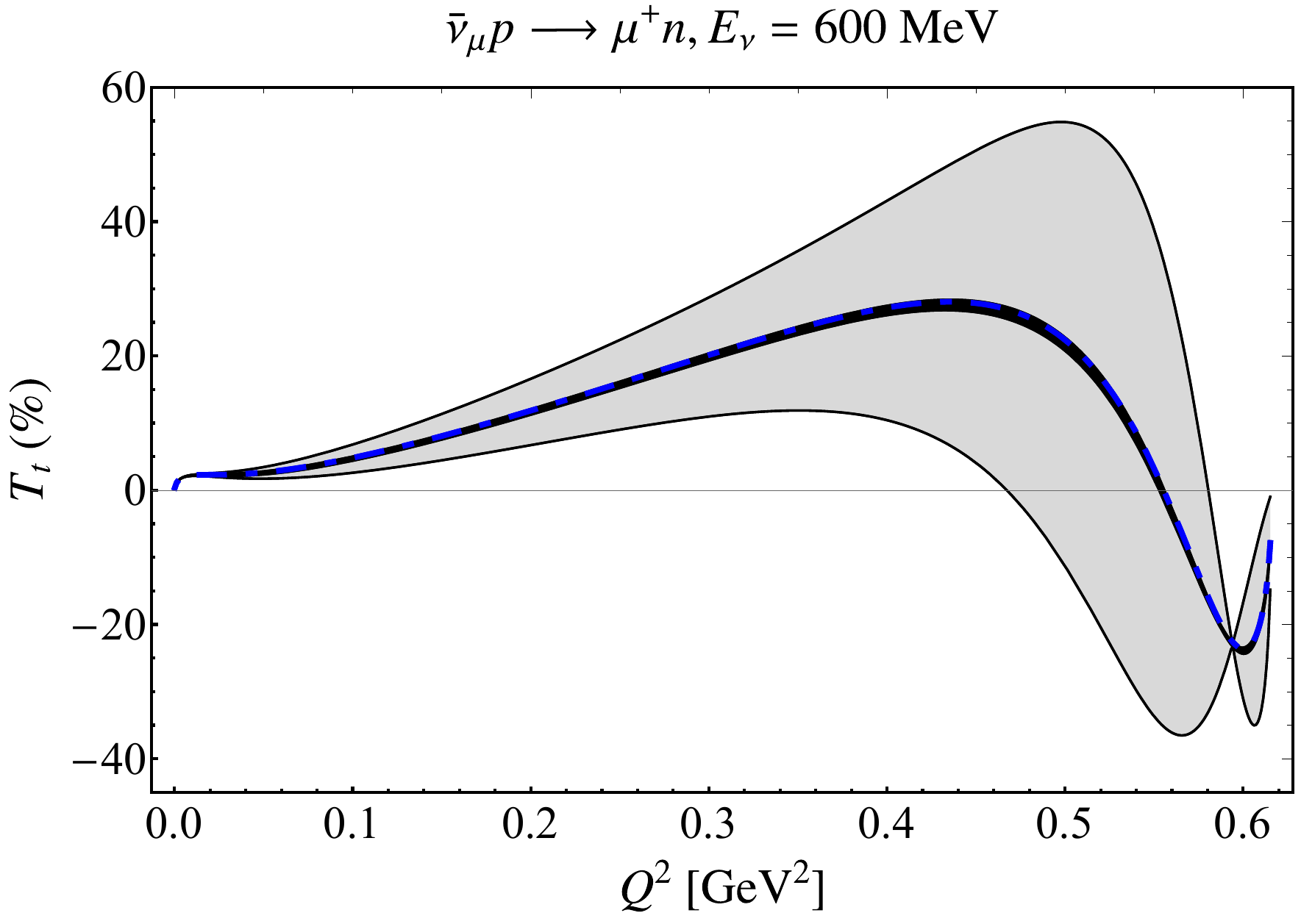}
\includegraphics[width=0.4\textwidth]{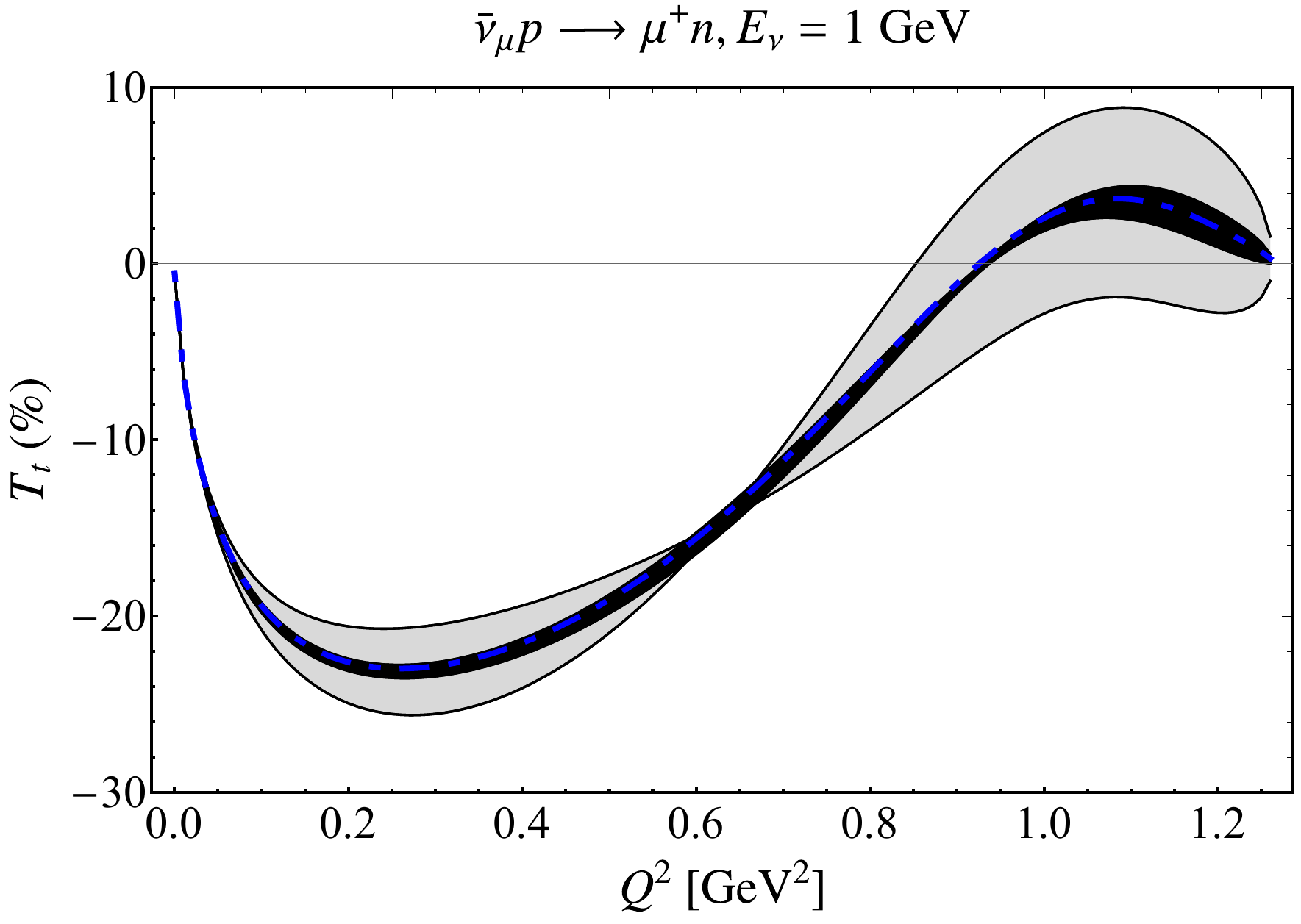}
\includegraphics[width=0.4\textwidth]{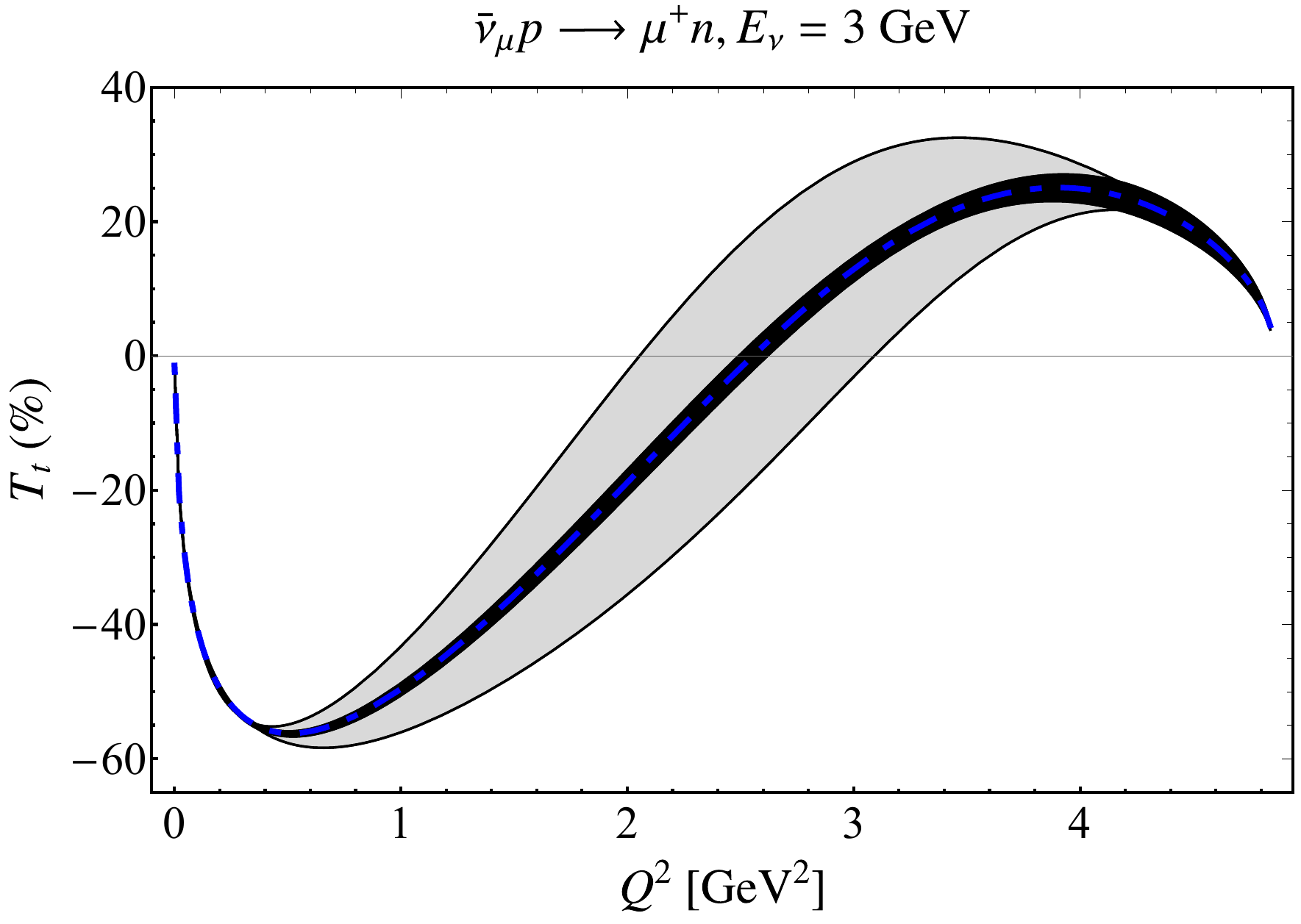}
\caption{Same as Fig.~\ref{fig:nu_Tt_radcorr} but for antineutrinos. \label{fig:antinu_Tt_radcorr}}
\end{figure}

\begin{figure}[H]
\centering
\includegraphics[width=0.4\textwidth]{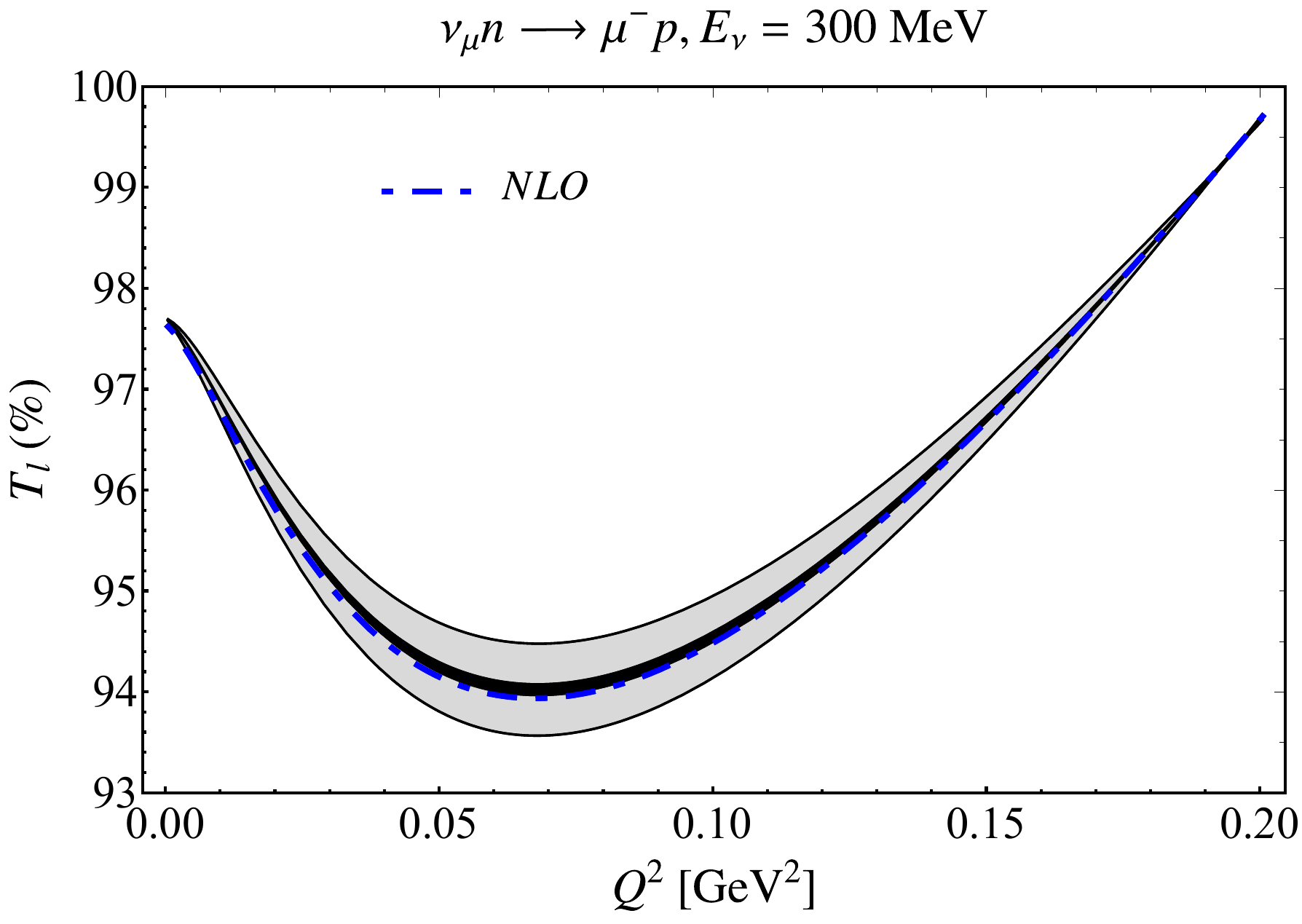}
\includegraphics[width=0.4\textwidth]{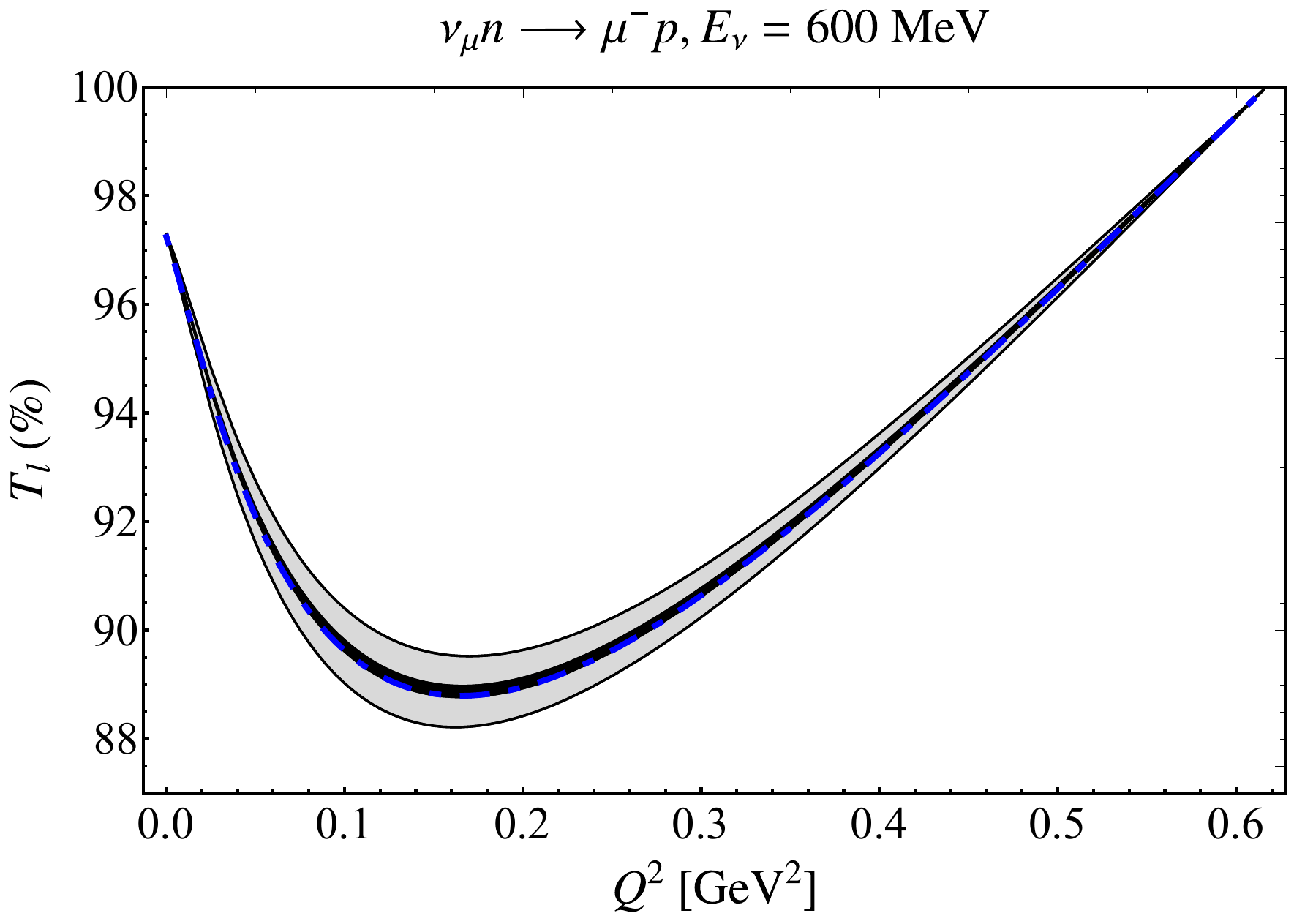}
\includegraphics[width=0.4\textwidth]{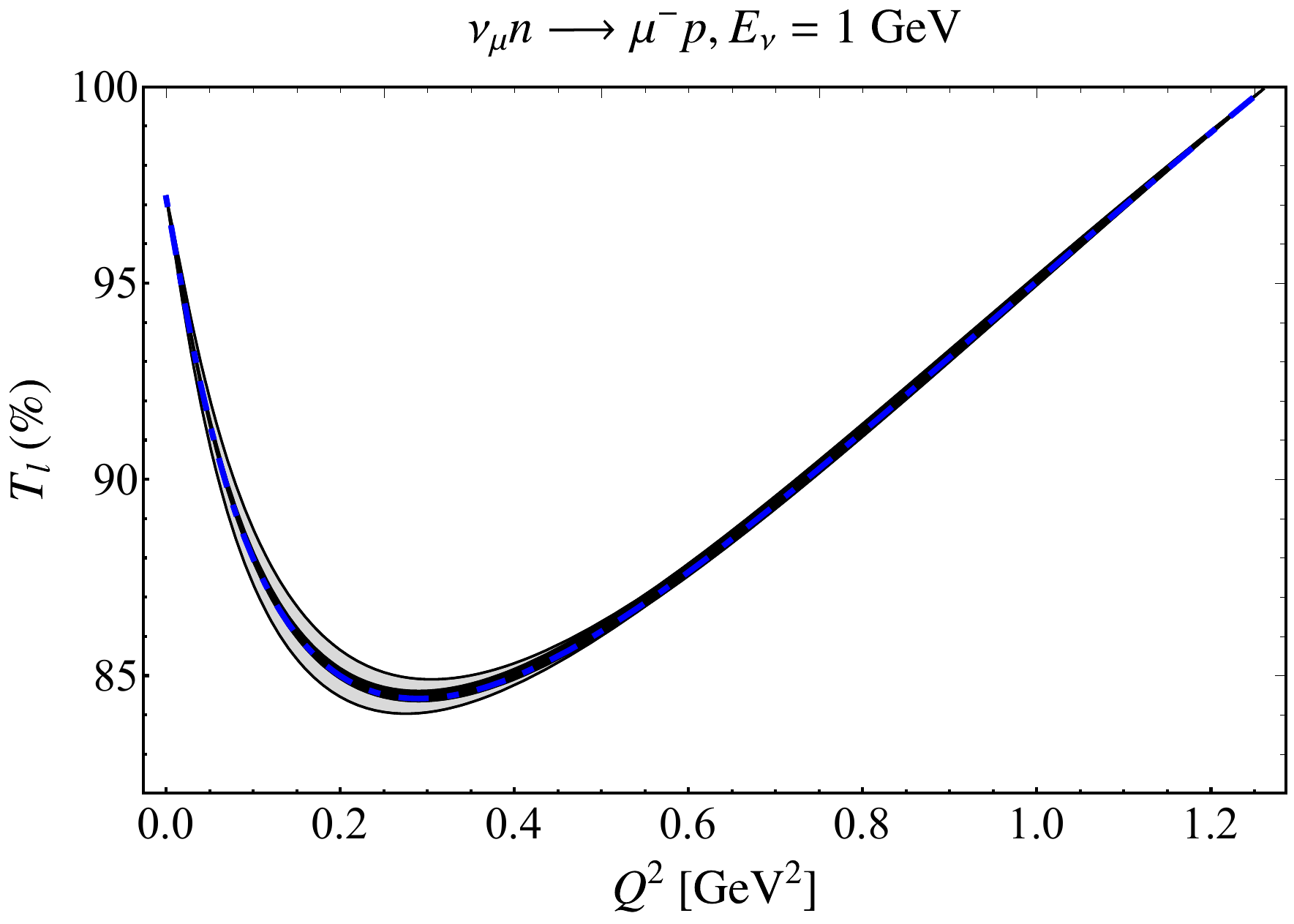}
\includegraphics[width=0.4\textwidth]{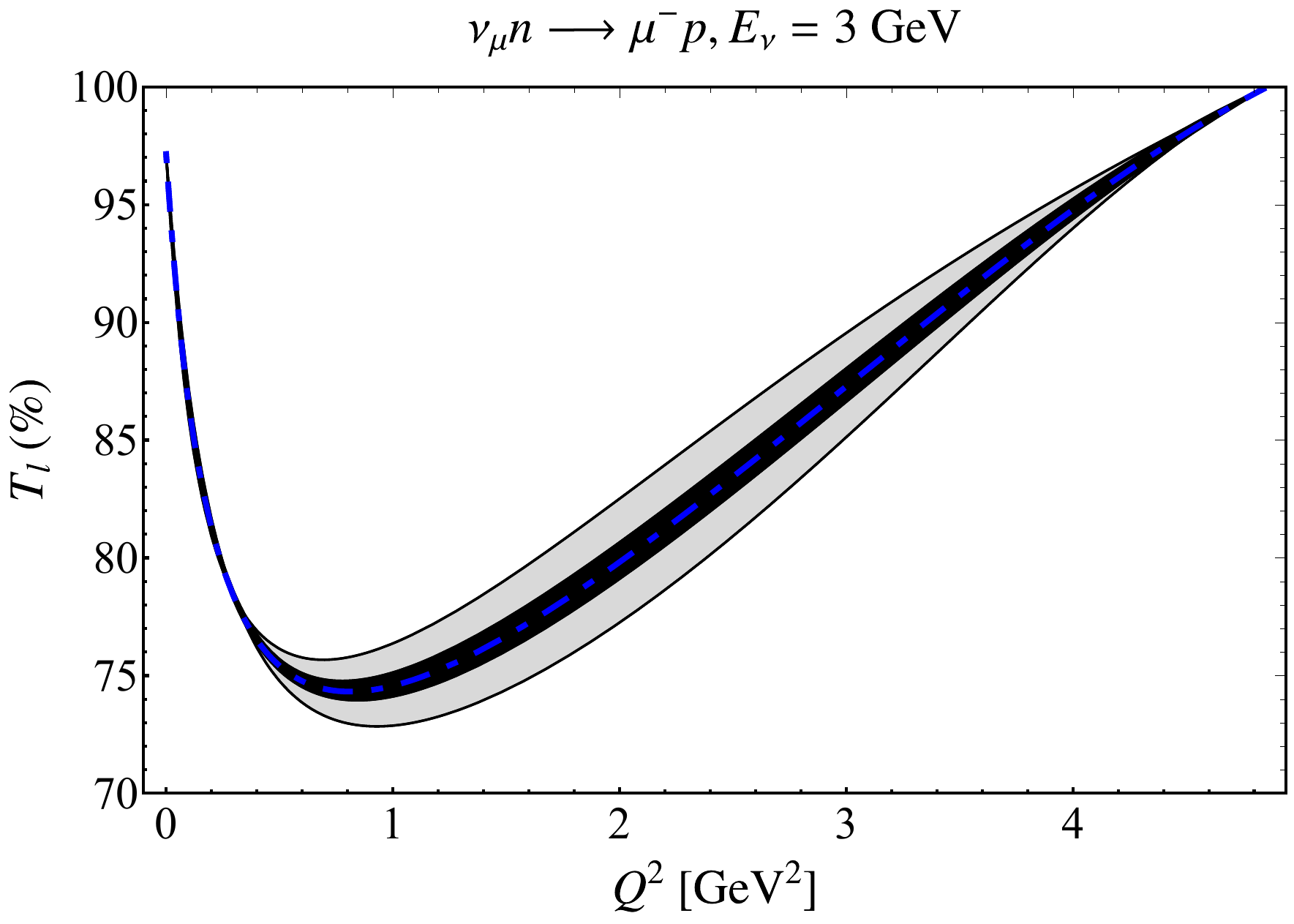}
\caption{Same as Fig.~\ref{fig:nu_Tt_radcorr} but for the longitudinal polarization observable $T_l$. \label{fig:nu_Tl_radcorr}}
\end{figure}

\begin{figure}[H]
\centering
\includegraphics[width=0.4\textwidth]{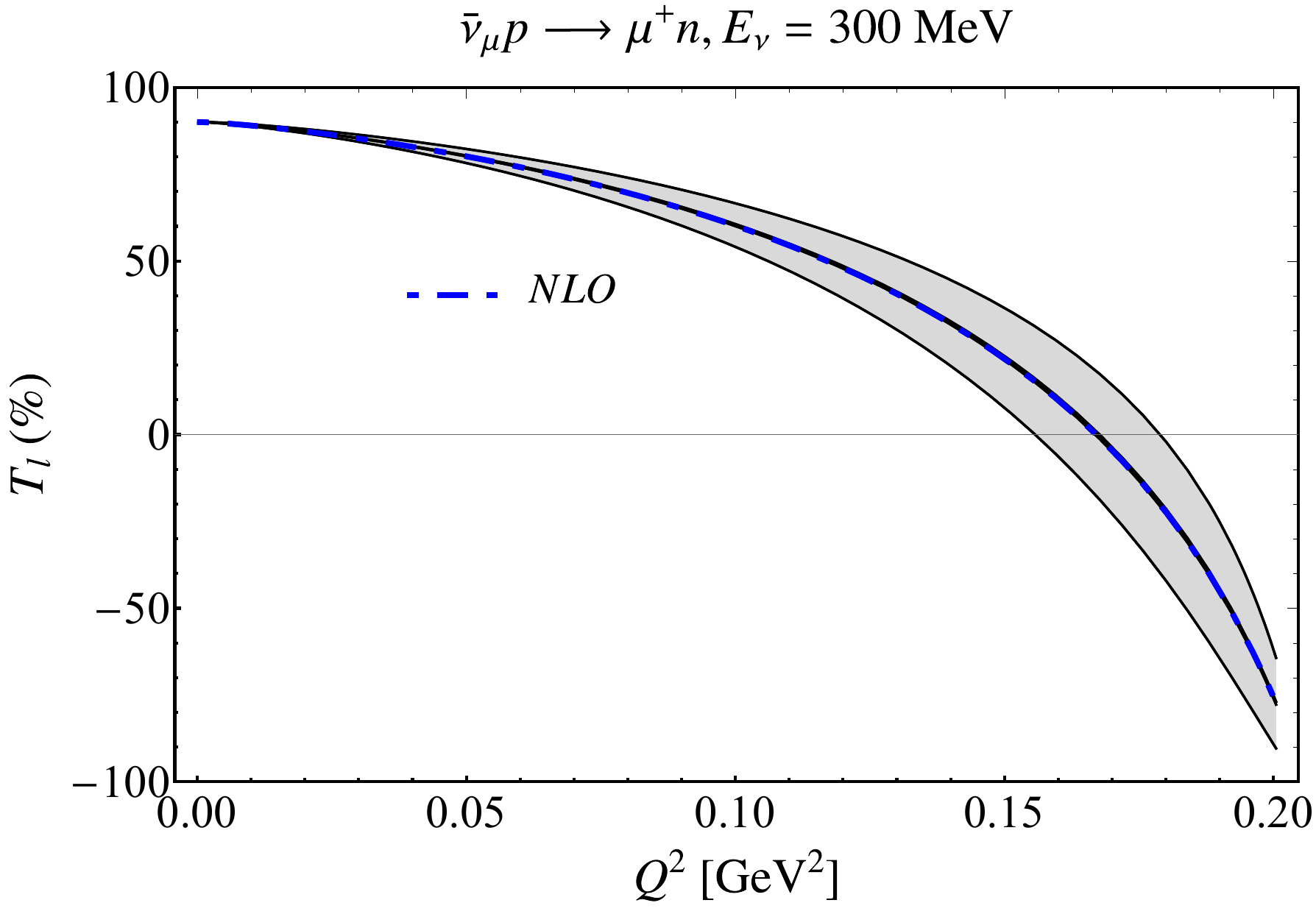}
\includegraphics[width=0.4\textwidth]{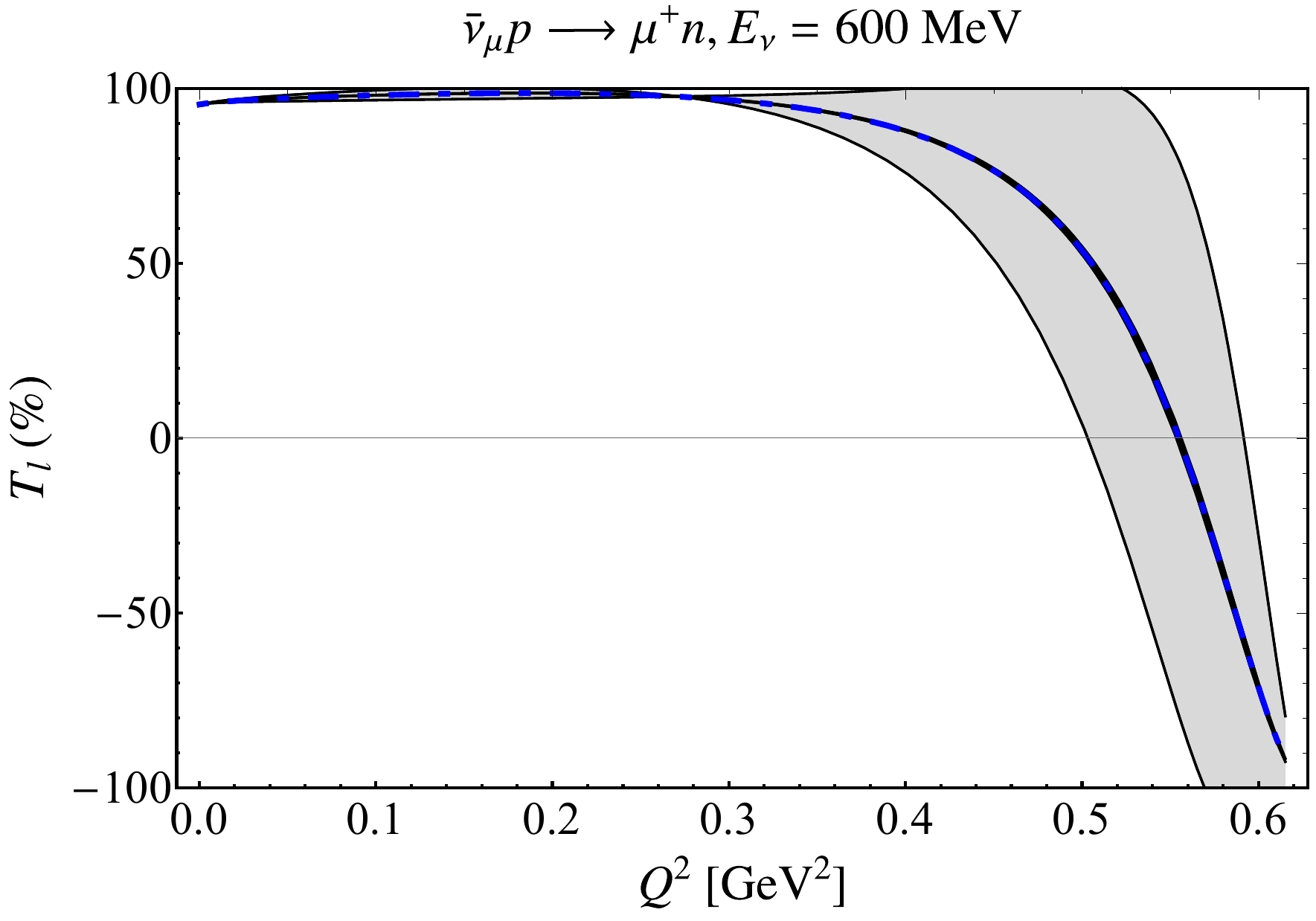}
\includegraphics[width=0.4\textwidth]{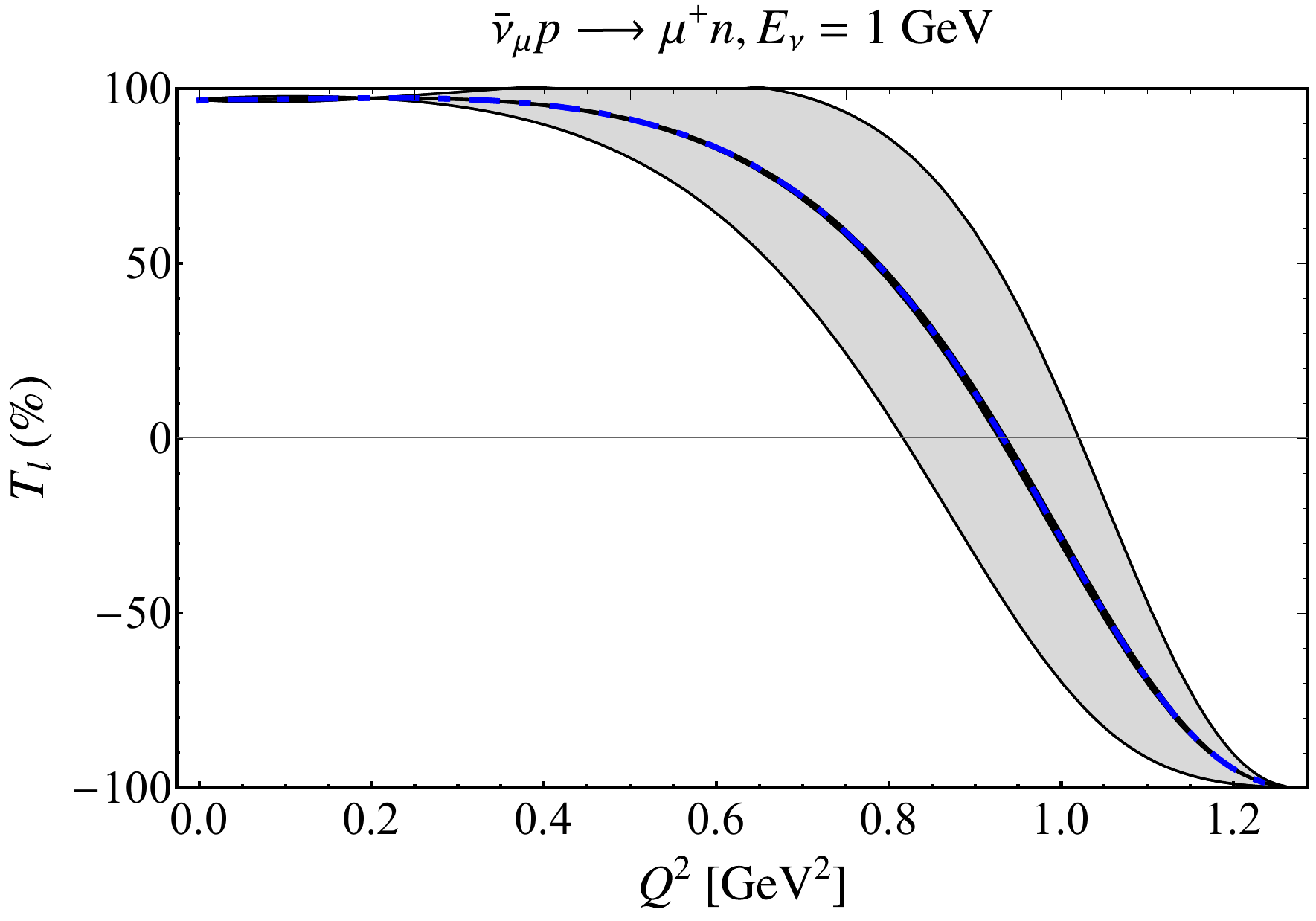}
\includegraphics[width=0.4\textwidth]{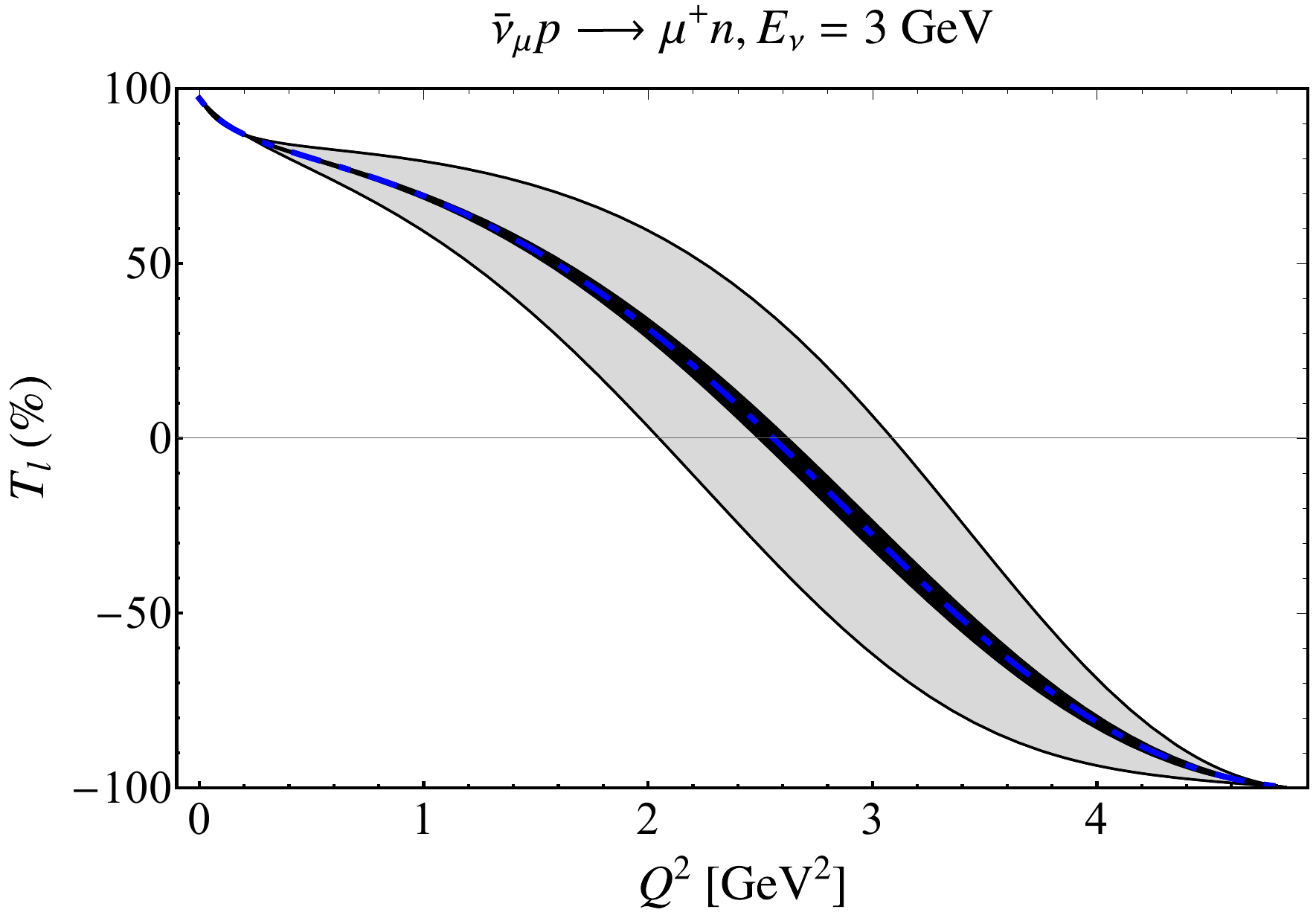}
\caption{Same as Fig.~\ref{fig:antinu_Tt_radcorr} but for the longitudinal polarization observable $T_l$. \label{fig:antinu_Tl_radcorr}}
\end{figure}

\begin{figure}[H]
\centering
\includegraphics[width=0.4\textwidth]{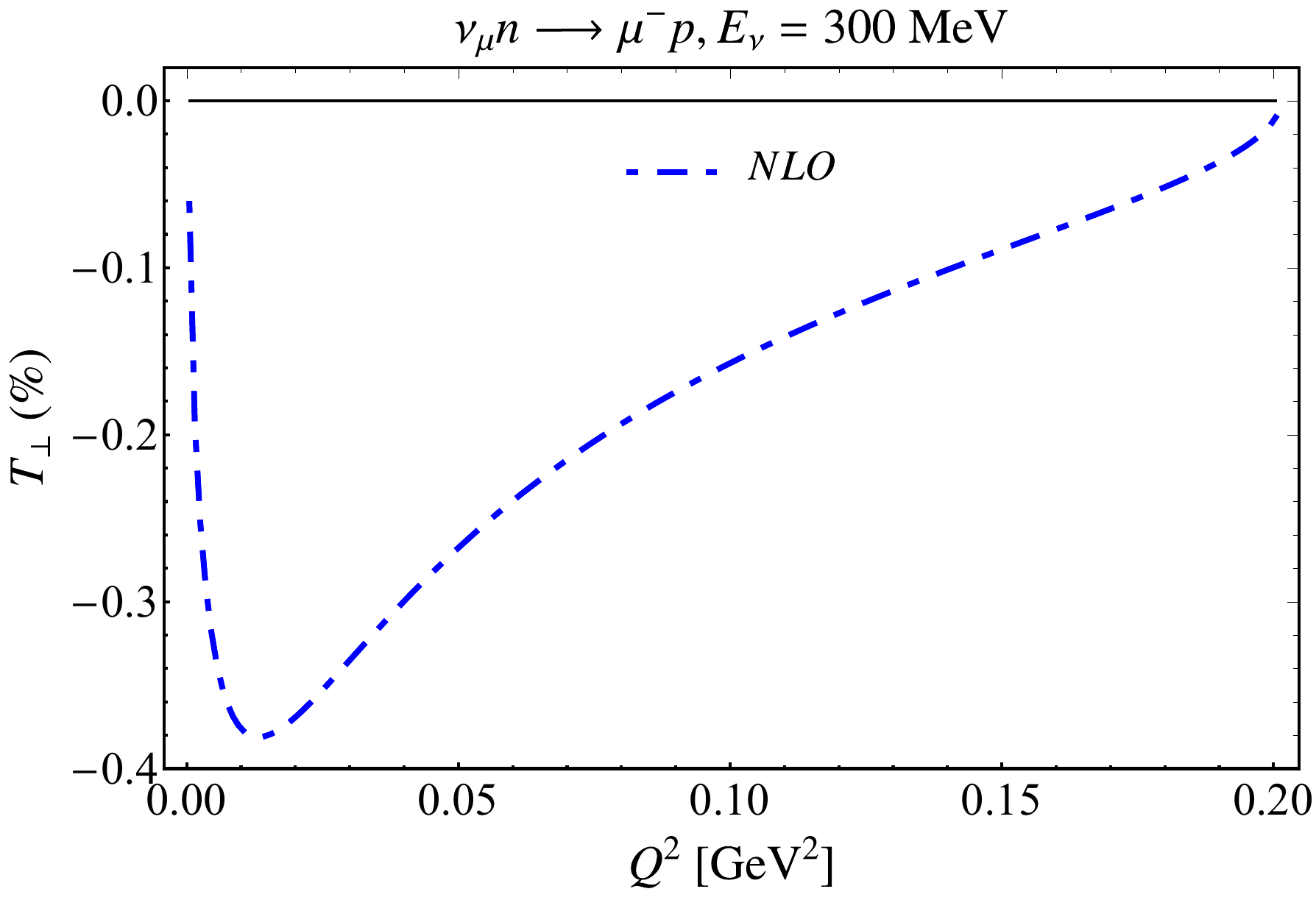}
\includegraphics[width=0.4\textwidth]{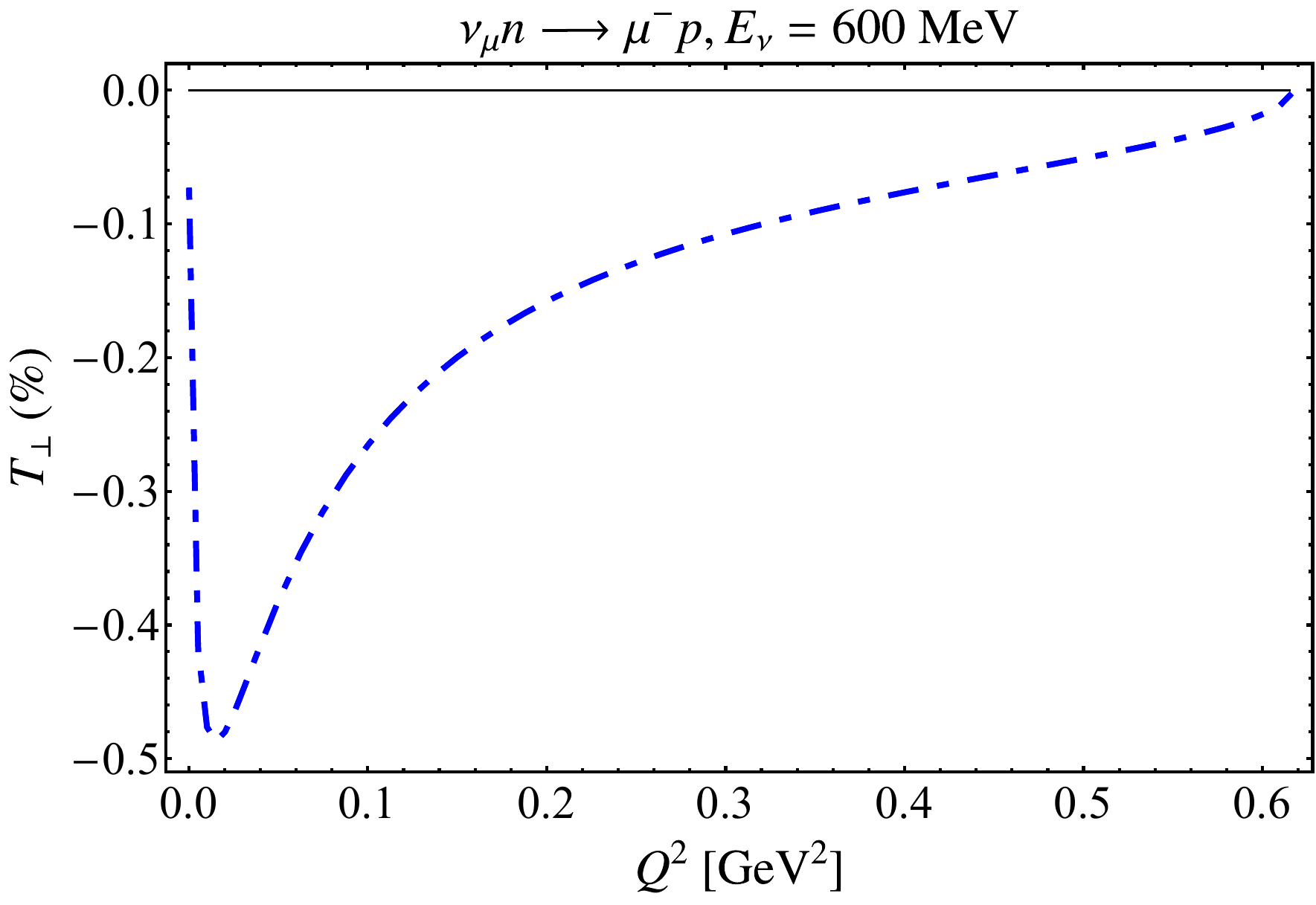}
\includegraphics[width=0.4\textwidth]{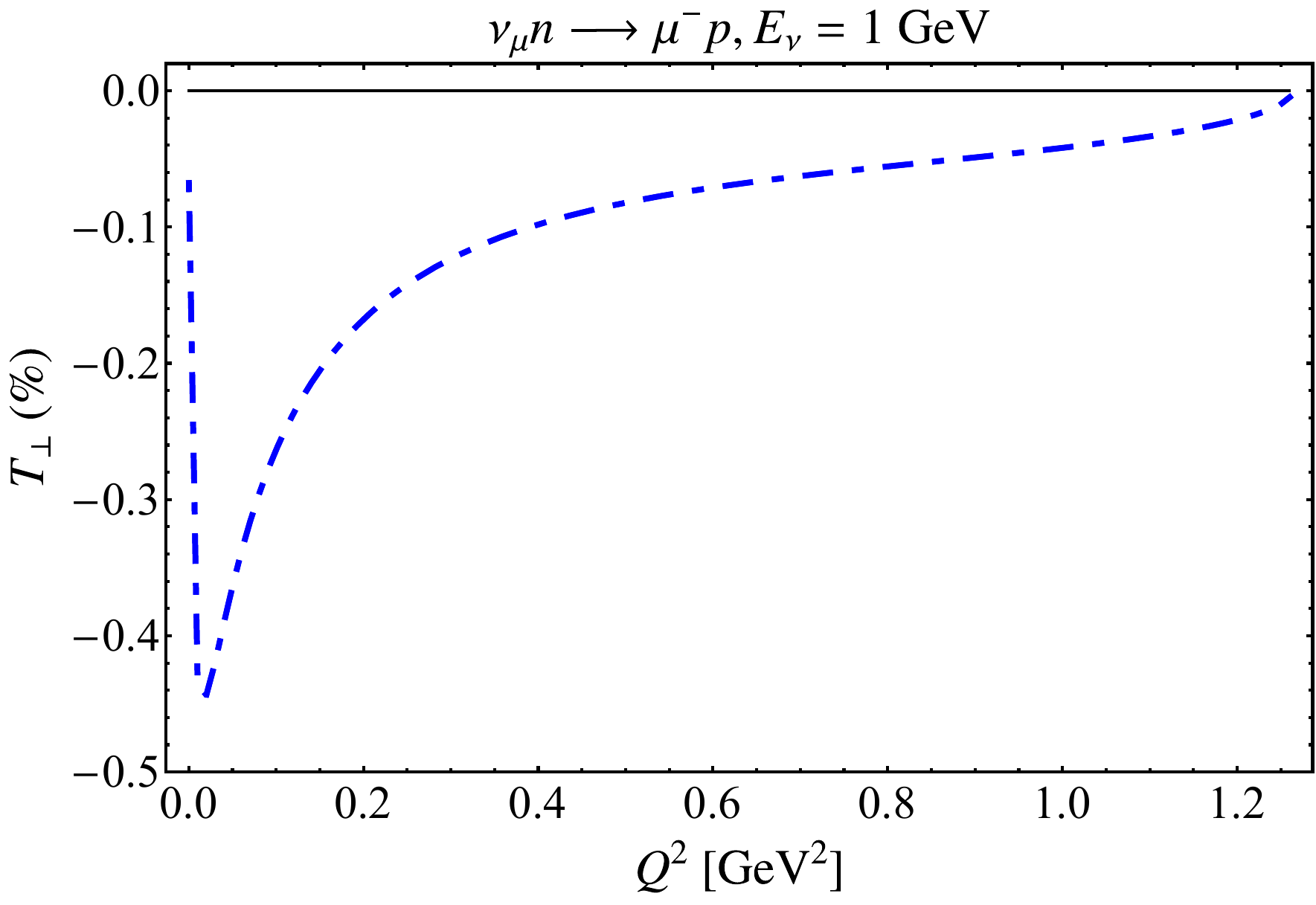}
\includegraphics[width=0.4\textwidth]{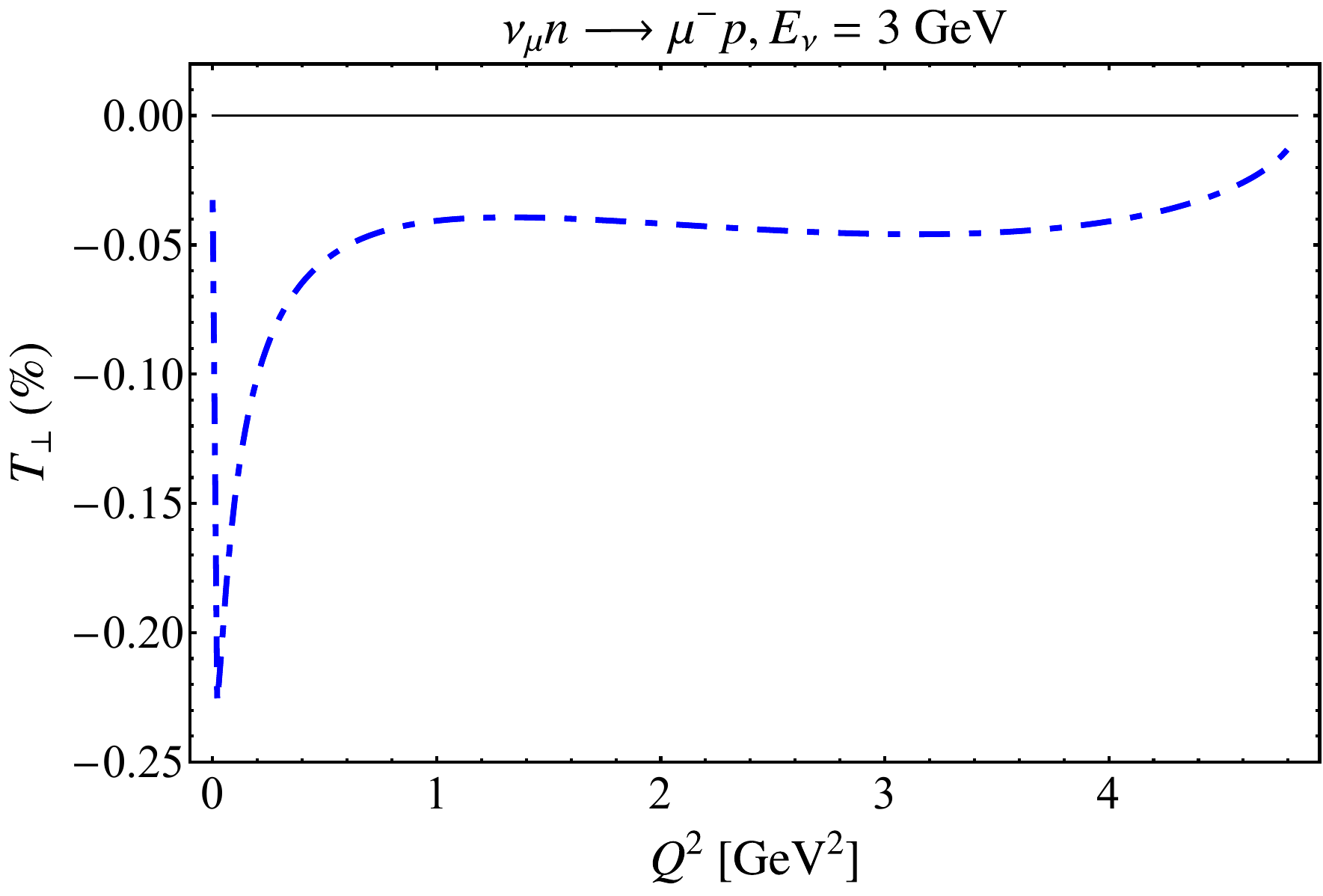}
\caption{Same as Fig.~\ref{fig:nu_Tt_radcorr} but for the transverse polarization observable $T_\perp$. \label{fig:nu_TTT_radcorr}}
\end{figure}

\begin{figure}[H]
\centering
\includegraphics[width=0.4\textwidth]{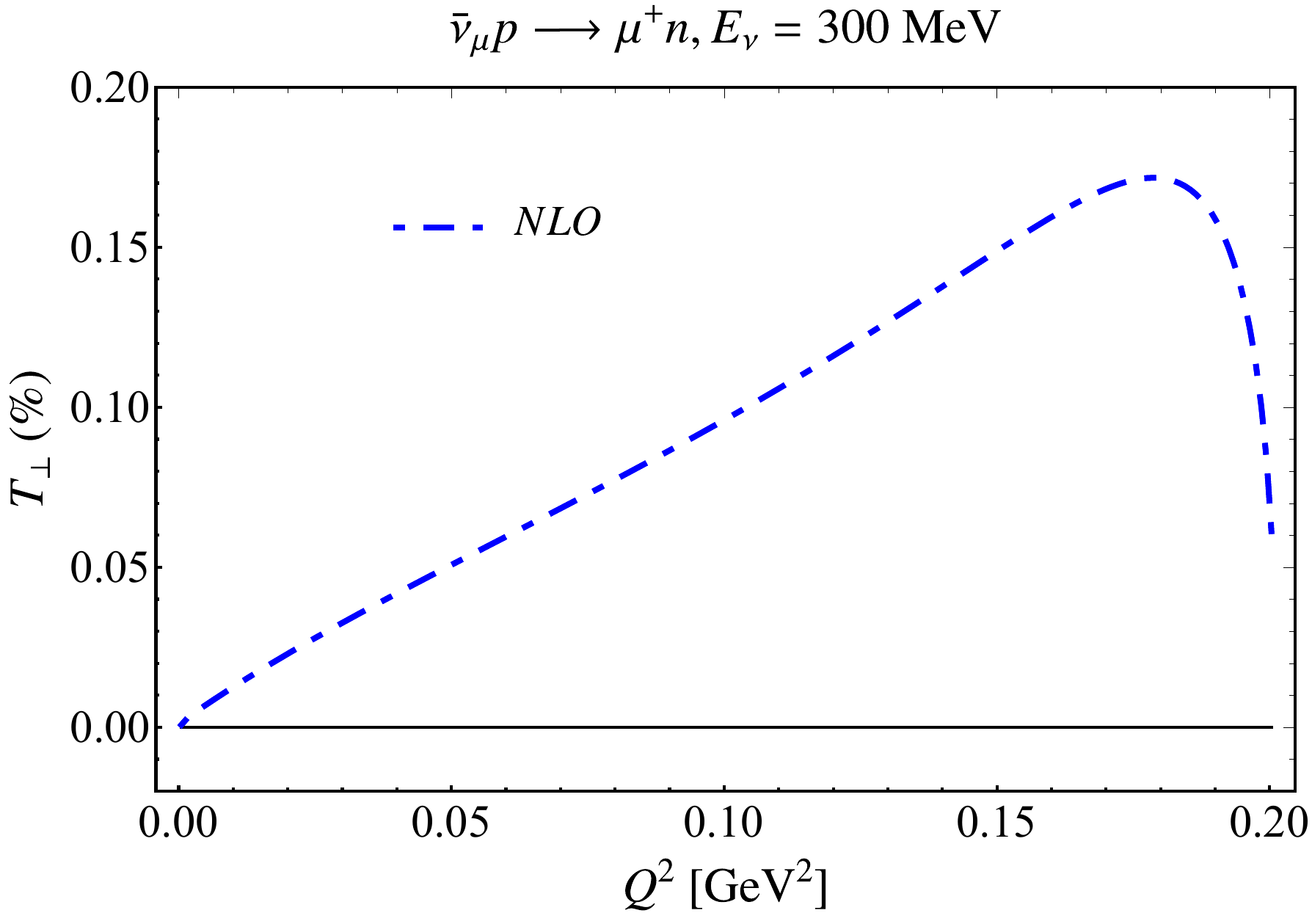}
\includegraphics[width=0.4\textwidth]{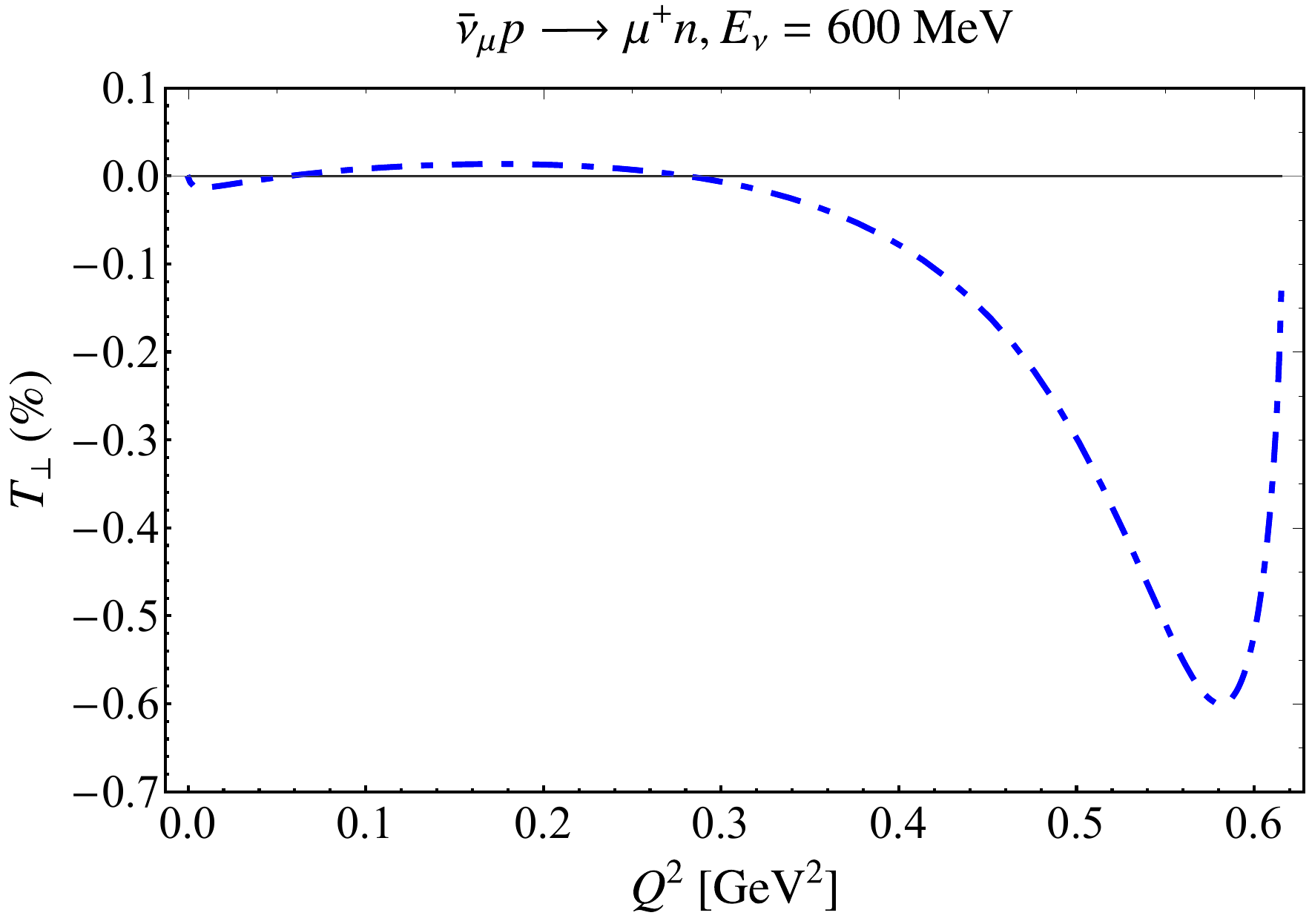}
\includegraphics[width=0.4\textwidth]{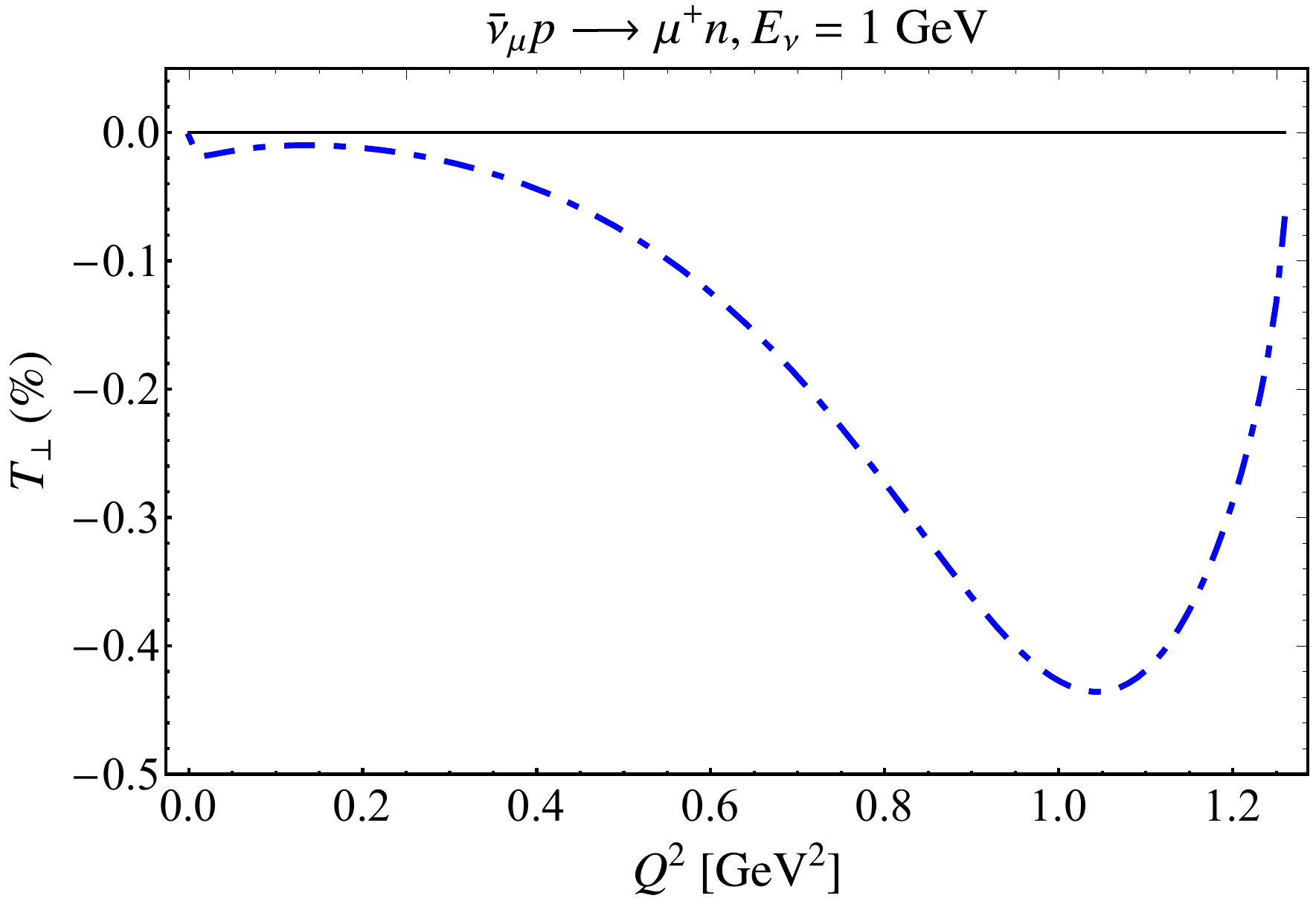}
\includegraphics[width=0.4\textwidth]{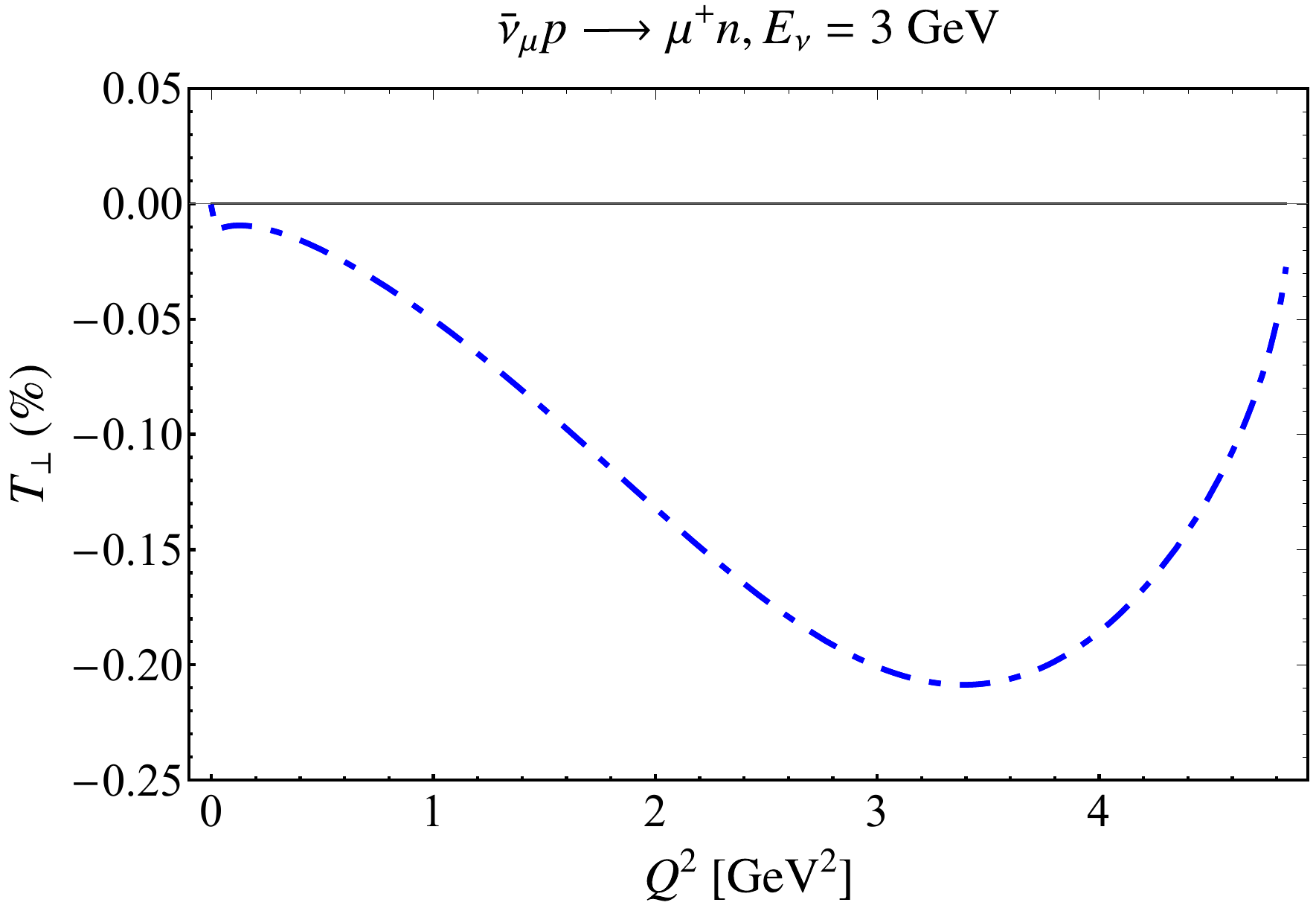}
\caption{Same as Fig.~\ref{fig:antinu_Tt_radcorr} but for the transverse polarization observable $T_\perp$. \label{fig:antinu_TTT_radcorr}}
\end{figure}

\begin{figure}[H]
\centering
\includegraphics[width=0.4\textwidth]{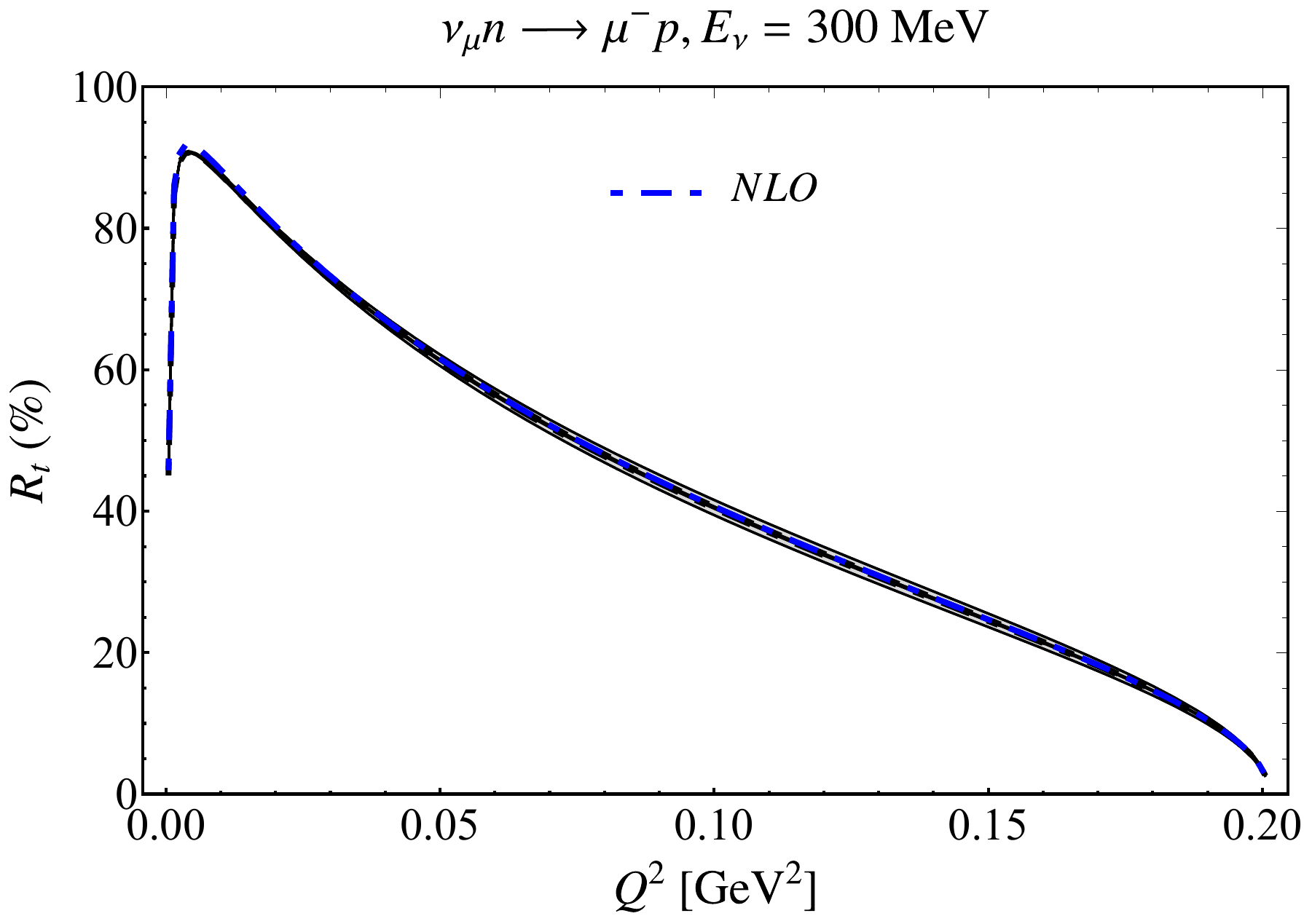}
\includegraphics[width=0.4\textwidth]{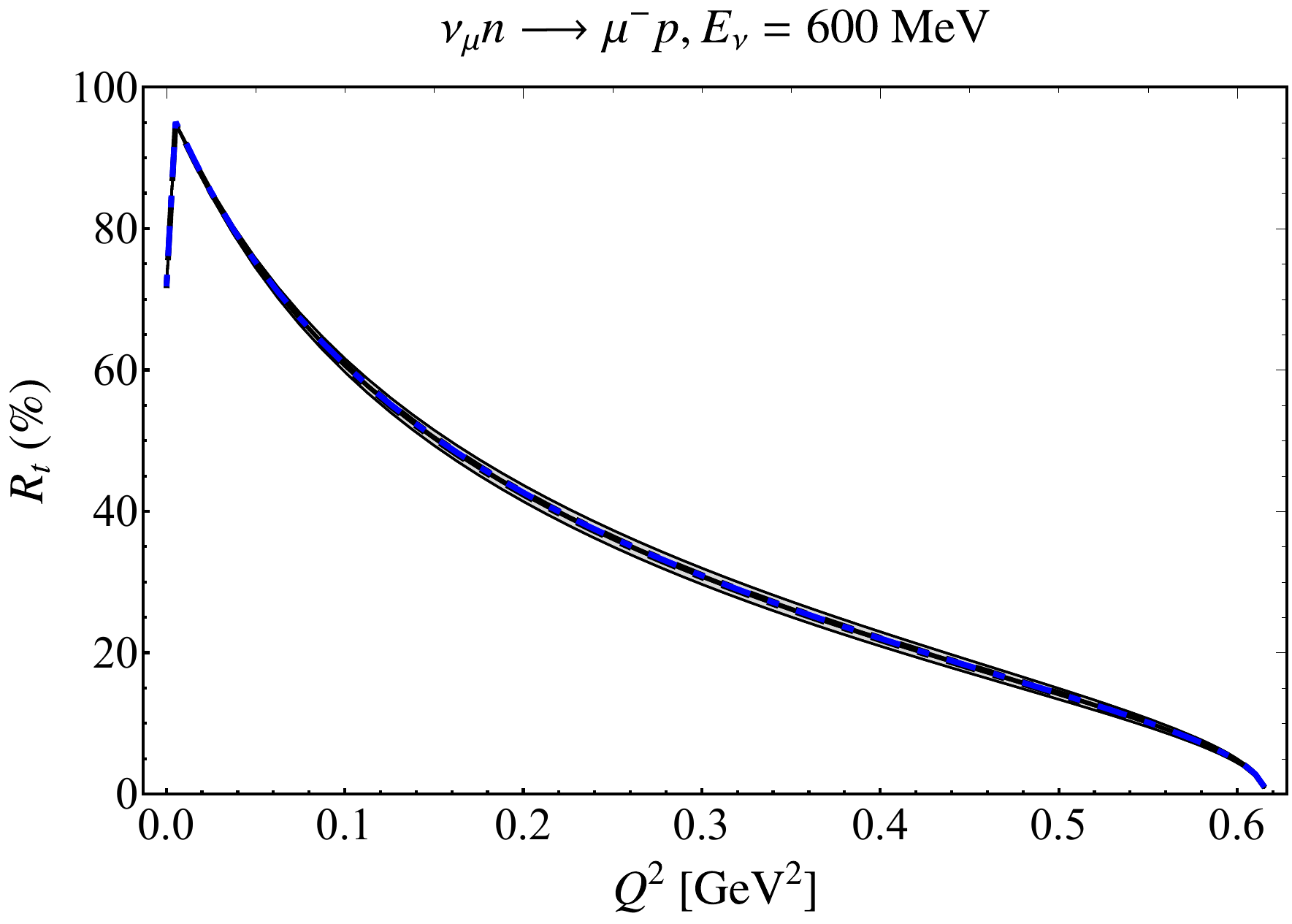}
\includegraphics[width=0.4\textwidth]{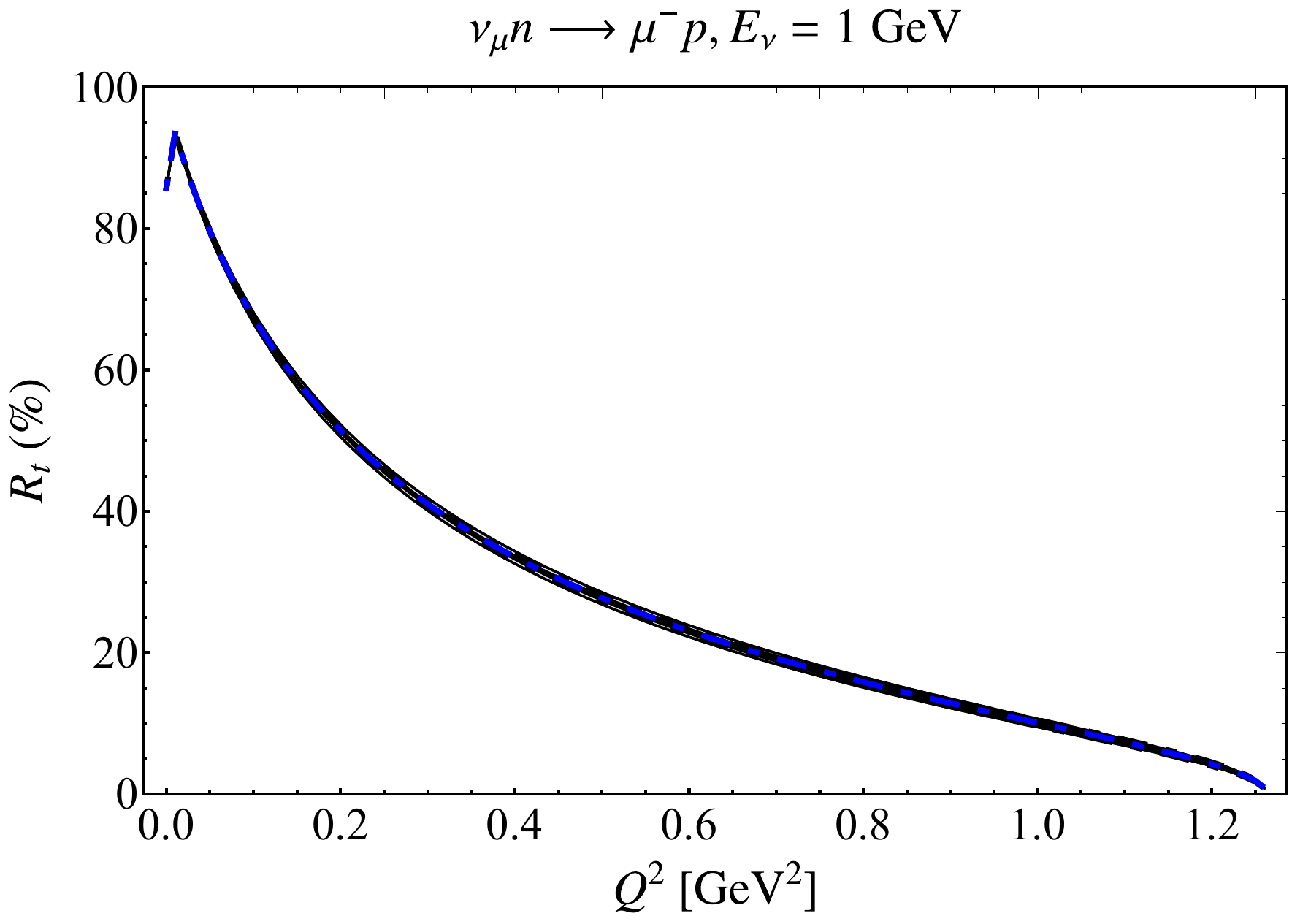}
\includegraphics[width=0.4\textwidth]{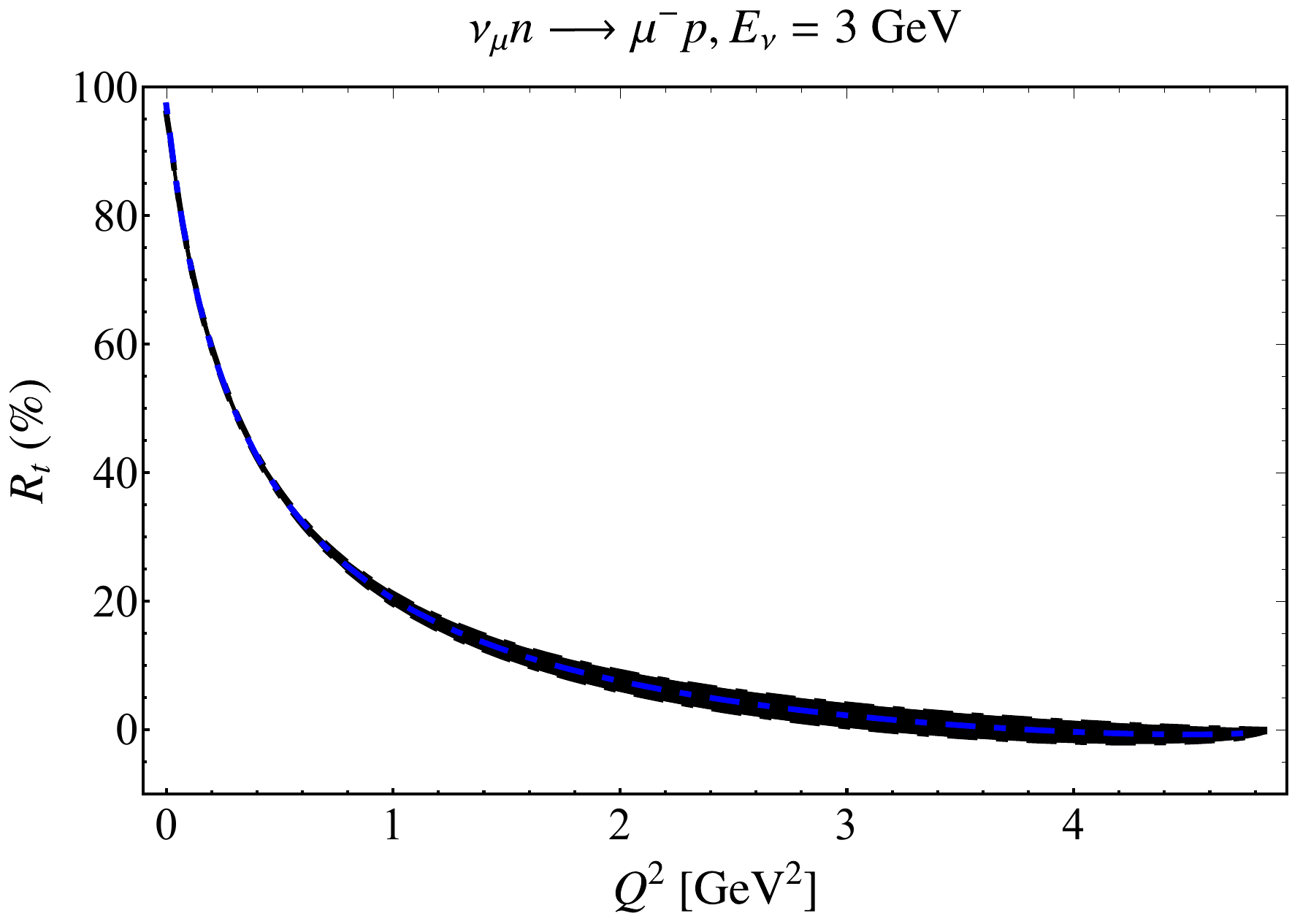}
\caption{Same as Fig.~\ref{fig:nu_Tt_radcorr} but for the transverse polarization observable $R_t$. \label{fig:nu_Rt_radcorr}}
\end{figure}

\begin{figure}[H]
\centering
\includegraphics[width=0.4\textwidth]{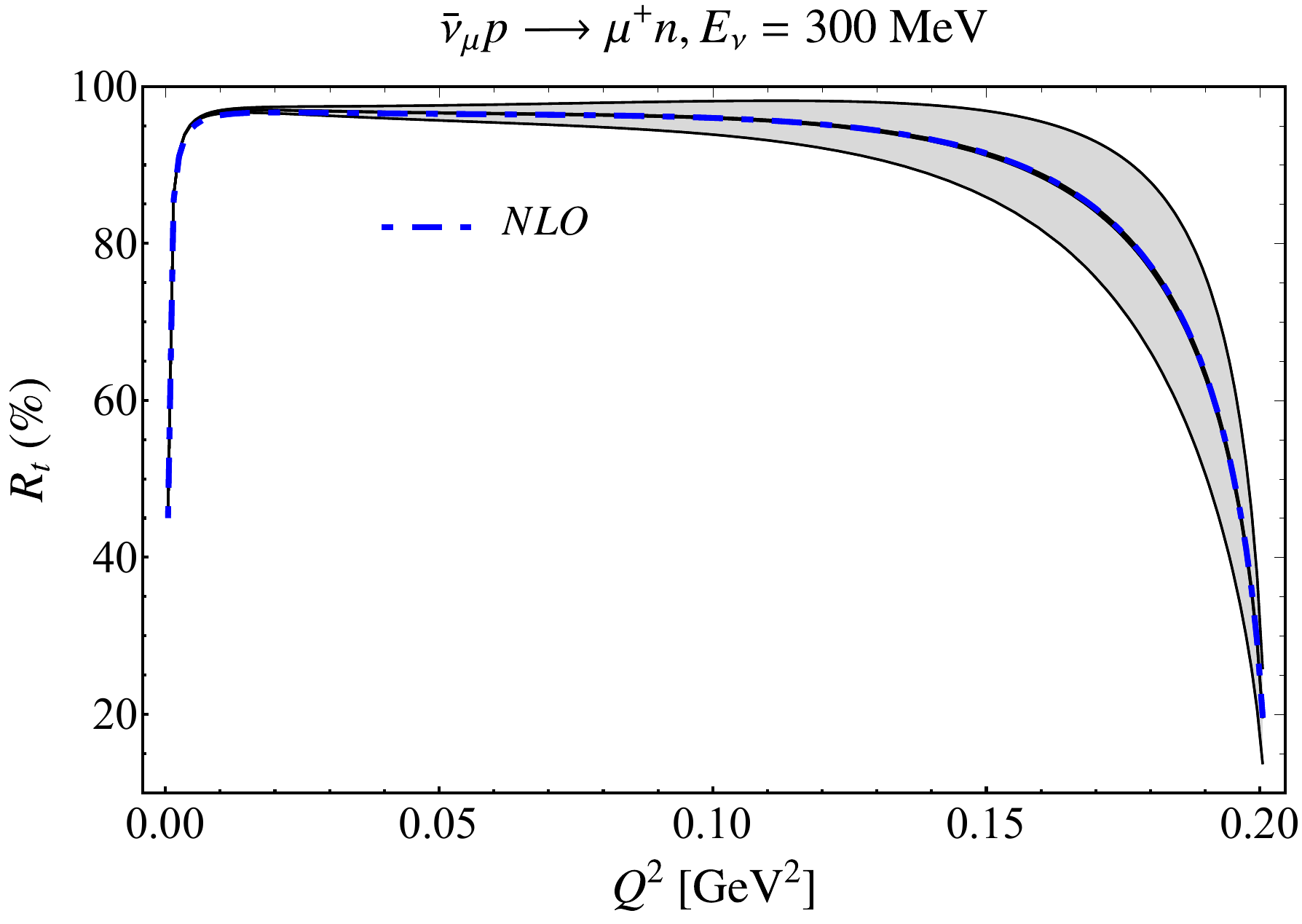}
\includegraphics[width=0.4\textwidth]{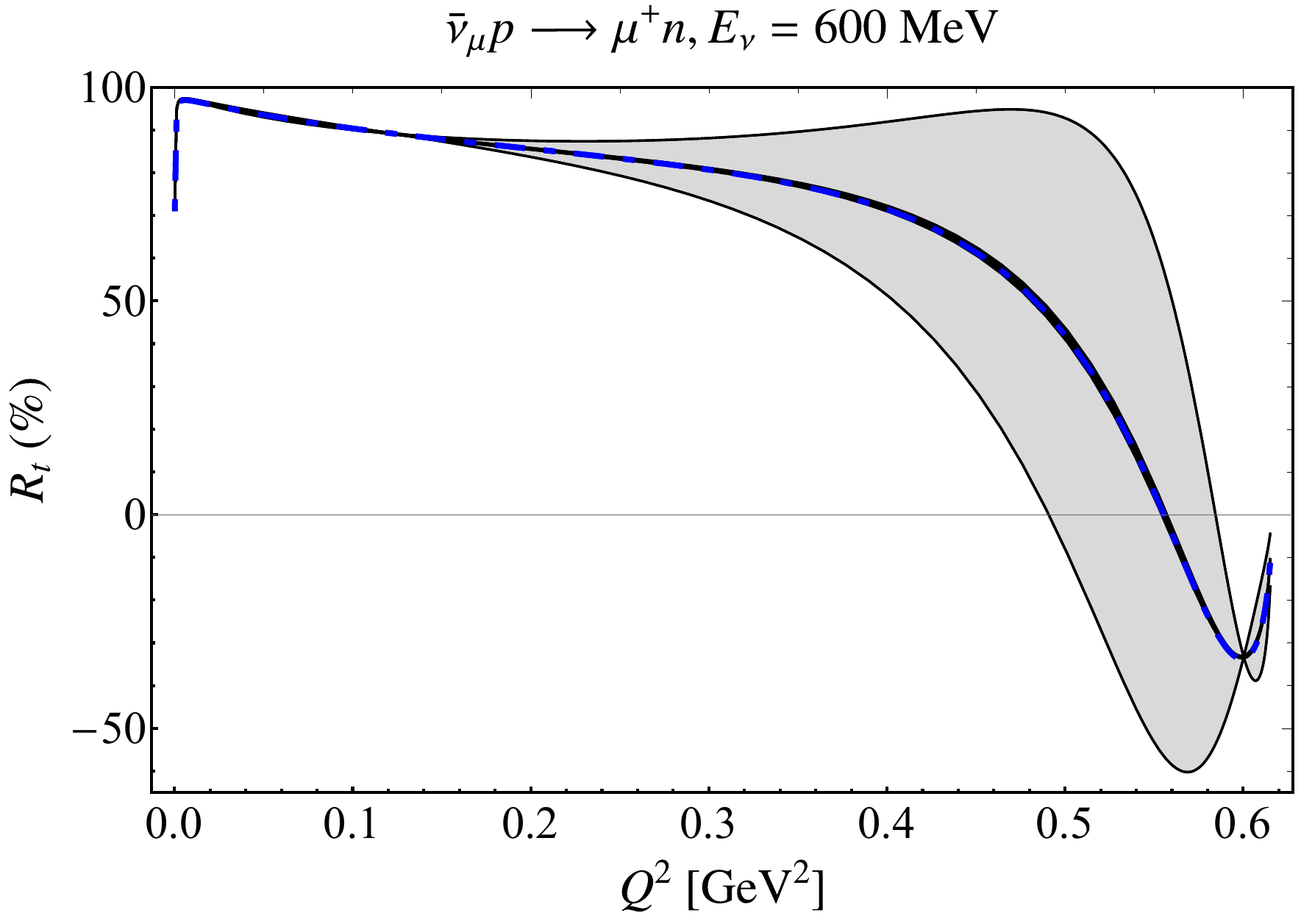}
\includegraphics[width=0.4\textwidth]{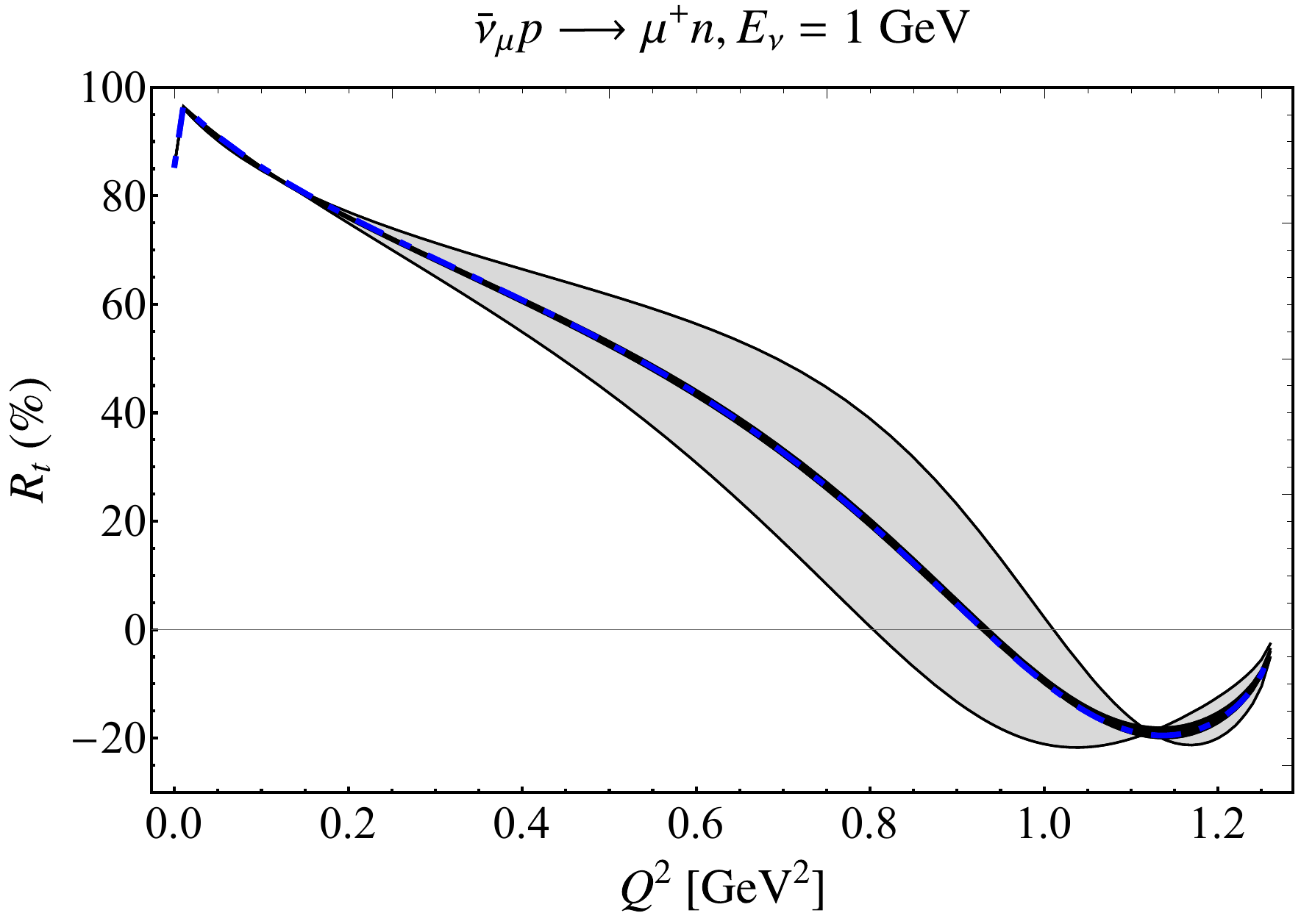}
\includegraphics[width=0.4\textwidth]{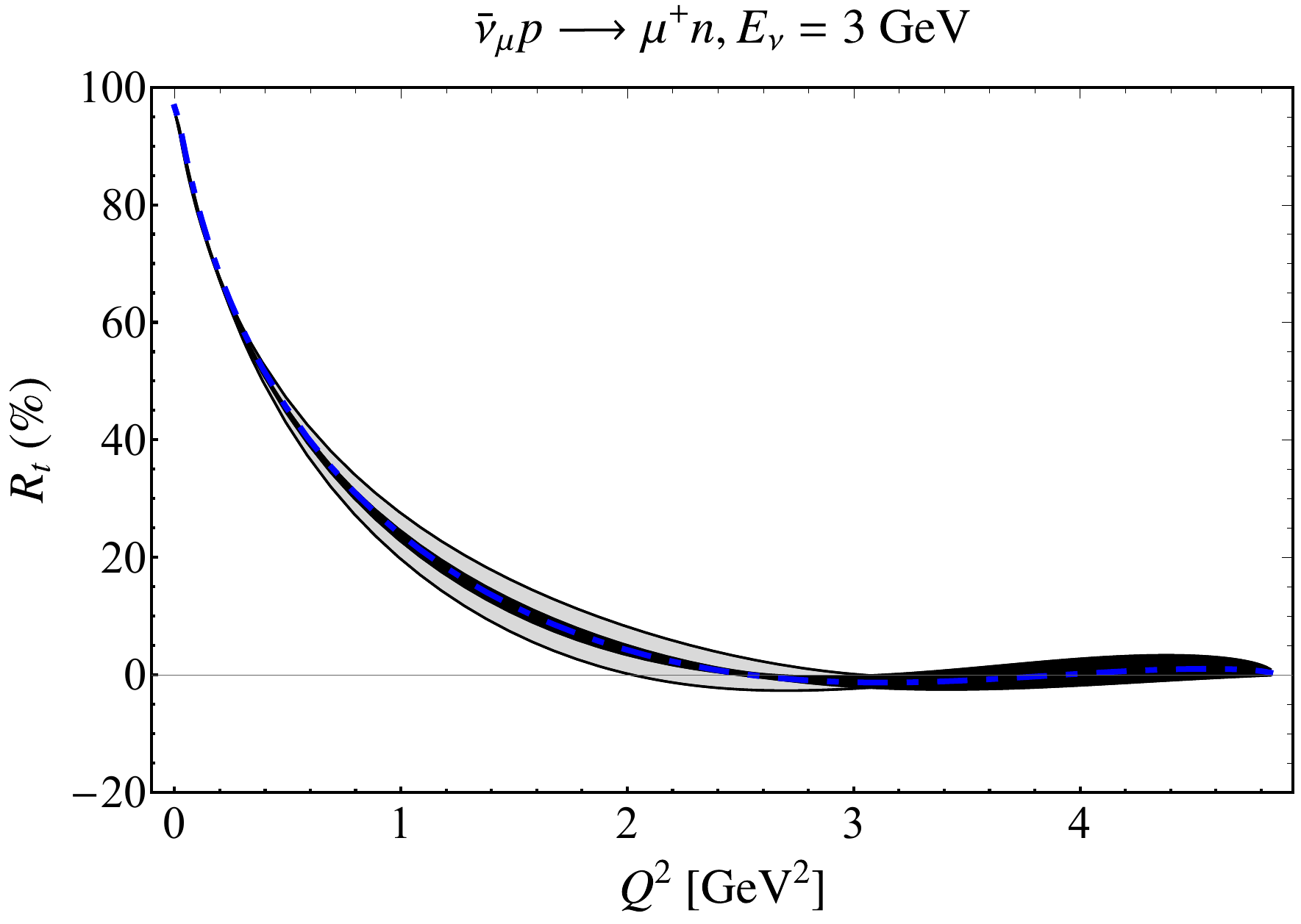}
\caption{Same as Fig.~\ref{fig:antinu_Tt_radcorr} but for the transverse polarization observable $R_t$. \label{fig:antinu_Rt_radcorr}}
\end{figure}

\begin{figure}[H]
\centering
\includegraphics[width=0.4\textwidth]{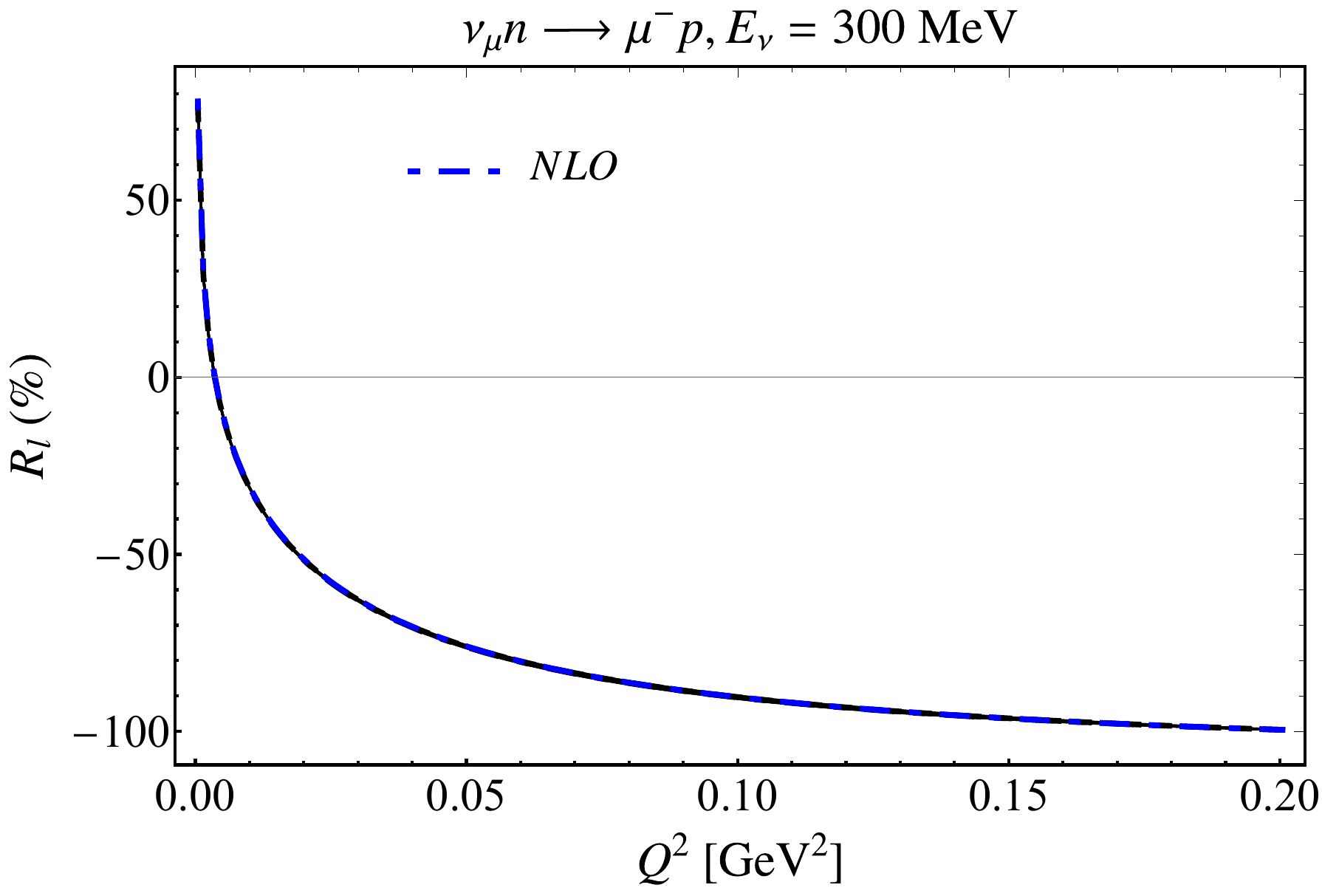}
\includegraphics[width=0.4\textwidth]{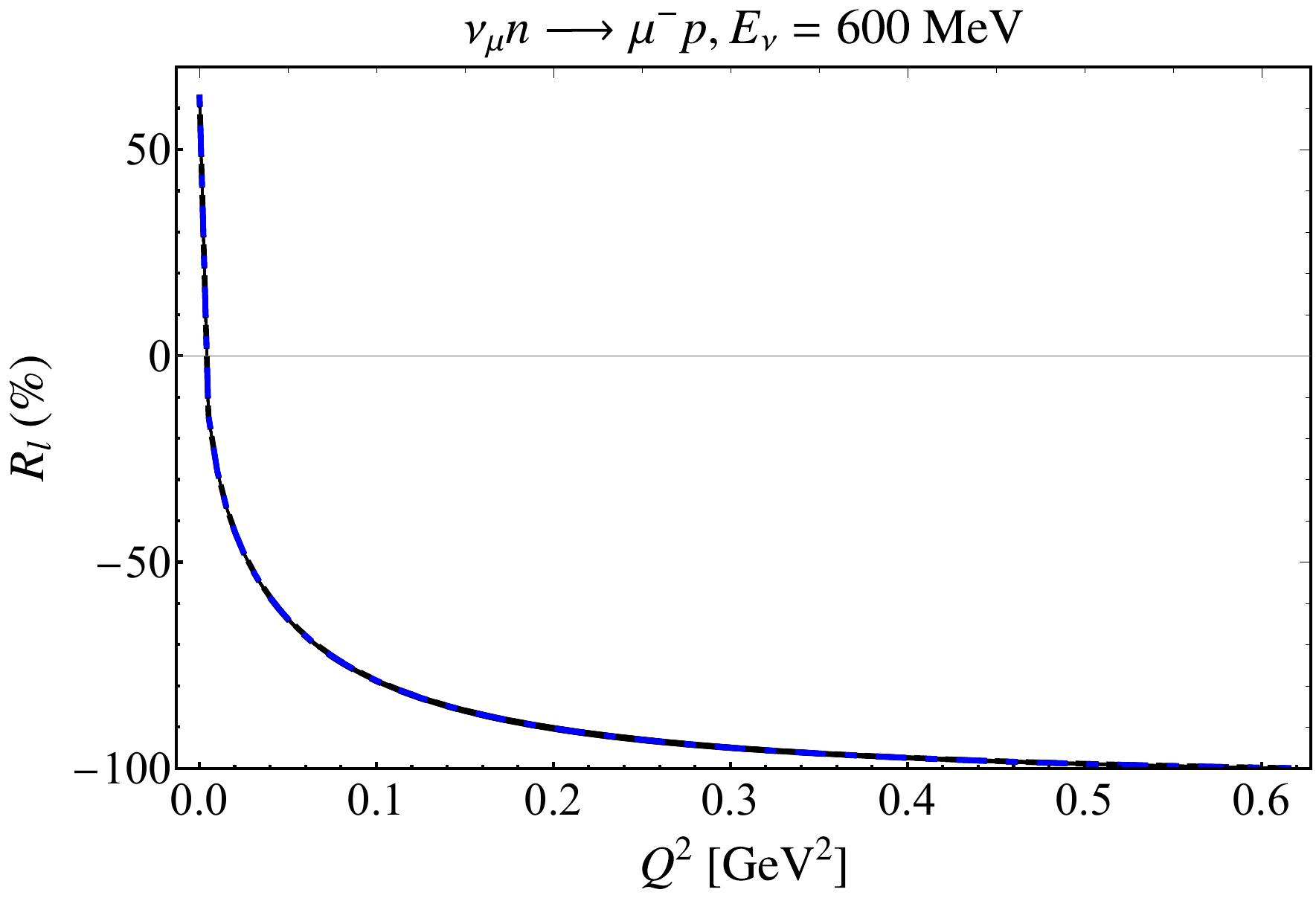}
\includegraphics[width=0.4\textwidth]{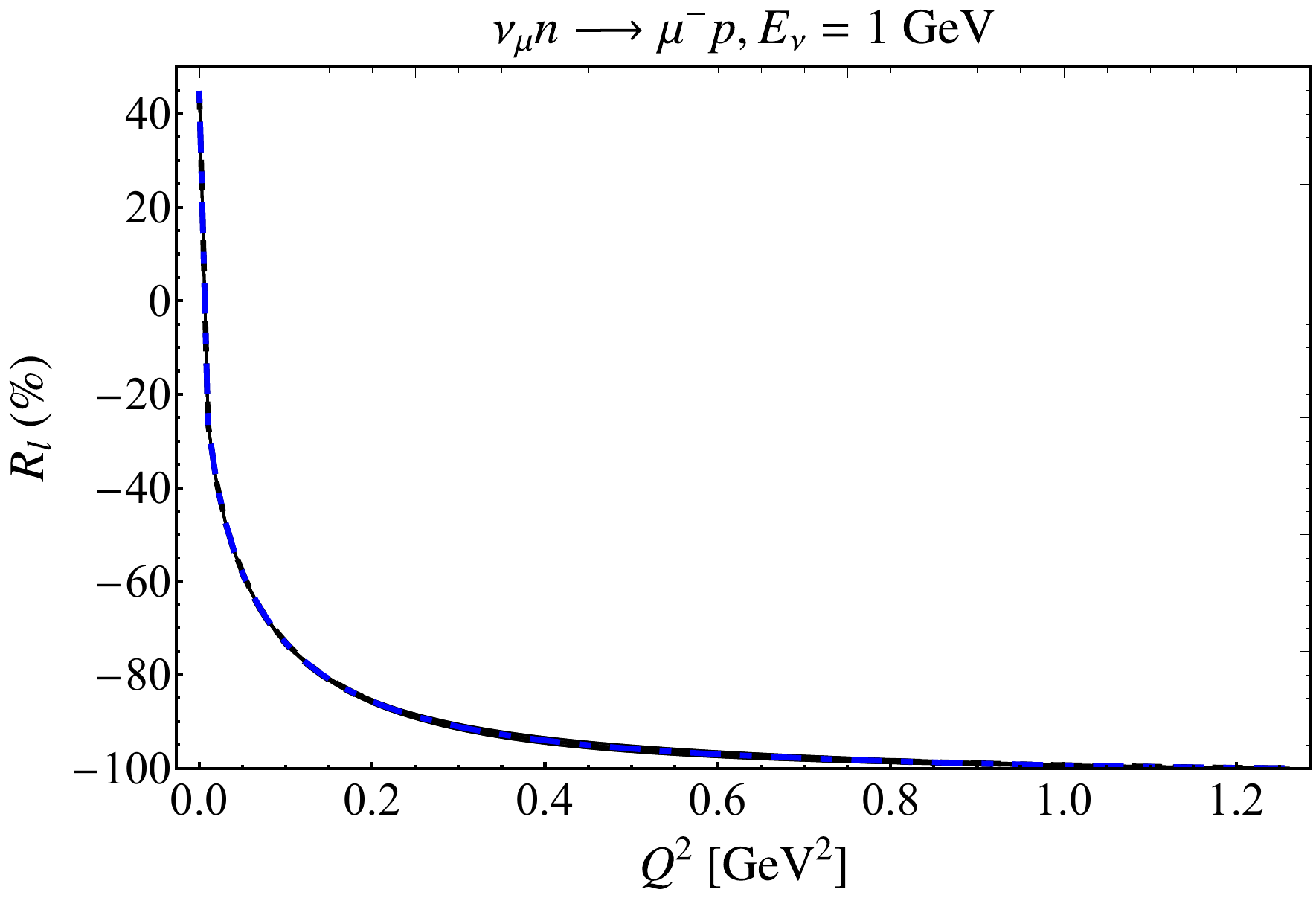}
\includegraphics[width=0.4\textwidth]{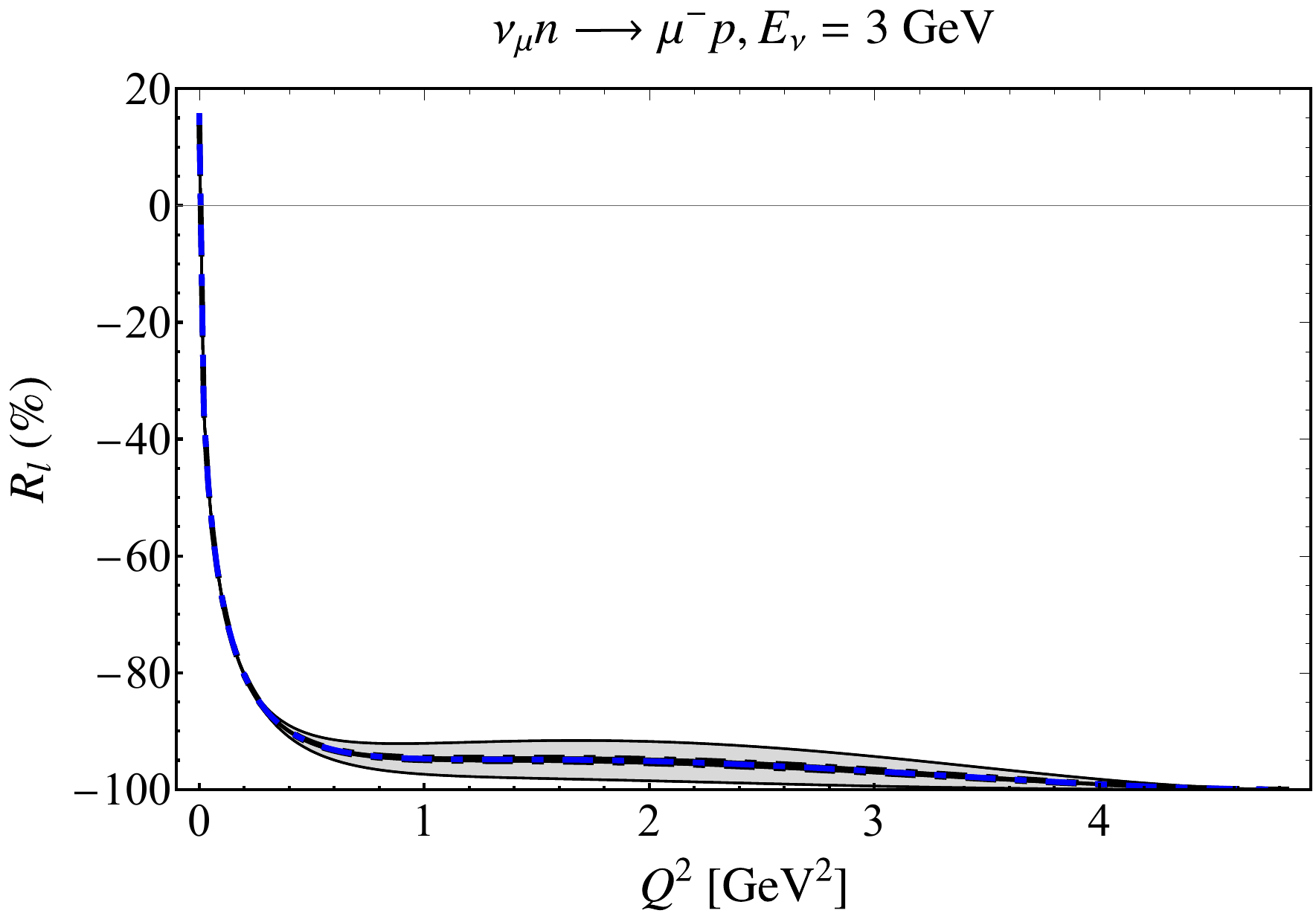}
\caption{Same as Fig.~\ref{fig:nu_Tt_radcorr} but for the longitudinal polarization observable $R_l$. \label{fig:nu_Rl_radcorr}}
\end{figure}

\begin{figure}[H]
\centering
\includegraphics[width=0.4\textwidth]{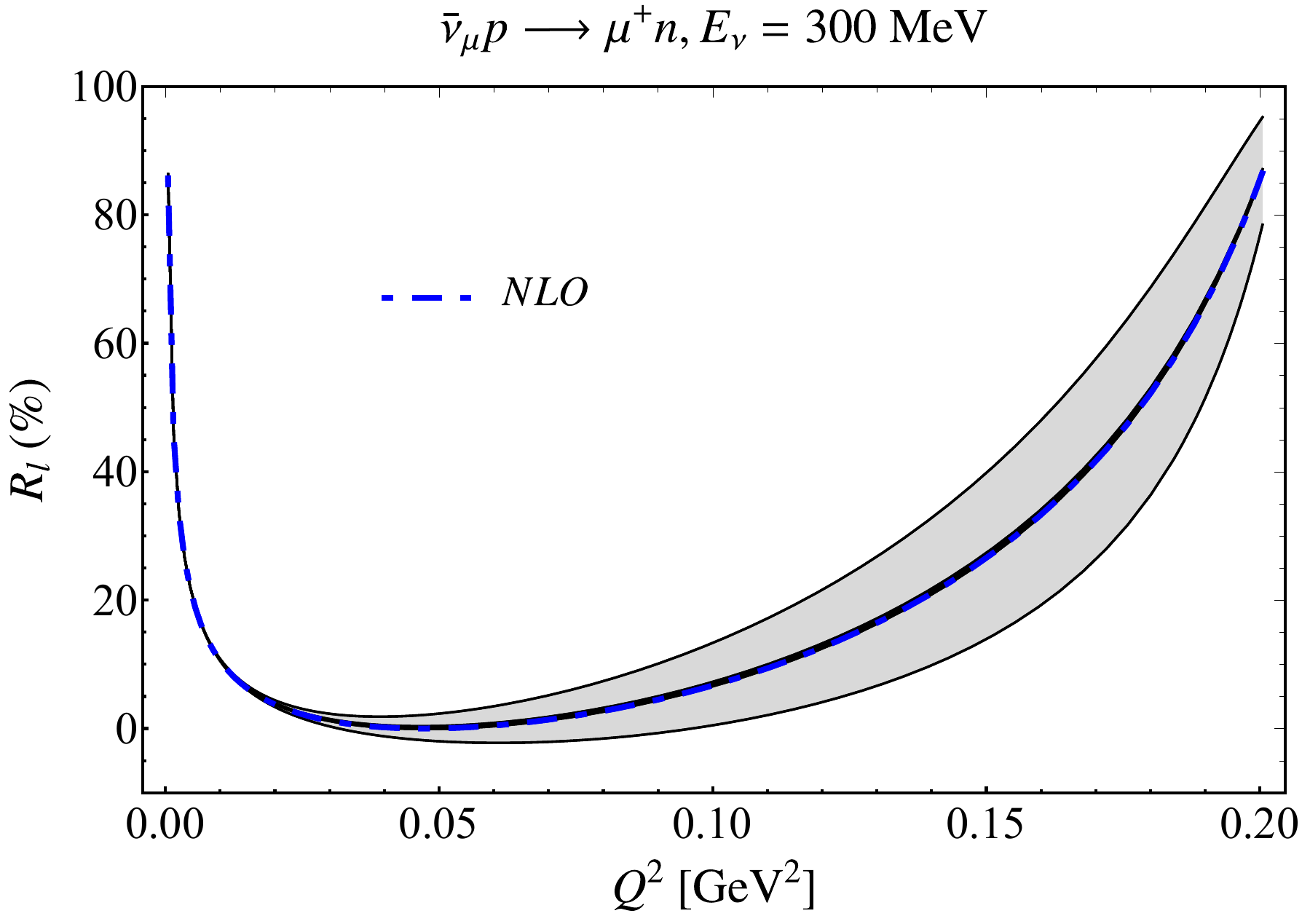}
\includegraphics[width=0.4\textwidth]{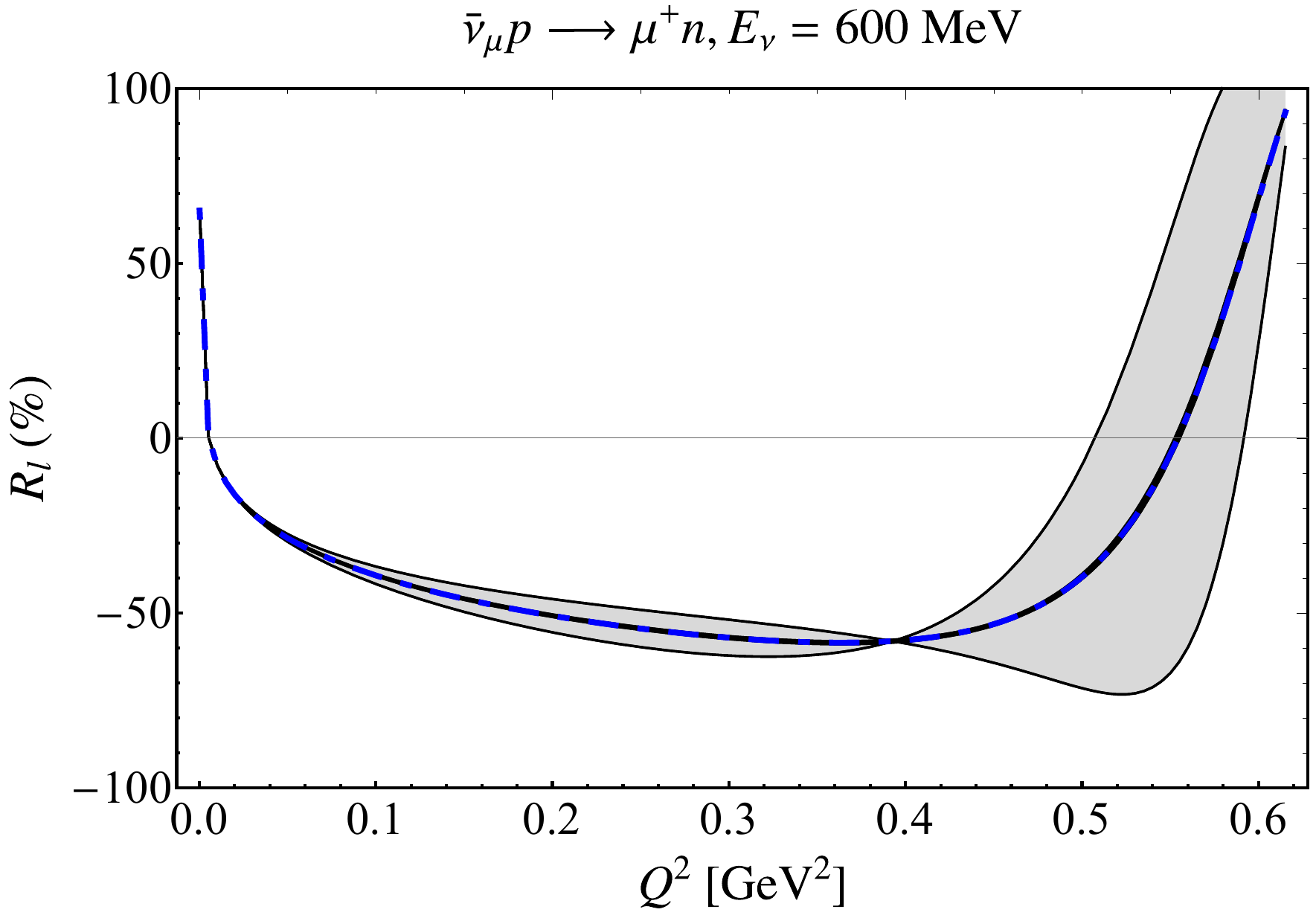}
\includegraphics[width=0.4\textwidth]{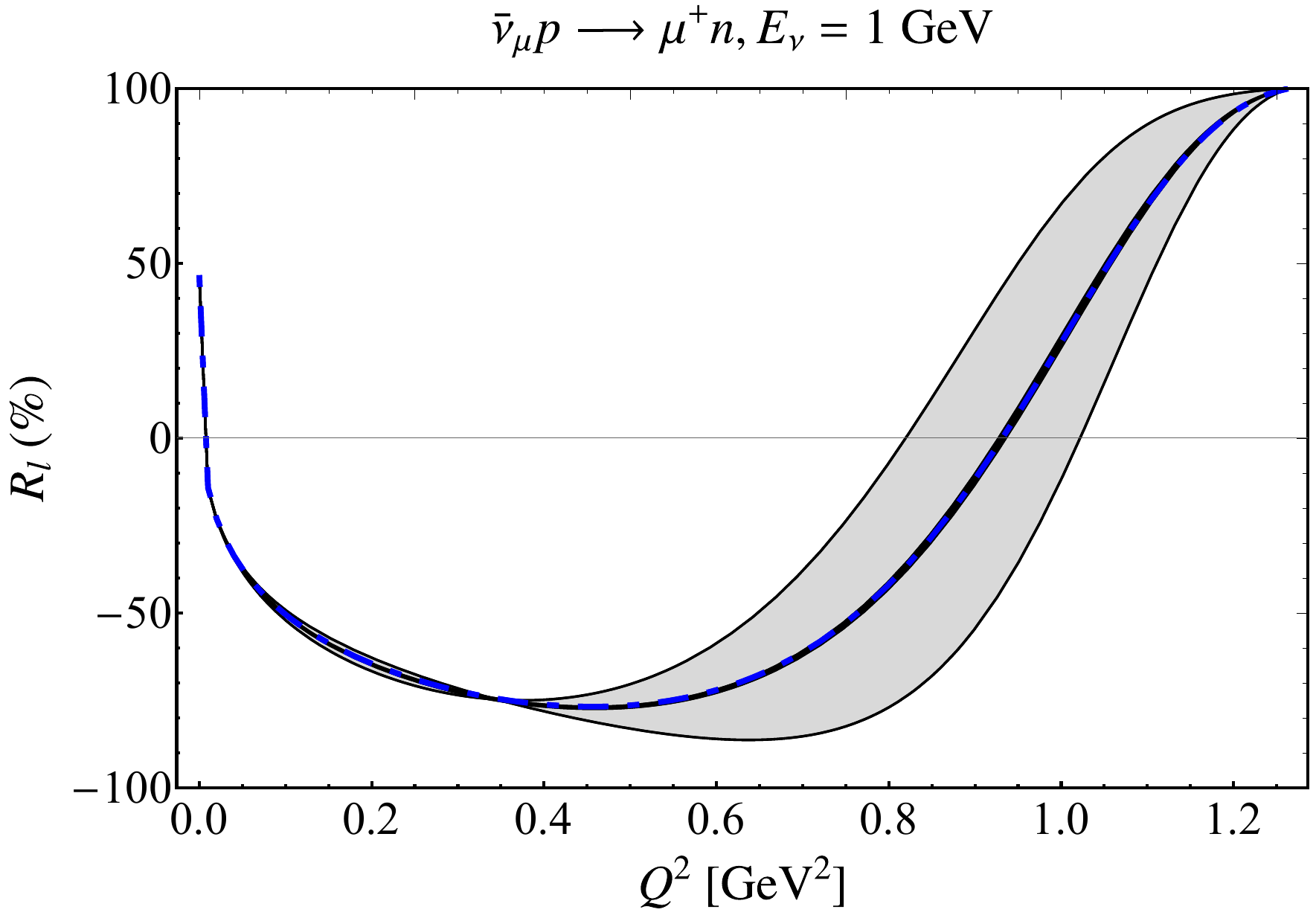}
\includegraphics[width=0.4\textwidth]{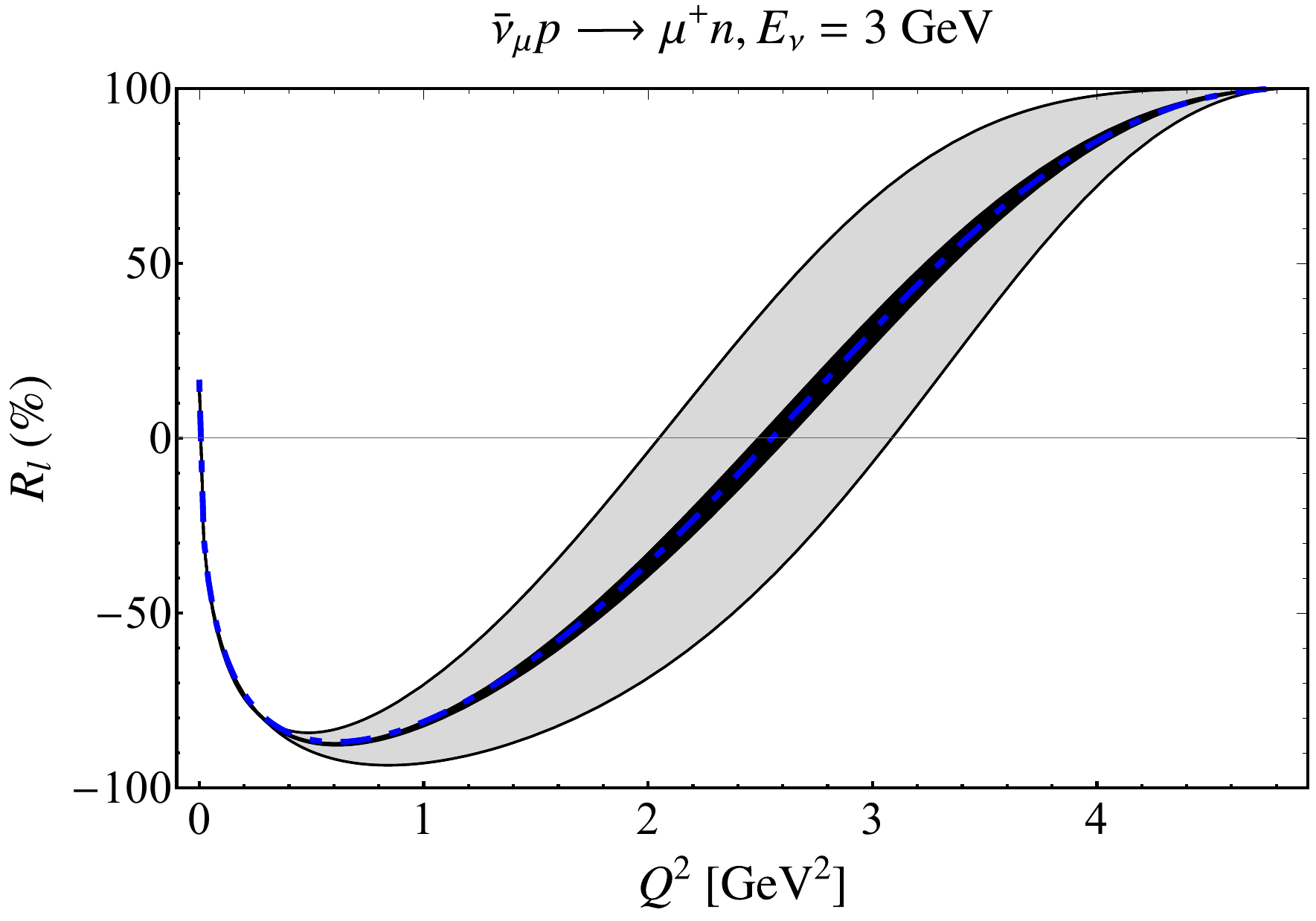}
\caption{Same as Fig.~\ref{fig:antinu_Tt_radcorr} but for the longitudinal polarization observable $R_l$. \label{fig:antinu_Rl_radcorr}}
\end{figure}

\begin{figure}[h]
\centering
\includegraphics[width=0.4\textwidth]{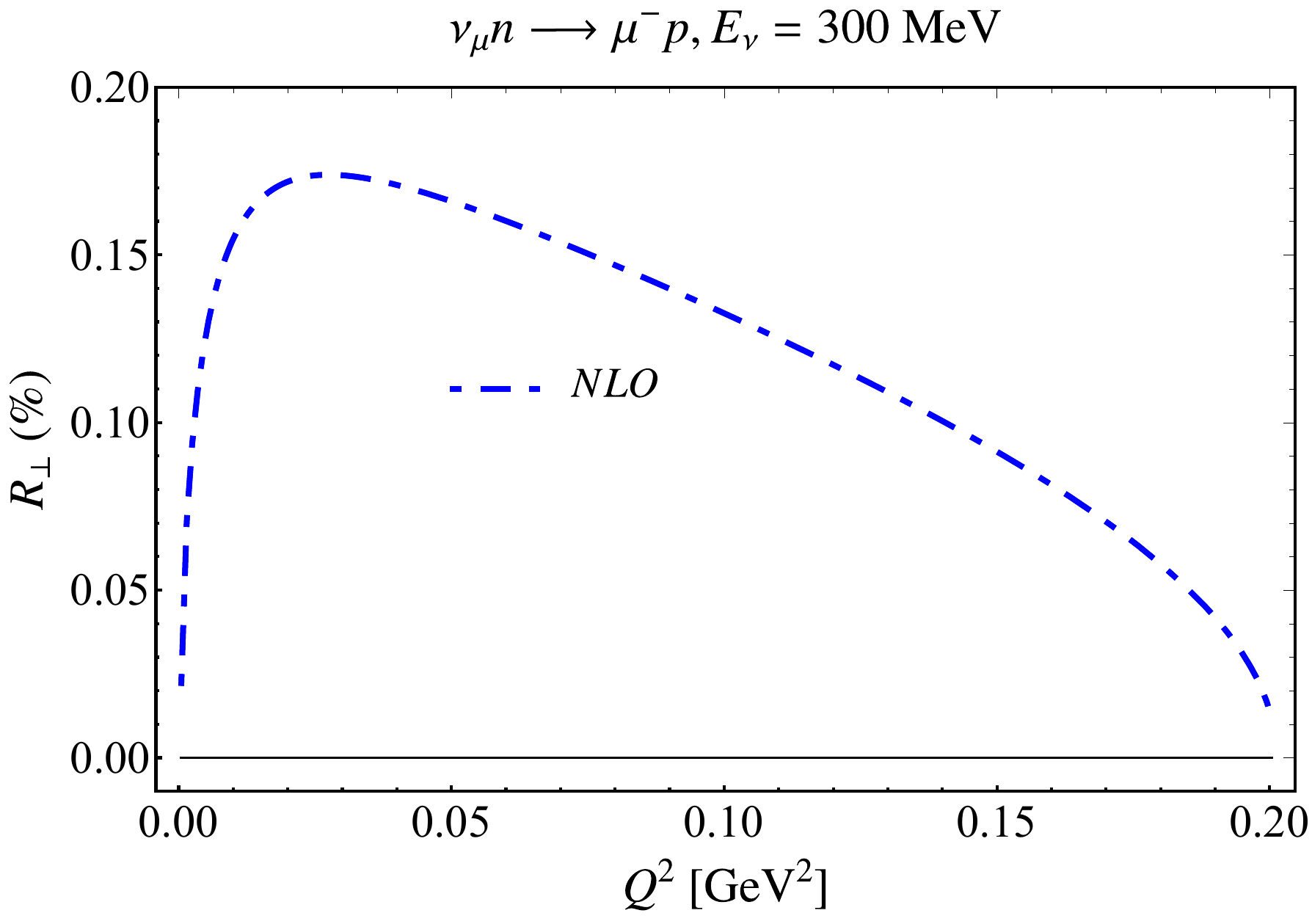}
\includegraphics[width=0.4\textwidth]{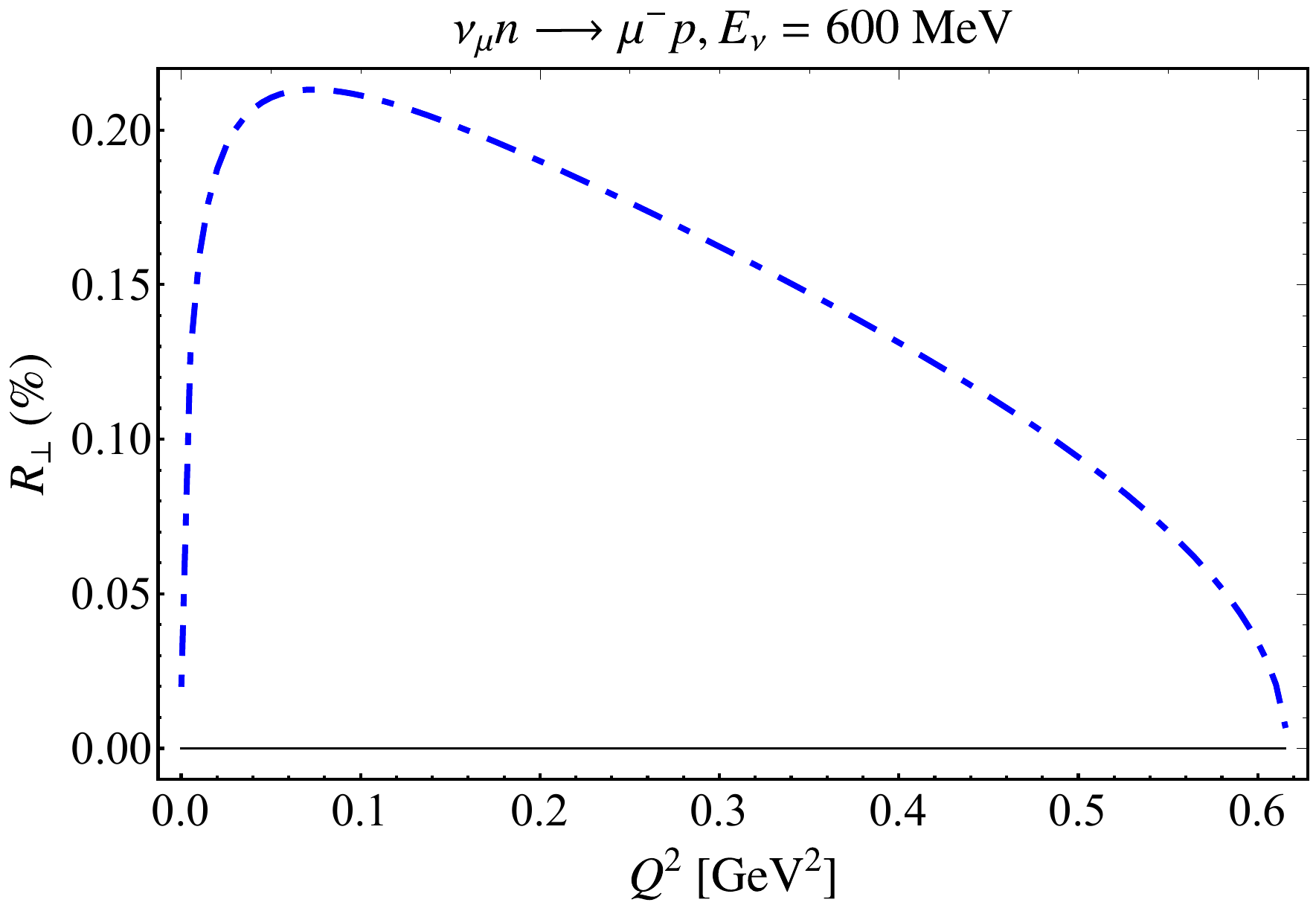}
\includegraphics[width=0.4\textwidth]{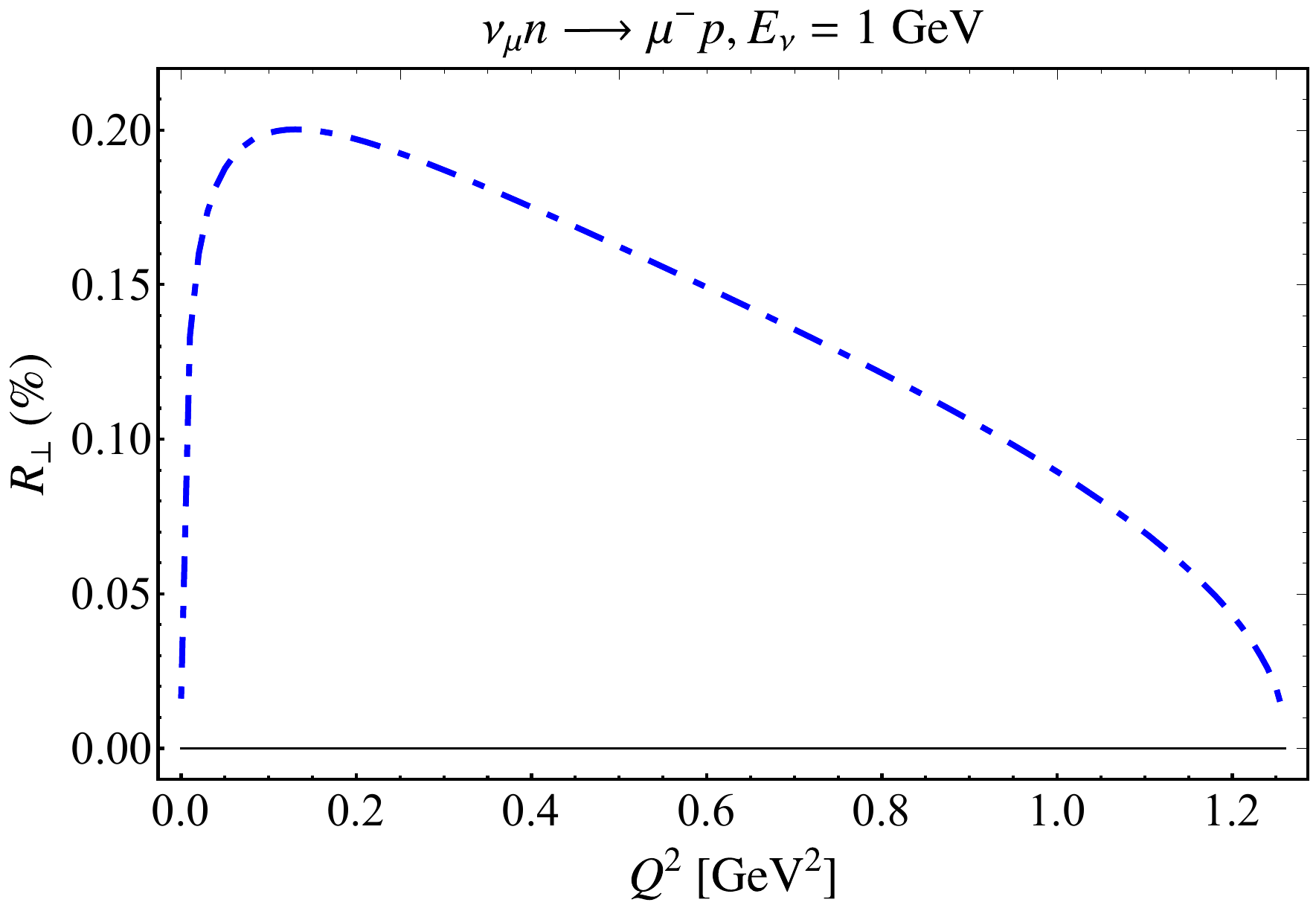}
\includegraphics[width=0.4\textwidth]{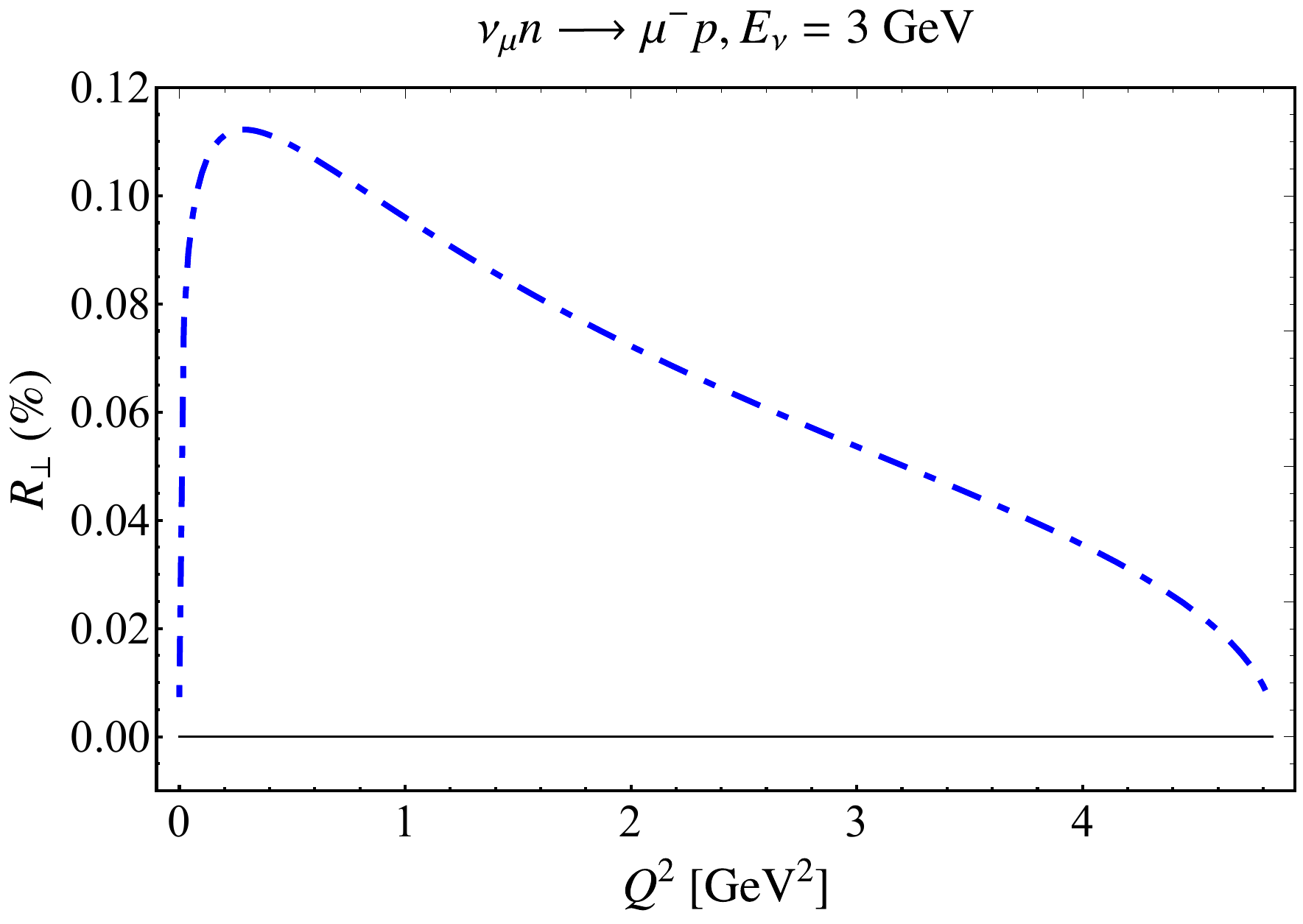}
\caption{Same as Fig.~\ref{fig:nu_Tt_radcorr} but for the transverse polarization observable $R_\perp$. \label{fig:nu_RTT_radcorr}}
\end{figure}

\begin{figure}[H]
\centering
\includegraphics[width=0.4\textwidth]{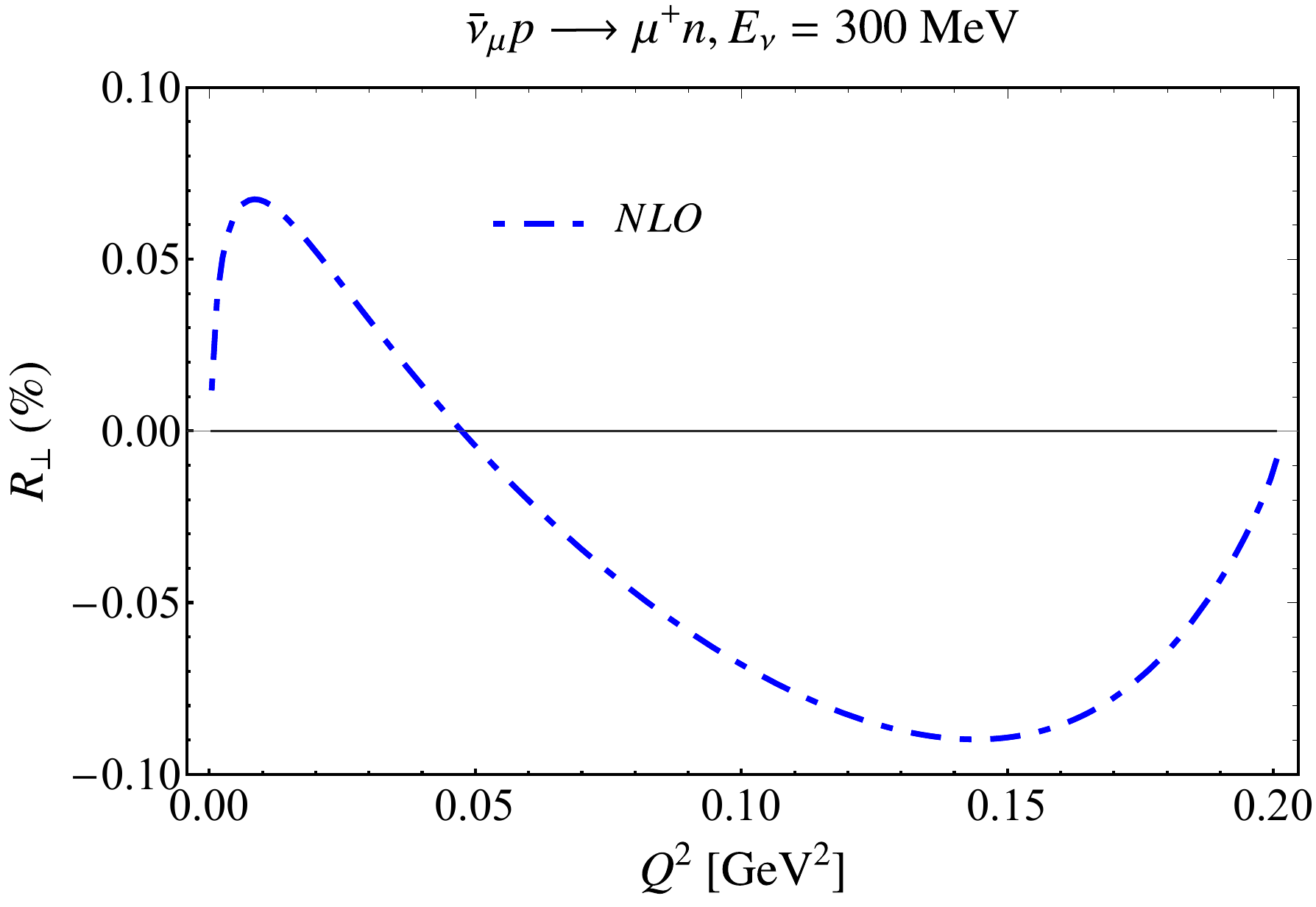}
\includegraphics[width=0.4\textwidth]{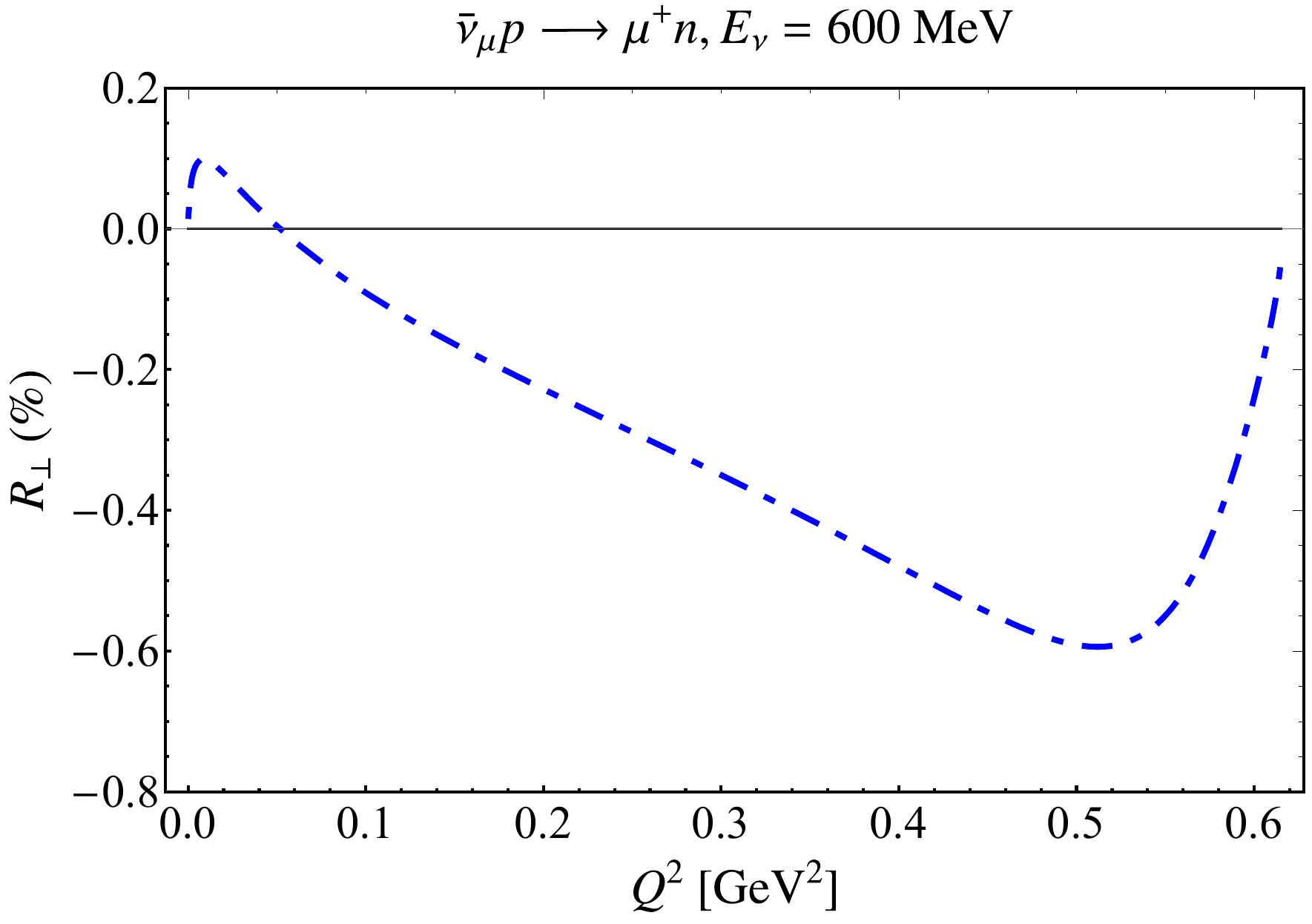}
\includegraphics[width=0.4\textwidth]{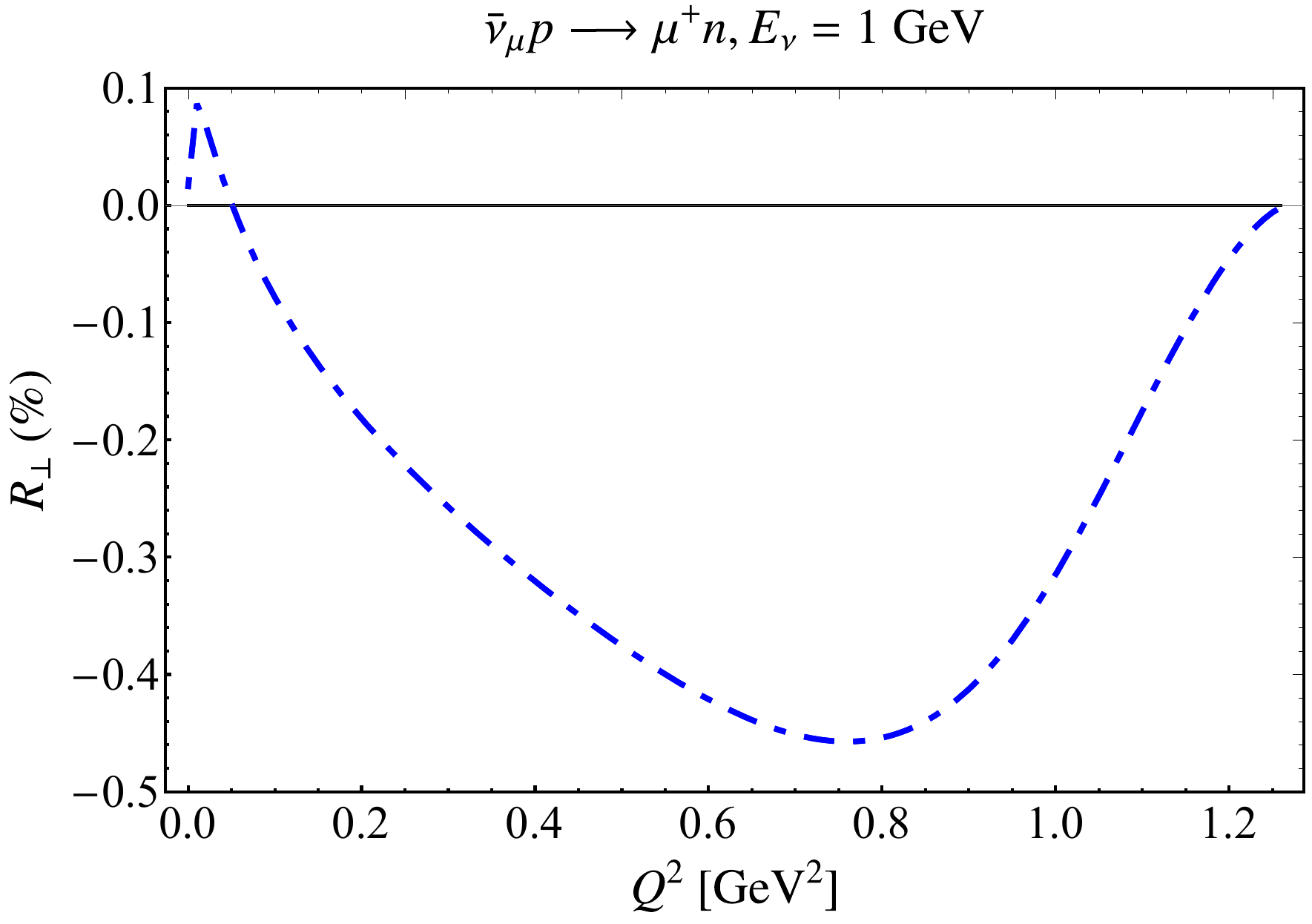}
\includegraphics[width=0.4\textwidth]{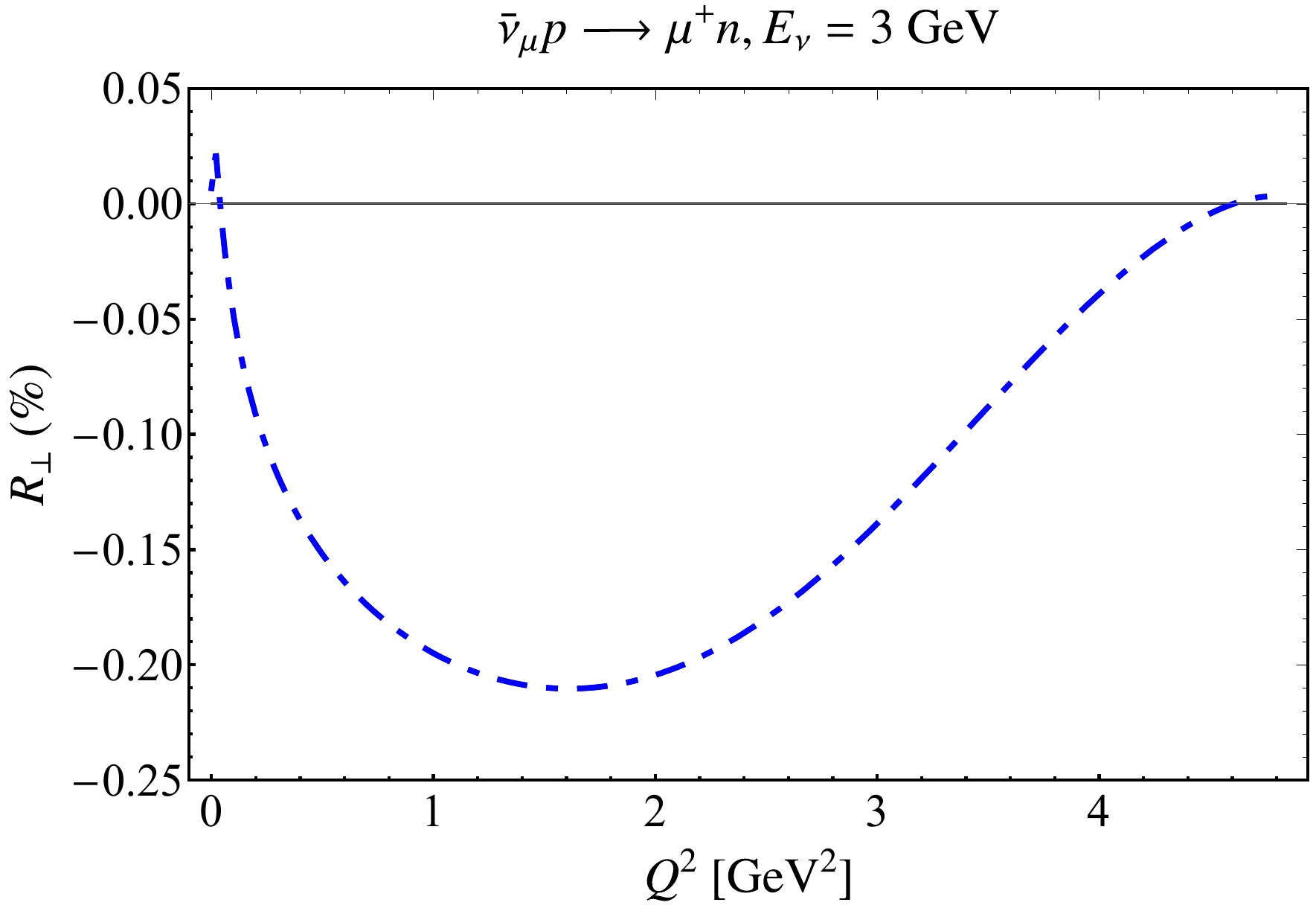}
\caption{Same as Fig.~\ref{fig:antinu_Tt_radcorr} but for the transverse polarization observable $R_\perp$. \label{fig:antinu_RTT_radcorr}}
\end{figure}

\begin{figure}[H]
\centering
\includegraphics[width=0.4\textwidth]{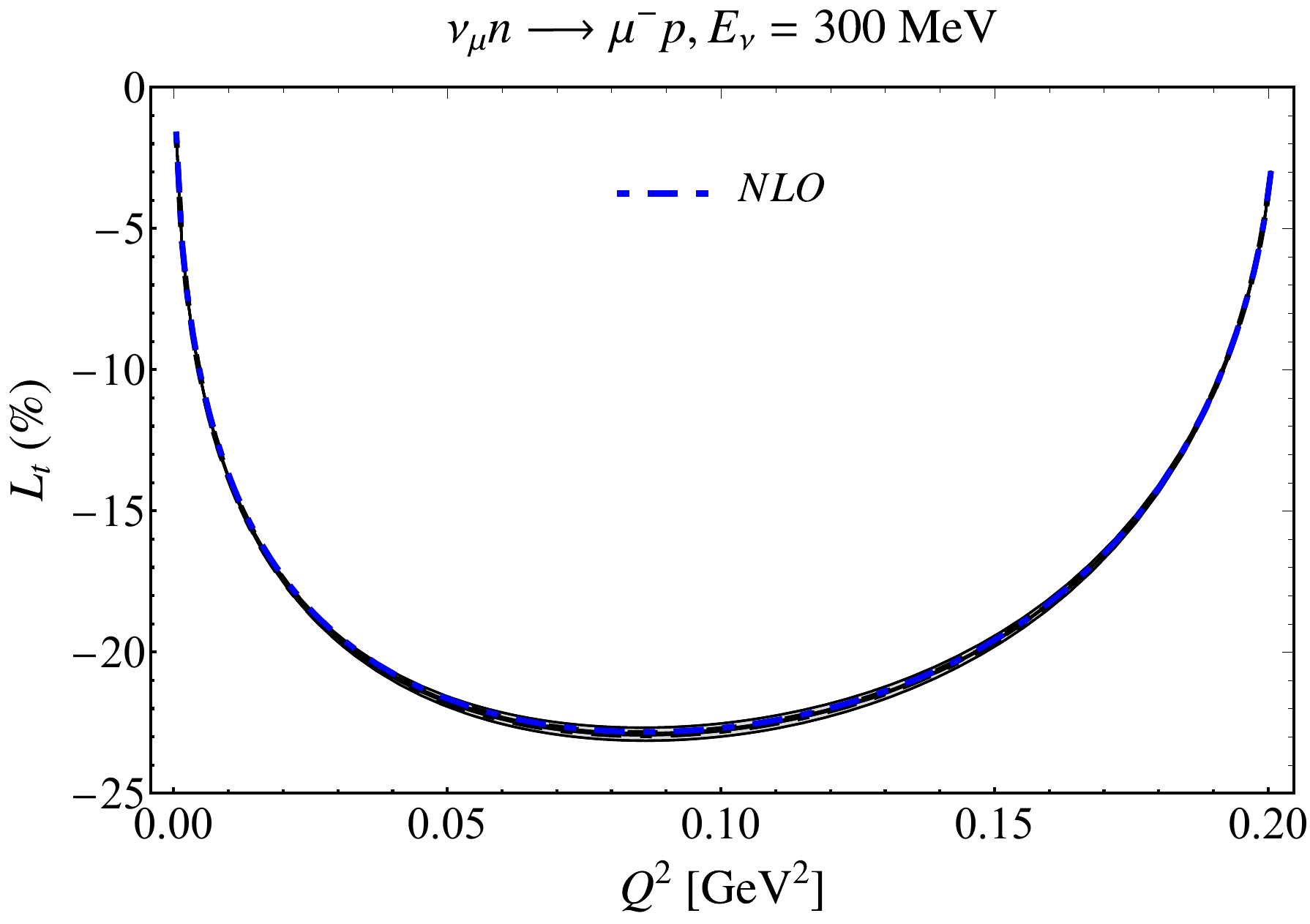}
\includegraphics[width=0.4\textwidth]{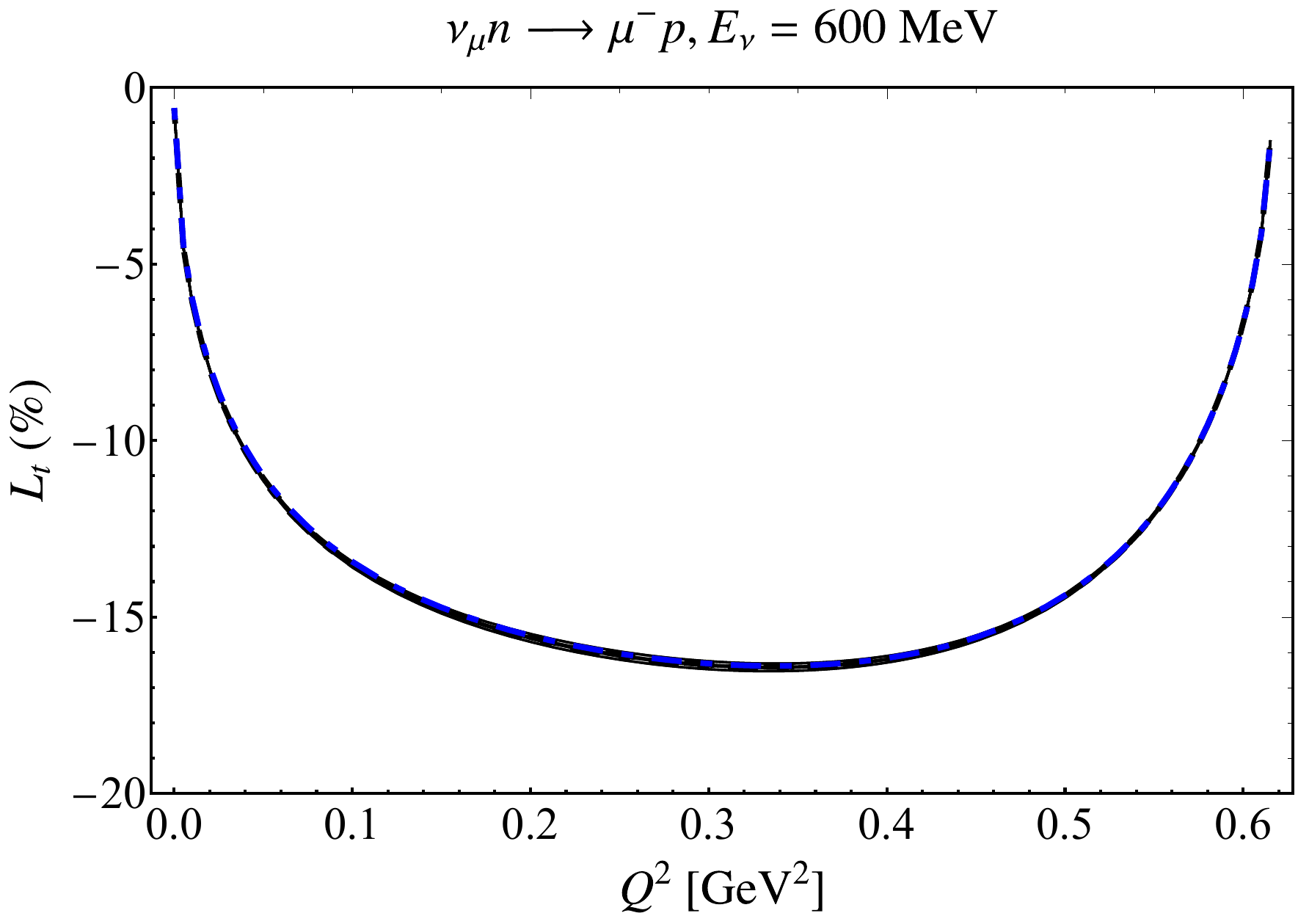}
\includegraphics[width=0.4\textwidth]{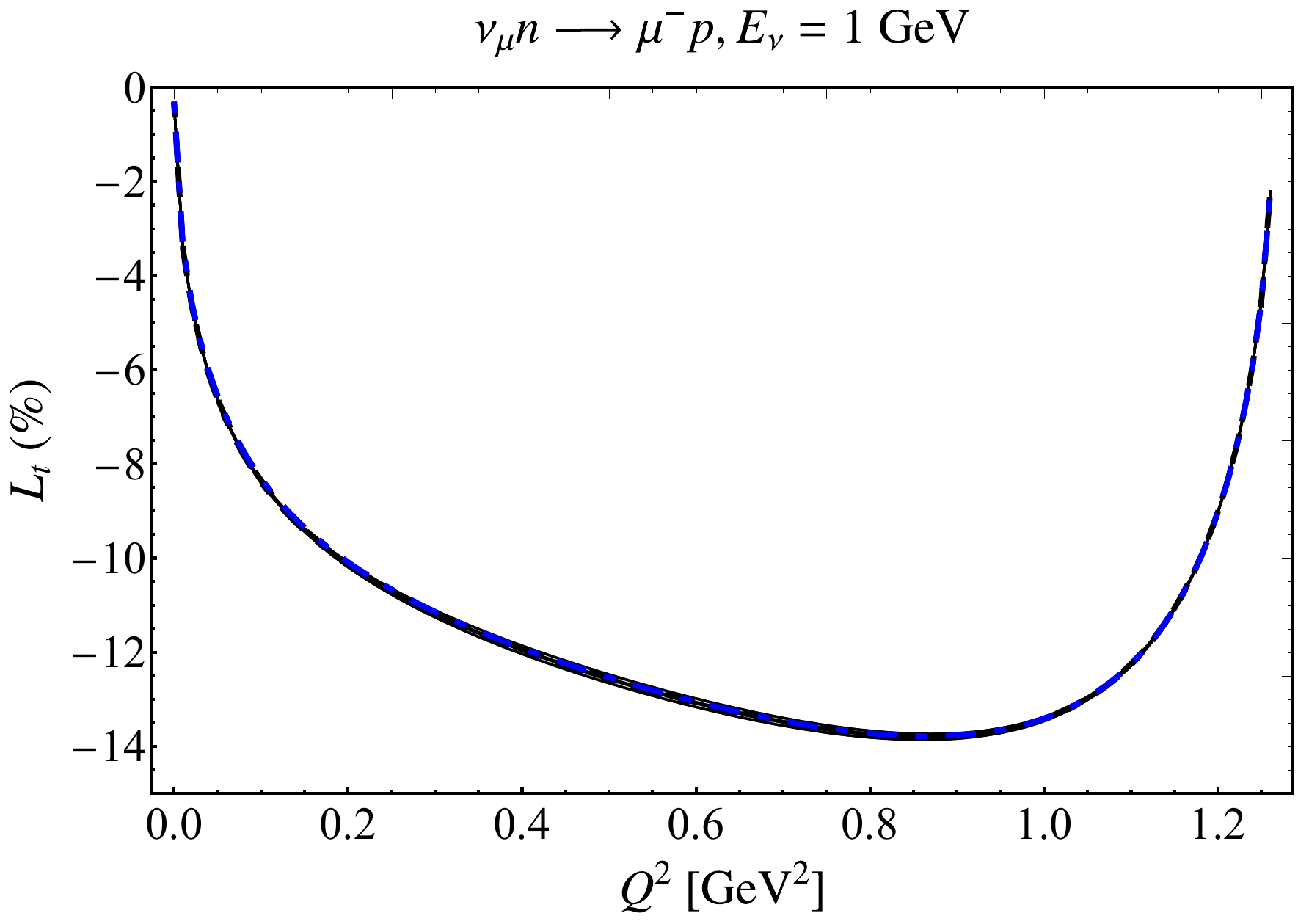}
\includegraphics[width=0.4\textwidth]{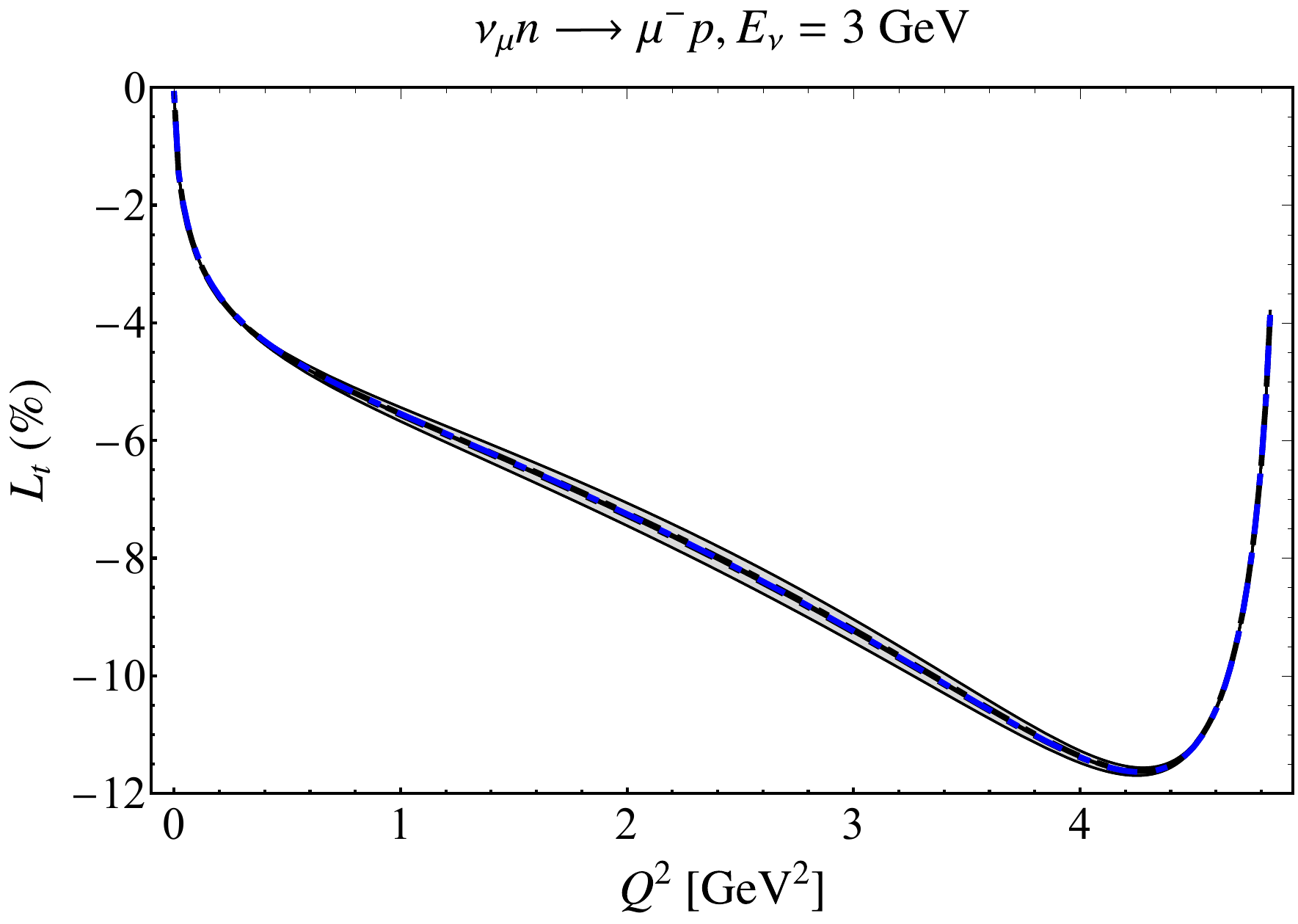}
\caption{Same as Fig.~\ref{fig:nu_Tt_radcorr} but for the transverse polarization observable $L_t$. \label{fig:nu_Lt_radcorr}}
\end{figure}

\begin{figure}[H]
\centering
\includegraphics[width=0.4\textwidth]{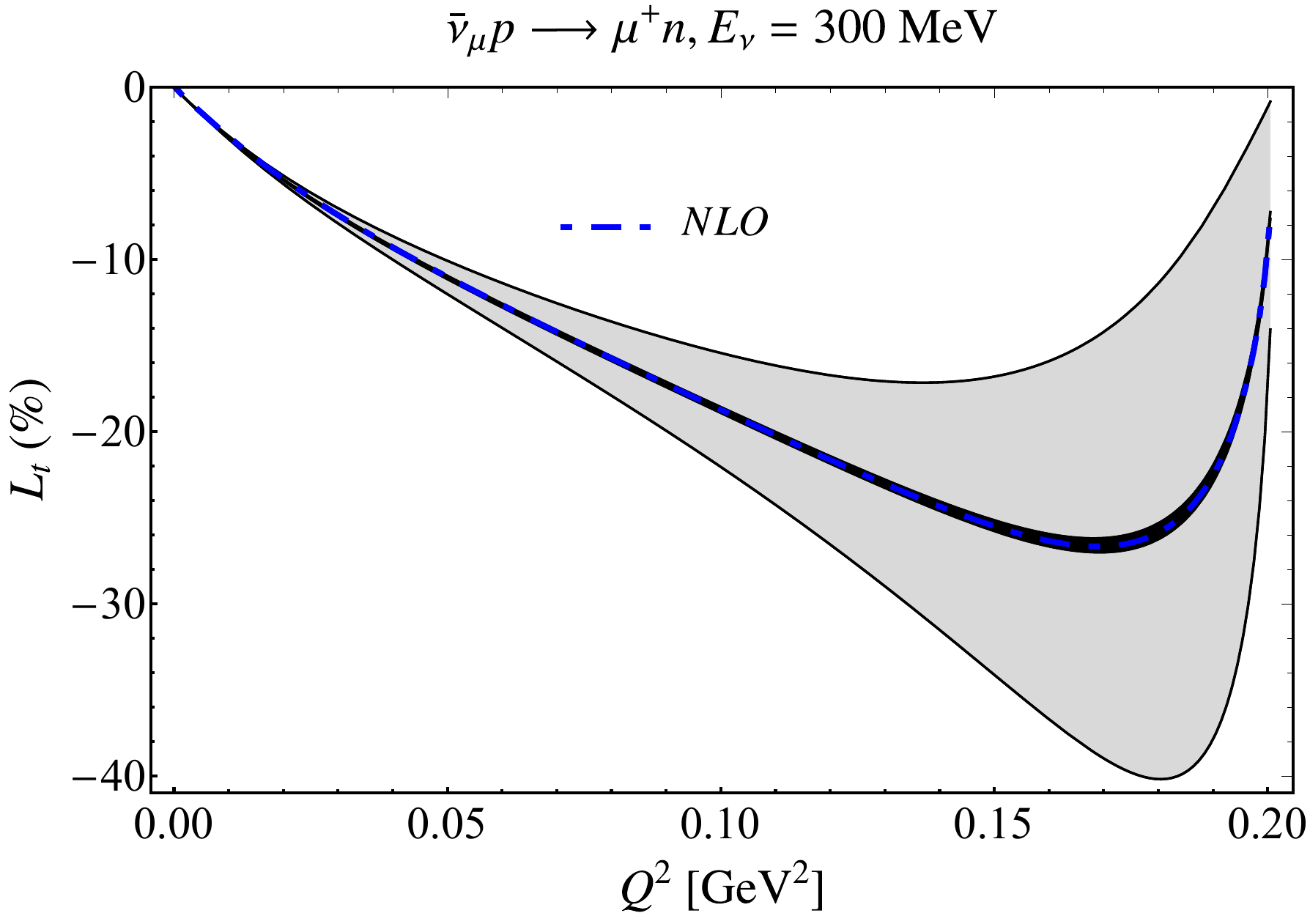}
\includegraphics[width=0.4\textwidth]{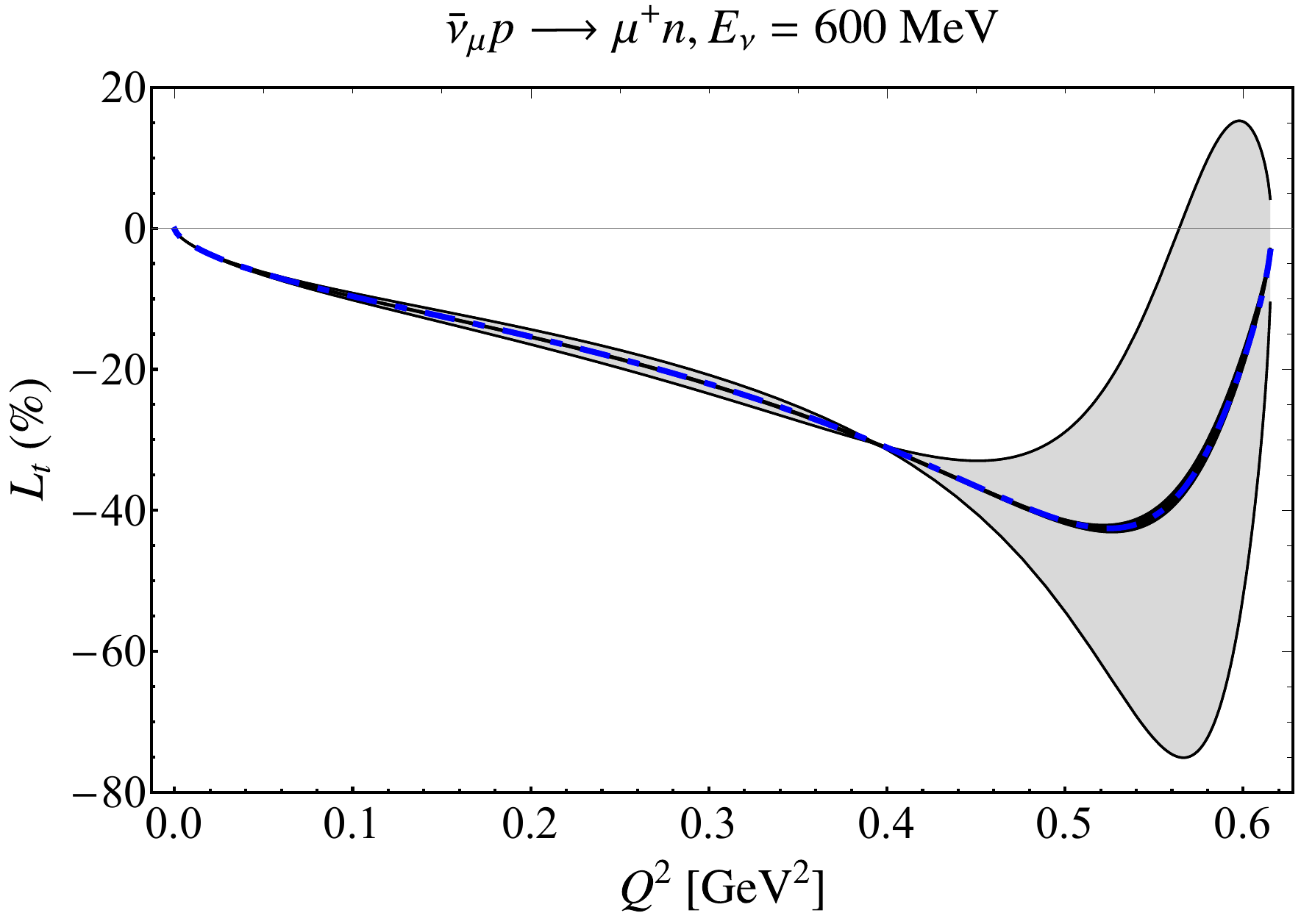}
\includegraphics[width=0.4\textwidth]{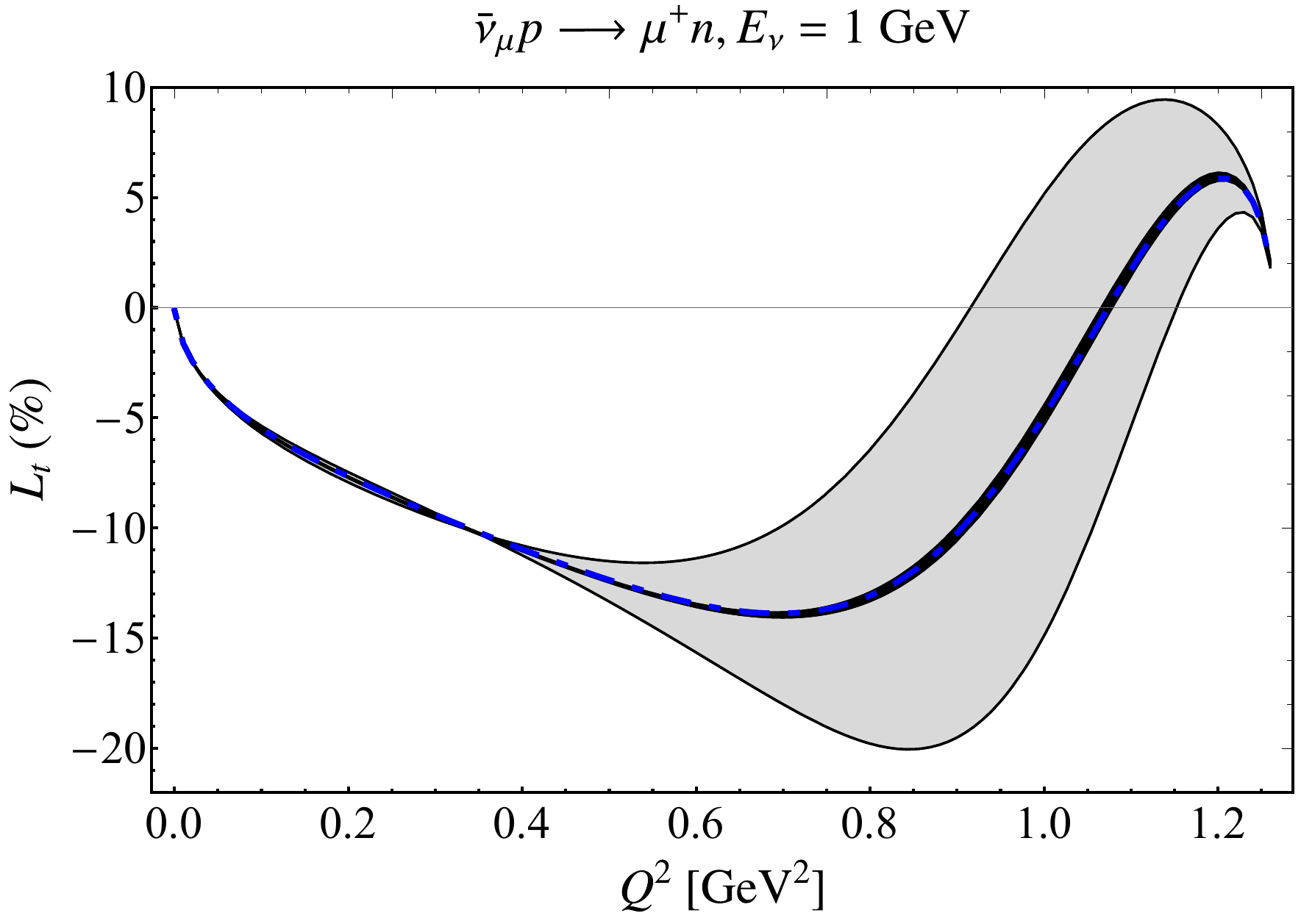}
\includegraphics[width=0.4\textwidth]{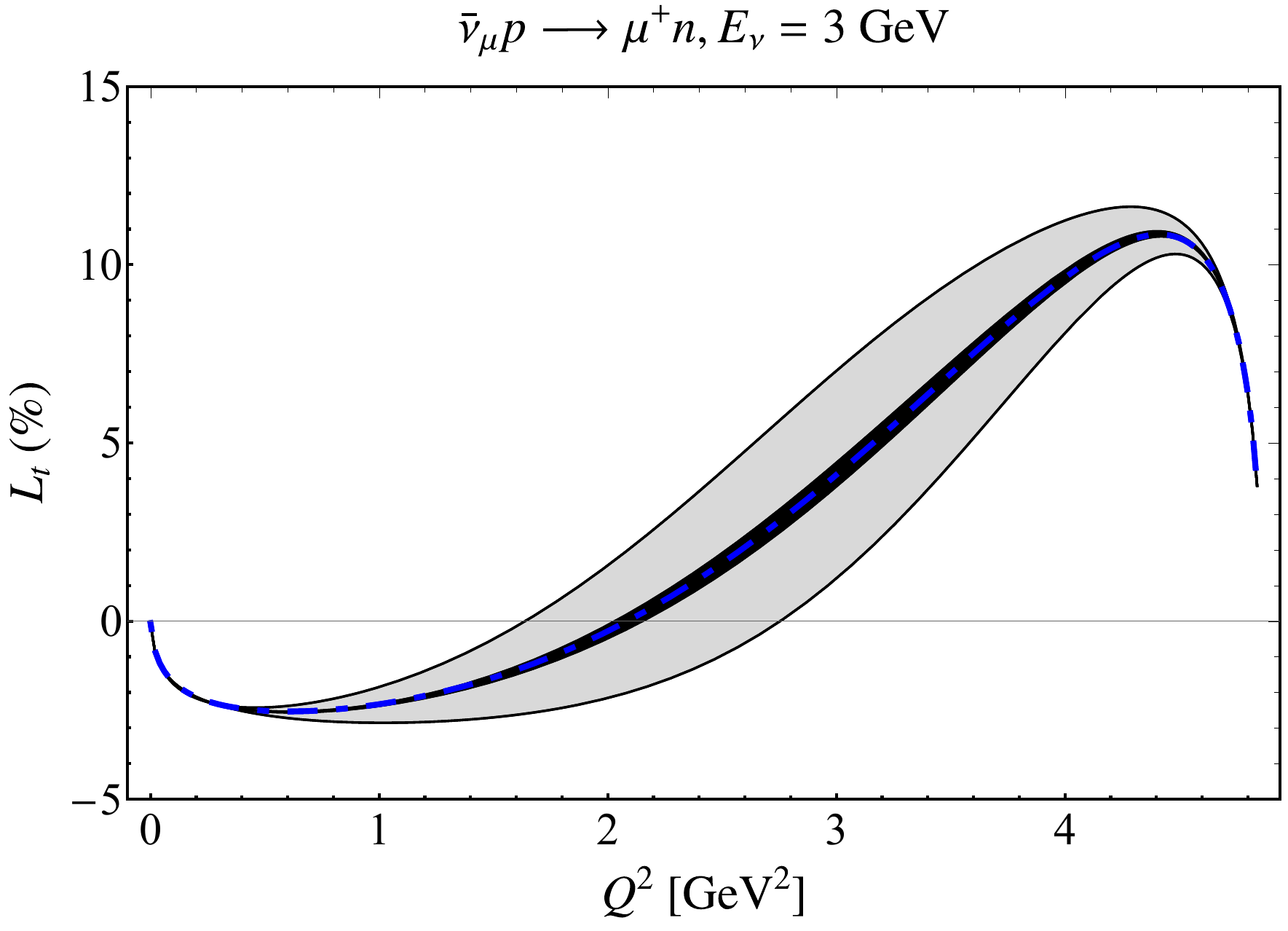}
\caption{Same as Fig.~\ref{fig:antinu_Tt_radcorr} but for the transverse polarization observable $L_t$. \label{fig:antinu_Lt_radcorr}}
\end{figure}

\begin{figure}[H]
\centering
\includegraphics[width=0.4\textwidth]{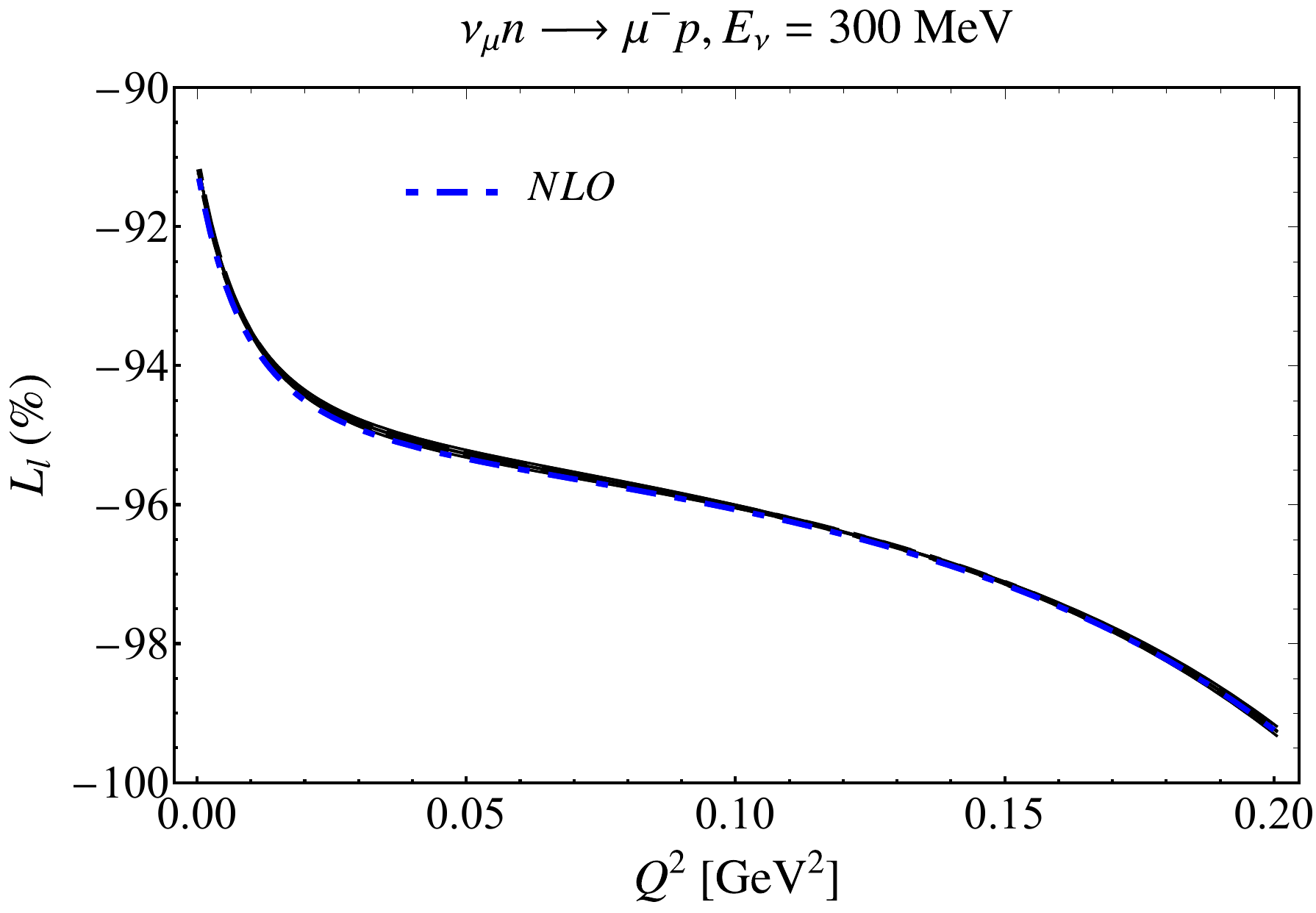}
\includegraphics[width=0.4\textwidth]{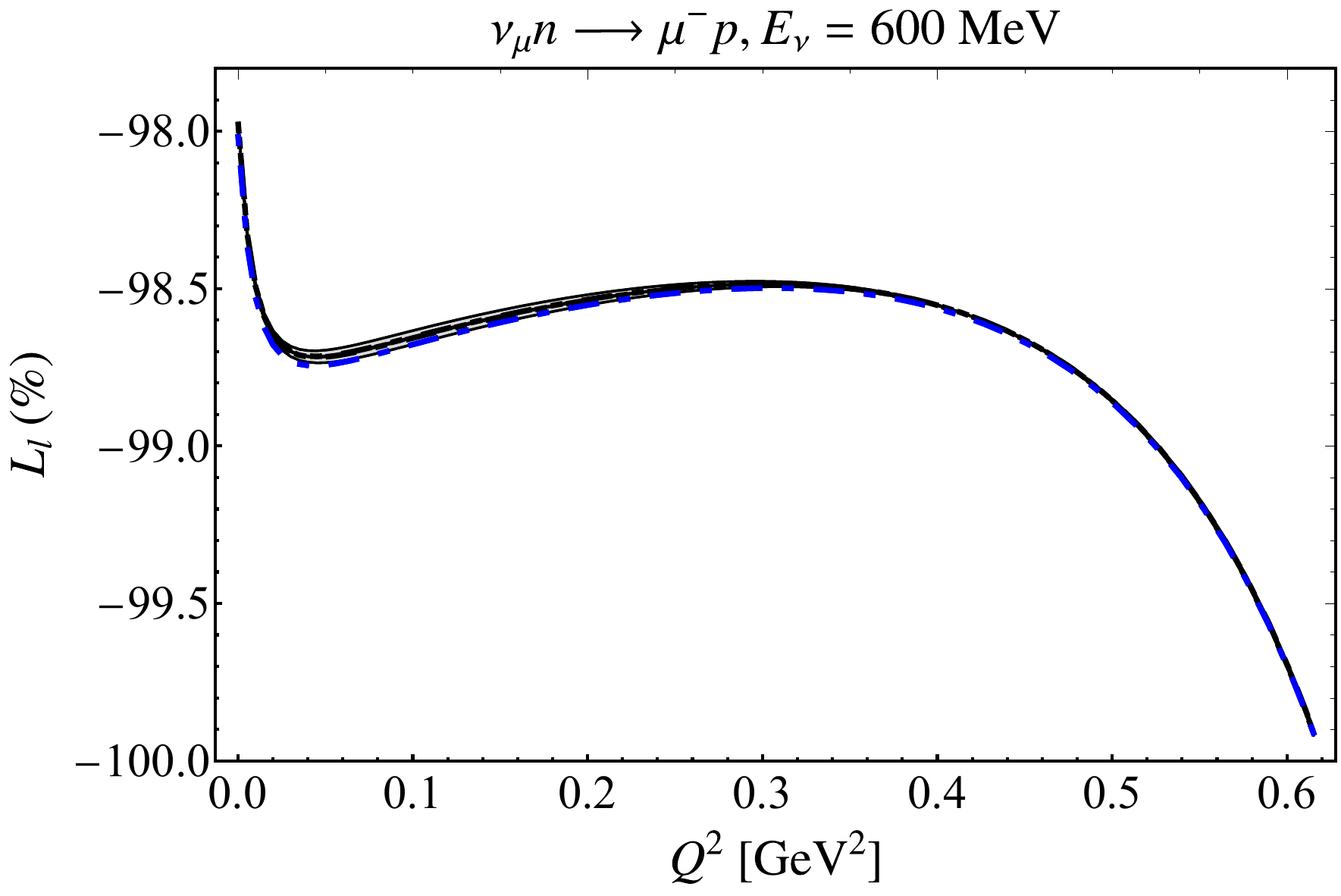}
\includegraphics[width=0.4\textwidth]{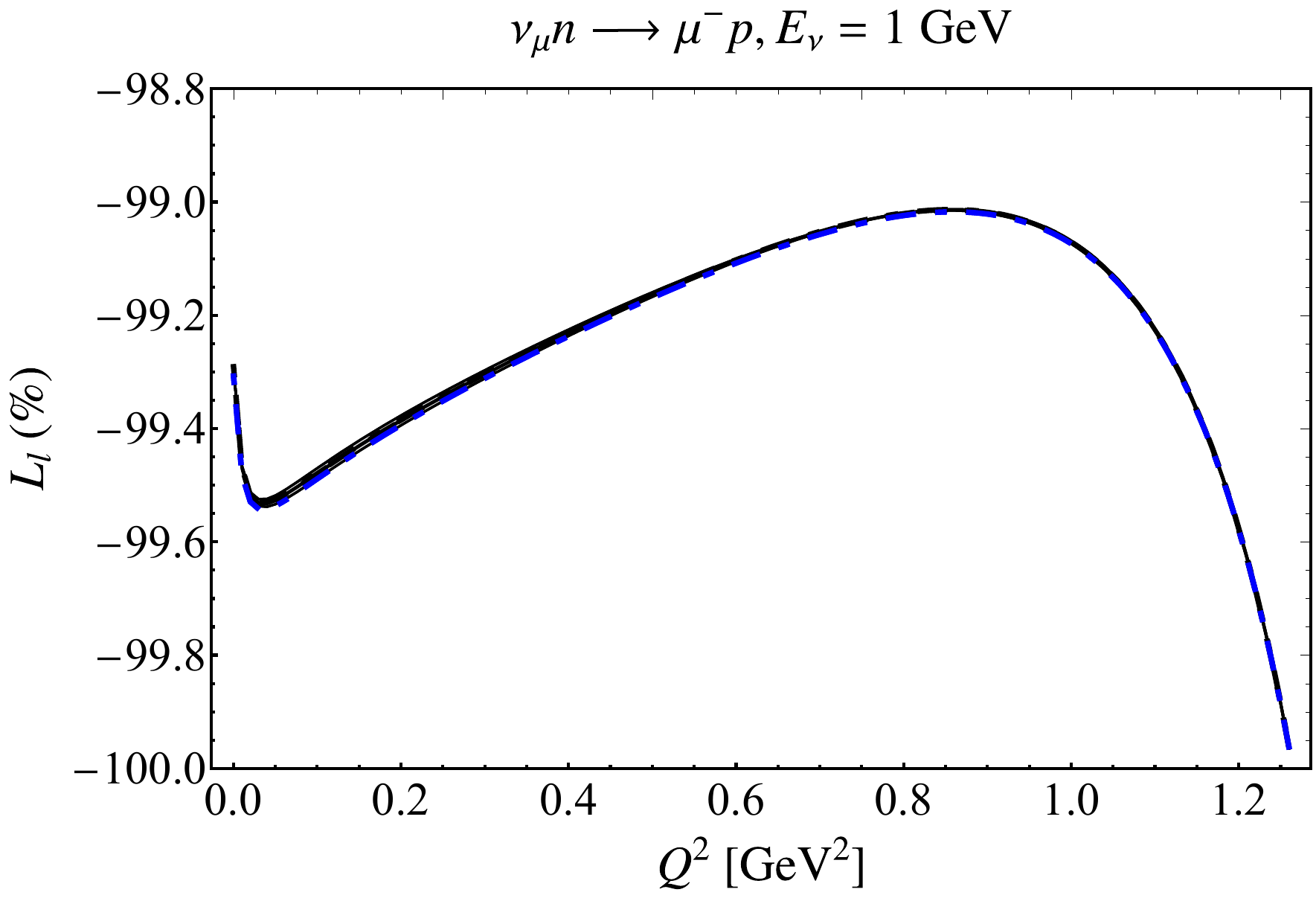}
\includegraphics[width=0.4\textwidth]{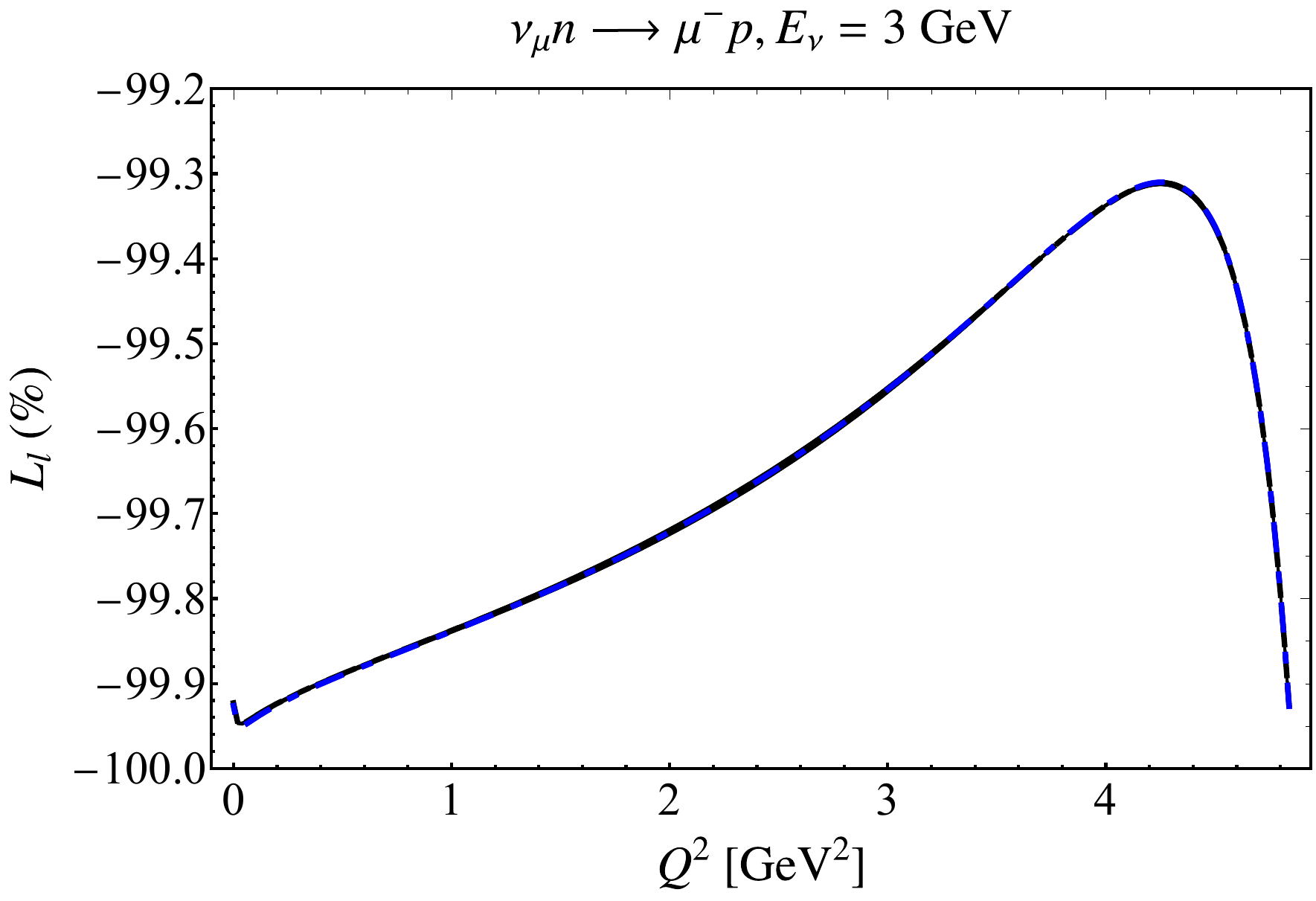}
\caption{Same as Fig.~\ref{fig:nu_Tt_radcorr} but for the longitudinal polarization observable $L_l$. \label{fig:nu_Ll_radcorr}}
\end{figure}

\begin{figure}[H]
\centering
\includegraphics[width=0.4\textwidth]{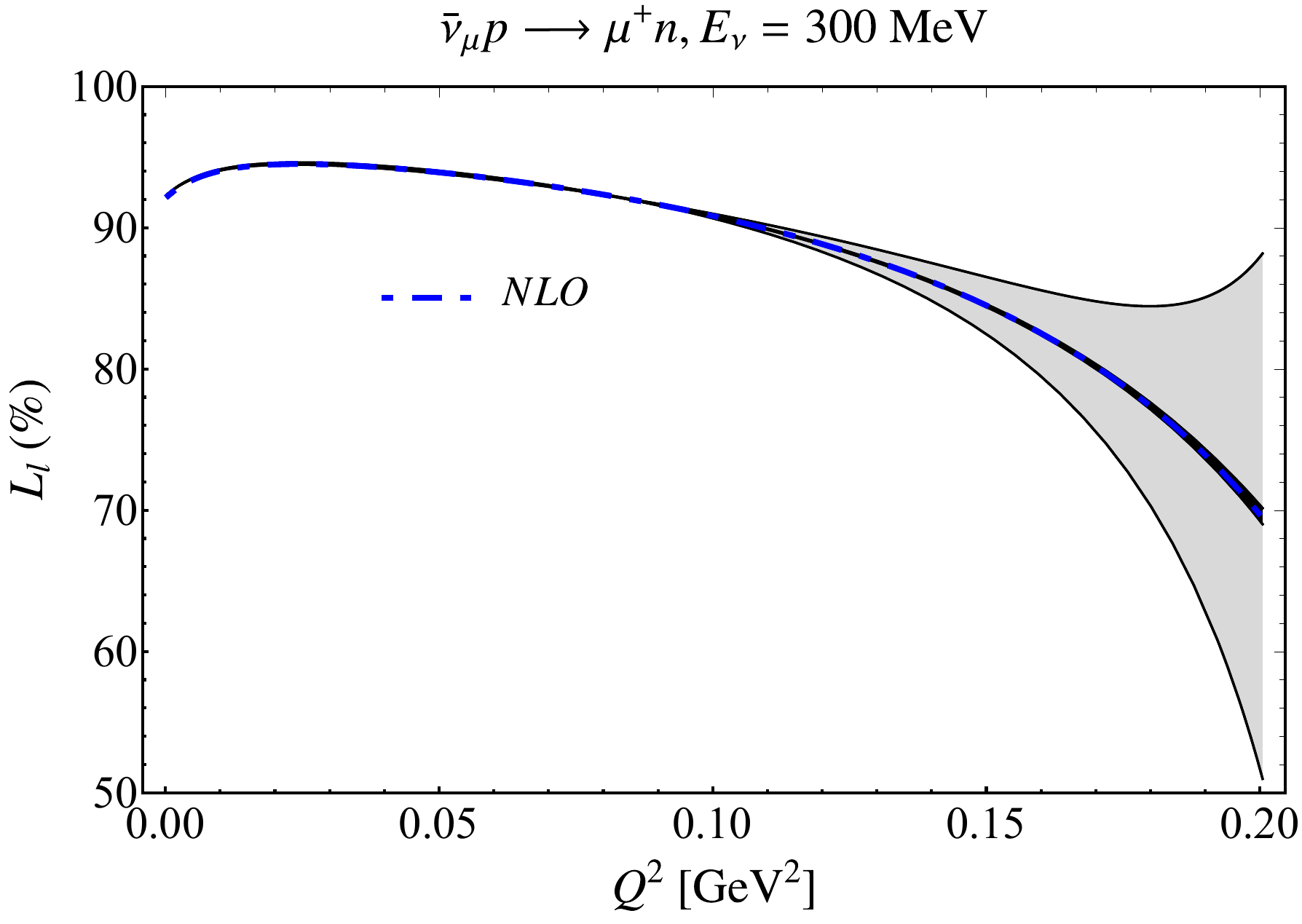}
\includegraphics[width=0.4\textwidth]{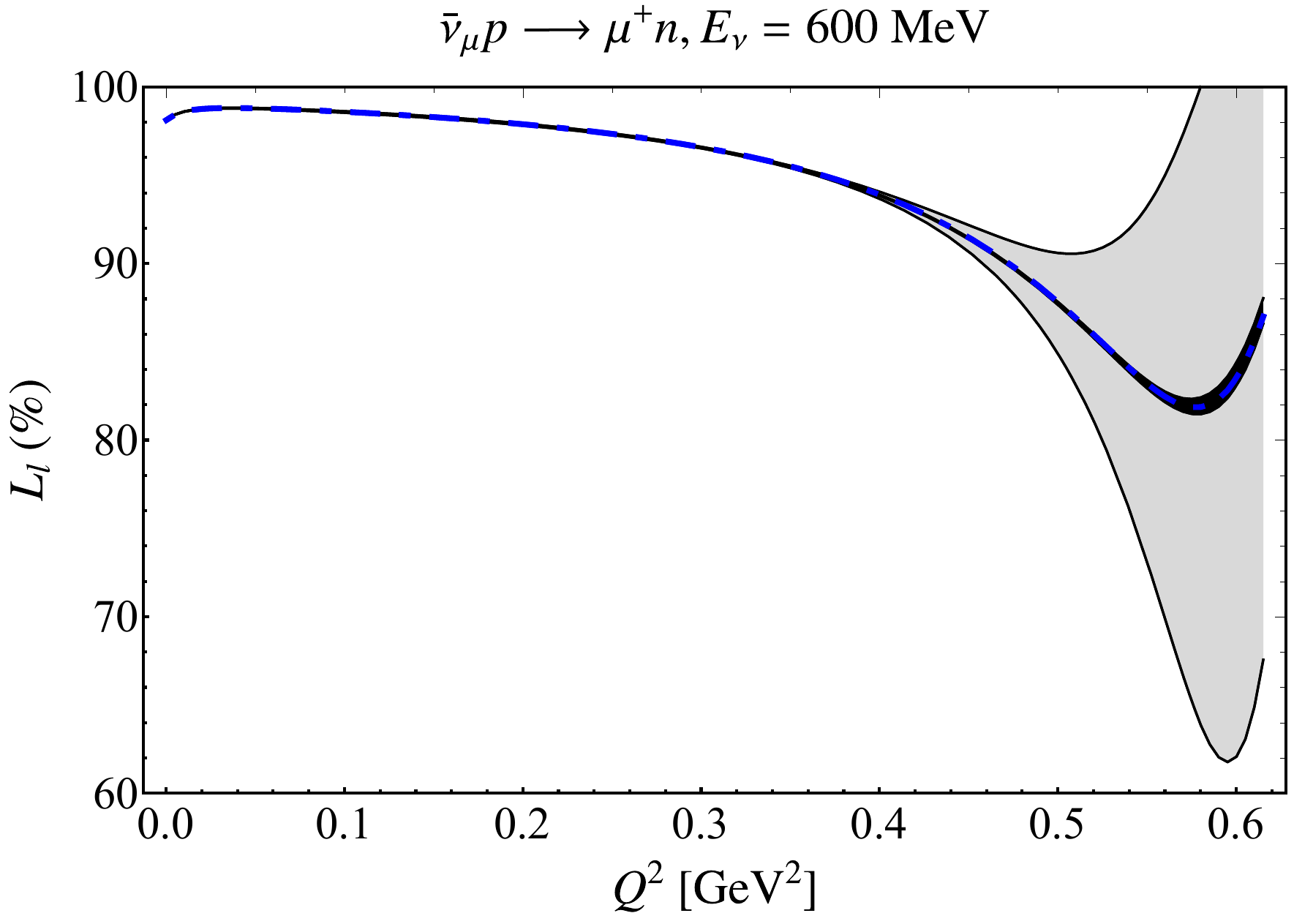}
\includegraphics[width=0.4\textwidth]{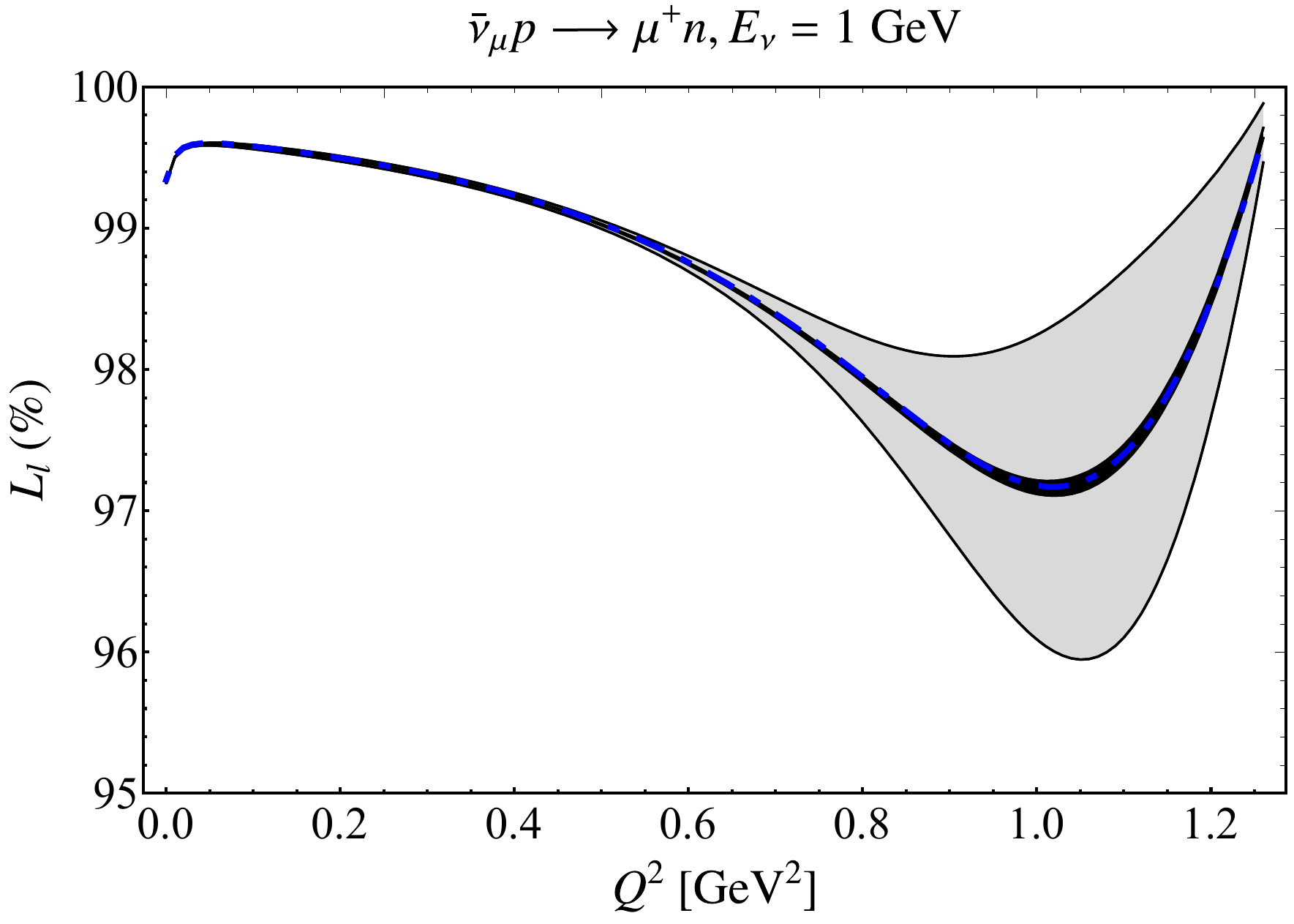}
\includegraphics[width=0.4\textwidth]{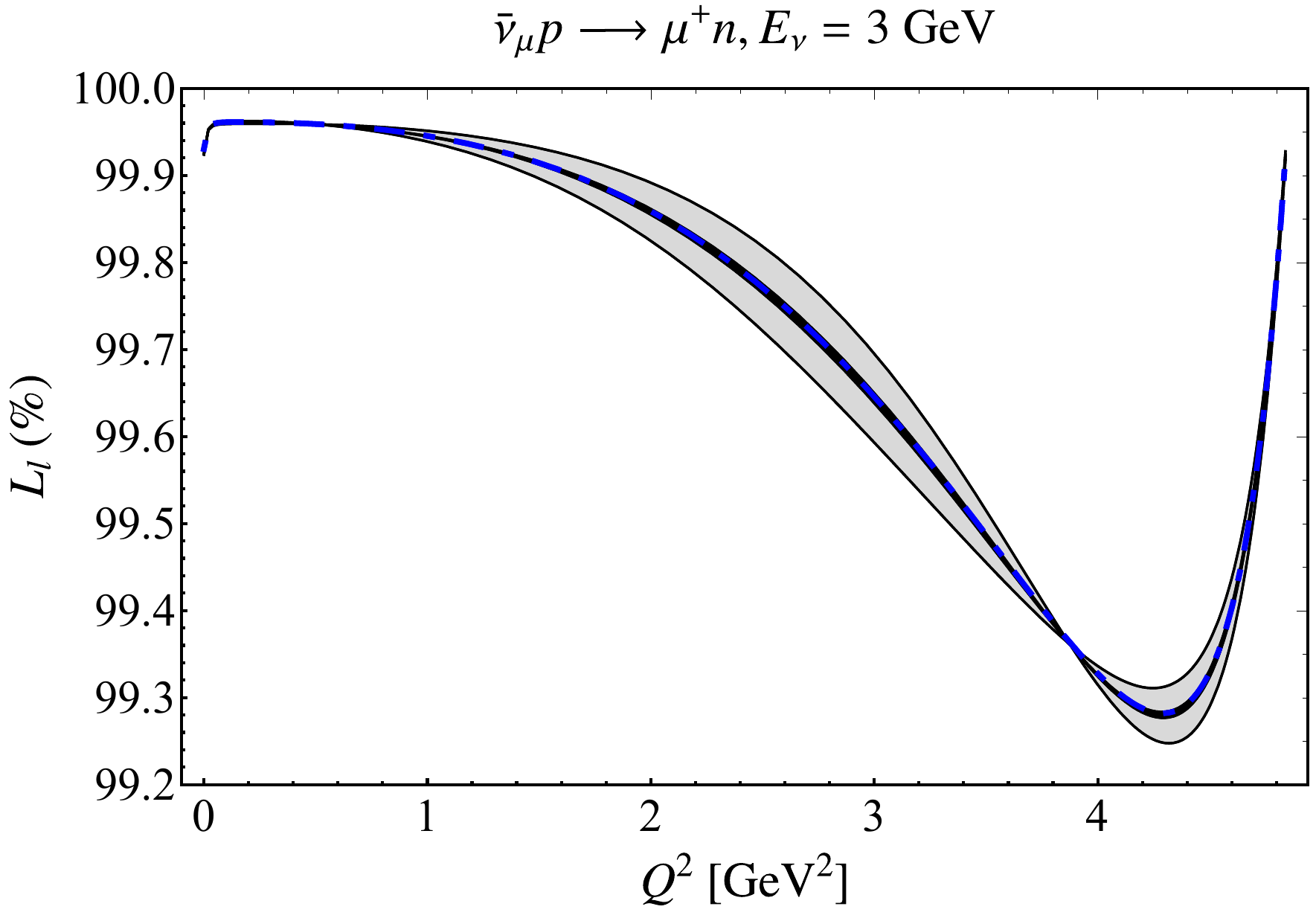}
\caption{Same as Fig.~\ref{fig:antinu_Tt_radcorr} but for the longitudinal polarization observable $L_l$. \label{fig:antinu_Ll_radcorr}}
\end{figure}

\begin{figure}[H]
\centering
\includegraphics[width=0.4\textwidth]{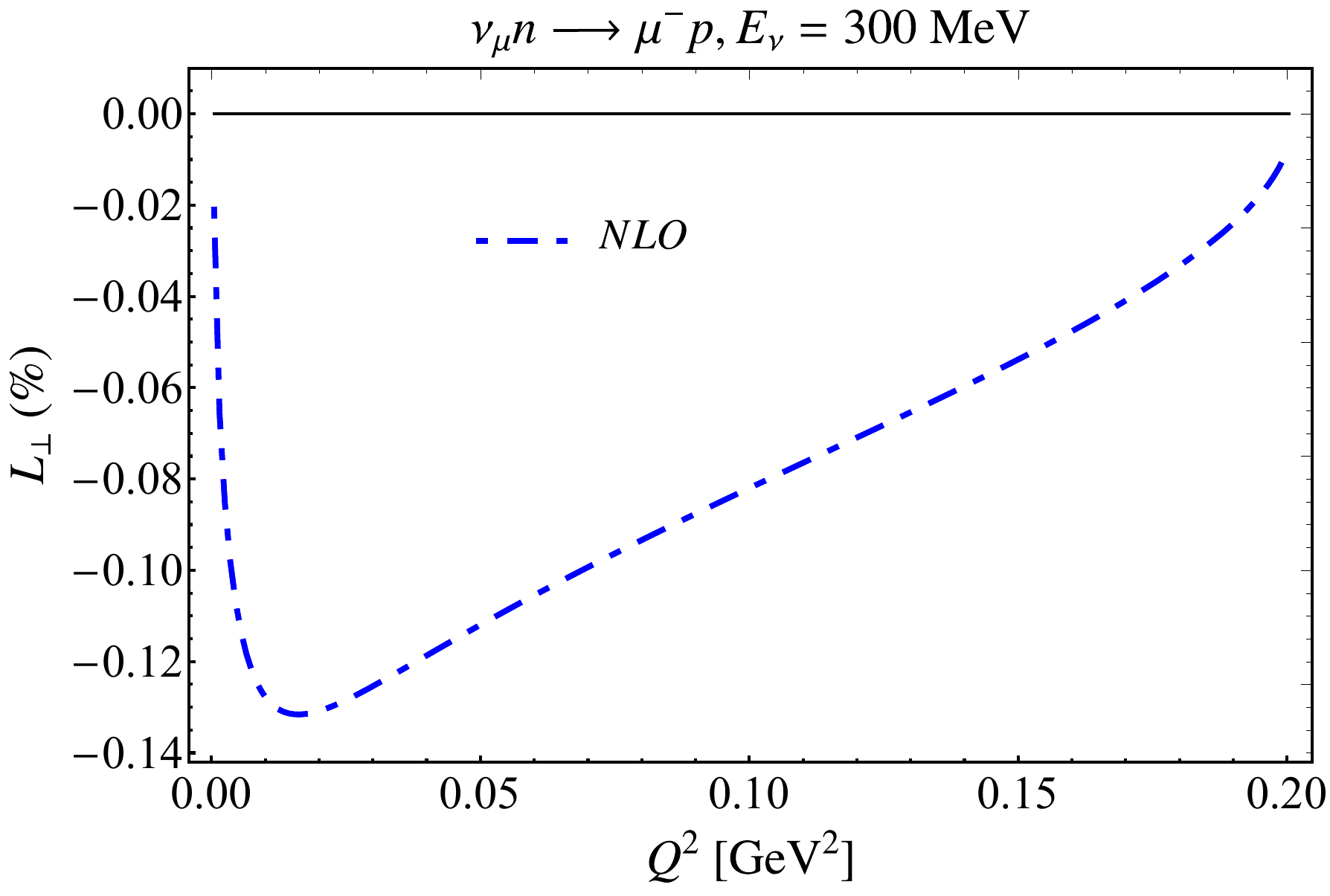}
\includegraphics[width=0.4\textwidth]{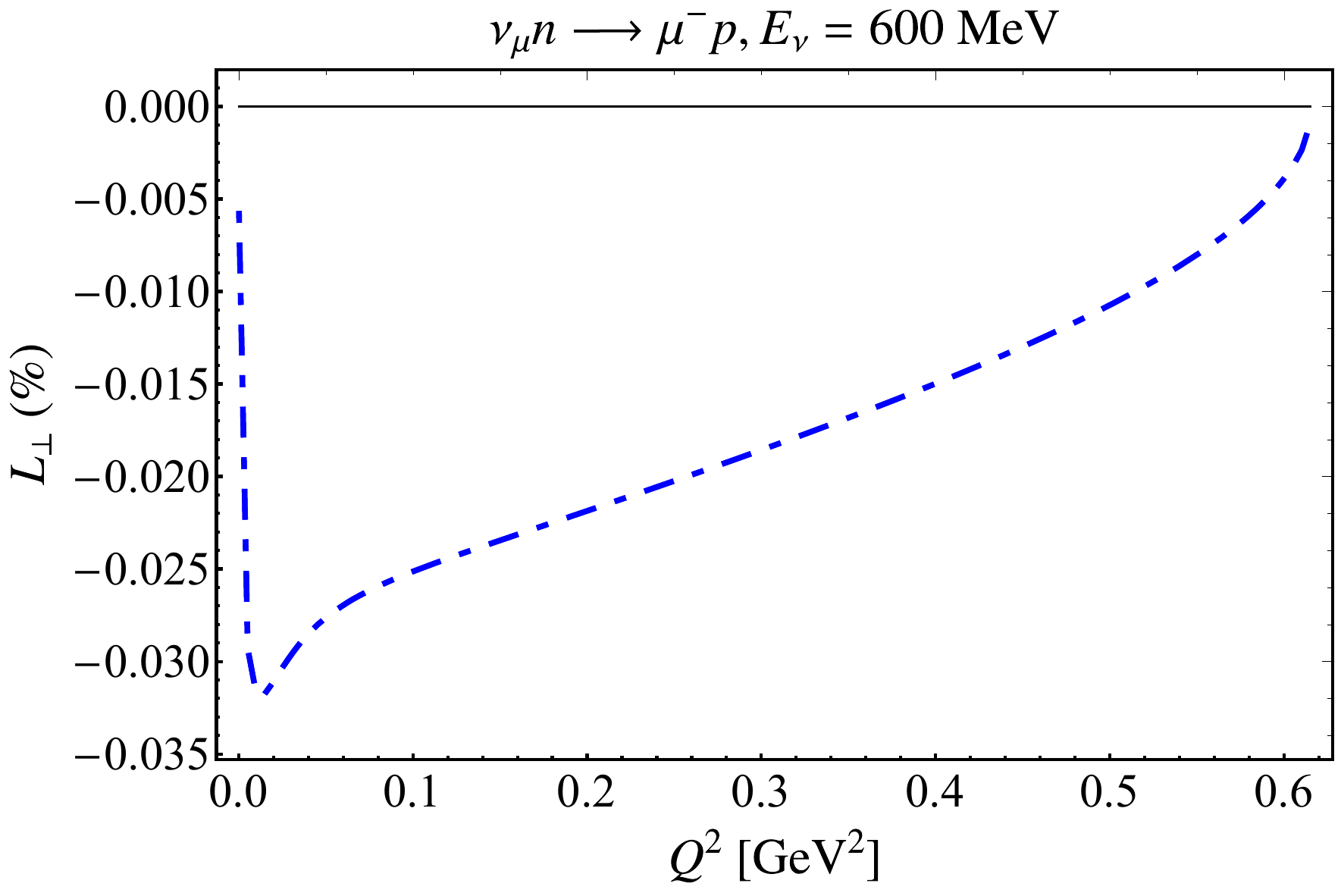}
\includegraphics[width=0.4\textwidth]{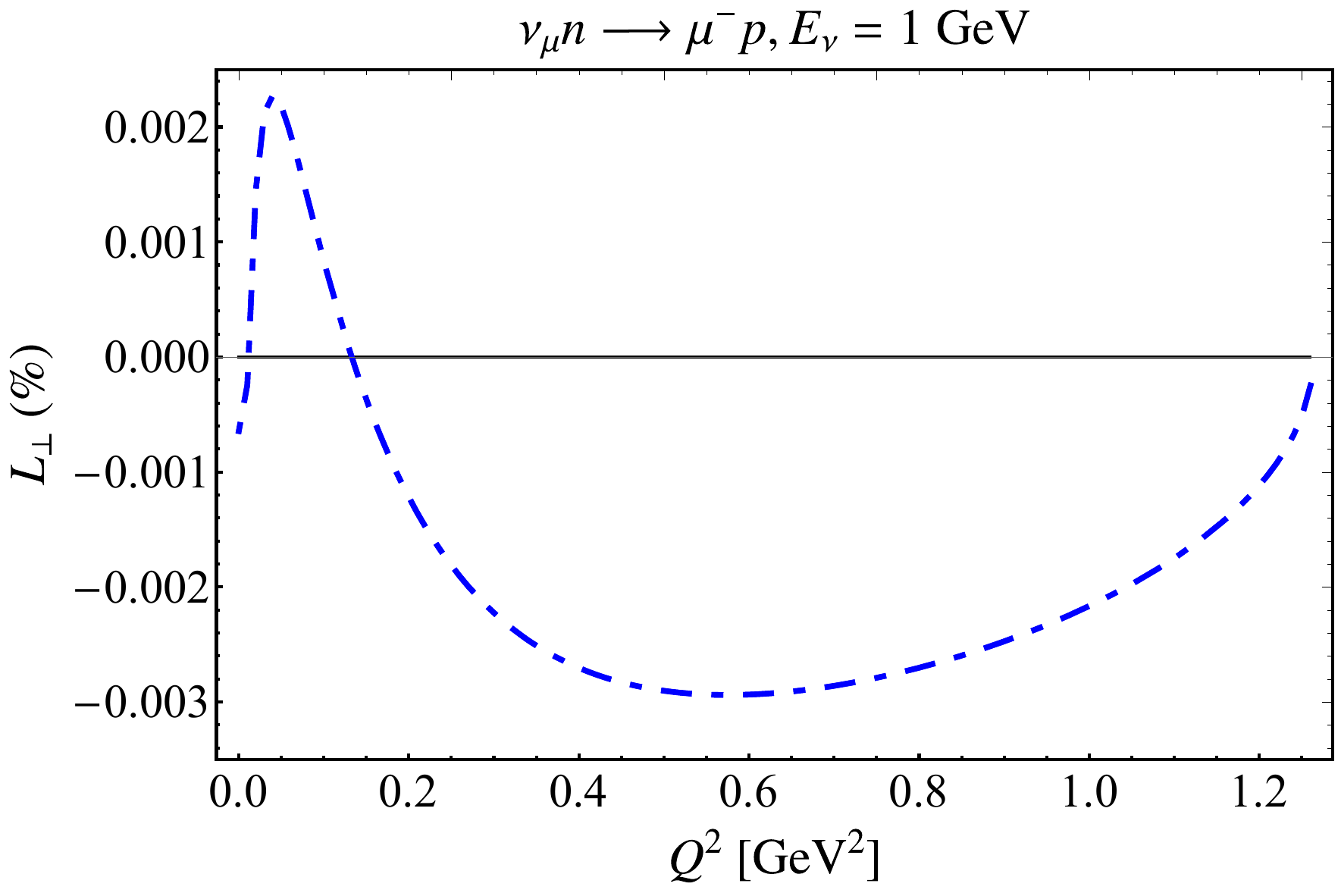}
\includegraphics[width=0.4\textwidth]{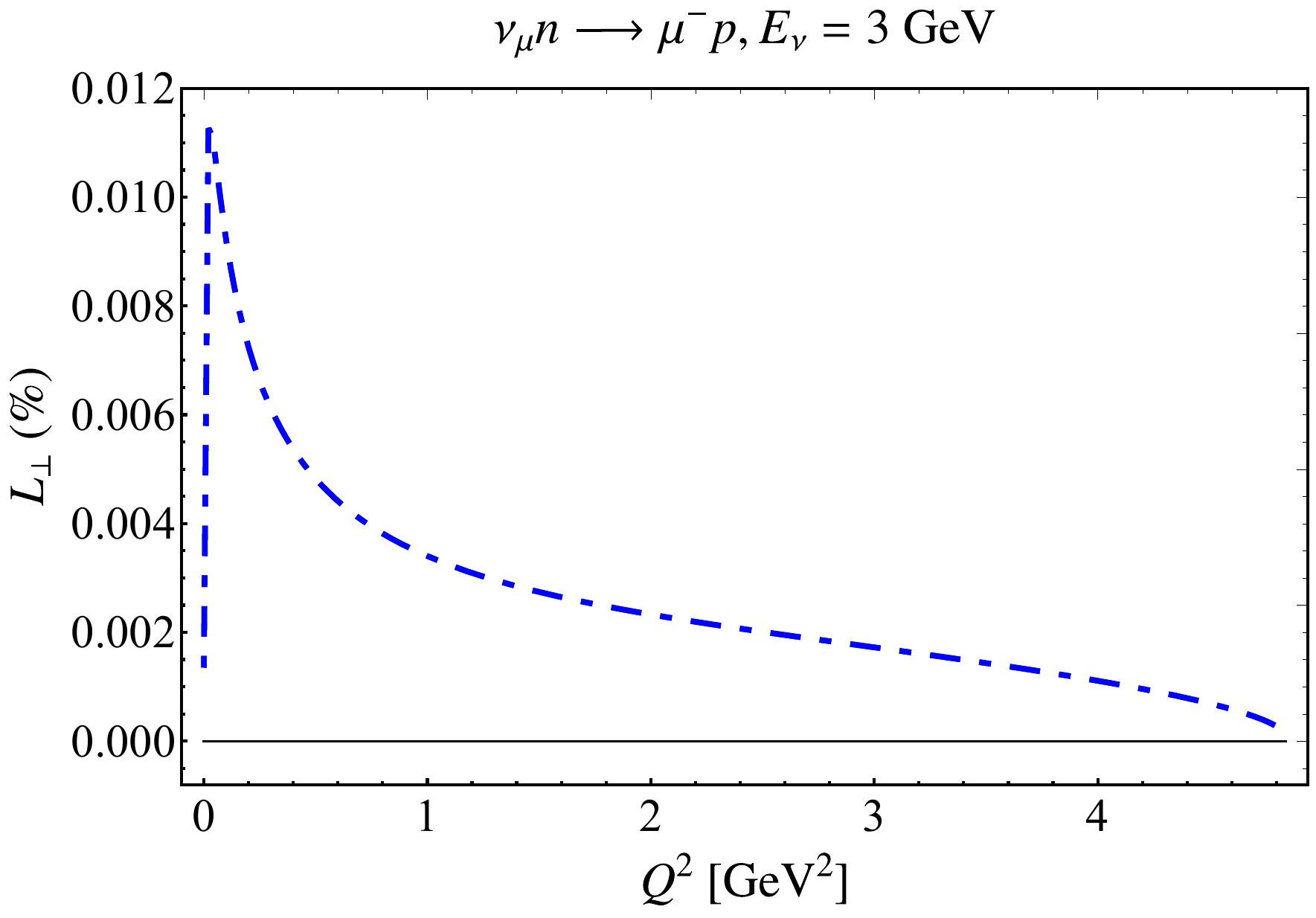}
\caption{Same as Fig.~\ref{fig:nu_Tt_radcorr} but for the transverse polarization observable $L_\perp$. \label{fig:nu_LTT_radcorr}}
\end{figure}

\begin{figure}[H]
\centering
\includegraphics[width=0.4\textwidth]{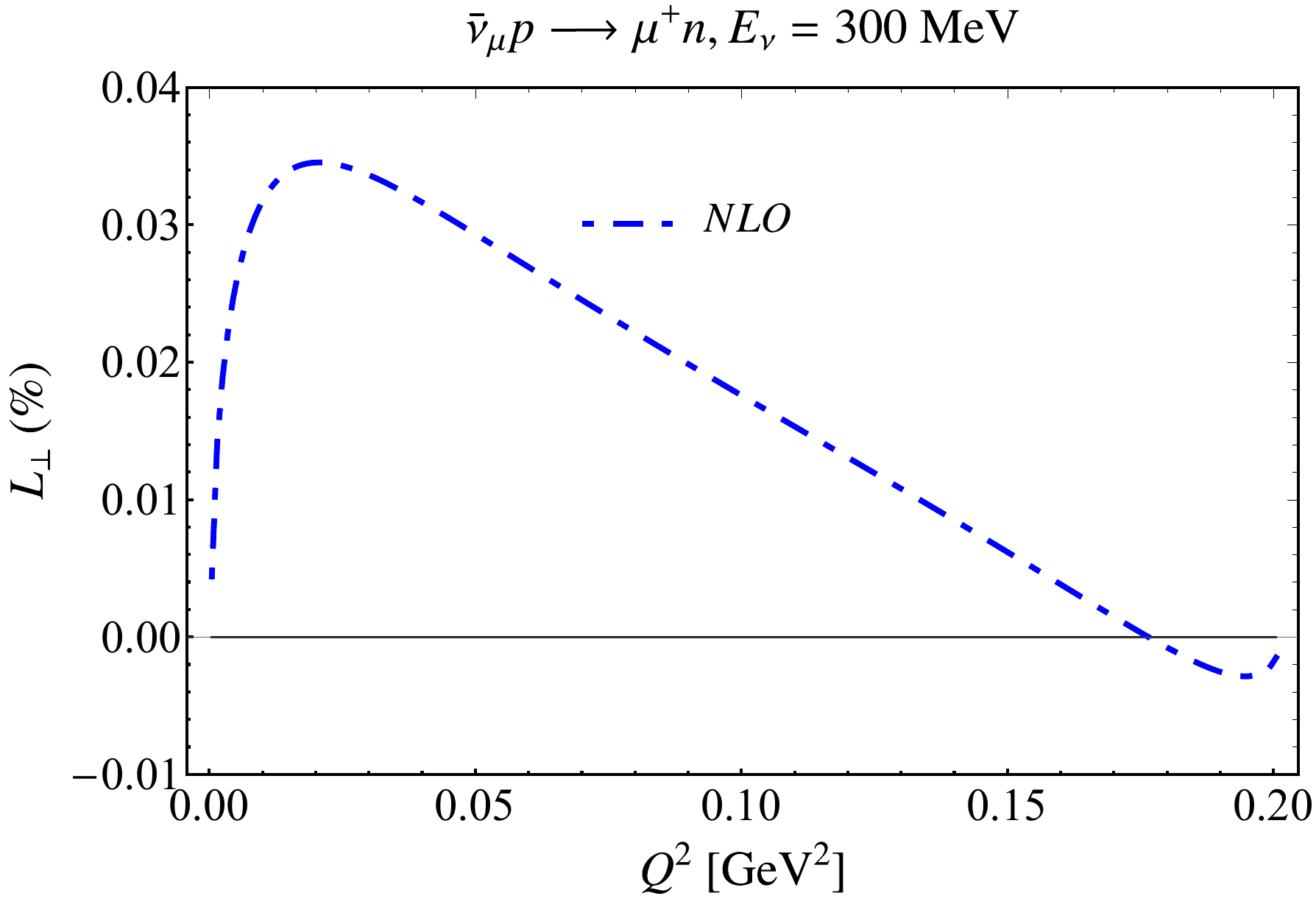}
\includegraphics[width=0.4\textwidth]{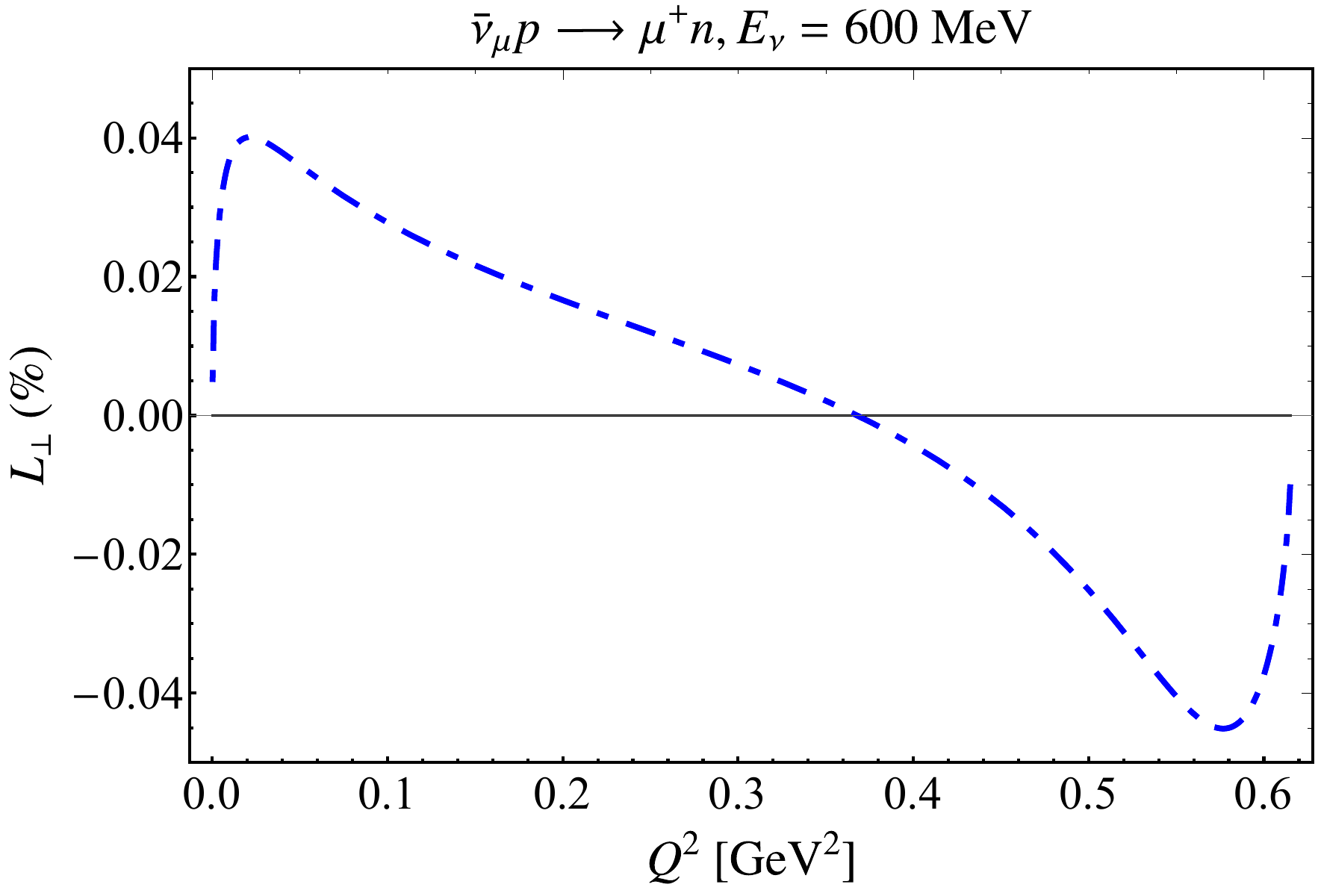}
\includegraphics[width=0.4\textwidth]{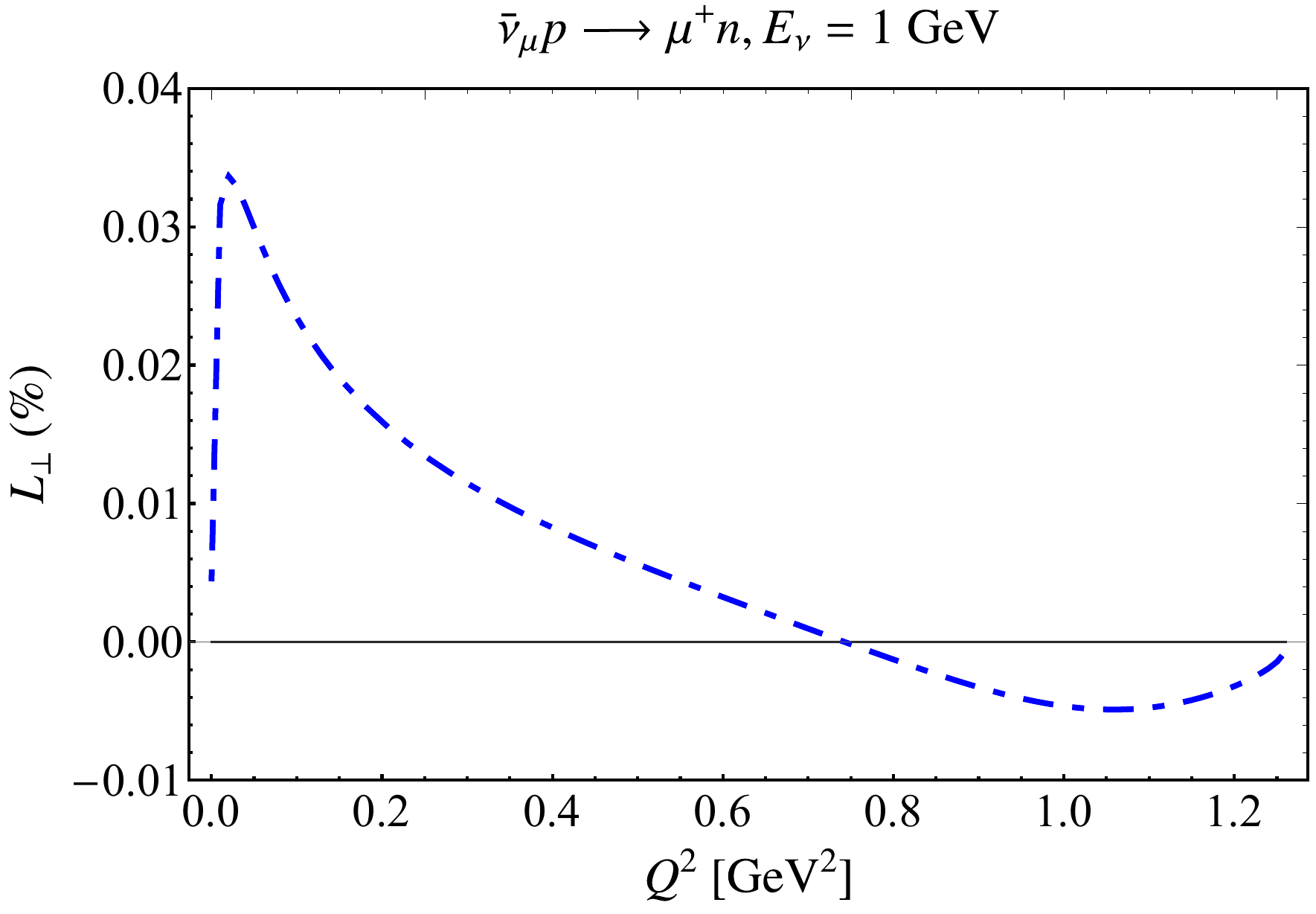}
\includegraphics[width=0.4\textwidth]{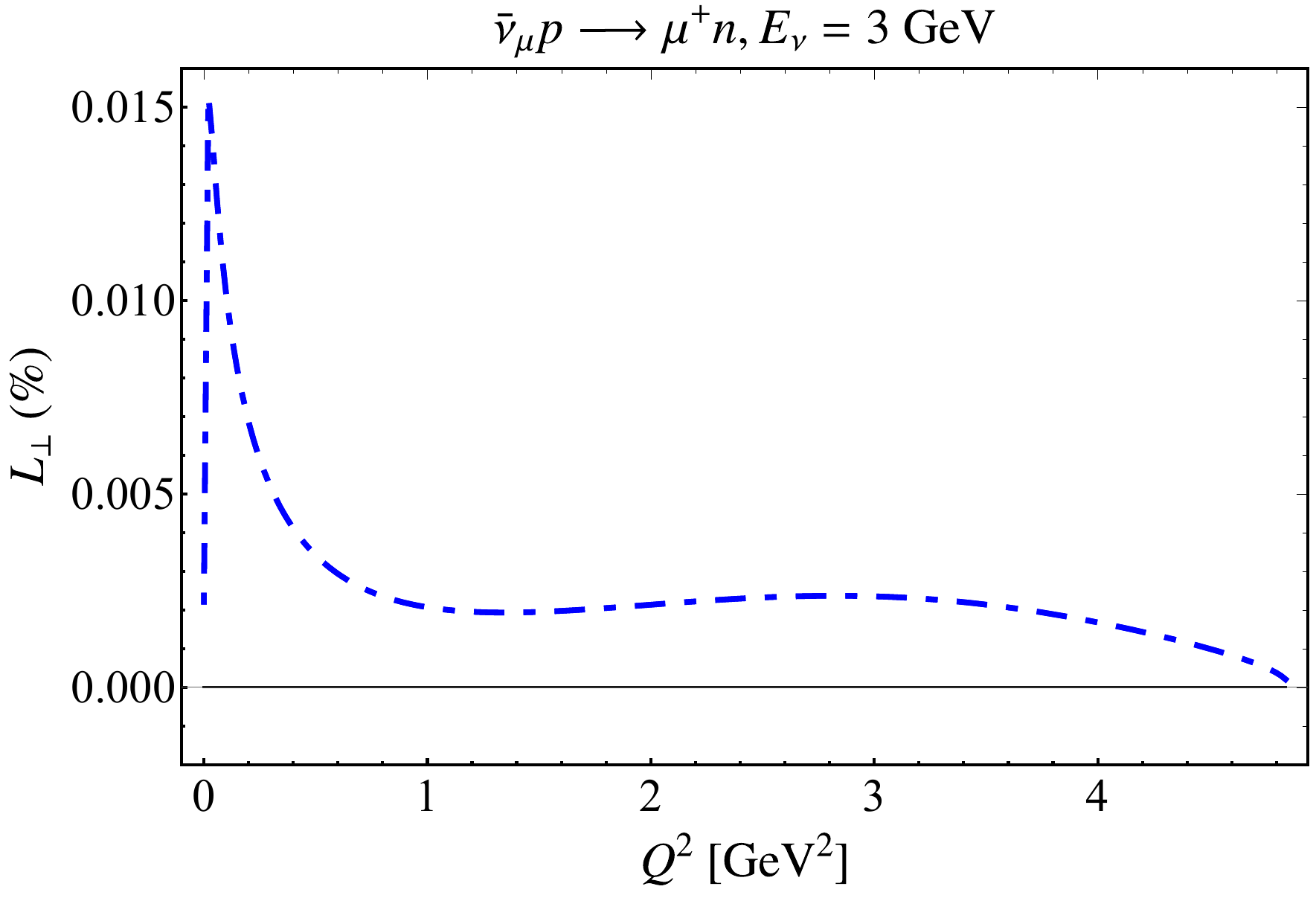}
\caption{Same as Fig.~\ref{fig:antinu_Tt_radcorr} but for the transverse polarization observable $L_\perp$. \label{fig:antinu_LTT_radcorr}}
\end{figure}

\newpage

\subsubsection{Radiative corrections to polarization asymmetries, tau (anti)neutrino}

In this Section, we present single-spin asymmetries for tau neutrinos and antineutrinos, including radiative corrections as described in Sec.~\ref{sec:radiative_corrections}. We consider neutrino energies $E_\nu = 5$~GeV, $7$~GeV, $10$~GeV, and $15$~GeV, and compare to the uncertainty from vector and axial-vector form factors from Sec.~\ref{sec:observables}. 

\begin{figure}[H]
\centering
\includegraphics[width=0.4\textwidth]{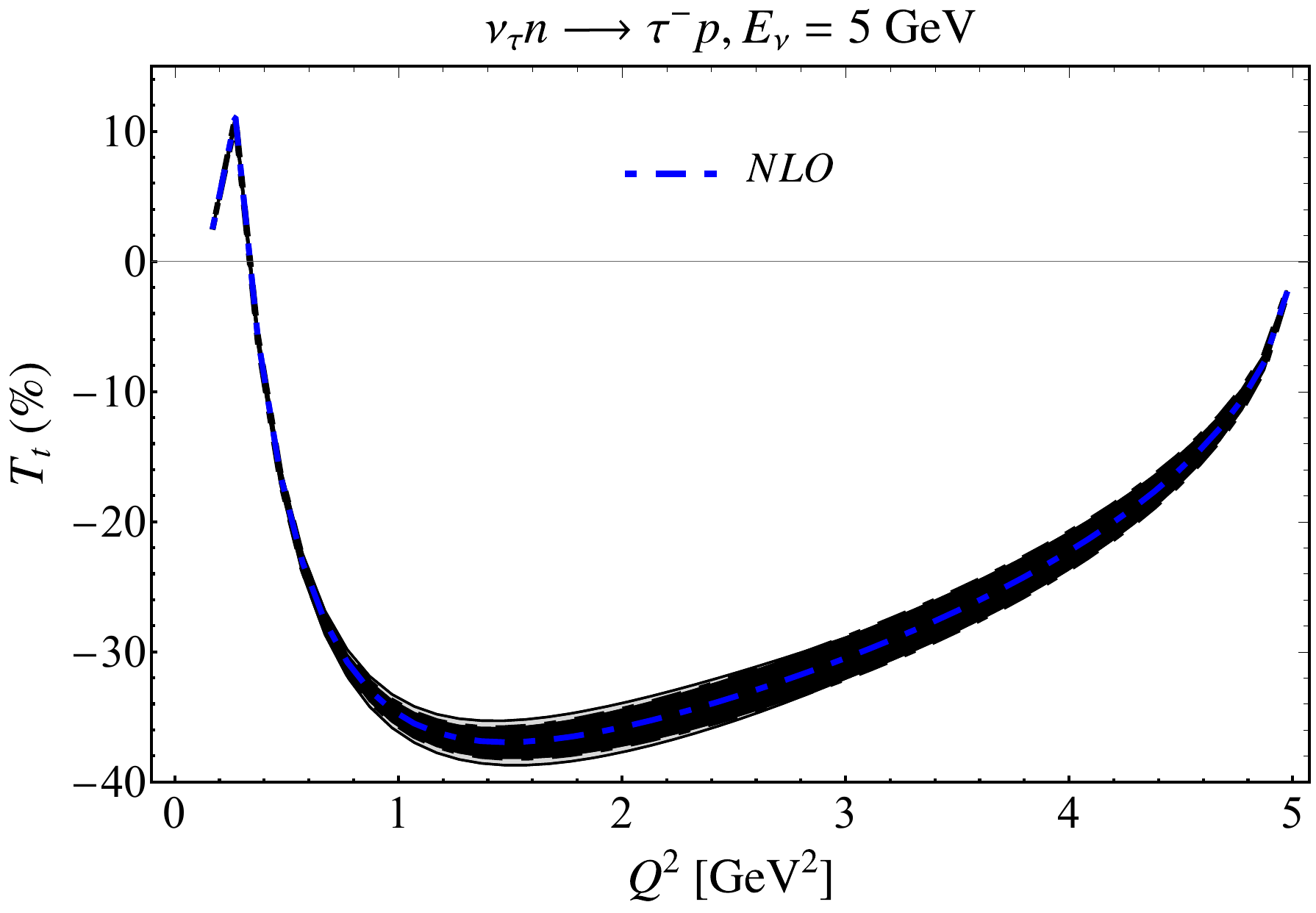}
\includegraphics[width=0.4\textwidth]{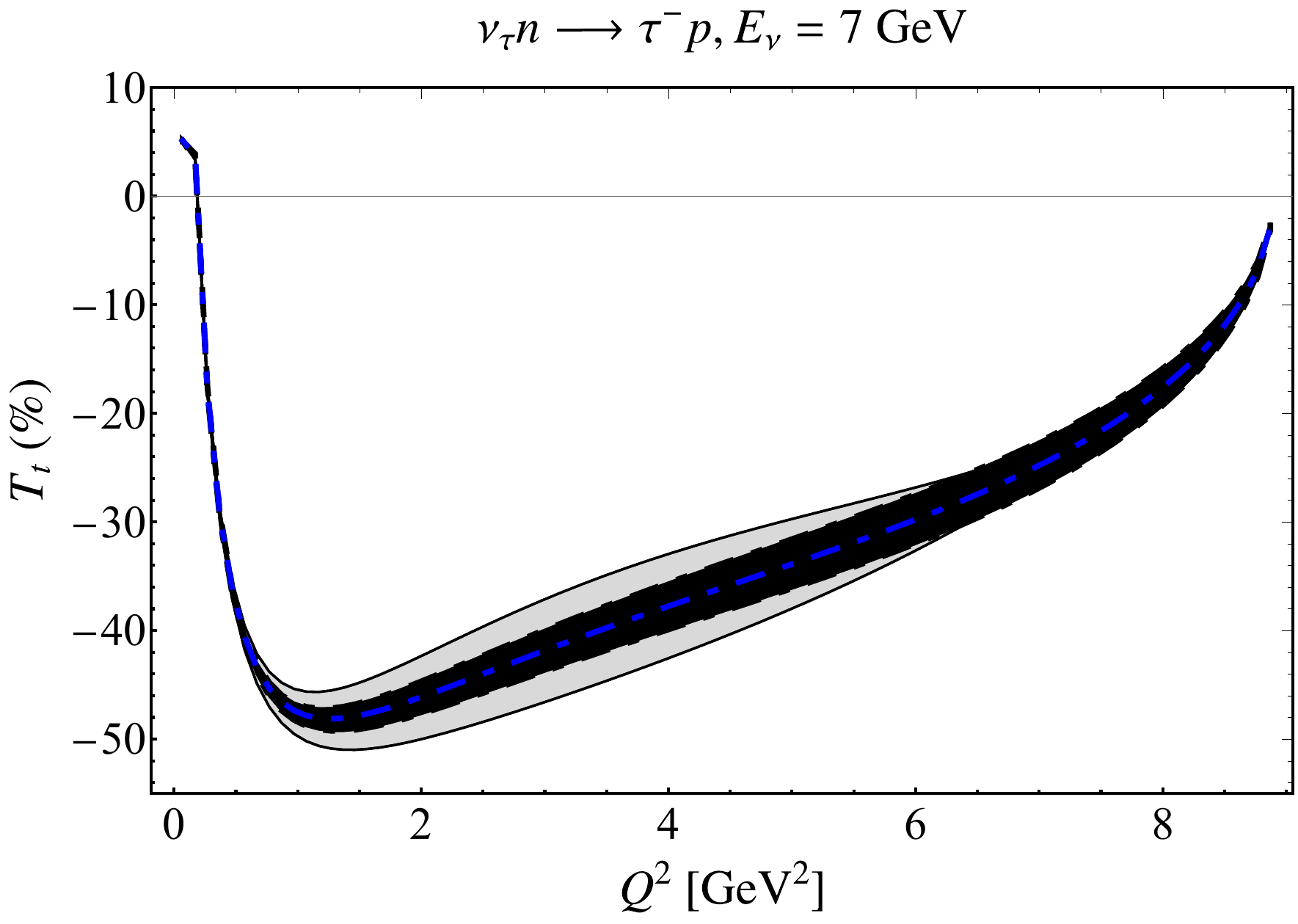}
\includegraphics[width=0.4\textwidth]{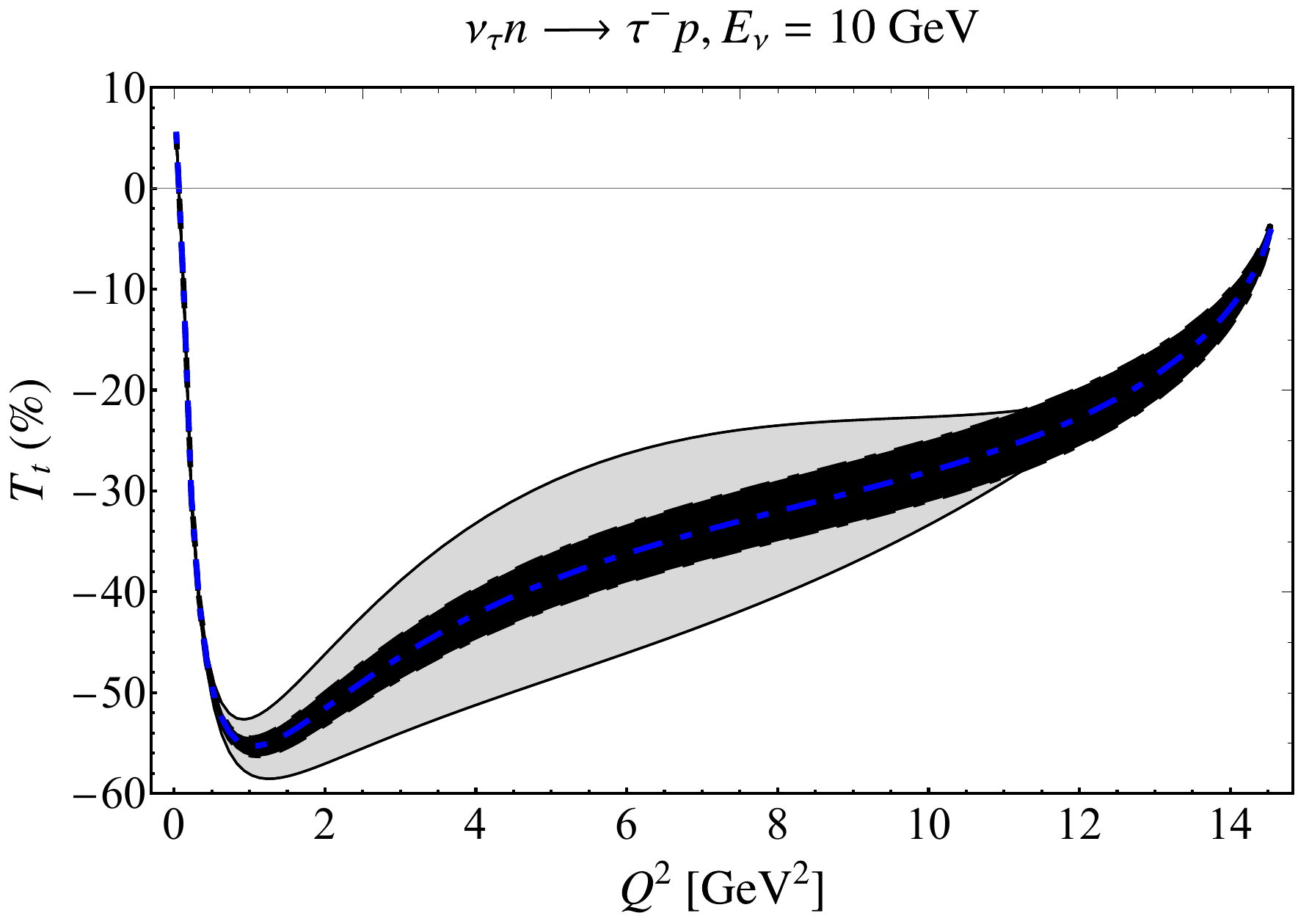}
\includegraphics[width=0.4\textwidth]{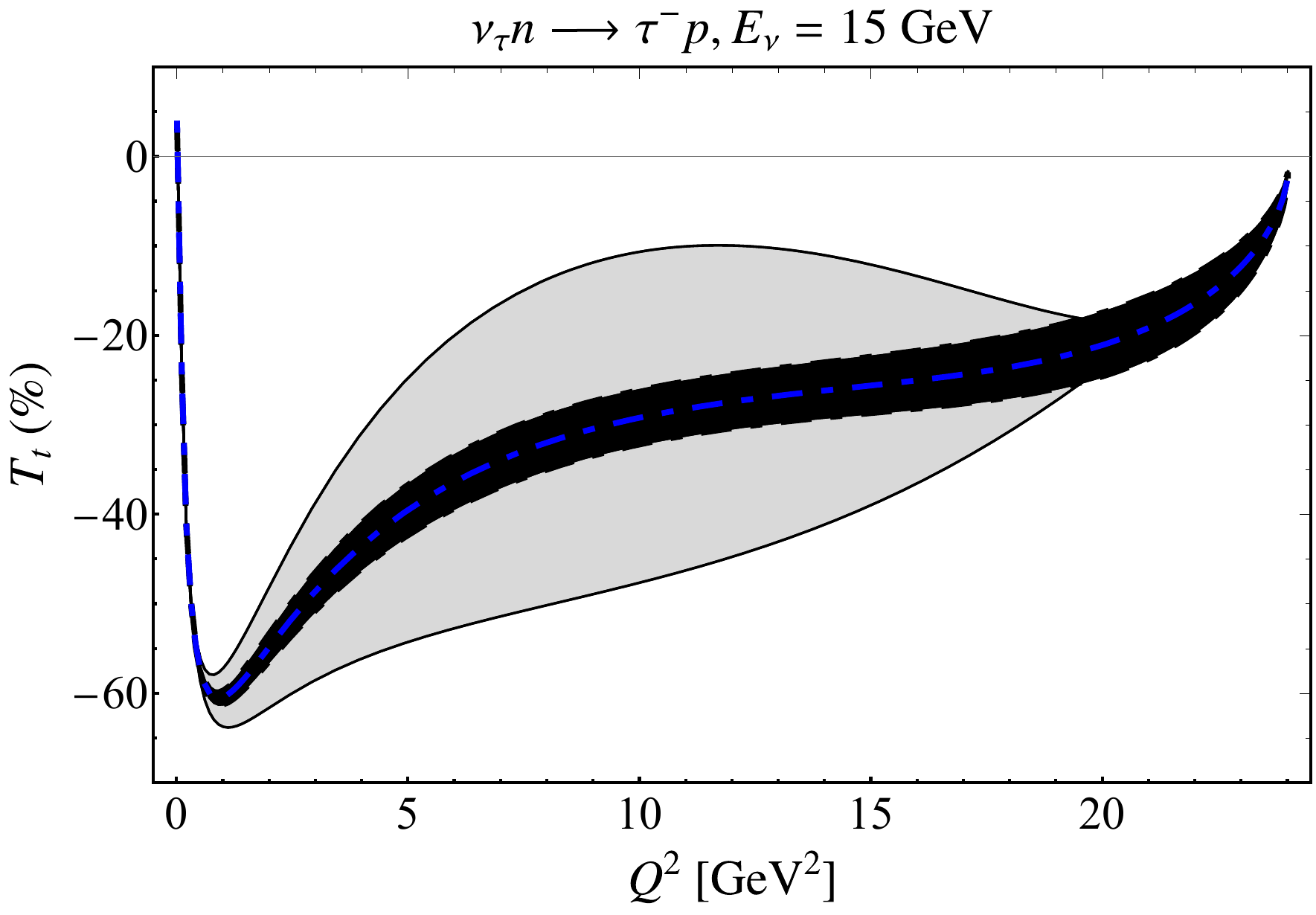}
\caption{Radiative correction to the transverse polarization observable $T_t$, at fixed tau neutrino energies $E_\nu = 5$~GeV, $7$~GeV, $10$~GeV, and $15$~GeV is illustrated. Radiatively-corrected observable, which includes virtual contributions and one real photon of energy below $10~\mathrm{MeV}$, is shown by the blue dashed-dotted line. The dark black and light gray bands correspond to vector and axial-vector form factor uncertainty. \label{fig:nu_Tt_radcorr_tau}}
\end{figure}

\begin{figure}[H]
\centering
\includegraphics[width=0.4\textwidth]{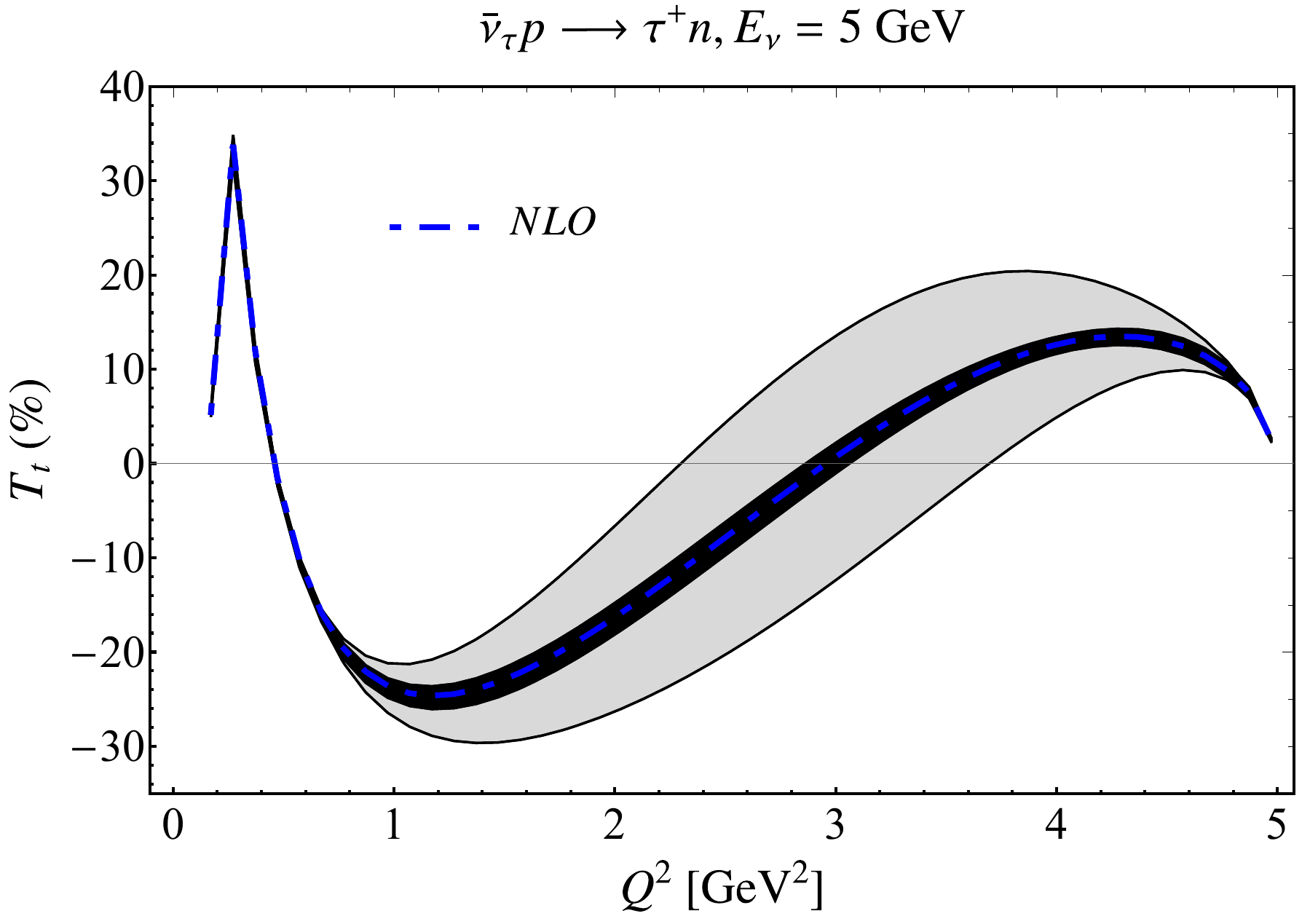}
\includegraphics[width=0.4\textwidth]{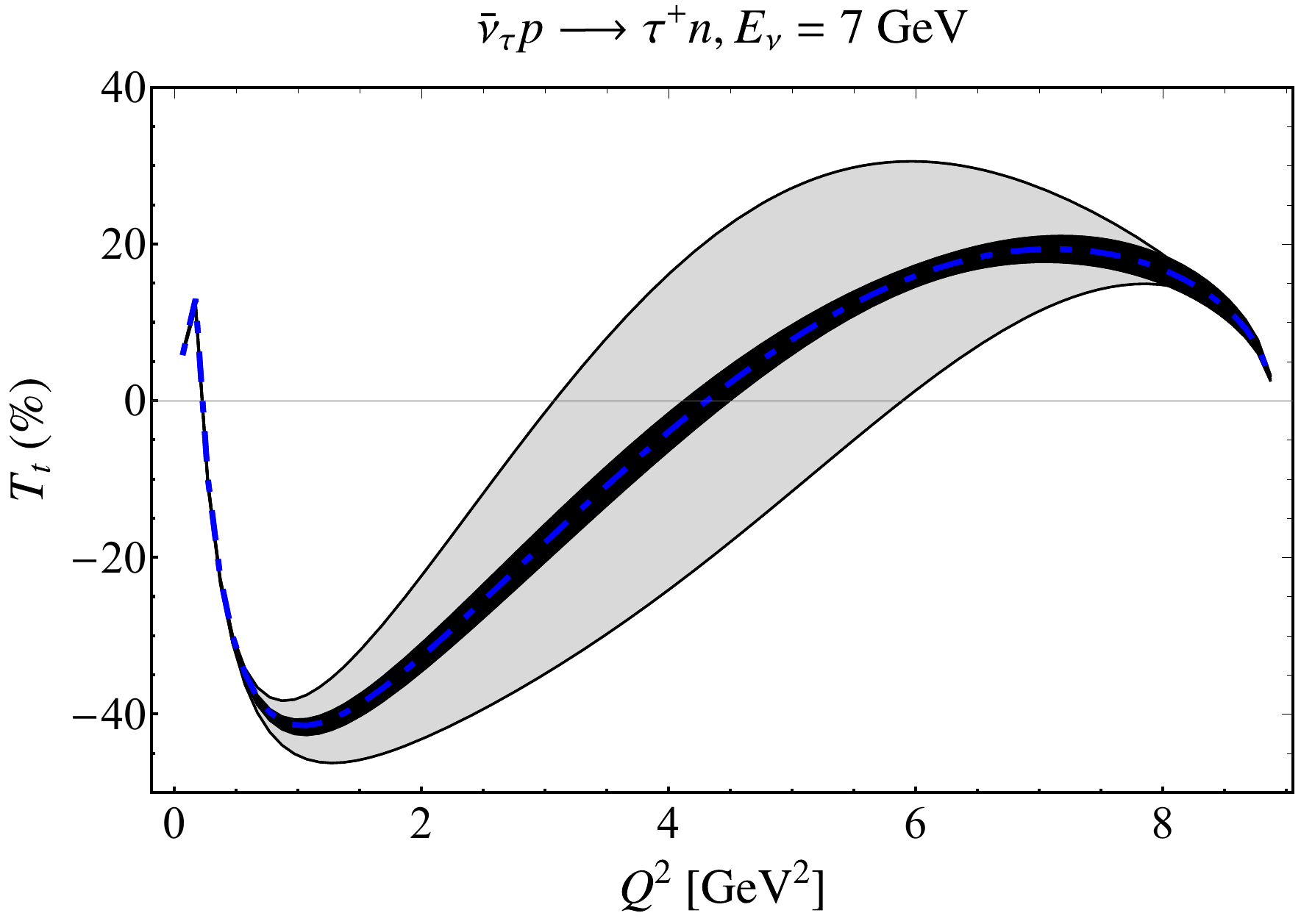}
\includegraphics[width=0.4\textwidth]{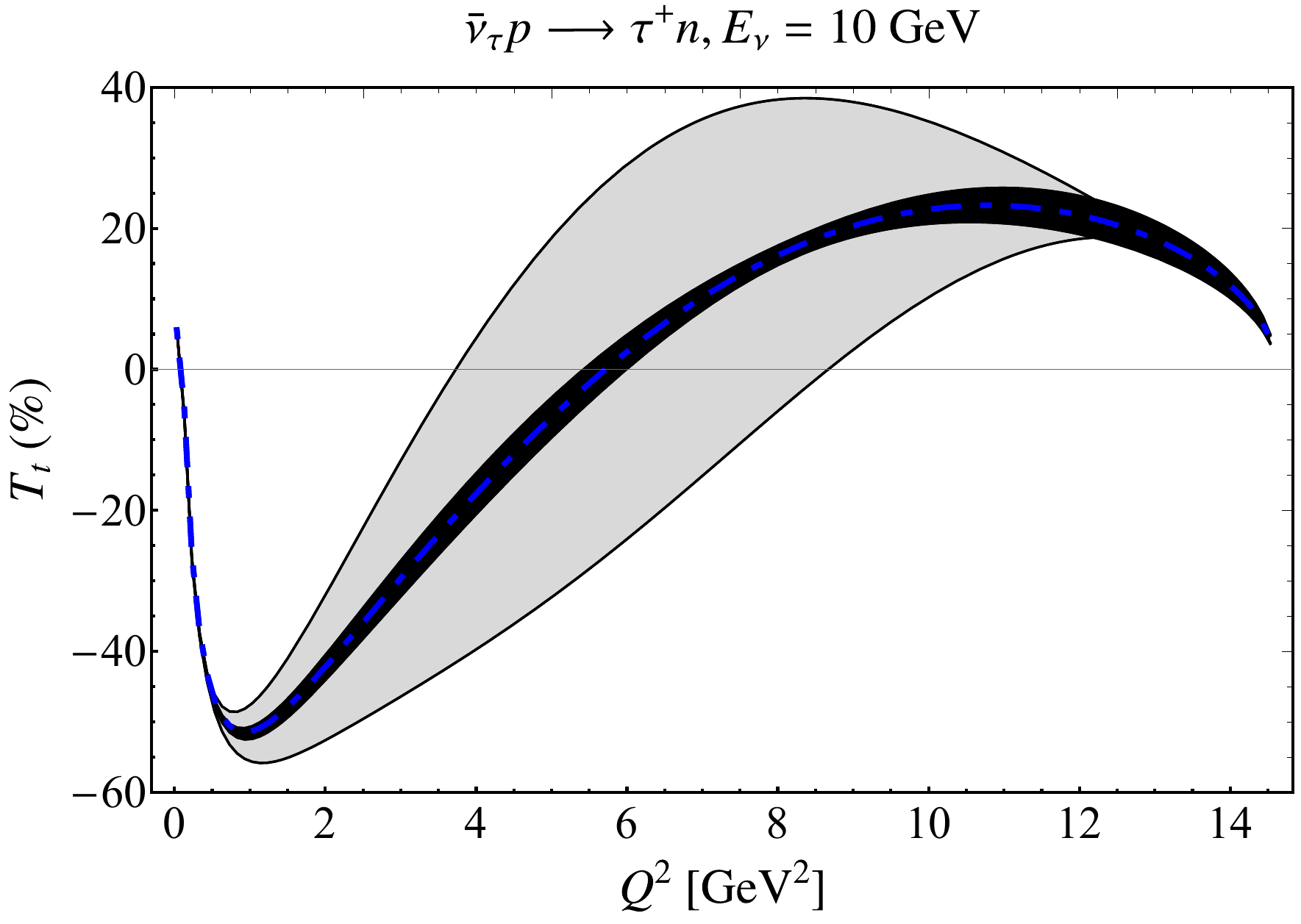}
\includegraphics[width=0.4\textwidth]{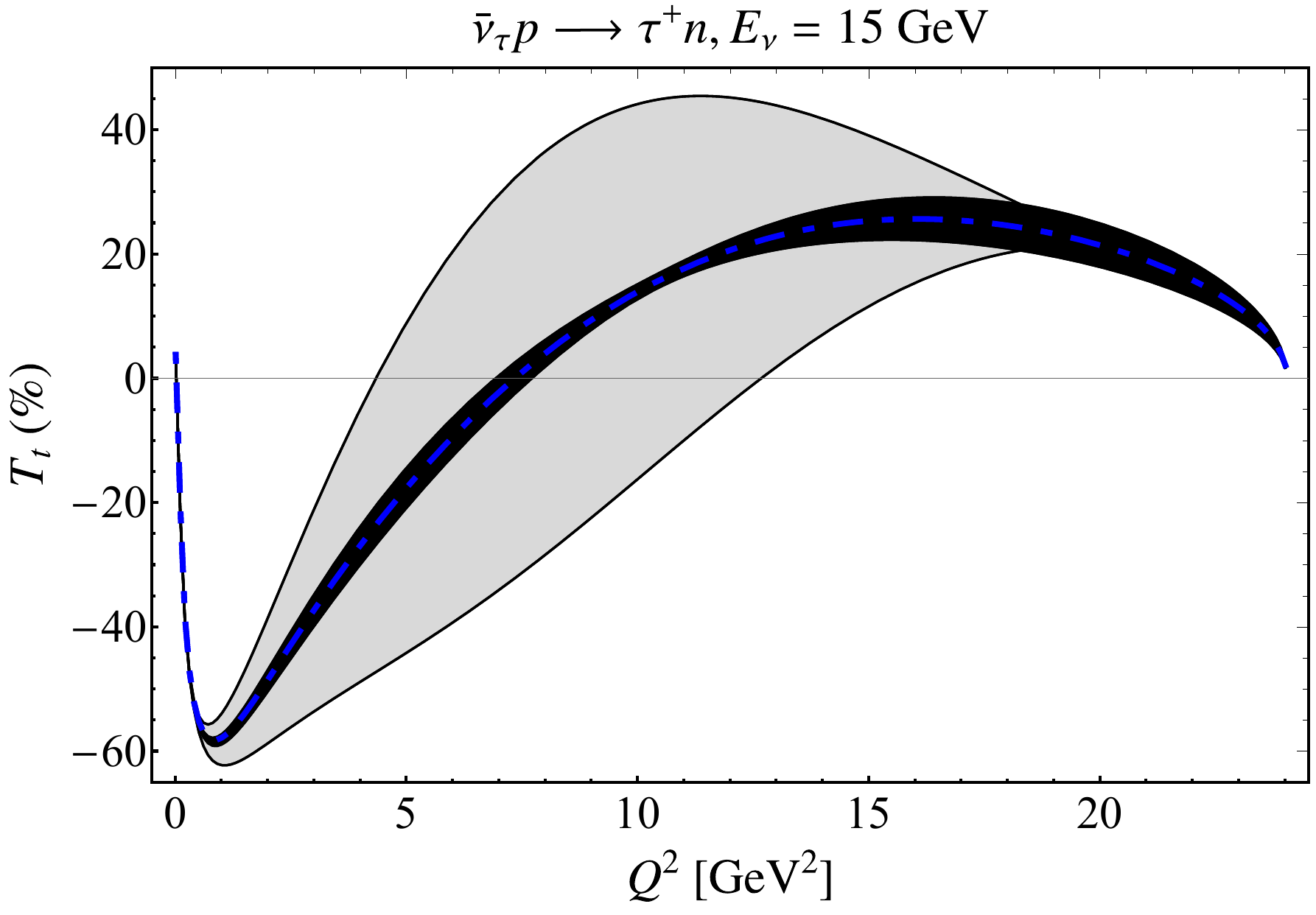}
\caption{Same as Fig.~\ref{fig:nu_Tt_radcorr_tau} but for tau antineutrinos. \label{fig:antinu_Tt_radcorr_tau}}
\end{figure}

\begin{figure}[H]
\centering
\includegraphics[width=0.4\textwidth]{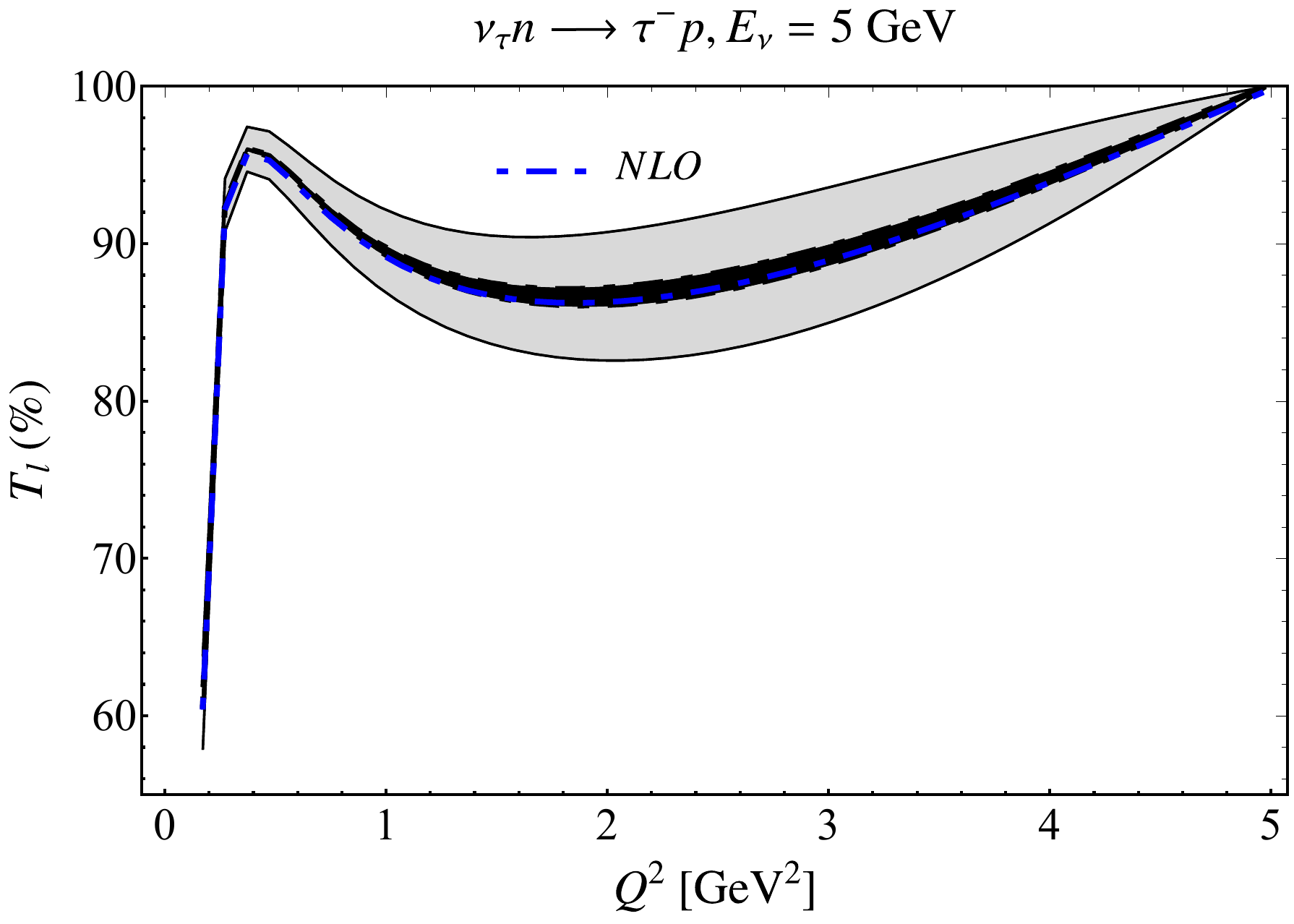}
\includegraphics[width=0.4\textwidth]{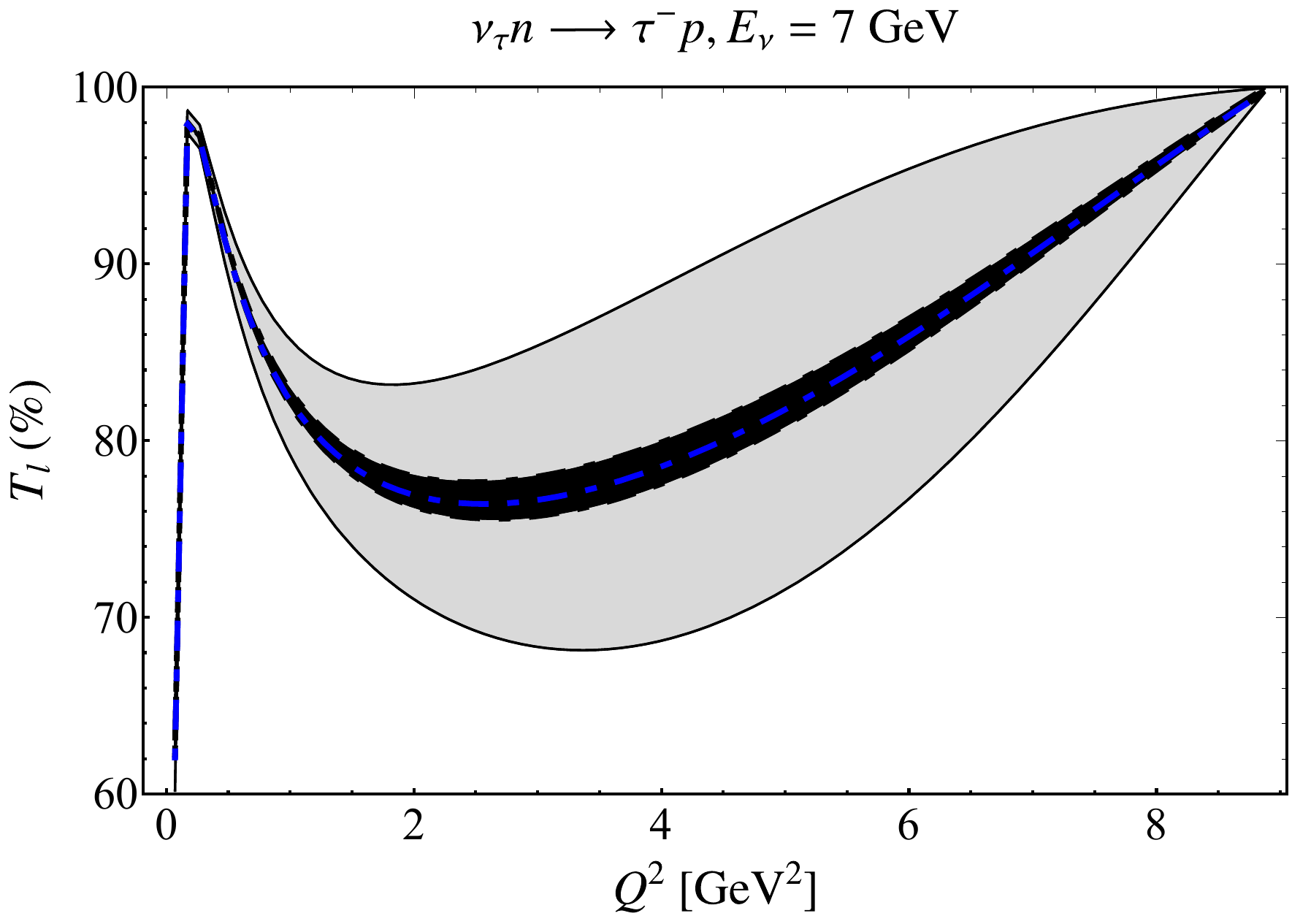}
\includegraphics[width=0.4\textwidth]{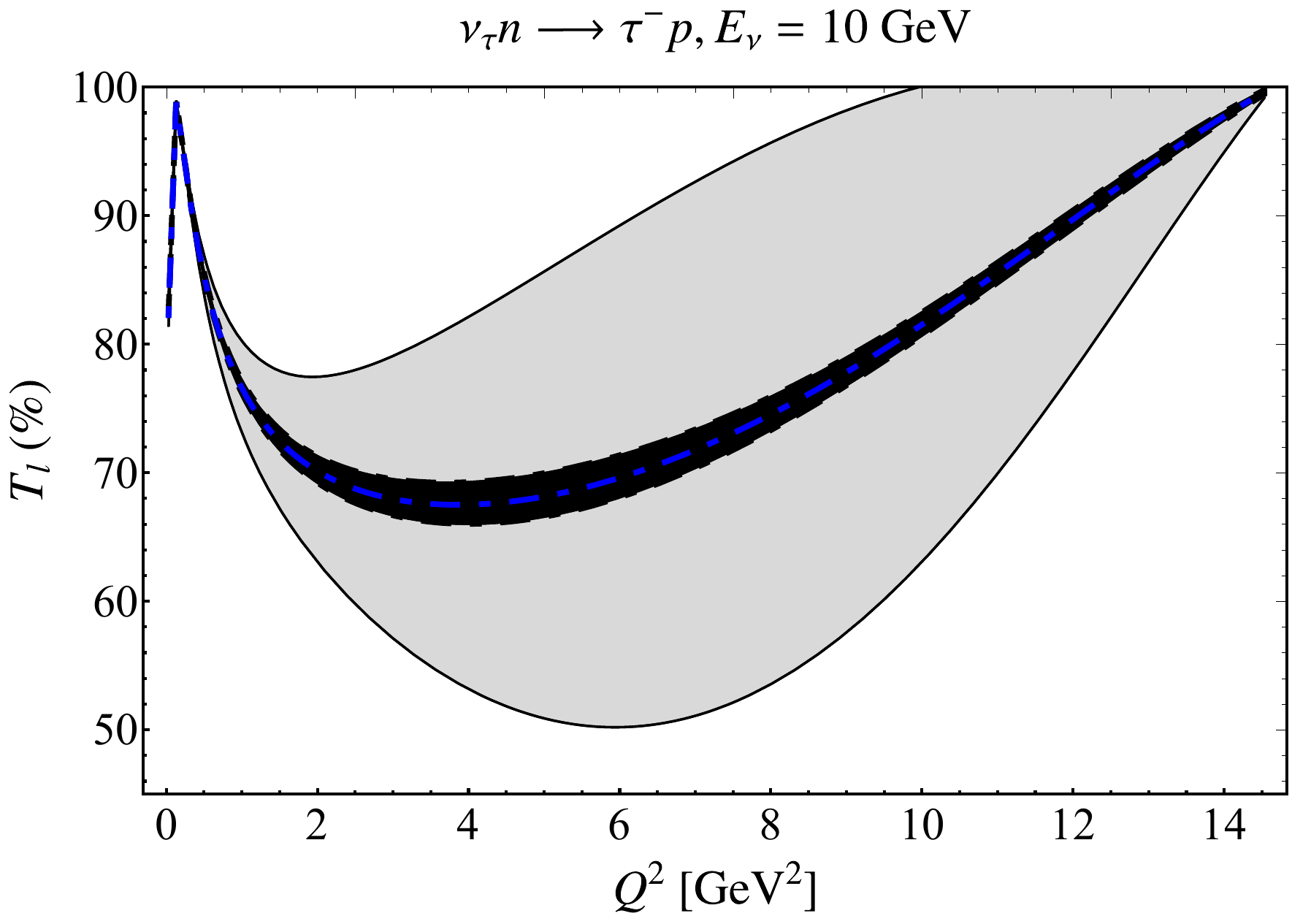}
\includegraphics[width=0.4\textwidth]{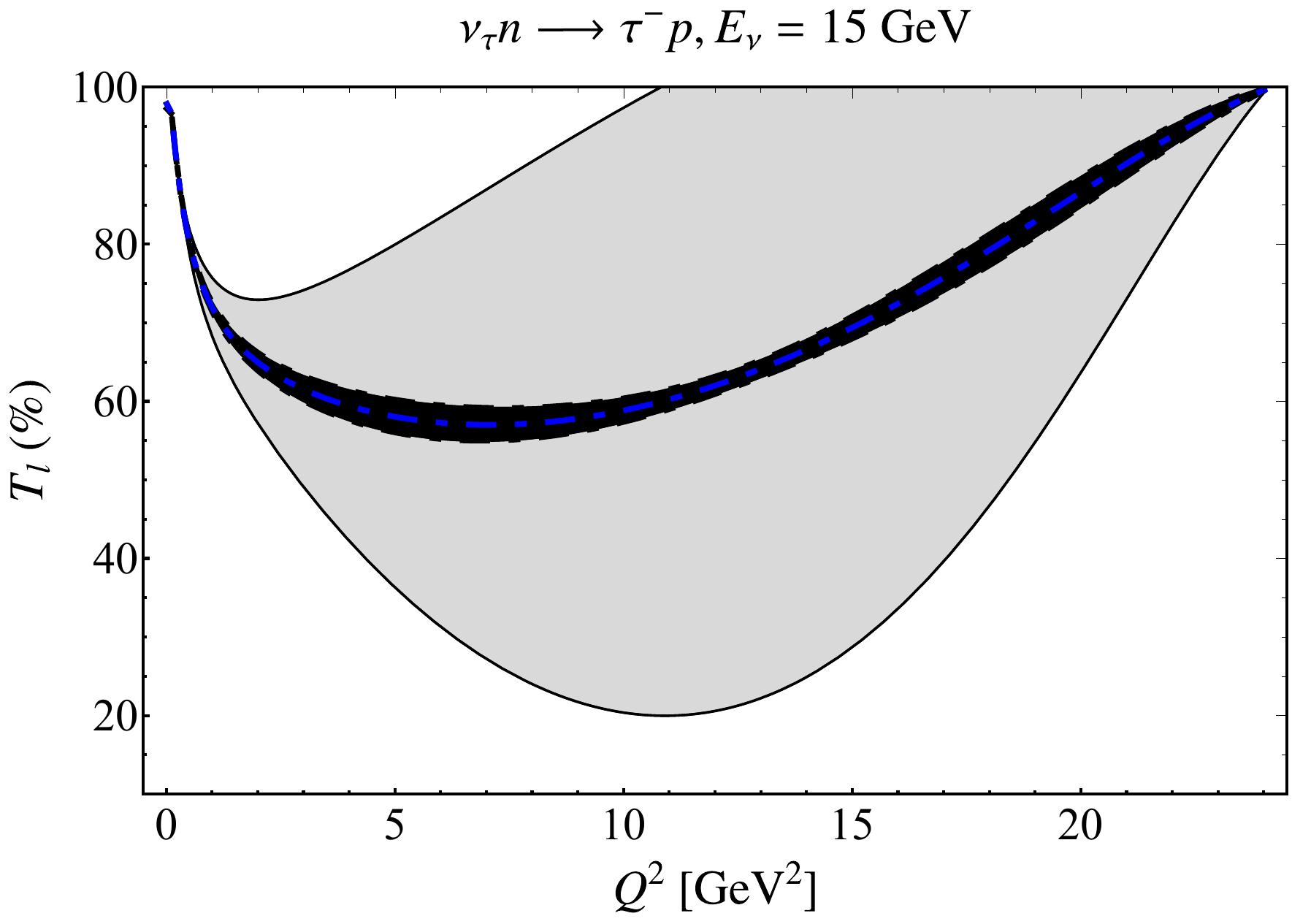}
\caption{Same as Fig.~\ref{fig:nu_Tt_radcorr_tau} but for the longitudinal polarization observable $T_l$. \label{fig:nu_Tl_radcorr_tau}}
\end{figure}

\begin{figure}[H]
\centering
\includegraphics[width=0.4\textwidth]{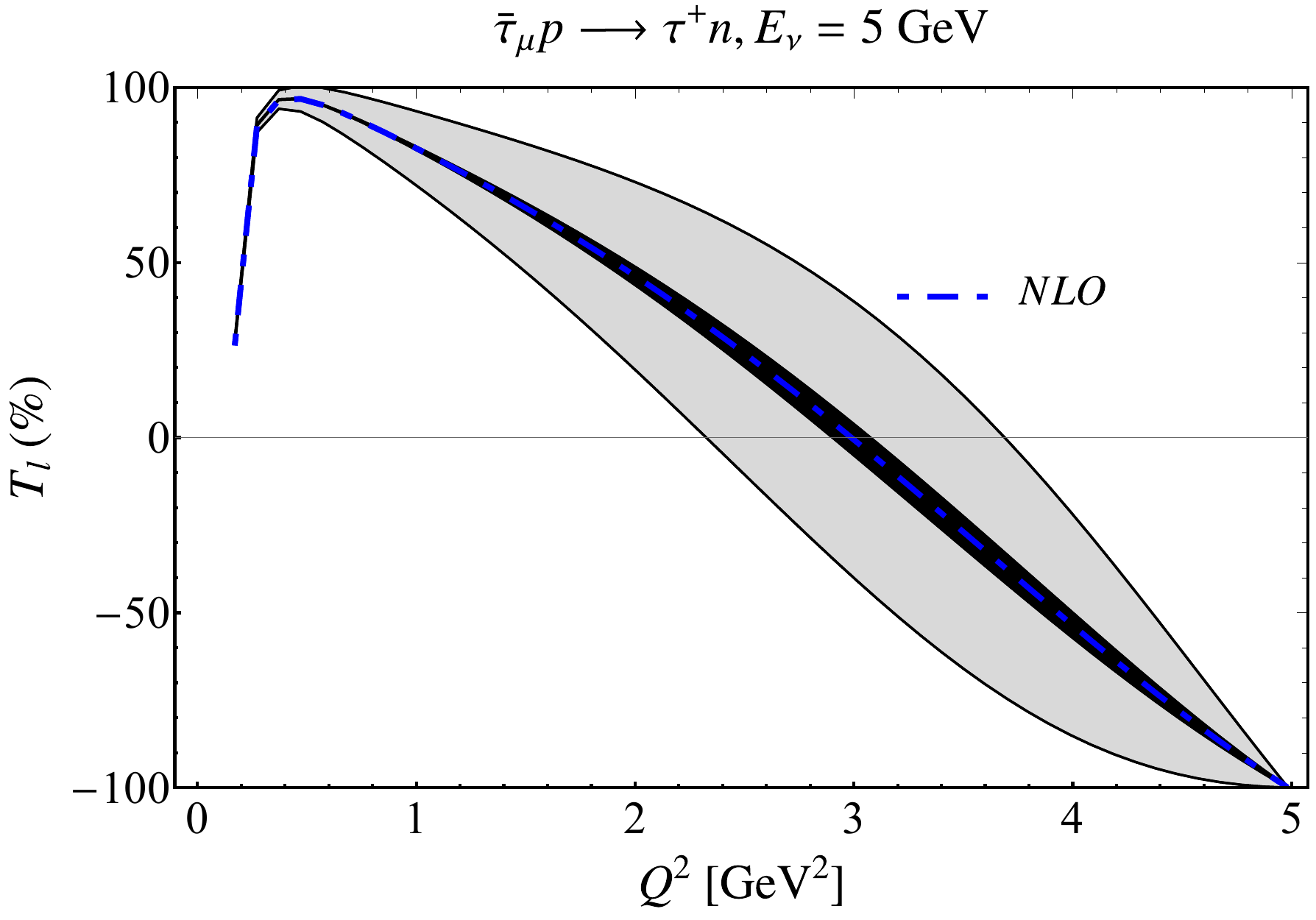}
\includegraphics[width=0.4\textwidth]{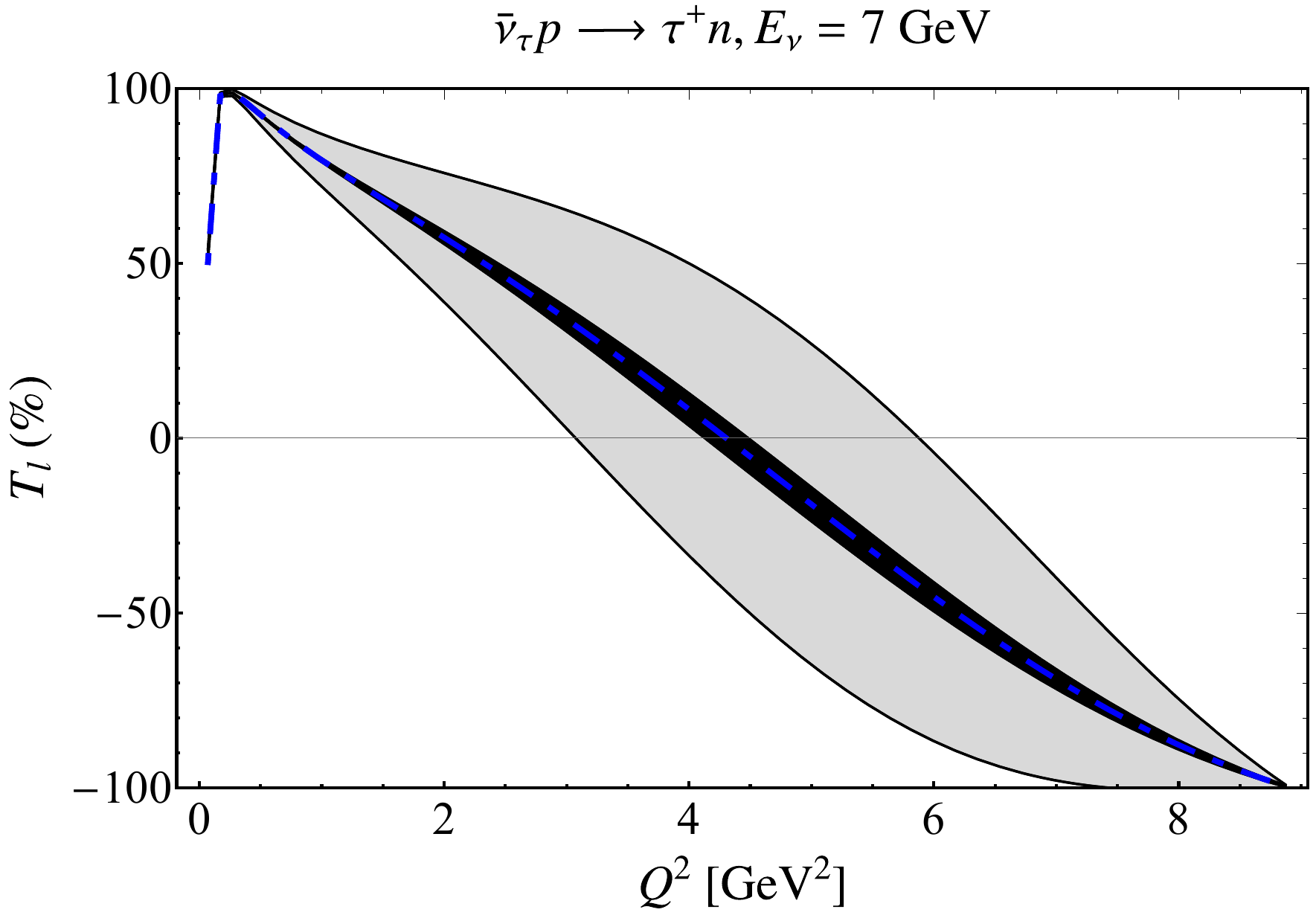}
\includegraphics[width=0.4\textwidth]{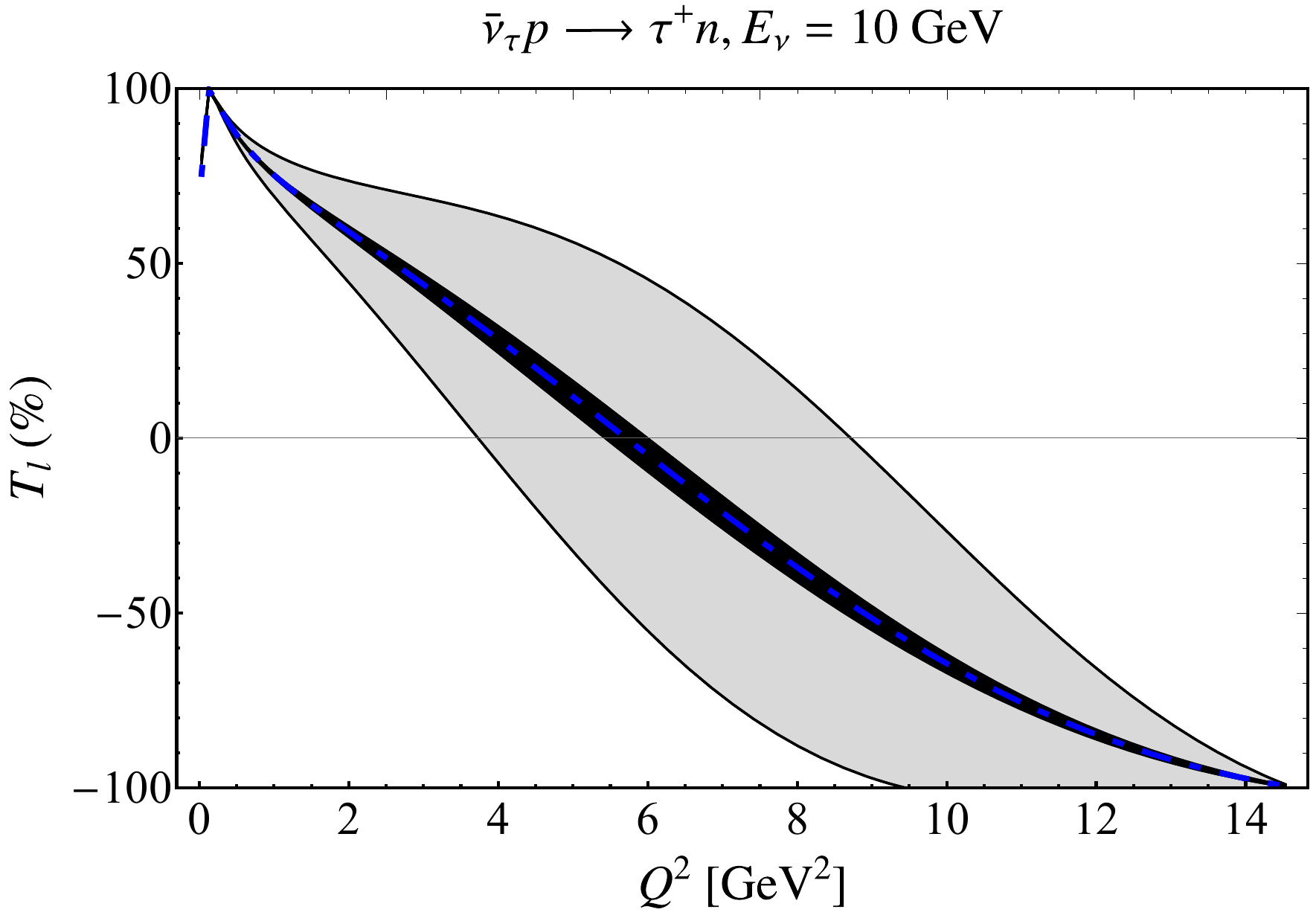}
\includegraphics[width=0.4\textwidth]{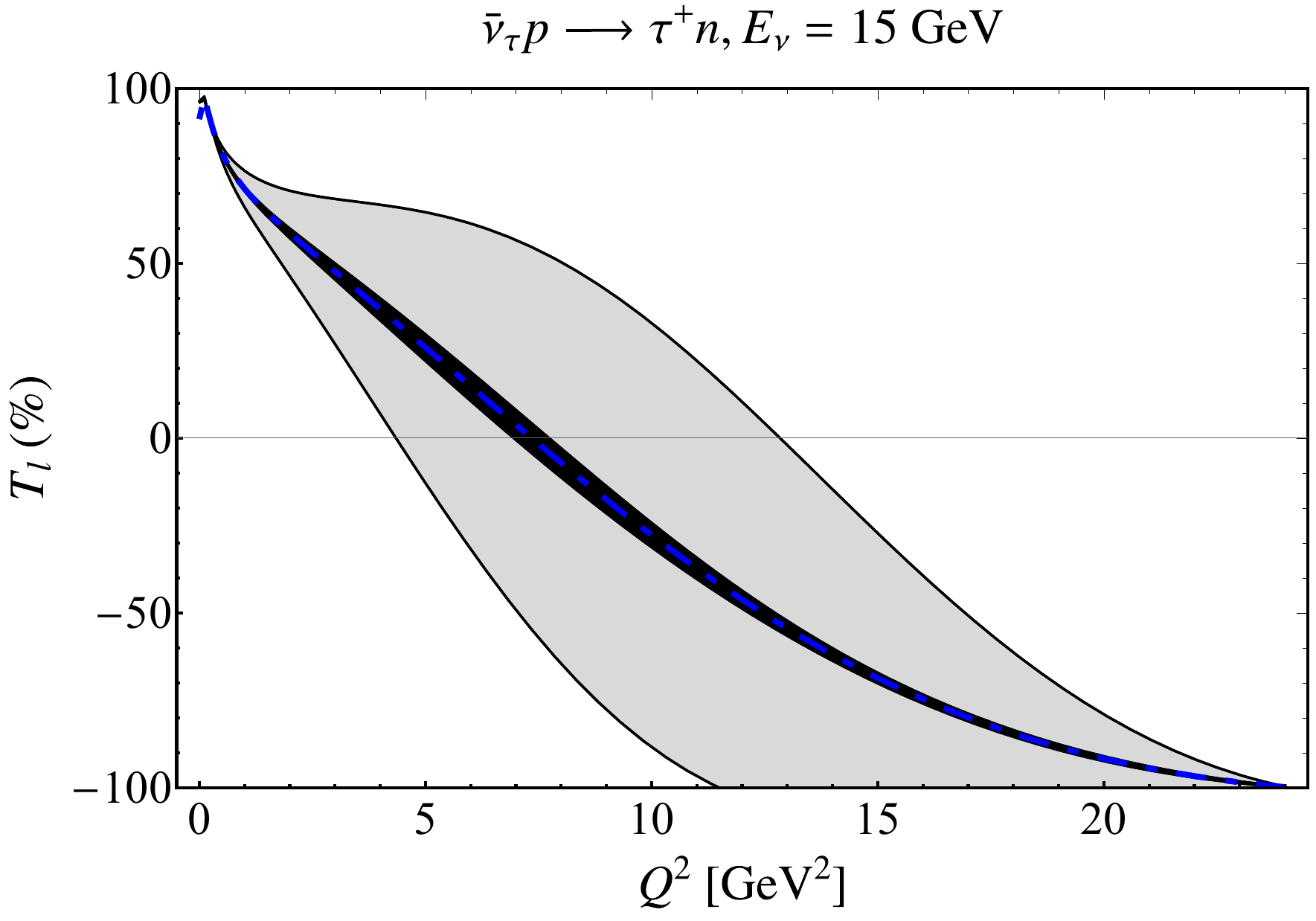}
\caption{Same as Fig.~\ref{fig:antinu_Tt_radcorr_tau} but for the longitudinal polarization observable $T_l$. \label{fig:antinu_Tl_radcorr_tau}}
\end{figure}

\begin{figure}[H]
\centering
\includegraphics[width=0.4\textwidth]{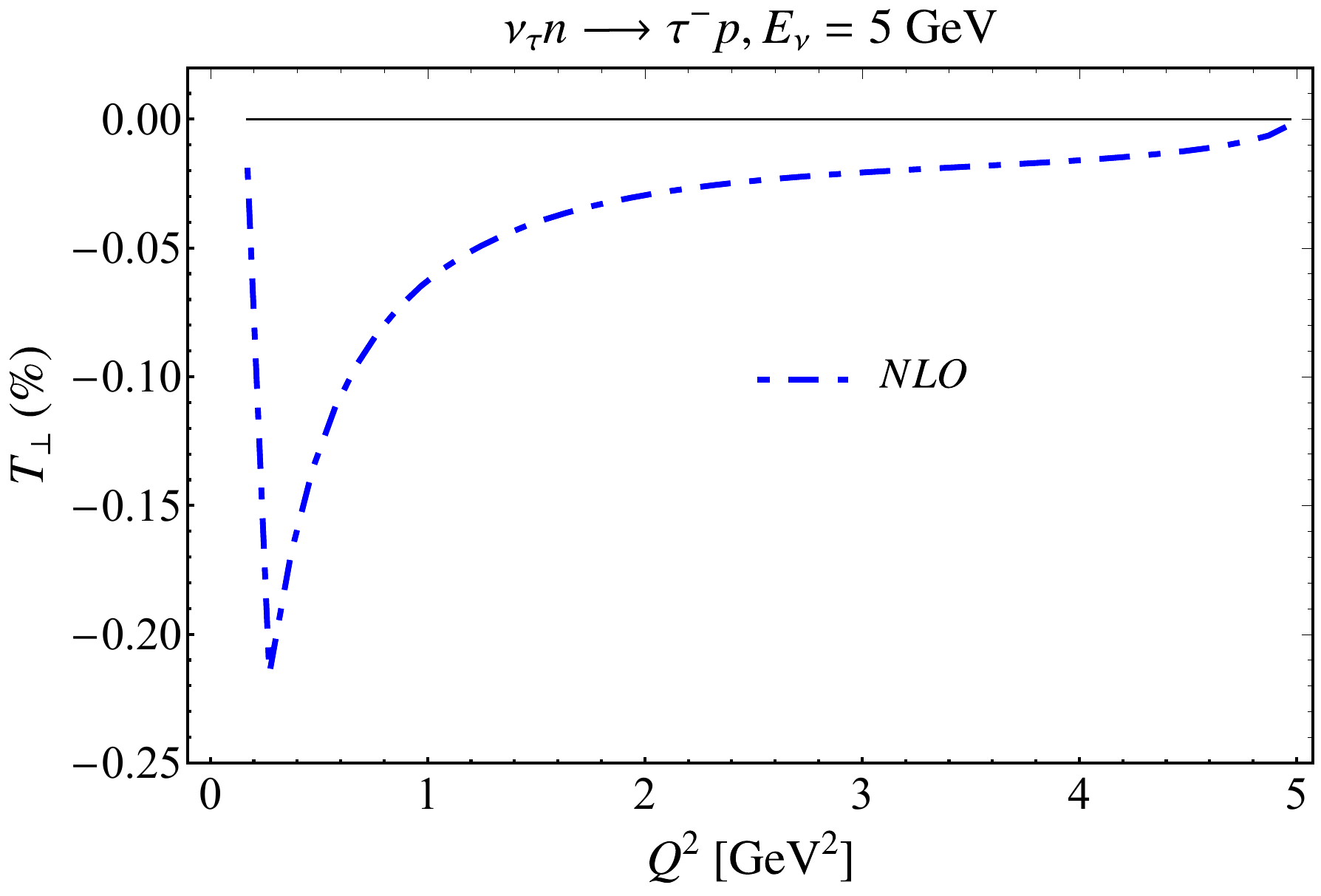}
\includegraphics[width=0.4\textwidth]{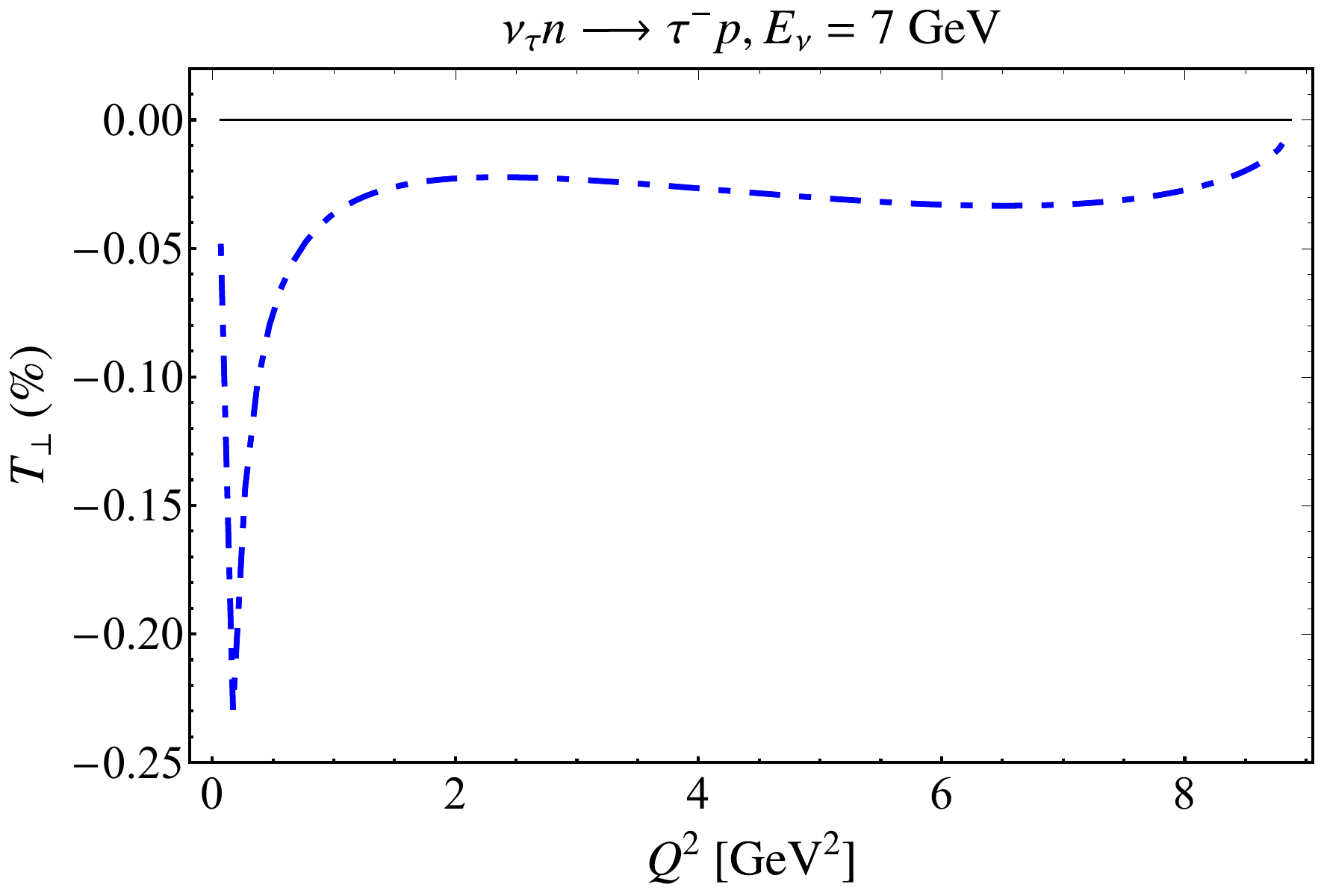}
\includegraphics[width=0.4\textwidth]{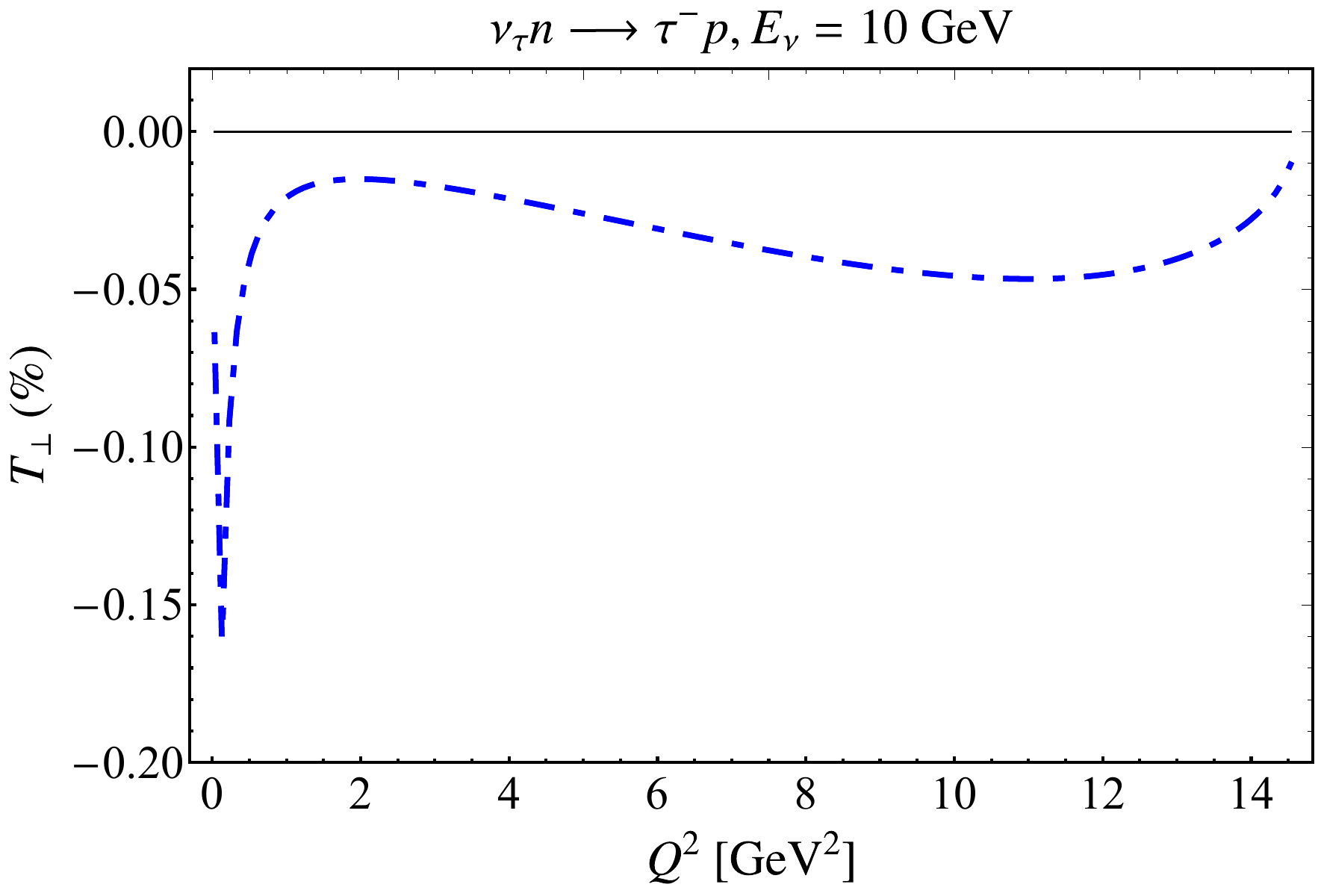}
\includegraphics[width=0.4\textwidth]{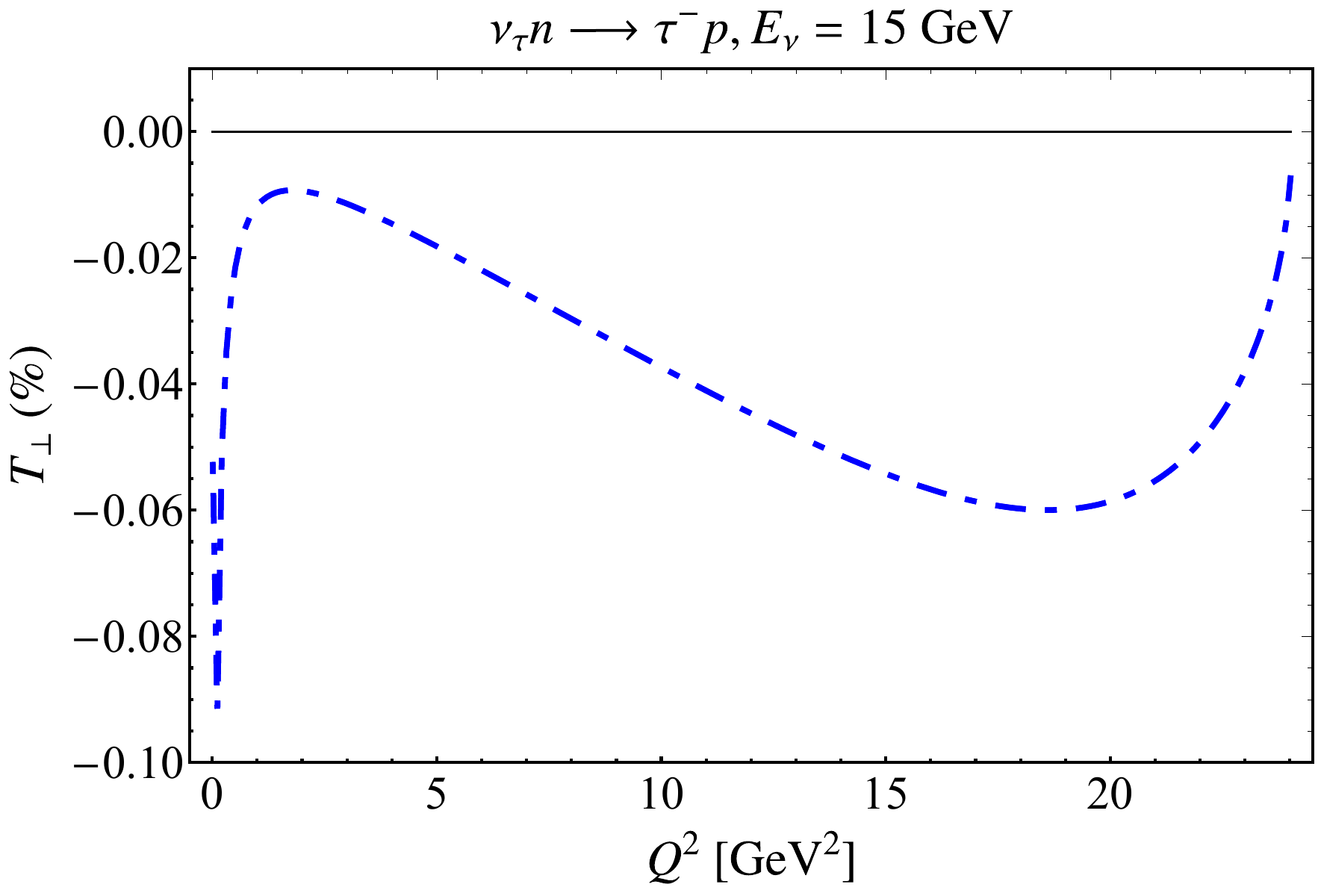}
\caption{Same as Fig.~\ref{fig:nu_Tt_radcorr_tau} but for the transverse polarization observable $T_\perp$. \label{fig:nu_TTT_radcorr_tau}}
\end{figure}

\begin{figure}[H]
\centering
\includegraphics[width=0.4\textwidth]{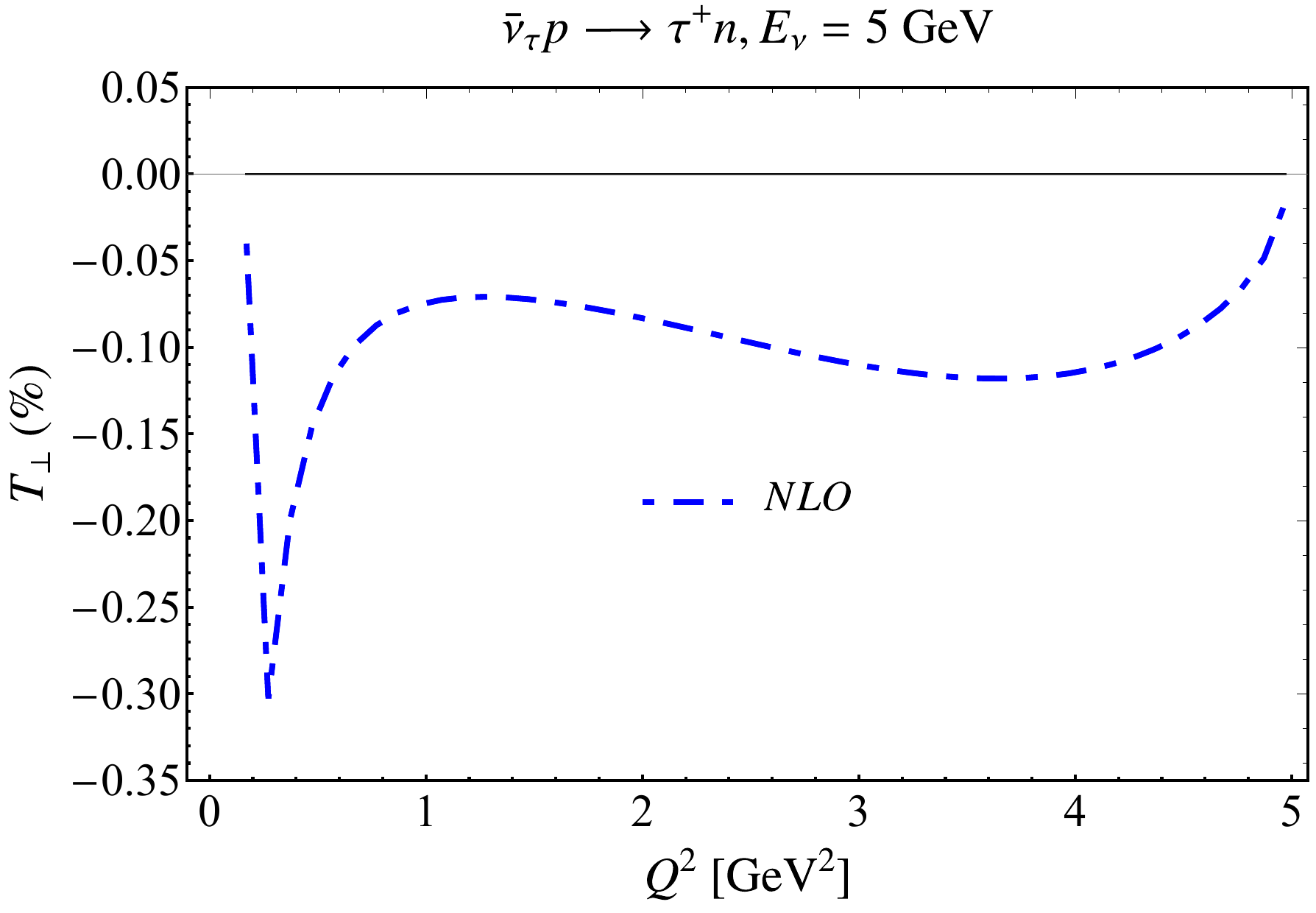}
\includegraphics[width=0.4\textwidth]{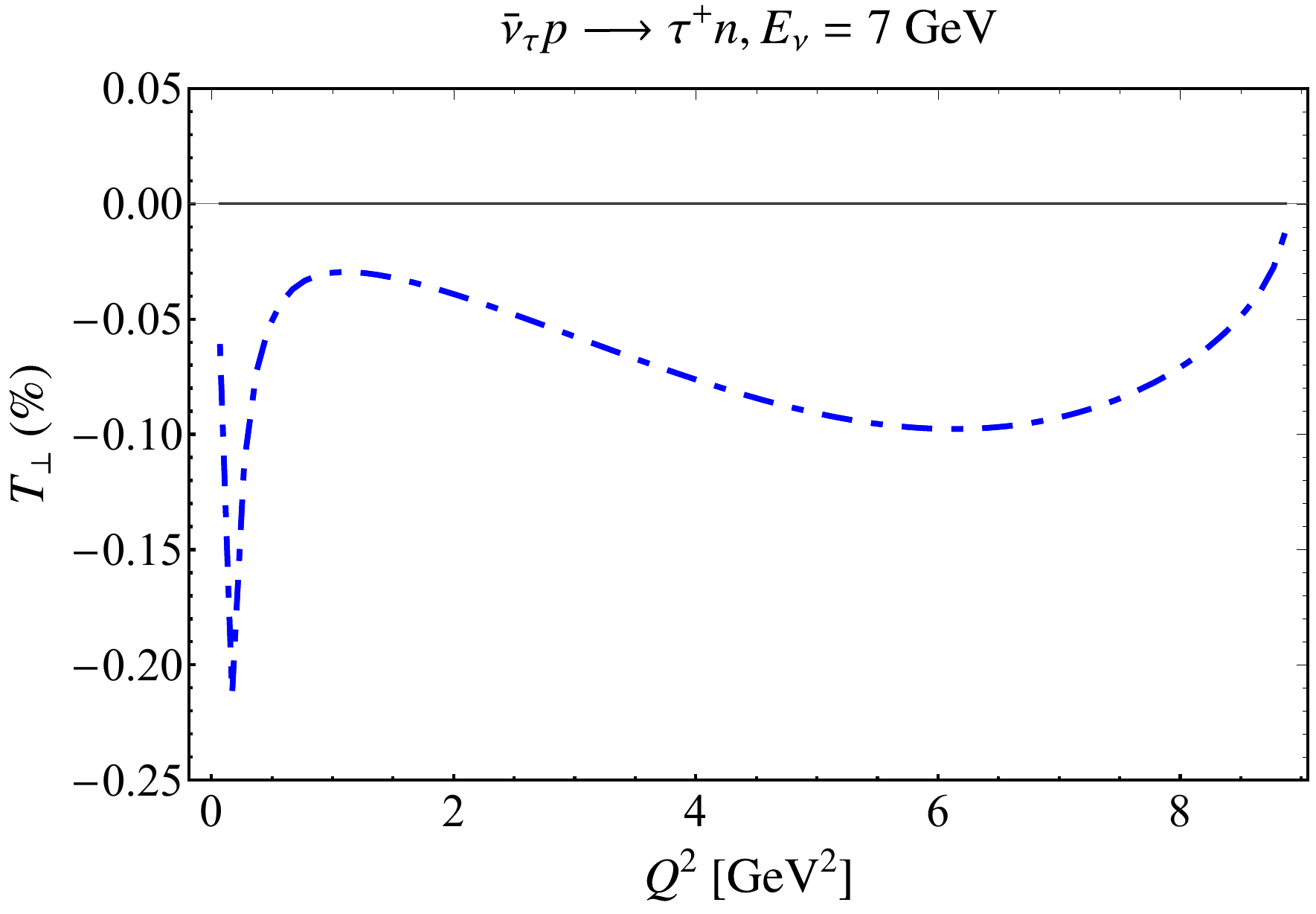}
\includegraphics[width=0.4\textwidth]{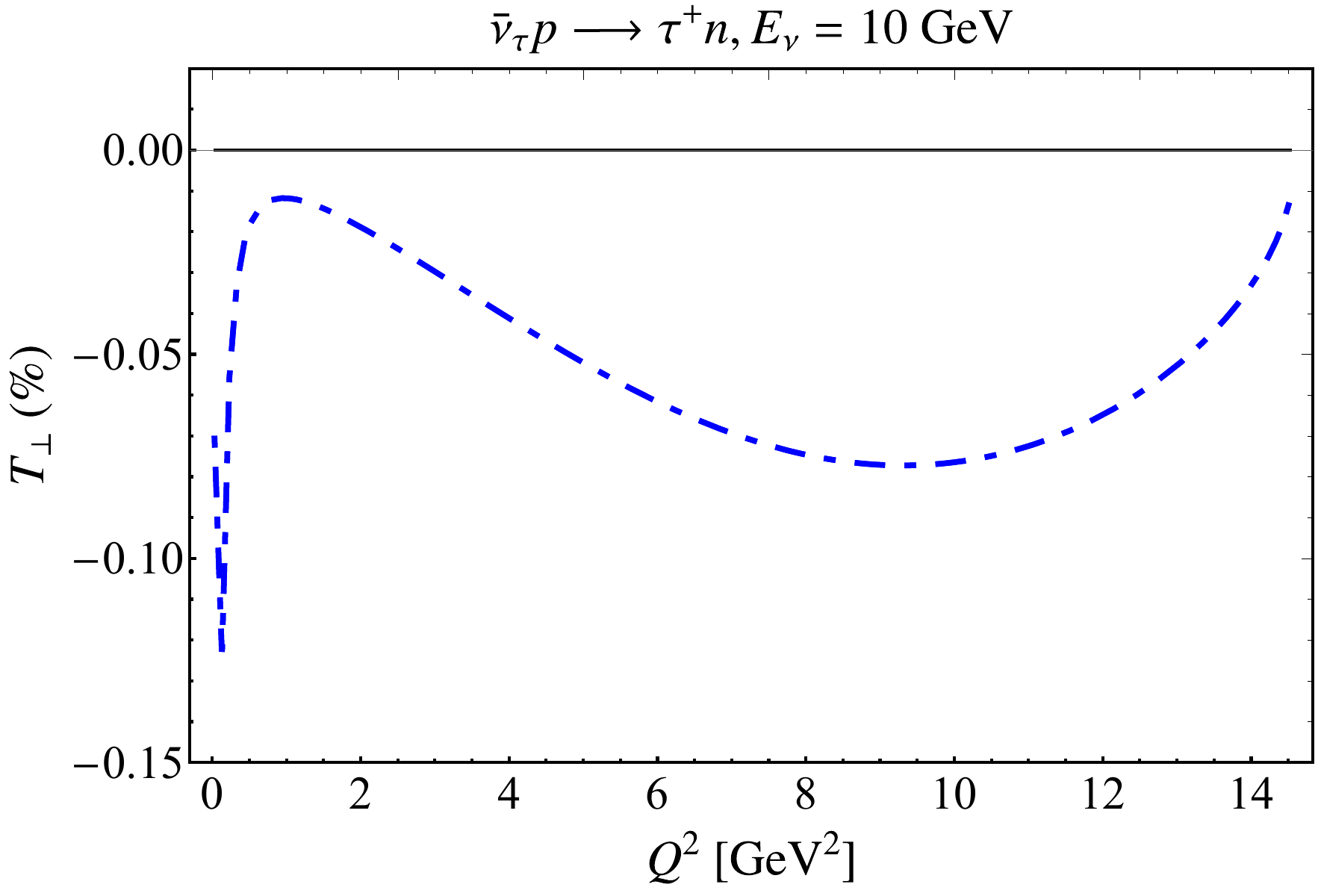}
\includegraphics[width=0.4\textwidth]{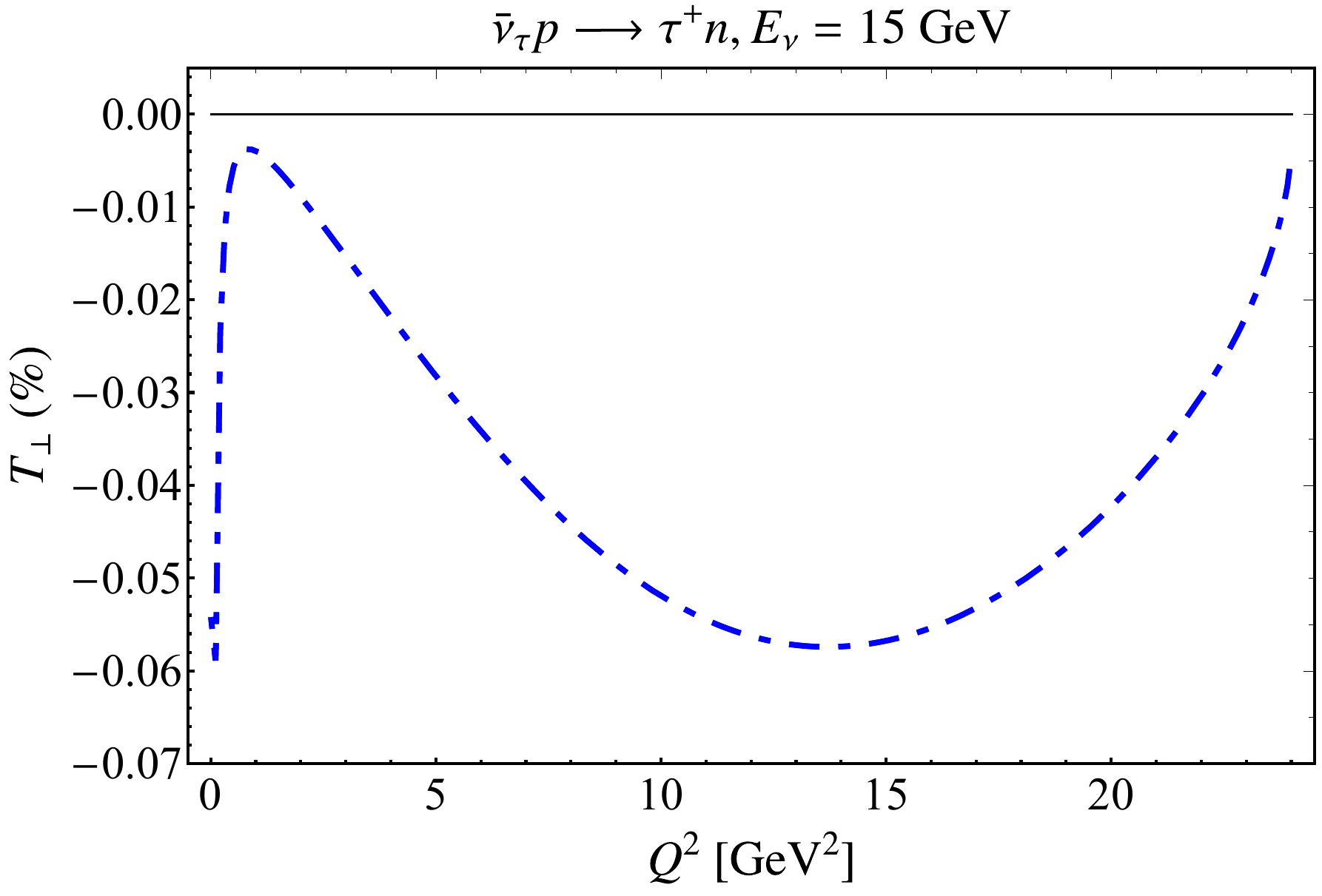}
\caption{Same as Fig.~\ref{fig:antinu_Tt_radcorr_tau} but for the transverse polarization observable $T_\perp$. \label{fig:antinu_TTT_radcorr_tau}}
\end{figure}

\begin{figure}[H]
\centering
\includegraphics[width=0.4\textwidth]{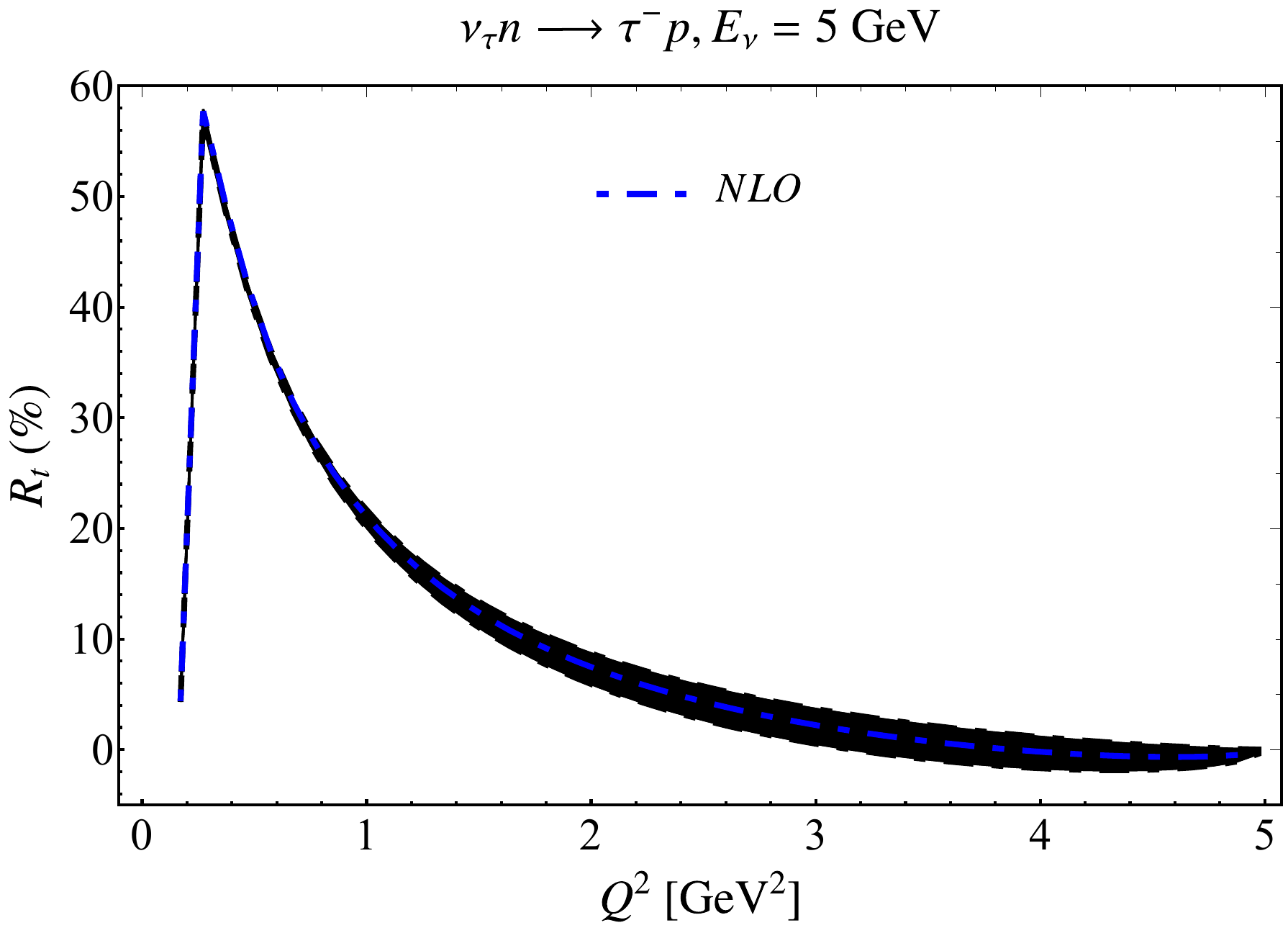}
\includegraphics[width=0.4\textwidth]{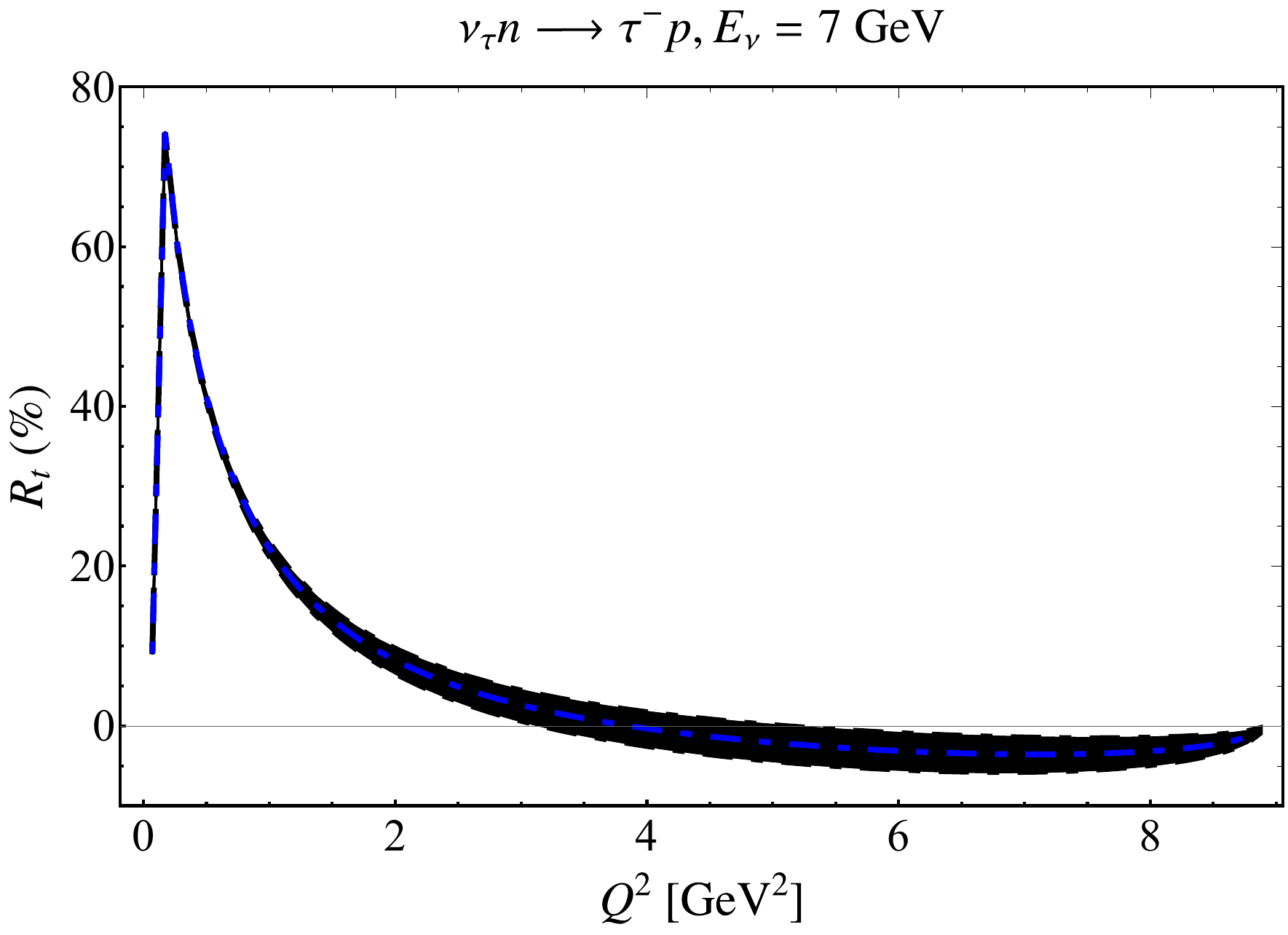}
\includegraphics[width=0.4\textwidth]{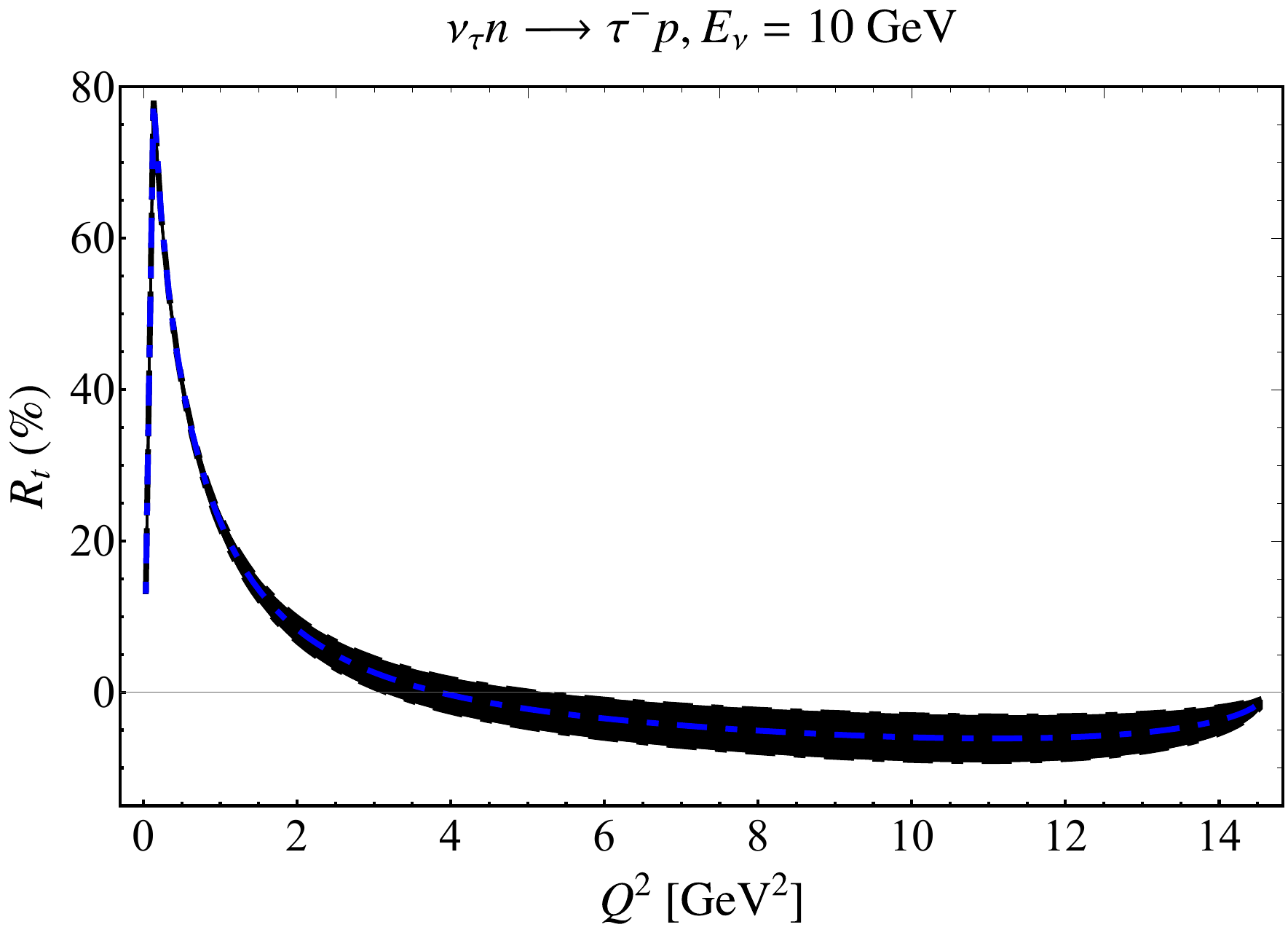}
\includegraphics[width=0.4\textwidth]{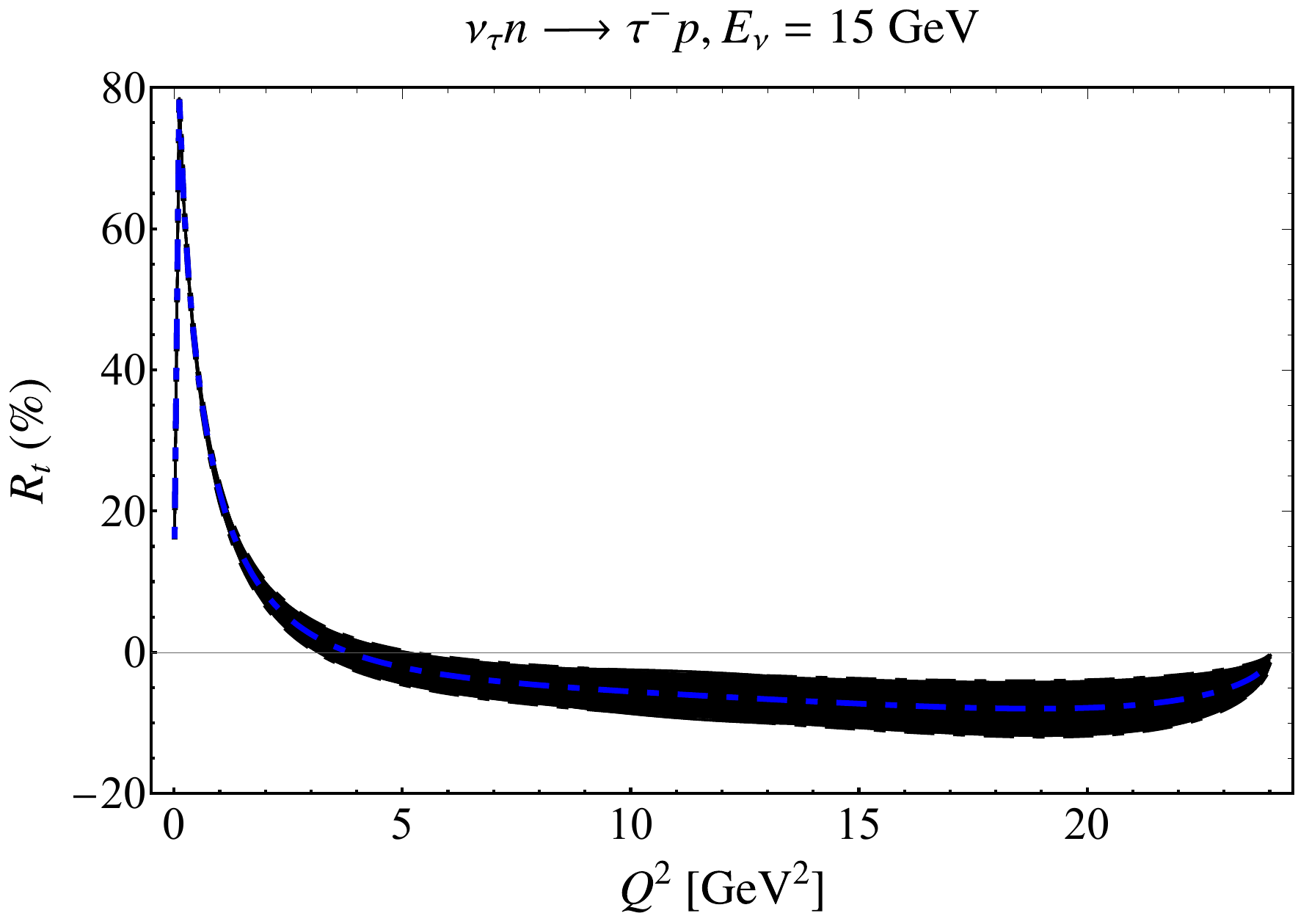}
\caption{Same as Fig.~\ref{fig:nu_Tt_radcorr_tau} but for the transverse polarization observable $R_t$. \label{fig:nu_Rt_radcorr_tau}}
\end{figure}

\begin{figure}[H]
\centering
\includegraphics[width=0.4\textwidth]{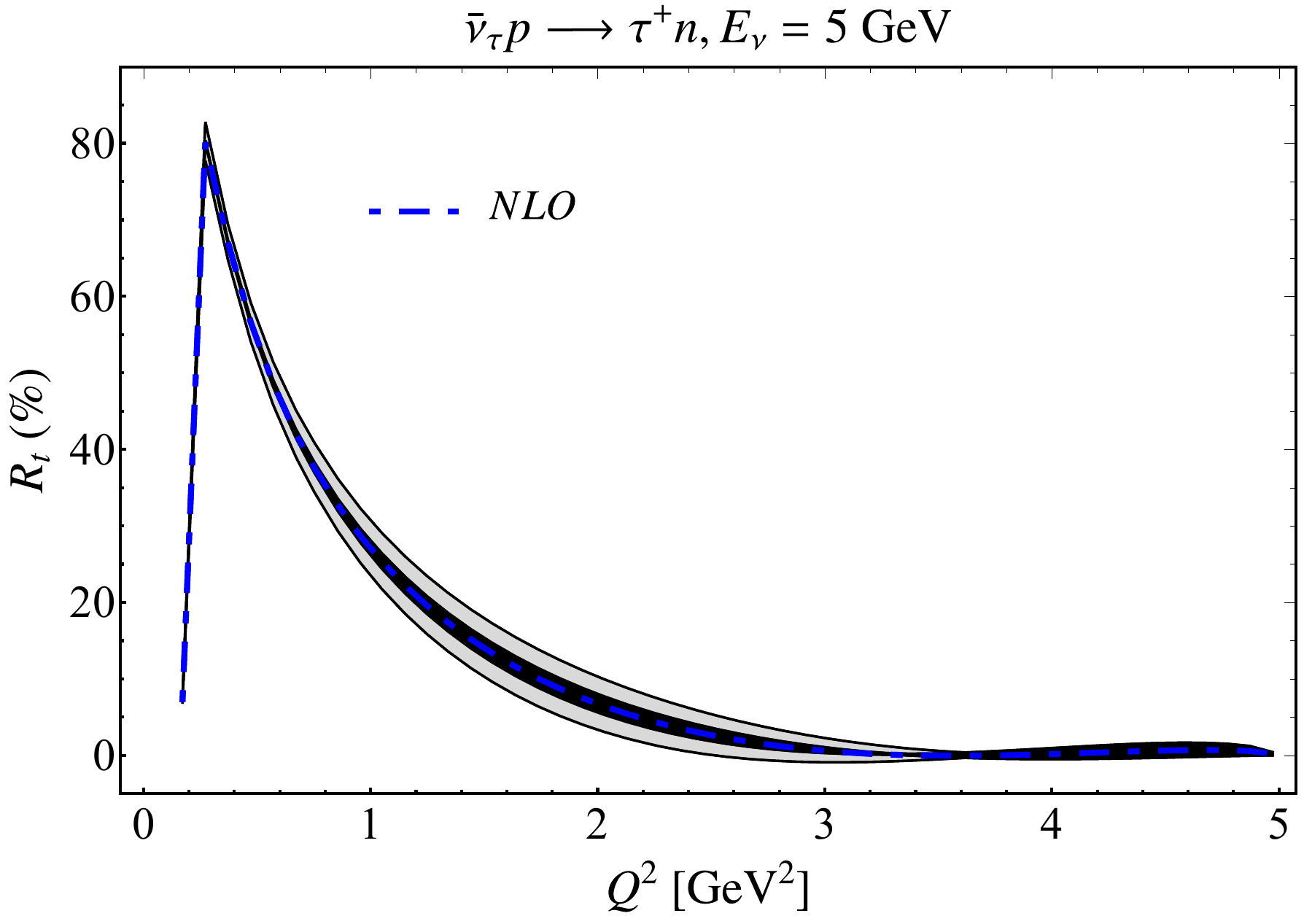}
\includegraphics[width=0.4\textwidth]{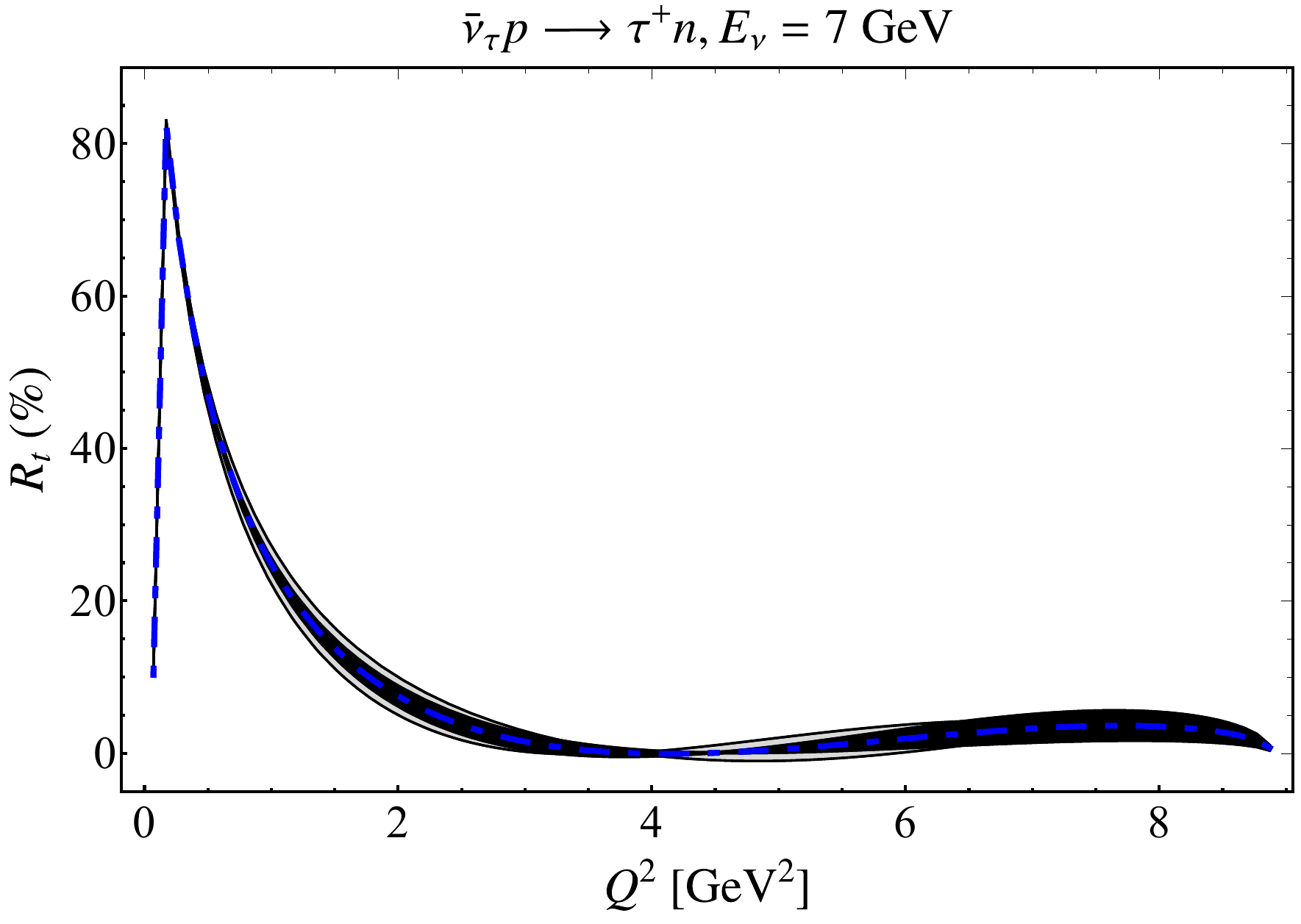}
\includegraphics[width=0.4\textwidth]{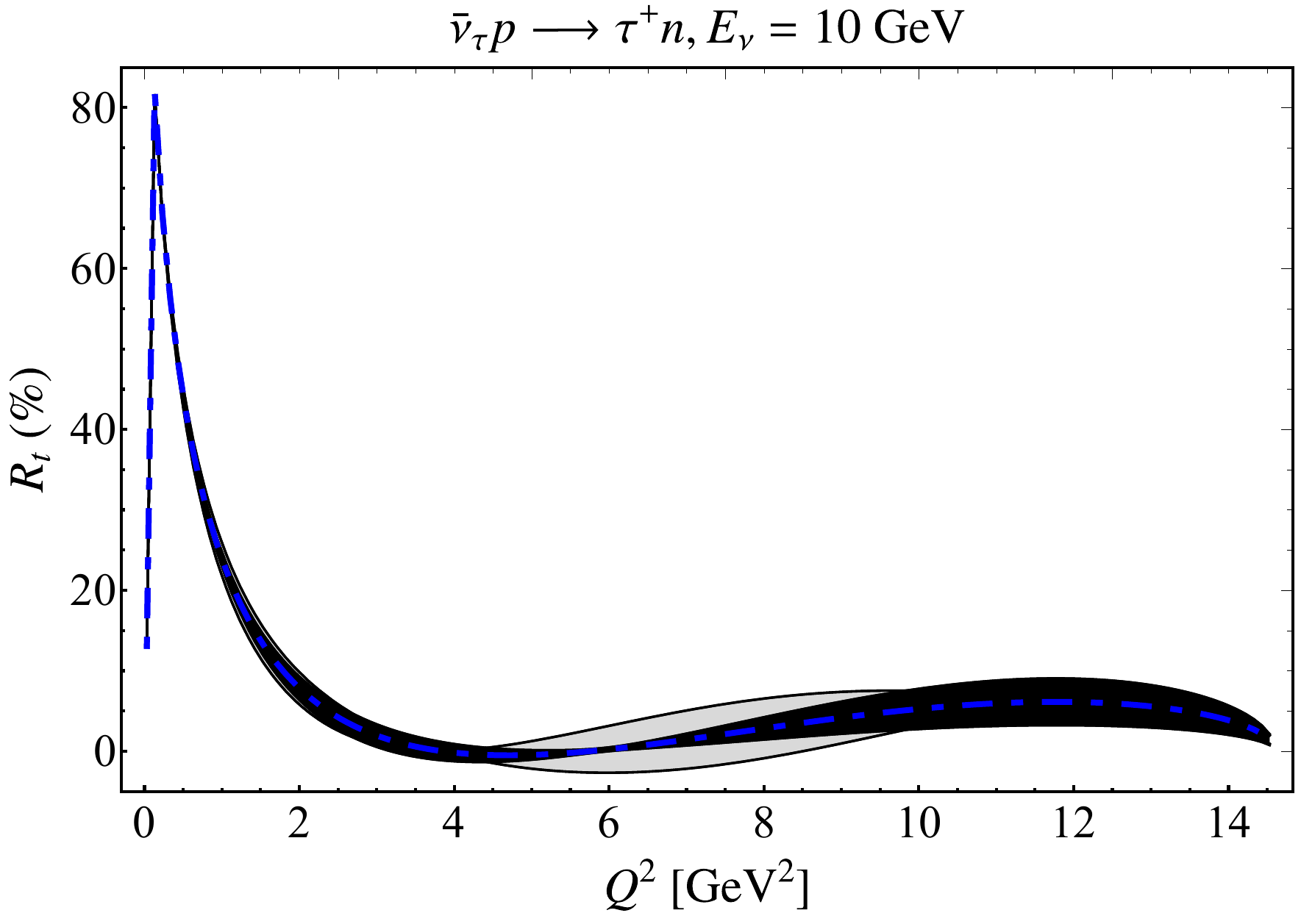}
\includegraphics[width=0.4\textwidth]{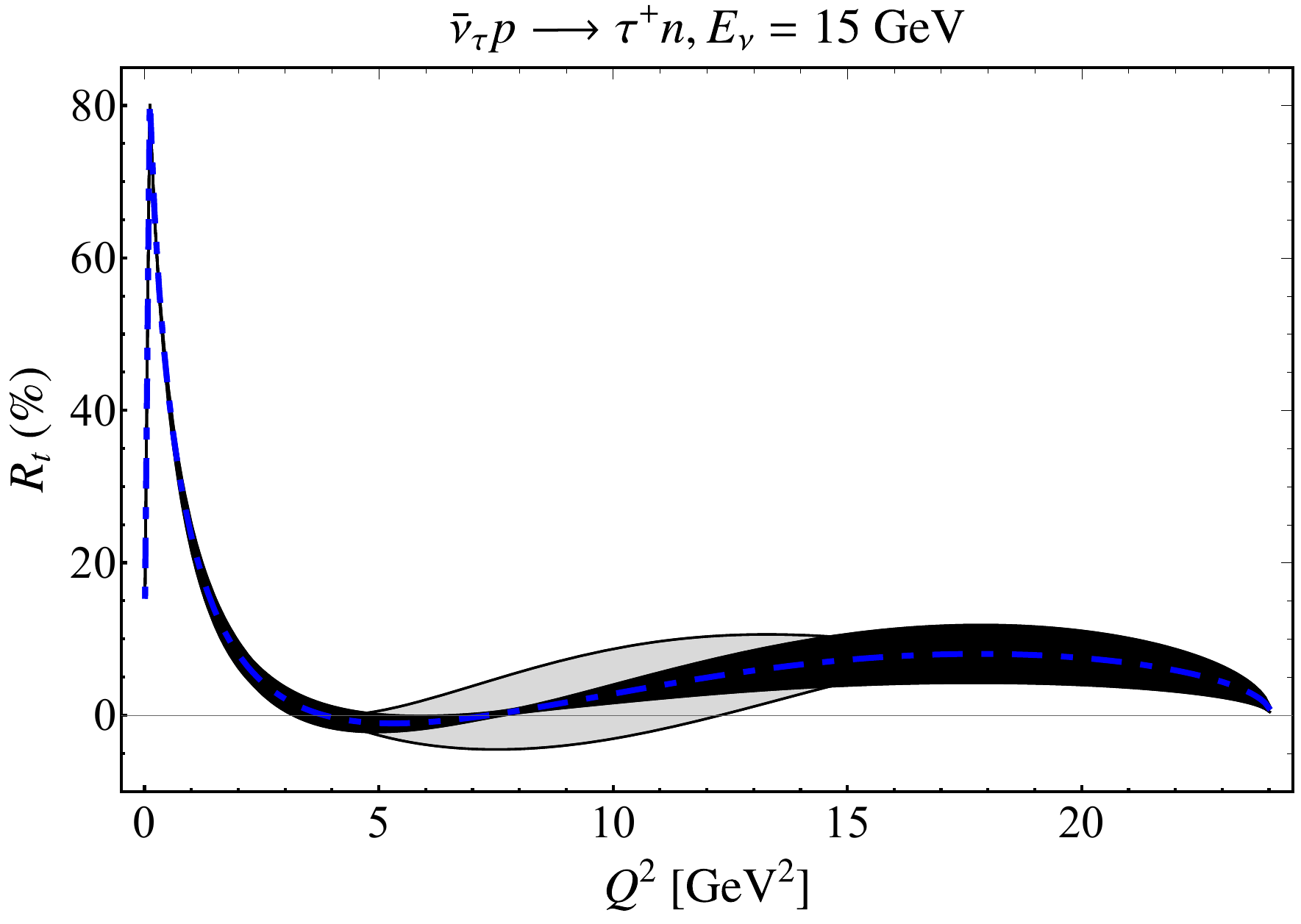}
\caption{Same as Fig.~\ref{fig:antinu_Tt_radcorr_tau} but for the transverse polarization observable $R_t$. \label{fig:antinu_Rt_radcorr_tau}}
\end{figure}

\begin{figure}[H]
\centering
\includegraphics[width=0.4\textwidth]{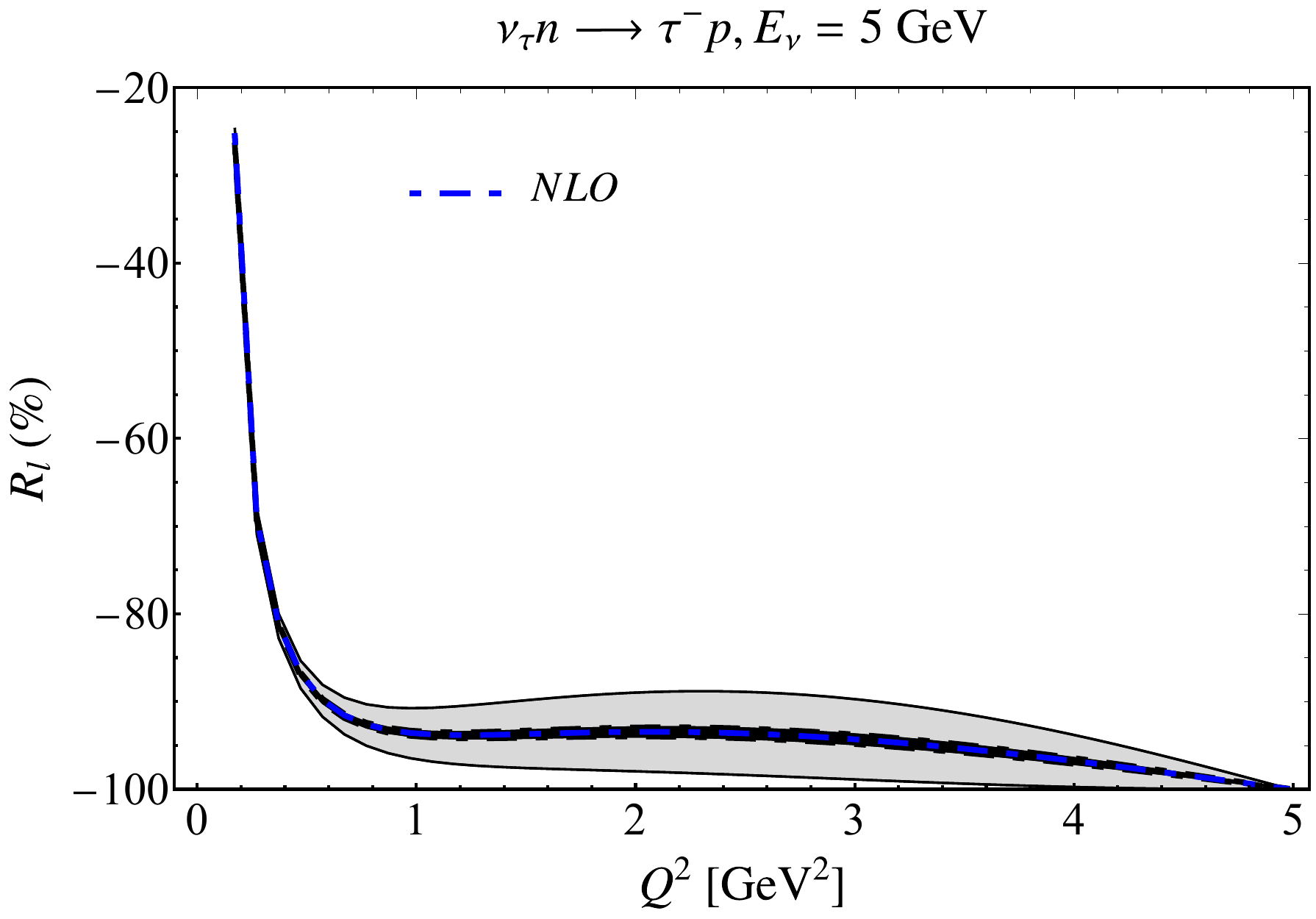}
\includegraphics[width=0.4\textwidth]{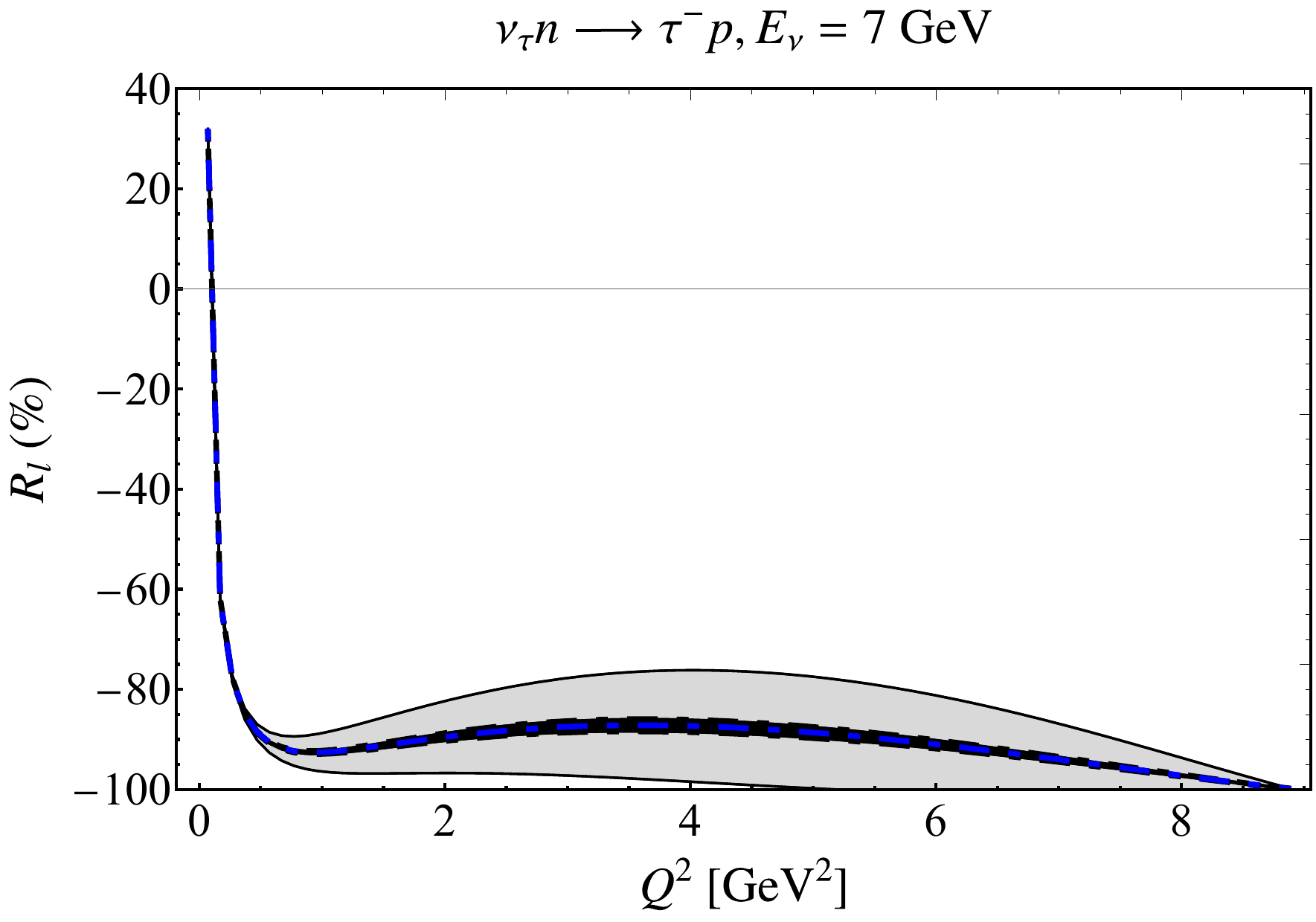}
\includegraphics[width=0.4\textwidth]{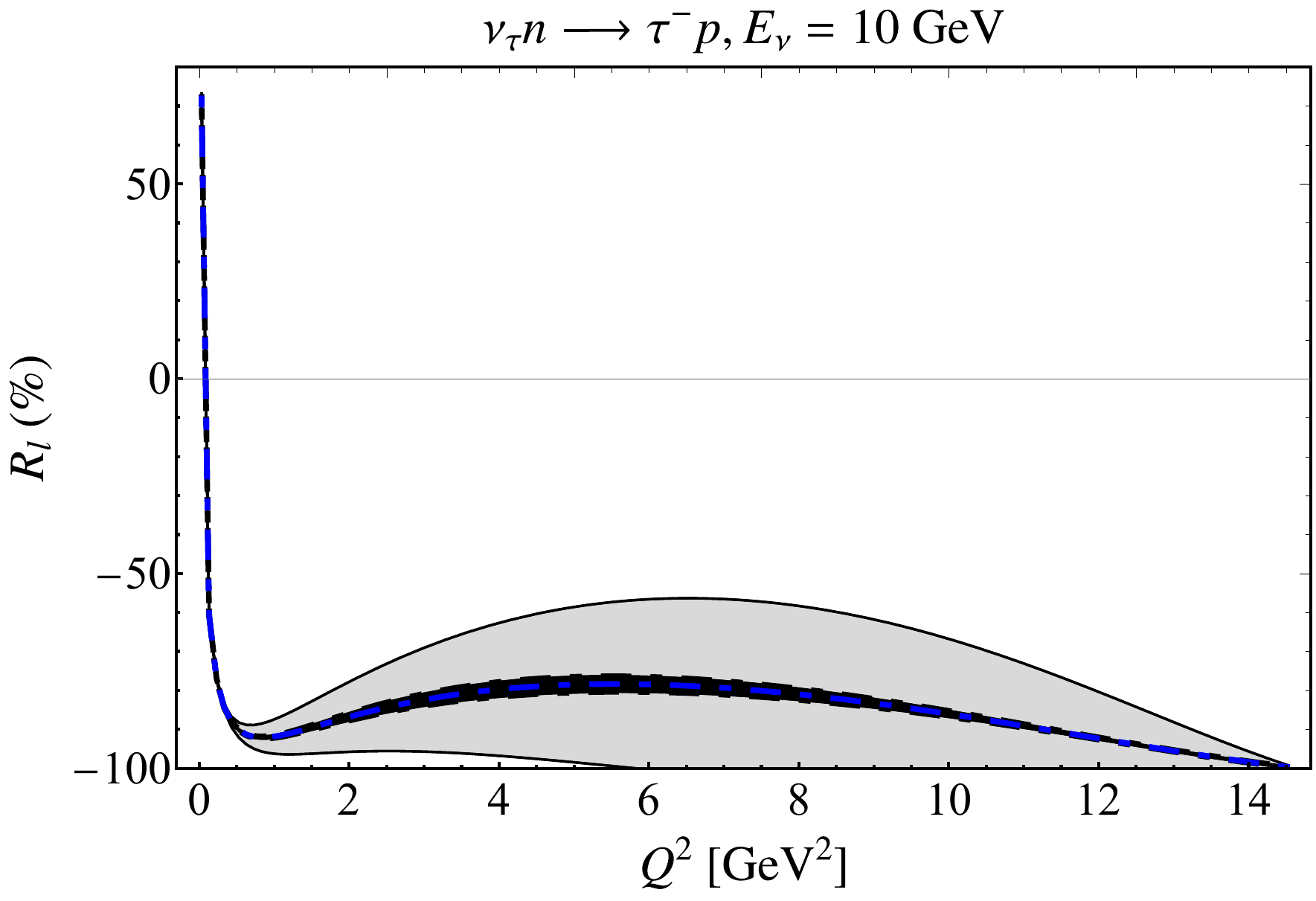}
\includegraphics[width=0.4\textwidth]{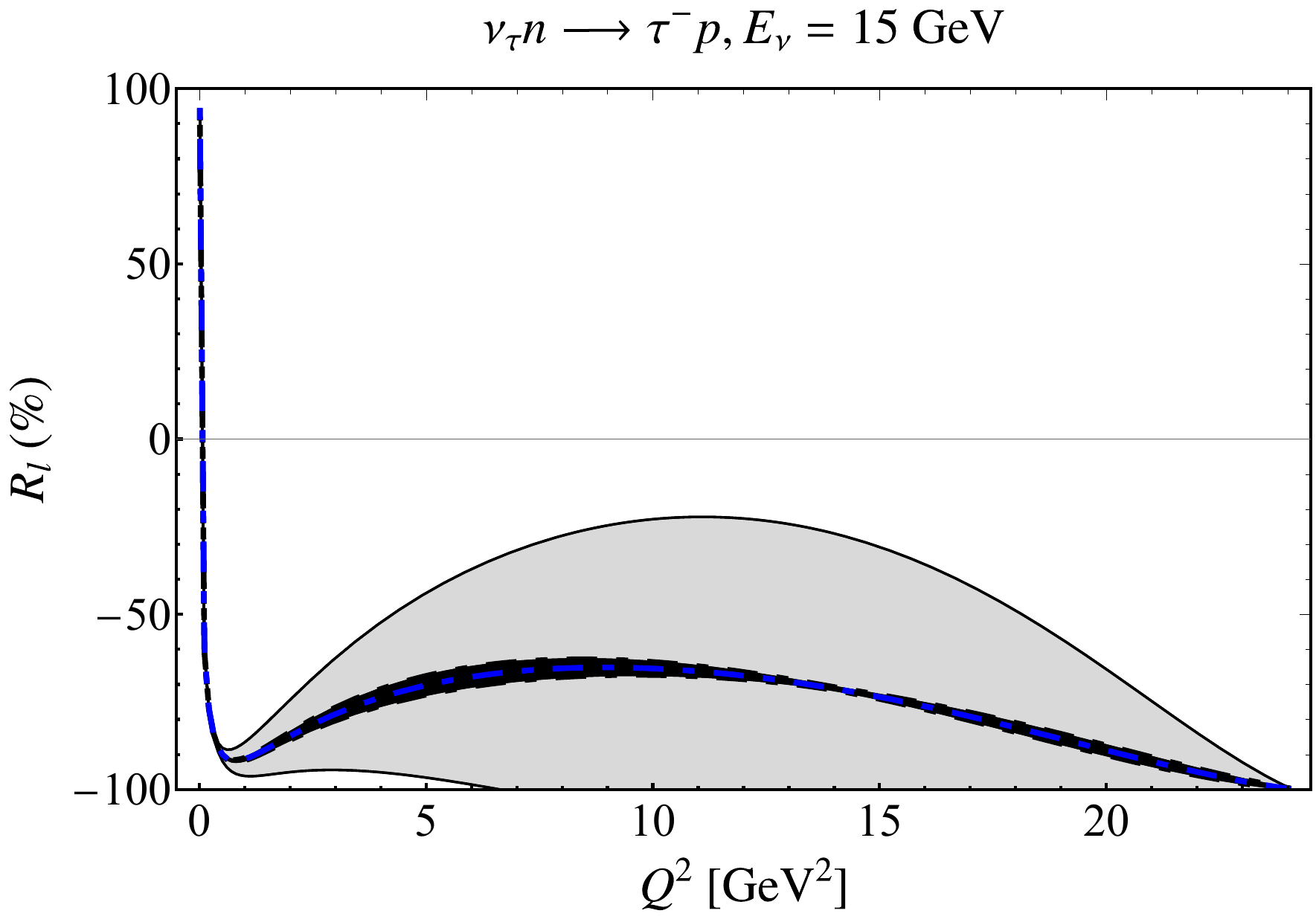}
\caption{Same as Fig.~\ref{fig:nu_Tt_radcorr_tau} but for the longitudinal polarization observable $R_l$. \label{fig:nu_Rl_radcorr_tau}}
\end{figure}

\begin{figure}[H]
\centering
\includegraphics[width=0.4\textwidth]{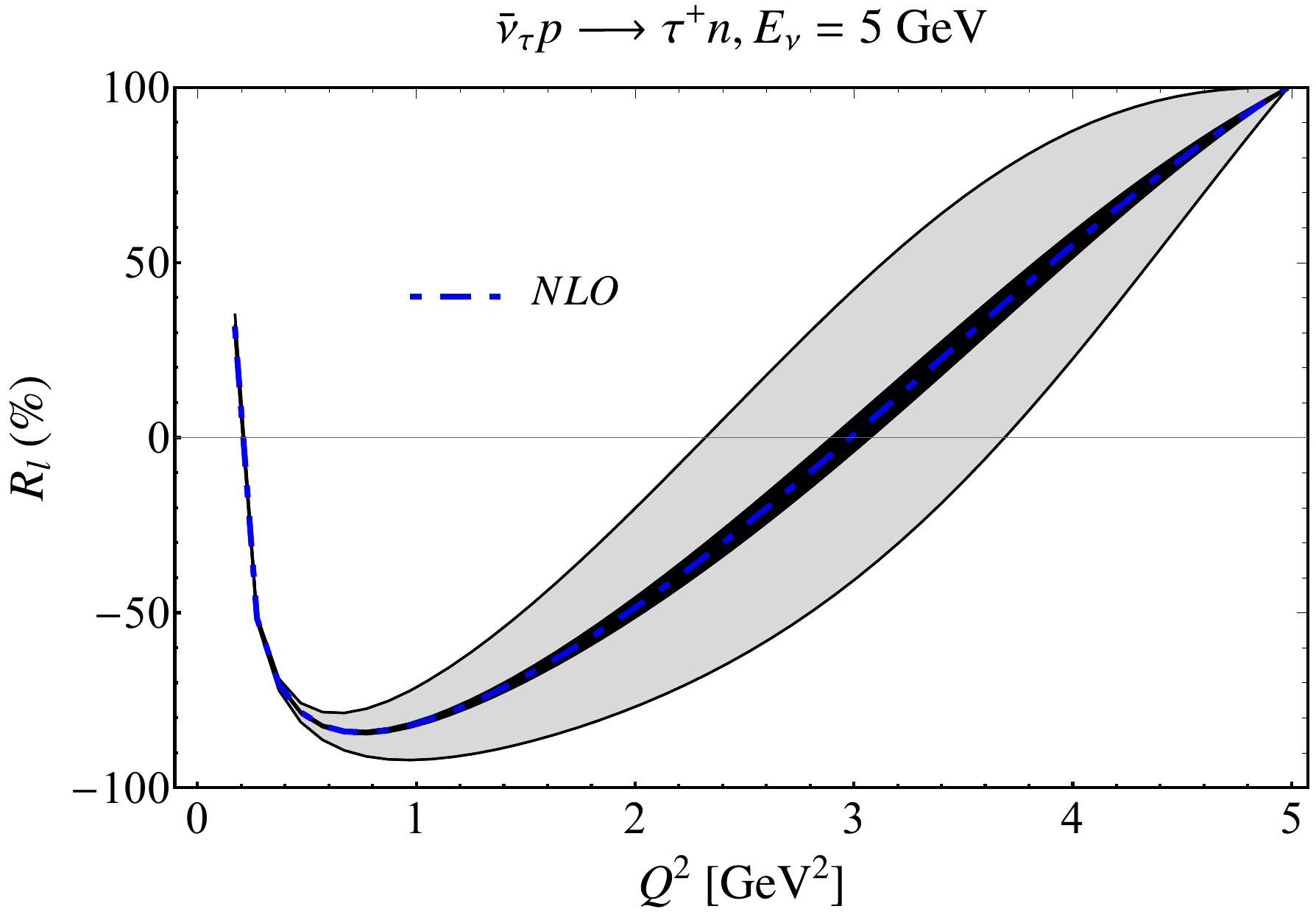}
\includegraphics[width=0.4\textwidth]{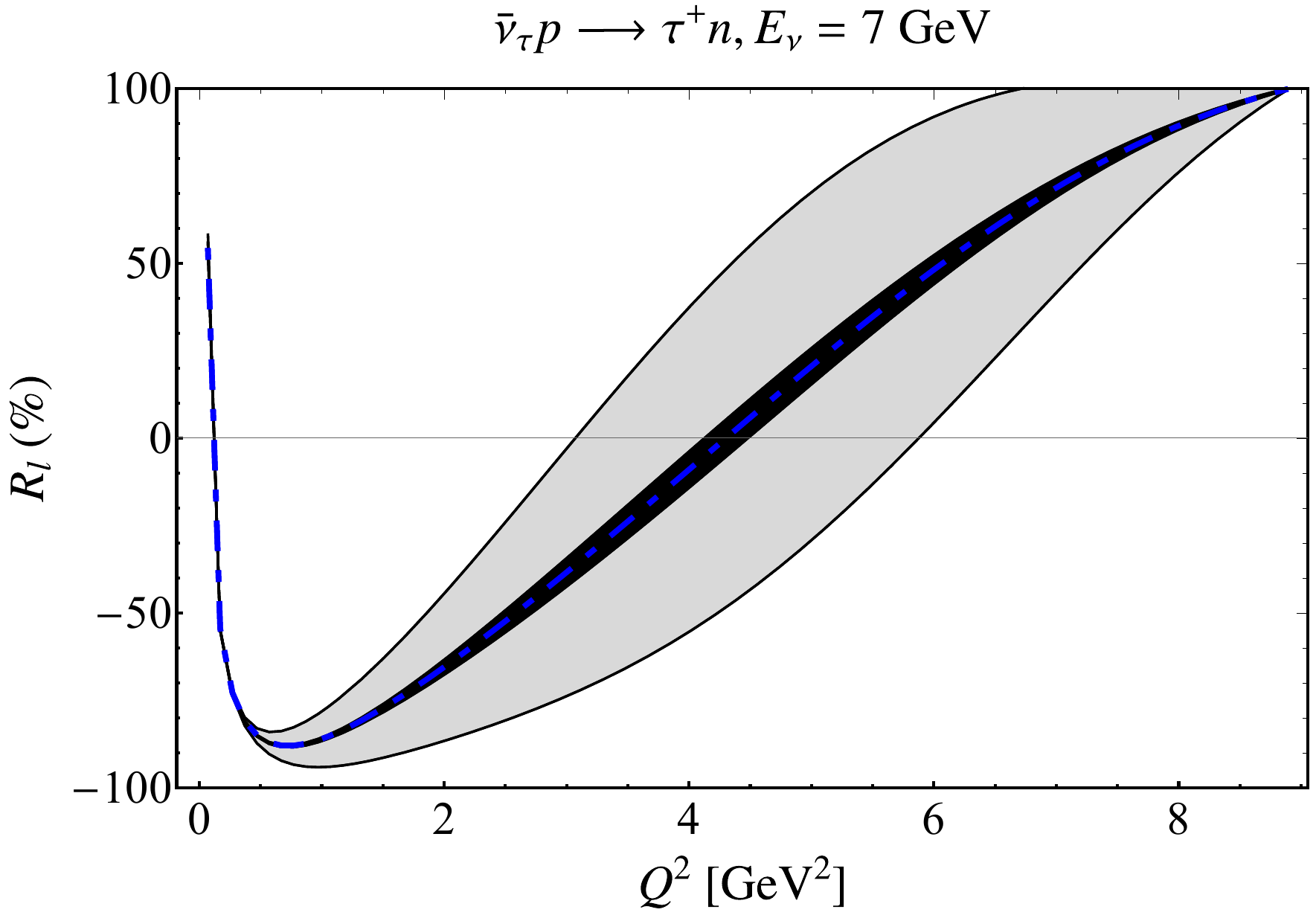}
\includegraphics[width=0.4\textwidth]{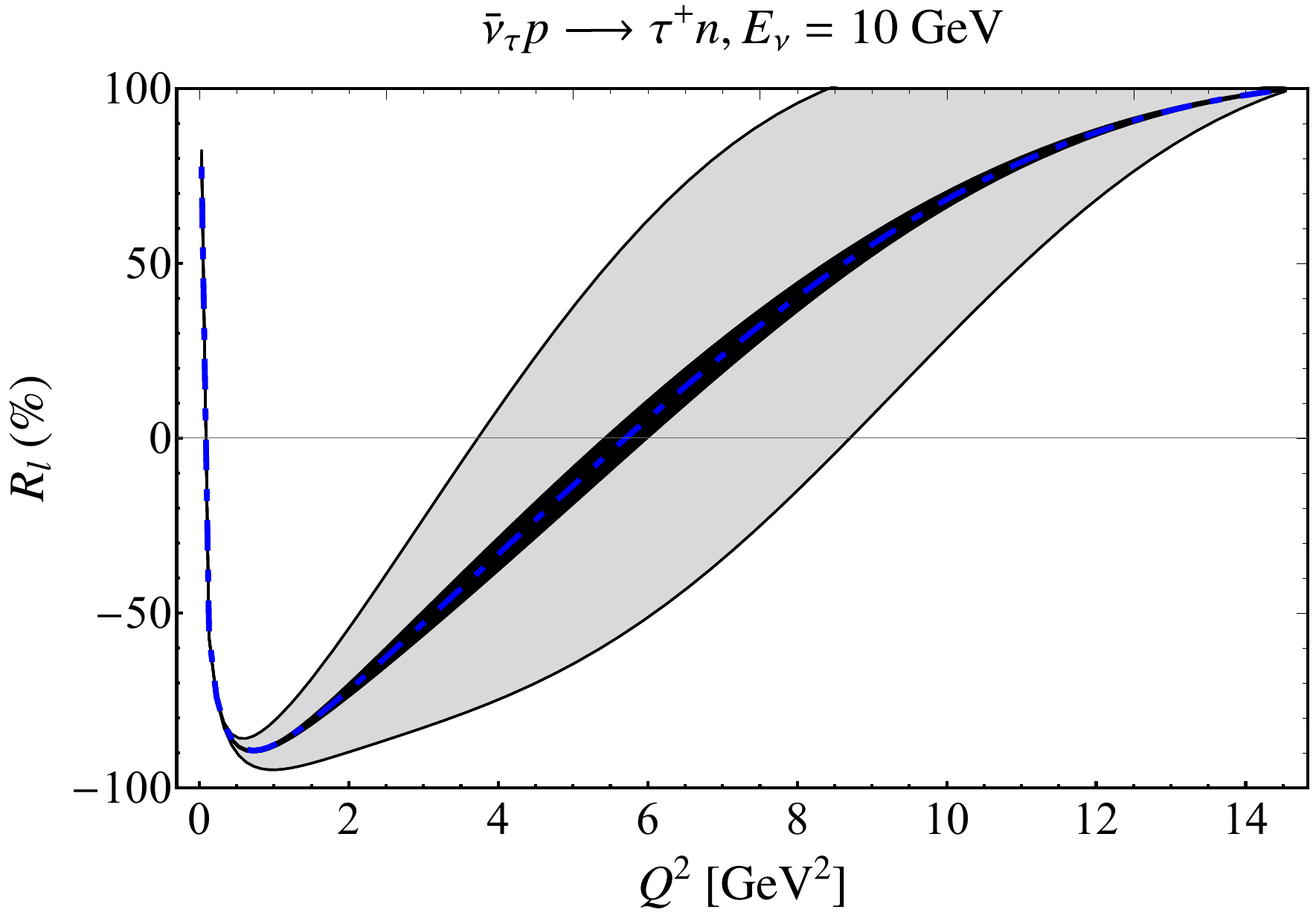}
\includegraphics[width=0.4\textwidth]{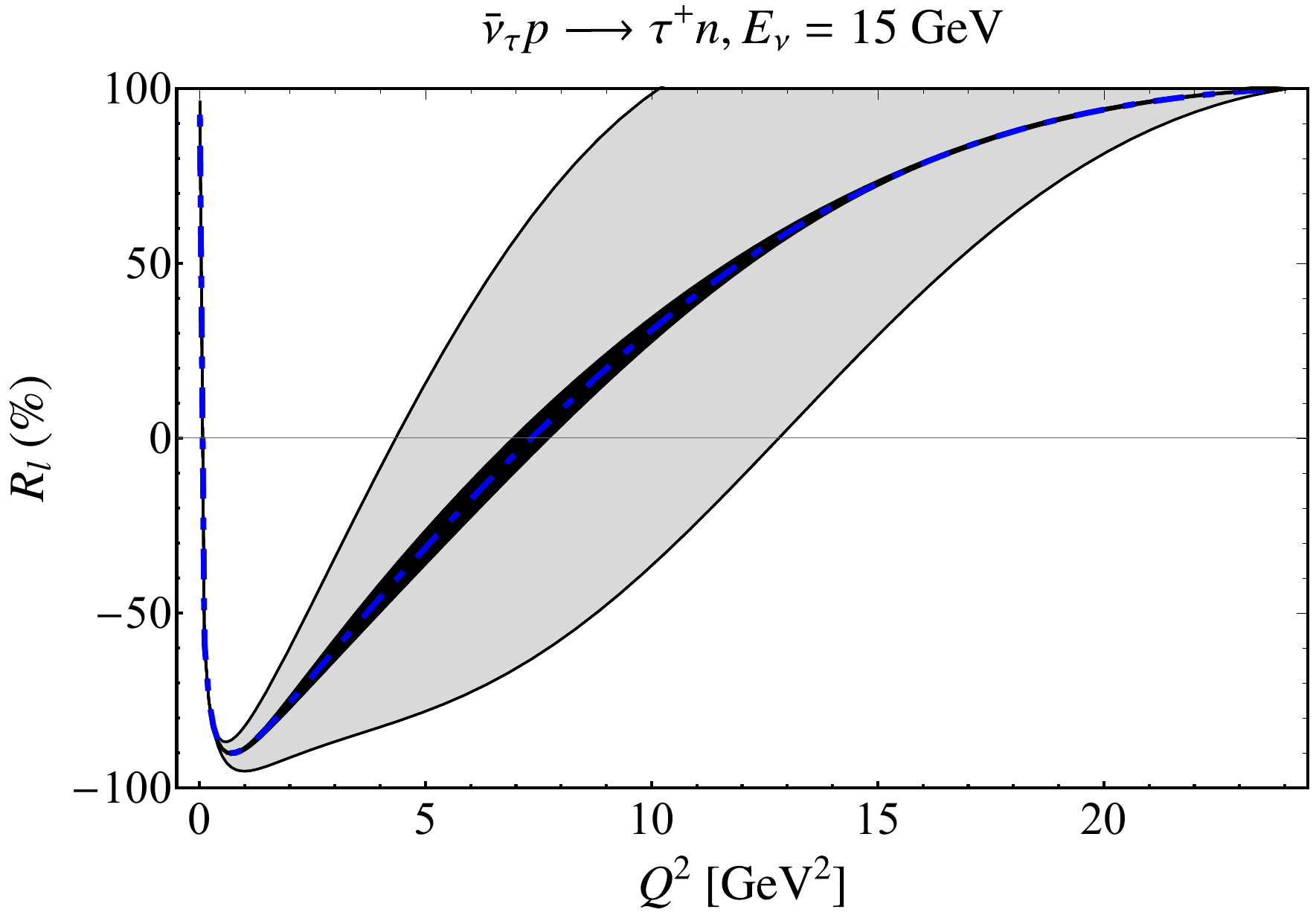}
\caption{Same as Fig.~\ref{fig:antinu_Tt_radcorr_tau} but for the longitudinal polarization observable $R_l$. \label{fig:antinu_Rl_radcorr_tau}}
\end{figure}

\begin{figure}[H]
\centering
\includegraphics[width=0.4\textwidth]{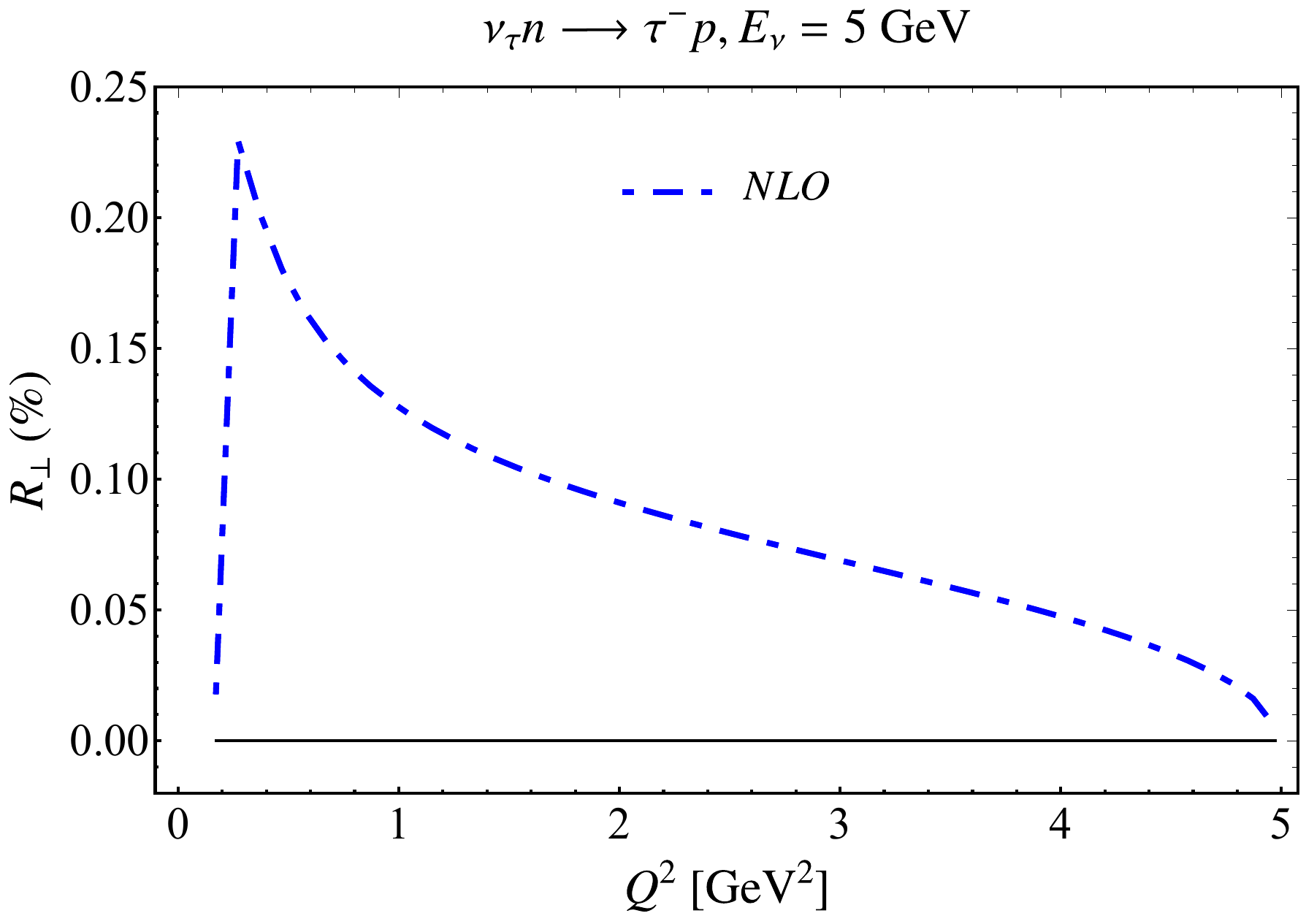}
\includegraphics[width=0.4\textwidth]{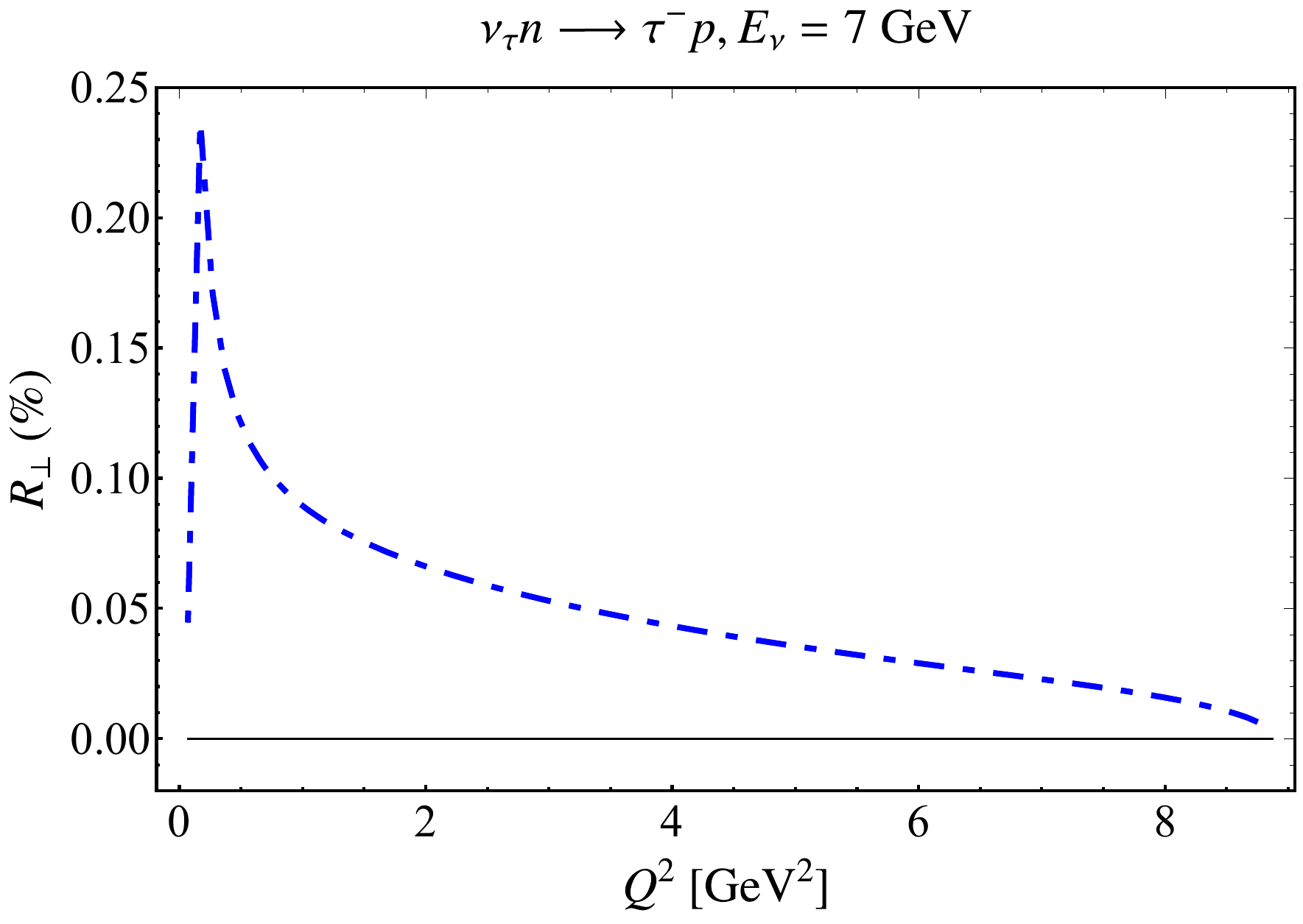}
\includegraphics[width=0.4\textwidth]{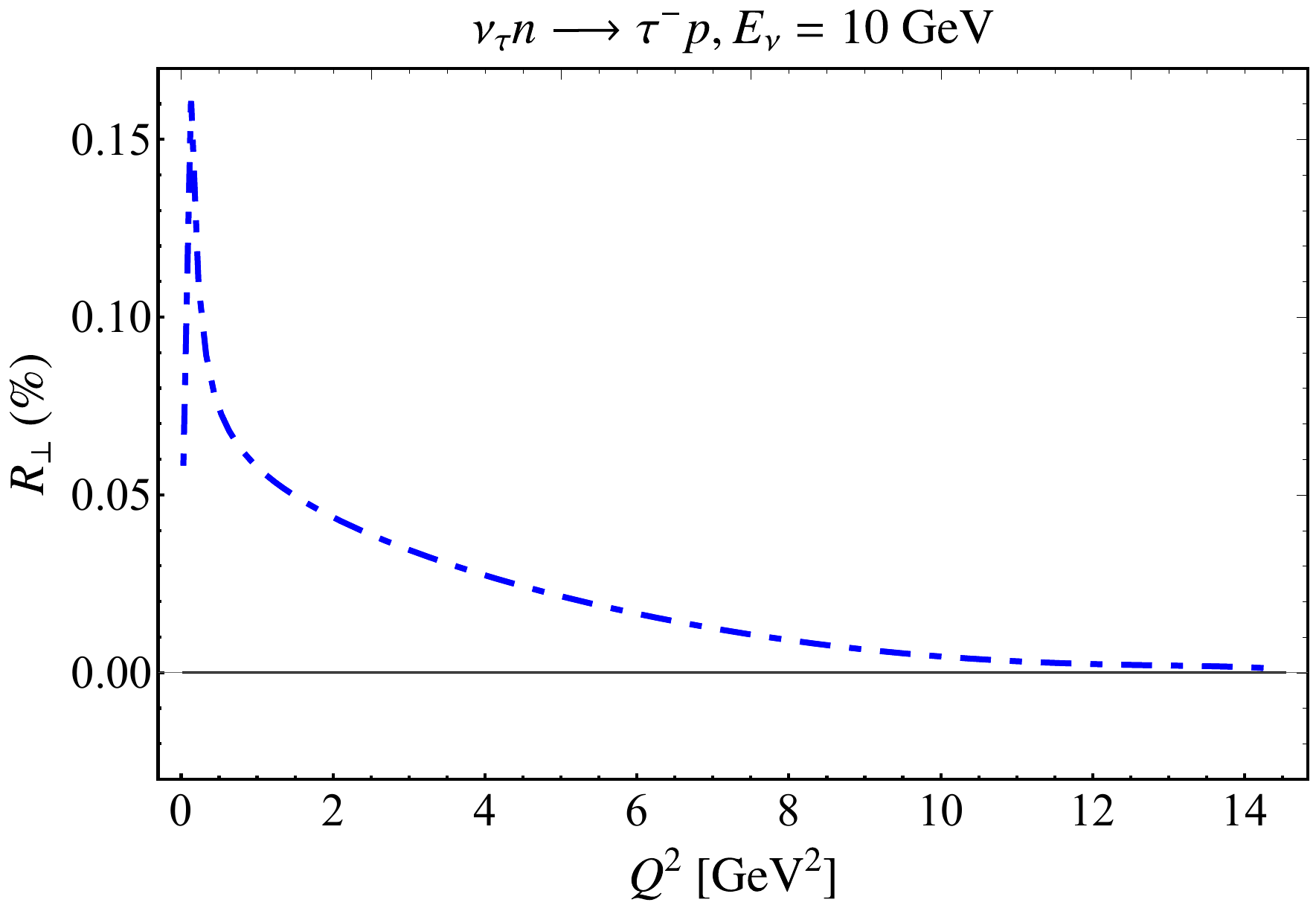}
\includegraphics[width=0.4\textwidth]{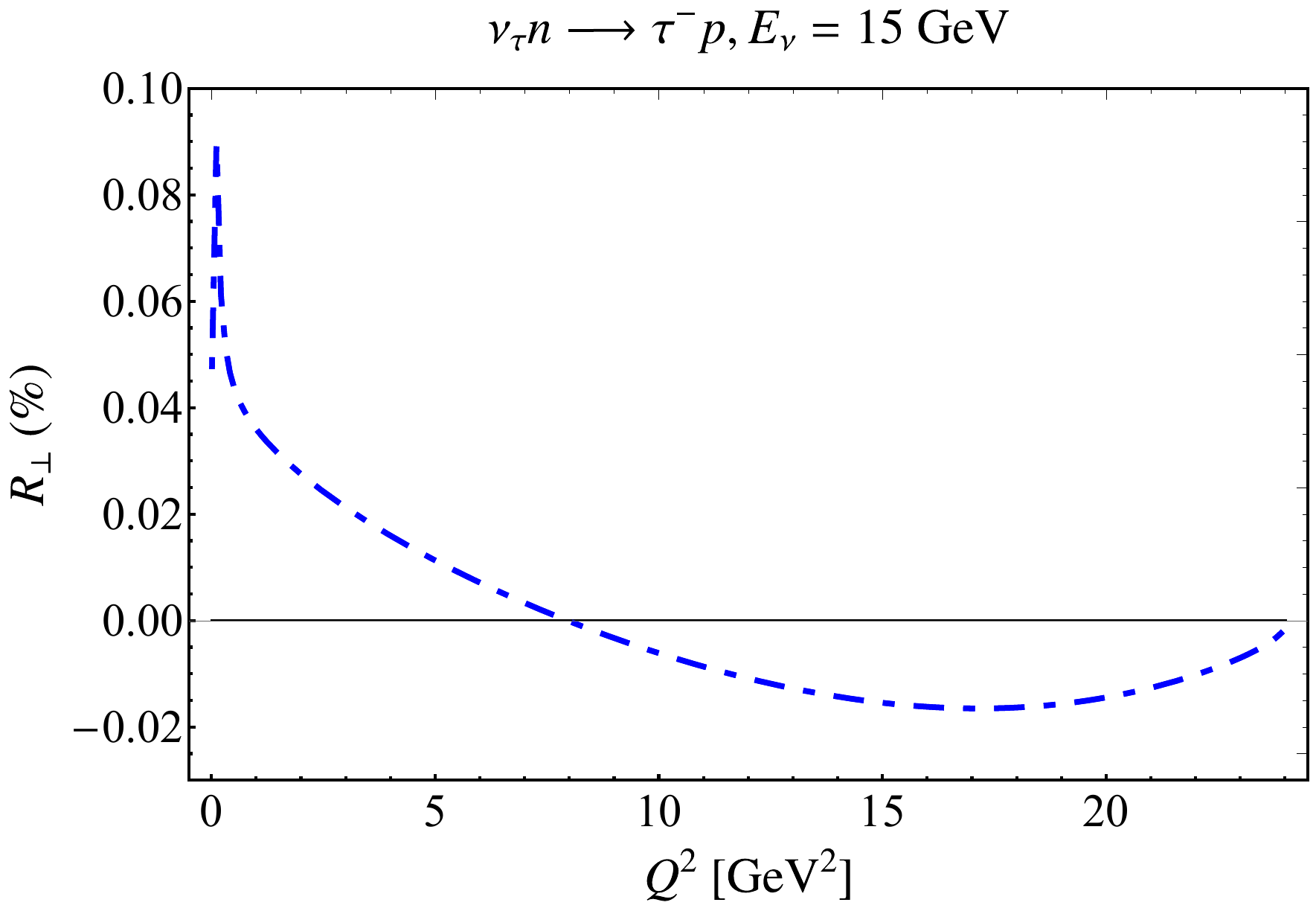}
\caption{Same as Fig.~\ref{fig:nu_Tt_radcorr_tau} but for the transverse polarization observable $R_\perp$. \label{fig:nu_RTT_radcorr_tau}}
\end{figure}

\begin{figure}[H]
\centering
\includegraphics[width=0.4\textwidth]{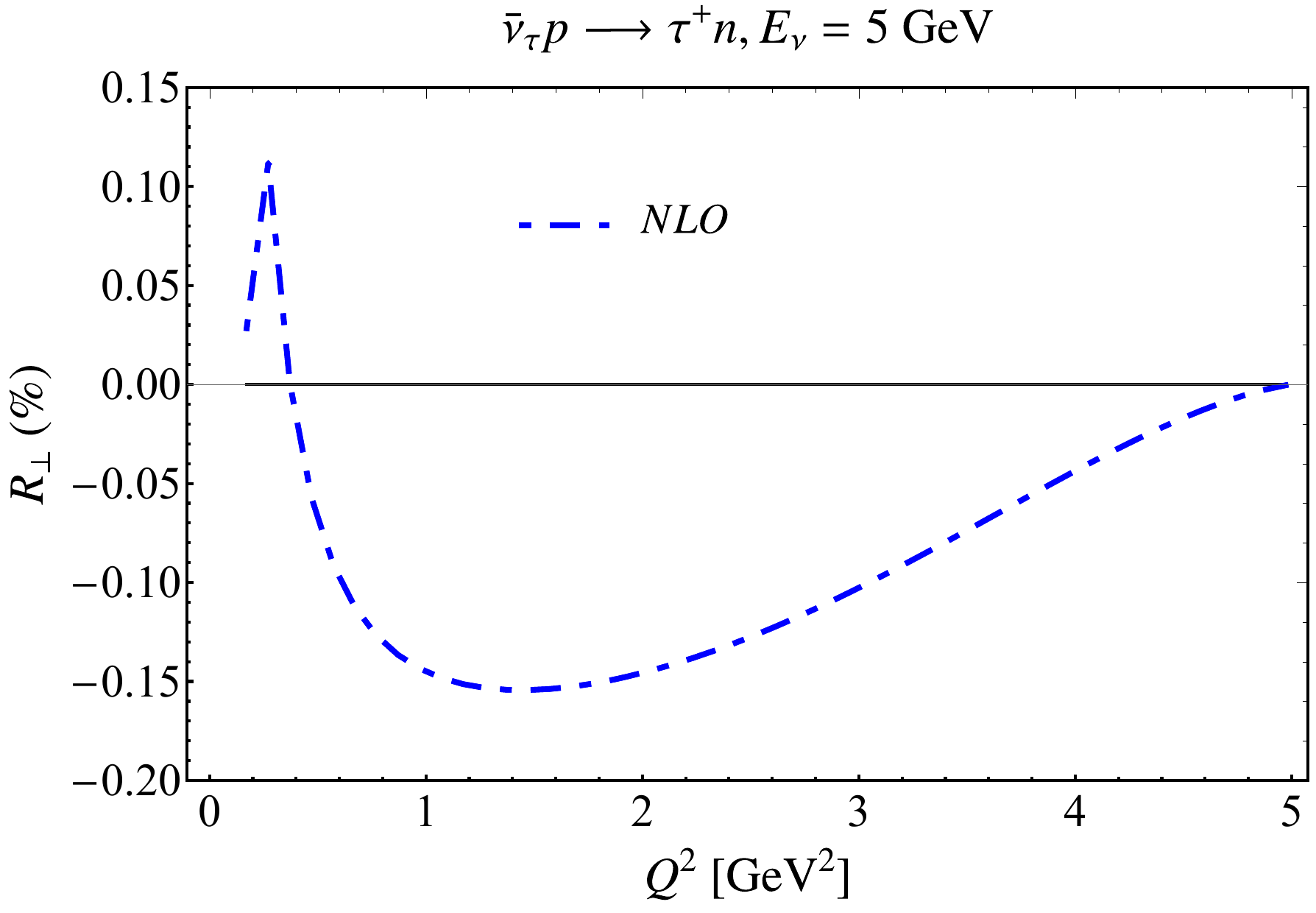}
\includegraphics[width=0.4\textwidth]{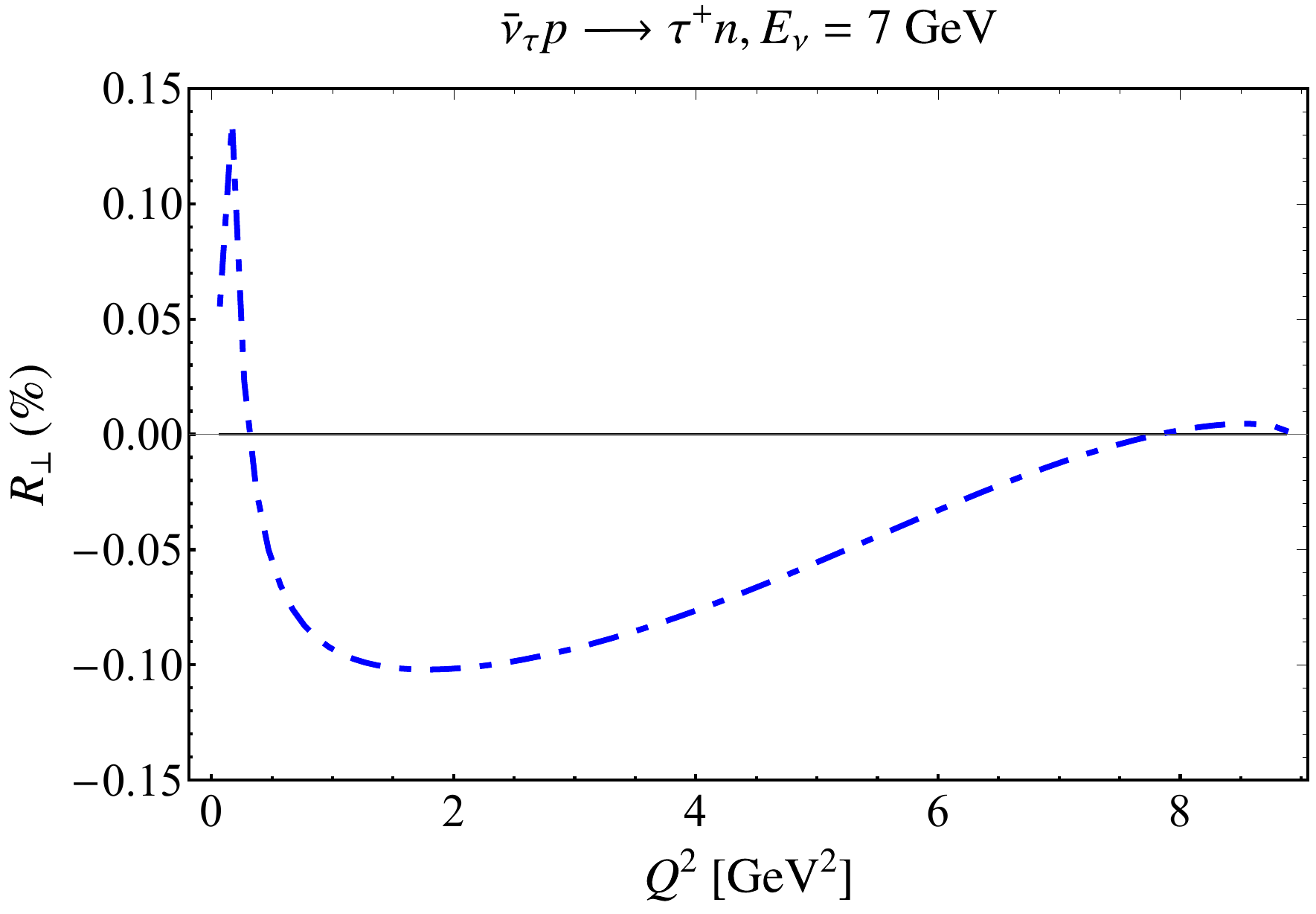}
\includegraphics[width=0.4\textwidth]{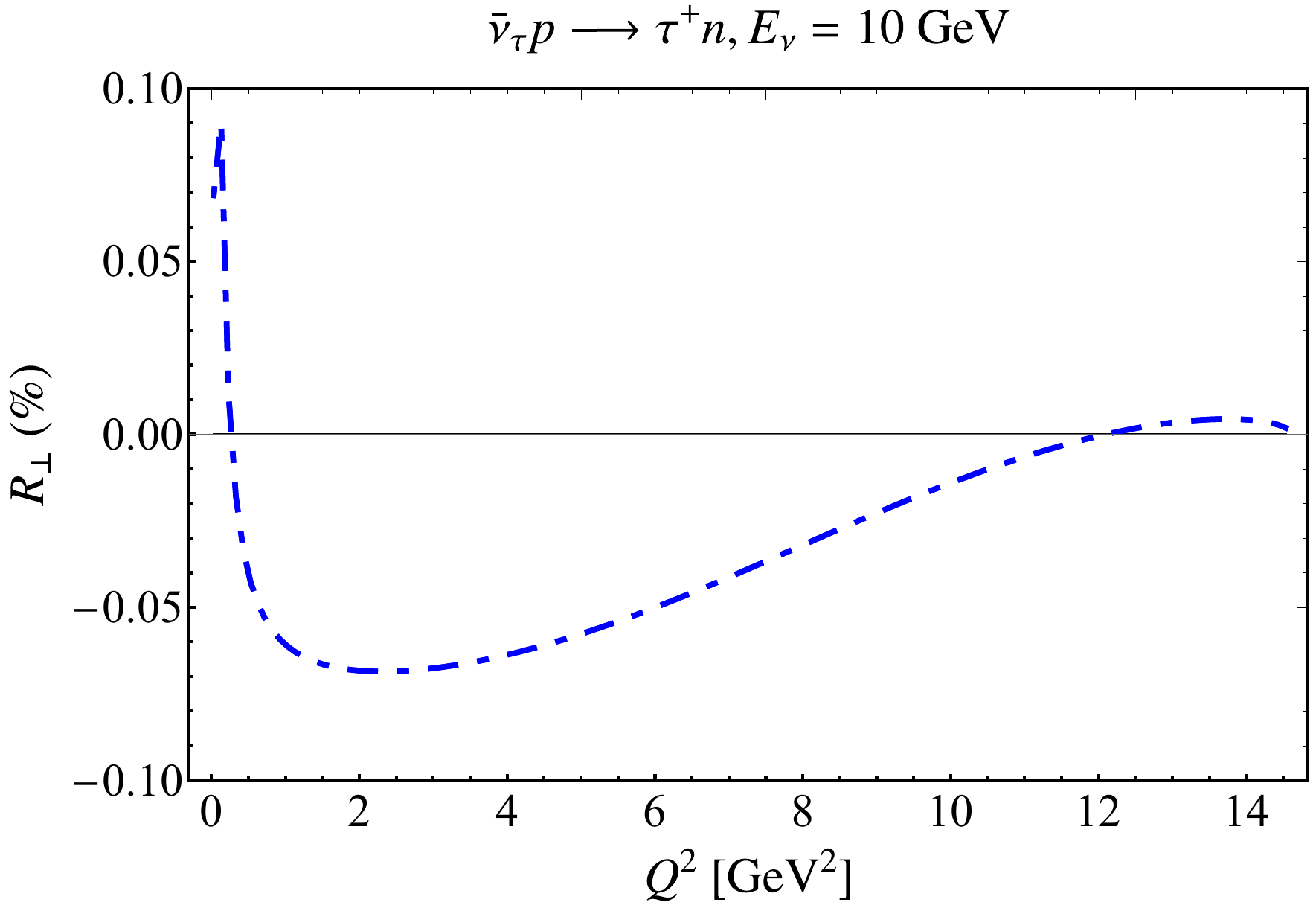}
\includegraphics[width=0.4\textwidth]{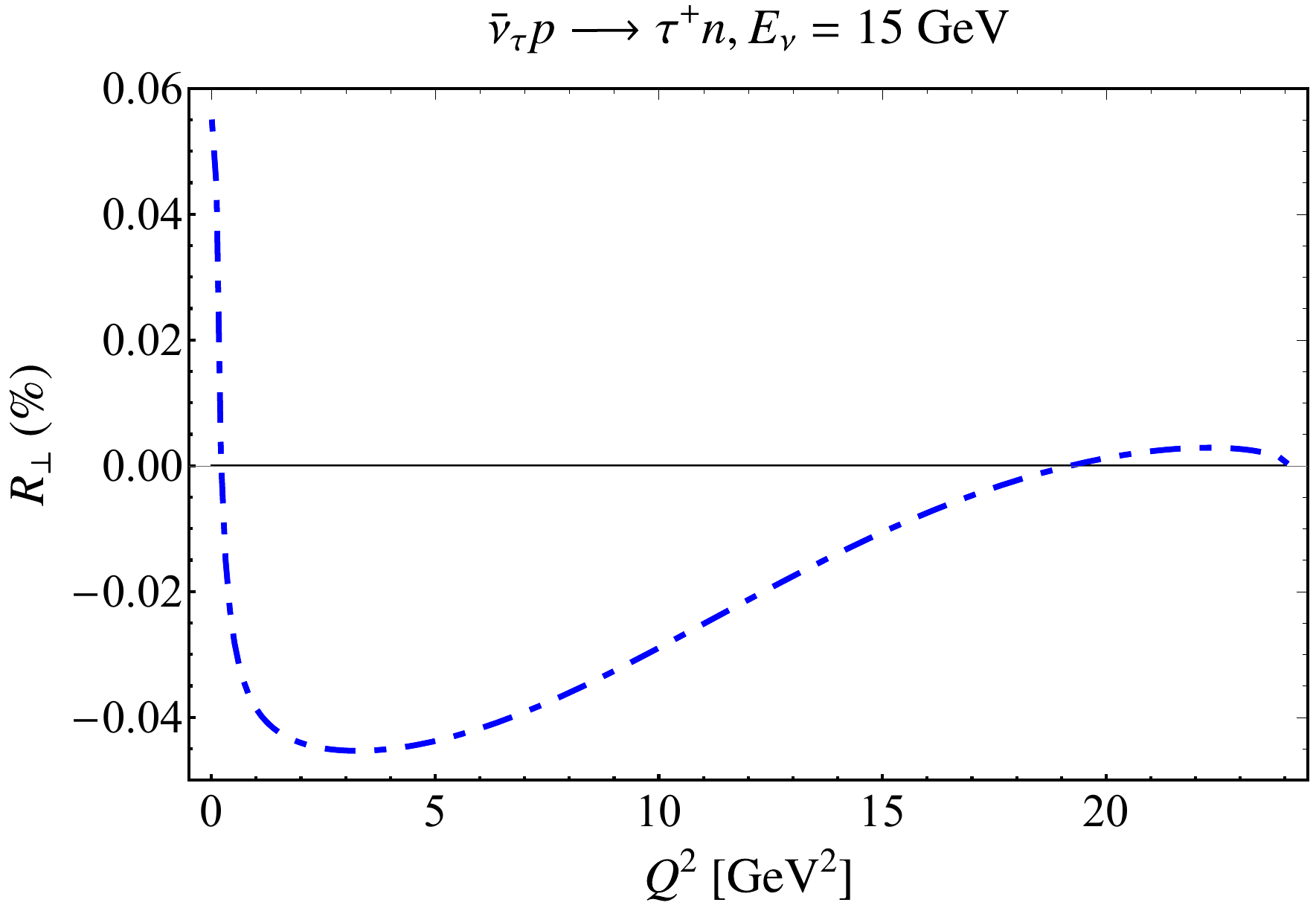}
\caption{Same as Fig.~\ref{fig:antinu_Tt_radcorr_tau} but for the transverse polarization observable $R_\perp$. \label{fig:antinu_RTT_radcorr_tau}}
\end{figure}

\begin{figure}[H]
\centering
\includegraphics[width=0.4\textwidth]{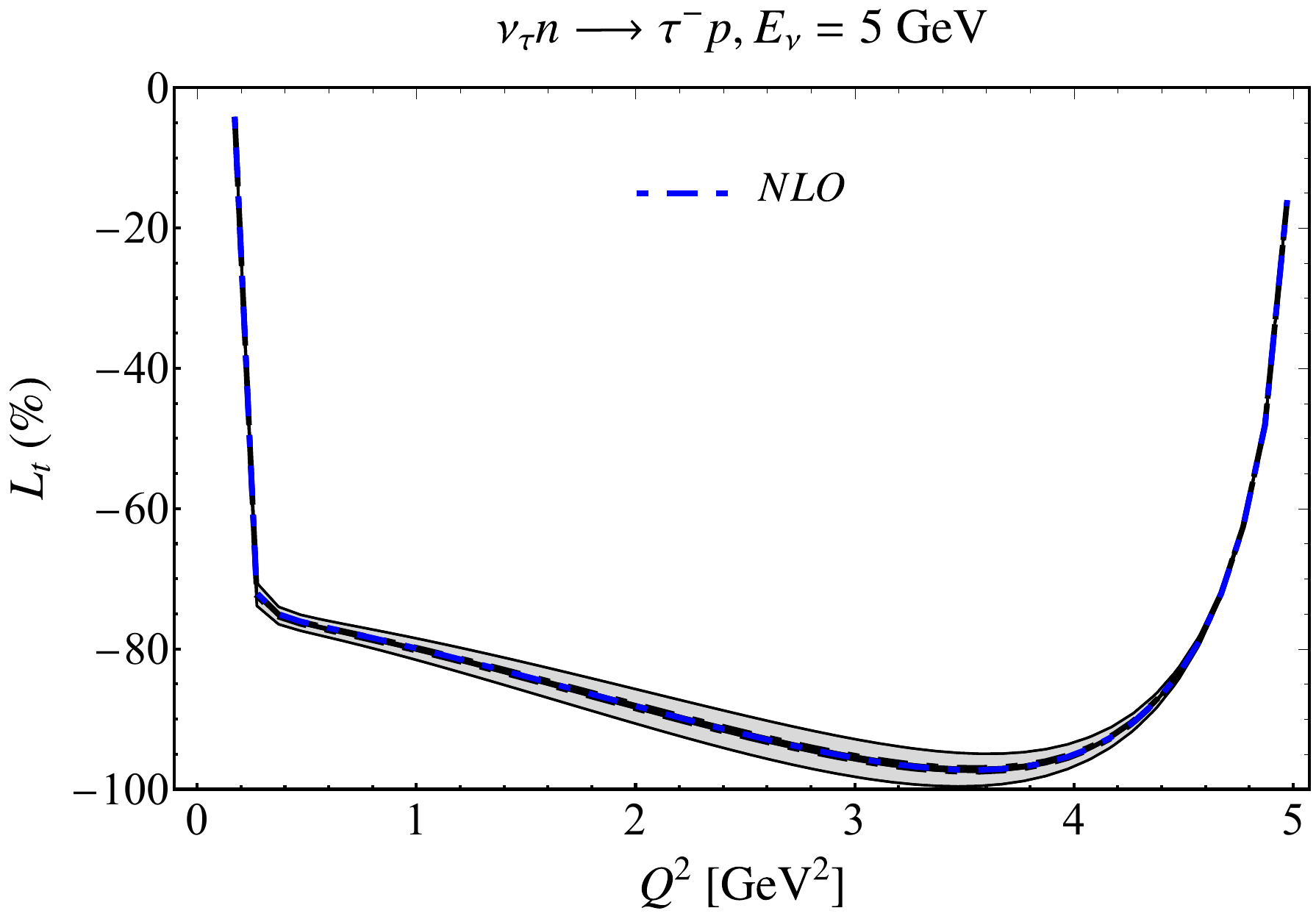}
\includegraphics[width=0.4\textwidth]{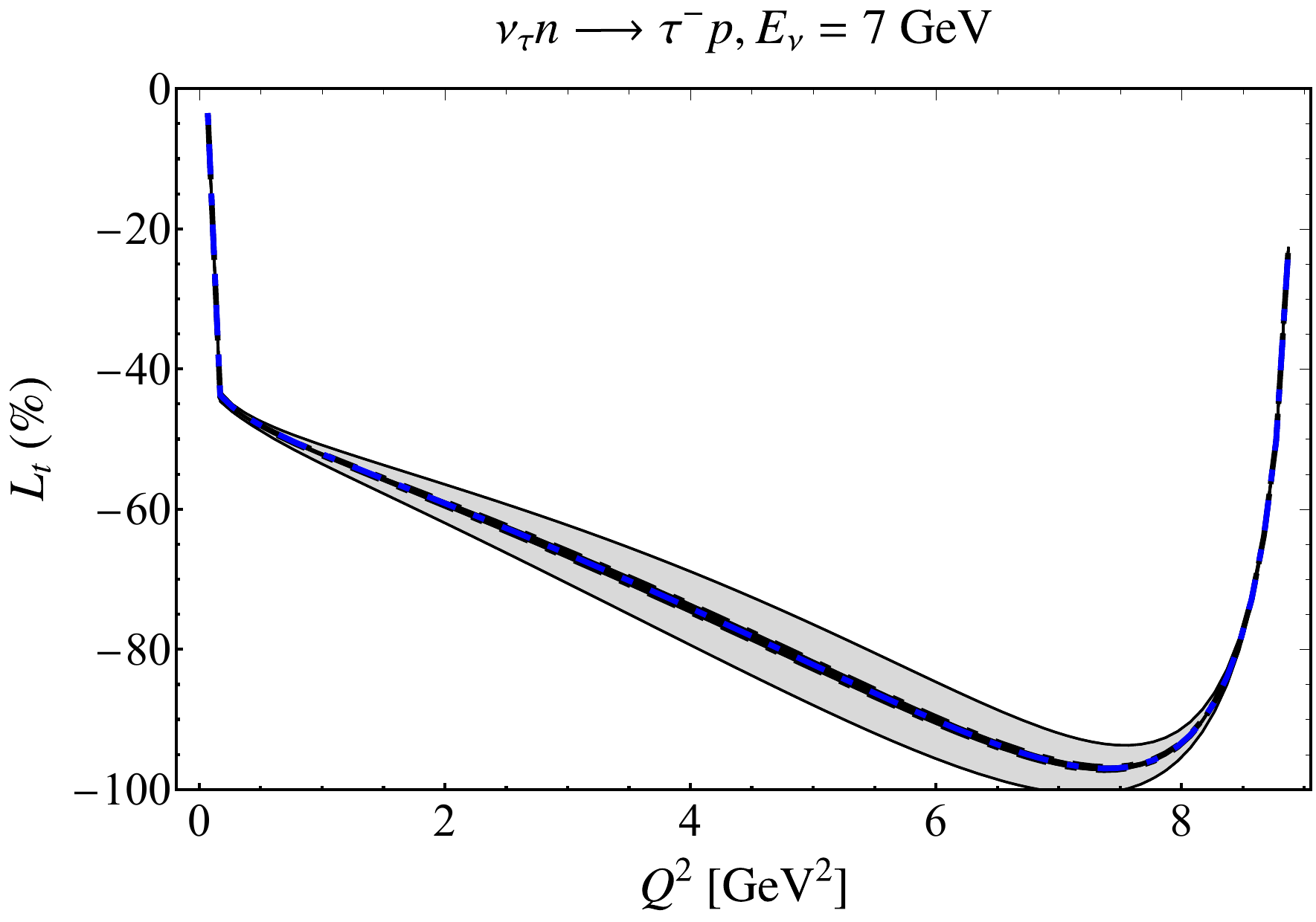}
\includegraphics[width=0.4\textwidth]{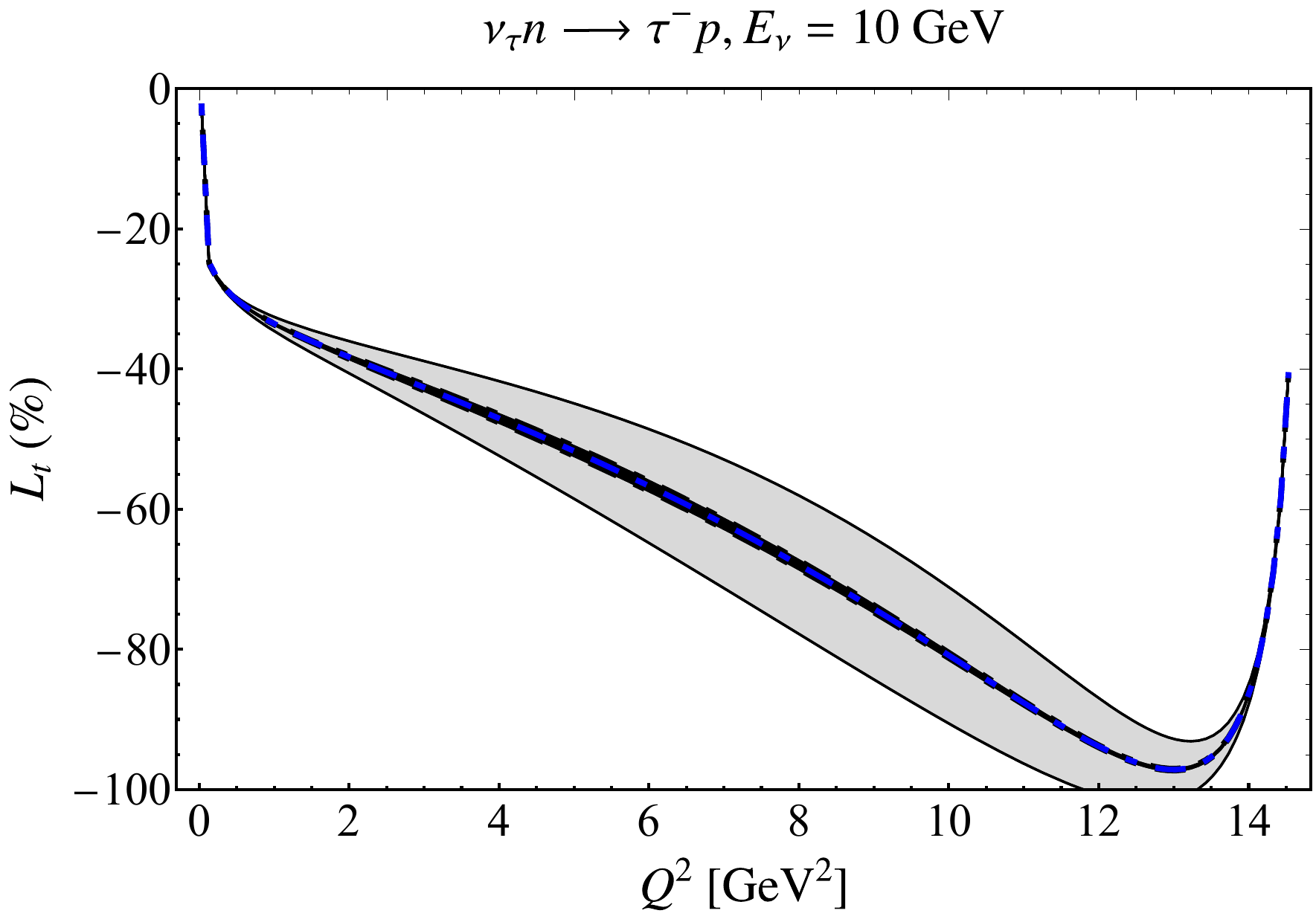}
\includegraphics[width=0.4\textwidth]{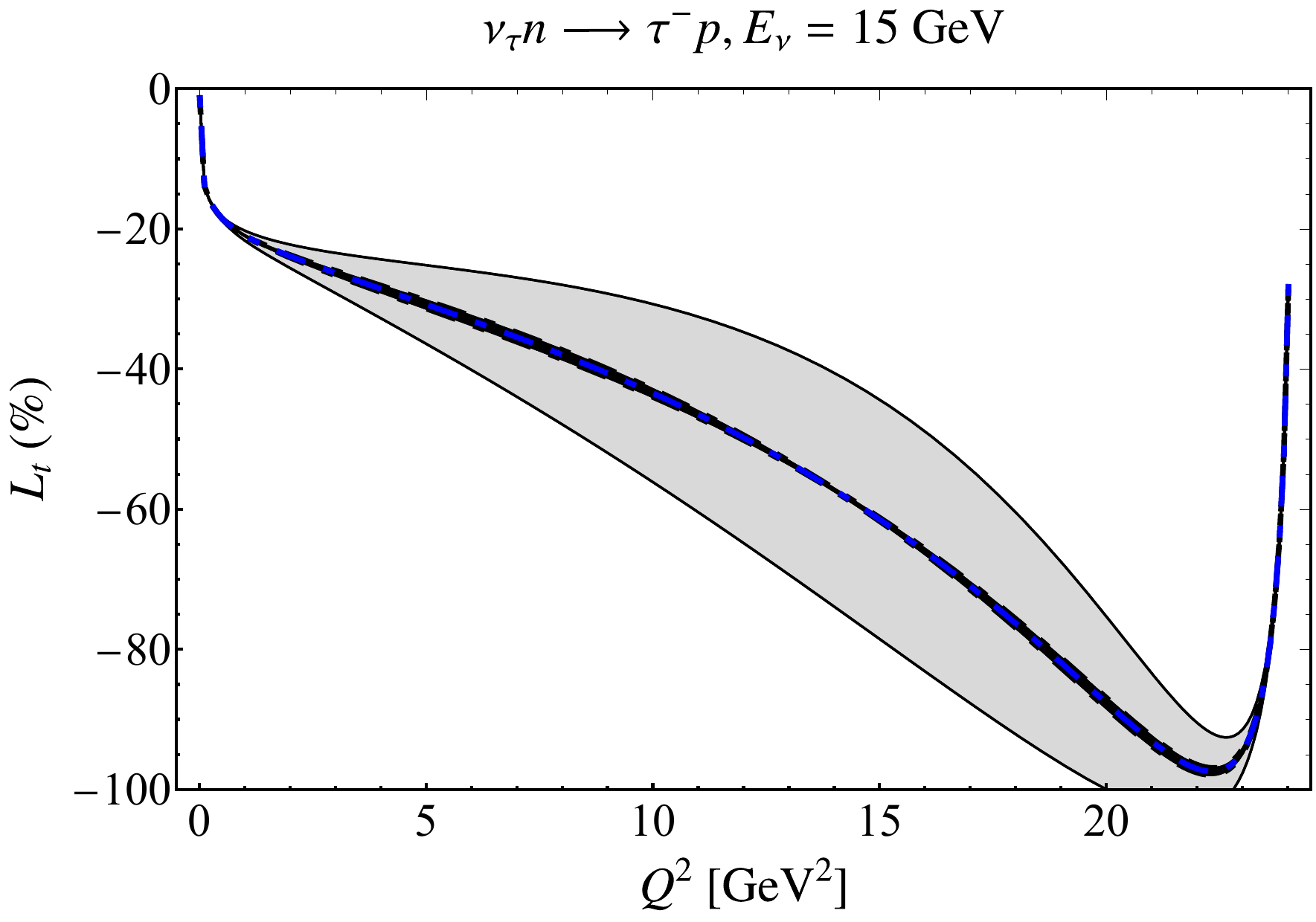}
\caption{Same as Fig.~\ref{fig:nu_Tt_radcorr_tau} but for the transverse polarization observable $L_t$. \label{fig:nu_Lt_radcorr_tau}}
\end{figure}

\begin{figure}[H]
\centering
\includegraphics[width=0.4\textwidth]{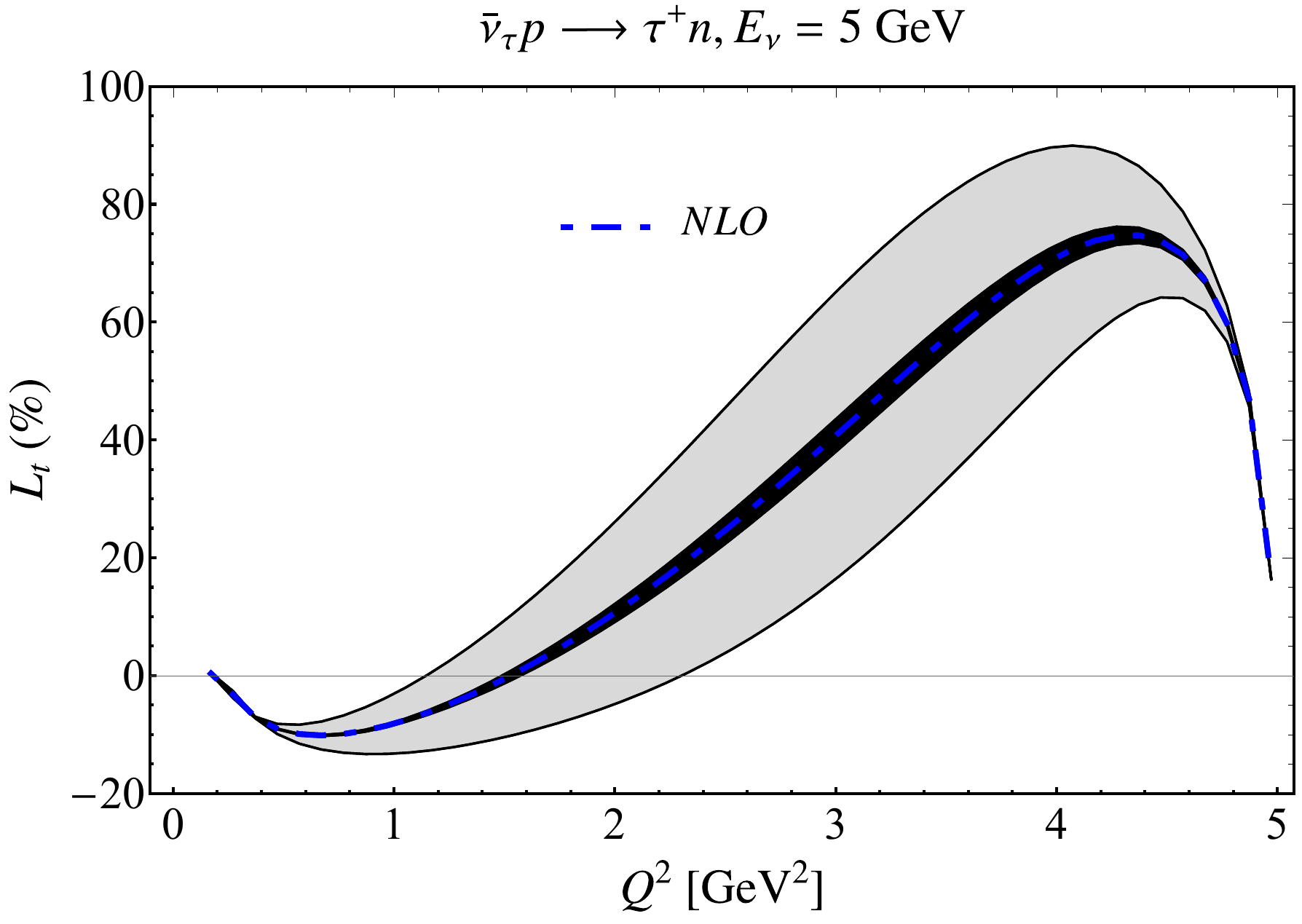}
\includegraphics[width=0.4\textwidth]{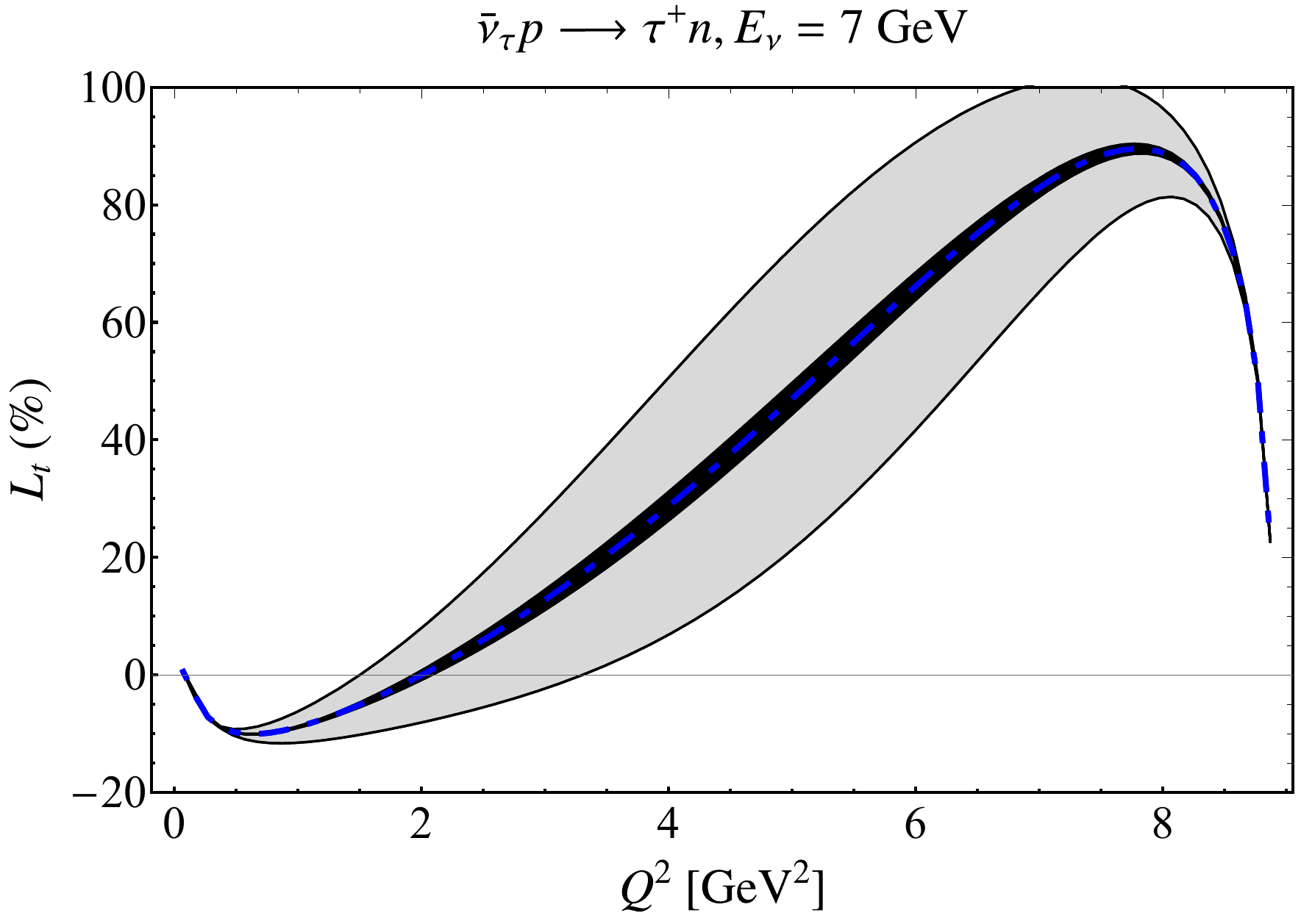}
\includegraphics[width=0.4\textwidth]{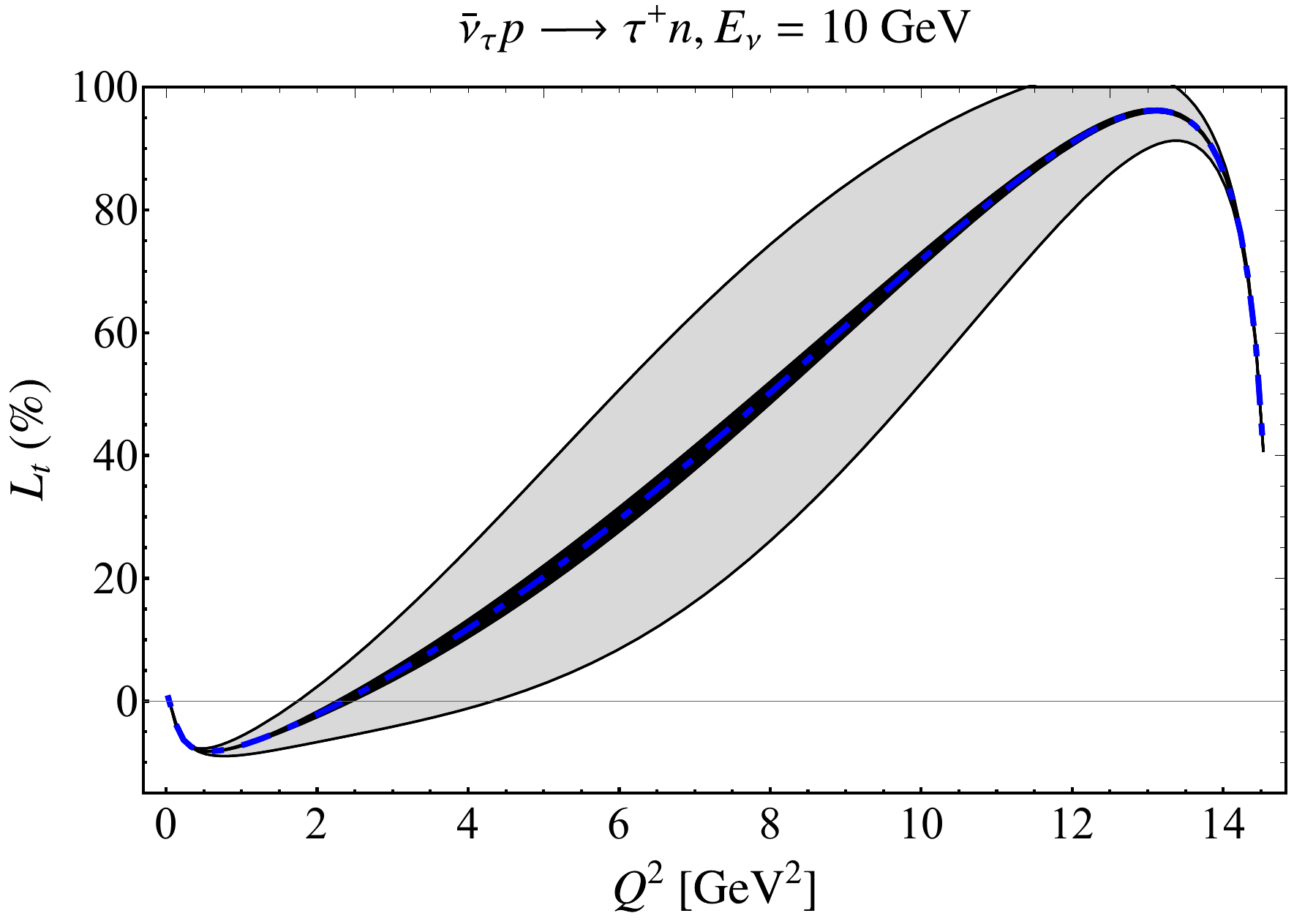}
\includegraphics[width=0.4\textwidth]{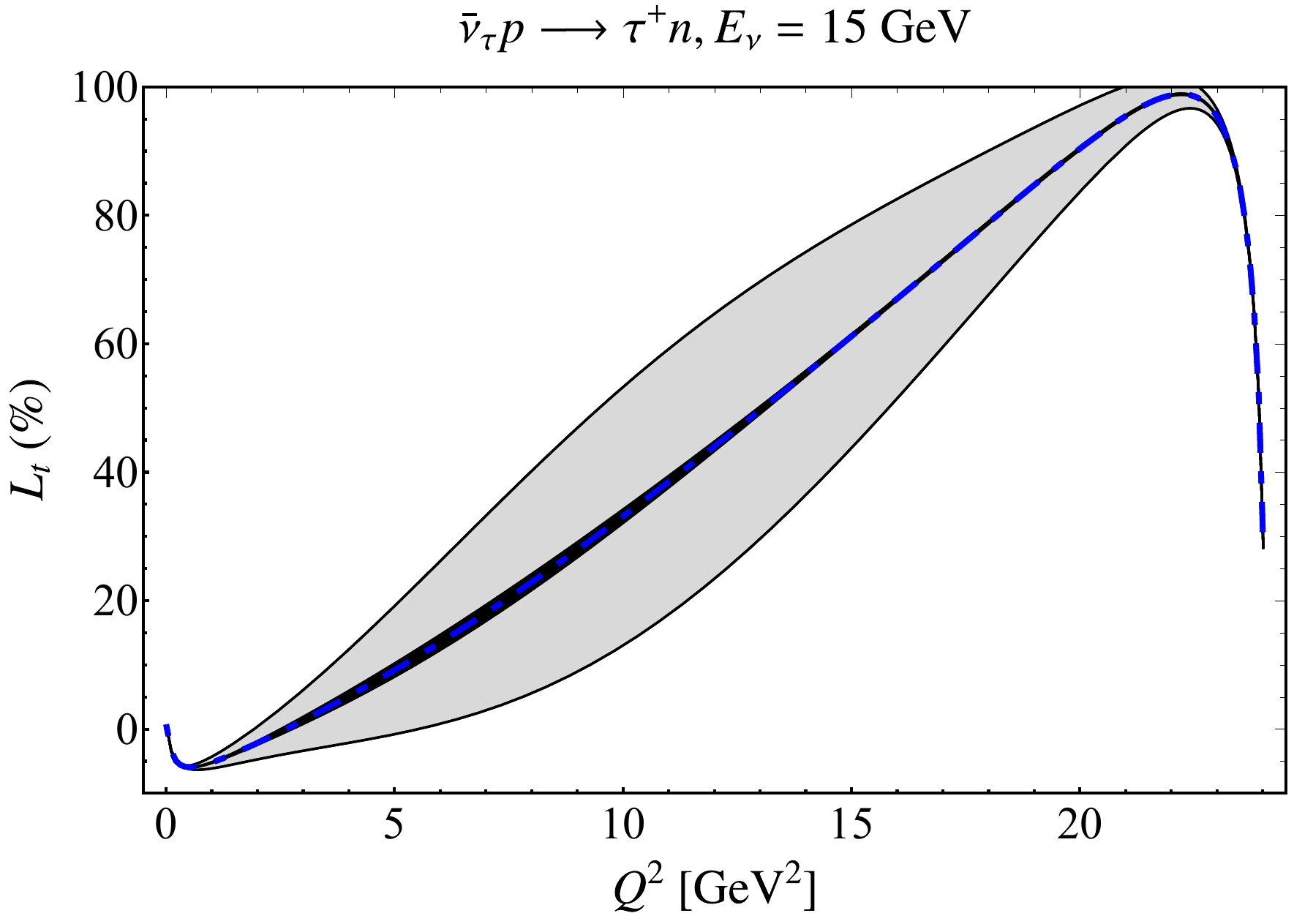}
\caption{Same as Fig.~\ref{fig:antinu_Tt_radcorr_tau} but for the transverse polarization observable $L_t$. \label{fig:antinu_Lt_radcorr_tau}}
\end{figure}

\begin{figure}[H]
\centering
\includegraphics[width=0.4\textwidth]{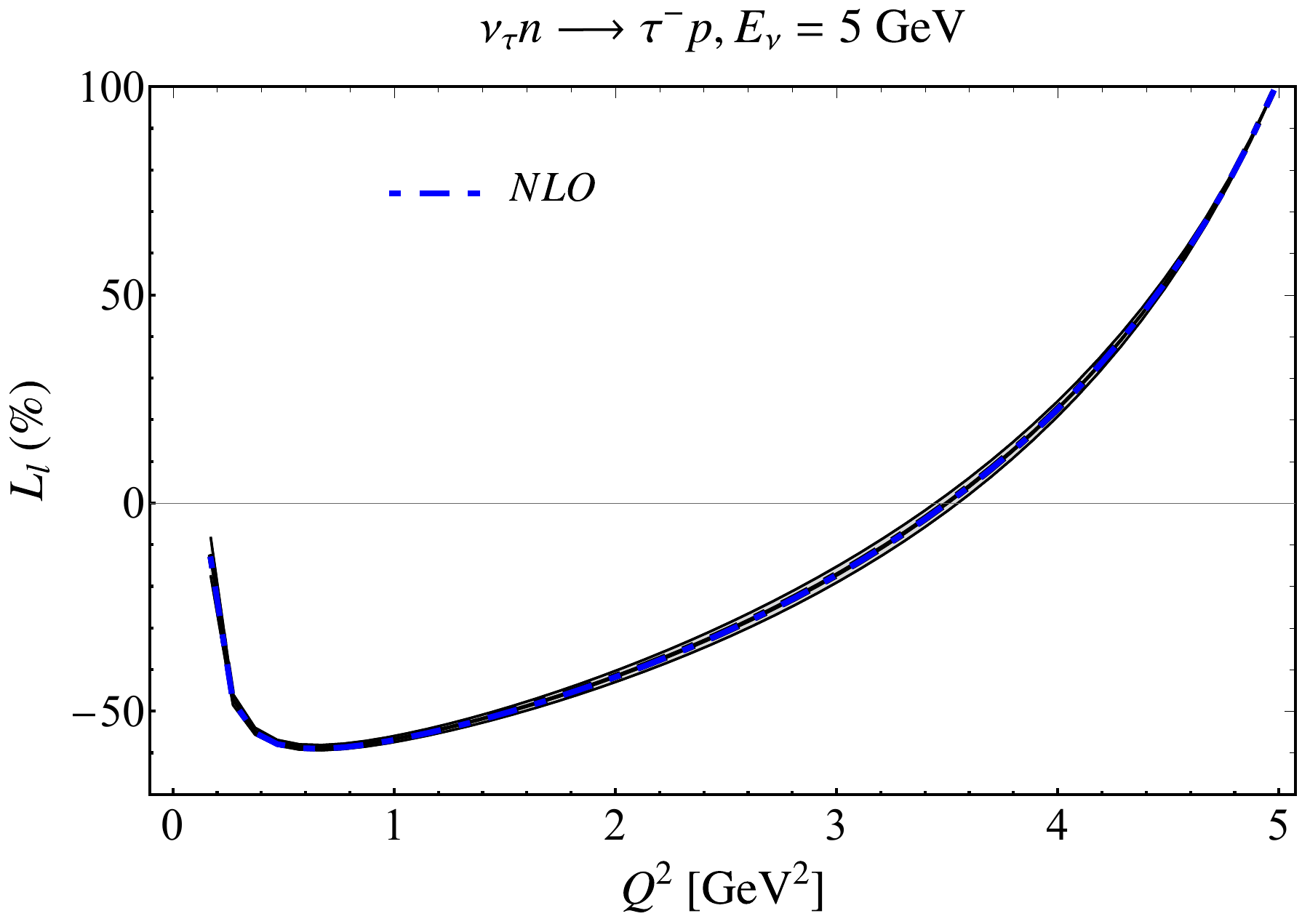}
\includegraphics[width=0.4\textwidth]{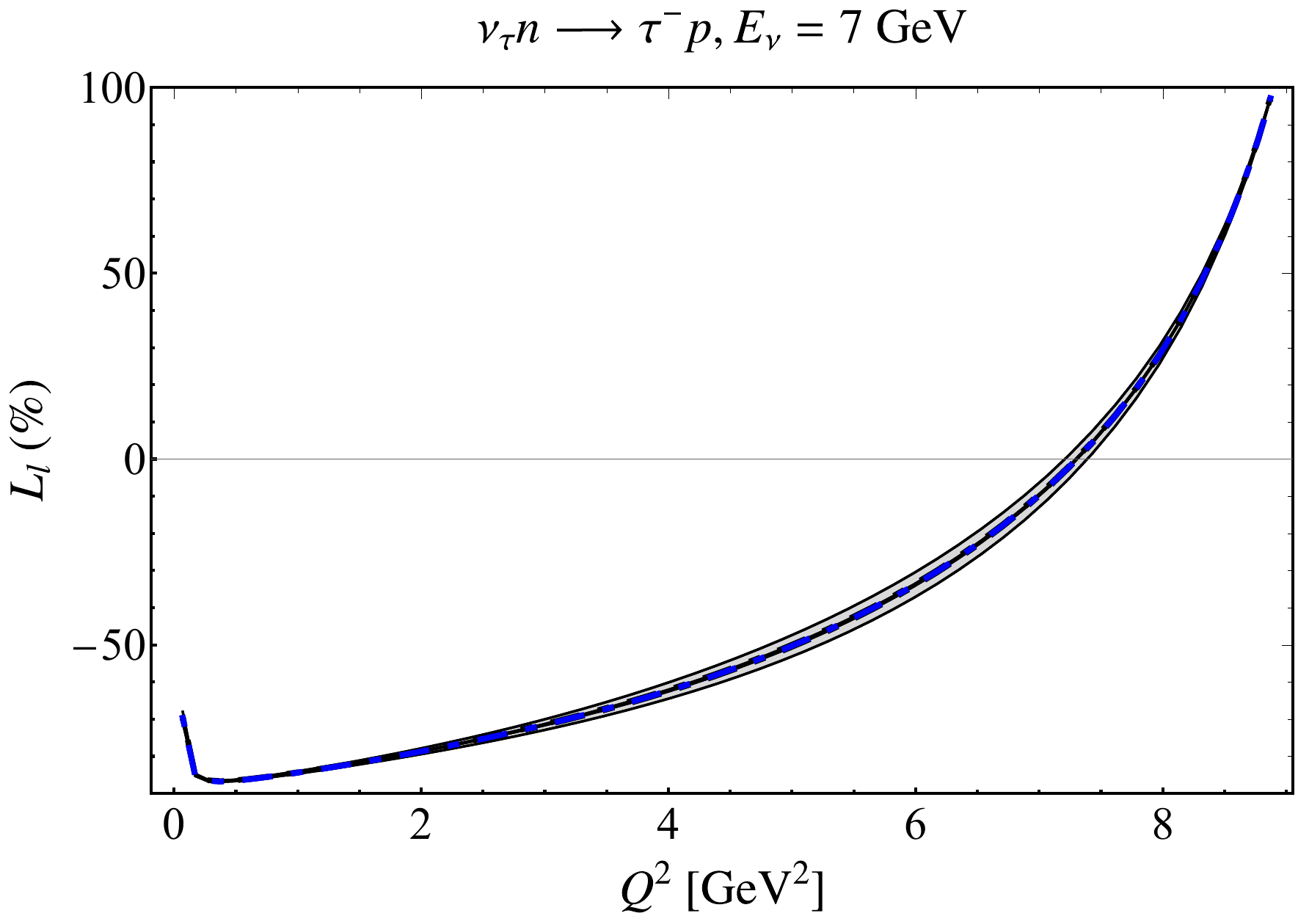}
\includegraphics[width=0.4\textwidth]{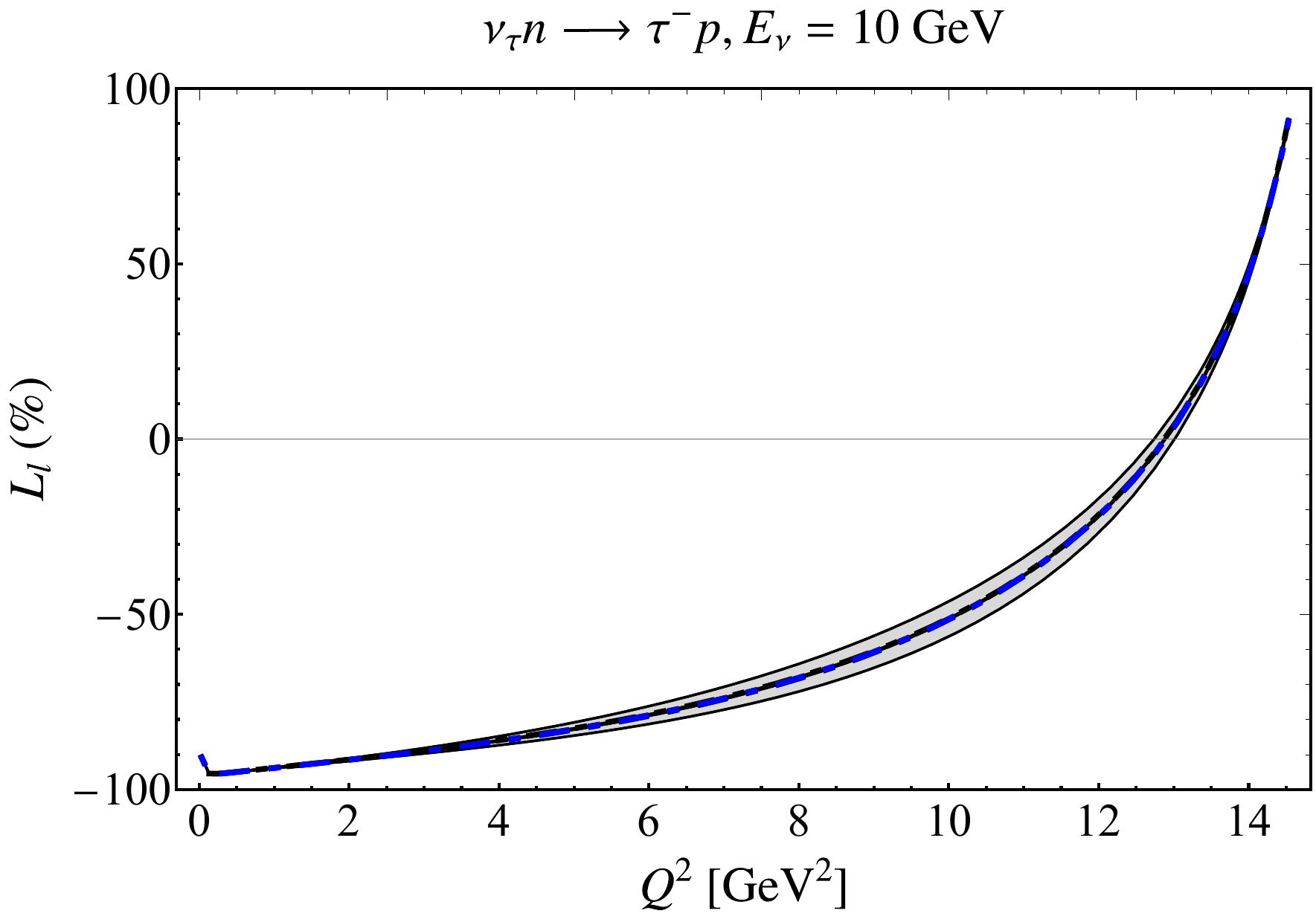}
\includegraphics[width=0.4\textwidth]{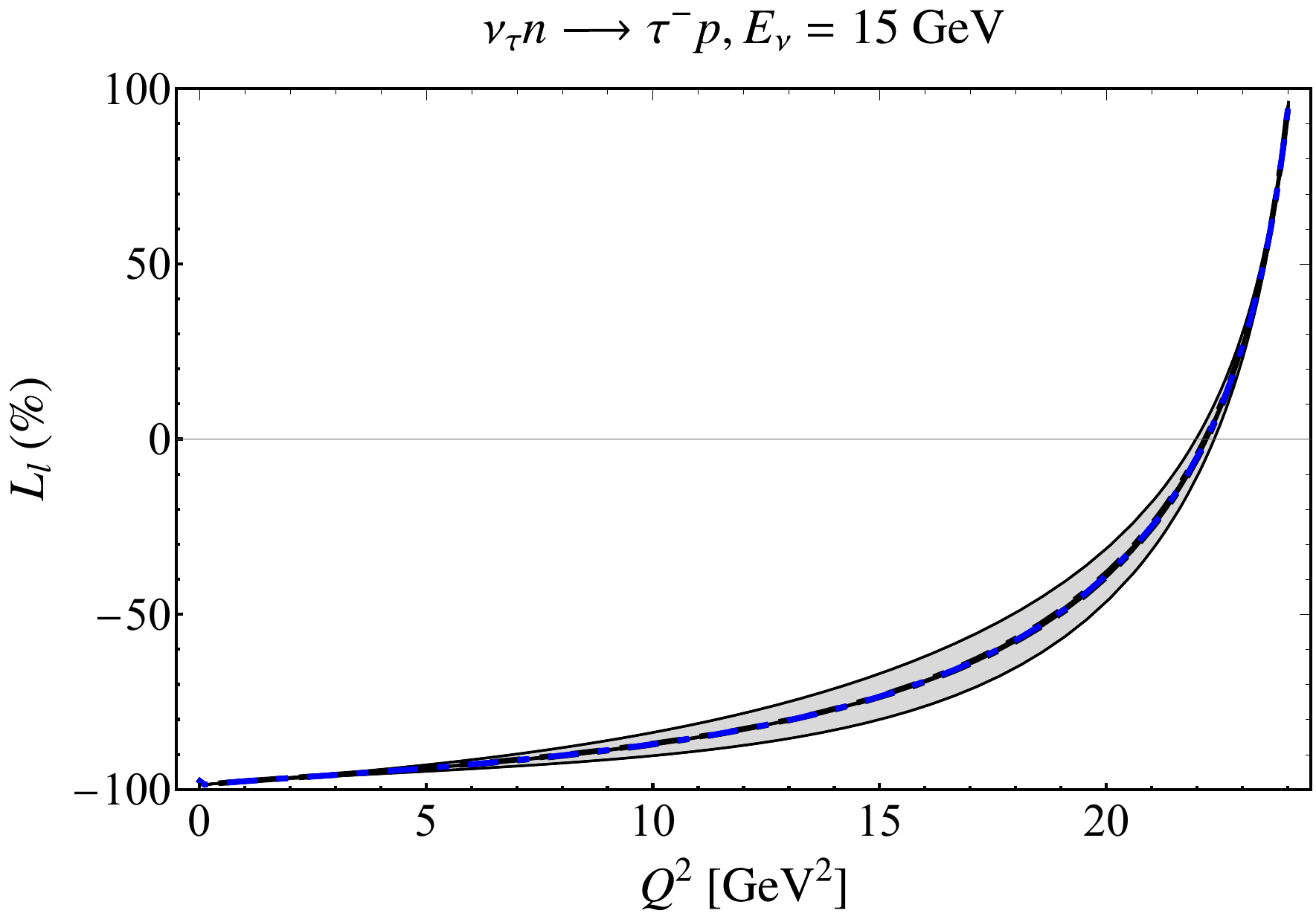}
\caption{Same as Fig.~\ref{fig:nu_Tt_radcorr_tau} but for the longitudinal polarization observable $L_l$. \label{fig:nu_Ll_radcorr_tau}}
\end{figure}

\begin{figure}[H] 
\centering
\includegraphics[width=0.4\textwidth]{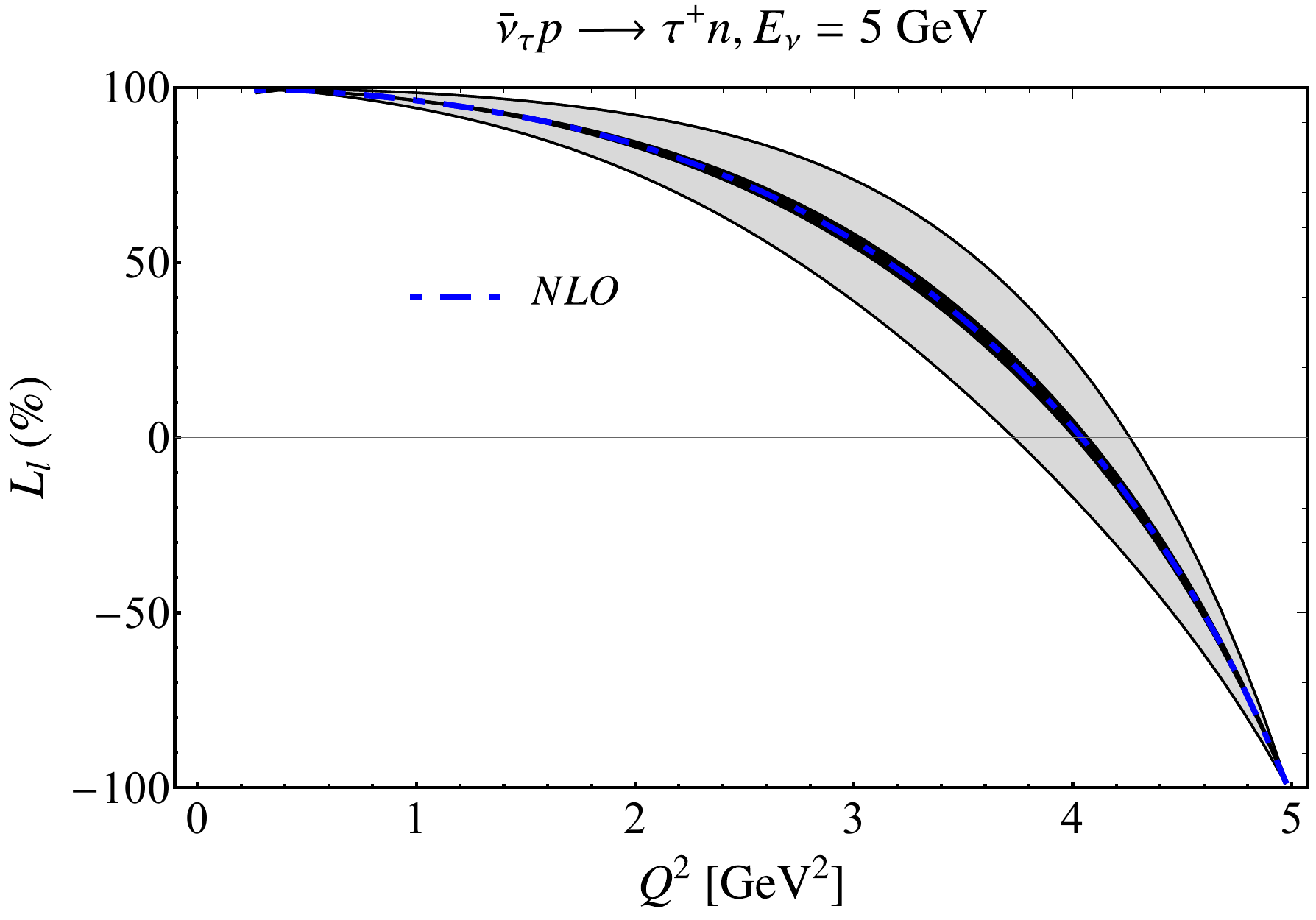}
\includegraphics[width=0.4\textwidth]{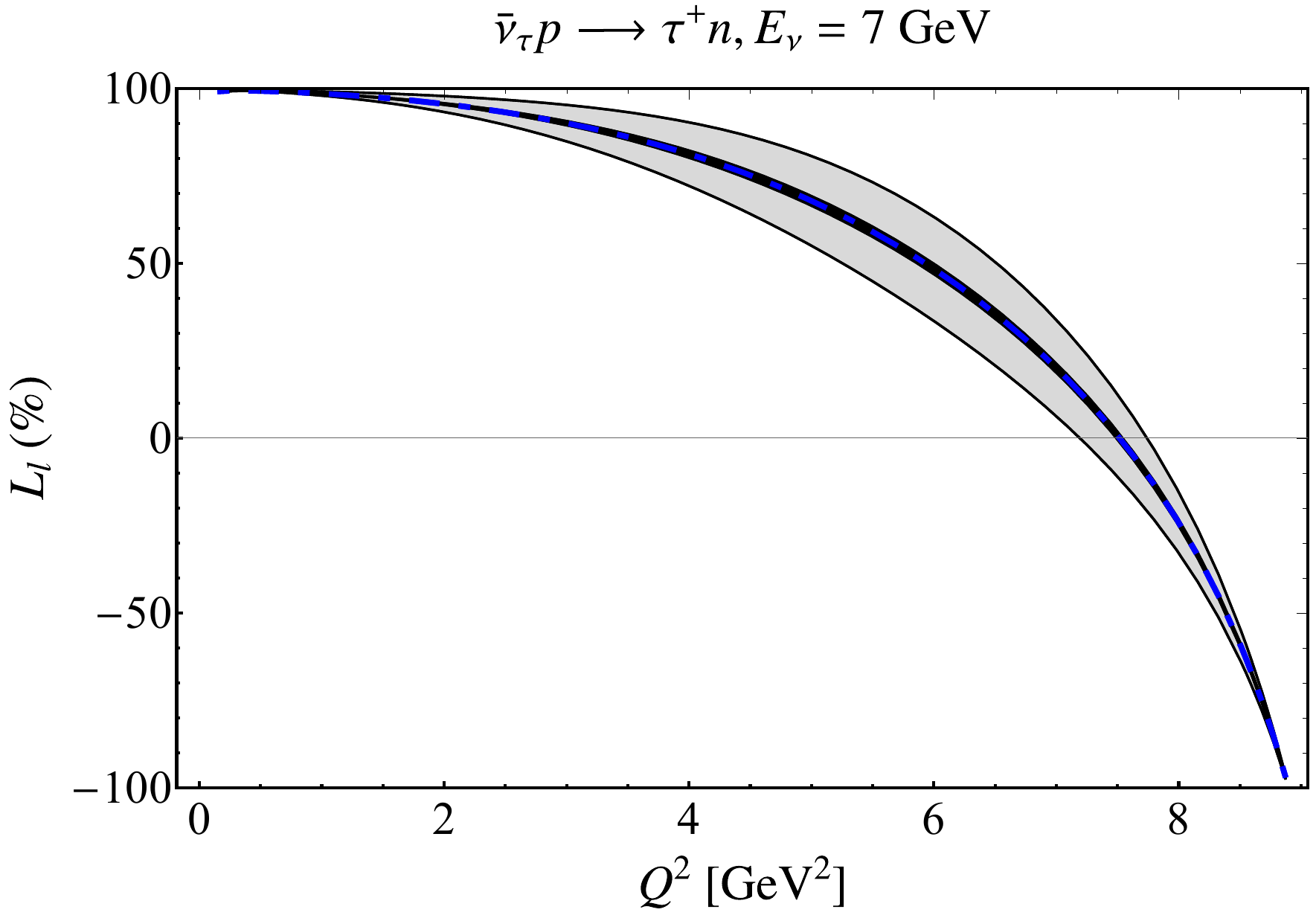}
\includegraphics[width=0.4\textwidth]{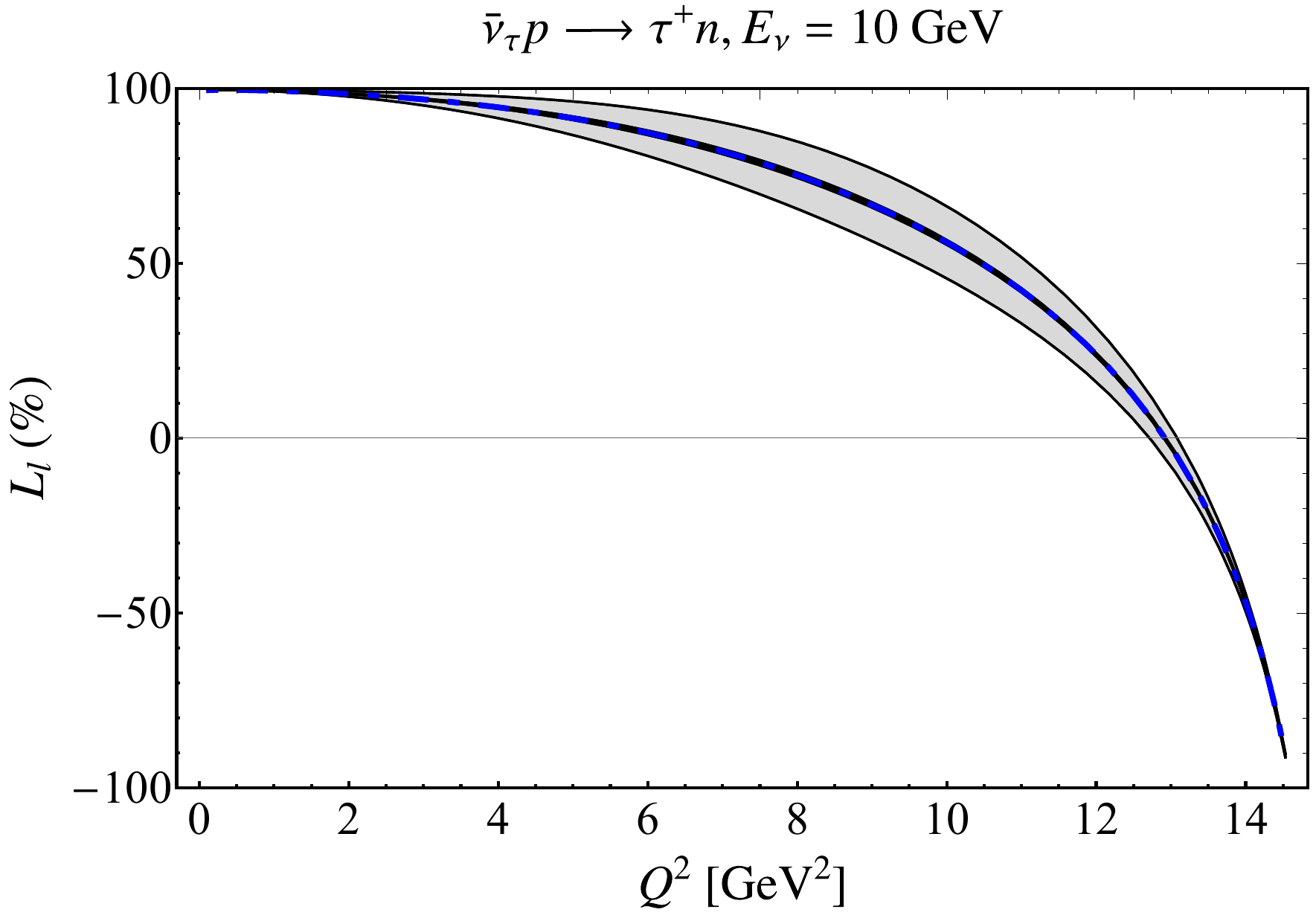}
\includegraphics[width=0.4\textwidth]{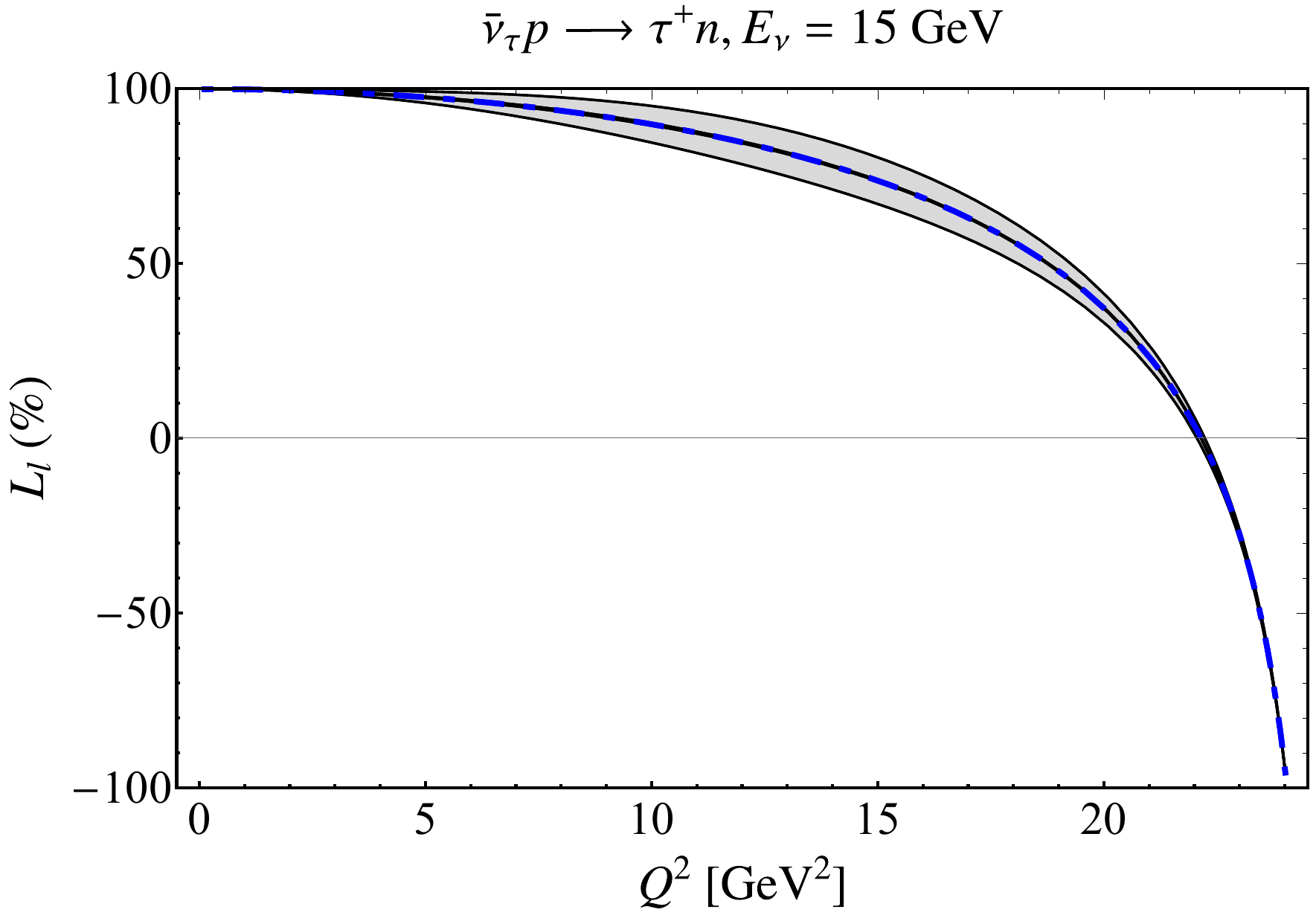}
\caption{Same as Fig.~\ref{fig:antinu_Tt_radcorr_tau} but for the longitudinal polarization observable $L_l$. \label{fig:antinu_Ll_radcorr_tau}}
\end{figure}

\begin{figure}[H]
\centering
\includegraphics[width=0.4\textwidth]{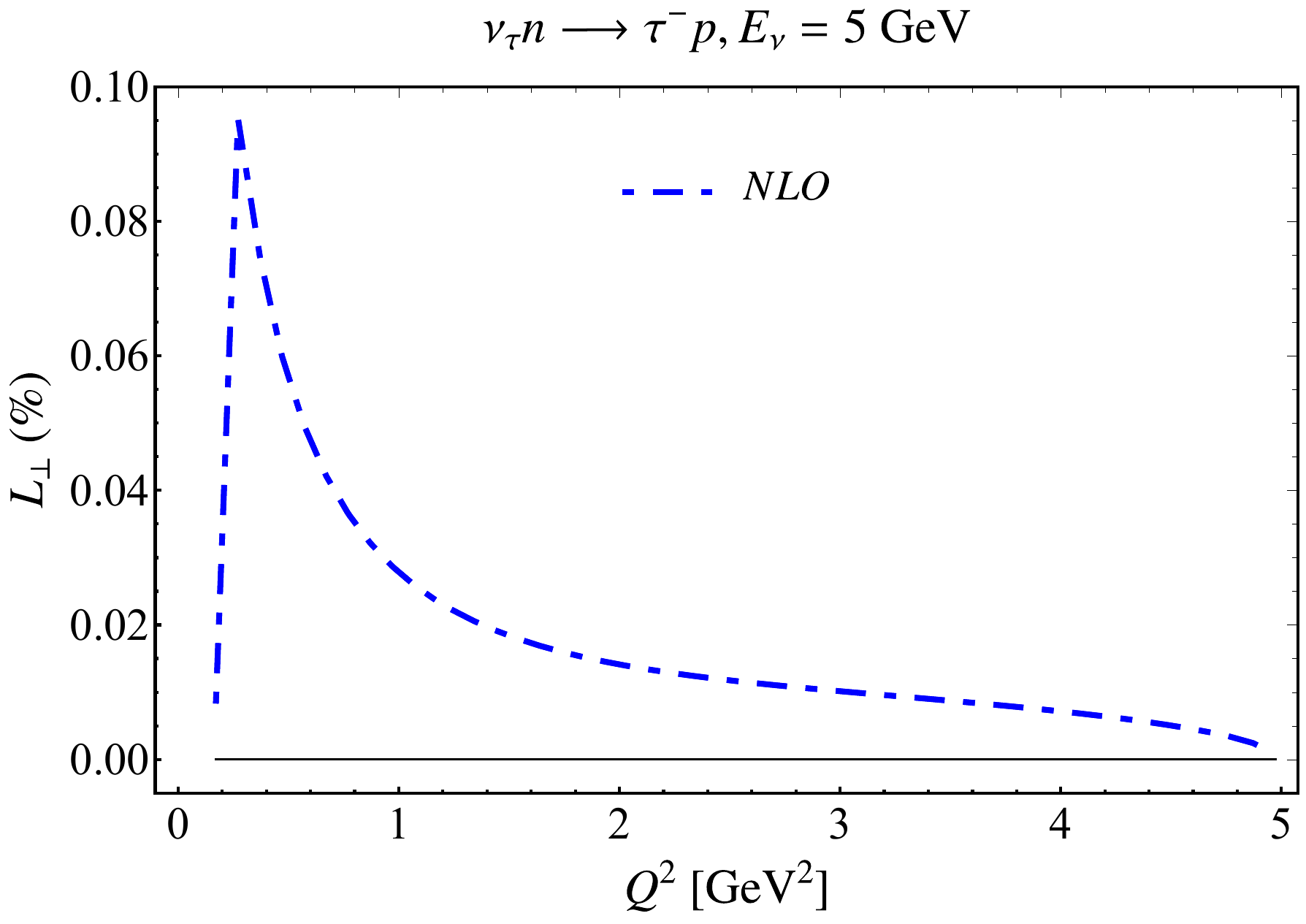}
\includegraphics[width=0.4\textwidth]{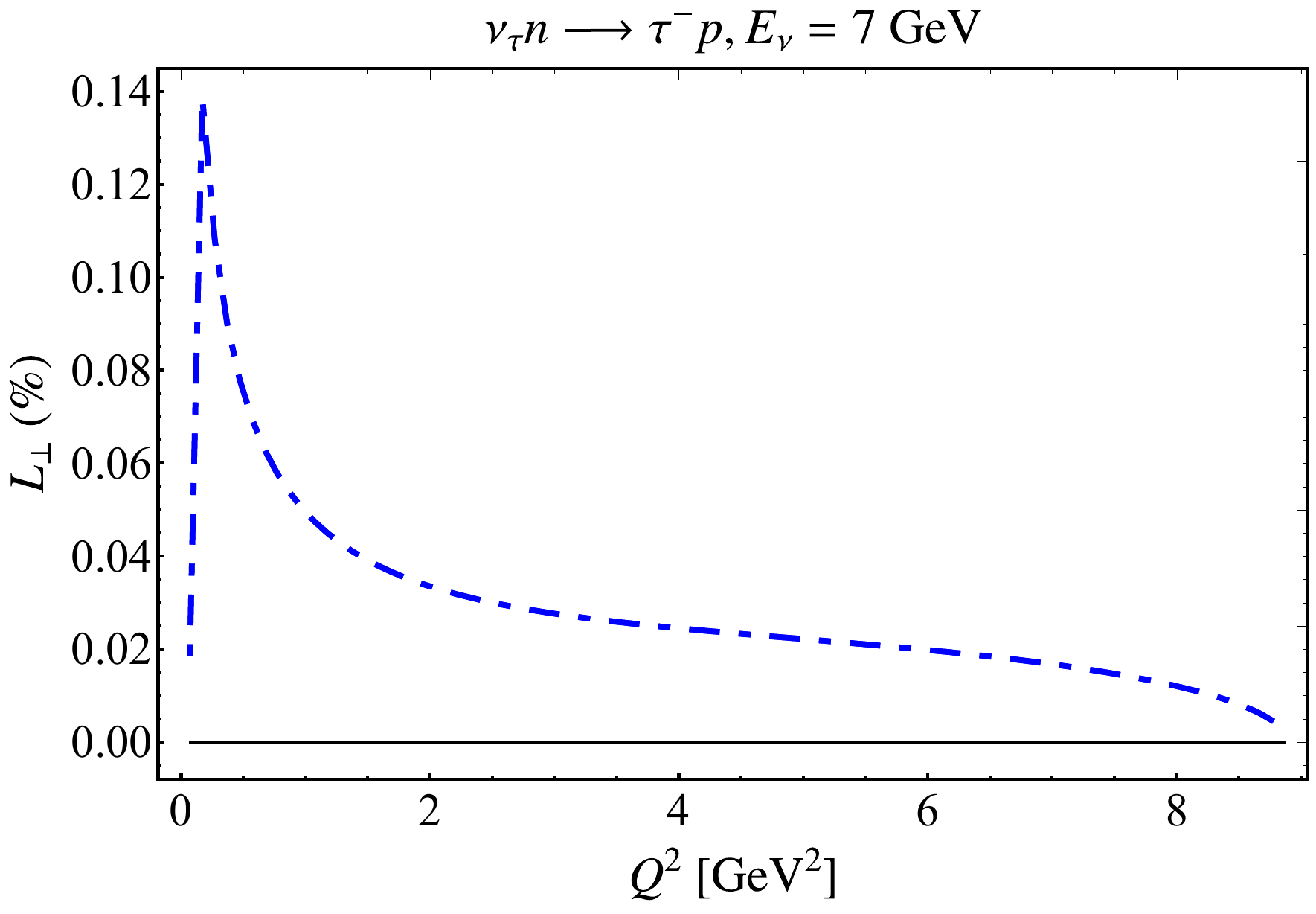}
\includegraphics[width=0.4\textwidth]{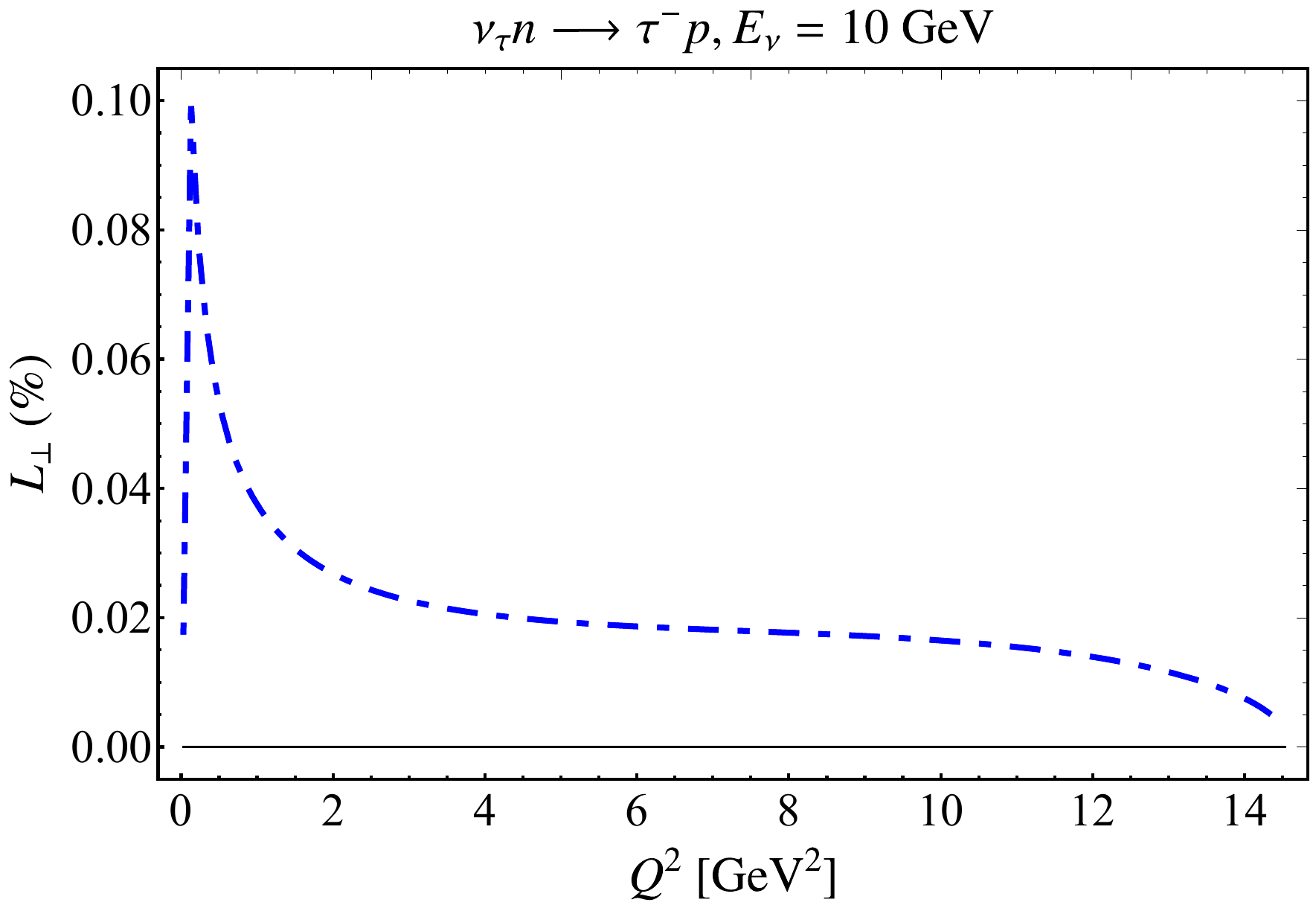}
\includegraphics[width=0.4\textwidth]{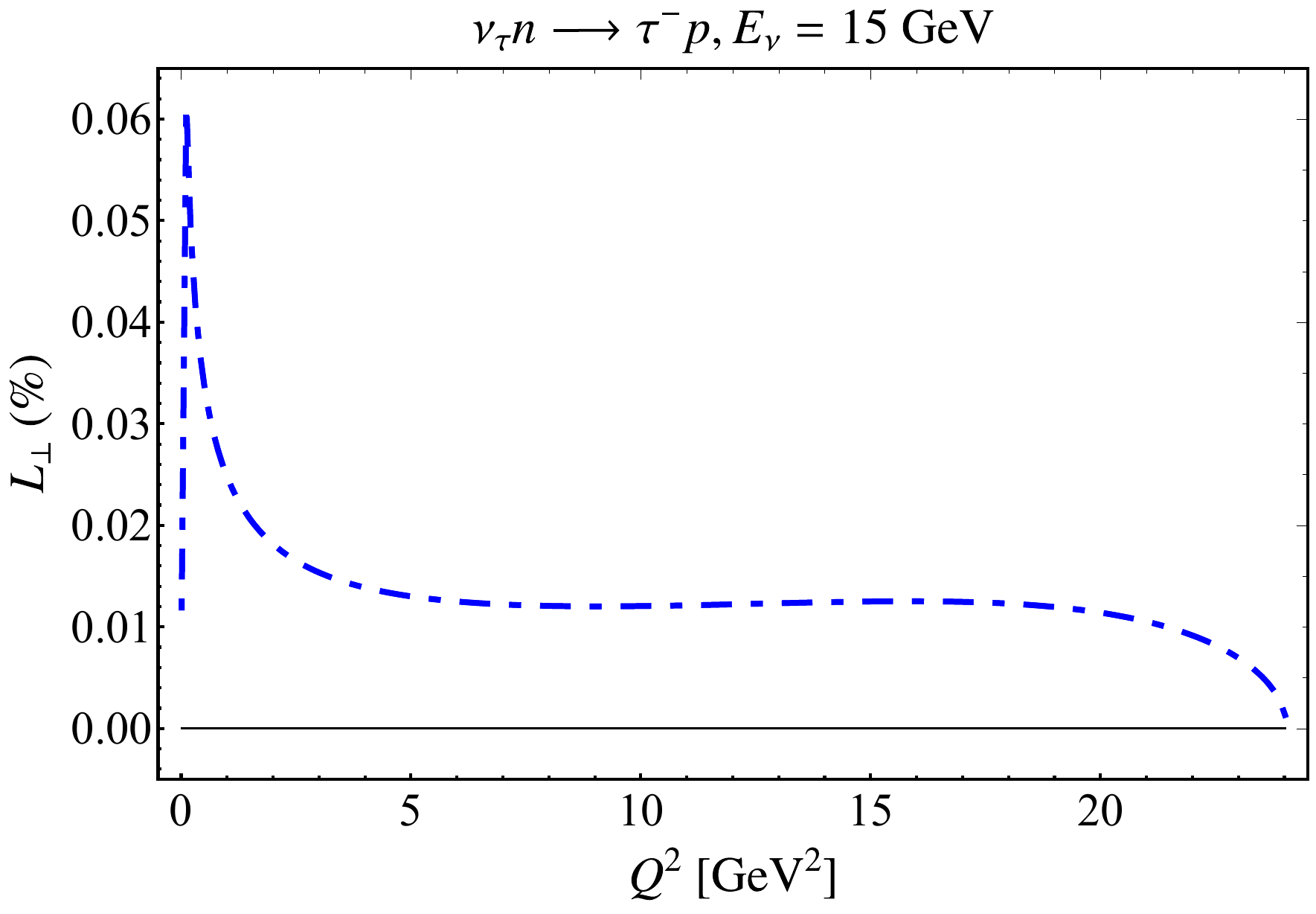}
\caption{Same as Fig.~\ref{fig:nu_Tt_radcorr_tau} but for the transverse polarization observable $L_\perp$. \label{fig:nu_LTT_radcorr_tau}}
\end{figure}

\begin{figure}[H]
\centering
\includegraphics[width=0.4\textwidth]{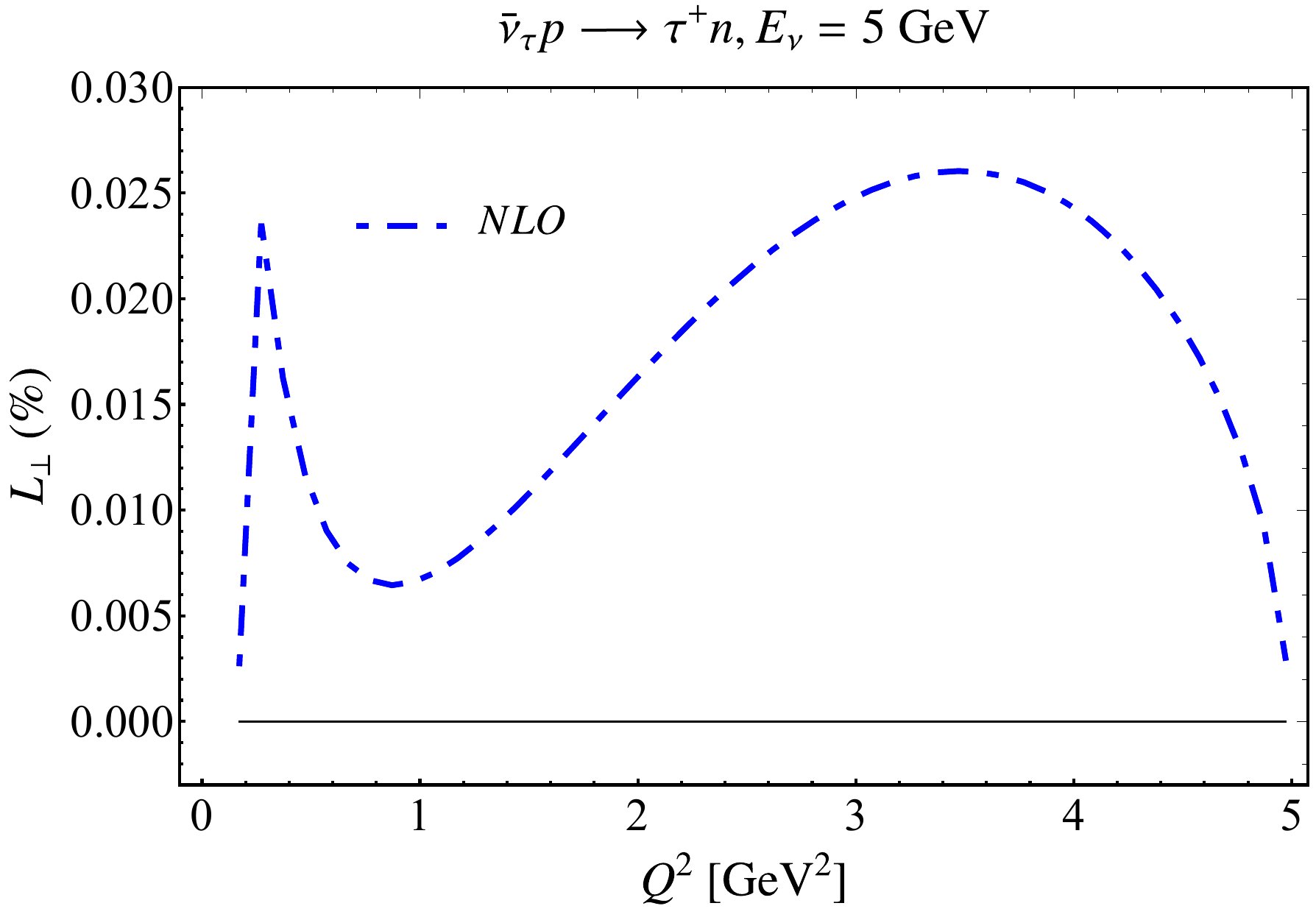}
\includegraphics[width=0.4\textwidth]{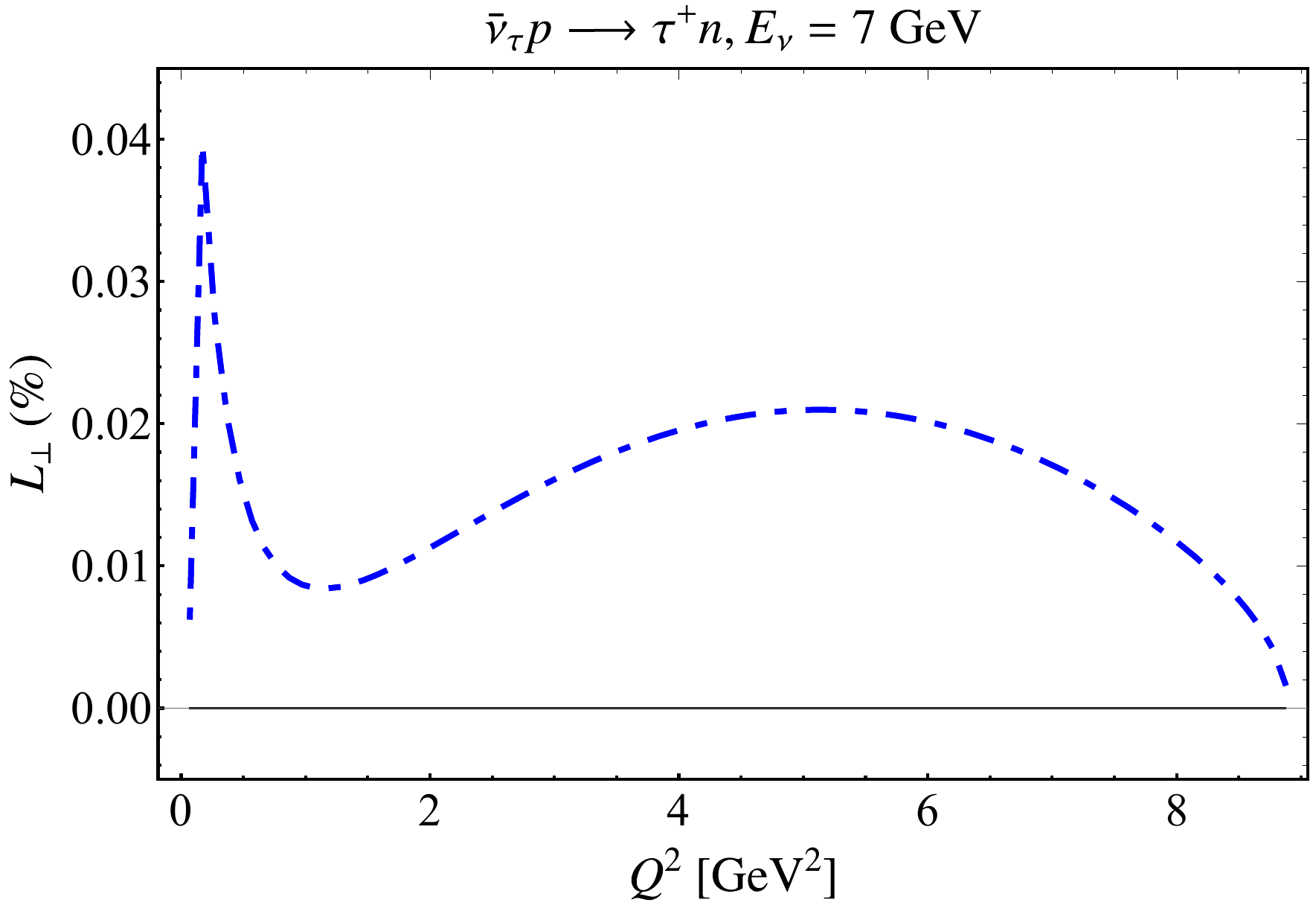}
\includegraphics[width=0.4\textwidth]{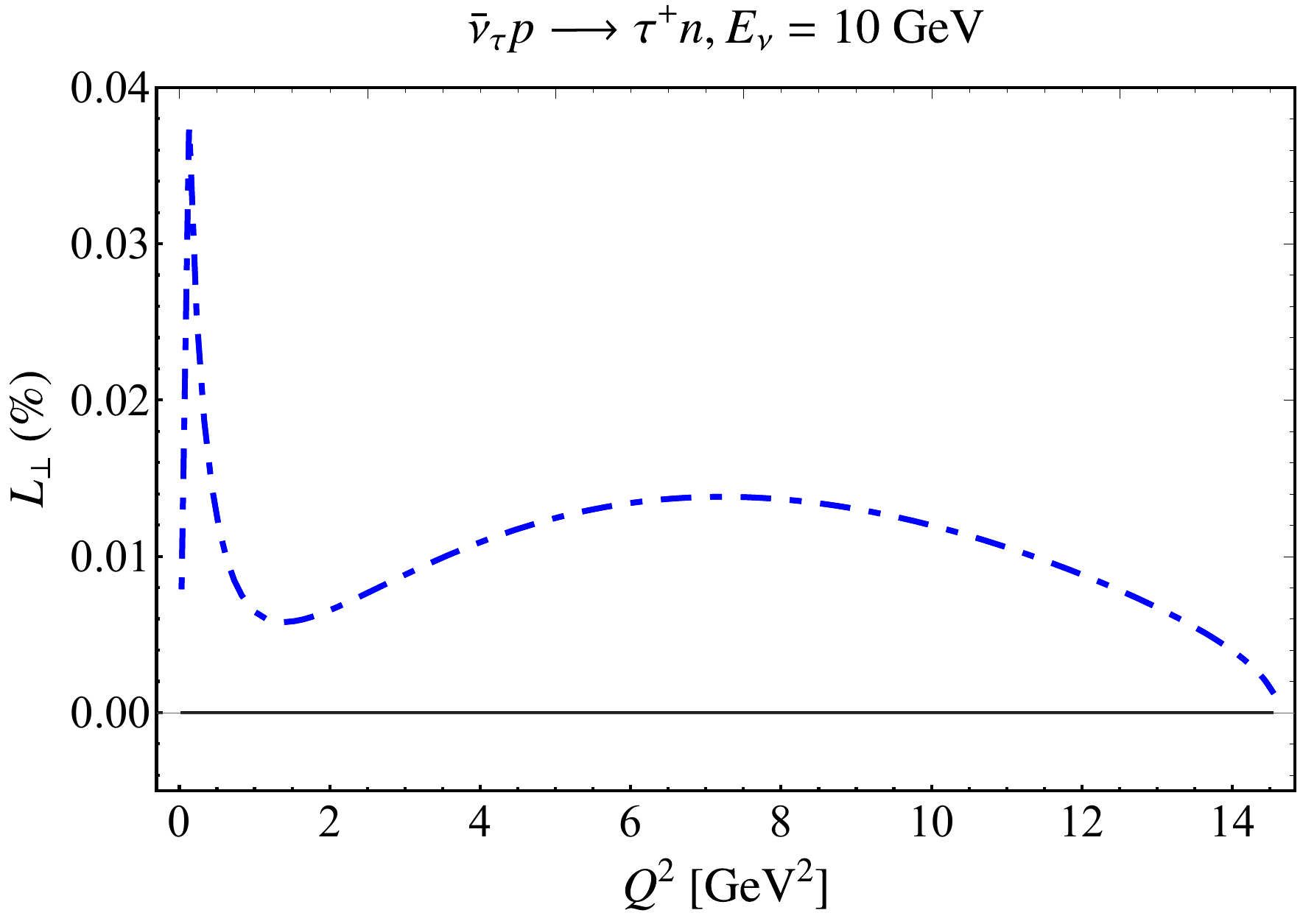}
\includegraphics[width=0.4\textwidth]{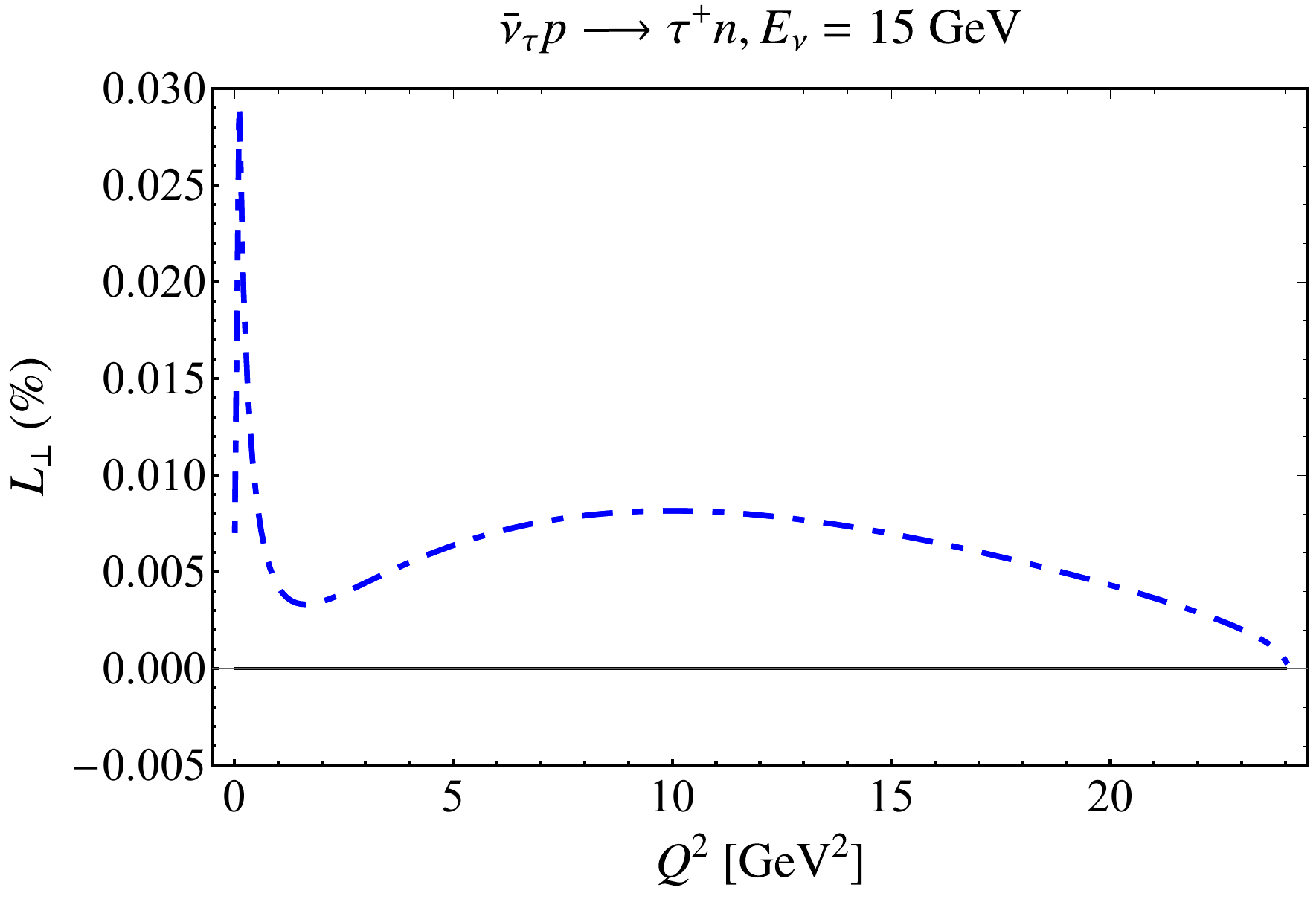}
\caption{Same as Fig.~\ref{fig:antinu_Tt_radcorr_tau} but for the transverse polarization observable $L_\perp$. \label{fig:antinu_LTT_radcorr_tau}}
\end{figure}
    
\end{appendix}

\newpage

\bibliography{secondclasscurrents}

\end{document}